\documentclass[11pt]{article}
\usepackage{amsmath,amssymb}
\usepackage{graphicx}
\pdfoutput=1
\font\ros=wncyr10
\font\rit=wncyi10
\makeatletter
\def\X#1#2{\mathopen{\hbox{$\left#2\vbox to\ifcase#1\or9.5\p@\or11.5\p@\or
14.5\p@\or17.5\p@\fi{}\right.\n@space$}}}
\def\Y#1#2{\mathclose{\hbox{$\left.\vbox to\ifcase#1\or9.5\p@\or11.5\p@\or
14.5\p@\or17.5\p@\fi{}\right#2\n@space$}}}
\def\Chr#1#2{\left\{{#1\@@atop#2}\right\}}
\def\pod#1#2{\mathrel{\mathop{\kern\z@#2}\limits_{#1}}}
\makeatother
\let\ti\textit
\def\atw#1 {\advance\textheight #1cm}
\let\ing\includegraphics

\numberwithin{equation}{section}
\let\t\times
\let\pa\partial
\def\pp#1#2{\frac{\pa #1}{\pa #2}}
\def\pz#1#2{\frac{d #1}{d #2}}
\def\ex#1{\exp(#1)}
\def\exd#1{\exp\X2(#1\Y2)}
\let\q\quad
\def\qh#1{\quad\hbox{#1}\quad}
\let\ov\overline
\let\ul\underline
\let\wt\widetilde
\let\wh\widehat
\def\hb#1{\wh{\ov{#1}}}
\let\iy\infty
\let\a\alpha
\let\b\beta
\let\d\delta
\let\D\Delta
\let\ve\varepsilon
\let\g\gamma
\let\G\varGamma
\let\k\kappa
\let\la\lambda
\let\La\Lambda
\let\o\omega
\let\O\varOmega
\let\vf\varphi
\let\vF\varPhi
\let\Ps\varPsi
\let\si\sigma
\let\th\theta
\let\vt\vartheta
\let\z\zeta
\def\m{{\mu\nu}}
\def\F{\mathcal F}
\def\cL{\mathcal L}
\def\cP{\mathcal P}
\def\R{\mathbb R}
\def\H{\mathbb H}
\def\E{\mathbb E}
\def\C{\mathbb C}
\def\dg #1,#2,#3,{{{#1}_{#2}}^{\!#3}}
\def\gd #1,#2,#3,{{{#1}^{#2}}_{\!#3}}
\let\sq\bigsqcap
\def\sq{\mathop{\hbox{\LARGE$\sqcap$}}}
\def\nad#1#2{\overset{\rm #1}{#2}}
\def\falg{{\mathrel{\lower5pt\hbox{${\scriptstyle\sim}$}\hskip-5pt g}}}
\def\Li{\operatorname{Li}\nolimits}
\def\Re{\operatorname{Re}\nolimits}
\def\Arg{\operatorname{Arg}\nolimits}
\def\cosech{\operatorname{cosech}\nolimits}
\def\arccot{\operatorname{arccot}\nolimits}
\def\Erf{\operatorname{erf}\nolimits}
\def\Int{\operatorname{int}\nolimits}
\def\Im{\operatorname{Im}\nolimits}
\def\crt{\operatorname{crt}\nolimits}
\def\eff{_{\rm eff}}
\def\sgn{\operatorname{sgn}\nolimits}
\def\res{\operatorname{res}\nolimits}
\def\rAB{\text{AB}}
\def\rE{\text{E}}
\def\U{\text{U}}
\def\SO{\text{SO}}
\def\SU{\text{SU}}
\def\pl{_{\rm pl}}
\let\er\eqref
\def\(#1){{(#1)}}
\def\[#1]{{[#1]}}
\def\ql#1{q_{\lower2pt\hbox{$\scriptstyle#1$}}}
\def\obraz#1{\refstepcounter{figure}\label{#1}}
\def\podpis#1{{\small \noindent Figure \thefigure #1}}

\def\beq{\begin{equation}\label}
\def\bea{\begin{eqnarray}\label}
\def\bml{\begin{multline}\label}
\def\bg{\begin{gather}\label}
\let\bal\aligned \let\eal\endaligned
\let\bga\gathered \let\ega\endgathered
\let\nn\nonumber
\def\mt#1{\left(\begin{matrix}#1\end{matrix}\right)}
\let\dsl\displaylines
\let\dss\displaystyle

\def\up#1{\uppercase{#1}}
\def\eu{\expandafter\up}
\def\ac{acceleration}
\def\an{anomalous}
\def\ap{approximat}
\def\cf{coefficient}
\def\cfn{confinement}
\def\cn{connection}
\def\ct{constant}
\def\cd{coordinate}
\def\co{cosmolog}
\def\dv{derivative}
\def\di{dimension}
\def\ef{effective}
\def\elm{electromagneti}
\def\e{equation}
\def\f{function}
\def\gr{gravitation}
\def\hy{hyperbolic}
\def\iv{invariant}
\def\JT{(Jordan--Thiry)}
\def\lg{lagrangian}
\def\nos{nonsymmetric}
\def\NK#1{Nonsymmetric Kaluza--Klein #1Theory}
\def\ph{perihelion}
\def\P{Pioneer 1}
\def\pt{potential}
\def\qu{quaternion}
\def\rl{relativistic}
\def\SS{Solar System}
\def\so{solution}
\def\sp{spacecraft}
\def\spt{space-time}
\def\st{such that }
\def\s{symmetric}
\let\TM\texttrademark
\def\Un{Universe}
\def\wrt{with respect to }
\def\YM{Yang--Mills}

\author{M. W. Kalinowski\\The Center of Excellence
BioExploratorium\\e-mail: markwkal@bioexploratorium.pl\\
Pracownia Bioinformatyki, Instytut Medycyny Do\'swiadczalnej i Klinicznej PAN\\
ul. Pawi\'nskiego 5, 02-106 Warszawa, Poland\\
e-mail: mkalinowski@imdik.pan.pl}
\title{Pioneer 10 and 11 Spacecraft Anomalous Acceleration\\
in the light of the Nonsymmetric Kaluza--Klein\\
(Jordan--Thiry) Theory}
\date{}
\begin{document}
\maketitle

\vskip-10pt
\setbox0=\hbox{\sl an indefatigable researcher in gravitation physics}
\rightline{\vbox{\hsize\wd0 \small
\sl \hbox{To John William Moffat,}
\copy0
\vskip5pt
\leftline{\rm Motto:}
\hbox{There is not an ultimate}
\hbox{theory of the Universe.}
\rightline{John William Moffat\indent}}\indent}

\begin{abstract}
The \NK{\JT\ } leads to a model of a modified \ac\ that can fit an
\an\ \ac\ experienced by the \P0 and 11 \sp. The future positions of those
\sp s are predicted using distorted hyperbolic orbit. A \cn\
between an \an\ \ac\ and a Hubble \ct\ is solved in the theory together
with a relation to a \co ical \ct\ in CDM$\La$ model.

In the paper we consider an exact \so\ of a point mass motion in the \SS\ under an
influence of an \an\ \ac. We find two types of orbits: periodic and chaotic.
Both orbits are bounded. This means there is no possibility to escape from
the \SS. Some possibilities to avoid this conclusion are considered. We
resolve also a coincidence between an \an\ \ac\ and the \co
ical \ct\ using a paradigm of modern \co y. Relativistic effects and a \co
ical drifting of a \gr al \ct\ are considered. The model of an \an\ \ac\ does
not cause any contradiction with \SS\ observations. We give a full
statistical analysis of the model. We consider also a full formalism of the
\eu\nos\ Jordan--Thiry Theory for the problem and present a relativistic
model of an \an\ \ac. We consider the model for General Relativity \ap ion,
i.e.\ $g_\m\ne \eta_\m$ ($g_\m=g_{\nu\mu}$).

In this model there are no contradictions with General Relativity tests
in the  \SS. \P0/11 \sp s will come back in $10^6$ years (a~time scale of
our periodic \so s is $10^6$ years). Moreover, almost
\rl\ or \rl\ \sp s can escape from the \SS. We consider also a model of a
\rl\ \ac\ which is more complicated, with $g_\m\ne g_{\nu\mu}$ taken into
account.
\end{abstract}

\newpage
\section{Introduction}
The paper deals with the issue of the Pioneer Anomaly: it proposes a \gr al
\so\ to it in terms of a modified model of gravity. In the centennial
anniversary of GR the Pioneer Anomaly must be put into an appropriate context
of \pt\ astrometric anomalies in the \SS\ (see Ref.~\cite{B1}). About possible
anomalies see Ref.~\cite{B2}, also Ref.~\cite{B3}. There are some anomalies reported in
several papers (see Refs \cite{B4,B5}). There is an interesting problem of
planetary flybys and the energy transfer (see Refs \cite{B6}--\cite{B11}).
Some problems with a Newton \gr al \ct\ are reported in Refs \cite{B12,B13}.
Constraints on non-standard gravitomagnetism by the \an\ \ph\ precession of
the planets are considered in Ref.~\cite{B14}.

Let us quote a list of anomalies in the \SS. They are: possible \an\ advances
of planetary perihelia, unexplained orbital residuals of recently discovered
moon of Uranus--Mab (from Queen Mab), the lingering unexplained secular
increase of the eccentricity of the orbit of Moon, the so-called Faint Young
Sun Paradox (FYSP), the secular decrease of the mass parameter of the Sun,
the Flyby Anomaly, the Pioneer Anomaly (PA), the \an\ secular increase of the
astronomical unit (see Ref.~\cite{B2}). The General Theory of Relativity
(GTR or~GR) has been created 100~years ago as not only a theory of \gr al
field but also as a theory of \spt. In the theory \gr al field is strictly
connected to a geometry of \spt. In this way according to Albert Einstein we
can summarize the theory of relativity in one sentence: \ti{time and space and
\gr\ have no separate existence from matter}.
The theory has been confirmed by many
observations and experiments in the \SS\ on the Earth and even beyond the \SS.
The theory is so established that it is a base for a \cd\ system in the \SS\
(BCRS---Barycentric Celestial Reference System) and on the Earth
(GCRS---Geocentric Celestial Reference System) (see Section~\ref{disturb}).
Moreover, there is a place for any alternative theory of \gr, which is
geometric, generally covariant valuable---passing all the tests in the \SS\
and beyond. This theory must be also a theory of gravity and a theory of a
\spt.

Thus any anomalies in the \SS, especially a PA-anomaly, can be a source of
such investigations in the field of alternative theories of \gr. Let us
reconsider a FYSP. According to the modern ideas a model of the Sun is well
established and an evolution of the luminosity (an energy output) of the Sun
is very well known. However, it happens that at a beginning of Archean era
(3.8 to 2.5~Gyr ago) a Solar \ct\ was insufficient to maintain a liquid water
on a surface of the Earth (see Ref.~\cite{B2} and references cited therein).
Moreover, we have strong independent evidence that at that time our planet
was covered by liquid water. Due to this fact a life flourished on the Earth.
This evident paradox can be easily resolved if we agree that our Earth was
closer to the Sun (the~radius of an orbit of our planet was smaller than now).
This fact can be explained if we suppose that \gr al attraction of the Sun
was stronger. The stronger attraction of the Sun can be obtained from an
alternative theory of \gr, where the \gr al \ct~$G$ is connected to a scalar
field changing in a space and in a time. In this way a \gr al ``\ct'' was
larger in Archean era and the Earth was closer to the Sun ($G=G(t)$ is
changing in a \co ical (maybe only in a geological) time).

In this paper we try to explain an \an\ \ac\ of \P0/11 (see Refs \cite1,
\cite2, \cite3, \cite{TurT}) using \NK{\JT\ }. It has been reported that \P0/11 is under an
influence of an \an\ \ac\ directed to the Sun of order $a_P=8\times10^{-10}\frac{\rm
m}{{\rm s}^2}$ on distances from 20\,{\rm AU} to 70\,AU.
Anderson et~al.\ reported an \an\ frequency shift derived from about ten years
of radio-metric data from Pioneer 10/11. The principle of observation is a
two way Doppler tracking. An emitter on the Earth sends a signal of a
frequency~$f_0$ which is seen by a spacecraft as a frequency
\beq{1.1}
f_1=\frac1{\sqrt{1-v^2_P/c^2}}\,\X2(1-\frac{v_P}c\Y2) f_0.
\end{equation}
The spaceship emits the signal back (with a slight offset). The receiver on
the Earth measures the frequency
\beq{1.2}
f_2=\frac1{\sqrt{1-v^2_P/c^2}}\,\X2(1-\frac{v_P}c\Y2) f_1.
\end{equation}
We can compare the sent and received frequency and get a velocity of a \sp
\beq{1.3}
z=\frac{f_2-f_0}{f_0}\simeq -2\frac{v_P}c\,.
\end{equation}
The observed value $z_1$ can be compared with the calculated value $z_2$
(from a~model).
Thus one gets
\beq{1.4}
\pz{}t(z_1-z_2)\simeq -2\,\frac{\ov a_P}c\,.
\end{equation}
This gives an \ac\ of the \sp.

This effect is small
and nonrelativistic. However, it is of nonnewtonian nature.

The \NK{\JT\ }
has been developed. It unifies the gauge invariance principle with the coordinate
invariance principle but in more than four-dimensional space-time. In
particular in the case of \elm c and \gr al interactions in 5-dimensional theory
(Ref.~\cite4).

A general nonabelian Yang--Mills fields have been unified with gravity in
$(n+4)$-\di al space-time ($n$---a \di\ of gauge group) (see Ref.~\cite5).
The theory uses a non\s\ metric defined on a metrized (in a non\s\
way) principal fibre bundle over a \spt\ with a structural group $\U(1)$ in an
\elm c case and in general case nonabelian semi-simple compact group~$G$. The
\cn\ on \spt\ and on a metrized principal fibre bundle is compatible with
this metric. This \cn\ is similar to a \cn\ from Einstein's Unified Field
Theory (Ref.~\cite6), however we use its higher \di al analogue. This \cn\ is
right-\iv\ \wrt an action of the group~$G$ (a~gauge group).

In the \elm c case the metric and the \cn\ are bi-\iv\ \wrt the group~$\U(1)$.
One can find all the details in Ref.~\cite7. The theory has been developed to
include a scalar field leading to an effective \gr al constant and \spt\
dependent cosmological terms. It is possible to extend the theory to include
Higgs' fields and spontaneous symmetry breaking of a gauge group to get
massive vector boson fields. Some exact \so\ of the fields equations has been
found (see~\cite2) and a proposition  to solve a \cfn\ of color in QCD has
been posed (see Refs~\cite8, \cite{Eur}, \cite{Novel}).

In Ref.~\cite{KalCEAS} we give some preliminary results of our approach to the
Pioneer \an\ \ac\ problem placing it on a wider background of a research
including more explanations by different researchers.

The theory is fully relativistic and unifies \elm c field, gauge fields,
Higgs' field and scalar forces with NGT (Non\s\ \eu\gr\ Theory)
in a nontrivial way (see Ref.~\cite9 for details on NGT).
By `in a nontrivial way' we mean that we get from the theory
something more than NGT, ordinary Kaluza--Klein \JT\ Theory, classical
electrodynamics, Yang--Mills' field theory with Higgs' field and spontaneous
symmetry breaking. These new features are some kind of ``interference
effects'' between all of them.
This theory unifies two important approaches in higher-dimensional
philosophy: Kaluza--Klein principle and a \di al reduction principle (see
Refs \cite{4}, \cite{5}, \cite{7} and the references quoted inside).

The beautiful theories such as Kaluza--Klein theory (a~Kaluza miracle) and
its descendents should pass the following test if they are treated as real
unified theories. They should incorporate chiral fermions. Since the
fundamental scale in the theory is a Planck's mass, fermions should be
massless up to the moment of spontaneous symmetry breaking. Thus they should
be zero modes. In our approach they can obtain masses on a \di al reduction
scale. Thus they are zero modes in $(4+n_1)$-\di al case. In this way
$(n_1+4)$-\di al fermions are not chiral (according to very well known
Witten's argument on an index of a Dirac operator). Moreover, they are
not zero modes after a \di al reduction, i.e., in 4-\di al case. It means we
can get chiral fermions under some assumptions.

We expect some nonrelativistic effects leading
to nonnewtonian gravity. Some applications of scalar fields to inflationary
scenario can be found in Ref.~\cite{10}.

In the paper we derive a model for an \an\ \ac\ using a scalar field from the
\NK{\JT\ }. The scalar field leads to an \ef\ \gr al \ct.
A~regularized difference between an ordinary newtonian \ac\ and the \ac\ with
an \ef\ \gr al \ct\ defines the \an\ \ac. The scalar field satisfies a
differential \e\ of the second order which we solve and substitute a \so\ to
the \ef\ \gr al \ct. We use our model to fit Anderson et al.\ data and we
proceed a full statistical analysis with two parameters. Both parameters are
expressed by initial conditions of the \e\ of the scalar field. One of them
has a \di\ of an \ac, the second $\frac1{{\rm AU}^2}$ (AU---Astronomical Unit
$= 149.500 \times 10^3\,{\rm km}$).
The order of the \ac\ \ct\ is the same as Anderson et al. \an\ \ac. The
second \ct\ can be expressed as a length scale of about 4\,AU. We connect the
first (the \ac\ \ct) to second scale of length getting about $1.5\times
10^3\,\rm AU$. Our fitting procedure is in half nonlinear fit.

Afterwards we use our \ac\ model to predict \ph\ movement of planets (caused
by the model) and distortion of elliptic orbits. We find no contradiction
with available data. Moreover, \P0/11 move along a \hy-like orbit. We use our
model to find a distortion of a \hy\ orbit up to $10^3$\,AU from the Sun
together with a distortion of a position on the orbit in comparison to
ordinary \hy\ motion. The same has been derived for a parabolic motion. We
obtain results under some \ap ion procedure working up to $10^3$\,AU.
Moreover, in the epoch of VLBI (Very Long Base Interferometry) it is possible
to consider more precise \ap ion. We proceed it describing the distorted \hy\
orbit using dilogarithm \f\ ($\Li_2(z)$). Moreover, it is not enough and we
are forced to introduce a new special \f. We find some interesting properties
of this \f. These results has been written in Appendix~B. Moreover, it is
interesting from pure theoretical point of view to find exact \so\ of \e s of
motion with \an\ \ac\ taken into account.

We find the \so\ in two cases: periodic and chaotic orbits, expressing them
by elliptic integrals of the first, second and third orders. Both orbits are
bounded. This means that an \an\ \ac\ forbids to escape from the \SS. In the
case of periodic orbits a period is given by a period of $\cP$-Weierstrass
\f. We consider also an influence of the \co ical \ct\ on the motion. The \co
ical \ct\ influence cannot remove a point mass (a~body) from the \SS. In the
case of \P0/11 a range is about $1.5\t10^5$\,AU and a time to come back
$10^6$\,yr.
We give these results in Appendix~A.

The important point in our research is a practical application of the results
for \P0/11 future. In order to do this we consider a movement of them along
distorted \hy\ orbit in the \SS\ using reference systems recommended by IAU.
We consider also a possibility to use a different \gr al theory than GR
(General Relativity), i.e.\ NGT (\eu\nos\ \eu\gr al Theory), Einstein--Cartan
Theory and Einstein--Cartan--Moffat Theory if it is necessary. We give
some suggestions to connect a reference frame to Cosmic Microwave
Background Radiation frame and to take into account a Coriolis force during a motion of a
barycenter of the \SS\ around a center of the Galaxy. In future measurements
of \sp s positions we suggest to use a next generation atomic clocks. In
order to get a precise predictions of positions of the \sp s we derive \e s
for perturbation of \hy\ orbits elements which includes all point masses with
known orbits.  Due to these procedures we hope to find parameters for
our model with the highest precision. This gives new lights for astrodynamics
and space navigation in the \SS.

We consider a mysterious connection between an \an\ \ac\ and the Hubble \ct\
within our model of \ac. We find a \so\ for the scalar field in a \co ical
background. We use a paradigm of model \co y using CDM$\La$ model for our
contemporary epoch, neglecting a radiation density. In this way we find that
the \ac\ \ct\ from our model is proportional to $cH_0$ with a parameter close
to~$1$ (but not exactly). We proceed a full statistical analysis for this
parameter. We find also a dependence of the second parameter of the model on
the Hubble \ct\ finding that a ratio of the two length scales (found by us)
is \ct\ in time. We consider also a second mysterious coincidence between the
\an\ \ac\ and the \co ical \ct. Some researchers claim that both coincidences
cannot be true simultaneously, for the first (this with the Hubble \ct)
depends on time, and the second (with the \co ical \ct) does not depend on
time. Moreover, this is not the case, for the second coincidence depends, in
our approach, on time in such a way that for our epoch (i.e.\ $14.5 \times
10^9$ yr) this dependence can be considered as an asymptotic. We find also an
interesting scale of a distance about $1.2\times 10^{26}$\,km (an order of
size of the visible Universe) and connect it to the \co ical \ct.

We consider relativistic effects on a distorted \hy\ orbit finding that they
have no significant influence in GR and NGT cases. A~time dependence on
$G\eff$ (an effective \gr al \ct) has been considered in our model. We have not
any profound contraction with observational data in the \SS. We consider the
full formalism of the \eu\nos\ Jordan--Thiry \e s applied to the problem of an
\an\ \ac. We present a relativistic model of an \an\ \ac\ using this
formalism.
We prove that the model passes General Relativity tests in the \SS.
Moreover, we need a fine tuning of initial conditions to get a full success.
We consider orbits as in the case of \P0/11 \sp s and we find that they are
bounded. Moreover, in the case of almost \rl\ or \rl\ \sp s they are
unbounded.
All the lengthy derivations we put in Appendix~A and Appendix~D.
In the paper we use Mathematica~7 in order to proceed some tedious
numerical and symbolic calculations.

The paper is organized as follows. In Section 2 we give some elements of
\NK{\JT\ }. In Section~3 we develop a model of an \an\ \ac\ and apply it to
the data of Anderson et al.\ (see Refs \cite1, \cite2, \cite3). Section~4 is
devoted to an influence of the model on perihelion shifts of elliptic orbit
and to a distortion of such orbits in the Solar System. We prove that our model
does not contradict observations.

In Section 5 we develop an application of the model to hyperbolic orbits
finding under some approximation procedure an \e\ of an orbit in polar
coordinates and an analogue of hyperbolic Kepler \e\ in our case. In
Section~6 we develop the same procedure for a parabolic orbit. This section
serves as an illustration of a model application because parabolic orbits are
rare. Moreover, \e s are simpler and it is possible to solve them exactly.
The results from Section~5 can be directly applied to the future movements of
Pioneers. In Section~7 we consider the distorted hyperbolic orbit in the Solar
System, consider perturbation of planets in order to compare predicted
positions with observation to fit parameters of an orbit plus a \ct~$b$.
This could lead to more refined models, which may be useful in astrodynamics
and for a navigation in the Solar System. Thus we need a feedback from
observational data. In Section~8 we find a \cn\ between an \an\ \ac\ \ct\ $b$
and a Hubble \ct. In Section 9 we consider relativistic effects.
In Section~10 we consider an effective \gr al \ct\ in the \SS.
In Section~11 we consider a relativistic model of an \an\ \ac\ based on the
full formalism of the \eu\nos\ Jordan--Thiry Theory.
We find that in the case of \P0/11 orbits are bounded with the size about
$10^5$\,AU and a time to come back to the center of the \SS\ is about $10^6$\,yr.
In Appendix~A we derive exact \e s for $r(\vf)$ and
$r(t)$ ($\rho(\vf)$,~$\rho(t)$ in the notation in the Appendix) in a plane of
motion for $r>30\,{\rm AU}$. We consider also numerical results, relativistic
extension of a model (relativistic \sp s) and a full \eu\nos\ Jordan--Thiry
Theory formalism for an \an\ \ac.

In Appendix~B we give more precise formulae for a hyperbolic (a distorted
hyperbola) and parabolic (a distorted parabola) motion. In Appendix~C we give
a programme to calculate $f(x,y)$ with a~result.
In Appendix~D we consider full \eu\nos\ Jordan--Thiry \e s in order to
define a \spt\ with an \an\ \ac, i.e.\ in a stationary, spherically \s\ case.
We give \e s and we solve them for a problem connecting to the \SS. We
find preliminary orbits.
We consider a match of a \spt\ with an \an\ \ac\ to a \co ical \so. We
consider a bending of light in this \spt, \an\ \ph\ movement, Shapiro effect,
using various parametrizations of elements of the metric. We consider in
details initial value problems to include the \SS\ parameters and Anderson et
al.\ data.
In Appendix~E we give listings of some programmes written in Mathematica~7
applied for problems from Appendix~D and Section~11.

Let us give some remarks concerning figures in the paper. In the case where
we have to do with orbits the length scale is 1\,AU (Astronomical Unit) or
$r_0\simeq 4$\,AU (see Eq.\ \er{3.35} in Section~3). In the text following or
preceding the figure it is explained which unit is considered. It seems that
the figures are more readable with such a convention. Sometimes we have to do
with \di less quantities and any unit is not supposed. If we have to do with
time dependence the velocity of light is equal to~1. Sometimes we use a very
big unit of length $L\sim 10$\,Mpc. In such a case this is also explained in
the text (below or above). Thus we have no misunderstandings. In general we
put all programs written in Mathematica~7 to Appendix~E. Moreover, from time
to time some short listings appear in the text. This is only for some
illustrations for our calculations and will not cause any misunderstandings.

\section{Elements of the \NK{\JT\ }}
The \NK{\JT\ } (see Refs \cite4, \cite5, \cite7) unifies the Nonsymmetric
Gravitational Theory (NGT) and gauge fields (Yang--Mills' fields) including
spontaneous symmetry breaking and the Higgs' mechanism with scalar
forces connected to the gravitational constant and cosmological terms
appearing as the so-called quintessence. The theory is
geometric and unifies tensor-scalar gravity with massive gauge theory
using a multidimensional manifold in a Jordan--Thiry manner.
We use a nonsymmetric version of this theory (see \cite5,
\cite7). The general scheme is the following. We introduce
the principal fibre bundle over the base $V = E \times G/G_{0}$ with the
structural group $H$, where $E$ is a space-time, $G$ is a compact semisimple
Lie group, $G_{0}$ is its compact subgroup and $H$ is a semisimple compact
group. The manifold $M = G/G_{0}$ has an interpretation as a ``vacuum states
manifold" if $G$ is broken to $G_{0}$ (classical vacuum states). We define on
the space-time $E$, the nonsymmetric tensor $g_{\alpha\beta}$ from NGT, which is
equivalent to the existence of two geometrical objects
\begin{eqnarray}
\ov{g}&=& g_{(\alpha \beta )}\overline{\theta }^\alpha \otimes \overline{\theta}^\beta
\label{2.1}\\
\ul{g}&=& g_{[\alpha \beta ]}\overline{\theta }^\alpha \wedge \overline{\theta}^\beta
\label{2.2}
\end{eqnarray}
the symmetric tensor $\ov{g}$ and the 2-form $\ul{g}$. Simultaneously we introduce
on $E$ two connections from NGT $\ov{W}{}_{\beta \gamma}^\alpha $ and
$\widetilde{\overline{\varGamma}}{}_{\beta \gamma}^\alpha $. On the homogeneous space $M$
we define the nonsymmetric metric tensor
\beq{2.3}
g_{\tilde{a}\tilde{b}} = h^{0}_{\tilde{a}\tilde{b}}
+ \zeta k^{0}_{\tilde{a}\tilde{b}}
\end{equation}
where $\zeta $ is the dimensionless constant, in a geometric way. Thus we
really have the nonsymmetric metric tensor on or $V = E \times
G/G_{0}$.
\beq{2.4}
\gamma _{\rAB} =
\left(\vcenter{\offinterlineskip\tabskip0pt
\halign{\strut\tabskip5pt plus10pt minus10pt
\hfill#\hfill&\vrule height10pt#&\hfill#\hfill\tabskip0pt\cr
$\vrule height0pt depth6pt width0pt g_{\alpha \beta }$ && $0$\cr
\noalign{\hrule}
$\vrule height9pt depth4pt width0pt 0$ &&
$r^2g_{\tilde{a}\tilde{b}}$\cr}}\right)
\end{equation}
$r$ is a parameter which characterizes the size of the manifold $M =
G/G_{0}$. Now on the principal bundle $\ul{P}$ we define the connection $\omega $, which
is the 1-form with values in the Lie algebra of $H$.

After this we introduce the nonsymmetric metric on $\ul{P}$
right-invariant with respect to the action of the group $H$, introducing
scalar field $\rho $ in a Jordan--Thiry manner.
The only difference is that
now our base space has more dimensions than four. It is
$(n_{1}+4)$-dimensional, where $n_{1} = \dim (M) = \dim (G)-\dim (G_{0})$. In other
words, we combine the nonsymmetric tensor $\gamma _{\rAB}$ on $V$ with the
right-invariant nonsymmetric tensor on the group $H$ using the connection
$\omega $ and the scalar field $\rho $. We suppose that the factor $\rho $ depends on a
space-time point only. This condition can be abandoned and we consider
a more general case where $\rho =\rho (x,y)$, $x\in E$, $y\in M$ resulting in a tower of
massive scalar field $\rho _{k}, k=1,2\ldots .$ This is really the Jordan--Thiry
theory in the nonsymmetric version but with $(n_{1}+4)$-dimensional
``space-time". After this we act in the classical manner. We introduce the linear connection
which is compatible with this nonsymmetric metric. This connection is
the multidimensional analogue of the connection
$\widetilde{\overline{\varGamma}}{}^{\alpha}_{\beta \gamma }$ on the space-time
$E$. Simultaneously we introduce the second connection $W$. The
connection $W$ is the multidimensional analogue of the $\ov{W}$-connection from
NGT and Einstein's Unified Field Theory. It is the same as the
connection from Refs.\ \cite5,~\cite7. Now we calculate the Moffat--Ricci
curvature scalar $R(W)$ for the connection $W$ and we get the following
result. $R(W)$ is equal to the sum of the Moffat--Ricci curvature on the
space-time $E$ (the gravitational lagrangian in Moffat's theory of
gravitation), plus $(n_{1}+4)$-dimensional lagrangian for the Yang--Mills'
field from the Nonsymmetric Kaluza--Klein Theory plus the Moffat--Ricci
curvature scalar on the homogeneous space $G/G_{0}$ and the Moffat--Ricci
curvature scalar on the group $H$ plus the lagrangian for the scalar
field~$\rho $. The only difference is that our Yang--Mills' field is defined
on $(n_{1}+4)$-dimensional ``space-time" and the existence of the
Moffat--Ricci
curvature scalar of the connection on the homogeneous space
$G/G_{0}$.
All of these terms (including $R(\ov{W})$) are multiplied by some
factors depending on the scalar field~$\rho $.

This lagrangian depends on the point of $V = E \times G/G_{0}$ i.e.\ on the
point of the space-time $E$ and on the point
of $M = G/G_{0}$. The curvature
scalar on $G/G_{0}$ also depends on the point of~$M$.

We now go to the group structure of our theory. We assume $G$ invariance of
the connection $\omega $ on the principal fibre bundle $\ul{P}$, the so
called Wang condition. According to the Wang theorem the connection $\omega $
decomposes into the connection $\widetilde{\omega }_{\rE}$ on the principal
bundle $Q$ over space-time $E$ with structural group $G$ and the multiplet of
scalar fields ${\varPhi} $. Due to this decomposition the multidimensional
Yang--Mills' lagrangian decomposes into: a 4-dimensional Yang--Mills'
lagrangian with the gauge group $G$ from the Nonsymmetric Kaluza--Klein
Theory, plus a polynomial of 4th order with respect to the fields ${\varPhi}
$, plus a term which is quadratic with respect to the gauge derivative of
${\varPhi}$ (the gauge derivative with respect to the connection
$\widetilde{\omega }_{\rE}$ on a space-time $E$) plus a new term which is of
2nd order in the ${\varPhi} $, and is linear with respect to the Yang--Mills'
field strength. After this we perform the dimensional reduction procedure for
the Moffat--Ricci scalar curvature on the manifold $\ul{P}$. We average
$R(W)$ with respect to the homogeneous space $M = G/G_{0}$.  In this way we
get the lagrangian of our theory. It is the sum of the Moffat--Ricci
curvature scalar on $E$ (gravitational lagrangian) plus a Yang--Mills'
lagrangian with gauge group~$G$ from the Nonsymmetric Kaluza--Klein Theory
(see \cite7), plus a kinetic term for the scalar field~${\varPhi} $, plus a
potential $V({\varPhi} )$ which is of 4th order with respect to ${\varPhi} $,
plus ${\cal L}_{\Int}$ which describes a nonminimal interaction between the
scalar field ${\varPhi} $ and the Yang--Mills' field, plus cosmological
terms, plus lagrangian for scalar field $\rho $. All of these terms
(including $\ov{R}(\ov{W})$) are multiplied of course by some factors
depending on the scalar field $\rho $. We redefine tensor $g_{\mu \nu }$ and
$\rho$ and pass from scalar field $\rho $ to ${\varPsi} $
\beq{2.5}
\rho = \ex{-{\varPsi} }
\end{equation}
After this we get lagrangian which is the sum of gravitational lagrangian,
Yang--Mills' lagrangian, Higgs' field lagrangian, interaction term ${\cal
L}_{\Int}$ and lagrangian for scalar field ${\varPsi} $ plus cosmological
terms. These terms depend now on the scalar field ${\varPsi} $. In this way
we have in our theory a multiplet of scalar fields $({\varPsi} ,{\varPhi} )$.
As in the Nonsymmetric-Nonabelian Kaluza--Klein Theory we get a polarization
tensor of the Yang--Mills' field induced by the skewsymmetric part of the
metric on the space-time and on the group $G$.  We get an additional term in
the Yang--Mills' lagrangian induced by the skewsymmetric part of the metric
$g_{\alpha \beta }$. We get also ${\cal L}_{\Int}$, which is absent in the
dimensional reduction procedure known up to now.  Simultaneously, our
potential for the scalar---Higgs' field has more complicated structure, due
to the skewsymmetric part of the metric on $G/G_{0}$ and on $H$.
This structure offers two kinds of critical points for
the minimum of this potential: ${\varPhi} ^{0}_{\crt }$ and ${\varPhi}
^{1}_{\crt }$. The first is known in the classical, symmetric dimensional
reduction procedure and corresponds to the trivial Higgs' field (``pure
gauge"). This is the ``true" vacuum state of the theory.  The second,
${\varPhi} ^{1}_{\crt }$, corresponds to a more complex configuration.  This
is only a local (no absolute) minimum of~$V$. It is a ``false" vacuum. The
Higgs' field is not a ``pure" gauge here. In the first case the unbroken
group is always $G_{0}$. In the second case, it is in general different and
strongly depends on the details of the theory: groups $G_{0}$, $G$, $H$,
tensors $\ell _{ab}$, $g_{\tilde{a}\tilde{b}}$ and the constants $\zeta$,
$\xi $. It results in a different spectrum of mass for intermediate bosons.
However, the scale of the mass is the same and it is fixed by a constant $r$
(``radius" of the manifold $M = G/G_{0})$. In the first case $V({\varPhi}
^{0}_{\crt }) =0$, in the second case it is, in general, not zero
$V({\varPhi} ^{1}_{\crt }) \neq 0$. Thus, in the first case, the cosmological
constant is a sum of the scalar curvature on $H$ and $G/G_{0}$, and in the
second case, we should add the value $V({\varPhi} ^{1}_{\crt })$. We proved
that using the constant $\xi$ we are able in some cases to make the
cosmological constant as small as we want (it can change the sign).  Here we
can perform the same procedure for the second term in the cosmological
constant using the constant $\zeta $. It can change the sign too.

The interesting point is that there exists an effective scale of masses,
which depends on the scalar field~${\varPsi} $.

Using Palatini variational principle we get an equation for fields
in our theory. We find a gravitational equation from N.G.T. with
Yang--Mills', Higgs' and scalar sources (for scalar field ${\varPsi} )$ with
cosmological terms. This gives us an interpretation of the scalar
field ${\varPsi} $ as an effective gravitational constant
\beq{2.6}
G\eff = G_{N}\ex{-(n+2){\varPsi} }\qh{or}
G\eff = \ov G_{0}\ex{-(n+2){\varPsi} },
\end{equation}
where $\ov G_0$ is a \ct\ connected to $G_N$ (Newton \gr al \ct) via some
cosmological circumstances (see Ref.~\cite{10}).

We get an equation for this scalar field ${\varPsi} $. Simultaneously we get
equations for Yang--Mills' and Higgs' field. We also discuss the change
of the effective scale of mass, $m\eff$ with a relation to the change of
the gravitational constant~$G\eff$.

In the ``true" vacuum case we get that the scalar field ${\varPsi} $ is
massive and has Yukawa-type behaviour. In this way the weak
equivalence principle is satisfied. In the ``false" vacuum case the
situation is more complex. It seems that there are possible some
scalar forces with infinite range. Thus the two worlds constructed
over the ``true" vacuum and the ``false" vacuum seem to be completely
different: with different unbroken groups, different mass spectrum for
the broken gauge and Higgs' bosons, different cosmological constants
and with different behaviour for the scalar field ${\varPsi} $. The last point
means that in the ``false" vacuum case the weak equivalence principle
could be violated and the gravitational constant (Newton's constant)
would increase in distance between bodies.

In this paper we are interested in properties of the scalar field $\Ps$ which
is a source of an inconstancy of an effective \gr al \ct\ (Eq.~\er{2.6}).
Thus we consider a \lg\ of this field, neglecting \YM' field and Higgs' fields
from the full theory. A kinetic part of a \lg\ of the field $\Ps$ looks:
\bea{2.7}
\cL_{\rm scal}^{\rm kin}(\Ps)&=&\bigl(\ov M g^\(\g\nu)+n^2g^\[\m]
g_{\d\mu}\wt g^\(\d\g)\bigr)\Ps_{,\nu}\Ps_\g \\
\ov M&=&\bigl(l^\[dc]l_\[dc]-3n(n-1)\bigr) \label{2.8}
\end{eqnarray}
where
\bg{2.9}
\wt g^\(\d\g)g_\(\d\a)=\gd \d,\g,\a,\\
l_{ab}=h_{ab}+\mu k_{ab}. \label{2.10}
\end{gather}
This field couples to \co ical \ct s in the theory: $\wt R(\wt\G)$---a scalar
curvature of a \cn\ $\wt\G$ on group manifold~$H$, and to
\beq{2.11}
\wt{\ul P}=\frac 1{V_2}\int_{G/G_0} d^{n_1}x\, \hb R(\hb \G),
\end{equation}
where $V_2$ is a volume of a manifold $M=G/G_0$ and $\hb R(\hb\G)$ is a
scalar curvature of a \cn\ $\hb\G$ defined on this manifold (see for details
Ref.~\cite4), and a full \lg\ for a field $\Ps$ looks
\beq{2.12}
L=\cL_{\rm scal}^{\rm kin}(\Ps)+\la _c,
\end{equation}
where
\beq{2.13}
\la_c=\a_s^2\,\frac{\ex{(n+2)\Ps}}{l\pl^2}\,\wt R(\wt\G)+\frac{\ex{n\Ps}}{r^2}\,\wt{\ul P},
\end{equation}
$l\pl=\sqrt{\frac{G_N\hbar}{c^3}}\simeq 10^{-33}$\,cm is a Planck's length, and
$\a_s$ is a \di less coupling \ct. Due to \nos ity of a \cn\ $\wt\G$ and
$\hb\G$, $\wt R(\wt \G)$ and $\wt{\ul P}$ are \f\ of \ct s $\mu$ and $\z$ and can change
the signs. Explicit examples are for $G=\SU(2)$ and $M=S^2$. $\la_c$ can be
written in a different way
\beq{2.14}
\la_c=\ex{(n+2)\Ps}\,\frac{\a_s^2}{l\pl^2}\,\wt R(\wt\G)
+\ex\Ps\,\frac{m_{\tilde A}^2}{\a_s^2}\Bigl(\frac c\hbar\Bigr)^2 \wt{\ul P},
\end{equation}
where $m_{\tilde A}$ is a scale of mass of broken gauge bosons. Moreover, we
get an \e\ for a scalar field~$\Ps$
\bml{2.15}
2\biggl[\Bigl((n^2+2\ov M)\wt g^\(\a\mu)-n^2g^{\m}g_{\d\nu}
\wt g^\(\a\d)\Bigr)\pp{^2\Ps}{x^\a \pa x^\mu}\\
{}+\frac1{\sqrt{-g}}\,\pa_\mu\biggl\{\sqrt{-g}\Bigl[n^2\wt g^\(\a\mu)
-\frac{n^2}2\,g_{\d\nu}\bigl(g^{\nu\a}\wt g^\(\mu\d)+g^{\nu\mu}\wt g^\(\mu\a)\bigr)
-2\ov M\wt g^\(\mu\a)\Bigr]\biggr\}\pp\Ps{x^\a}\biggr]\\
{}-\frac n{r^2}\,\ex{n\Ps}\wt{\ul P}
-\frac{(n+2)\a_s^2}{l\pl^2}\,\ex{(n+2)\Ps}\wt R(\wt\G)=0.
\end{multline}
We neglect in Eq.\ \er{2.15} terms involving \YM' fields and Higgs' fields.

In the paper we are interested in a propagation of this field in Riemannian
geometry. Thus Eq.~\er{2.15} simplifies. The \co ical terms in the full \lg\
of the theory can be considered as selfinteraction potential of a scalar
field~$\Ps$
\beq{2.16}
U(\Ps)=\ex{(n+2)\Ps}\,\frac{\a_s^2\wt R(\wt\G)}{l\pl^2}+
\frac{\wt P}{r^2}\,\ex{n\Ps}=\g \ex{n\Ps}+\b \ex{(n+2)\Ps}.
\end{equation}
The \co ical ``\ct'' is equal to
\beq{2.17}
\la_{\rm co}(\Ps)=-\frac\g2\,\ex{n\Ps}-\frac\b2 \ex{(n+2)\Ps}.
\end{equation}
The simplified \e\ for $\Ps$ looks:
\beq{2.18}
2\ov M \wt\nabla_\a(g^{\a\b}\pa_\b \Ps)-(n+2)\ex{(n+2)\Ps}\b
-n\ex{n\Ps}\g=0.
\end{equation}

In order to find a \co ical \ct\ in the theory we should minimize a
selfinteraction potential \wrt field~$\Ps$. One gets that for the value
\beq{2.19}
\ex{\Ps_0}=x_0=\sqrt{\frac{n|\g|}{(n+2)\b}}
\end{equation}
we get a minimum for $U$. In this way
\beq{2.20}
\la_{\rm co}(\Ps_0)=\frac{x_0^n|\g|}{(n+2)}
\end{equation}
($\g<0$, $\b>0$).

Let us notice the following fact. If we write $\Ps=\Ps_0+\vf$ where we have
redefined field $\Ps$ we get
\beq{2.21}
2\wt\nabla_\a \bigl(g^{\a\b} \pa_\b \vf)-\frac{x_0^n|\g|n}{\ov M}\,
\ex{n\vf}(\ex{2\vf}-1)=0.
\end{equation}
Using Eq.\ \er{2.20} we get
\beq{2.22}
\wt\nabla_\a \bigl(g^{\a\b}\pa_\b \vf)-\frac{n(n+2)}{2\ov M}\,
\la_{\rm co}\ex{n\vf}(\ex{2\vf}-1)=0.
\end{equation}
Taking a known value of a contemporary \co ical \ct\ as $\la_{\rm co}$
($\La=10^{-52}\frac1{\rm m^2}$) we
come to the \e
\beq{2.22a}
\wt\nabla_\a \bigl(g^{\a\b}\pa_\b \vf\bigr)+\wt\ve \ex{n\vf}(\ex{2\vf}-1)=0,
\end{equation}
where $\wt\ve=\sgn\ov M$. In Eq.~\er{2.20} we use natural scales of space and
time \cd s.
\beq{2.23}
L=\sqrt{\frac{n(n+2)\la_{\rm co}}{2|\ov M|}}
\simeq 10\,{\rm Mpc}, \q T=\frac Lc\simeq32\times10^6{\rm yr.}
\end{equation}
$T$ is of order of a geological time or a propagation time of temperature
perturbations from a center of the Sun to its surface (see Ref.~\cite{11n}).

In a \co ical background, spatially flat one gets
\beq{2.24}
\frac1{a^3(t)}\,\pp{}t\Bigl(a^3(t)\pp\vf t\Bigr)
-\frac1{a^2(t)}\,\D\vf+\wt\ve \ex{n\vf}(\ex{2\vf}-1)=0,
\end{equation}
where $a(t)$ is a scale factor and $\D$ is a Laplace operator in $\R^3$.
This can be rewritten as
\beq{2.25}
\ov\D\vf-\frac1{a^3(t)}\,\pp{}t\Bigl(a^3(t)\pp\vf t\Bigr)
-\wt\ve \ex{n\vf}(\ex{2\vf}-1)=0
\end{equation}
where new space \cd s are defined: $\ov x=a(t)x$, $\ov y=a(t)y$, $\ov
z=a(t)z$, and $\ov \D$ is written in these \cd s.

Let us consider Eq.\ \er{2.2} in flat \spt\ for a static, spherically-\s\
case. One gets
\beq{2.26}
\frac1{x^2}\,\pz{}x\Bigl(x^2\,\pz\vf x\Bigr)-\wt\ve \ex{n\vf}(\ex{2\vf}-1)=0.
\end{equation}
$\vf=\vf(x)$ and $x$ is a radial \cd\ measured in $L$.

\section{An \an\ \ac}
Let us consider Eq.\ \er{2.26} in the following form in the Solar System
application.
\beq{3.1}
x\,\pz{^2y}{x^2}+2\,\pz yx-\wt\ve x\exp(ny)(\exp(2y)-1)=0.
\end{equation}
Let us remind to the reader that for a scale of length (about 10~Mpc)
in the Solar System $x$ is of $10^{-12}$ of this scale, that in the first
order of \ap ion we get
\beq{3.2}
2\,\pz{\ov y}x=0, \q \ov y=C_1={\rm const.}
\end{equation}
Let us consider a small deviation of this first order \so\
\beq{3.3}
y=C_1+y_1,
\end{equation}
where $y_1=y_1(x)$ is small. One gets
\beq{3.4}
x\,\pz{^2y_1}{x^2}+2\,\pz{y_1}x -\wt\ve x\ex{nC_1}(1+y_1)
\bigl(\ex{2C_1}(1+y_1)-1\bigr)=0
\end{equation}
or
\beq{3.5}
x\,\pz{^2y_1}{x^2}+2\,\pz{y_1}x-\wt\ve x\ex{nC_1}\bigl((\ex{2C_1}-1)
+y_1(2\ex{2C_1}-1)+\ex{2C_1}y_1^2\bigr)=0.
\end{equation}

We \ap e an exponential \f\ by a sum of the first and the second terms in a
power expansion. Neglecting a term with $y_1^2$ we get an \e
\beq{3.6}
\frac1{x^2}\,\pz{}x\Bigl(x^2\,\pz{y_1}x\Bigr)-\wt\ve \ex{nC_1}
(2\ex{2C_1}-1)y_1=\wt\ve \ex{nC_1}(\ex{2C_1}-1).
\end{equation}
Moreover, in our simple model we can neglect also a linear term. Eventually
we get
\beq{3.7}
\frac1{x^2}\,\pz{}x\Bigl(x^2\,\pz{y_1}x\Bigr)
=\wt\ve \ex{nC_1}(\ex{2C_1}-1).
\end{equation}

Let us solve this \e. One finds
\beq{3.8}
y_1=\frac q6\,x^2-\frac{C_2}x+C_3,
\end{equation}
where $q=\wt\ve \ex{nC_1}(\ex{2C_1}-1)$. Taking $C_2=0$ (a solution is singular
at $x=0$) we get finally
\beq{3.9}
y=\ov C_2+\frac{\wt\ve \ex{nC_1}(\ex{2C_1}-1)}6\,x^2,
\end{equation}
where $\ov C_2=C_1+C_3$ is another \ct. In this way we get an \ap e \so\ of
Eq.~\er{3.1} depending on two \ct s of integration.

Let us remind to the reader that an \ef\ \gr al \ct\ from the \eu\nos\
Jordan--Thiry Theory reads
\beq{3.10}
G\eff=G_N\exp\bigl(-(n+2)y\bigr).
\end{equation}
Thus we get
\beq{3.11}
G\eff=G_N\exp\Bigl(-(n+2)C_2-\wt\ve\,\frac{n+2}6\,\ex{nC_1}(\ex{2C_1}-1)x^2
\Bigr).
\end{equation}
$C_2$ and $C_1$ are arbitrary and finally we have the following shape of
$G\eff$:
\beq{3.12}
G\eff=G_N\exp(ar^2+b_1),
\end{equation}
where $a$ and $b_1$ are arbitrary and are easily expressed by $C_1$ and $C_2$
and a scale of 10\,Mpc and $r$ is measured in the Solar System units
of length (AU or km). Now we calculate an \an\ \ac\ which is the difference
between an \ac\ with $G\eff$ and a newtonian one.
\beq{3.13}
a_p(r)=-\frac{G\eff M_\odot}{r^2}+\frac{G_NM_\odot}{r^2},
\end{equation}
where $M_\odot$ is a mass of the Sun. One gets
\beq{3.14}
a_p(r)=-\frac{k^2}{r^2}\bigl(\exp(ar^2+b_1)-1\bigr)
\end{equation}
where $k^2=G_NM_\odot$.

However, this \ac\ is singular at $r=0$ which is unreasonable.

We should regularize it. First of all, we subtract a pole at $r=0$ and a
value at $r=0$
$$
a_p^{\rm reg}(r)=a_p(r)+\frac{k^2(\ex{b_1}-1)}{r^2}+k^2\ex{b_1}a
$$
getting
\beq{3.15}
a_p^{\rm reg}(r)=-\frac{k^2}{r^2}\,\ex{b_1}\bigl(\exp(ar^2)-1-ar^2\bigr).
\end{equation}
Denoting $k^2\ex{b_1}=b$ we get a model for an \an\ \ac
\beq{3.16}
a_p(r)=-\frac b{r^2}\bigl(\exp(ar^2)-1-ar^2\bigr).
\end{equation}

Regularization applied here is similar in spirit to any regularization known
in quantum theory or statistical physics where we are going to remove poles
(and only poles). Otherwise the regularization procedure is not unique.
Similar singularities appear also in perturbation calculus in celestial
mechanics and recently also in perturbative expansion in General Relativity
(a~novel approach to EIH method).

Let us notice that $b>0$. In this way we can write a model as
\beq{3.17}
a_p(r)=-\ov b g(ar^2)
\end{equation}
where
\beq{3.18}
g(x)=\frac1x(\ex x-1-x)
\end{equation}
and
$$
\ov b=ba.
$$
Let us notice that a \f\ $g$ is nonnegative for $x>0$, for $\ex x=1+x+
\frac{x^2}2+\ldots>1+x$. In this way $a_p(r)<0$ for $a>0$. Moreover, we are
interested also in the case $a<0$. Writing
\beq{3.19}
a_p(r)=-\frac b{r^2}\bigl(\exp(-ar^2)-1+ar^2\bigr)
\end{equation}
we should prove that $a_p(r)<0$. In order to prove it let us consider a \f\
$g_1(x) =\ex{-x}-1+x$ for $x\ge0$. One gets $g_1(0)=0$ and
\beq{3.20}
\pz{g_1}x = -\ex{-x}+1>0 \qh{for} x>0.
\end{equation}
Thus this \f\ is increasing (even $\pz{g_1}x(0)=0$) and cannot be negative.
Thus in all cases we have $a_p(r)<0$, $r>0$.

Let us consider a model \er{3.17}. We get
\beq{3.21}
a_p(r)=-\frac{\ov b}{ar^2}\bigl(\ex{ar^2}-1-ar^2\bigr).
\end{equation}
In the case of $a<0$ we can write
\beq{3.22}
a_p(r)=\frac{\ov b}{|a|r^2}\bigl(\ex{-|a|r^2}-1+|a|r^2\bigr)
\end{equation}
and a \ct\ $\ov b<0$.

It is easy to see that
\beq{3.23}
\ov b=-b|a|.
\end{equation}
We meet this situation below.
Let us use the model \er{3.17} to fit the data of Anderson et al.\ for an
\an\ \ac\ of \P0/11 (see Refs \cite1, \cite2, \cite3). In order to do this we
apply the least square method and define a \f
\beq{3.24}
M(a,\ov b)=\frac12 \sum_{i=1}^n \frac{\bigl(\ov bg(ax_i^2)-y_i\bigr)^2}
{\si_i^2}\,.
\end{equation}
According to the mentioned method we get
\beq{3.25}
\bal
\pp Ma&=b\sum_{i=1}^n \frac{x_i^2}{\si_i^2}\bigl(bg(ax_i^2)-y_i\bigr)
g'(ax_i^2)=0\cr
\pp Mb&=b \sum_{i=1}^n \frac{g(ax_i^2)}{\si_i^2}-\sum_{i=1}^n
\frac{y_ig(ax_i^2)} {\si_i^2}=0,
\eal
\end{equation}
where $x_i$ is a position (in AU) of \P0/11 and $y_i$ a residual \an\ \ac\ of
them, $\si_i$ are standard deviations. In the case of Anderson et al.\ data
$n=19$. Our fit is in half linear (a~linear dependence of~$\ov b$). Supposing
$\ov b\ne0$ we get the following \e s for $a$ and $\ov b$
\refstepcounter{equation}\label{3.27}
$$
\dsl{
\hfill DA-BC=0 \hfill (\theequation\rm a)\cr
\hfill \ov b=\frac BA \hfill (\theequation\rm b)}
$$
where
\beq{3.28}
\bal
A&=\sum_{i=1}^n \frac{g^2(ax_i^2)}{\si_i^2}\cr
B&=\sum_{i=1}^n \frac{y_ig^2(ax_i^2)}{\si_i^2}\cr
C&=\sum_{i=1}^n \frac{x_i^2g^2(ax_i^2)g'(ax_i^2)}{\si_i^2}\cr
D&=\sum_{i=1}^n \frac{y_ix_i^2g'(ax_i^2)}{\si_i^2}
\eal
\end{equation}

From Eq.\ (\ref{3.27}a) we get $a$ and after that $\ov b$ from Eq.\
(\ref{3.27}b). Eq.~(\ref{3.27}a) can be solved using a Newton method with a
starting point $a_1=-0.01661$, which is a value of a parameter $a$ for a
fitted model without weights $\frac1{\si_i^2}$ (in order to avoid spurious
\so s). We get
\beq{3.29}
\bal
a&=-0.05938\cr
b&=-8.52151.
\eal
\end{equation}
In order to estimate statistical errors we calculate a matrix
\beq{3.30}
W=\mt{
\dss \pp{^2M}{a^2}&\ &\dss \pp{^2M}{a\pa\ov b}\\ \noalign{\vskip5pt}
\dss\pp{^2M}{\ov b\pa a}&\ &\dss\pp{^2M}{\ov b{}^2}
}
\end{equation}
and an inverse matrix $E=\mt{\si_a^2&\ &V(a,\ov b)\\
V(a,\ov b)&&\si_{\bar b}^2}=W^{-1}$, which is an asymptotic covariance matrix
for obtained parameters, $\si_a$ and $\si_{\bar b}$ are standard deviations of $a$
and~$\ov b$ and $V(a,\ov b)$ means a covariance of them.

\begin{figure}
\centerline{\rotatebox{90}{\ing{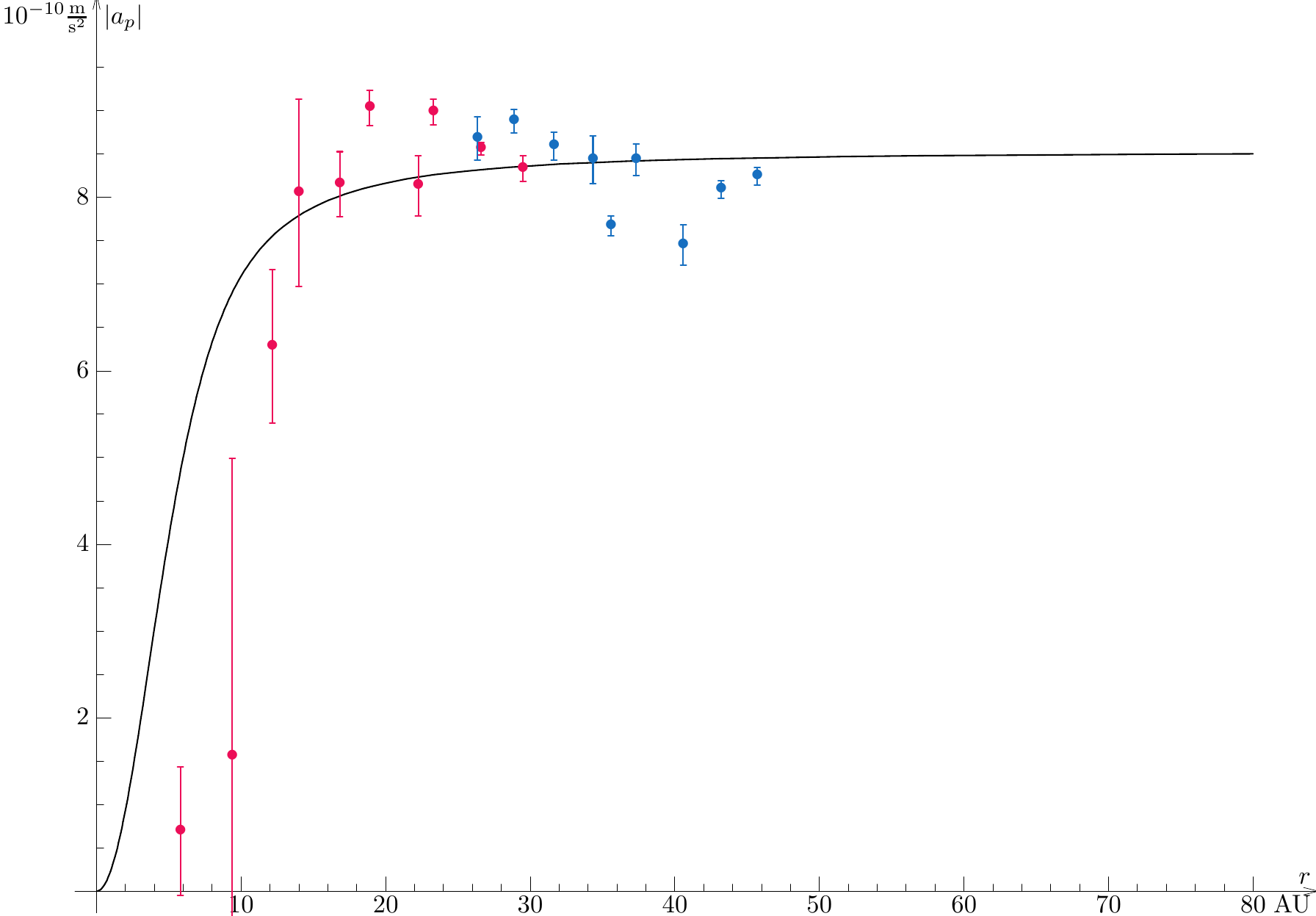}}}
\caption{A fit of \an\ \ac\ $|a_p|$ against a distance from the Sun with
Anderson et al. data and a prediction of a model up to 80\,AU. Red points
mean \P1 data, blue \P0 data (extracted from Fig.~4 of Ref.~\cite3).}\label{f1}
\end{figure}

One gets
\bg{3.31}
W=\mt{3896.28&\ &505.498\\\noalign{\vskip2pt}
505.498&&181.938}\\
E=\mt{0.000401315&\ &-0.00111502\\\noalign{\vskip2pt}
-0.00111502&&0.00859434}. \label{3.32}
\end{gather}
Thus we get
\beq{3.33}
\bal
|a|&=(0.05938\pm 0.02032)\,{\rm AU}^{-2}\cr
|\ov b|&=(8.52151\pm 0.0927)\times10^{-10}\,\tfrac{\rm m}{\rm s^2}
\eal
\end{equation}
and a correlation parameter
\beq{3.34}
\rho=\frac{{\rm cov}(a,\ov b)}{\si_a\si_{\bar b}}=-0.60039.
\end{equation}
It is interesting to notice that parameters $a$ and $\ov b$ induce two scales
in the \SS
\bea{3.35}
r_0=\frac1{\sqrt{|a|}}&=&(4.103\pm 0.069)\,{\rm AU}\\
R=\sqrt{\frac{k^2}{|\ov b|}}&=&(2.640\pm0.0143)\times10^3\,{\rm AU}\label{3.36}
\end{eqnarray}
with the same correlation parameter.
$R$ is the order of the size of the \SS.

\begin{figure}
\centerline{\ing{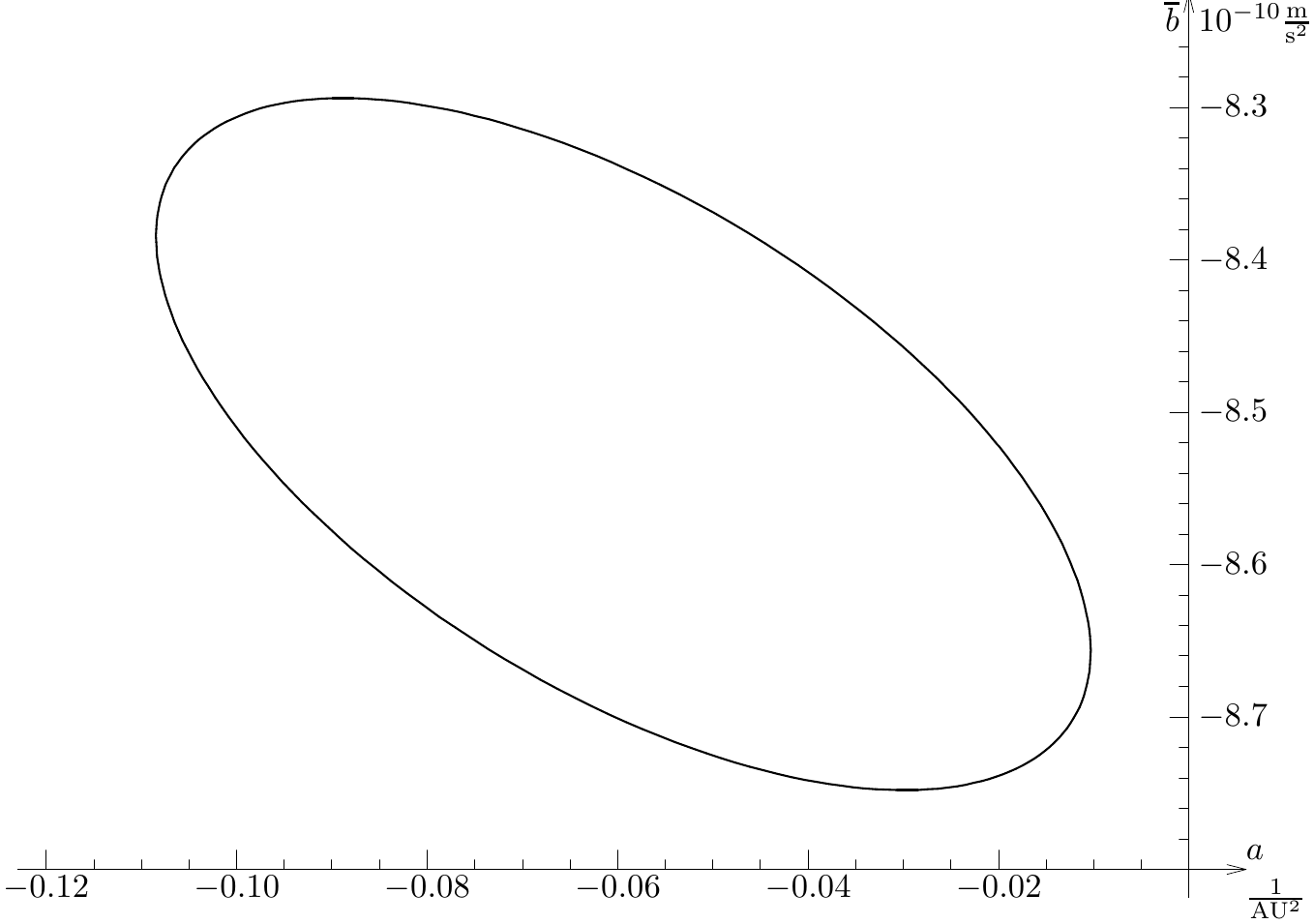}}
\caption{An ellipse of 95\% confidence for $a$ and $\ov b$ parameters.}\label{f2}
\end{figure}

The model for an \an\ \ac\ has been fitted on Fig.~\ref{f1}; on Fig.~\ref{f2} we fitted an
ellipse of 95\% confidence. Let us notice the following points. From this
model we get two \ct s: $\ov b$---an \ac\ and $r_0$---a~length. Probably in
a more refined model $r_0$ can be connected to some kind of cosmogonic scale.
The \ct~$\ov b$ is more interesting. Let us remind to the reader the
following facts from \co y. The Hubble \ct\ for our contemporary epoch is
equal to
\beq{3.38}
H_0=H(t_0)=h\cdot \ov H_0
\end{equation}
where
$$
H_0=\frac{100\,\frac{\rm km}{\rm s}}{1\,{\rm Mpc}}
$$
and $h$ is a dimensionless parameter.

\begin{figure}
\centerline{\ing{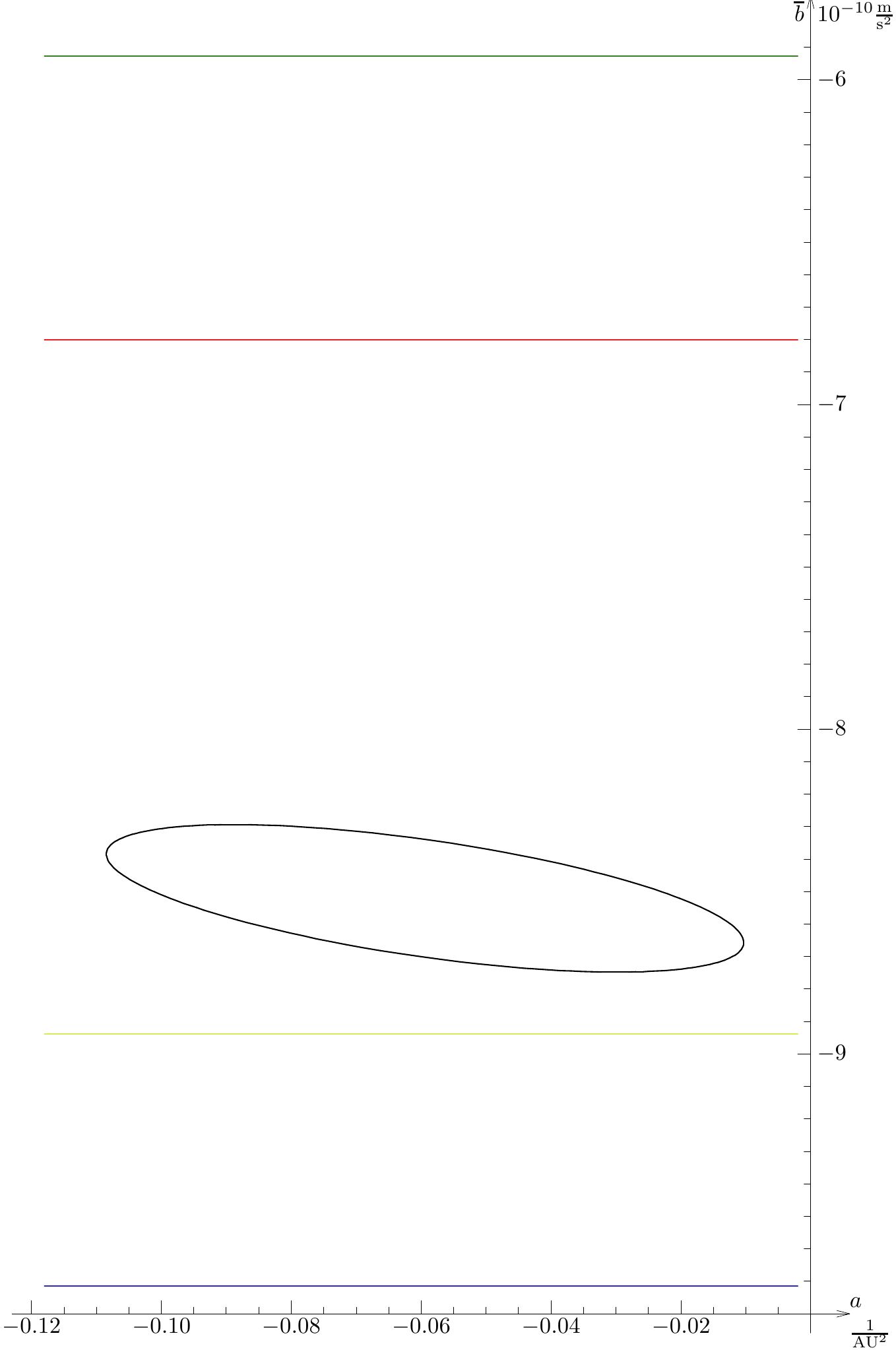}}
\caption{An ellipse of 95\% confidence of parameters $a$ and $\ov b$ among
values of \ac\ (with minus sign) obtained from several methods of obtaining
Hubble \ct\ (see text).}\label{f3}
\end{figure}

From various data we have different values of $h$. From WMAP data we have
$$
h=0.732\begin{matrix}{}+0.031\\{}-0.032\end{matrix}
$$
(see Ref.\ \cite{11}). This value has been obtained under some theoretical
assumptions (spatially flat Freedman Universe and a specific scenario of the
early Universe (a cold dark matter)). From Ia Supernovae $h=0.71\pm0.06$ (see
Ref.\ \cite{12}).
Moreover, recently we get from Planck's satellite the following value for~$H$
\bea{3.37n}
H&=&(67.4\pm1.4)\,\frac{\rm km/s}{\rm 1\,Mpc}\\
\hbox{or}\quad H&=&(68.0\pm1.4)\,\frac{\rm km/s}{\rm 1\,Mpc},\label{3.38n}
\end{eqnarray}
see Ref.~\cite{Planck} which is inconsistent with Ref.~\cite{12}.

Tully--Fisher relation $0.71\pm0.03\pm0.07$ (first statistical and second
systematic errors, Refs~\cite{13}, \cite{14}); $0.78\pm0.05\pm0.09$---Faber--Jackson
(see Ref.~\cite{15}, also first statistical and second
systematic errors); $0.81\pm0.06$---Tonry et al.\ (see Ref.~\cite{16}). Maybe
in the future all the values will be closer. Moreover, we calculate a value
$c\ov H_0$ ($c$~is a velocity of light). We get
\beq{3.39}
\ov b_0=c\ov H_0=9.716\times 10^{-10}\,\tfrac{\rm m}{\rm s^2}\,.
\end{equation}

Thus we have to do with some kind of an \ac\ of \co ical origin equal to
\beq{3.40}
b=hb_0.
\end{equation}
If we compare the value of our \ct~$\ov b=8.52\times 10^{-10}\,\frac{\rm
m}{\rm s^2}$, we easily notice that $\ov b$ can be equal to $hb_0$ for some
value of~$h$ and can be found in an interval of error. We plot our confidence
ellipse together with some extreme values of $hb_0$ in Fig.~\ref{f3}. The \ac\ \ct\
from Milgrom phenomenological theory (see Ref.~\cite{17}) is of the same
order. Finally we define a ratio of two scales
\beq{3.41}
\eta=\frac {\,\ov R\,}{r_0}=(0.643\pm0.014)\times10^{3}.
\end{equation}
In this way our model can be written in four ways:
\beq{3.42}
\bal
a_p&=\frac{|b|}{|a|r^2}\bigl(\ex{-ar^2}-1+ar^2\bigr)\cr
&=\frac{k^2}{\ov R{}^2}\Bigl(\frac{r_0}r\Bigr)^2\biggl(\exd{-\X2(\frac r{r_0}\Y2)^2}
-1+\Bigl(\frac r{r_0}\Bigr)^2\biggr)\cr
&=\frac{k^2}{\eta^2 r^2}\biggl(\exd{-\X2(\frac r{r_0}\Y2)^2}
-1+\Bigl(\frac r{r_0}\Bigr)^2\biggr)\\
&=\frac{{k'}^2}{r^2}\biggl(\exd{-\X2(\frac r{r_0}\Y2)^2}
-1+\Bigl(\frac r{r_0}\Bigr)^2\biggr)
\eal
\end{equation}
where ${k'}^2=\frac{k^2}{\eta^2}$.

\begin{figure}[h]
\hbox to \textwidth{\includegraphics[width=0.45\textwidth]{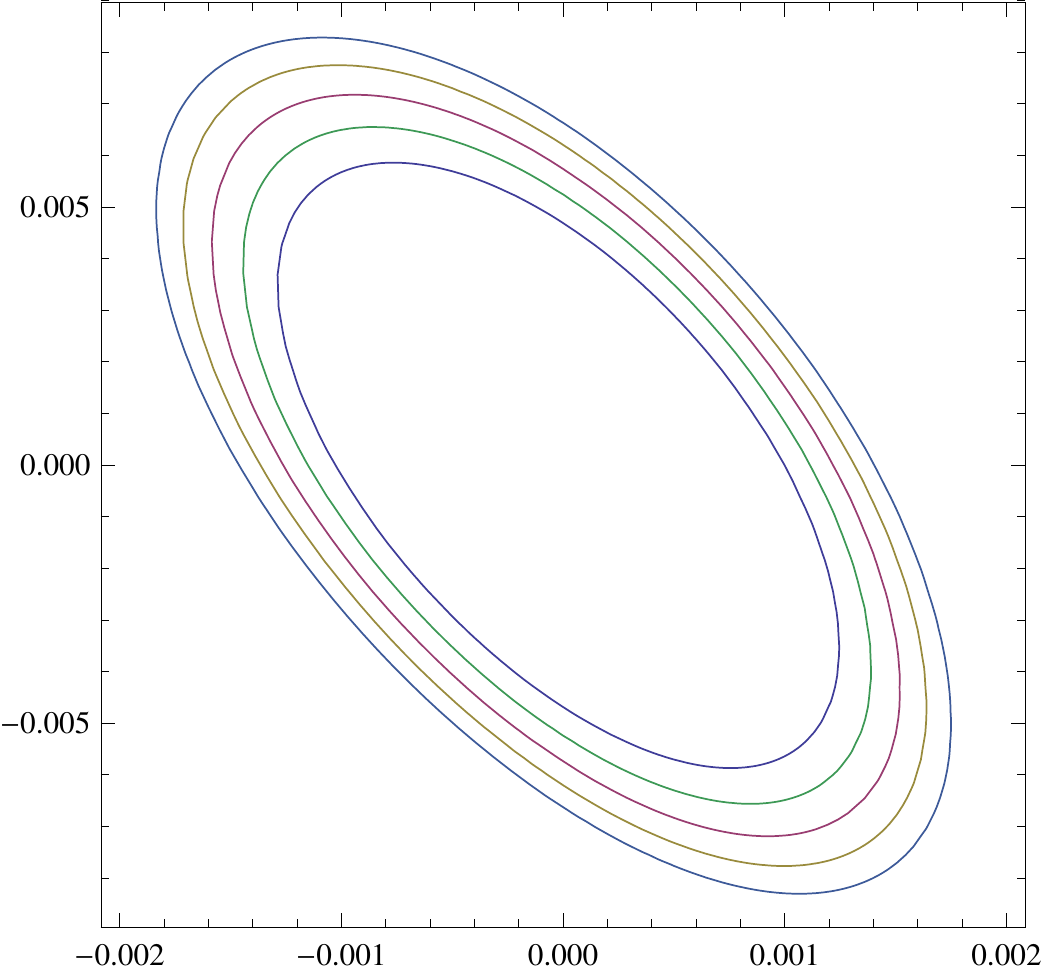}\hfil
\includegraphics[width=0.45\textwidth]{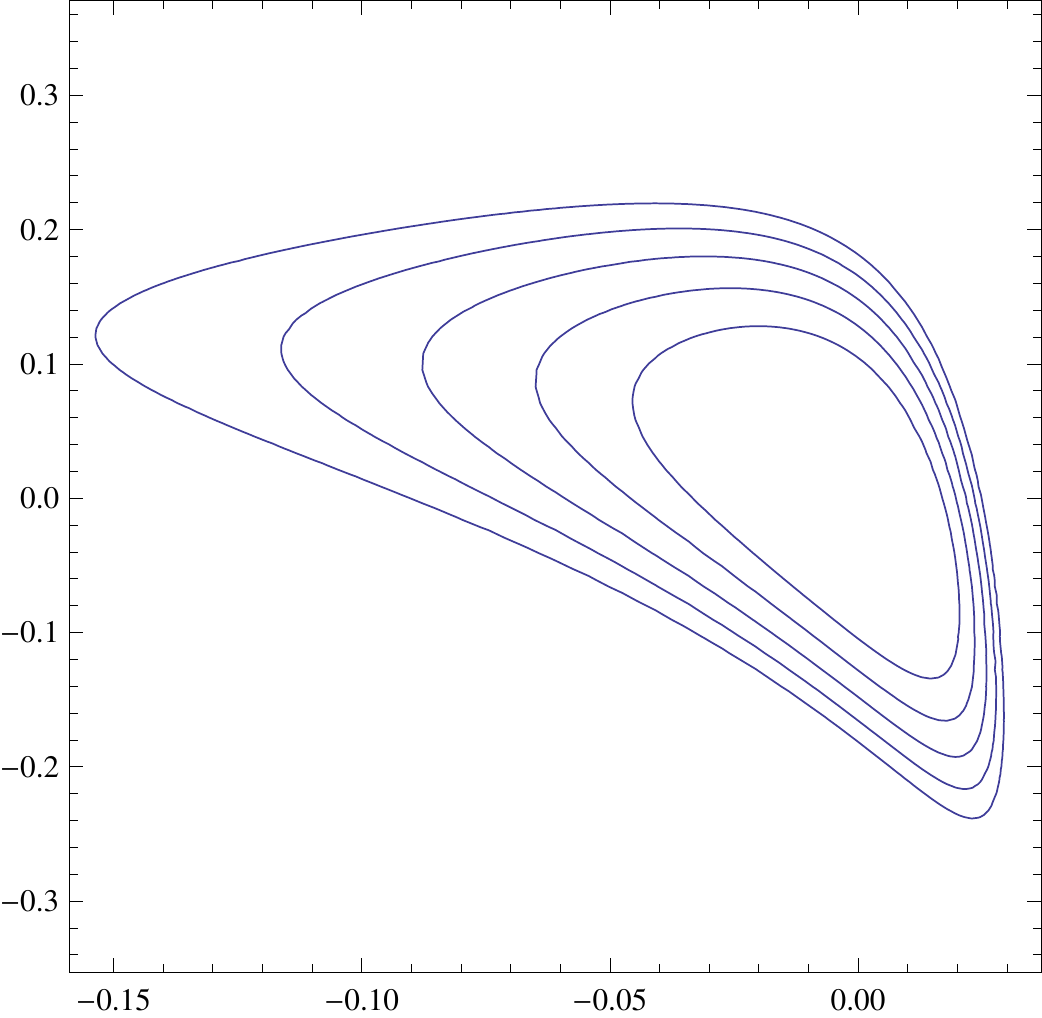}}
\hbox to \textwidth{\hbox to 0.45\textwidth{\hfil(A)\hfil}\hfil
\hbox to 0.45\textwidth{\hfil(B)\hfil}}
\caption{Contour plots of $f(x,y)=c$. (A)---small parameters, (B)---large
parameters (see a text).}\label{f8}
\end{figure}

Let us consider the \f
\beq{3.41a}
f(x,y)=2\X1(M(a+x,\ov b+y)-M(a,\ov b)\Y1)
\end{equation}
where $a=-0.0593$, $\ov b=-8.52$ (see Eq.~\er{3.29}). Let us calculate a
contour plot of $f(x,y)=c>0$ for several values of the parameter~$c$. On
Fig.~\ref{f8} we give contour plots for small values of $c=0.004, 0.005,
0.006, 0.007, 0.008$ (A) and for large values of~$c=1.991, 2.991, 3.991,
4.991, 5.991$ (B). In the first case they are almost ellipses. In the second
case they are distorted. Moreover, if we use a contour plot for $c=5.9914$,
we can still read errors for $a$ and $\ov b$ of the same order as for an
asymptotic case (an ellipse of 95\%~CL) even the region of 95\% confidence is
different and an error for~$a$ is asymmetric. One gets
\bea{3.42a}
|a|&=&\X3(0.05913\bal{}&+0.015\\{}&-0.071\eal\Y3)\tfrac1{\rm AU^2}\\
|\ov b|&=&\X1(8.52151\pm0.09\Y1)\times 10^{-10}\tfrac{\rm m}{\rm s^2}.
\label{3.43a}
\end{eqnarray}

95\% CL corresponds to $2\si_a$ and $2\si_{\bar b}$ and is consistent
with an asymptotic covariance analysis (see Fig.~\ref{f2}). To be honest, we
give a programme written in Mathematica~7 to calculate $f(x,y)$ with a result
of calculations in Appendix~C.

Let us consider $2M_{\min}=2M(a,\ov b)$ for $a=-0.05938$, $\ov b=-8.52151$.
One gets $2M_{\min}=46.35$. The last value is a $\chi^2$ for $n=19-2=17$
degrees of freedom,
\beq{3.44a}
\frac{2M_{\min}}{17}=\frac{\chi_{n=17}^2}{17}=2.73, \q
\sqrt{\frac{\chi_{n=17}^2}{17}}=1.6.
\end{equation}
The last value is greater than 1 and it means that mean fluctuations of
Anderson et al.\ data are quite large. Let us notice that
$E(\chi_{n=17}^2)=17<46.35$ ($E$ is an expectation value). One can easily
calculate a probability:
\bml{3.44}
P(\chi_{n=17}^2\ge 2M_{\min})=\frac1{\G(17/2)2^{17/2}}
\int_{2M_{\min}}^\iy u^{-17/2}e^{-u/2}\,du\\
{}=1-\frac1{\G(17/2)2^{17/2}}
\int^{2M_{\min}}_0 u^{-17/2}e^{-u/2}\,du\simeq 0.000153.
\end{multline}
Thus we get large fluctuations of data and low probability of a
statistical hypothesis. It means we need more data to conclude. In this way
we need a feedback from observations.

Moreover, Anderson et al.\ data cannot be considered as measurements
and because of it a low probability of a statistical hypothesis is not conclusive
for a model. They are residual \ac s.

Let us notice that the model has some limits of an application. This is a
length scale 10\,Mpc. Moreover, in order to keep an \ap  ion on a reasonable
level we should apply the model up to $10^{10}\,\rm AU$ only. For such a length
scale an \ap ion procedure for Eq.~\er{3.1} should work correctly.

In all sections starting from Section 4 (except Section~8) we use for
simplicity in place of $|\ov b|$ simply $b$.

The problem of an \an\ \ac\ of \P0/11 \sp s has a vast literature to obtain
an explanation using several approaches. They are conventional, non-\gr al
(see Refs \cite{N20}--\cite{N25}). In all papers cited above the authors are
trying to evaluate the force acting on a \sp\ coming from thermal radiation.
The idea is very simple. Every \elm\ radiation emitted from sources carries
energy and momentum. Thus it is enough to find such sources in the \sp\ and
evaluate the force afterwards to compare it with an \an\ \ac. However, the
problem is very complex. First of all it is necessary to divide the \sp\ into
some small elements (it means, to use thermal finite element method).
Secondly to model ways of rays. It means to use a ray-tracing algorithm. This
procedure demands knowledge of some details of an interaction of a radiation
with several surfaces in the \sp. The last point is very crucial for we need
to estimate a momentum transfer. They are using some phenomenological
approaches known in computer graphics which have no physical justification
(I~write on it in Conclusions). For this they cannot obtain
from their analysis (which is correct except the last point) reliable results
on \ac\ caused by thermal effects.

The Pioneer effect (Pioneer Anomaly---PA) should have an influence on the
motion on major bodies in the \SS\ (see Refs \cite{N26}--\cite{N28}. Moreover
the motion of those bodies are not so well known with sufficient precision
during a sufficiently long time in the region between 20 and 100~AU. Even the
inverse square law of gravity is poorly probed by experimental tests at
distances of $\sim10$~AU. Thus there is a place for an \an\ \ac\ for in any
application we are using Gauss \e s to get secular changing of orbital
elements of an elliptic orbit of a body in the Sun \gr al field (Newtonian
one). Even a second Newton law of dynamics is poorly probed for a very small
\ac. This is a place
for another approach to the problem---MOND dynamics \cite{17}. In the case of
small \ac\ even minor planets cannot help us (see \cite{N29}). Thus some
arguments from Refs \cite{N30}--\cite{N35} can be applied after a full
simulation of major bodies in the \SS\ with sufficient precision during a
long time not only for future but also in past. I~do not mean here a
perturbation calculus in celestial mechanics but a real computer simulation
of \e s of motion with all important interactions. Only such an approach can
falsificate an \an\ \ac\ as some artefacts or different effects connecting to
an interpretation of Doppler tracking data (see Ref.~\cite{N36}) in a
conformal \co y coming from conformal gravity with different field \e s.
In Ref.~\cite{N36} the author considers an effect of conformal \co y obtained
from conformal gravity which forced us to reinterpret radio-tracking data as
an additional conformal Doppler effect without an \an\ \ac. It
is also possible to reinterpret Doppler tracking data using nonlinear
electrodynamics (see Ref.~\cite{N37}). Let us notice that if we have
nongeodesic motion of \sp s, it means, an \an\ \ac\ is a non-metric effect in
relativistic theory of \gr, we can avoid some conclusions from \cite{N35} for
in our approach (see Section~9) it is an external (non-geodesic) effect. In
Refs \cite{N38}--\cite{N39} it is proved that in standard \co y based on
General Relativity it is impossible to get an \an\ \ac\ of \P0/11 from \co
ical expansion.

Let us notice that an estimation of a Lense--Thirring effect (\cite{N40}) cannot rule out
the \gr al origin of \P0/11 effect for a reason that we need a full
simulation of bodies in the \SS\ with position results (the \ph\ movement of
Uranus is not measurable---only a position is measurable). In Ref.~\cite{N41}
the author considers orbital effects of a time-dependent Pioneer-like \an\
\ac\ using new radio-tracking data. Moreover, a different interpretation is
possible which we quoted in Conclusions.

In Ref.~\cite{N42} the author raised an interesting problem to consider the
Neptunian system of satellites under an influence of the Pioneer \an\ \ac. In
our approach it would be necessary to find an analogous model of \ac\ for the
Neptunian system  as for the \SS, which is beyond the scope of our paper (it
would be necessary to add both \ac s). The problem raised in Ref.~\cite{N43}
is also very interesting (i.e.\ velocity dependent forces). Moreover, in this
case we should also give a full numerical simulation of movements with
calculated real (measurable) positions. In our case, our forces are velocity
independent.

For thermal effects explanation see also recent Ref.~\cite{N45}. For EPM and
Relativity see Ref.~\cite{N46}.

\section{A \ph\ shift and a distortion of elliptic orbits}
Let us apply an \an\ \ac\ to the planet motion in the \SS. In order to do
this we consider a Binet formula (see Refs \cite{18}, \cite{19}) for a total
\ac, i.e.\ a \gr al newtonian \ac\ caused by an attraction of the Sun plus an
\an\ \ac. The additional \ac\ corresponds to a central force and a curve of
motion is planar.

One gets in polar \cd s in a plane of motion
\beq{4.1}
a_r=-\frac{A^2}{r^2}\Bigl(\pz{^2}{\vf^2}\Bigl(\frac1r\Bigr)+\frac1r\Bigr),
\end{equation}
where $A=r^2\dot\vf={\rm const.}$,
\beq{4.2}
a_r=-\frac{k^2}{r^2}-\frac b{ar^2}\bigl(\ex{-ar^2}-1+ar^2\bigr).
\end{equation}
We consider a motion in two different sectors for $r<20\,{\rm AU}$ and for
$r>20\,{\rm AU}$. In order to do this we expand an exponential \f\ in an \an\
\ac
\beq{4.3}
a_p=-\frac b{ar^2}\X2(1-ar^2+\frac{(ar^2)^2}2+\ldots-1+ar^2\Y2)
\simeq -\frac{bar^2}2\,.
\end{equation}
We find an asymptotic behaviour of $a(r)$ and we get
\beq{4.4}
a_p=-b.
\end{equation}
We apply formula \er{4.3} for $r<20\,{\rm AU}$ and formula \er{4.4} for
$r>20\,{\rm AU}$. It is easy to see that they are very good \ap ion in these
regions (see Fig.~\ref{f1}). One gets from Binet formula in both cases.
\refstepcounter{equation}\label{4.5}
$$
\dsl{
\hfill \pz{^2u}{\vf^2}+u=\frac{k^2}{A^2}+\frac{ba}{2A^2u^4} \hfill
(\theequation\rm a)\cr
\hfill \pz{^2u}{\vf^2}+u=\frac{k^2}{A^2}+\frac{b}{A^2u^2} \hfill
(\theequation\rm b)}
$$
where $u=\frac1r$.

The \ct\ $b$ is very small and for this we can consider a motion as a
keplerian motion plus a small distortion, i.e.
\beq{4.6}
u=u_0+\D u, \q u_0=\frac{k^2}{A^2}(1+e\cos\vf),
\end{equation}
where $e$ is an eccentricity of an elliptic orbit ($e<1$) and $\D u$ is a
deviation from the orbit caused by an \an\ \ac. For $|\D u|\ll1$ we get in
both cases
\refstepcounter{equation}\label{4.7}
$$
\dsl{
\hfill \pz{^2\D u}{\vf^2}+\D u=B_1(1+e\cos\vf)^{-4}\hfill
(\theequation\rm a)\cr
\hfill \pz{^2\D u}{\vf^2}+\D u=B_2(1+e\cos\vf)^{-2}\hfill
(\theequation\rm b)}
$$
where
\beq{4.8}
B_1=\frac{baA^6}{2k^8}, \q B_2=\frac{bA^2}{k^4}\,.
\end{equation}
In both cases we use an \e\ for undisturbed orbit
\beq{4.9}
\pz{^2u_0}{\vf^2}+u_0=\frac{k^2}{A^2}
\end{equation}
and we put $u=u_0$ on the right-hand side of \e s.

In order to solve both \e s we develop both right-hand sides into Fourier
series, i.e.
\refstepcounter{equation}\label{4.10}
$$
\dsl{
\hfill (1+e\cos\vf)^{-2}=\frac{b_{02}}2 + \sum_{n=1}^\iy b_{n2}\cos n\vf
\hfill (\theequation\rm a)\cr
\hfill (1+e\cos\vf)^{-4}=\frac{b_{01}}2 + \sum_{n=1}^\iy b_{n1}\cos n\vf
\hfill (\theequation\rm b)}
$$
where
\refstepcounter{equation}\label{4.11}
$$
\dsl{
\hfill b_{n2}=\frac1\pi \int_{-\pi}^\pi \frac{\cos n\vf}{(1+e\cos\vf)^2}
\,d\vf=\frac2\pi \int_0^\pi \frac{\cos n\vf}{(1+e\cos\vf)^2}
\,d\vf \hfill (\theequation\rm a)\cr
\hfill b_{n1}=\frac1\pi \int_{-\pi}^\pi \frac{\cos n\vf}{(1+e\cos\vf)^4}
\,d\vf=\frac2\pi \int_0^\pi \frac{\cos n\vf}{(1+e\cos\vf)^4}
\,d\vf \hfill (\theequation\rm b)}
$$
$n=0,1,2,\dots$ (both \f s are odd).

For every component of a Fourier expansion we get in both cases
\bea{4.12}
\pz{^2u_{ni}}{\vf^2}+u_{ni}&=&K_{ni}\cos n\vf, \q n=1,2,\dots\\
K_{ni}&=&B_i\cdot b_{ni}\nn\\
\pz{^2u_{0i}}{\vf^2}+u_{0i}&=&K_{0i}, \label{4.13}\\
K_{0i}&=&\frac{b_{0i}}2\,B_i,\q i=1,2,\nn
\end{eqnarray}
where
\beq{4.14}
u_i=\frac{u_{0i}}2+\sum_{n=1}^\iy u_{ni}\cos n\vf.
\end{equation}
We are interested only in a particular integral of Eqs (\ref{4.7}ab).

One easily gets particular integrals for Eqs (\ref{4.12}--\ref{4.13}):
\bea{4.15}
u_{ni}&=&\frac{K_{ni}}{1-n^2}\cos n\vf, \q n=2,3,\dots\\
u_{0i}&=&K_{0i},\q i=1,2.\label{4.16}
\end{eqnarray}
However, the case $n=1$ has a peculiar character. In this case one gets
\beq{4.17}
u_{1i}=\frac{B_ib_{1i}}2\,\vf\sin\vf.
\end{equation}
The \so\ \er{4.17} gives a \ph\ shift of the orbit and Eqs
(\ref{4.15}--\ref{4.16}) lead to a distortion of the orbit.

One gets
\bea{4.18}
b_{11}&=&-\frac{e(4+e^2)}{2(1-e^2)^{7/2}}\\
b_{12}&=&-\frac e{2(1-e^2)^{3/2}}\label{4.19}
\end{eqnarray}
and
\bea{4.20}
u_{1i}&=&\frac{k^2}{A^2}\,e\a_i\vf\sin\vf\\
\a_1&=&-\frac{4+e^2}{8(1-e^2)^{7/2}}\,\frac{baA^8}{k^{10}}\label{4.21}\\
\a_2&=&-\frac b{8(1-e^2)^{3/2}}\,\frac{A^4}{k^8}\,.\label{4.22}
\end{eqnarray}

A \ph\ shift is given by the formula
\beq{4.23}
\D\vf_i=2\pi e\a_i.
\end{equation}
Thus one gets
\refstepcounter{equation}\label{4.24}
$$
\dsl{
\hfill \D\vf_1=-\frac{\pi e(4+e^2)baA^8}{4(1-e^2)^{7/2}k^{10}}
=-\frac{\pi e(4+e^2)baa_e^4(1-e^2)^{1/2}}{8k^2} \hfill (\theequation\rm a)\cr
\hfill \D\vf_2=-\frac{\pi eb}{2(1-e^2)^{3/2}}\,\frac{A^4}{k^6}
=-\frac{\pi eb(1-e^2)^{1/2}a_e^2}{k^2} \hfill (\theequation\rm b)}
$$
where we use the formula
\beq{4.25}
A^2=pk^2=a_e(1-e^2)k^2,
\end{equation}
$a_e$ is a major semiaxis of an ellipse.

A relativistic \ph\ shift is given by the formula (see Ref.~\cite{20})
\beq{4.26}
\D\vf_{\rm rel}=\frac{6\pi k^2}{a_e(1-e^2)c^2}\,.
\end{equation}
Let us compare our \ph\ shift with the relativistic one:
\beq{4.27}
\eta_i=\frac{\D\vf_i}{\D\vf_{\rm rel}}\,.
\end{equation}
One gets
\refstepcounter{equation}\label{4.28}
$$
\dsl{
\hfill \eta_1=-\frac{(4+e^2)ea_e^5(1-e^2)^{3/2}bac^2}{48k^4}
\simeq -1.6 e(4+e^2)(1-e^2)^{3/2}a_e^5\times10^{-2}
\hfill (\theequation\rm a)\cr
\hfill \eta_2=-\frac{(1-e^2)^{3/2}ea_e^3bc^2}{12k^4}
=-1.2e(1-e^2)^{3/2}a_e^3 \hfill (\theequation\rm b)}
$$
where $a_e$ is measured in AU.

A \ph\ shift by an \an\ \ac\ has reverse sign in comparison to the
relativistic one. It is a regress, not advance, i.e.
\beq{4.49}
\dot \o_i=-\dot\o_{\rm rel}|\eta_i|, \q i=1,2.
\end{equation}
If we take the greatest relativistic \ph\ advance in the \SS---\ph\ advance
of the Mercury---we get that
\beq{4.50}
\bga
|\dot\o_i|<\dot\o_{\rm Mercury}|\eta_i|\cr
\dot\o_{\rm Mercury}=43''\hbox{ per century.}
\ega
\end{equation}
It means that $\dot\o_i$ is unmeasurable using contemporary techniques.

It is interesting to find a value of an \an\ \ph\ movement for Saturn. Using
Eqs (\ref{4.24}a), (\ref{4.28}a) ($a_e<20\,{\rm AU}$) one gets
\beq{4.31}
\dot\o_{\rm Saturn}=-3.84''\hbox{ per century}
\end{equation}
with an error estimation more than 50\%. If we compare it with the recent
result for Saturn (see Ref.~\cite{Jono}, also Ref.~\cite{26*}), it is too big.
A~recent data reanalysis yields a new uncertainty of 0.43 miliseconds per
century.

Now let us consider a distortion of an elliptic orbit in the \SS\ caused by
an \an\ \ac. We can write
\beq{4.51}
u_i=u_0+\frac{B_ib_{0i}}2+B_i\sum_{n=2}^\iy \frac{b_{ni}}{1-n^2}\cos n\vf
\end{equation}
where $u_0$ is given by the formula \er{4.6}
\refstepcounter{equation}\label{4.52}
$$
\dsl{
\hfill b_{01}=\frac{2+3e^2}{2(1-e^2)^{7/2}} \hfill (\theequation\rm a)\cr
\hfill b_{02}=\frac1{(1-e^2)^{3/2}} \hfill (\theequation\rm b)}
$$

One gets
\refstepcounter{equation}\label{4.53}
$$
\dsl{
\hfill |b_{n1}|=\X3|\frac2\pi \int_0^\pi \frac{\cos n\vf}{(1+e\cos\vf)^4}
\,d\vf\Y3| \le \frac2\pi \int_0^\pi \frac{\left|\cos n\vf\right|}{(1+e\cos\vf)^4}\,d\vf
\le \frac2\pi \int_0^\pi \frac{d\vf}{(1+e\cos\vf)^4}=b_{01}
\hfill (\theequation\rm a)\cr
\hfill |b_{n2}|\le \frac2\pi \int_0^\pi \frac{\left|\cos n\vf\right|}{(1+e\cos\vf)^2}\,d\vf
\le b_{02}.\hfill (\theequation\rm b)}
$$
Thus
\bg{4.54}
|b_{ni}|\le b_{0i}\\
u_i=\frac{k^2}{A^2}\X3(1+e\cos\vf+\frac{A^2B_ib_{0i}}{2k^2}
+\frac{A^2}{k^2}\sum_{n=2}^\iy \frac{b_{ni}}{1-n^2}\cos n\vf\Y3).\label{4.55}
\end{gather}
One obtains
\bml{4.57}
\X3|\sum_{n=2}^\iy \frac{b_{ni}}{1-n^2}\cos n\vf\Y3| \le
\sum_{n=2}^\iy \frac{|b_{ni}|}{n^2-1}\left|\cos n\vf\right|\le
\sum_{n=2}^\iy \frac{b_{0i}}{n^2-1}\\ {}\le
\sum_{n=2}^\iy \frac{3b_{0i}}{n^2}<3b_{0i}\sum_{n=1}^\iy \frac1{n^2}
=b_{0i}\frac{\pi^2}2
\end{multline}
and
\bml{4.58}
\frac{B_iA^2}{k^2}\X3|\frac{b_{0i}}2+\sum_{n=2}^\iy \frac{b_{ni}\cos n\vf}
{1-n^2}\Y3|
\le \frac{B_iA^2}{k^2}\X3(\frac{b_{0i}}2
+\sum_{n=2}^\iy \frac{b_{0i}\left|\cos n\vf\right|}{n^2-1}\Y3)
\le \frac{B_iA^2b_{0i}(\pi^2+1)}{2k^2}=\ve_i,\\ \q i=1,2.
\end{multline}

For $i=1$ one gets
\refstepcounter{equation}\label{4.59}
$$
\dsl{
\hfill \ve_1=\frac{\pi^2+1}8 \sqrt{1-e^2}\,(2+3e^2)a_e^4 \,\frac{ba}{k^2}
\simeq 1.02(2+3e^2)\sqrt{1-e^2}\,a_e^4\times 10^{-8}.
\hfill (\theequation\rm a)}
$$
For $i=2$ one finds
$$
\dsl{
\hfill \ve_2=\frac{\pi^2+1}2\sqrt{1-e^2}a_e^2\,\frac b{k^2}\simeq 7.6a_e^2
\sqrt{1-e^2}\times 10^{-7}.
\hfill (\theequation\rm b)}
$$
One finally obtains
\refstepcounter{equation}\label{4.60}
$$
\dsl{
\hfill |u_1-u_0|<1.02(2+3e^2)\sqrt{1-e^2}\,a_e^4\times 10^{-8}
\hfill (\theequation\rm a)\cr
\hfill |u_2-u_0|<7.6a_e^2 \sqrt{1-e^2}\times 10^{-8}.
\hfill (\theequation\rm b)}
$$

In this way we can get a distortion of the major semiaxis $\D a_e$ in both
cases. We calculate
\bg{4.61}
a_{e\,\rm distorted}=\frac{r_{\max}+r_{\min}}2 \simeq \frac{a_e(1-e^2)}2
\X2[\frac1{1-e\pm \ve_i}+\frac1{1+e\pm\ve_i}\Y2]=a_e(1+\ve_i)\\
\D a_e=a_{e\,\rm distorted}-a_e=a_e\ve_i\nn\\
\frac{\D a_e}{a_e}=\ve_i=
\begin{cases}
21(2+3e^2)\sqrt{1-e^2}\,a_e^4\ (\mu\rm arcs), & i=1,\\
156.8\sqrt{1-e^2}\,a_e^2\ (\mu\rm arcs), & i=2.
\end{cases} \label{4.62}
\end{gather}

In this way we prove that an influence of an \an\ \ac\ on planetary orbits in
the \SS\ is negligible.

Some authors consider some additional contribution to the \ph\ precession and
some possibilities to measure it (see Refs \cite{N47}--\cite{N51}), i.e.\
Lense--Thirring effect, Solar oblateness etc. Using Reissner--Nordstr\"om
\so\ in the place of Schwarzschild \so, it is possible to take under
consideration an electric charge of the Sun and a planet (see Refs
\cite{N52}--\cite{N54}). However, it would be necessary to consider charged
bodies in astrophysics and electrostatic interactions which are stronger than
\gr al ones. It happens that some authors consider more exotic
effect caused by extra-\ac s of order $r^k$, $k=2$, similar to our \er{4.3}
and \er{4.4} which they obtained perturbatively (see Refs
\cite{N55}--\cite{N56}). There are also some papers for a \ph\ precession
caused by central extra-forces (see Refs \cite{N57}, \cite{N58}). Some authors
calculated also \ph\ precession caused by Pioneer-like extra-\ac\ such as
Eq.~\er{4.4} (see Refs \cite{N26}, \cite{N59}). All these results can be
compared also to Ref.~\cite{N61} where the authors consider the possibility
to obtain an \an\ \ac\ from dark matter in the \SS\ versus a modification of
\gr al inverse-square law. Our method to calculate a \ph\
advance (or regress) is different and in agreement with methods cited here.
We have a different model of an \an\ \ac. We calculated also a distortion of
the orbit in terms of changing of parameters of an elliptic orbit (i.e.\
semimajor axis). It does not mean that the energy is not conserved. Our
orbits are not ellipses and everybody knows that orbits in astronomy are not
ellipses too. Our additional force is a central force and does not change any
elements of a Kepler orbit except a shape which can be translated to \ph\
movement and a distortion.

Let us continue our considerations about \SS s, their experimental bounds
and some theoretical predictions, either rooted in~GR and in alternative
models. In Refs \cite{N63}--\cite{N64a} we have a state of arts in observational
and experimental status. In Refs \cite{N65a}--\cite{N68a} the authors consider
some alternative models with Yukawa parametrization, logarithmic correction,
a power-low parametrization getting some upper limits. In Ref.~\cite{N69a}
a gravitomagnetic field of the Sun gives new constraints on a Yukawa-like
fifth force from planetary data. Also some ingredients from OPERA superluminal
neutrons phenomenology (see Ref.~\cite{N70a}) can give some constraints on
orbital motion around the Earth. GRACE geopotential models can give novel
considerations about the error budget of the LAGEOS-based tests of
frame-dragging (see Ref.~\cite{N71a}). In Ref.~\cite{N72a} the authors consider
Weyl conformastatic perihelion advance. Ref.~\cite{N73a} is devoted to some
bounds in \SS\ for an extra \ac\ of $f(R,T)$, where $f(R,T)$ is a
generalization of GR Lagrangian with $R$ being a Ricci scalar and $T$ a trace
of an energy momentum tensor. The body problem with the \co ical \ct\
is also considered giving some observational constraints (see Ref.~\cite{N74a}).

There is a body of work in this domain which could be
incorporated in our approach (see Ref.~\cite
{N26}). We mean a problem of
Neptune satellite motion. An \an\ \ac\ according to our investigations can be
applied also for systems different than the \SS. It means, for a satellite of
Jupiter, Saturn, Uranus, Neptune. Thus we can investigate an \an\ \ac\ model
for these systems. Moreover, in this case values of parameters $a_s$ and $b_s$
are uncertain ($s$ means the system $(\rm J,S,U,N)$).

Moreover, let us notice the following fact. In Eq.\ \er{3.15}, \er{3.42} and
\er{4.3} we have a mass
of the Sun $M_\odot$. Moreover, in a definition of the \ct\ $b$ this quantity
has been absorbed by an integration \ct\ (initial conditions). Moreover, we
can pose a conjecture that $b$ is proportional to the mass of the Sun
(a~central body). In this way if we apply an \an\ \ac\ model for systems of
satellites of planets in the \SS\ supposing that an \an\ \ac\ in the system
is proportional to a mass of a central body. In this way we get
\beq{4.43}
\bal
b_J&=\frac{M_J}{M_\odot}\,b_\odot&&=9.54 \t 10^{-4}b_\odot\cr
b_S&=\frac{M_S}{M_\odot}\,b_\odot&&=2.85 \t 10^{-4}b_\odot\cr
b_U&=\frac{M_U}{M_\odot}\,b_\odot&&=4.35 \t 10^{-4}b_\odot\cr
b_N&=\frac{M_N}{M_\odot}\,b_\odot&&=5.17 \t 10^{-4}b_\odot
\eal
\end{equation}
where $b_\odot=b$ and $M_J,M_S,M_U,M_N$ are masses of Jupiter, Saturn, Uranus
and Neptun. Thus a conjecture connects initial conditions for the field $\vf
(r,t)$ in the \SS\ (with the Sun as a central body) to the initial conditions
of some subsystems in the \SS. In the case of the length scale $a_s$
(or~$r_{0s}$)---an analogue of~$a$ in our model of an \an\ \ac---there are no
suggestions and the simplest conjecture tells us: the \an\ scale parameter in these
systems is \ct, $a_s=a_\odot=a$. This working conjecture should be considered for a motion of
satellites in the mentioned systems. Isaac Newton said: \ti{Hypotheses non
fingo}. Moreover, this is only a working conjecture.

Let us consider a different conjecture. In this conjecture we put
\beq{4.44}
\bal
r_J&=\sqrt{\frac{M_J}{M_\odot}}\,r_\odot=3.08\t 10^{-2}r_\odot\cr
r_S&=\sqrt{\frac{M_S}{M_\odot}}\,r_\odot=1.68\t 10^{-2}r_\odot\cr
r_U&=\sqrt{\frac{M_U}{M_\odot}}\,r_\odot=2.08\t 10^{-2}r_\odot\cr
r_N&=\sqrt{\frac{M_N}{M_\odot}}\,r_\odot=2.27\t 10^{-2}r_\odot
\eal
\end{equation}
where $r_\odot=r_0$ is a model scale for the Sun.

This conjecture can be applied for the satellites of gigant planets.
Moreover, the simplest conjecture keeps $b$ and $a$ \ct. In this way we get
using our model on a border of their satellites region (see Ref.~\cite{71})
\beq{4.45}
\bal
b_J&=1.6\t 10^{-3}b_\odot\cr
b_S&=1.1\t 10^{-3}b_\odot\cr
b_U&=0.61\t 10^{-3}b_\odot\cr
b_N&=1.7\t 10^{-3}b_\odot
\eal
\end{equation}
$b_\odot=b$ (for the Sun).

\section{An influence of an \an\ \ac\ on a \hy\ orbit}
In the case of a central potential we can reduce a motion of a point mass to
two \e s in a plane of motion. In polar \cd s (see \cite{18}, \cite{19}) one
gets
\bea{5.1}
\pz r\vf &=& \pm \frac{r^2}A \sqrt{2U(r)+2H -\frac{A^2}{r^2}}\\
\pz rt &=& \pm \sqrt{2U(r)+2H -\frac{A^2}{r^2}}\label{5.2}
\end{eqnarray}
where as before $A=r^2\vf$, $r$ and $\vf$ are polar \cd s in a plane of
motion. $H$~is a total energy per unit mass.

In our case
\beq{5.3}
U(r)=-\frac{k^2}r +\ov U(r)
\end{equation}
where $\ov U(r)$ is a potential corresponding to an \an\ \ac
\bml{5.4}
\ov U(r)=-\int_0^r \frac b{a\rho^2} \X1(\ex{-a\rho^2}-1+a\rho^2\Y1)\,d\rho+c\\
=-\frac b{ar}\X1(\ex{-ar^2}+1+ar^2+\sqrt\pi \Erf(\sqrt a\,r)\Y1)+c,
\end{multline}
where $\Erf$ is an error \f\
$$
\Erf(z)=\frac2{\sqrt \pi}\int _0^z \ex{-t^2}\,dt
$$
and $c$ is a \ct\ which we put to zero, $\ov U(0)=0$.

In the case of a \hy\ orbit we consider the case $\rho>20\,{\rm AU}$. In our
terminology from Section~4 it means $i=2$ $(\frac\pi2<\vf<\arccos(-\frac1e)$).
Thus in this case
\beq{5.5}
\ov U(r)=-br.
\end{equation}

Let us consider Eq.\ \er{5.1}. Let us suppose that we have to do with a \hy\
motion ($e>1$) distorted by an additional potential, which is small. One gets
\beq{5.6}
r=r_0+r_1
\end{equation}
where $r_0$ is a \so\ for a \hy\ motion and $r_1$ is a small perturbation.

One gets
\bea{5.7}
\pz r\vf&=&\frac{r^2}A \sqrt{-\frac{2k^2}r+2H-\frac{A^2}{r^2}}\,
\X3(1-\frac{br}{2(-\frac{2k^2}r+2H-\frac{A^2}{r^2})}\Y3)\\
\pz {r_1}\vf&=&-\frac{r_0^2b(r_0+r_1)}{2A\sqrt{-\frac{2k^2}{r_0}+2H
-\frac{A^2}{r_0^2}}}\label{5.8}\\
\pz{r_0}\vf&=&\frac{r_0^2}A{\sqrt{-\frac{2k^2}{r_0}+2H-\frac{A^2}{r_0^2}}}
\label{5.9}
\end{eqnarray}
Eq.\ \er{5.9} is an \e\ for unperturbed \hy\ orbit. Thus we get
\beq{5.10}
\pz{r_0}\vf\cdot\frac A{r_0^2}=\sqrt{-\frac{2k^2}{r_0}+2H
-\frac{A^2}{r_0^2}}\,.
\end{equation}
Using Eq.\ \er{5.10} we get the \e\ for a perturbation
\beq{5.11}
\pz{r_1}\vf = -\frac{br_0^5}{2A^2\pz{r_0}\vf}-\frac{br_0^4}
{2A^2\pz{r_0}\vf}\,r_1.
\end{equation}
For $|r_1|\ll 1$ we neglect the second term in Eq.\ \er{5.11}.

We have a \so\ for unperturbed orbit
\bea{5.12}
r_0&=&\frac p{1+e\cos\vf}\\
\pz {r_0}\vf&=&\frac{ep\sin\vf}{(1+e\cos\vf)^2} \q (e>1).\label{5.13}
\end{eqnarray}
Thus we get
\beq{5.14}
\pz{r_1}\vf=-\frac{bp^4}{2eA^2}\,\frac1{(1+e\cos\vf)^3\sin\vf}\,.
\end{equation}
Using relations $p=a_h(e^2-1)$ and $A=\sqrt{k^2a_h(e^2-1)}$ we get
\beq{5.15}
r_1=-\frac{ba_h^3(e^2-1)^3}{4ek^2}\,f(\vf)
\end{equation}
where
\bml{5.16}
f(\vf)=\frac{2e(e^2+3)}{(e^2-1)^2}\log\frac{1+e\cos\vf}e
-\frac{\log(1-\cos\vf)}{(e+1)^3}-\frac{\log(1+\cos\vf)}{(e-1)^3}\\
{}+\frac{4e}{(1+e\cos\vf)(e^2-1)^2}-\frac e{(1+e\cos\vf)^2(e^2-1)}\,.
\end{multline}
$e$ is an eccentrity of a hyperbola, $p$ a \ti{similatus rectum}, $a_h$ a
semi-major axis of a hyperbola.

In this way an \e\ of an orbit is
\beq{5.17}
r(\vf)=\frac{a_h(e^2-1)}{1+e\cos\vf}-\frac{ba_h^3(e^2-1)^3}{4ek^2}\,f(\vf).
\end{equation}

In order to find a dependence $\vf$ on time $t$ we use a relation $A=r^2\dot\vf$,
i.e.
\bg{5.18}
A\,dt=r^2\,d\vf\\
r^2=(r_0+r_1)^2 \simeq r_0^2+2r_0r_1
=\frac{a_h^2(e^2-1)^2}{(1+e\cos\vf)^2}-\frac{ba_h^2(e^2-1)^4}{2ek^2}
\,\frac{f(\vf)}{1+e\cos\vf}\,. \label{5.19}
\end{gather}
We get
\beq{5.20}
\frac k{a_h^{3/2}(e^2-1)^{3/2}}\,dt = d\vf \X3(
\frac1{(1+e\cos\vf)^2}-\frac{ba_h^2(e^2-1)^2}{2ek^2}\,
\frac{f(\vf)}{1+e\cos\vf}\Y3).
\end{equation}
We use a substitution
\bg{5.21}
\eta=\tan\vf/2, \q \vf=2\arctan\eta\\
d\vf=\frac{2\,d\eta}{1+\eta^2} \label{5.22}\\
\cos\vf=\frac{1-\eta^2}{1+\eta^2} \label{5.23}\\
\g=\frac{1-e}{1+e}, \q e>0, \q \g<0. \label{5.24}
\end{gather}
In order to do an integration we rewrite a \f\ $f(\vf)$ as
\bml{5.25}
f(\vf)=-\frac{2e(e^2+3)\log e}{(e^2-1)^2}+\frac{2e(e^2+3)}{(e^2-1)^2}
\log(1+e\cos\vf)-\frac{\log(1-\cos\vf)}{(e+1)^3}\\
{}-\frac{\log(1+\cos\vf)}{(e-1)^3}-\frac{4e}{(e^2-1)^2(1+e\cos\vf)}
-\frac{e}{(e^2-1)(1+e\cos\vf)^2}\,.
\end{multline}
We make the following simplifications:
\bea{5.26}
\log(1+e\cos\vf)&\simeq&e\cos\vf-\frac{e^2\cos^2\vf}2\\
\log(1+\cos\vf)&\simeq&\cos\vf \label{5.27}\\
\log(1-\cos\vf)&\simeq&-\cos\vf. \label{5.28}
\end{eqnarray}
Finally
\bml{5.29}
f(\vf)=\frac{e(e^2+3)\log e}{(e^2-1)^2}\,(1-e^2\cos^2\vf)
+\frac{2\bigl(e^2(e^2+3)(e^2-1)\log e+3e^2+1\Y1)}{(e^2-1)^3}\cos\vf\\
{}-\frac{3e(e^2+3)\log e}{(e^2-1)^2}
+\frac{4e}{(e^2-1)^2}\,\frac1{(1+e\cos\vf)}
-\frac e{(e^2-1)}\,\frac1{(1+e\cos\vf)^2}\,.
\end{multline}
Now we can do an integration
\bml{5.30}
F(\vf)=\int \frac{f(\vf)}{1+e\cos\vf}\,d\vf=
B(e)\log\tan\X2(\frac\pi4+\frac q2\Y2)
+C(e)\tan q+A(e)\vf\\
{}-\frac{e^2}{4(e^2-1)^3}\sin\vf
\X3[4e^2(e^2+3)(e^2-1)\log e+\frac{e^2-4-3e\cos\vf}{(1+e\cos\vf)^2}\Y3]
\end{multline}
where
\bg{5.31}
A(e)=\frac{2\X1(e^2\log e(e^2+3)(e^2-1)(2-e^2)+3e^2+1\Y1)}{e(e^2-1)^3}\\
B(e)=-\frac{(e^2+1)^{1/2}(e^3+6e^2+6e+2)+e(e^2+3)(e^2-1)^{1/2}
\log e(6e^2-3e^4-1)}{(e^2-1)^{7/2}(e^2+1)^{1/2}} \label{5.32}\\
C(e)=\frac{4e}{(e^2-1)^{5/2}(e-1)^2} \label{5.33}\\
\tan\frac q2=\sqrt{-\g}\,\eta=\sqrt{\frac{e-1}{e+1}}\tan\frac\vf2\,.
\end{gather}
Finally we integrate Eq.\ \er{5.20} getting
\beq{5.35}
\frac k{a_h^{3/2}(e^2-1)^{3/2}}\,(t-T) = e\tan q -\log\X2(\tan\X2(
\frac\pi4+\frac q2\Y2)\Y2) - \frac{ba_h^2(e^2-1)(e-1)}{ek^2}\,F(\vf).
\end{equation}
For a \hy\ motion the following \e\ is satisfied:
\beq{5.36}
\frac k{a_h^{3/2}(e^2-1)^{3/2}}\,(t-T) = e\tan q_0 -\log\X2(\tan\X2(
\frac\pi4+\frac {q_0}2\Y2)\Y2).
\end{equation}
From Eq.\ \er{5.36} one gets $q_0=q_0(t)$ and
\beq{5.37}
\vf_0(t)=2\arctan\X3(\sqrt{\frac{e+1}{e-1}}\tan\frac{q_0(t)}2\Y3).
\end{equation}

In order to find $\vf=\vf(t)$ for a perturbed \hy\ orbit we should solve
Eq.~\er{5.35} treating an additional term as a small perturbation. Let
\beq{5.38}
q(t)=q_0(t)+\D q(t)
\end{equation}
where $q_0(t)$ is a \so\ of an unperturbed problem and $\D q(t)$ is a small
perturbation. For $\D q(t)$ is small we can keep only first order terms in
Eq.~\er{5.35} calculating all \dv s at a point $q_0(t)$. An additional term
with $F(\vf)$ is calculated at $\vf=\vf_0$. Under these assumptions one gets
\bea{5.39}
\D q&=&\frac{ba_h^2(e^2-1)(e-1)\cos^2q_0}{ek^2(e+\cos q_0)}\,F(\vf_0)\\
\D \vf&=&\frac{\sqrt{e^2-1}}{(e+1)-2\cos^2(q_0/2)}\,\D q \label{5.40}
\end{eqnarray}
where $\D \vf$ is a deviation for $\vf$ corresponding to $\D q$.

Finally one gets
\bml{5.41}
\D\vf=-\frac b{k(e+1-2\cos^2\frac{q_0}2)(e+\cos q_0)}
\X3[\ov B(e)\sqrt{a_h}\cos^2q_0(t-T)-
\frac{a_h^2}k \X2[\ov D(e)\sin2q_0\\
{}+\cos^2q_0\X2[\ov A(e)\vf_0-\ov C(e)\sin\vf_0\X2(\ov E(e)
+\frac{e^2-4-3e\cos\vf_0}{(1+e\cos^2\vf_0)^2}\Y2)\Y2]\Y2]\Y3]
\end{multline}
where
\bea{5.42}
\ov A(e)&=&\frac{(e-1)\X1(e^2(2-e^2)(e^2+3)(e^2-1)\log e+3e^2+1\Y1)}
{e(e^2-1)^{3/2}}\\
\ov B(e)&=&-\frac{(e^2+1)^{1/2}(e^3+6e^2+6e+2)+e(e^2+3)(e^2-1)^{1/2}
(6e^2-3e^4-1)\log e}{e(e^2+1)^{1/2}(e^2-1)^{5/2}(e+1)} \q
\label{5.43}\\
\ov C(e)&=&\frac{e(e-1)}{4(e^2-1)^{3/2}}
\label{5.44}\\
\ov D(e)&=&\frac{(e^2+1)^{1/2}(2-e^3-6e^2-6e)+e(e^2+3)(e^2-1)^{1/2}
(3e^4-6e^2+1)\log e}{2(e^2-1)(e^2+1)^{1/2}(e+1)}
\label{5.46}\\
\ov E(e)&=&4e^2(e^2+3)(e^2-1)\log e.
\label{5.47}
\end{eqnarray}
Finally
\bea{5.48}
\vf(t)&=&2\arctan \X3(\sqrt{\frac{e-1}{e+1}}\tan\frac{q_0(t)}2\Y3)+\D\vf(t)\\
\vf_0(t)&=&2\arctan \X3(\sqrt{\frac{e-1}{e+1}}\tan\frac{q_0(t)}2\Y3).
\label{5.49}
\end{eqnarray}

Let us come back to Eq.\ \er{5.17}. We get
\bml{5.50}
r(t)=r(\vf(t))=r(\vf_0(t)+\D\vf(t)) \simeq r(\vf_0(t))+\pz{r_0}\vf\,\D\vf\\
{}=\frac{a_h^2(e^2-1)}{1+e\cos\vf_0(t)}
-\frac{ba_h^3(e^2-1)^3}{4ek}\,f(\vf_0(t))
+\frac{a_he(e^2-1)\sin\vf_0(t)}{(1+e\cos\vf_0(t))^2}\,
\D\vf(t)
\end{multline}
and $f(\vf_0(t))$ is given by Eq.\ \er{5.29}.

Summing up we get a perturbed motion of a point mass where Eq.\ \er{5.17}
gives us an \e\ of an orbit and Eqs~\er{5.49}, \er{5.50} a dependence on time
in polar \cd s. According to our \ap ion it breaks for $r>R_0 \simeq
1.5\times10^3$\,AU. For such large distances an \an\ \ac\ is equal to ordinary
\gr al (newtonian) of the Sun.

\section{An influence of an \an\ \ac\ on a parabolic orbit}
Let us consider a parabolic orbit disturbed by an additional  central force
caused by an \an\ \ac. According to Section~5 we can derive a similar \e\ for
$e=1$ (see Refs \cite{18}, \cite{19}). Thus we have
\bg{6.1}
r_0=\frac p{1+\cos\vf}\\
\pz{r_1}\vf=-\frac{bp^4}{2A^2}\,\frac1{(1+\cos\vf)^3\sin\vf}\,, \label{6.2}
\end{gather}
where $p$ is a parameter of a parabola (a {\it semilatus rectum}).

One gets
\beq{6.3}
r_1=-\frac{bp^4}{4A^2}\X3[\frac1{1+\cos\vf}-\log\sqrt{\frac{1+\cos\vf}
{1-\cos\vf}}\Y3]
\end{equation}
and
\bea{6.4}
r&=&r_0+r_1=\frac p{1+\cos\vf}-
\frac{bp^3}{4k^2}\X3[\frac1{1+\cos\vf}-\frac12 \log\frac{1+\cos\vf}{1-\cos\vf}
\Y3]\\
r_1&=&-\frac{bp^4}{4A^2}f_p(\vf) \label{6.5}\\
f_p(\vf)&=&\frac1{1+\cos\vf}-\log\sqrt{\frac{1+\cos\vf}{1-\cos\vf}}, \q
\frac\pi2<\vf<\pi
\label{6.6}\\
F_p(\vf)&=&\int\frac{f_p(\vf)}{1+\cos\vf}\,d\vf. \label{6.7}
\end{eqnarray}
Using an \ap ion
\bea{6.8}
\log(1+\cos\vf)\simeq \cos\vf-\frac{\cos^2\vf}2\\
\log(1-\cos\vf)\simeq -\cos\vf-\frac{\cos^2\vf}2\label{6.9}
\end{eqnarray}
one gets
\beq{6.10}
F_p(\vf)=\frac12 \X2(-\vf+\sin\vf+\frac32 \tan\frac\vf2\Y2).
\end{equation}
From the \e
\beq{6.11}
r^2\,d\vf=A\,dt
\end{equation}
we get
\beq{6.12}
r_0^2\,d\vf-\frac{bp^4r_0}{2A^2}\,f_p(\vf)=A\,dt
\end{equation}
and
\beq{6.12a}
\frac{d\vf}{(1+\cos\vf)^2}-\frac{k^4bp^5f_p(\vf)}{2A^6(1+\cos\vf)}
=\frac{k^4}{A^3}\,dt.
\end{equation}

\begin{figure}[h]
\hbox to \textwidth{\ing[width=0.45\textwidth]{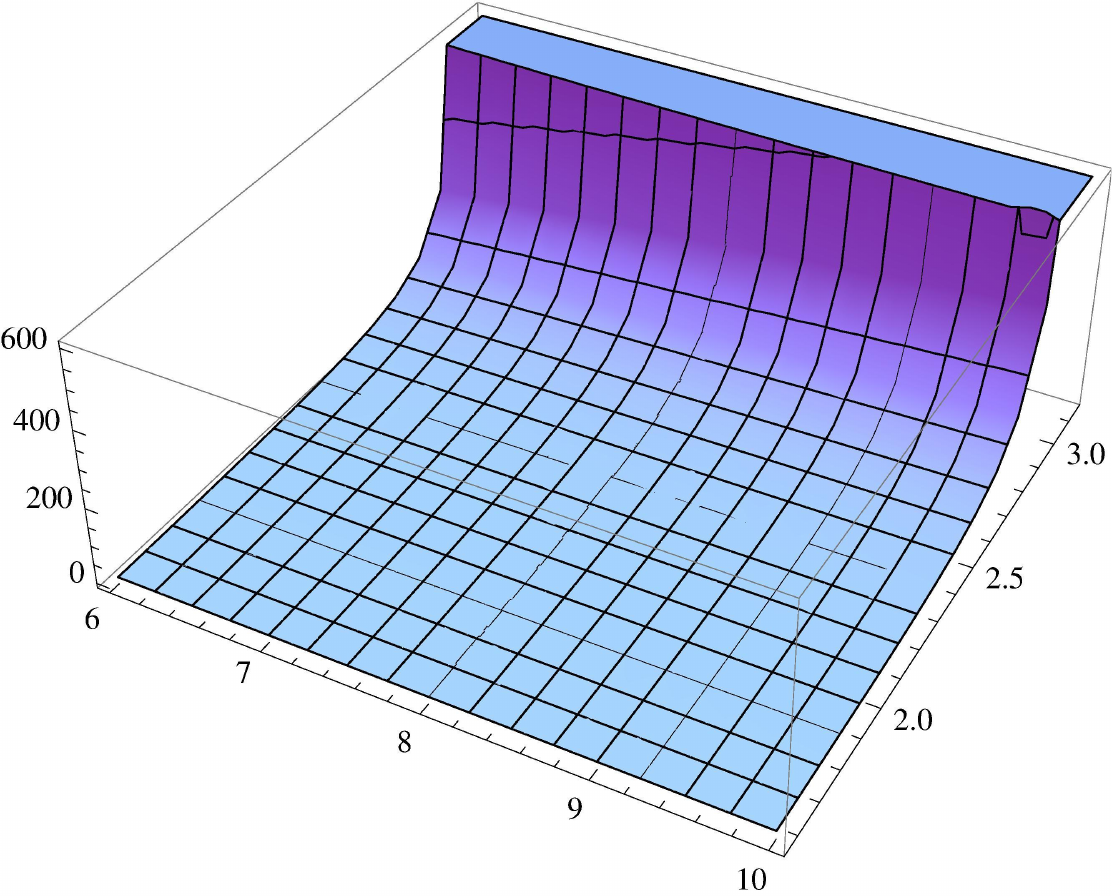}\hfil\ing[width=0.45\textwidth]{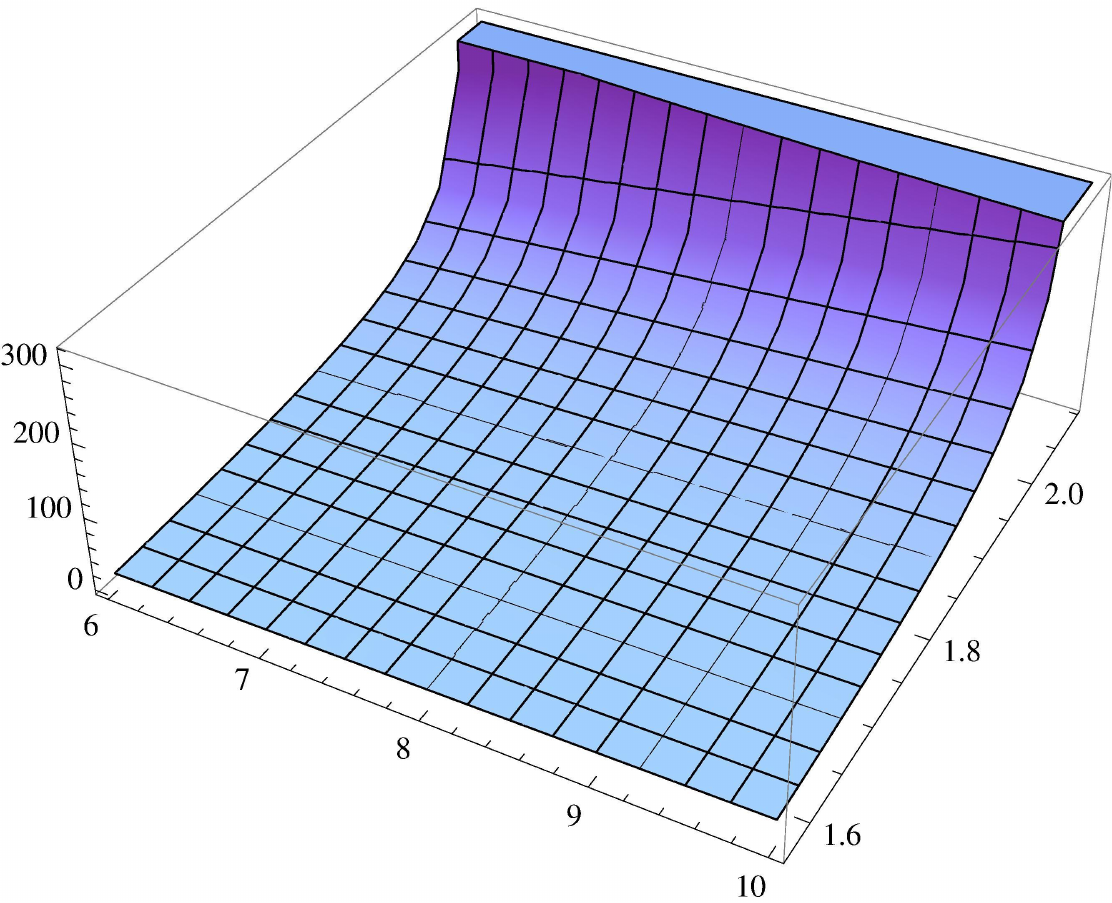}}
\hbox to \textwidth{\hbox to 0.45\textwidth{\hfil(A)\hfil}\hfil
\hbox to 0.45\textwidth{\hfil(B)\hfil}}
\hbox to \textwidth{\ing[width=0.45\textwidth]{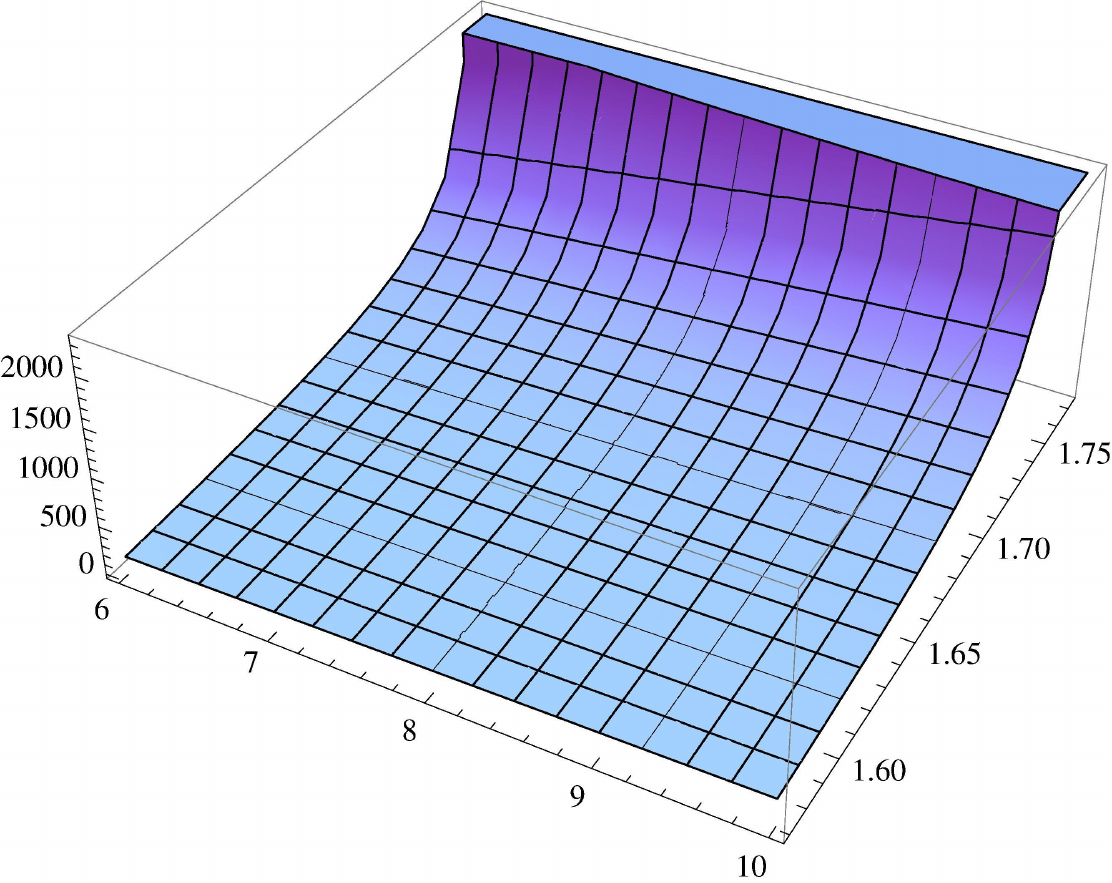}\hfil \ing[width=0.45\textwidth]{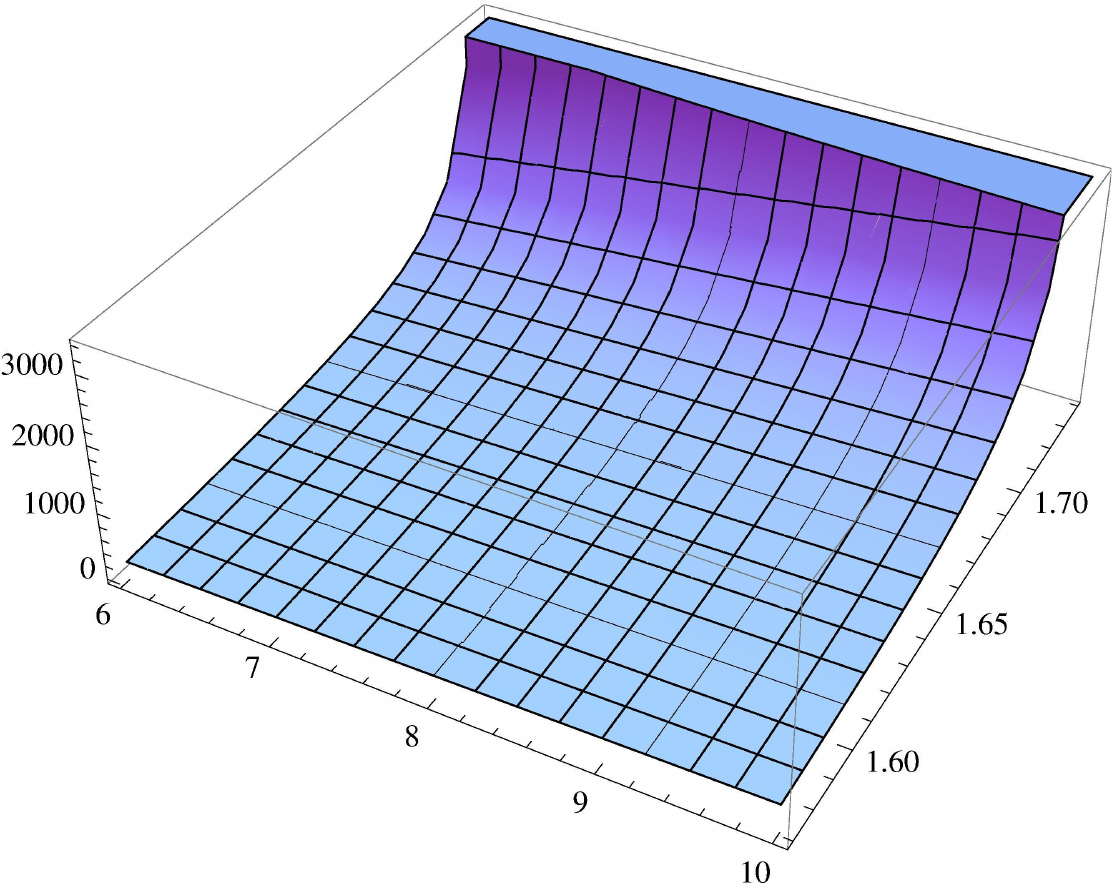}}
\hbox to \textwidth{\hbox to 0.45\textwidth{\hfil(C)\hfil}\hfil
\hbox to 0.45\textwidth{\hfil(D)\hfil}}
\caption{(A) 3D plot of distorted parabolic orbits, (B)~3D plot of distorted
\hy\ orbits for $e=2$, (C)~3D plot of distorted
\hy\ orbits for $e=5$, (D)~3D plot of distorted
\hy\ orbits for $e=6$ (see text).}\label{WK}
\end{figure}

After integration one gets
\beq{6.13}
\tan\frac\vf2+\frac13\tan^3\frac\vf2-\frac{k^4bp^5}{A^6}
\X2(-\vf+\sin\vf+3\tan\frac\vf2\Y2)=\frac{2k^4}{A^3}(t-T)
\end{equation}
or
\beq{6.14}
\tan\frac\vf2+\frac13\tan^3\frac\vf2-\frac{bp^2}{2k^2}
\X2(-\vf+\sin\vf+3\tan\frac\vf2\Y2)=\frac{2k}{p^{3/2}}(t-T).
\end{equation}
Let us suppose as in Section 5:
\beq{6.15}
\vf=\vf_0+\D\vf
\end{equation}
where $\vf_0$ satisfies an unperturbed \e
\beq{6.16}
\tan\frac{\vf_0}2+\frac13\tan^3\frac{\vf_0}2=
\frac{2k}{p^{3/2}}(t-T)
\end{equation}
and $\D\vf$ is a small perturbation. Using the same \ap ion procedure as in
Section~5 (linearization \wrt $\D\vf$) one gets
\beq{6.17}
\D\vf=-\frac{bp^2}{2k^2}\X2(\vf_0-\sin\vf_0-3\tan\frac{\vf_0}2\Y2)
\cos^4\X2(\frac{\vf_0}2\Y2)
\end{equation}
where
\bg{6.18}
\eta+\frac13\,\eta^3-\frac{2k}{p^{3/2}}(t-T)=0\\
\vf_0=2\arctan\eta. \label{6.19}
\end{gather}
We can solve \er{6.18} \wrt $\eta$ using Cardano method. In this case the
discriminant of the cubic \e~\er{6.18}
\beq{6.20}
D=1+\frac{9k^2(t-T)^2}{p^3}>0
\end{equation}
is positive (we have only one real root of Eq.~\er{6.18}).

One gets
\beq{6.21}
\eta=\frac1{p^{1/2}}\X3[\root3\of{3k(t-T)+\sqrt{9k^2(t-T)^2+p^3}}
+\root3\of{3k(t-T)-\sqrt{9k^2(t-T)^2+p^3}}\Y3]\,.
\end{equation}
Finally
\beq{6.22}
\vf=\vf_0+\D\vf=2\arctan\eta
\X3(1-\frac{bp^2}{2k^2(1-\eta^2)^2}\Y3)
+\frac{bp^2}{2k^2}\,\frac\eta{(1-\eta^2)^2}\X2(\frac2{1+\eta^2}-3\Y2).
\end{equation}

In this way we get a \so: \e\ of an orbit in polar \cd s \er{6.4} and a \f\
$\vf(t)$ given by Eqs~(\ref{6.21}--\ref{6.22}). It is easy to see that $\vf=0$
for $t=T$.

Let us calculate a distortion of~$p$ from Eq.~\er{6.4}. One gets
\bg{6.23}
p\eff=p\X2(1-\frac{bp^2}{4k^2}\Y2)\\
\frac{\D p}p=-\frac{bp^2}{4k^2}\,. \label{6.24}
\end{gather}

On Fig.~\ref{WK} we summarize distorted parabolic ($e=1$) and \hy\
($e>1$) orbits obtained in Sections 6 and 5, using 3d plots. For $e=2,5,6$
and $e=1$ (parabolic case) we plot $r(a_h,\vf)$, where $a_h$ is a semi-major axis
and $\frac\pi2 < \vf <\arccos(-1/e)$, $6\le a_h\le 10$ (for a distorted
hyperbola (Eq.~\er{5.17}) and $r(p,\vf)$, $\frac\pi2<\vf<\pi$, $6\le p\le
10$, where $p$ is a \ti{semilatus rectum} (for a distorted parabola,
Eq.~\er{6.4}). $r(a_h,\vf)$, $r(p,\vf)$, $a_h,p$ are measured in AU.
They look similar even they represent different types of distorted orbits:
parabolic and \hy.

\section{A disturbed \hy\ orbit in the \SS}\label{disturb}
\begin{figure}[ht]
\centerline{\ing{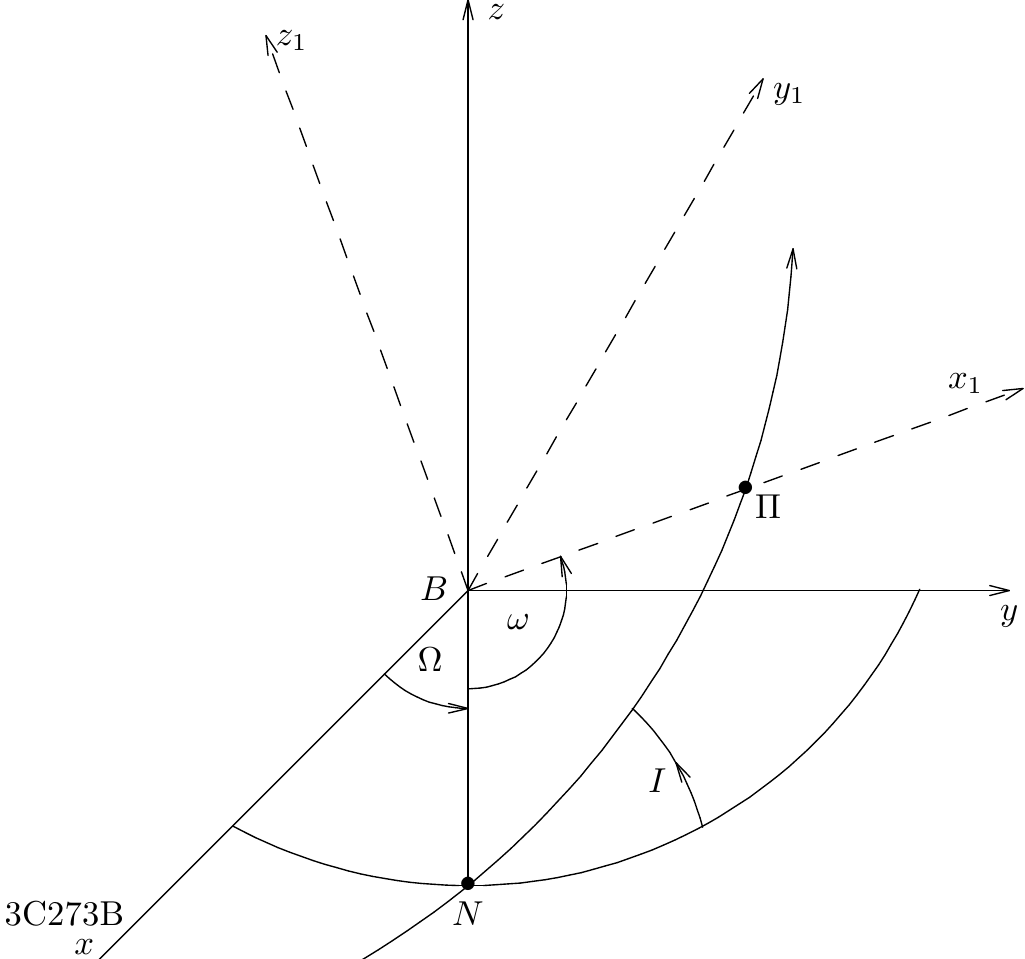}}
\caption{A disturbed \hy\ orbit in a cartesian \cd\ system connected to the
\SS\ BCRS. All elements of the orbit are quoted (see a text). $\Pi$ is a
perihelium point. $N$~is an ascending node.}\label{f4}
\end{figure}

In this section we consider a perturbed \hy\ orbit in BCRS \cd s (see the
text below)
in order to obtain elements of an orbit from observation and to
introduce perturbations from planets. Let us introduce cartesian \cd s on a
plane of motion (see Refs \cite{18}, \cite{19}),
\beq{7.1}
\bal
x_1&=r\cos\vf\cr
y_1&=r\sin\vf\cr
z_1&=0.
\eal
\end{equation}
In cartesian BCRS \cd s we get
\beq{7.2}
\bal
x(t)&=r(t)\X1(\cos(\vf(t)+\o)\cos\O-\sin(\vf(t)+\o)\sin\O\cos I\Y1)\cr
y(t)&=r(t)\X1(\cos(\vf(t)+\o)\sin\O+\sin(\vf(t)+\o)\cos\O\cos I\Y1)\cr
z(t)&=r(t)\sin(\vf(t)+\o)\sin I
\eal
\end{equation}
where $\o,\O$ and $I$ have usual meaning for elements of an orbit (see
Fig.~\ref{f4}). $B$~is a barycenter of the \SS, $t$ is a time for BCRS (see a text
below), i.e., $t={\rm TCB}$ (Barycentric \eu\cd\ Time).

Let us notice the following fact. According to GR (General Relativity) and
alternative theories of \gr\ (e.g.~NGT) a time $t$ as an argument in \e s of
motion should be changed to a proper time $\tau$ in a path along which a material point moves.
It means
$$
\pz{^2x^\mu}{\tau^2}+\G_{\a\b}^\mu\,\pz{x^\a}\tau\,\pz{x^\b}\tau=F^\mu
$$
where $\G_{\a\b}^\mu$ are \cf s of a connection (Christoffel symbols in~GR)
and $F^\mu$ is an external force. In our case relativistic analogue of an
\an\ \ac. Thus we have
$$
\pz{^2t}{\tau^2}+\G_{\a\b}^4\,\pz{x^\a}\tau \, \pz{x^\b}\tau=F^4 \qh{and}
t\ne \tau.
$$
Moreover, using PN \ap ion one can prove that a difference between $t$
and~$\tau$ is negligible if we are far from the Sun (20\,AU is enough, see
Ref.~\cite{21}).

We can also use spherical \cd s \st
\beq{7.2a}
\bal
x(t)&=r(t)\sin\b(t)\cos\la(t)\cr
y(t)&=r(t)\sin\b(t)\sin\la(t)\cr
z(t)&=r(t)\cos\b(t).
\eal
\end{equation}
Working in cartesian BCRS \cd s we can find elements of an orbit
measuring \cd s and velocity of an object. Let us suppose for a moment that
we have to do with a real \hy\ orbit. In this case we have
\bea{7.3}
2H&=&v_0^2-\frac{2k^2}{r_0}-2br_0\\
v_0^2&=&v_{0x}^2+v_{0y}^2+v_{0z}^2 \label{7.4}\\
r_0^2&=&x_0^2+y_0^2+z_0^2 \label{7.5}\\
A^2&=&\frac1{k^2}\X2(v_0^2r_0^2-\X1(x_0v_{0x}+y_0v_{0y}+z_0v_{0z}\Y1)^2\Y2)
\label{7.6}
\end{eqnarray}
where $\vec r_0=(x_0,y_0,z_0)$ are \cd s of an object and $\vec v=(v_{0x},
v_{0y},v_{0z})$ its velocity at the same time $t_0$. From Eqs~\er{7.3}
and~\er{7.6} we find $e$ and $a_h$:
\bea{7.7}
e&=&\sqrt{\frac{2HA^2}{k^4}+1}, \q H>0,\\
a_h&=&\frac{A^2}{k^2(e^2-1)}=\frac{k^2}{2H}\,.\label{7.8}
\end{eqnarray}

Let us define the following quantities:
\bg{7.9}
\bal
\ov A&=\frac{x_0A}{r_0e}\,(2Hr_0+k^2)-r_0\sin\vf_0v_{0x}\cr
\ov B&=\frac{y_0A}{r_0e}\,(2Hr_0+k^2)-r_0\sin\vf_0v_{0y}\cr
\ov C&=\frac{z_0A}{r_0e}\,(2Hr_0+k^2)-r_0\sin\vf_0v_{0z}
\eal\\
\bal
\ov D&=r_0v_{0x}\cos\vf_0+\frac{x_0k^2\sin\vf_0}A\cr
\ov E&=r_0v_{0y}\cos\vf_0+\frac{y_0k^2\sin\vf_0}A\cr
\ov F&=r_0v_{0z}\cos\vf_0+\frac{z_0k^2\sin\vf_0}A
\eal \label{7.10}
\end{gather}
where
\beq{7.11}
\vf_0=\arccos\X2(\frac{A^2-k^2r_0}{r_0k^2e}\Y2).
\end{equation}

From Eqs \er{7.9}, \er{7.10} we get
\bea{7.12}
\o&=&\arctan\X3(\frac{\ov C}{\,\ov F\,}\Y3)\\
I&=&\arcsin\X3(\frac{\ov C}{\ov A\sin\o}\Y3) \label{7.13}\\
\O&=&\arctan\X3(\frac{\ov B-\ov F\tan\o}{\ov A-\ov C\tan\o}\Y3). \label{7.14}
\end{eqnarray}
We have
\beq{7.15}
q_0=2\arctan\X3(\sqrt{\frac{e-1}{e+1}}\tan\frac{\vf_0}2\Y3)
\end{equation}
and
\beq{7.16}
\ov T=t_0-\frac{a_h^{3/2}(e^2-1)^{3/2}}k \X3[e\tan q_0 - \log\X2(\tan\X2(\frac{q_0}2
+\frac\pi4\Y2)\Y2)\Y3].
\end{equation}

Let us notice that we use a total potential energy to define~$e$ and
unperturbed \hy\ \e~\er{7.16} to define~$\ov T$.
According to a standard notion $n$~is a mean anomaly, $I$~an inclination,
$\O$~a longitude of the ascending node, $\o$~an argument of a peribarycenter,
$\ov T$~a time of a peribarycenter (perihelium).
The parameters $\O$, $\o$
and $\ov T$ are quite artificial for the orbit is considered for $r(\vf)
>20\,{\rm AU}$. Moreover, they are very useful. In order to calculate $\ov T$
we can use also Eqs \er{5.41}--\er{5.49} or even \er{B.1}--\er{B.3}.

Moreover, it is necessary to include an influence of planets in the \SS\ on
the orbit. In order to do this we define a perturbation \f
\beq{7.17}
P=k^2\sum_{A=1}^n m_A\X3\{\frac1{\X1[(x-x_A)^2+(y-y_A)^2+(z-z_A)^2\Y1]}
-\frac{xx_A+yy_A+zz_A}{r_A^2}\Y3\}
\end{equation}
where $m_A$, $A=1,2,\dots,n$, are masses of planets, $x_A,y_A,z_A$ their
cartesian BCRS \cd s, and $r_A=\sqrt{x_A^2+y_A^2+z_A^2}$. Using this \f\ we can
write \e\ for a slow changing of elements of the orbit.
\bea{7.18}
\pz{a_h}t&=&\frac2{na_h}\,\pp P\ve\\
\pz{e}t&=&-\frac{(e^2-1)^{1/2}}{na_h^2e}\X1(1-(e^2-1)^{1/2}\Y1)\pp P\ve
-\frac{(e^2-1)^{1/2}}{na_h^2e}\,\pp P{\ov \pi} \label{7.19}\\
\pz{I}t&=&-\frac{\tan(I/2)}{na_h^2(e^2-1)^{1/2}}\X2(\pp P\ve+\pp P{\ov \pi}
\Y2)-\frac1{na_h^2(e^2-1)^{1/2}\sin I}\,\pp P\O \label{7.20}\\
\pz{\ve}t&=&-\frac2{na_h}\,\pp P{a_h}+\frac{(e^2-1)^{1/2}}{na_h^2\ve}
\X1(1-(e^2-1)^{1/2}\Y1)\pp P\ve+\frac{\tan(I/2)}{na_h^2(e^2-1)^{1/2}}
\pp PI \label{7.21}\\
\pz\O t&=&\frac1{na_h^2(e^2-1)^{1/2}\sin I}\,\pp PI \label{7.22}
\end{eqnarray}
where $x,y,z$ are given by Eq.~\er{7.2},
\bea{7.23}
n&=&ka_h^{-3/2}(e^2-1)^{-1/2}\\
\ve&=&\o+\O-n\ov T \label{7.24}\\
\ov\pi&=&\o+\O \label{7.25}
\end{eqnarray}
and a time evolution is given by Eqs (\ref{5.49}--\ref{5.50}) for a perturbed
hyperbola. In this way we can have something similar to a tangent disturbed
hyperbola at any instant of time (Eqs~(\ref{7.18}--\ref{7.22}) are exact).
One gets similar results as Eqs~(\ref{7.18}--\ref{7.22}) in Ref.~\cite{23d}.
The rate of change of elements of perturbed hyperbola is slower than a motion
of a point mass on the orbit and for this we improve calculated position
changing elements of an orbit.

Let us consider a kinematic \cd\ system recommended by IAU (see
Ref.~\cite{22}), called ICRS/ICRF (International Celestial Reference System /
International Celestial Reference\break Frame). (Since $1^{\rm st}$~January
2010 we are obliged to use ICRF2 in place of ICRF according to the
recommendation of IAU.)
ICRS is defined by positions of 212
extragalactical radiosources (quasars) of a precision 0.4\,marcs. ICRF is a
catalogue of equator \cd s of 608 extragalactical radiosources obtained from
VLBI observations. The origin point of ICRS is a barycenter of the \SS. There
is not any rotation of ICRS in an inertial space. There is not any influence
on this frame coming from precession--nutation model. Some individual
movement of radiosources caused by changing their structure have no
influence on the model.

Let us notice the following fact. The most natural quasi-inertial frame in
the Universe is the frame connecting to a black body radiation background. It
is not an aether for an aether \ti{does not exist\/}. Moreover, a barycenter of
the \SS\ moves \wrt this frame. The movement is very well known (Cobe, WMAP,
Planck's satellite). Thus the transformation (a~general covariant
transformation) from this frame to BCRS can be easily done. It would be very
natural to do it and to use as a standard of IAU.

The motion of the barycenter of the \SS\ is not linear in its orbit around
the center of the Galaxy. There is therefore a Coriolis-like force, which
gives us galactic geodesic precession. This is not included in a ICRS
definition. Thus we should in principle add this additional force to a
definition of BCRS. There is also a possibility of such an \ac\ (force) \wrt
black body radiation frame (a~linear too). There is no global rotation of the
Universe. It means there is not a global Coriolis \ac.

The fundamental directions $x,y,z$ in this system are defined in such a way
that definitions of axes are the same as directions from FK5 catalogue for an
epoch 2000.0. It means that an IAU pole of an epoch J2000.0 is based on a
precession model IAU1976 and a nutation model IAU1980. An origin of a
rectascence is defined by a rectascence $\a_{\rm FK3}$ (a fundamental
catalogue FK3) of a radiosource 3C273B transformed to FK5. All directions of
axes are considered as convections and are fixed. Moreover, they are close to
J2000.0 ecliptic plane and an equinox point of an epoch of 2000.0. From this
moment ICRS and ecliptical \cd s are disconnected. Moreover, it is possible
to do some calculations based on ecliptical \cd s of J2000.0 if the precision
is not so high (Ref.~\cite{23}). In our case (see Fig.~\ref{f4}) $x,y,z$ are defined
as \cd\ system of ICRS/ICRF. The directions $x,y,z$ are in some sense
generalizations of a celestial sphere from spherical astronomy. Moreover, it
is necessary to define a reference system based on ICRS/ICRF which includes
General Relativity. It means we need the fourth axis---a time. Thus we define
a \spt\ metric tensor of GR (General Relativity) according to the
recommendation of IAU (see Ref.~\cite{22}). It is based on EIH
(Einstein--Infeld--Hoffmann) method (see Ref.~\cite{24}), developed by
T.~Damour et al.\ (see Refs \cite{25}, \cite{26}, \cite{27}, \cite{28} and
\cite{29}). It is a celestial relativistic mechanics up to a certain order of
$\X1(\frac1c\Y1)$ (see also Ref.~\cite{30}).

According to the mentioned recommendation we define a metric tensor
\beq{7.27}
ds^2=g_{\mu\nu}\,dx^\mu\,dx^\nu, \q \mu,\nu=1,2,3,4,
\end{equation}
where
\beq{7.28}
\bal
g_{44}&=1-\frac{2U}{c^2}\\
g_{4k}&=g_{k4}=0\\
g_{ik}&=-\X2(1+\frac{2U}{c^2}\Y2), \q i,k=1,2,3.
\eal
\end{equation}
In these formulae $U$ is a potential of all \gr al fields in the \SS\ in a
postnewtonian \ap ion. We use a convention of a signature of the
\spt\ different from that in Ref.~\cite{22}. It means $(---+)$ as in our
papers. 
It is supposed that $U(x)$ is going to zero far away from the origin of the
\SS. This system is called BCRS (Barycentric Celestial Reference System).
It reads in a more expanded and exact form
\beq{7.29}
\bal
g_{44}&=1-\frac{2w(t,\vec x)}{c^2}+\frac{2w(t,\vec x)^2}{c^4}\\
g_{4i}&=-\frac4{c^3}\,w^i(t,\vec x)\\
g_{ij}&=-\d_{ij}\X2(1+\frac2{c^2}\,w(t,\vec x)\Y2)
\eal
\end{equation}
where
\bg{7.30}
w(t,\vec x)=G_N \int d^3\vec x{}'\,\frac{\si(t,\vec x{}')}{\|\vec x-\vec x{}'\|}
+\frac{2G_N}{c^2}\,\pp{^2}{t^2}\X3(\int d^3\vec x{}'\,\si(t,\vec x{}')
\|\vec x-\vec x{}'\|\Y3)\\
w^i(t,\vec x)=G_N \int d^3\vec x{}'\,\frac{\si^i(t,\vec x{}')}{\|\vec
x-\vec x{}'\|}\label{7.31}
\end{gather}
$\si$ and $\si^i$ are density of active \gr al masses and densities of their
currents. The formulae for potential $w$ and $w^i$ define $g_{44}$ up to
$\frac1{c^3}$, $g_{4i}$ up to $\frac1{c^5}$ and $g_{ij}$ up to $\frac1{c^4}$
(according to the EIH method). $\si$ and~$\si^i$ are defined by a tensor of
energy--momentum of a matter of all bodies of the \SS\ (see Refs \cite{25},
\cite{26}, \cite{27}, \cite{28}, \cite{29}, \cite{30}, \cite{31}).
Moreover, now we need a transformation from the BCRS to
GCRS (Geocentric Celestial Reference System). (In this case we have a
celestial sphere connecting to Earth, a geocentric celestial sphere.)
In order to do this we should
define a metric tensor for GCRS. It is
\bea{7.32}
G_{44}&=&1-\frac2{c^2}\,W(T,\vec X)+\frac2{c^4}\,W(T,\vec X)^2\\
G_{4a}&=&-\frac4{c^3}W^a(T,\vec X) \label{7.33}\\
G_{ab}&=&-\d_{ab}\X2(1+\frac2{c^2}W(T,\vec X)\Y2), \label{7.34}
\end{eqnarray}
$a,b=1,2,3$.

Let us give the following comment. Metric tensors $g_{\mu\nu}$ and
$G_{\mu\nu}$ are given in a harmonic system of coordinates, i.e.\ $x^i$
and~$X^i$ satisfy a harmonic gauge condition. They satisfy also a time
harmonic gauge. In this a formalism is slightly different than in~\cite{30}.
This is not a de~Donder condition which can be satisfied in BCRS according to
Fock suggestion in the case of Schwarzschild \so. This condition cannot be
satisfied in GCRS for Schwarzschild \so. Moreover, harmonic gauge condition in
this case is not exactly a de Donder condition (even this condition is
usually called harmonic) and the contradiction does not take place.

Both geocentric potentials $W(T,\vec X)$ and $W^a(T,\vec X)$ are a sum of all
potentials coming from \gr al interactions of the Earth $W_\oplus$ and $W^a_\oplus$,
and external potentials $W_{\rm ext}$ and $W^a_{\rm ext}$ caused by tidal and
uninertial sources.
\bea{7.35}
W(T,\vec X)&=&W_\oplus(T,\vec X)+W_{\rm ext}(T,\vec X)\\
W^a(T,\vec X)&=&W_\oplus^a(T,\vec X)+W^a_{\rm ext}(T,\vec X). \label{7.36}
\end{eqnarray}

It is supposed that external potentials are zero at a barycenter of mass of
the Earth and they are power series in positive \cf s in $\vec X$. Potentials
$W_\oplus$, $W^a_\oplus$ are defined in the same way as $w$ and~$w^i$, but with
arguments calculated in GCRS with integration for all the Earth. According
to the resolution IAU (see Ref.~\cite{22}) it is necessary to develop a
GCRS postnewtonian potential of the Earth outside of the Earth
\bml{7.38}
W_\oplus(T,\vec X)=\frac{M_\oplus G_N}r \,\X3[1+
\sum_{l=2}^\iy \sum_{m=0}^l \X2(\frac{\,\ov a\,}r\Y2)^l
P_{lm}(\sin\psi)\X2(C_{\oplus lm}(T)\cos m\la\\
{}+S_{\oplus lm}(T)\sin m\la\Y2)\Y3]
\end{multline}
where $C_{\oplus lm}$ and $S_{\oplus lm}$ are (up to sufficient precision)
equivalent to postnewtonian multipole moments, $\theta=\frac\pi2-\psi$ and $\la$ are
spherical angles corresponding to (cartesian) \cd s $\vec X$ in GCRS.
$M_\oplus$ is a mass of the Earth, $\ov a$ its equator radius and
$r=\|\vec X\|$ (see the first approach to this formalism Ref.~\cite{Tul}).

The mentioned resolution recommends that a vector potential outside of the
Earth, being a source of the Lense--Thirring effect, expressed by a \f\ of a
total angular momentum of the Earth $\vec S_\oplus$ is in a form
\beq{7.39}
W^a_\oplus(T,\vec X)=-\frac{G_N}2\,\frac{(\vec X\t \vec S_\oplus)^a}{R^3}\,.
\end{equation}
External potentials $W_{\rm ext}$ and $W_{\rm ext}^a$ are expressed as a sum
of tidal and uninertial terms
$$
\bal
W_{\rm ext}&=W_{\rm tidal}+W_{\rm inert}\\
W^a_{\rm ext}&=W^a_{\rm tidal}+W^a_{\rm inert}\,.
\eal
$$
The potential $W_{\rm tidal}$ is a general relativistic (up to a certain
order) generalization of newtonian tidal potential (see Refs \cite{28},
\cite{30}, \cite{31}). Potentials $W_{\rm inert}$, $W_{\rm
inert}^a$ are linear in~$X^a$. The first of them is expressed by some
expressions connecting nonsphericity of the Earth with external potentials.
In rotating GCRS, $W^a_{\rm inert}$ describes a Coriolis force caused by
a geodetic precession (this precession---a~de~Sitter precession---is a
relativistic effect of a motion of a barycenter of the system Earth--Moon).
Let us come back to the relation between BCRS and GCRS. Let us notice the
following facts. In order to go from one metric tensor on a \spt\ to the
second one we should perform the following transformation
\beq{7.40}
g_{\mu\nu}=\pp{x^{\a'}}{x^\mu}\,\pp{x^{\b'}}{x^\nu}\,G_{\a'\b'}
\end{equation}
where $x^{\a'}(x^\mu)$ is a \f\ of a transformation of \cd\ system
$x^{\a'}=(X^a,T)$, $x^\a=(x^i,t)$ (in general mixing
space and time \cd s).

In our case we have (up to a certain order of $\frac1c$)
\bg{7.41}
T=t-\frac1{c^2}\X1(A(t)+v_\oplus^ir_\oplus^i\Y1)+
\frac1{c^4}\X1[B(t)+B'(t)r_\oplus'+B^i(r)r_\oplus^i
+B^{ij}(t)r_\oplus^ir_\oplus^j+C(t,\vec x)\Y1]\hskip0.1\textwidth \nn\\
\hskip0.6\textwidth{}+O\X2(\frac1{c^5}\Y2)\\
X^a=\d_{ai}\X2[r_\oplus^i+\frac1{c^2}\X2(\frac12v_\oplus^iv_\oplus^j
r_\oplus^j+W_{\rm ext}(\vec X_\oplus)r_\oplus^i+r_\oplus^i
+r_\oplus^ia_\oplus^jr_\oplus^j-\frac12a_\oplus^jr_\oplus^2\Y2)\Y2]+
O\X2(\frac1{c^4}\Y2) \label{7.42}
\end{gather}
where
\bea{7.43}
{}\kern-15pt \pz{}t\,A(t)&=&\frac12\,v_\oplus^2+W_{\rm ext}(\vec X_\oplus)\\
{}\kern-15pt \pz{}t\,B(t)&=&-\frac18\,v_\oplus^4-\frac32v_\oplus^2W_{\rm ext}(\vec X_\oplus)
+4v_\oplus^i W_{\rm ext}^i(\vec X_\oplus)+\frac12\,W_{\rm ext}^2(\vec X_\oplus)
\label{7.44}\\
{}\kern-15pt B^i(t)&=&-\frac12\,v_\oplus^2v_\oplus^i+4W_{\rm ext}^i(\vec X_\oplus)
-3v_\oplus^iW_{\rm ext}(\vec X_\oplus) \label{7.45}\\
{}\kern-15pt B^{ij}(t)&=&-v_\oplus^i\d_{aj}Q^a+2\,\pp{}{x_j}\X1(W_{\rm ext}^i(\vec
X_\oplus)\Y1)-v_\oplus^i\,\pp{}{x^j}\X1(W_{\rm ext}(\vec X_\oplus)\Y1)
+\frac12\,\d^{ij}\,\pz{}t\X1(W_{\rm ext}(\vec X_\oplus)\Y1) \label{7.46}\\
{}\kern-15pt C(t,\vec x)&=&-\frac1{10}\,r_\oplus^2\X2(\pz{}t\,a_\oplus^ir_\oplus^i\Y2)
\label{7.47}
\end{eqnarray}
where $\vec X_\oplus,\vec v_\oplus,\vec a_\oplus$ are barycentric position,
velocity and \ac\ of the Earth and
\bg{7.48}
Q^a=\d_{ai}\X2[\pp{}{x^i}W_{\rm ext}(\vec X_\oplus)-a_\oplus^i\Y2]\\
W_{\rm ext}=\sum_{A\ne \oplus}W_A, \q W_{\rm ext}^i=\sum_{A\ne \oplus}W_A^i. \label{7.49}
\end{gather}
It means a sum for all bodies in the \SS\ except the Earth.

Let us sum up. We have a reference system (ICRS) and in this system a \spt\
is defined by a metric tensor for BCRS and for GCRS separately. Both systems
BCRS and GCRS are dynamical and have different \cd\ times $T=\rm TCG$
(Geocentric \eu\cd\ Time), $t=\rm TCB$ (Barycentric \eu\cd\ Time). In \e s
$$
x^i=\vec x(t)=\X1(x(t),y(t),z(t)\Y1)=\X1(x^1(t),x^2(t),x^3(t)\Y1)
$$
of our orbit $t$ means TCB. In this way we transform an orbit from BCRS to
GCRS and the \e s are given by
\beq{7.50}
X^a=\vec X(T)=\X1(X(T),Y(T),Z(T)\Y1)=\X1(X^1(T),X^2(T),X^3(T)\Y1)
\end{equation}
and $T=\rm TCG$. Let us notice that in the case of Newton theory $T=t$ and
$X^a=x^a-x^a_\oplus(t)$ where $x_\oplus^a(t)$ means a barycentric position of
a geocenter.

If we define a metric tensor $G_{\m}^{\rm rot}$ as a metric tensor on a
rotating Earth (a~geoid) we get
$$
G_{\m}=\pp{x^{\a'}}{x^\mu}\cdot\pp{x^{\b'}}{x^\nu}\,G_{\a'\b'}^{\rm rot},
\q x^{\a'}=(\vec x_{\rm ITRS},\rm TT)
$$
(see a text below).

Let us notice that in a frame on a surface of the Earth (on a geoid) a metric
tensor is different, i.e.\
$$
G_{44}^{\rm rot}=G_{44}-\frac{\ov\o{}^2r^2}{c^2}\,\cos^2\psi,
$$
where $\ov\o$ is an angular velocity.

It is interesting to notice that if it were necessary it would be possible to
include some elements of alternative theories of \gr, i.e., different from
GR, moreover, with GR as a limit (according to the Bohr correspondence
principle). The most promising theory is NGT (Ref.~\cite9). This theory uses
a \nos\ metric tensor. Thus in this case $g_{\mu\nu}\ne g_{\nu\mu}$ and
$G_{\mu\nu}\ne G_{\nu\mu}$, also $G_{\m}^{\rm rot}\ne G_{\nu\mu}^{\rm rot}$,
$G_{\m}^{\rm loc}\ne G_{\nu\mu}^{\rm loc}$.
The transformation \er{7.40} can be done in a
similar way by using a postnewtonian \ap ion in this theory (Refs \cite{32},
\cite{33}, \cite{34}, \cite{35}). This theory has a well defined Cauchy
initial problem (Ref.~\cite{36}) and a consistent PN formalism. It uses
additional phenomenological material sources (a~fermion charge current). The
PN \e\ in NGT differs from that of GR, especially for perihelia movement
(Ref.~\cite{37}). The Einstein--Cartan theory (see Ref.~\cite{38}) can also
be incorporated in the formalism with phenomenological spin sources
(a~Mathisson spin). We can also try to use a unification of NGT and
Einstein--Cartan theory---the so called Einstein--Cartan--Moffat theory (see
Ref.~\cite{39}). Moreover, in the last case the PN formalism has not been
developed. This would be done if necessary.

In all mentioned systems we are using SI second and SI meter as units of time
and length. The quantities extended for use with any system should be called
system-compatible (e.g., GCRS-compatible).

Let us introduce TRS (Terrestial Reference System). This system rotates with
the Earth. In the system positions of points on a surface of the Earth are
defined by \cd s which change not significally with time due to geophysical
effects. The system is called CTRS (Conventional Terrestial Reference System)
and it is a quasi-cartesian system defined by a space rotation \wrt GCRS.
(GCRS does not rotate.) The time for CTRS is TCG (see Ref.~\cite{40},
\cite{41}, \cite{42}, \cite{42k}, \cite{42.1}, \cite{44n}, \cite{44o}). An
origin of CTRS is a barycenter of the Earth with the oceans and the
atmosphere taken into account. CTRS does not globally rotate \wrt any
horizontal motion of the Earth (even residually). CTRS investigated by IERS
(International Earth Rotation Service)
has been called ITRS (International Terrestial Reference System). ITRS is a
geocentric system with a length unit 1\,m\,SI
and a time unit 1\,s SI. A~transition from a celestial
(quasi-inertial) system to a terrestial (observational) system is achieved
via an intermediate system (see Ref.~\cite{40}). It is a Guinot idea.
In particular we have a transformation
\beq{7.51}
\vec X_{\rm GCRS}=Q(t_1)R(t_1)W(t_1)\vec X_{\rm ITRS}
\end{equation}
where $\vec X_{\rm ITRS}$ is a position vector in ITRS and $\vec X_{\rm
GCRS}$ its image in GCRS ($\vec X_{\rm GCRS}=\vec X$).

The transformation \er{7.51} is an isometry. All matrices given below are
orthogonal.
$W(t_1)$, $R(t_1)$ and $Q(t_1)$ are matrices of the transformation. $W(t_1)$
defines a motion of CIP (Celestial Intermediate Pole) \wrt the terrestial
system ITRS. $R(t_1)$ is a rotation of an intermediate system IRS
(Intermediate System) around an axis of CIP. $Q(t_1)$ is a rotation matrix
defining a motion of CIP \wrt the celestial system GCRS (defined below). The
parameter $t_1$ is a time (defined below---see \er{7.64}). In this way we
divide a physical precession--nutation of the Earth into two parts:
a~precession--nutation model and a motion of the pole. This division is an arbitrary
convention (in a frequency domain). The structure of rotation matrices is
standard.

Let
\refstepcounter{equation}\label{7.52}
$$
\dsl{
\hfill R_1(\vartheta_1)=\mt{1&\ &0&\ &0\\
0&&\cos\vartheta_1&&\sin\vartheta_1\\
0&&-\sin\vartheta_1&&\cos\vartheta_1}
\hfill (\theequation\rm a)\cr
\hfill R_2(\vartheta_2)=\mt{\cos\vartheta_2&\ &0&\ &-\sin\vartheta_2\\
0&&1&&0\\
\sin\vartheta_2&&0&&\cos\vartheta_2}
\hfill (\theequation\rm b)\cr
\hfill R_3(\vartheta_3)=\mt{\cos\vartheta_3&\ &\sin\vartheta_3&\ &0\\
-\sin\vartheta_3&&\cos\vartheta_3&&0\\
0&&0&&1}
\hfill (\theequation\rm c)\cr
\refstepcounter{equation}\label{7.53}
\hfill W(t_1)=R_3(-s')R_2(x_p)R_1(y_p)
\hfill (\theequation)}
$$
where $x_p,y_p$ are positions of CIP in ITRS for an epoch $t_1$ and are
defined as
\beq{7.54}
(x_p,y_p)=(x,y)_{\rm IERS}+(\D x,\D y)_{\rm tidal}+(\D x,\D y)_{\rm
nutation}\,.
\end{equation}
$(x,y)$ are \cd s of the pole (CIP) given by IERS (International Earth
Rotation Service). $(\D x,\D y)_{\rm tidal}$ are tidal corrections (oceans'
tides), $(\D x,\D y)_{\rm nutation}$ are nutation corrections, which are not
included into the precession--nutation model IAU2000. They have short periods
of an order of a day of lower. The quantity $s'$ is defined by a motion of
TEO (Terrestial Ephemeris Origin) on an equator of CIP \wrt ITRS caused by a
motion of the pole (CIP). Let us notice that now a pole (CIP) is not a north
pole of a celestial sphere (considered earlier) and a point to calculate a
rectascence (TEO) is not an equinox point. Those notions are disconnected.
Moreover, they are quite close
\bg{7.55}
s'(t_1)=\frac12 \int_{t_0}^{t_1}(x_p\dot y_p-\dot x_py_p)\,d\tau.\\
R(t_1)=R_3(-{\rm ERA}) \label{7.56}
\end{gather}
where ERA (Earth Rotation Angle) is an angle of the Earth rotation (see
Ref.~\cite{41}, \cite{42}).
\beq{7.57}
Q(t_1)=R_3(-E)R_2(-d)R_3(E)R_3(s)
\end{equation}

Let $\ov X,\ov Y,\ov Z$ be cartesian \cd s of CIP in GCRS,
\beq{7.58}
\bal
\ov X&=\sin d \cos E\\
\ov Y&=\sin d \sin E\\
\ov Z&=\cos d.
\eal
\end{equation}
The parameter $s$ describes a change of a position of CEO (Celestial
Ephemeris Origin) on an CIP equator caused by a motion of CIP,
\bg{7.59}
s(t_1)=-\frac12\X1[\ov X(t_1)\ov Y(t_1)-\ov X(t_0)\ov Y(t_0)\Y1]
+\int_{t_0}^{t_1}\dot{\ov X}(\tau)\ov Y(\tau)\,d\tau+s_0\\
t_0=J2000.0, \q s_0=-94\,\mu{\rm arcs} \nn
\end{gather}
($\tau$ means here $T$).

Let us find $\vec X_{\rm GCRS}$ for an orbit \er{7.56}. One gets
\beq{7.60}
\vec X_{\rm GCRS}=\X1(X(T),Y(T),Z(T)\Y1)
\end{equation}
where
\beq{7.61}
T={\rm TCG}.
\end{equation}
According to the Astronomical Almanach we have
\beq{7.62}
{\rm ERA}(T_u)=2\pi(0.7790572732640+1.00273781191135448\, T_u)
\end{equation}
where
\beq{7.63}
T_u=\X1[{\rm JD(UT1)}-2451545.0\Y1],
\end{equation}
UT1 is a universal time.
\beq{7.64}
t_1=\frac1{36525}\X1[{\rm JD(TT)}-{\rm2000\,January\,}1^d12^h\,{\rm TT}\Y1],
\end{equation}
TT is a Terrestial Time.
\beq{7.65}
{\rm TCG}-{\rm TT}=L_G\t ({\rm JD}-2443144.5)\times 86400.
\end{equation}
The time TT is a geocentric time on a surface of a geoid,
\beq{7.66}
L_G=\frac{U_G}{c^2}
\end{equation}
where $U_G$ is a potential of a geoid. TT is connected with a scale of atomic
time (TAI).

Let us notice that we should rescale space coordinates going from GCRS to
ITRS, i.e.\ $\vec X_{\rm ITRS}=(1-L_G)\vec X_{\rm GCRS}$ (see Eqs \er{7.51},
\er{7.66}).

Recently a revival of \qu\ approach to rotation in 3-\di al Euclidean space
has been observed. Let us remind some fundamental notions of \qu\ algebra.
\eu\qu s form a normed division algebra $\H$ over real numbers $\R$,
$\dim\H=4$. Thus they can be considered as vectors in a 4-\di al space \st
\beq{7.66k}
q=f_0+if_1+jf_2+kf_3 , \q f_0,f_1,f_2,f_3\in\R
\end{equation}
\st $i^2=j^2=k^2=ijk=-1$ and a product is defined as usual
\beq{7.67k}
qp=(f_0+if_1+jf_2+kf_3)(f_0'+if_1'+jf_2'+kf_3').
\end{equation}
This product is noncommutative, moreover, is associative. There is a unit
element of the product, $1\in\R$, and there is for any $q\ne0$ a unique
inverse $q'$ \st $qq'=1$. For $q'$ is unique we use a notation $q'=q^{-1}$.
One can define a norm of a \qu
\beq{7.68k}
\|q\|=\sqrt{f_0^2+f_1^2+f_2^2+f_3^2}.
\end{equation}
The most important in applications are \qu s of a unit norm $\|q\|=1$, known
also as unit \qu s. \eu\qu s can be described as a sum of a scalar and a
vector part
\beq{7.69k}
q=q_s+q_v,
\end{equation}
where
$$
q_s=f_0, \quad q_v=if_1+jf_2+kf_3.
$$
In this way every vector in a 3-\di al Euclidean space is equivalent to a
\qu\ with $q_s=0$. In this way R.~Hamilton realized his dream to find a
possibility to divide vectors in a 3-\di al space.

Every rotation in a 3-\di al space can be described by an orthogonal rotation
matrix $R$, $R^{-1}=R^T$, $\det R=1$, i.e.\ $\vec r{\hskip1pt}'=R\vec r$. Moreover, if
we represent vectors $\vec r, \vec r{\hskip1pt}'$ as \qu s with zero scalar part,
\beq{7.70k}
\bal q_{\vec r{\hskip0.5pt}'}&= ix'+jy'+kz',\\
q_{\vec r}&= ix+jy+kz,
\eal
\end{equation}
we can write an orthogonal transformation as
\beq{7.71k}
\ql{\vec r{\hskip0.5pt}'}=\ql R \ql{\vec r} \ql R^\ast,
\end{equation}
where $\ql R$ is a unit \qu\ corresponding to a rotation matrix and $\ql R^\ast$
is a conjugate \qu
\beq{7.72k}
q^\ast = f_0-if_1 -jf_2 -kf_3.
\end{equation}
In the case of unit \qu s $q^\ast=q^{-1}$. A norm of a \qu\ can be written as
$\|q\|^2=qq^\ast$ and an inverse \qu\ as $q^{-1}=\frac{q^\ast}{\|q\|^2}$.

Now we need a representation of unit \qu s by a rotation matrix and \ti{vice
versa}. One gets:
\beq{7.73k}
R=\left(\begin{matrix}
f_0^2+f_1^2-f_2^2-f_3^2 &\quad& 2(f_1f_2-f_0f_3) &\quad& 2(f_1f_3+f_0f_2) \\
2(f_1f_2+f_0f_3) && f_0^2-f_1^2+f_2^2-f_3^2 && 2(f_2f_3-f_0f_1)\\
2(f_1f_3-f_0f_2) && 2(f_2f_3+f_0f_1) && f_0^2 -f_1^2-f_2^2+f_3^2
\end{matrix}\right)^T
\end{equation}
where $f_0^2+f_1^2+f_2^2+f_3^2=1$.

Using standard rotation matrices (\ref{7.52}a,b,c) we can find \qu s
corresponding to the principal rotations given by angles $\vt_1,\vt_2,\vt_3$.
In this way we can write
\beq{7.74k}
\ql{\rm GCRS}=\ql{Q(t_1)} \ql{R(t_1)} \ql{W(t_1)}\ql{\rm ITRS}
 \ql{W(t_1)}^{-1}  \ql{R(t_1)}^{-1}  \ql{Q(t_1)}^{-1} .
\end{equation}
\refstepcounter{equation}\label{7.75k}
A \qu\ corresponding to a rotation by an angle $\vt$ reads:
$$
f_0=\cos\tfrac\vt2, \q f_i=\g_i \sin\tfrac\vt2, \q i=1,2,3,
$$
where $\g_i$ are cosines of an axis of a rotation.
\eu\qu s corresponding to rotation (\ref{7.52}a--c) read
$$
\dsl{
\hfill \ql{R_1}=\cos\tfrac{\vt_1}2+i\sin\tfrac{\vt_1}2
\hfill (\theequation\rm a)\cr
\hfill \ql{R_2}=\cos\tfrac{\vt_2}2+j\sin\tfrac{\vt_2}2
\hfill (\theequation\rm b)\cr
\hfill \ql{R_3}=\cos\tfrac{\vt_3}2+k\sin\tfrac{\vt_3}2
\hfill (\theequation\rm c)
}
$$
We can also use a not-unit \qu\ with a norm equal to $\sqrt{(1-L_G)}$.
Moreover, in this case we use the formula \er{7.71k}.

Vector \qu s $\ql{\rm GCRS}, \ql{\rm ITRS}$ correspond to vectors $\vec X_{\rm
GCRS}, \vec X_{\rm ITRS}$ and unit \qu s $\ql{Q(t_1)},\ql{R(t_1)},\ql{W(t_1)}$
to rotation matrices $Q(t_1),R(t_1),W(t_1)$. For an incredible importance of
\qu s in robotics, avionic and \sp\ navigation, molecular dynamics and
computer graphics there are a lot of computer programmes to proceed a
translation from \qu s to rotation matrices in several representations. They
can be easily used for our purposes. Eventually we give some useful formulae
(for an interested reader) to represent a \qu\ product. Let $p$ and $q$ be
two \qu s \st
\beq{7.76k}
p=p_s + \vec p_v, \q q=\ql s+{\vec q}_{\lower3pt\hbox{$\scriptstyle v$}},
\end{equation}
where $p_s,\ql s$ are scalar parts and $\vec p_v,\vec q{\lower3pt\hbox{$\scriptstyle v$}}$ are vector parts of
them.

In this notation a product is
\beq{7.77k}
pq=p_s\ql s - \vec p_v\cdot \vec q{\lower3pt\hbox{$\scriptstyle v$}}
+p_s\vec q{\lower3pt\hbox{$\scriptstyle v$}}+ \vec p_v \ql s
+\vec p_v\t \vec q{\lower3pt\hbox{$\scriptstyle v$}},
\end{equation}
where $\vec p_v\cdot \vec q{\lower3pt\hbox{$\scriptstyle v$}}$ means a scalar product of vector parts and
$\vec p_v\t \vec q{\lower3pt\hbox{$\scriptstyle v$}}$ means a vector product of two 3-\di al vectors
in~$\E^3$. \eu\qu s have a polar decomposition similar to complex numbers
(they can be considered as pairs of complex numbers with similar definitions
of a product as complex numbers being pairs of real numbers on a Hamilton
plane)
\beq{7.78k}
q=\|q\|\cdot U(q),
\end{equation}
where $U(q)$ is an element of $\SU(2)$ corresponding to \qu\ $q$. Unit \qu s
are elements of $\SU(2)$. In this way they form a universal covering space
$\SO(3)$ being elements of an algebra. This is very useful.

One can make the following identifications for $i,j,k$ and Pauli matrices
$$
i \to \si_1\si_2, \q j \to \si_3\si_1, \q k \to \si_2\si_3, \q 1 \to I,
$$
where
$$
\si_1=\mt{0 &\ & 1 \\ 1 && 0}, \q
\si_2=\mt{0 &\ & -\sqrt{-1} \\ \sqrt{-1} && 0}, \q
\si_3=\mt{1 &\ & 0 \\ 0 && -1}
$$
are Pauli matrices and
$$
I=\mt{1 &\ & 0 \\ 0 && 1}
$$
is a unit matrix $2\t 2$. The unit \qu s form a group isomorphic to $\SU(2)$.
This gives us another description of unit \qu s in terms of Pauli matrices
$$
U(q)=e^{\sqrt{-1}\,\a\si_1}e^{\sqrt{-1}\,\b\si_2}e^{\sqrt{-1}\,\g\si_3},
\q \|q\|=1, \q \a,\b,\g\in\R.
$$
Every \qu\ can be considered as a pair of complex numbers such that
$q=(z_1,z_2)$, $z_i\in \C$, $i=1,2$, according to the Cayley--Dickson
construction. Moreover, we can write any \qu\ $q$ as
$$
q=a+Jb, \q a,b\in\R,
$$
and $J$ is a \qu\ axis $J^2=-1$. \eu\qu s do not form an algebra over complex
numbers.

If we consider an \e\
$$
z^2=-1, \ z\in\H,
$$
we find that this \e\ has an infinite number of \so s in~$\H$. This \qu\ \so\
is a vector \qu\ with a unit norm $f_1^2+f_2^2+f_3^2=1$. These vectors form a
unit sphere $S^2\subset E^3$. Thus they correspond to direction in $E^3$
(axes). In this way we can consider \qu s as a sum of infinite copies of
complex planes for every established~$J$, \st $J^2=-1$. In other words,
$x+y\sqrt{-1}$ corresponds to $x+yJ$ with an identification for~$J$ and~$-J$.
Every non-real \qu\ belongs to a unique copy of $\C$. Using a decomposition
of $J=J_s+\vec J_v$ we can write
$$
J=J_s+\|\vec J_v\| U(\vec J_v)=U(\vec J_v).
$$
We can write a correspondence as
$x+y\sqrt{-1}$ corresponds to $x+yU(\vec J_v)$. There is also a decomposition
for every \qu\
$$
q=q_s+\|q_v\|U(q_v).
$$

Any details of
\qu\ theory one can find in Refs \cite{45k}, \cite{46k}, \cite{47k}.

We can also use another old notion of rotation cracovians (see
Ref.~\cite{48k}). They represent rotations. Moreover, a product is different.
If $A$ and $B$ are rotational matrices, a cracovian product is defined as
$A\ast B=A^TB$. In this way this product is noncommutative. It is also
nonassociative. They had some applications in astronometry. Now there is some
revival in aplications of astronomy in geography (see Ref.~\cite{49k}).

The question is why \qu s and why cracovians. In the case of \qu s the answer
is simple. The calculations using \qu s have a lower complexity and they
avoid a gimbal lock singularity. In the case of cracovians the calculations
are simpler too. T.~Banachiewicz used cracovians in his Ph.D. dissertation on
the Moon librations. However, cracovians cannot avoid a gimbal lock singularity.

In this way we reduce a motion on a distorted hyperbola to the observational
system on the Earth. Moreover, we should move from a barycenter of the Earth
to a surface of the Earth (a~surface of a geoid). In order to do this we write
\beq{7.67}
\X1(\wt x,\wt y,\wt z\Y1)=\vec{\wt r}=\vec X_{\rm ITRS}-\vec R(\ov\vf,\la)
\end{equation}
where $\vec r$ is a position vector (in cartesian \cd s) on a surface of a
geoid and $\vec R(\ov\vf,\la)$ a vector pointing from the barycenter of the
Earth to a point on a geoid surface. Simultaneously we have a reduction to
TT. The time UT1 is changed into a local time ${\rm UT1}+\la$, where $\la$ is
a geographical (astronomical) longitude of a point on a surface of the Earth.
$\ov\vf$ is a geographical (astronomical) latitude. At a point of an
observation with an elevation $\ov h$ \wrt a geoid we construct a celestial
sphere and we get
\beq{7.68}
(x_h,y_h,z_h)=\vec r_h=\vec{\wt r}-\ov h \vec n_1=r_h(\cos h\cos a,
\cos h \sin a,\sin h)
\end{equation}
($\ov h$ is along a normal to geoid), where $a$ is an azimuth and $h$ is a
zenithal height (attitude) (we can also use $\ov z$ such that $h+\ov z=\frac\pi2$),
\beq{7.69}
r_h=\|\vec X_{\rm ITRS}-\vec R-\ov h\vec n_1\|,
\end{equation}
$\vec n_1$ is a unit vector orthogonal to a geoid surface (a ``surface of the
Earth''). We should also correct a time to
\beq{7.69a}
{\rm TT_{loc}}={\rm TT}\X2(1-\frac{g\ov h}{c^2}\Y2)
\end{equation}
where $g$ is a \gr al \ac\ in a point of an observation on a geoid. In order
to define $\vec R$ we should introduce a reference ellipsoid and ellipsoidal
\cd \ system on it. In this way
\beq{7.70}
\vec R=\vec R_e+\vec n H
\end{equation}
where $H$ is an elevation of a geoid \wrt an ellipsoid of reference, and
$\vec n$ is a unit normal vector to it in a point of interest.
\beq{7.71}
\vec R_e=r_e \mt{\cos\psi\cos \la\\ \cos\psi\sin\la\\ \sin\psi}, \q
\vec n=\mt{\cos\ov\vf\cos \la\\ \cos\ov\vf\sin\la\\ \sin\ov\vf},
\end{equation}
$\la$ is a geographical (astronomical) longitude. We are using the same
spherical \cd s $\theta=\frac\pi2-\psi$, $\la$, $r$ in all cases,
\beq{7.72}
r_e=\ov a\sqrt{\frac{1-f^2}{1-f^2\cos^2\psi}}\,,
\end{equation}
$\ov a$ is a major semiaxis of an ellipsoid (an equator radius), $f$ is its eccentricity, and
$\psi$ is a geocentric latitude, which is connected to geographical
(astronomical) latitude by
\beq{7.73}
\tan\psi=(1-f^2)\tan\ov\vf.
\end{equation}
In general a potential of the Earth is given by the formula (see Eq.\ \er{7.73})
\bml{7.74}
U(r,\ov\vf,\la)=\sum_{n=0}^\iy \frac1{r^{n+1}} \sum_{k=0}^\iy
\X1(D_{nk}\cos k\la+E_{nk}\sin k\la\Y1)P_{nk}(\sin\psi(\ov\vf))\\
{}+\frac{\ov\o{}^2r^2}{3}\,\X1(1-P_{20}(\sin\psi(\ov\vf))\Y1).
\end{multline}
$\frac\pi2-\psi$ and $\la$ are spherical \cd s,
$P_{nk}$ are generalized Legendre \f s
\beq{7.75}
P_{nk}(t)=\frac1{2^nn!}(1-t^2)^{k/2}\,\frac{d^{n+k}(1-t^2)^n}{dt^{n+k}}\,.
\end{equation}
$D_{nk}$ and $E_{nk}$ are \cf s, $\ov\o$ is an angular velocity of the Earth
(see \er{7.38}, now $D_{nk}$ and $E_{nk}$ are \ct).

Centrifugal \pt, monopole \pt, quadrupole and higher multipole \pt\ terms cooperate to
give an equi\pt\ on the rotating Earth (on a geoid).

Eq.~\er{7.74} can be written in a different form
\bml{7.74a}
U(r,\ov\vf,\la)=\frac{M_\oplus G_N}r \X3\{1-\sum_{n=1}^\iy
\sum_{k=0}^n \X2(\frac{\,\ov a\,}r\Y2)^n\X1(I_{nk}\cos k\la
+K_{nk}\sin k\la\Y1)P_{nk}(\sin\psi(\ov\vf))\Y3\}\\
{}+\frac{\ov\o{}^2r^2}3\X1(1-P_{20}(\sin\psi(\ov\vf))\Y1)
\end{multline}
(see Eq.\ \er{7.73}). In this formula some $I_{nk}$ and $K_{nk}$ have simple physical
interpretation
$$
\bal
I_{10}&=I_{11}=K_{11}=I_{21}=K_{21}=0\\
I_{20}&=\frac{C-\frac12(A+B)}{M_\oplus {\ov a}^2}\\
I_{22}&=\frac{A-B}{4M_\oplus {\ov a}^2}\\
K_{22}&=-\frac F{2M_\oplus {\ov a}^2}
\eal
$$
where $A,B,C$ are principal inertial moments of the Earth and $F$ is a
deviation moment (an axis $Oz$ is along a direction of an axis of maximal
principal moment of inertia, and an origin is a barycenter of the Earth).
$\ov a$ is a major semiaxis of an ellipsoid of reference.

For a surface of a geoid is an equipotential surface, the geoid surface is
given by the \e
\beq{7.77}
U(r,\ov\vf,\la)=U_G={\rm const.}
\end{equation}
or
\beq{7.78}
r=r(\ov\vf,\la).
\end{equation}
$D_{nk}$ and $E_{nk}$ are known from satellites' motion up to higher order. In
practice it is enough to get only a few \cf s. For $k=0$ (ordinary Legendre
polynomials) a potential depends only on~$\ov\vf$. For $n=2l$ Legendre \f s
are symmetric \wrt an equator, for $n=2l+1$ they describe some nonregularity
on north and south hemispheres. We need harmonics with $k\ne0$. For a geoid
\er{7.77} we have of course an ellipsoid of reference (important also from
cartographical point of view).

Eq.\ \er{7.74} can be simplified by taking only terms with $k=0$. In this way
we get
\beq{7.79}
\ov U(r,\ov\vf)=\frac{G_NM_\oplus}r \,\X3[1-\sum_{n=2}^\iy
\X2(\frac{\,\ov a\,}r\Y2)^n I_nP_{n0}(\sin\psi(\ov\vf))\Y3]
+\frac{\ov\o{}^2r^2}3 \,(1-P_{20}(\sin\psi(\ov\vf)))
\end{equation}
where we reparametrize Eq.\ \er{7.74} and drop a dependence on~$\la$ (see
Eq.\ \er{7.73}). Usually
it is enough to take $I_2,I_3,I_4,I_5,I_6$. For this a geoid looks like a
pear.

Let us notice that in order to calculate a correction $\frac{g\ov h}{c^2}$ to~TT
it is enough to use a formula for an ellipsoid of reference WGS-84
\beq{7.80}
g=9.780049(1+0.005317\sin^2\ov\vf+0.000007\sin^22\ov\vf)\tfrac{\rm m}{\rm
s^2}\,.
\end{equation}
We have also different formulae
\beq{7.80a}
g=g_0(1+a_1\sin^2\ov\vf+a_2\sin^22\ov\vf),
\end{equation}
where
\beq{7.80b}
\gathered
a_1=\frac52 \,q -f -\frac{17}{14}\,qf, \quad
a_2=\frac{f^2}8 - \frac58\,qf, \quad
q=\frac{\ov \o{}^2\ov a}{g_0},\\
g_0=\frac{G_NM_\oplus}{\ov a{}^2}\X2(1+\frac{3I_2}2 -q\Y2), \quad
\ov\o=7.292115 \times 10^{-5}\tfrac{\rm rad}{\rm s},\\
I_2=0.001920 \simeq \frac f2,\\
g_0=9.780049 \,\tfrac{\rm m}{\rm s^2} \hbox{ ---an \ac\ on an equator},\\
\ov a=6378137.0\,{\rm m} \ \hbox{(WGS-84)}, \quad
G_NM_\oplus=k^2=3986005 \times 10^8 \hbox{ (WGS-84)},\\
g_0=9.780318(1+5.3024\times 10^{-3}\sin^2\ov\vf
-5.9\times 10^{-6}\sin^22\vf)\,\tfrac{\rm m}{\rm s^2}.
\endgathered
\end{equation}
$$
\frac {dg}{d\ov h}= -0.0000030877 (1-1.39\times 10^{-3}\sin^2\ov\vf)\tfrac1{\rm
s^2},
$$
where $\ov h$ is an elevation over a geoid.
\beq{7.80c}
\gathered
g_{\bar\vf}(\ov h)=g_{0\bar\vf}-(g_3+g_4\cos2\ov\vf)\ov h,\\
g_{0\bar\vf}=g_0+g_1\cos2\ov\vf+g_2\cos^22\ov\vf
\endgathered
\end{equation}
where
$$
\gathered
g_0=9.80611\,\tfrac{\rm m}{\rm s^2}, \quad g_1=0.0026373g_0, \quad
g_2=0.0000059g_0,\\
g_3=3.085462\times 10^{-6}\tfrac1{\rm s^2}, \quad
g_4=2.27 \times 10^{-9}\tfrac1{\rm s^2}
\endgathered
$$
(see Refs \cite{45n}--\cite{47n}).

In this case we have only the \cf s $I_2,I_4,I_6,I_8$,
\beq{7.81}
U_G=6263686.0850\times10\tfrac{\rm m^2}{\rm s^2}\,.
\end{equation}

The problem, which is not important for us here, is to use a difference
\beq{7.82}
U(r,\ov\vf,\la)-\ov U(r,\ov \vf)=W(r,\la,\ov\vf)
\end{equation}
in order to find a real shape of the Earth (it means, to find a real
elevation of a surface of the Earth over a geoid). This is the so called
Stokes idea (Stokes' parameters). An important point of our considerations is
to connect UT1 to~TT.  It is solved by introducing a $\D T$ correction ${\rm
UT1}={\rm TT} +\D T$.

We can also proceed in a following way, introducing UTC \st
$$
\gathered
{\rm TAI}-{\rm UTC}=32\,{\rm s}\\
\hbox{and \ }{\rm UT1}={\rm UTC}+[{\rm UT1}-{\rm UTC}]_{\rm IERS}
\endgathered
$$
where a correction $[{\rm UT1}-{\rm UTC}]_{\rm IERS}$ is given by IERS
(International Earth Rotating Service). UTC is not measured by any clock on
the Earth. UTC (\eu\cd d Universal Time) is calculated by the Bureau
International des Poides et Measures (BIPM). In particular, UTC(NIST) is a
time generated and distributed by NIST (National Institute of Standard and
Technology).

Let us consider $\D T$ as a \f\ of a terrestial time TT. In this way one gets
\beq{7.83}
{\rm UT1}={\rm TT}+\D T={\rm TAI} +32.184\,{\rm s}+\D T
\end{equation}
where TAI (International Atomic Time, Temps Atomique International (in French))
is a time measured by atomic clocks on the Earth. $\D T$ can be expressed by
TT or TAI as a polynomial \f
\beq{7.84}
\D T=\D T(T)=W(T-T_0)=W({\rm TAI}-T_0')=\sum_{n=0}^N a_nu^n
\end{equation}
where $u=T-T_0={\rm TAI}-T_0'$ for several different periods of time (see
\cite{42a}, \cite{43}, \cite{44}). It is possible to get these corrections
from the International Earth Rotations and Reference Systems Service (IERS)
(\cite{43}, \cite{45}). In those corrections it is possible to include
earthquakes and atmospheric phenomena as El Ni\~no and La Ni\~na. All of
these effects can change Stokes parameters.
Using this expression,
$\theta={\rm ERA}$ is a complicated \f\ of an
atomic time TAI (or GPS time, ${\rm GPST}={\rm TAI}+19^{\rm s}+{\rm Co}$).
\beq{7.85}
\ov \o =\frac{d\theta}{d({\rm TT})}=\pz\theta{\rm(TAI)}=\ov \o_0+\D\ov \o({\rm TAI})
\end{equation}
where $\ov\o_0$ is a \ct\ and $\D\ov\o$ is a correction. Thus one gets from
\er{7.74}
\beq{7.86}
U_G \simeq U_{G_0}(\ov\o_0)+\frac23\,\ov\o_0\D\ov\o r^2(1-P_{20}(\sin\psi(\ov\vf)))
\end{equation}
where $r=r(\la,\ov\vf)$ (see Eq.\ \er{7.78}).

$\D\ov\o$ gives an additional correction to $\rm TT_{loc}$,
\beq{7.86a}
{\rm TT_{loc}}={\rm TT}\X3(1-\frac{U_G(\ov\o_0)}{c^2}-\frac23\,\frac{\ov\o_0\D\ov\o}{c^2}
\, r^2\X1(1-P_{20}(\sin\psi(\ov\vf))\Y1)-\frac{g\ov h}{c^2}\Y3)
\end{equation}
where $\D\ov\o=\D\ov\o({\rm TT})$. $U_G(\ov\o_0)$ is a \pt\ \er{7.74a} or
\er{7.79} with $\ov\o_0$ in a place of~$\ov\o$. $U_G(\ov\o_0)$ can be
considered as $U_G$ (see Eq.~\er{7.81}).

Let us consider Eq.\ \er{7.69a}---a time correction for an elevation $h$ over
a surface of a geoid. We change the term $\frac{gh}{c^2}$ into
\beq{7.87}
\int_0^h \frac{\vec n\vec g(\la,\ov\vf,x)}{c^2}\,dx
\end{equation}
or into
$$
\frac{W(r,\la,\ov\vf)}{c^2}\,.
$$
($\vec g(\la,\ov\vf,x)$ is the Earth \gr al \ac).

We can define a next metric tensor $G_\m^{\rm loc}$ for a frame of
observation. In this case
$$
G_{\a\b}^{\rm rot}=\pp{x^{\mu'}}{x^\a}\,\pp{x^{\nu'}}{x^\b}\,G_{\a\b}^{\rm
loc}, \q x^\a=(\ov\vf,\la,\ov h,{\rm TT_{loc}}).
$$

The proposed corrections \er{7.86a} and \er{7.87} are very small. Moreover,
in an advent of a new generation of atomic clocks (based on Einstein--Bose
condensates, see Refs \cite{48n}, \cite{49n}) they could be measured in a next
future leading to local time variations \wrt an elevation,
geographic latitude and longitude. These technologies are progressing very
rapidly (see Refs \cite{50n}--\cite{54n}).

See also \cite{kappa} for cold $^{88}{\rm Sr}$ and $^{87}{\rm Sr}$
applications in atomic clock metrology. A~new possibility to use atomic
nucleus energy levels in metrology, in particular $^{229}{\rm Th}^{3+}$ ion,
see Ref.~\cite{kappa1}. In the last case we use an energy transition
of $7.6\,\rm eV$ ($160\,\rm nm$) getting an accuracy $10^{-19}$.

There is a significant progress on a portable rubidium fountain frequency
standard (see \cite{57n}). Due to this we can measure $W(r,\la,\ov\vf)$ very
precisely.

Thus for an orbit (a distorted \hy\ orbit) we get at a point of observation
\beq{7.88}
\bal
a&=a({\rm TT_{loc}},\la,\ov\vf)\\
h&=h({\rm TT_{loc}},\la,\ov\vf)
\eal
\end{equation}
and the time $\rm TT_{loc}$ is given by atomic clock at a point of
observation. The only one thing which we should do is to take under
consideration an atmospheric refraction at a point of observation. Taking
many such observations with a great precision we can measure the orbit
parameters $a_h$, $e$, $J$, $\ve$, $\O$, $\ov\pi$, $\ov T$ and the parameter
$b$. This gives us a feedback to new observation and a value of a
parameter~$b$ can be precisely predicted.

\section{The \ct\ $\ov b$ and the Hubble \ct}
Let us consider Eq.\ \er{2.25} in the following form
\beq{8.1}
\X3[\frac1{x^2}\,\pp{}x \X2(x^2\,\pp\vf x\Y2)-\wt\ve \ex{n\vf}(\ex{2\vf}-1)\Y3]
\frac1{R^2(t)}+ \frac1{R^3(t)}\X3[\pp{}t\X2(R^3(t)\,\pp\vf t\Y2)\Y3]=0
\end{equation}
where $R(t)$ is a scale factor of a spatially flat \co ical model. It means that
we consider spherically \s\ case, but in a \co ical background. Thus
$\vf=\vf(x,t)$. Let us suppose that the first term in quadratic brackets in
Eq.~\er{8.1} is zero. Simultaneously, the second term in quadratic brackets
is zero, too.

In this way we get
\bg{8.2}
\frac1{r^2}\,\pp{}r \X2(r^2\,\pp\vf r\Y2)-\wt\ve \ex{n\vf}(\ex{2\vf}-1)=0\\
\pp{}t\X2(R^3(t)\,\pp\vf t\Y2)=0 \label{8.3}
\end{gather}
where we change a scale from $x$ to $r$ (see Section~3). In this way we
separated space and time \cd s. Let us find a \so\ of Eq.~\er{8.3}. One gets
\beq{8.4}
\vf=C_1 \int \frac{dt}{R^3(t)}+C_2,
\end{equation}
where $C_1=C_1(r)$, $C_2=C_2(r)$ and $\vf$ depends parametrically on~$r$.
Simultaneously we get a \so\ of \er{8.2} (see Section~3):
\beq{8.5}
\vf(r,t)=\ov C_2+\frac q6\,r^2
\end{equation}
where $\ov C_2=\ov C_2(t)$, $q=q(t)$ depend parametrically on~$t$. Let us
consider \er{8.4} for a Lema\^\i tre--Freedman model filled by a dust with
\co ical \ct\ greater than zero. This model can be considered as a model of our
contemporary Universe for we can neglect a radiation energy density and
consider ``dark matter'' as a dust (CDM$\La$-model).
We have (see Refs \cite{21},~\cite{46})
\bg{8.6}
R(t)=\X3(\frac{1-\O_{\La,0}}{\O_{\La,0}}\Y3)^{1/3}
\sinh^{2/3}\X3(\frac32\sqrt{\frac{\La c^2}3\, t}\Y3)\\
\O_{\La,0}>0 \nn\\
H(t)=\sqrt{\frac{\La c^2}3}\,\coth\X3(\frac32\sqrt{\frac{\La c^2}3}\,t\Y3)
\label{8.7}\\
\O_{\La,0}=\frac{\La c^2}{3H_0^2} \label{8.8}\\
\rho_m(t)=\frac{\La c^2}{8\pi G_N} \cosech^2\X3(\frac32\sqrt{\frac{\La
c^2}3\, t}\Y3). \label{8.9}
\end{gather}
$\rho_m(t)$ is a density of matter in the \Un, $H(t)$ a ``Hubble \ct''. Using
Eq.~\er{8.6} one gets from \er{8.4}
\beq{8.10}
\vf(r,t)=C_3(r)H(t)+C_2(r)
\end{equation}
where $C_3=C_3(r)$ depends parametrically on $r$.

Let us compare \er{8.10} and \er{8.5}. If we want to identify this
representation of the \so\ we should put
\bg{8.11}
\frac{q(t)}6=\xi_1 H(t), \q \xi_1={\rm const.}\\
C_1(r)=r^2. \label{8.12}\\
C_2(r)=\ov C_2(t)={\rm const.}=b_1. \label{8.13}
\end{gather}
One gets
\bg{8.14}
\vf(r,t)=b_1+H(t)\xi_1r^2\\
a=\xi_1H(t). \label{8.15}
\end{gather}
In this way the ``\ct'' $\ov b$ equals
\bg{8.16}
\ov b=-b|a|=-\xi_1 k^2\ex{b_1}H(t)=c\xi H(t), \q \xi={\rm const.}\\
\ov b(t)=\xi c^2\sqrt{\frac13}\coth\X3(\frac32\,c\sqrt{\frac\La3}\,t\Y3).
\label{8.17}
\end{gather}

Let us estimate the \ct\ $\xi$, which is dimensionless, using the results
from Section~3. In order to do this, we write Eq.~\er{8.17} for our
contemporary epoch
\beq{8.18}
\ov b=c\xi h\ov H_0.
\end{equation}
For we are using a specific model of the \Un, we should use data from
Ref.~\cite{11}. They suppose the same model. One gets
\beq{8.19}
\xi=\frac1h\,\frac{c\ov H_0}{\ov b}=1.557
\end{equation}
and finally
\beq{8.20}
\xi=1.557\begin{matrix}+0.083\\-0.081\end{matrix}\,.
\end{equation}
Let us consider both scales of length obtained in Section~3, $r_0$ and $R$,
\beq{8.21}
r_0=\frac1{\sqrt{\xi_1}}\,H^{-1/2}(t)=\xi_2H^{-1/2}(t)
\end{equation}
and let us estimate the \ct\ $\xi_2$ using results from Section~3. One gets
\beq{8.22}
\xi_2=\X2(0.631\begin{matrix}+0.0239\\-0.0243\end{matrix}\Y2)
\frac{\rm AU}{\sqrt s}\,.
\end{equation}
For $\ov R=\sqrt{\frac{k^2}{|\ov b|}}$ we get
\beq{8.23}
\ov R=\sqrt{\frac{k^2}{c\xi H(t)}}=\frac1c\sqrt{\frac{k^2}\xi}\tanh^{1/2}
\X3(\frac32\,c\sqrt{\frac\La3}\,t\Y3).
\end{equation}
One easily gets
\beq{8.24}
\eta=\frac{\,\ov R\,}{r_0}=\frac{\sqrt{k^2}}{\xi_2\sqrt{c\xi}}={\rm const.}
\end{equation}
Thus the ratio of both length scales is \ct\ on a \co ical time scale.

Let us notice that we obtain the same formula for $\ov b$ and $a$ (also for
$r_0$ and~$\ov R$) using a model of Freedman spatially flat \Un\ filled with
matter (without a \co ical \ct). However, we cannot obtain such formulae for
de~Sitter \Un\ and for radiation filled \Un. In our case $\La$ is equal to
$\la_{\rm co}$ (see Section~2).

One can easily resolve also a mysterious \cn\ of an \an\ \ac\ to a \co ical
\ct, i.e.,
\beq{8.25}
\ov b\simeq c^2\X2(\frac \La3\Y2)^{1/2}.
\end{equation}
Using Eq.\ \er{8.7} we get
\beq{8.26}
\ov b=\xi c^2\X2(\frac \La3\Y2)^{1/2}\coth\X3(\frac32\sqrt{\frac{\La c^2}3}\,t
\Y3)\simeq \xi c^2\X2(\frac \La3\Y2)^{1/2}
\end{equation}
because for a large argument $\coth x$ is close to one.

Let us consider a \co ical scale of length caused by a \ct\ $\ov b$
\beq{8.27}
R_u=\frac{c^2}{|\ov b|}\simeq 1.2\times 10^{26}{\rm km}.
\end{equation}
The estimated size of the visible Universe is $5\times 10^{26}$\,km.

Using Eq.\ \er{8.6} one finds
\beq{8.27a}
R_u(t)=\frac{c^2}{b(t)}=\frac1\xi \sqrt{\frac3\La}\tanh\X3(\frac32\,c
\sqrt{\frac\La3}\,t\Y3)\ne R(t).
\end{equation}
For $\tanh x$ is close to one for a large argument,
\beq{8.28}
R_u(t_0)\simeq\frac1\xi\sqrt{\frac3\La}
\end{equation}
and the scale \er{8.27a} is connected to the \co ical \ct.

In this way we solve a problem of mysterious \cn\ between a Hubble \ct, a \co
ical \ct\ and
an \an\ \ac. The \so\ is \ap e for we separate spatial and time \cd s. This
is rational for we do not see any coupling between \co ical and the \SS\
degrees of freedom.

Finally, we give the following comment. We use the Freedman--Lema\^\i tre
model of the \Un\ according to a paradigm in modern \co y. Moreover, there
are some different approaches (see Ref.~\cite{47}) using Lema\^\i tre--Tolman
inhomogeneous models or even Szekers anisotropic models. If a paradigm
changes, the results from Section~3 will be still correct.

Let us notice that $\xi$ and $\xi_2$ depend on initial conditions. In a
system different from the \SS\ they can be different. This is reasonable.
According to A.~Wigner, initial conditions do not belong to Physics, they belong
to Geology, Cosmogony or to \eu\co y, and this is a case.

Different initial conditions should be applied if we want to consider a model
of an \an\ \ac\ to e.g. Neptun satellites system.

Several authors consider local orbital effects forced by a Hubble \ct\ (H)
(see Refs \cite{N64}--\cite{N70}). In all of these papers the effect of a \co
ical expansion on a planetary motion in the \SS\ is negligible or zero (see
Ref.~\cite{N67}). In order to find some post-Newtonian effects in such a case
it is necessary to develop a post-Newtonian expansion in a Robertson--Walker
\Un\ \wrt a Hubble \ct\ which is under construction (see Ref.~\cite{N70}).
None of those approaches concern us for in our case we get a \cn\ between our
\ct~$b$ (which is of order of an \an\ \ac) and a Hubble \ct. In this way we
get a possibility of a \co ical evolution of~$b$.

Many authors consider an influence of a \co ical \ct\ on planetary motion in
the \SS\ and a light bending. They obtained very interesting results getting
a negligible influence of a \co ical \ct\ via Kotler \so\ of Einstein \e s
and {\it non-zero} bending of light, even Kotler \so\ is not asymptotically
flat (see Refs \cite{N56}, \cite{N72}--\cite{N83}). These results have nothing to do with
our approach to connect a \co ical \ct\ to~$b$.

Let us mention that the embedding of a Schwarzschild \so\ in
Robertson--Walker \Un\ has been started by A.~Einstein and afterwards
developed by MacVittie. The \so\ by MacVittie interpolates between
Schwarzschild \so\ and Robertson--Walker \so. However, there is no physically
reasonable \e\ of state for a matter
(see Ref.~\cite{N67}). In a conformal theory of \gr\
(a~quadratic \gr al Lagrangian in a curvature---a square of a Weyl tensor)
there is an osculating \so\ between Schwarzschild-like and a \co ical-like.
About possible local effects of Hubble parameter in \SS\ we also have in
Refs~\cite{N83a,N65}.

\section{Relativistic effects}
In this section we consider relativistic effects with an anomalous \ac. Let
us consider an \e\ of motion in GR (also in NGT) of a point massive particle in
presence of an external \pt\ $\ov U(x)$, $x\in\R^4$. One gets
\beq{9.1}
u^\b \nabla_\b u^\mu=-\pp {\ov U}{x^\mu}=F^\mu
\end{equation}
where $F^\mu$ is an external four-force (four-\ac).

$\nabla_\b$ is a covariant \dv\ \wrt a \cn\ on a \spt. This \cn\ is not
necessarily a Levi-Civit\`a \cn. $\ov U(x)$ is an external \pt\ responsible for
an \an\ \ac\ (see Eq.~\er{5.4}). $u^\mu=\pz{x^\mu}{\tau}$ is a four-velocity of
a material point.

Using \cf s of a \cn\ one gets
\beq{9.2}
\pz{^2x^\mu}{\tau^2}+\G_{\a\b}^\mu \,\pz{x^\a}\tau \cdot \pz{x^\b}\tau
+\pp {\,\ov U\,}{x^\mu}=0.
\end{equation}
In the case of GR $\G_{\a\b}^\mu=\Chr\mu{\a\b}$ (Christoffel symbols).

Let us consider a Schwarzschild \spt\ in spherical \cd s
$x^\mu=(r,\theta,\vf,t)$
\beq{9.3}
ds^2=c^2\X2(1-\frac{2k^2}{c^2r}\Y2)dt^2-\X2(1-\frac{2k^2}{c^2r}\Y2)^{-1}dr^2
-r^2\X1(d\theta^2+\sin^2\theta\,d\vf^2\Y1), \q k^2=M_\odot G_N.
\end{equation}
Eq.\ \er{9.2} for a Schwarzschild \so\ can be reduced to
\beq{9.4}
\pz{^2u}{\vf^2}+u=\frac{k^2}{A^2}+3\,\frac{k^2}{c^2}\,u^2+\frac{b}{A^2u^2}
\end{equation}
where $u=\frac1r$.

If we use a spherical stationary \so\ from NGT (see Refs \cite9, \cite{37}),
\bg{9.5}
ds^2=c^2\X2(1+\frac{\ell ^4}{r^4}\Y2)\X2(1-\frac{2k^2}{c^2r}\Y2)dt^2
-\X2(1-\frac{2k^2}{c^2r}\Y2)^{-1}dr^2-r^2\X1(d\theta^2+\sin^2\theta\,d\theta^2)
\\
g_{[14]}=\frac{\ell ^2}{r^2}\,, \label{9.6}
\end{gather}
one gets
\beq{9.7}
\pz{^2u}{\vf^2}+u=\frac{k^2}{A^2}+3\,\frac{k^2}{c^2}\,u^2
+\frac{2\ell ^4c^2}{A^2}\,u^3+\frac{b}{A^2u^2}\,.
\end{equation}
$\ell $ is a \ct\ from NGT equal to a fermion charge of a source. We are using an
\ap ion of an \an\ \ac\ for $r>20\,AU$.

Let us consider a \hy\ orbit. In this case one gets
\beq{9.8}
\pz{^2u}{\vf^2}+u=a_h(e^2-1)+3\,\frac{k^2}{c^2}\,u^2+
\frac{2\ell ^4c^2}{k^2a_h(e^2-1)}\,u^3+\frac{b}{u^2k^2a_h(e^2-1)}\,.
\end{equation}
GR limit can be obtained by putting $\ell =0$. Eqs \er{9.4} and \er{9.7} are
generalization of a Binet \e. Let us estimate a distance from the Sun where
an \an\ \ac\ term is equal to a relativistic correction from GR. One gets
\beq{9.9}
3\,\frac{k^2}{c^2}\,u_1^2=\frac{b}{u_1^2k^2a_h(e^2-1)}
\end{equation}
where $u_1=\frac1{R_1}$.

Finally one gets
\beq{9.10}
R_1=k\root4\of{\frac{a_h(e^2-1)}{bc^2}}
\end{equation}
or
\beq{9.11}
R_1\simeq 3.39\root4\of{a_h(e^2-1)}\times 10^{-4}\,\rm AU
\end{equation}
where $a_h$ is measured in AU. It is easy to see that for $r>20\,\rm AU$ a
relativistic effect is much smaller than an \an\ \ac\ effect. This justifies
our approach to \hy\ orbit in Section~5 (neglecting this effect). If we take
$a_h=6\,\rm AU$ and for $e=3,7,10$ we get $R_1(e=3)=8.01\t 10^{-4}\,\rm AU$,
$R_2(e=7)=14.02\t 10^{-4}\,\rm AU$, $R_3(e=10)=16.73\t10^{-4}\,\rm AU$. These
values correspond to the Pioneer case. In the case
of NGT correction a result is similar,
\bg{9.12}
\frac{2\ell ^4}{k^2a_h(e^2-1)}\,u_2^3=\frac{b}{u_2^2k^2a_h(e^2-1)}\\
u_2=\frac1{R_2}, \q R_2=\root5\of{\frac{2\ell ^4c^2}{b}}\,. \label{9.13}
\end{gather}
Taking $\ell =\ell _\odot=3.1\times 10^3\,\rm km$ (a fermion charge of the Sun, see
Ref.~\cite{37}), one gets
\beq{9.14}
R_2\simeq 10.2\times 10^7\,{\rm km}\simeq 0.68\,\rm AU.
\end{equation}
This is of course much smaller than 20\,AU. Moreover, the value $\ell _\odot$ can
be much more smaller. For such a value of $\ell _\odot$ (as from
Ref.~\cite{37}) a quadrupole moment of mass for the Sun should be quite big
in order to get an agreement with observation data concerning a \ph\ movement
of Mercury orbit ($J_2=(5.5\pm1.3)\times10^{-6}$, see Ref.~\cite{50}).
According to modern measurement, $J_2=(1.9\pm0.3)\times10^{-7}$ (see
Ref.~\cite{51}). Thus we can neglect
relativistic effects from GR and NGT on a \hy\ orbit with comparison to an
\an\ \ac. Let us notice that in the formula \er{9.13} we get a scale length
\beq{9.15}
R_u\simeq\frac{c^2}{b}\simeq1.2\times10^{26}\,\rm km.
\end{equation}

Let us consider an extension of spherical stationary \so\ in GR and NGT to a
case with a \co ical \ct. In this case it is enough to shift for both metrics
(Eqs~\er{9.3}, \er{9.5})
\beq{9.16}
1-\frac{2k^2}{c^2r}\to 1-\frac{2k^2}{c^2r}-\frac{\La r^2}3
\end{equation}
where $\La$ is a \co ical \ct. In this case an \e\ of motion reads
\beq{9.17}
\pz{^2u}{\vf^2}+u=\frac{k^2}{A^2}+3\,\frac{k^2}{c^2}\,u^2+\frac{2\ell ^4c^2}{A^2}
\,u^3+\frac{b}{A^2u^2}+\frac{\La c^2}{3u^3}\,.
\end{equation}
In the case of \hy\ orbit one gets
\beq{9.18}
\pz{^2u}{\vf^2}+u=\frac1{a_h(e^2-1)}+3\,\frac{k^2}{c^2}\,u^2
+\frac{2\ell ^4c^2}{k^2a_h(e^2-1)}\,u^3+\frac{b}{u^2k^2a_h(e^2-1)}
+\frac{2\la}{u^3}
\end{equation}
($\la=\frac{\La c^2}6$).

It is easy to prove that a total energy is conserved during a motion
\beq{9.19}
\pz{}\tau\X2(\frac12\,g_{\mu\nu}u^\mu u^\nu+\ov U(x)\Y2)=0.
\end{equation}
This is valid for GR and NGT.

Let us consider a movement of a photon. One has
\beq{9.20}
g_\m k^\mu k^\nu=0
\end{equation}
where $k^\mu$ is a four-wave vector of a photon. From \er{9.20} one gets
\beq{9.21}
k^\a g_\m\nabla_\a k^\mu k^\nu=0
\end{equation}
and finally
\beq{9.22}
\pz{k^\mu}{\la}+\G_{\a\b}^\mu k^\a k^\b=0,
\end{equation}
$\la$ means an affine parameter along a light ray.

It means that an \an\ \ac\ has no influence on a photon motion (photons are
not accelerated). This is valid for GR and NGT.

Some authors develop a formalism of a post-Newtonian perturbation theory
in which 1PN reference trajectory is adopted (see Refs \cite{N84,N85}).
Here we are doing similarly, treating our model of an \an\ \ac\ as a
perturbation to 1PN reference trajectory in a Schwarzschild or Kotler \spt\
(in both GR and NGT cases). We can also consider an influence of a \co ical
\ct\ to relativistic bending of light (see Ref.~\cite{N83}).

\section{The effective \gr al \ct\ in the \SS}
Let us consider the effective \gr al \ct\ in our model
\beq{10.1}
G\eff(r,t)=G_N\exp\X1(-(n+2)\vf(r,t)\Y1)
\end{equation}
where
\beq{10.2}
\vf(r,t)=b_1+H(t)\,\frac{r^2}{\xi_2^2}
\end{equation}
(see Section 8).

A rate of a change in time of $G\eff$ reads
\beq{10.3}
\frac{\D G\eff}{G\eff}\simeq \frac{\dot G\eff}{G\eff}=\pz{}t \log G\eff=
\dot H(t)\,\pz{}H\log G\eff=-(n+2)\dot H(t)\,\frac{r^2}{\xi_2}
\end{equation}
or
\beq{10.4}
\frac{\D G\eff}{G\eff}\simeq (n+2)\,\frac{r^2}{\xi_2}\,(q(t)+1)H^2(t)
\end{equation}
where $q(t)$ is a deceleration parameter
\beq{10.5}
q(t)=-\frac{\ddot R(t)R(t)}{\X1(\dot R(t)\Y1)^2}\,.
\end{equation}
In this way we get a rate of change $G\eff$ as a \f\ of a distance from the
Sun and we connect it to a value of~$n$, $n=\dim H$ (a~\di\ of a gauge
unification group). Taking for our contemporary epoch
\beq{10.6}
q=-0.58
\end{equation}
we get
\beq{10.7}
\frac{\D G\eff}{G\eff}=3.8(n+2)h^2r^2\times 10^{-27}/{\rm per\ centry}
\end{equation}
where $h$ is a \di less Hubble \ct\ (${}\simeq0.7$). The variation of the \gr
al \ct\ obtained from the \SS\ measurement is
\beq{10.8}
-4.2\t10^{-14}<\frac{\dot G}G<7.5\t10^{-14}\frac1{\rm yr}
\end{equation}
(see Ref.\ \cite{51}).

The value \er{10.8} is statistically nonzero and our estimate even it depends
on~$r$ agrees with measurement.

\def\nd{(n+2)}
\def\eb#1{e^{#1B(r)}}
\def\ef#1{e^{#1\vF(r)}}
\section{\eu\rl\ model of an \an\ \ac}
In this section we develop a \rl\ model of an \an\ \ac. The model is based on
a model of a \gr al field in the \SS\ described in Appendices A and~D. It
consists of a metric
\beq{11.1}
ds^2=c^2e^{2A(r)}\,dt^2- e^{2B(r)}\,dr^2 - r^2(d\th^2+\sin^2\th\,d\vf^2)
\end{equation}
and a scalar field $\vF(r)$. $A(r)$, $B(r)$ and $\vF(r)$ satisfy the \e s
\er{D.15}--\er{D.17} which we quote here
\bea{11.2}
\pz{A(r)}{r}&=&\frac{\eb2-1}{4r}\\
\pz{^2B(r)}{r^2}&=&\frac{9-10\eb2+\eb4}{8r^2}\label{11.3}\\
\pz{^2\wt\vF(r)}{r^2}&=&-\frac{\eb2-1}{4r^2}\,\eb2 \label{11.4}\\
\vF(r)&=&\frac{n+2}{\ov M}\,\wt\vF(r).\label{11.5}
\end{eqnarray}
We suppose that $n=120$, $\ov M=1$ (see Appendix~D).

Simultaneously we define initial conditions for $A(r)$, $B(r)$ and
$\wt\vF(r)$ in such a way that $r$ is measured in $r_0$ unit,
$4.103\div4.104$ (see Section~3). The initial conditions are as follows:
\beq{11.6}
\bal
B(643)&=4\t10^{-12}\\
\pz B r(643)&=0\\
A(643)&=-4\t10^{-12}\\
\wt\vF(643)&=0\\
\pz {\wt\vF} r(643)&=0.
\eal
\end{equation}
The value $643r_0$ corresponds to the second scale of our model from
Section~3, $R=2640$\,AU. Thus the initial Cauchy condition has been defined
far away from the Sun. The initial Cauchy condition has been defined properly
(see Appendix~D for details). In our model there is also a density of a dust
inside the \SS\ which we consider to be a dark matter consisting of skewon
and quintessence particles. This density $\ov\rho(r)$ has been calculated in
our model (see \er{D.5}). This density has been estimated to be lower than
interplanetary matter inside the \SS. The field $\vF(r)$ is connected to the
effective \gr al ``\ct'',
\beq{11.7}
G\eff=G_N \ef{-\nd}=G_N \ef{-\frac{\nd^2}{\ov M}\wt}.
\end{equation}
In this model we use the following two parametrizations (see Appendix D):
\bea{11.8}
e^{2A(r)}&=&1-\frac{r_s}r-2\ov V_1(r)\\
e^{2B(r)}&=&\frac1{1-\frac{r_s}r-2\ov V_2(r)} \label{11.8a}
\end{eqnarray}
and
\bea{11.9}
e^{2A(r)}&=&1-\frac{r_s}r \,\ef{-\nd}-\frac{\La r^2}3-2\wt V_1(r)\\
e^{2B(r)}&=&\frac1{1-\frac{r_s}r \,\ef{-\nd}-\frac{\La r^2}3-2\wt V_2(r)}
\label{11.10}
\end{eqnarray}
$\ov V_1(r)$, $\ov V_2(r)$ give a deviation from the Schwarzschild \so, $\wt
V_1(r)$, $\wt V_2(r)$ give a deviation from the Schwarzschild \so\ with a \co
ical \ct~$\La$. In this case we take into account a changing of the effective
\gr al ``\ct'' introducing a factor $\ef{-\nd}$ (see Appendix~D).

We define also \ac s connected with $\ov V_1(r)$, $\ov V_2(r)$, $\wt V_1(r)$,
$\wt V_2(r)$
\bg{11.11}
\bal b_1(r)&=\pz{\ov V_1}r(r)\\
b_2(r)&=\pz{\ov V_2}r(r)\eal \\
\bal \wt b_1(r)&=\pz{\wt V_1}r(r)\\
\wt b_2(r)&=\pz{\wt V_2}r(r)\eal \label{11.12}
\end{gather}

According to our investigations from Appendix D, an \an\ \ac\ which has an
important influence on a planetary motion and on a motion of \sp s is
$b_1(r)$. Simultaneously an \an\ \ac\ is given by the formula
\beq{11.13}
b=-\frac{c^2}2\,b_1(r)
\end{equation}
or in a proper system of units
\beq{11.14}
b=-7.33\t10^4b_1(r).
\end{equation}
In the case of a different parametrization
\beq{11.15}
b=-\frac{c^2}2\,\wt b_1(r)
\end{equation}
or
\beq{11.16}
b=-7.33\t10^4\wt b_1(r).
\end{equation}

Let us notice the following fact. The model considered here has been obtained
as a \so\ of the full field \e s from the \eu\nos\ Jordan--Thiry Theory (see
Ref.~\cite4) in spherically \s\ and stationary case (see Appendix~A and
Appendix~D). We consider \e s for $A(r)$, $B(r)$, $\vF(r)$ (after eliminating
$\ov\rho(r)$ from the \e s) in a limit $r\ll L$  where $L\simeq 10\xi$\,Mpc,
$\xi$~is a factor of order one. Thus in the \SS\ this \ap ion is pretty
satisfied. Let us notice that among the \e s \er{11.2}--\er{11.4} \e\
\er{11.3} for $B(r)$ is crucial and solving it we get $A(r)$ and $\vF(r)$ by
quadratures. Maybe it is possoble to solve Eq.\ \er{11.3} exactly using some
special \f s. This problem is still under investigations. Up to now we know
that around a zero the \so\ is nonsingular.

In our parametrization we use for $r_s$ a Schwarzschild radius for the Sun
$$
r_s\simeq 3\,{\rm km}, \q \La=10^{-52}\,{\rm\frac1{m^2}}
$$
(see Appendix D for details). \eu\e s \er{11.2}--\er{11.4} with initial
conditions \er{11.6} have been solved numerically using {\tt NDSolve}
instruction from Mathematica~7. In this way we can calculate
\beq{11.17}
\bal
A(7.03)&=-4.865794911570204 \t 10^{-12}\\
B(7.03)&=3.469214116529213 \t 10^{-13}\\
\pz Br(7.03)&=-2.7206955894132885 \t 10^{-14}
\eal
\end{equation}
We calculate also
\bea{11.18}
\ov V_1(7.03)&=&4.94562 \t 10^{-11} \\
b_1(7.03)&=&-3.42985 \t 10^{-10}.\label{11.19}
\end{eqnarray}
The value $7.03r_0$ corresponds to the value $28.88$\,AU. At this point the
value of an \an\ \ac\ from Anderson et al.\ data (see Refs \cite{1,2,3}) is
evaluated for \P0 to be
\beq{11.20}
b_A(28.88)=-8.88 \t 10^{-10}\,{\rm \tfrac m{s^2}}
\end{equation}
and an uncertainty for this value is quite small (see also Fig.~\ref{f1} and
Appendix~C). The value $b_A(7.03)$ in Eq.\ \er{11.19} corresponds to
$b_1(7.03)$
\beq{11.21}
b_1^A(7.03)=-1.21 \t 10^{-14}.
\end{equation}
In this way, using our parametrization for $e^{2A(r)}$ with $\ov V_1(r)$ one
gets
\beq{11.22}
e^{2A(7.03)}=1-\frac{r_s}{7.03}-2\ov V_1(7.03).
\end{equation}
For $A(7.03)$ is small we get
\beq{11.23}
A(7.03)=-\frac12\X2(\frac{r_s}{7.03} + 2\ov V_1(7.03)\Y2).
\end{equation}
However, our calculated value for $b_1(7.03)$ does match the Anderson et al.\
data value $b_1^A(7.03)$ (see Eqs \er{11.19} and \er{11.21}). Thus we
redefine $A(7.03)$ in the following way
\beq{11.24}
A_{\rm
re}(7.03)=-\frac12\X2(\frac{r_s}{7.03}+2\,\frac{b_1^A(7.03)}{b_1(7.03)} \ov
V_1(7.03)\Y2)
\end{equation}
getting
\beq{11.25}
A_{\rm re}(7.03) = -6.957012802275961 \t 10^{-10} (0.5 +
0.00012069886778807667)
\end{equation}
where $A_{\rm re}(7.03)$ is a corrected value of $A(7.03)$. This value will
be considered as an initial value for $A(r)$ at $r=7.03$ together with
calculated values for $B(7.03)$, $\pz Br(7.03)$ and $\wt\vF(7.03)$,
$\pz{\wt\vF}r(7.03)$. In this way we get the following initial conditions at
$7.03$: \er{11.17}, \er{11.25} and
\beq{11.26}
\bal
\wt\vF(7.03)&=-9.568184581692126 \t 10^{-13}\\
\pz{\wt\vF}r(7.03)&=-5.348841232880765 \t 10^{-15}.
\eal
\end{equation}

\def\<#1>{\left\langle #1\right\rangle}
Let us sum up. We get a model of the \gr al field in the \SS\ from the
\eu\nos\ Jordan--Thiry Theory considering the full field \e s in the
theory, i.e.\ Eqs \er{A.115a}--\er{A.121} in the case of \s\ metric $g_\m
=g_{\nu\mu}$. We consider these field \e s in a spherically \s, stationary
case, i.e.\ Eqs \er{A.124}--\er{A.126}. We prove consistency of these \e s
via Bianchi identities. After eliminating $\ov\rho(r)$ from the field \e s we
can solve \e s for $A(r)$, $B(r)$ and $\vF(r)$, \er{D.10}--\er{D.12}. In the
small $r$ limit $r\ll 1$ we derive Eqs \er{D.15}--\er{D.16}. Eqs
\er{D.10}--\er{D.12} and \er{D.15}--\er{D.16} have a well defined Cauchy
initial problem for $r\in\<a,b>$ where $0\notin\<a,b>$. Thus we define such
an initial problem at $r=643r_0$. At this point, according to the model from
Section~3 the Newtonian \gr al \pt\ of the Sun is equal to the  \pt\ of an
\an\ \ac. Simultaneously the first \dv\ of a sum of both \pt s is equal to
zero. We suppose that the value $\vF(r)$ and its first \dv\ $\pz\vF r(r)$ at
this point is equal to zero. We use these initial conditions to solve the
problem (see Appendix~D for details of this \so). Afterwards we use the \so\
to define a Cauchy initial problem at 7.03 calculating values of relevant
quantities at 7.03. Using Anderson et al.\ data we find a discrepancy of an
\an\ \ac\ from Anderson et al.\ data and our calculated value. We tune the
value of $A(7.03)$ calculated by us to a new value $A_{\rm re}(7.03)$ in
order to remove a discrepancy. In that moment we arrive to appropriate
initial conditions of our problem. Eqs \er{11.2}--\er{11.4} and initial
conditions \er{11.17}, \er{11.25} and \er{11.26} are our model of the \gr al
field in the \SS\ together with Eqs \er{11.5}, \er{11.1}, \er{11.17} and
\er{D.5}. In this way they are model of an \an\ \ac. This model is \rl. We
will consider a movement of a massive point body in a background of this
field using general \rl\ treatment according to the formalism of Appendix~D.
Let us solve numerically Eqs \er{11.2}--\er{11.4} with our initial
conditions. In all relevant formulae we put $n=120$, $\ov M=1$ (see
Appendix~D for details).

On Fig. \ref{pp} we plot our results for $A(r)$, $B(r)$, $e^{2A(r)}$,
$e^{2B(r)}$, $e^{2(A(r)+B(r))}$, $\pz{}r(e^{2A(r)})$,
$\pz{}r(e^{2B(r)})$, $\pz{}r(e^{2(A(r)+B(r))})$, $\wt\vF(r)$, $\pz{\wt\vF} r(r)$,
$\frac{G\eff}{G_N}$ for several regions of~$r$.

\bigskip

\begin{figure}[h]
\hbox to \textwidth{\ing[width=0.45\textwidth]{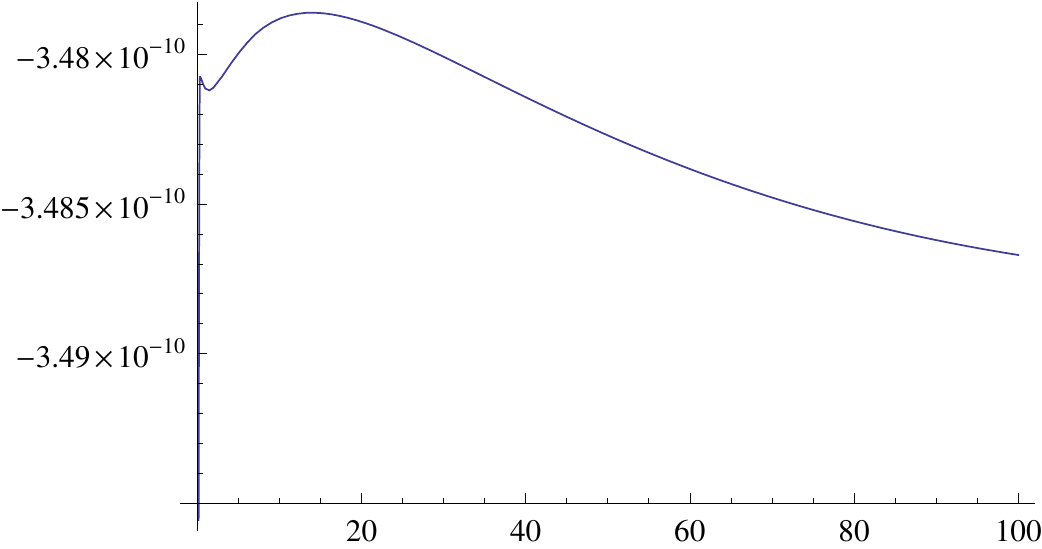}\hfil
\ing[width=0.45\textwidth]{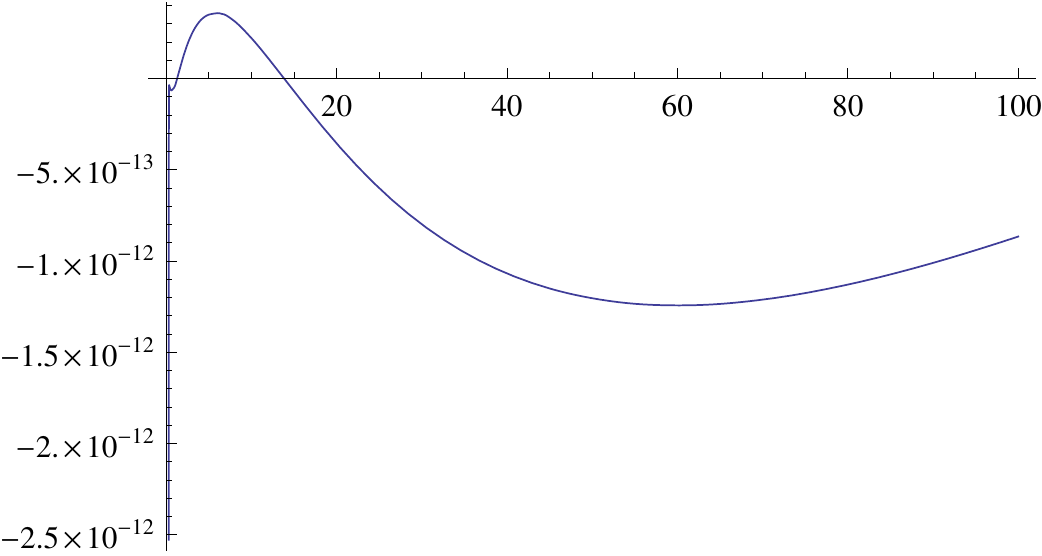}}
\hbox to \textwidth{\hbox to 0.45\textwidth{\hfil(A)\hfil}\hfil
\hbox to 0.45\textwidth{\hfil(B)\hfil}}
\hbox to \textwidth{\ing[width=0.45\textwidth]{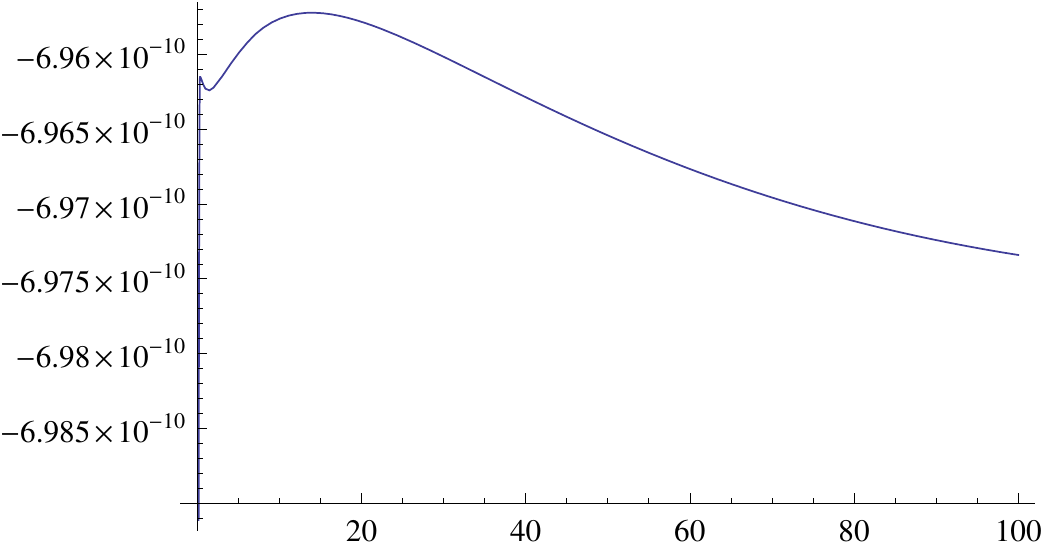}\hfil
\ing[width=0.45\textwidth]{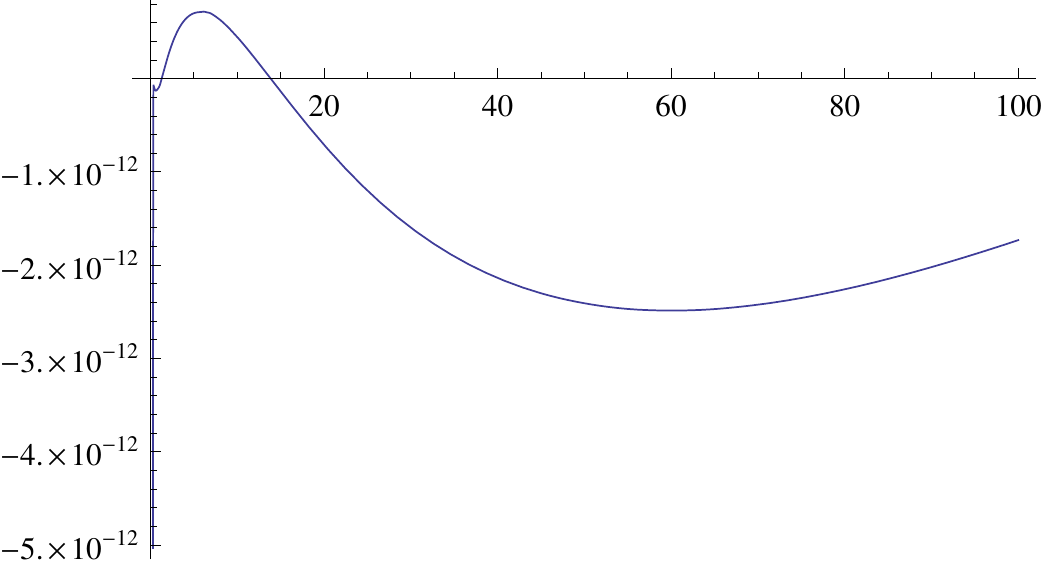}}
\hbox to \textwidth{\hbox to 0.45\textwidth{\hfil(C)\hfil}\hfil
\hbox to 0.45\textwidth{\hfil(D)\hfil}}
\hbox to \textwidth{\ing[width=0.45\textwidth]{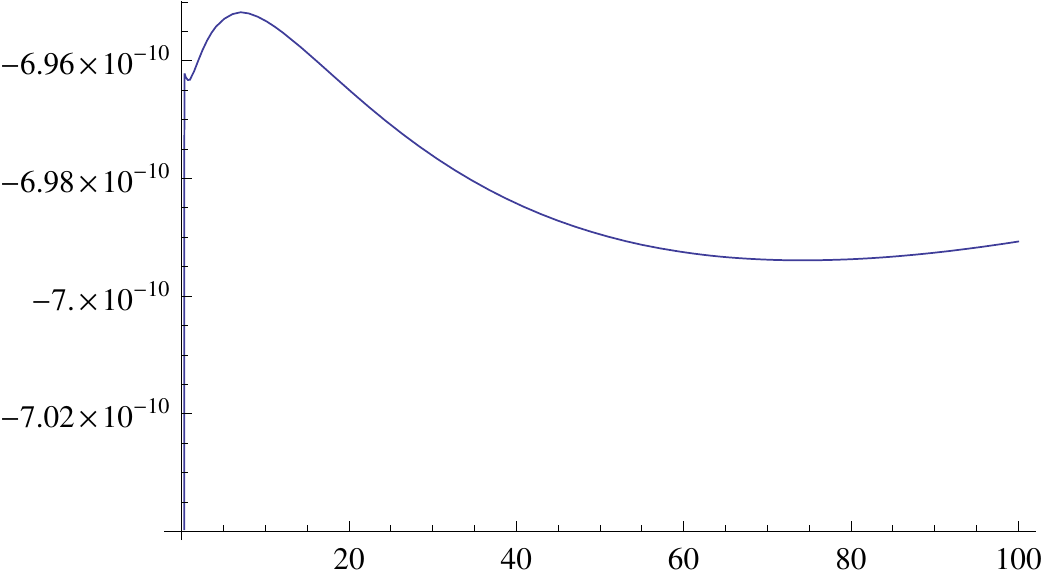}\hfil
\ing[width=0.45\textwidth]{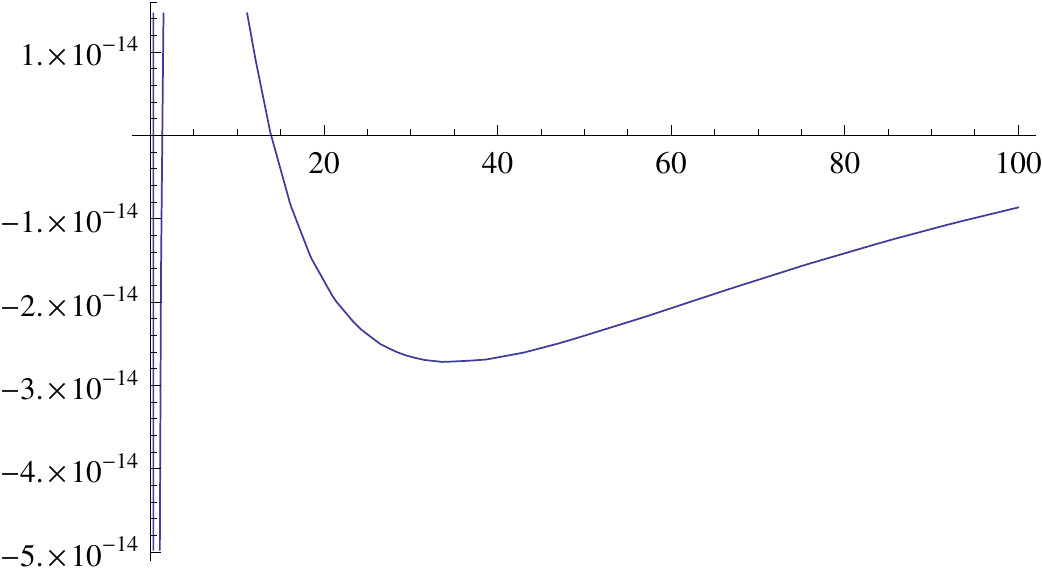}}
\hbox to \textwidth{\hbox to 0.45\textwidth{\hfil(E)\hfil}\hfil
\hbox to 0.45\textwidth{\hfil(F)\hfil}}
\medskip
\caption{%
(A)---a plot of $A(r)$ for $10^{-3}<r<100$,
(B)---a plot of $B(r)$ for $10^{-3}<r<100$.
(C)---a plot of $e^{2A(r)}-1$ for $10^{-3}<r<100$,
(D)---a plot of $e^{2B(r)}-1$ for $10^{-3}<r<100$,
(E)---a plot of $e^{2(A(r)+B(r))}-1$ for $10^{-3}<r<100$,
(F)---a plot of $\pz{}r e^{2A(r)}$ for $10^{-3}<r<100$.
}
\label{pp}
\end{figure}


\hbox to \textwidth{\ing[width=0.45\textwidth]{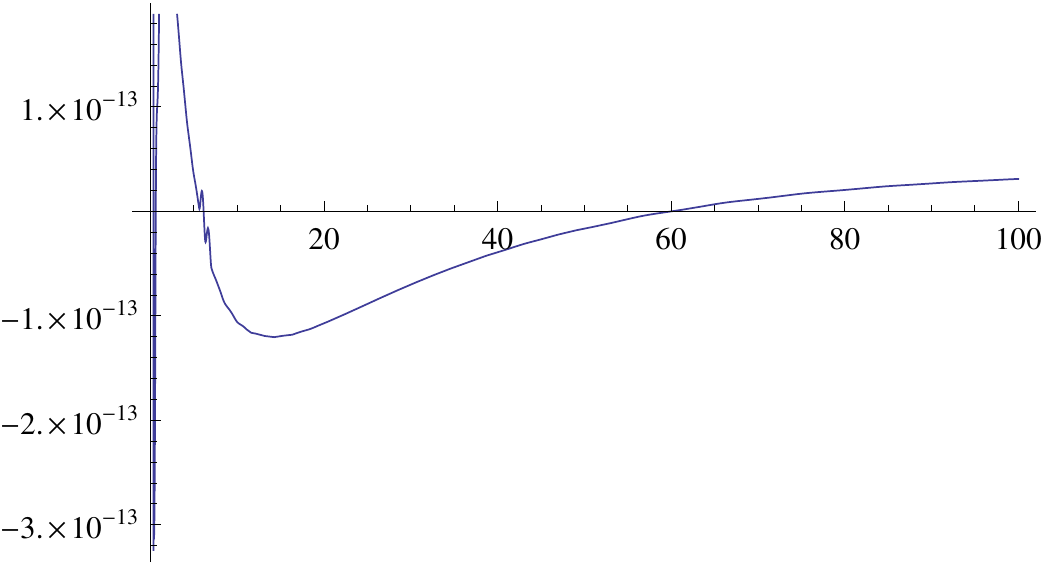}\hfil
\ing[width=0.45\textwidth]{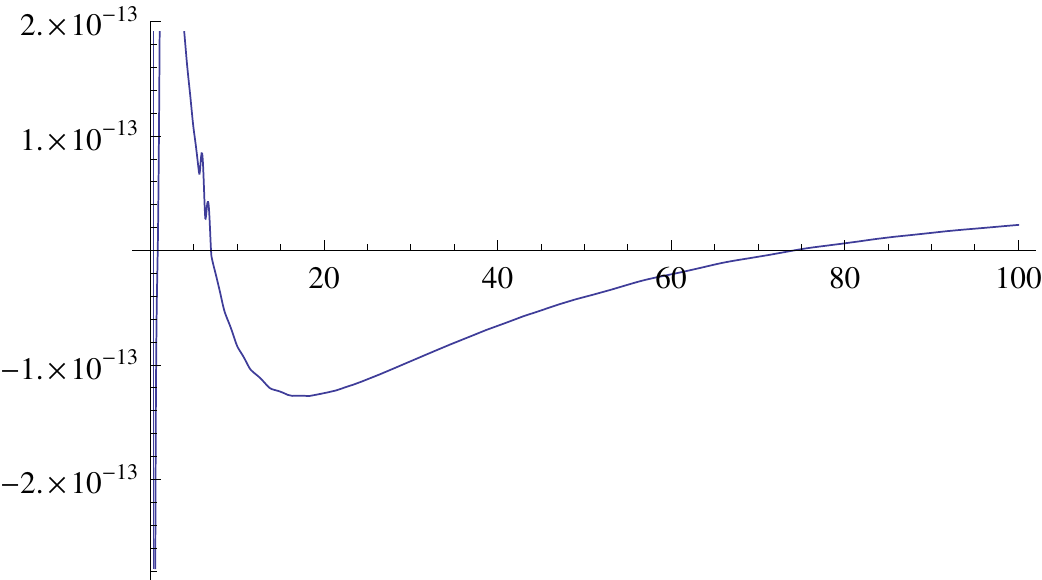}}
\hbox to \textwidth{\hbox to 0.45\textwidth{\hfil(G)\hfil}\hfil
\hbox to 0.45\textwidth{\hfil(H)\hfil}}
\hbox to \textwidth{\ing[width=0.45\textwidth]{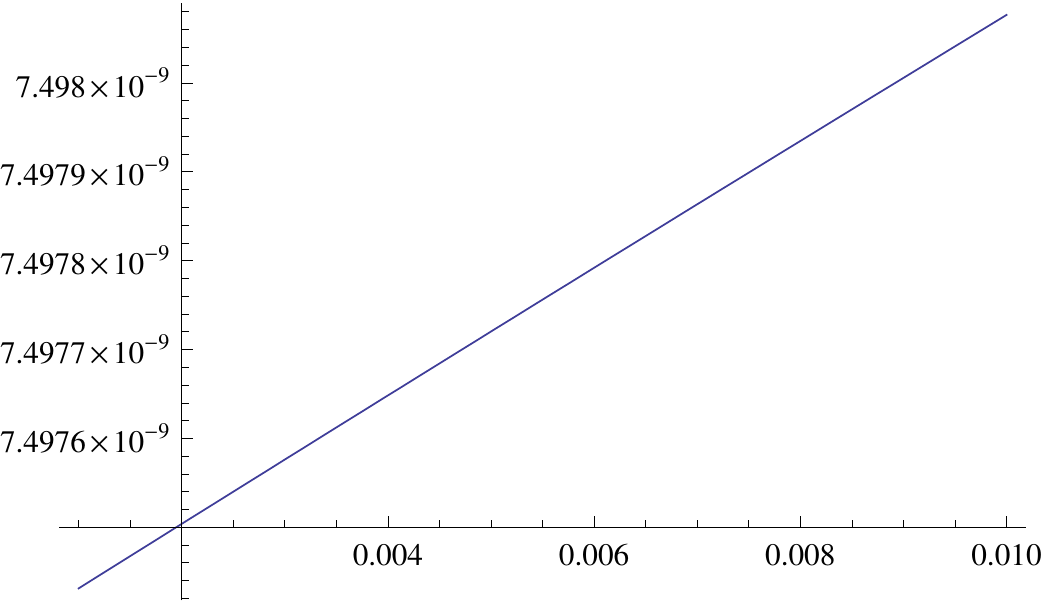}\hfil
 \ing[width=0.45\textwidth]{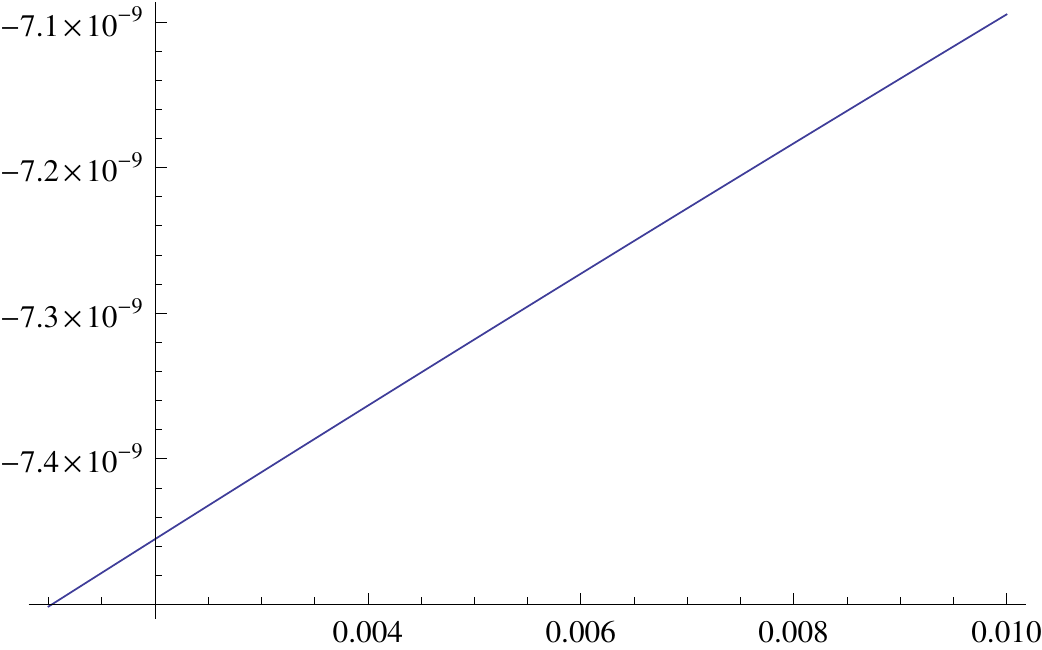}}
\hbox to \textwidth{\hbox to 0.45\textwidth{\hfil(I)\hfil}\hfil
\hbox to 0.45\textwidth{\hfil(J)\hfil}}
\hbox to \textwidth{\ing[width=0.45\textwidth]{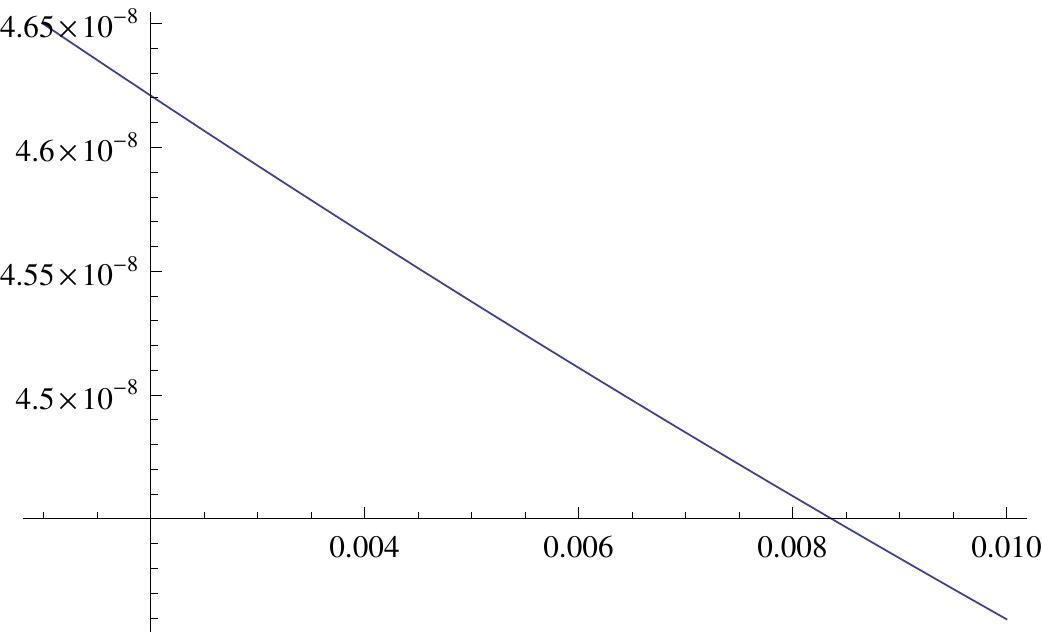}\hfil
\ing[width=0.45\textwidth]{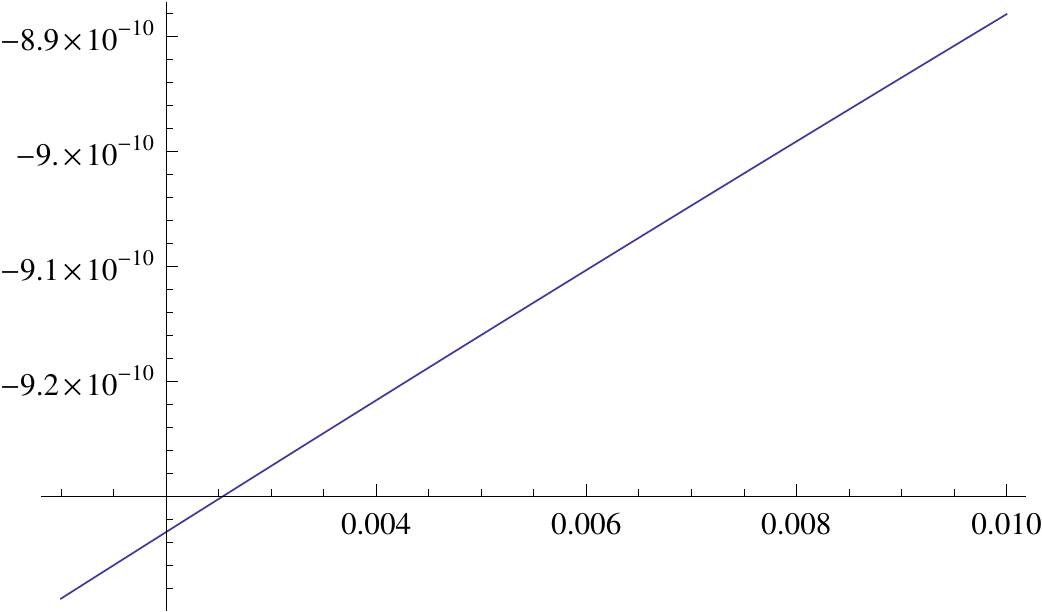}}
\hbox to \textwidth{\hbox to 0.45\textwidth{\hfil(K)\hfil}\hfil
\hbox to 0.45\textwidth{\hfil(L)\hfil}}
\hbox to \textwidth{\ing[width=0.45\textwidth]{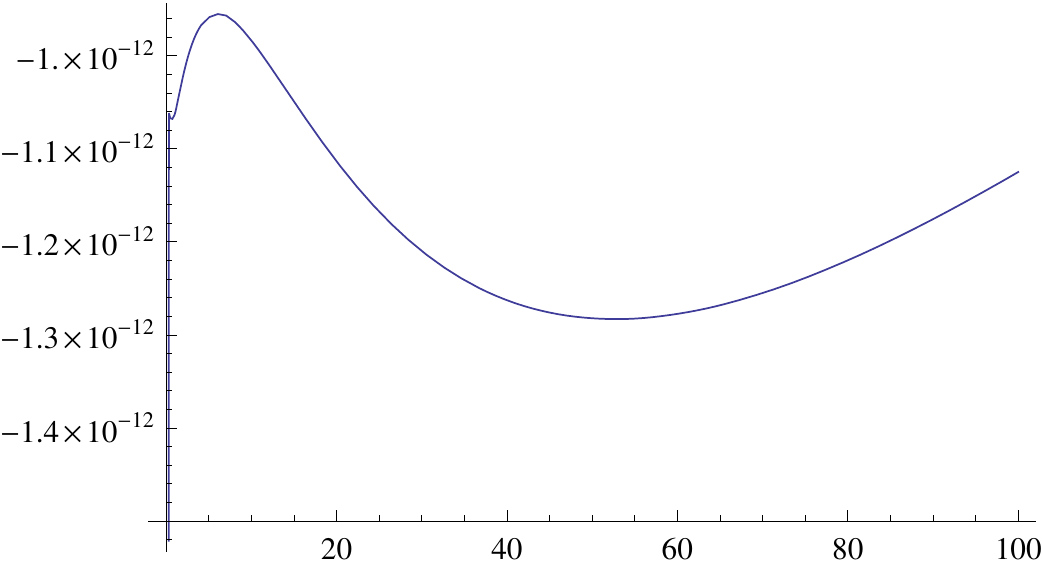}\hfil
\ing[width=0.45\textwidth]{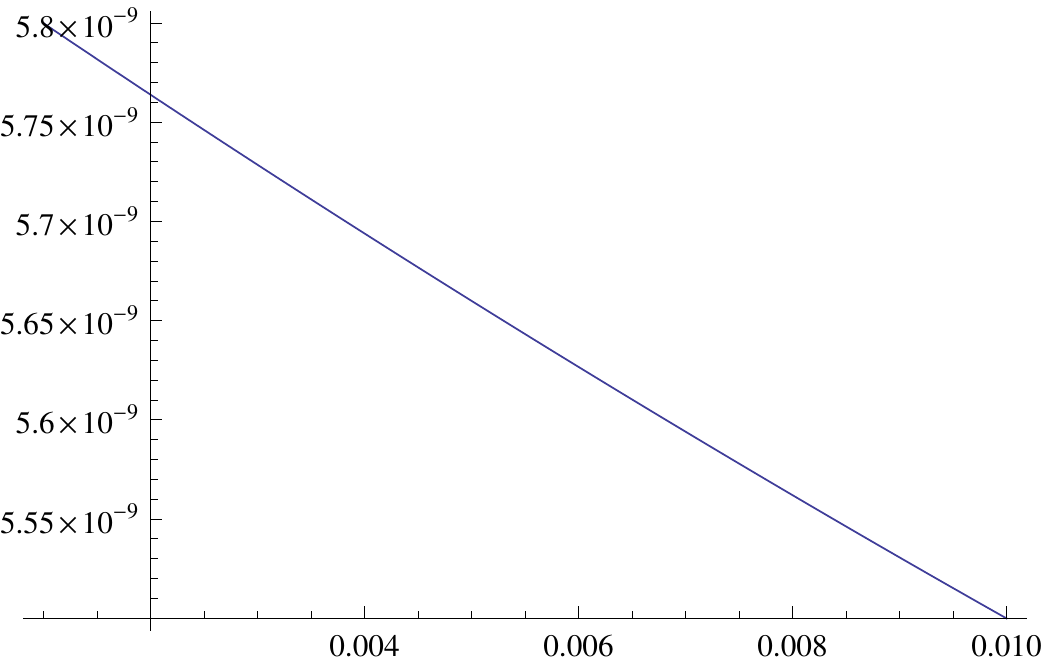}}
\hbox to \textwidth{\hbox to 0.45\textwidth{\hfil(M)\hfil}\hfil
\hbox to 0.45\textwidth{\hfil(N)\hfil}}
\medskip

{\small \noindent Figure \thefigure\ (cont.):
(G)---a plot of $\pz{}r e^{2B(r)}$ for $10^{-3}<r<100$,
(H)---a plot of $\pz{}r e^{2(A(r)+B(r))}$ for $10^{-3}<r<100$,
(I)---a plot of $e^{2A(r)}-1+8.20516 \t 10^{-9}$ for $10^{-3}<r<10^{-2}$,
(J)---a plot of $e^{2B(r)}-1$ for $10^{-3}<r<10^{-2}$,
(K)---a plot of $\pz{}r e^{2(A(r)+B(r))}$ for $10^{-3}<r<10^{-2}$,
(L)---a plot of $\wt\vF(r)$ for $10^{-3}<r<10^{-2}$,
(M)---a plot of $\wt\vF(r)$ for $10^{-3}<r<100$,
(N)---a plot of $\pz{\wt\vF}r(r)$ for $10^{-3}<r<10^{-2}$.}

\hbox to \textwidth{\ing[width=0.45\textwidth]{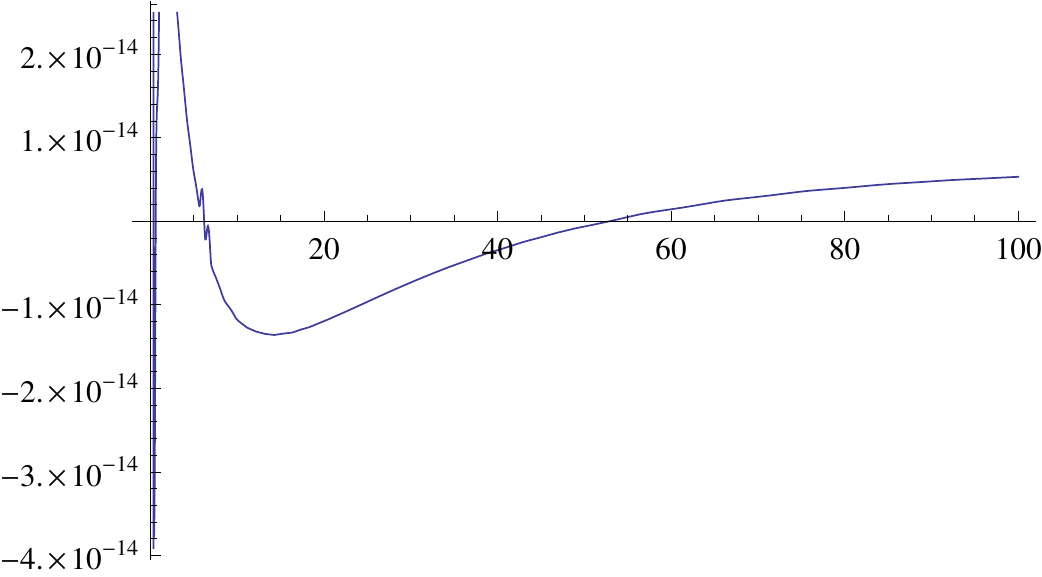}\hfil
\ing[width=0.45\textwidth]{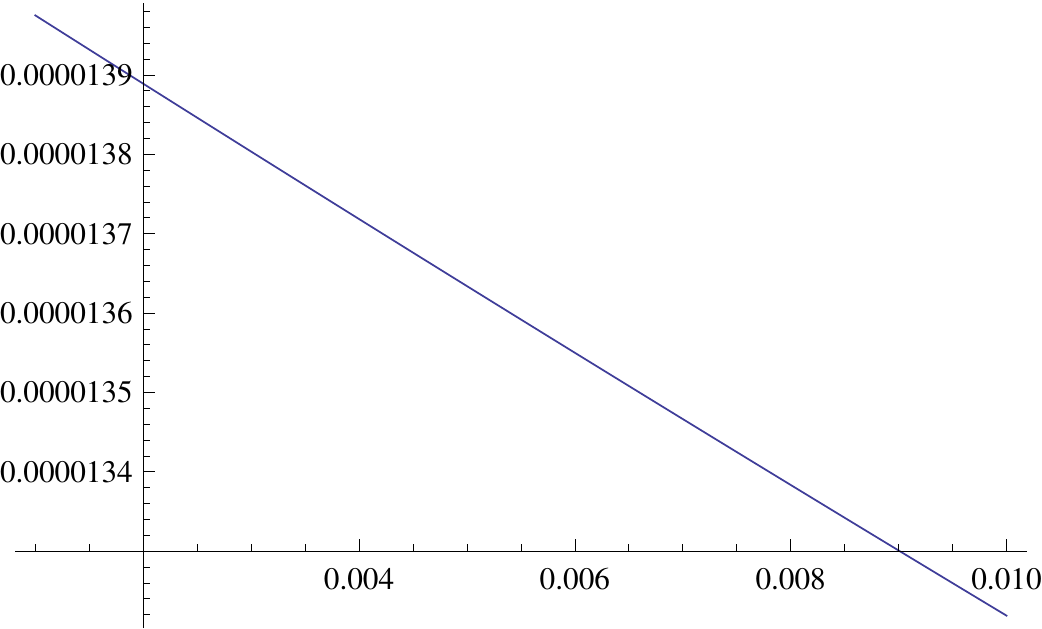}}
\hbox to \textwidth{\hbox to 0.45\textwidth{\hfil(O)\hfil}\hfil
\hbox to 0.45\textwidth{\hfil(P)\hfil}}
\hbox to \textwidth{\ing[width=0.45\textwidth]{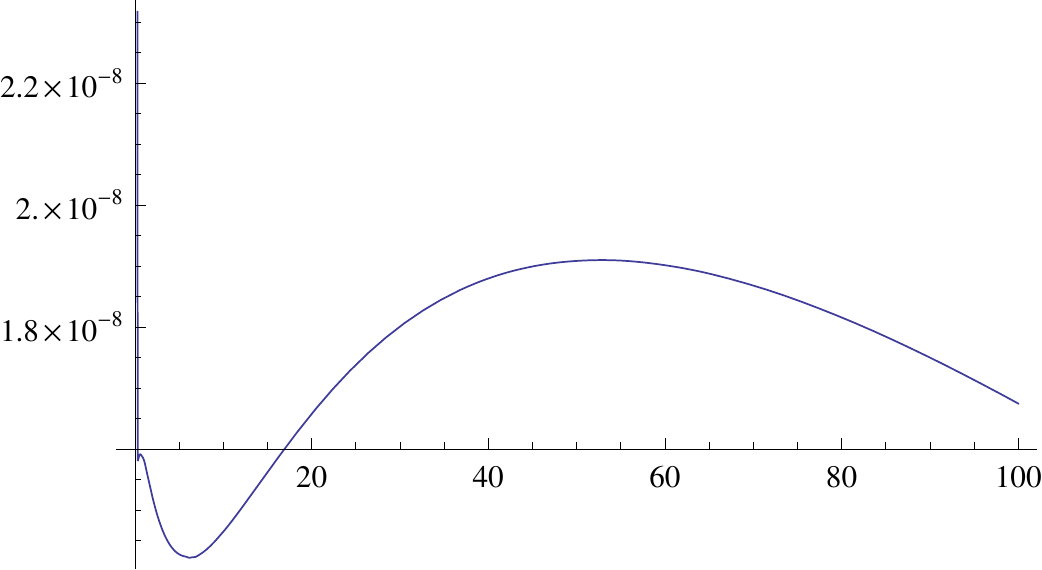}\hfil
\ing[width=0.45\textwidth]{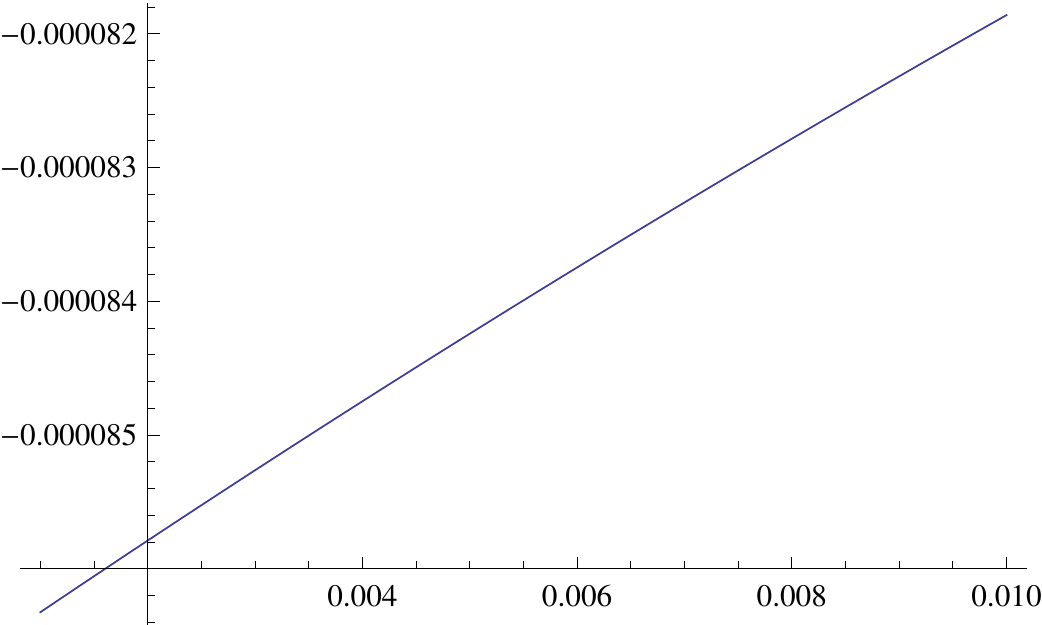}}
\hbox to \textwidth{\hbox to 0.45\textwidth{\hfil(Q)\hfil}\hfil
\hbox to 0.45\textwidth{\hfil(R)\hfil}}
\hbox to \textwidth{\hfil \ing[width=0.45\textwidth]{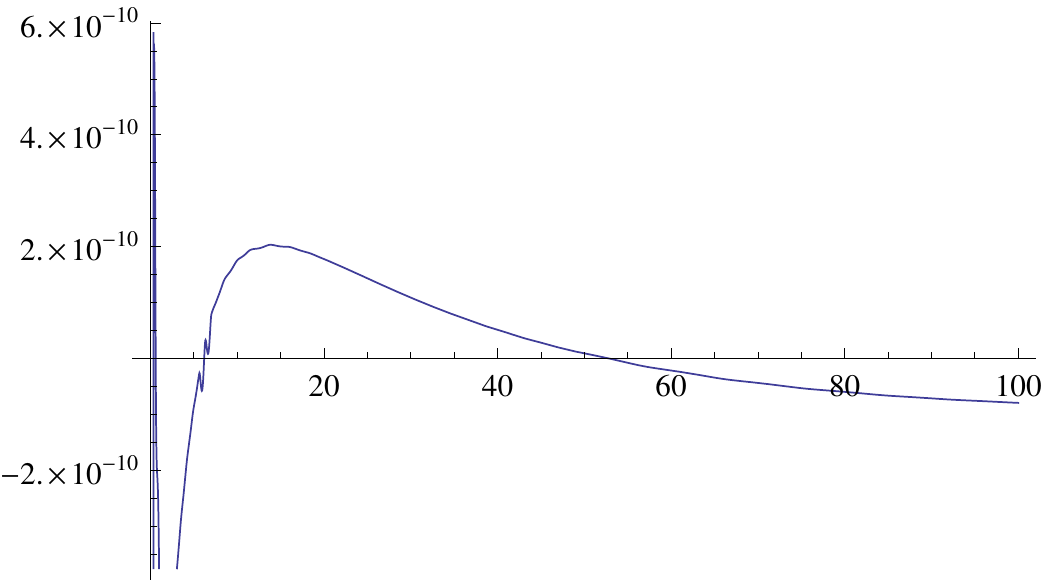}\hfil}
\hbox to \textwidth{\hfil\hbox to 0.45\textwidth{\hfil(S)\hfil}\hfil}
\medskip
{\small \noindent Figure \thefigure\ (cont.):
(O)---a plot of $\pz{\wt\vF}r(r)$ for $10^{-3}<r<100$,
(P)---a plot of $\frac{G\eff(r)}{G_N}-1$ for $10^{-3}<r<10^{-2}$,
(Q)---a plot of $\frac{G\eff(r)}{G_N}-1$ for $10^{-3}<r<100$,
(R)---a plot of $\pz{}r(\frac{G\eff(r)}{G_N})$ for $10^{-3}<r<10^{-2}$,
(S)---a plot of $\pz{}r(\frac{G\eff(r)}{G_N})$ for $10^{-3}<r<100$.}

\medskip

\eject
On Fig. \ref{mm} we plot our results for $\ov\rho(r)$
 for several regions of~$r$.

\medskip
\refstepcounter{figure}\label{mm}
\hbox to \textwidth{\ing[width=0.43\textwidth]{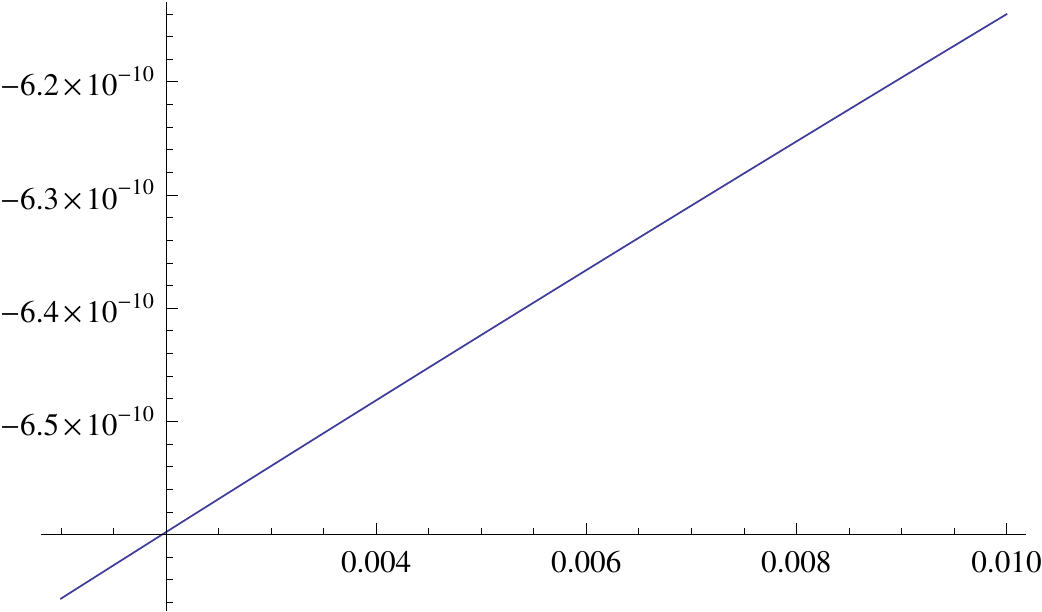}\hfil
\ing[width=0.43\textwidth]{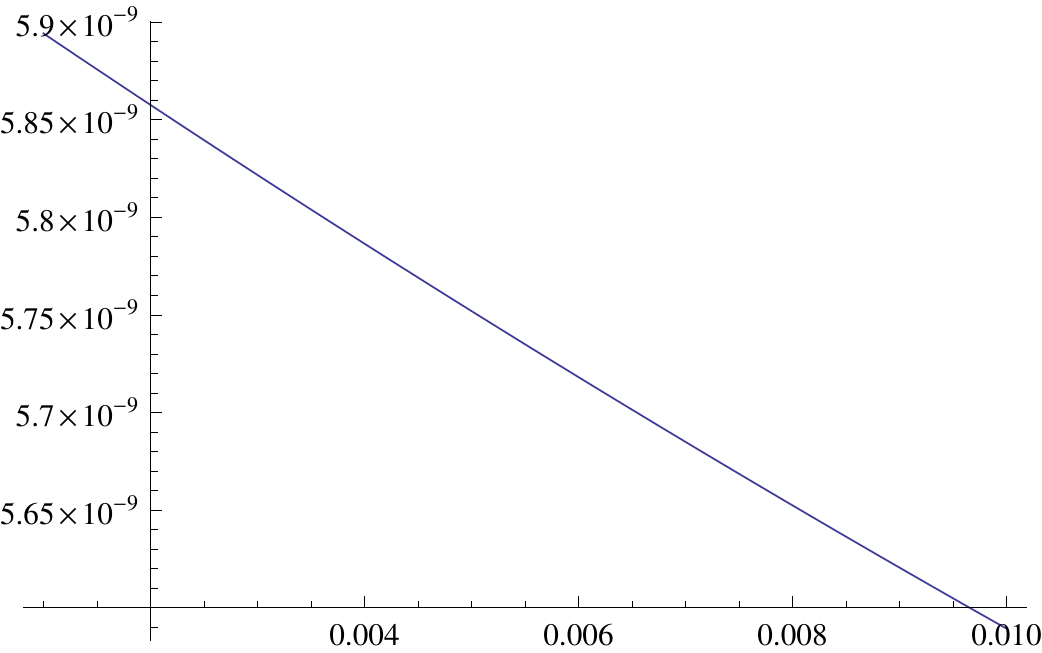}}
\hbox to \textwidth{\hbox to 0.43\textwidth{\hfil(A)\hfil}\hfil
\hbox to 0.43\textwidth{\hfil(B)\hfil}}
\hbox to \textwidth{\ing[width=0.43\textwidth]{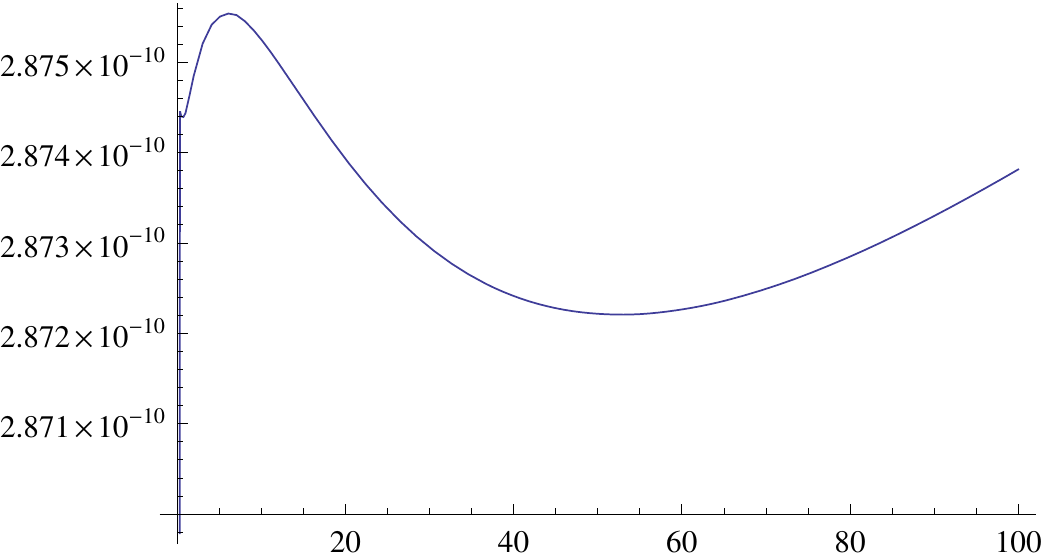}\hfil
\ing[width=0.43\textwidth]{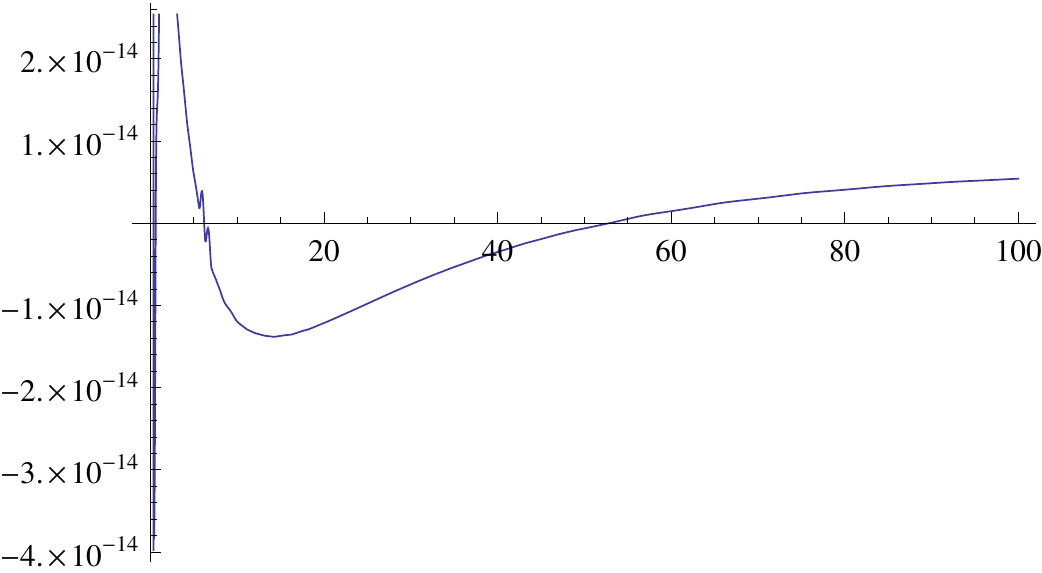}}
\hbox to \textwidth{\hbox to 0.43\textwidth{\hfil(C)\hfil}\hfil
\hbox to 0.43\textwidth{\hfil(D)\hfil}}
\hbox to \textwidth{\ing[width=0.43\textwidth]{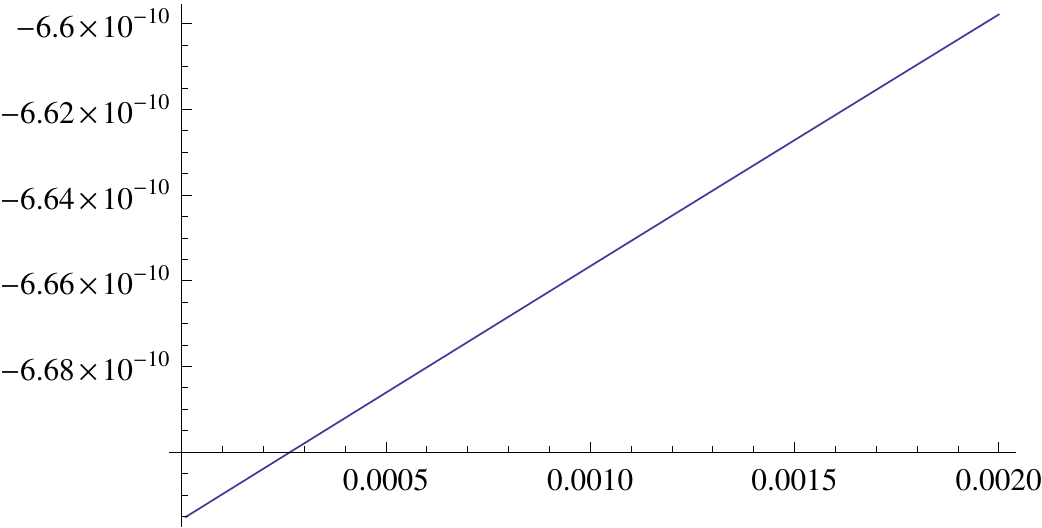}\hfil
\ing[width=0.44\textwidth]{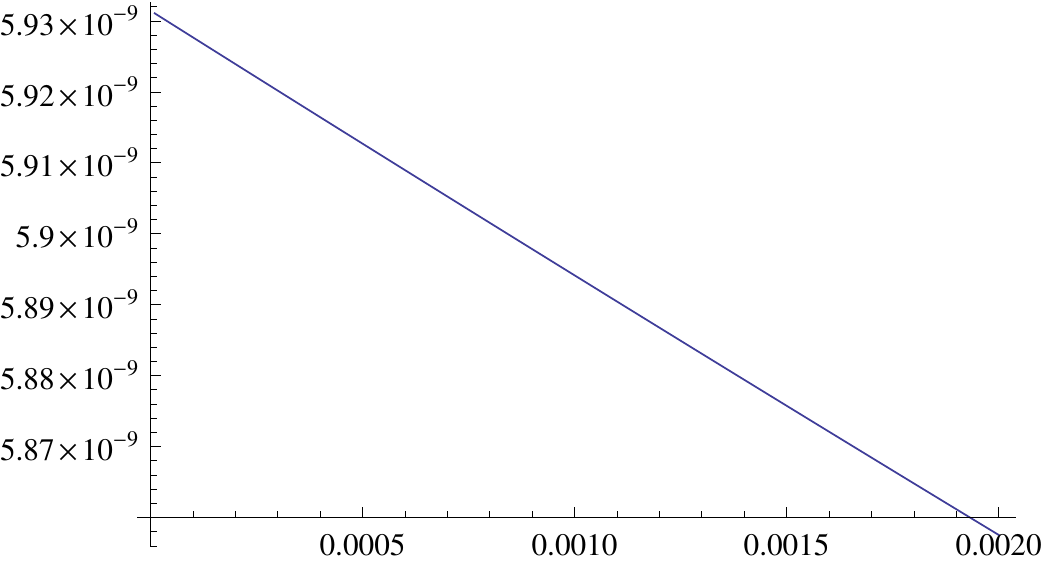}}
\hbox to \textwidth{\hbox to 0.43\textwidth{\hfil(E)\hfil}\hfil
\hbox to 0.43\textwidth{\hfil(F)\hfil}}
\medskip
{\small \noindent Figure \thefigure:
(A)---a plot of $\ov\rho(r)-0.0000682784$ for $10^{-3}<r<10^{-2}$,
(B)---a plot of $\pz{\ov\rho(r)}r$ for $10^{-3}<r<10^{-2}$,
(C)---a plot of $\ov\rho(r)-0.0000682784$ for $10^{-3}<r<100$,
(D)---a plot of $\pz{\ov\rho(r)}r$ for $10^{-3}<r<100$,
(E)---a plot of $\ov\rho(r)-0.0000682784$ for $10^{-5}<r<2\t10^{-3}$,
(F)---a plot of $\pz{\ov\rho(r)}r$ for $10^{-5}<r<2\t10^{-3}$.}

\eject
On Fig. \ref{cyfry} we plot our results for $V_1(r)$, $V_2(r)$, $\ov V_1(r)$,
$\ov V_2(r)$, $b_1(r)$, $b_2(r)$ for several regions of~$r$ and compare them
with $\ov U(r)$ and its \dv s plotting $\eta_1(r)$, $\eta_2(r)$.

\bigskip
\refstepcounter{figure}\label{cyfry}
\hbox to \textwidth{\ing[width=0.44\textwidth]{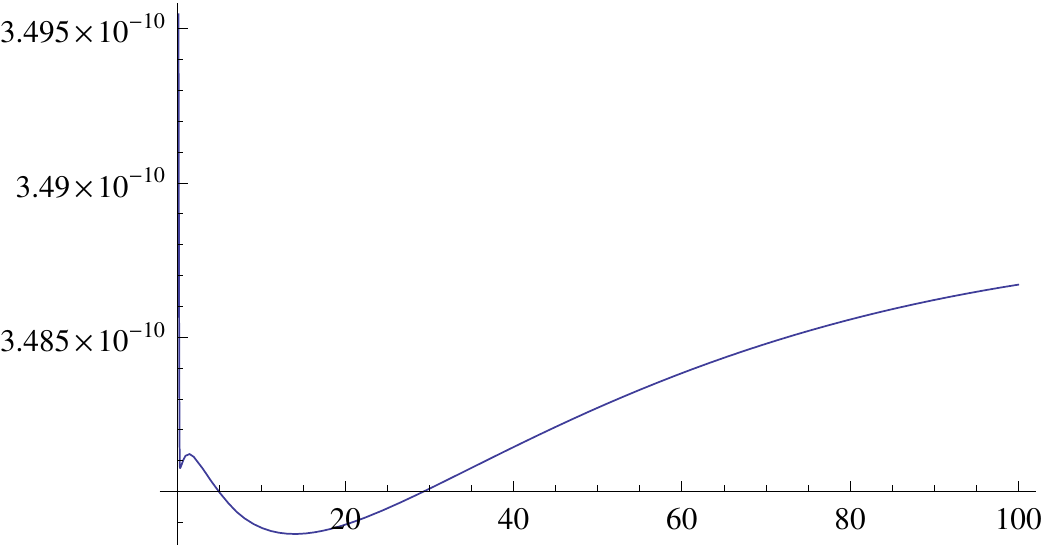}\hfil
\ing[width=0.44\textwidth]{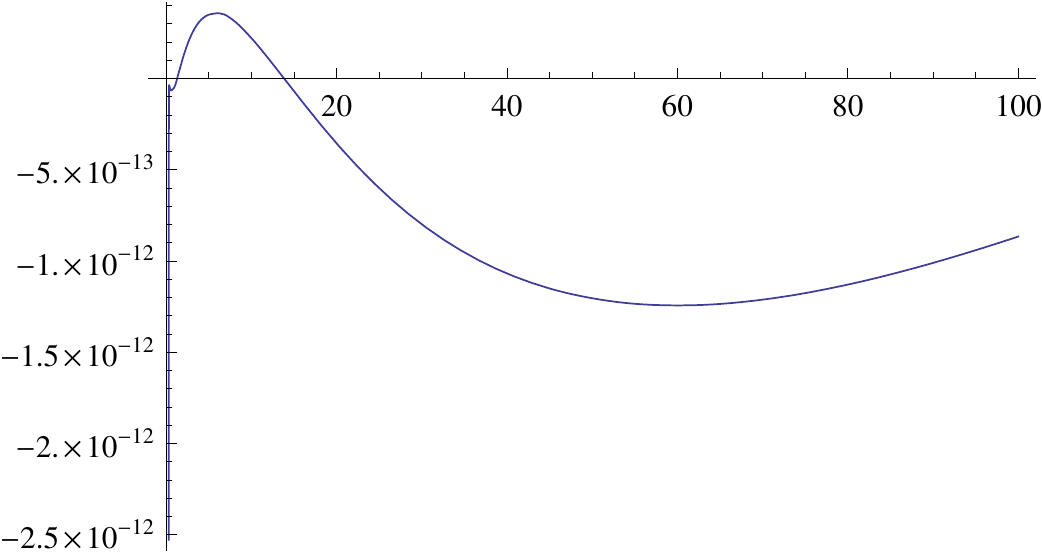}}
\hbox to \textwidth{\hbox to 0.44\textwidth{\hfil(A)\hfil}\hfil
\hbox to 0.44\textwidth{\hfil(B)\hfil}}
\hbox to \textwidth{\ing[width=0.44\textwidth]{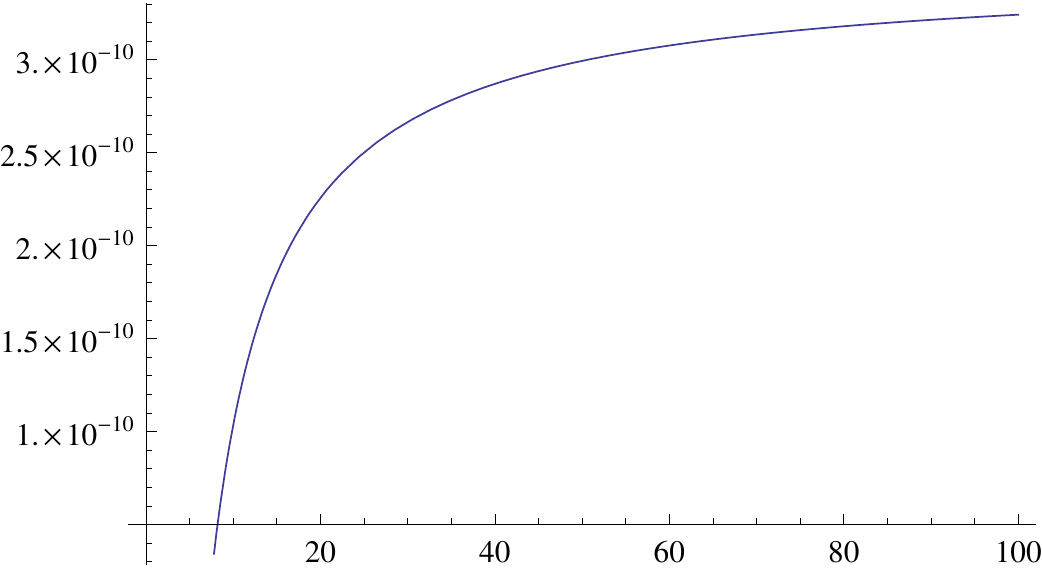}\hfil
\ing[width=0.44\textwidth]{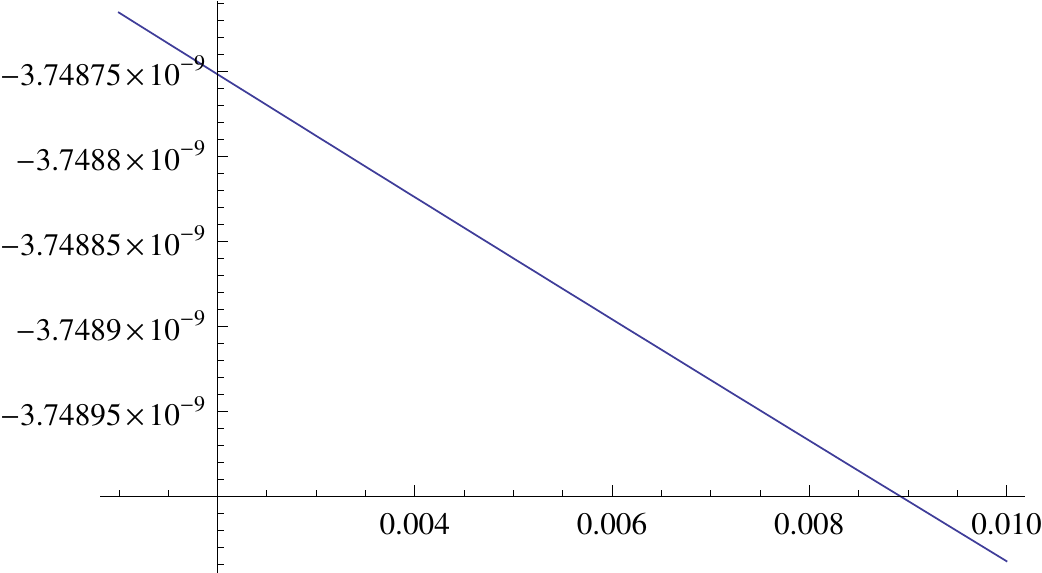}}
\hbox to \textwidth{\hbox to 0.44\textwidth{\hfil(C)\hfil}\hfil
\hbox to 0.44\textwidth{\hfil(D)\hfil}}
\hbox to \textwidth{\ing[width=0.44\textwidth]{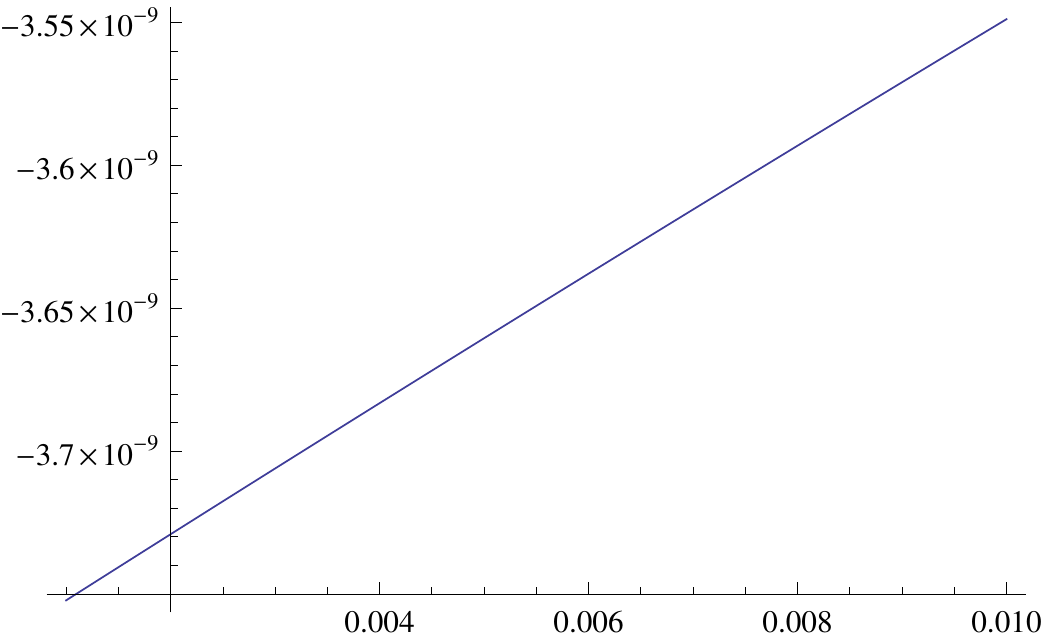}\hfil
\ing[width=0.44\textwidth]{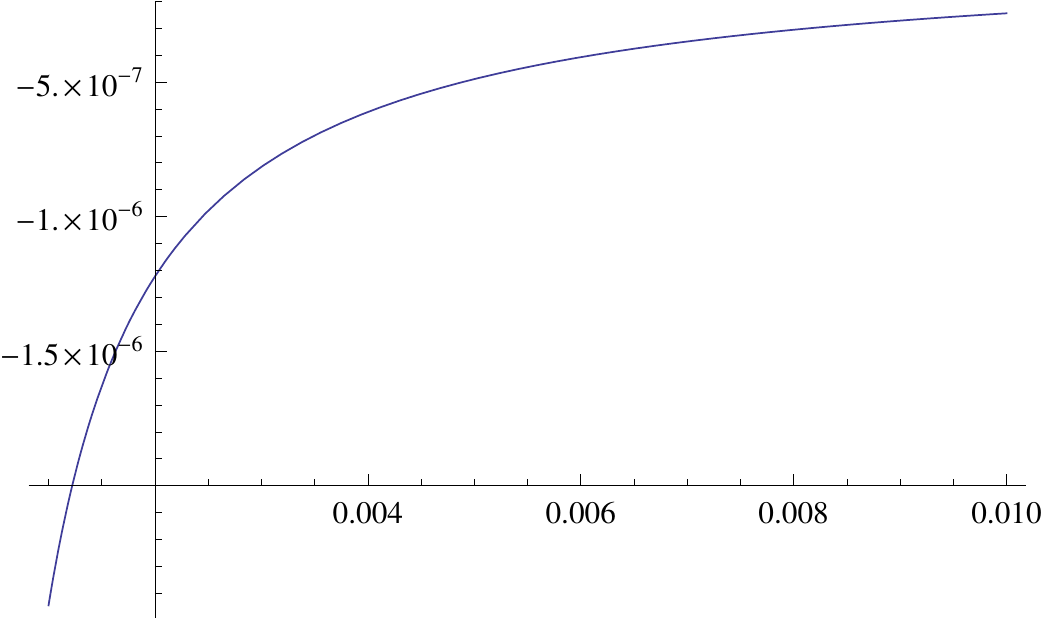}}
\hbox to \textwidth{\hbox to 0.44\textwidth{\hfil(E)\hfil}\hfil
\hbox to 0.44\textwidth{\hfil(F)\hfil}}
\hbox to \textwidth{\ing[width=0.44\textwidth]{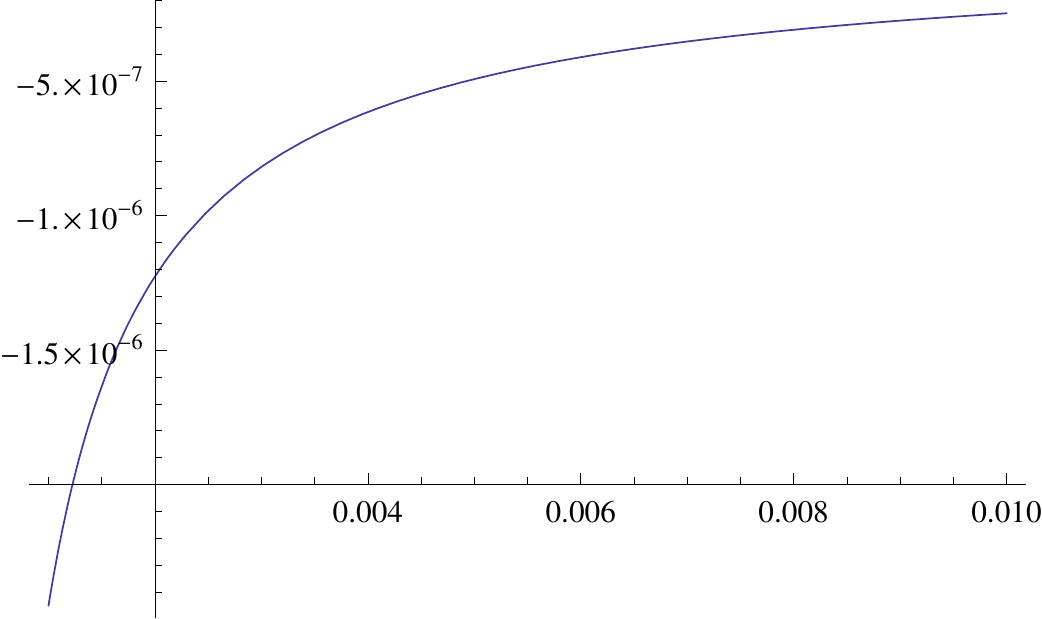}\hfil
\ing[width=0.44\textwidth]{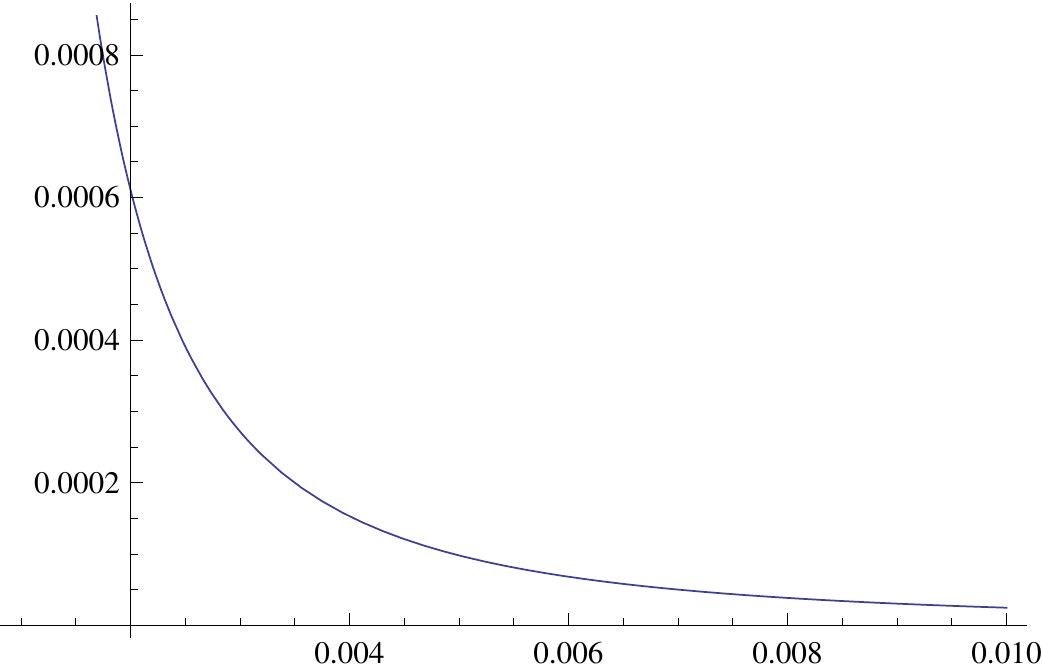}}
\hbox to \textwidth{\hbox to 0.44\textwidth{\hfil(G)\hfil}\hfil
\hbox to 0.44\textwidth{\hfil(H)\hfil}}
\smallskip \noindent
{\small Figure \thefigure:
(A)---a plot of $V_1(r)$ for $10^{-3}<r<100$,
(B)---a plot of $V_2(r)$ for $10^{-3}<r<100$,
(C)---a plot of $\ov V_1(r)$ for $10^{-3}<r<100$,
(D)---a plot of $V_1(r)-4.10258 \t 10^{-9}$ for $10^{-3}<r<10^{-2}$,
(E)---a plot of $V_2(r)-1.5\t10^{-12}$ for $10^{-3}<r<10^{-2}$,
(F)---a plot of $\ov V_1(r)$ for $10^{-3}<r<10^{-2}$,
(G)---a plot of $\ov V_2(r)$ for $10^{-3}<r<10^{-2}$,
(H)---a plot of $b_1(r)$ for $10^{-3}<r<10^{-2}$.}

\hbox to \textwidth{\ing[width=0.45\textwidth]{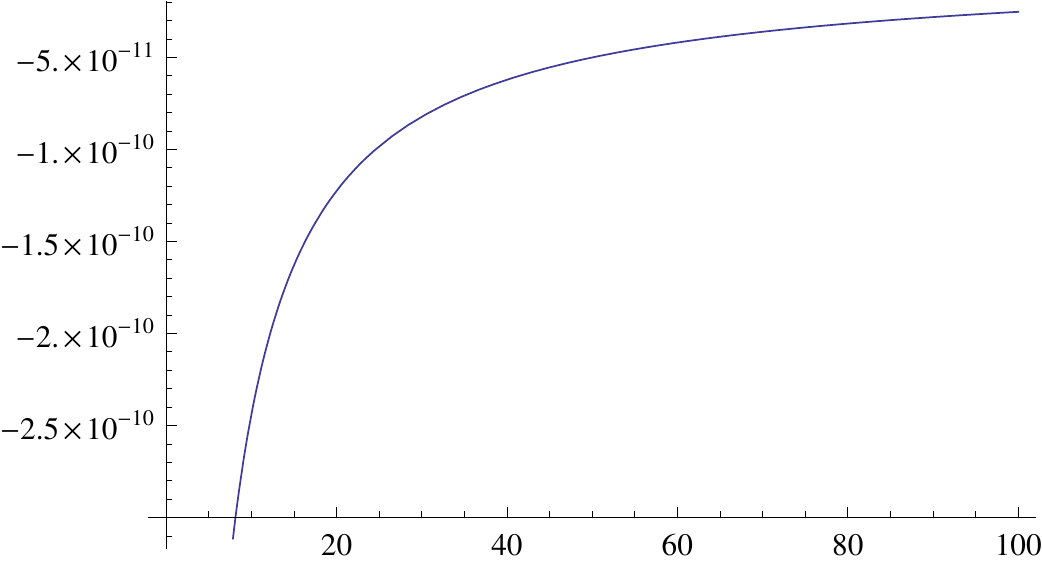}\hfil
\ing[width=0.45\textwidth]{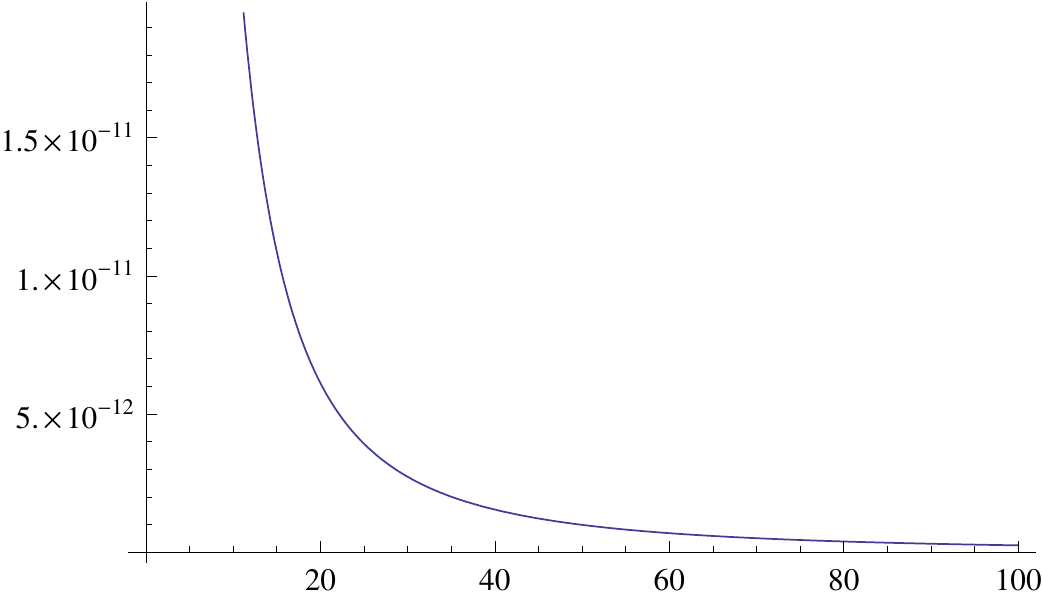}}
\hbox to \textwidth{\hbox to 0.45\textwidth{\hfil(I)\hfil}\hfil
\hbox to 0.45\textwidth{\hfil(J)\hfil}}
\hbox to \textwidth{\ing[width=0.45\textwidth]{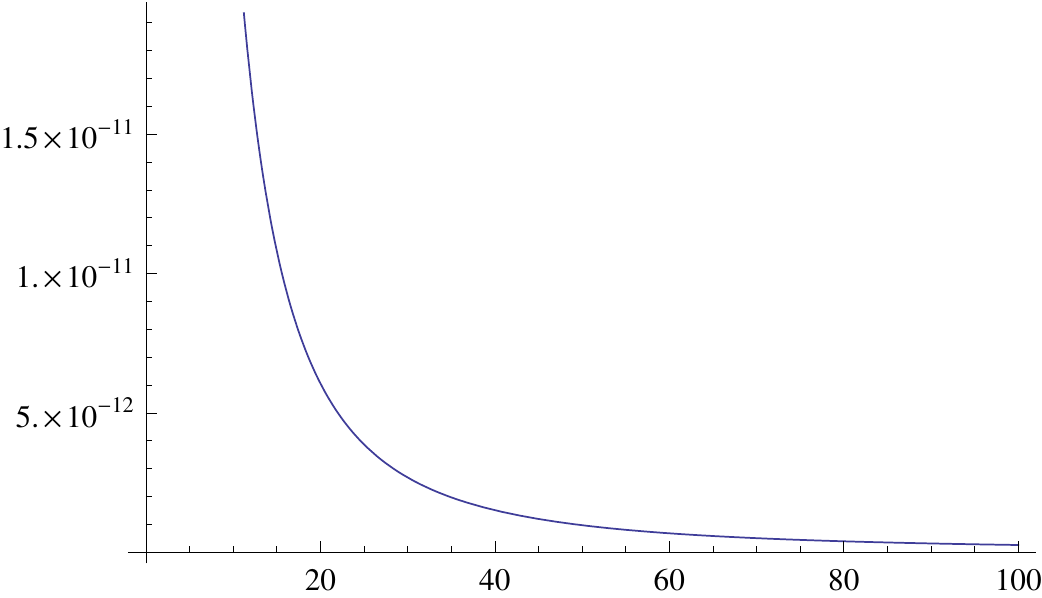}\hfil
\ing[width=0.45\textwidth]{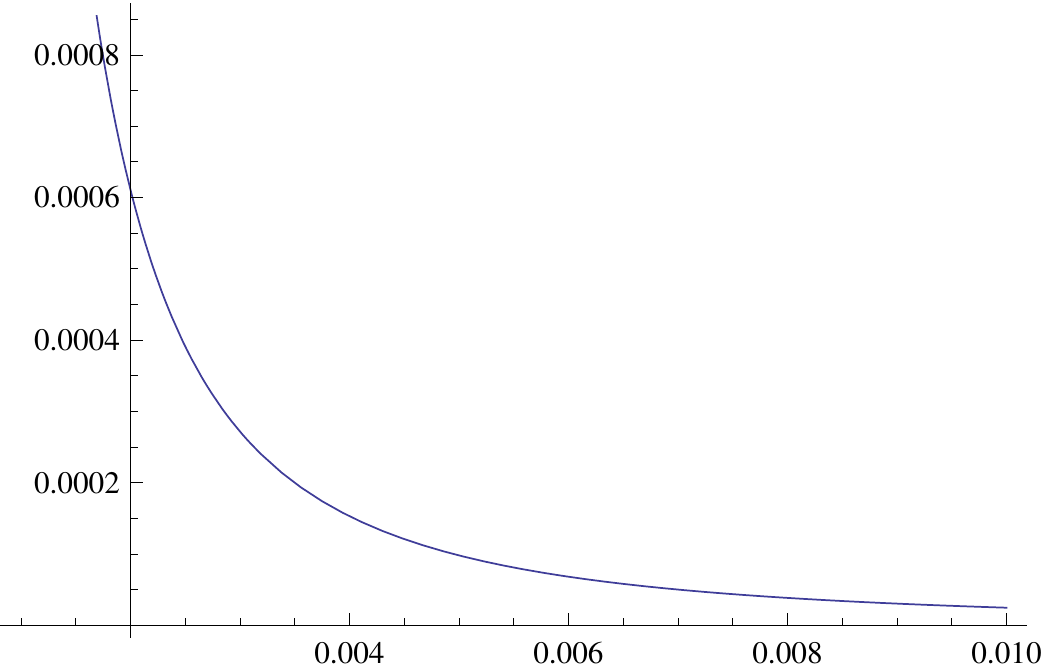}}
\hbox to \textwidth{\hbox to 0.45\textwidth{\hfil(K)\hfil}\hfil
\hbox to 0.45\textwidth{\hfil(L)\hfil}}
\hbox to \textwidth{\ing[width=0.45\textwidth]{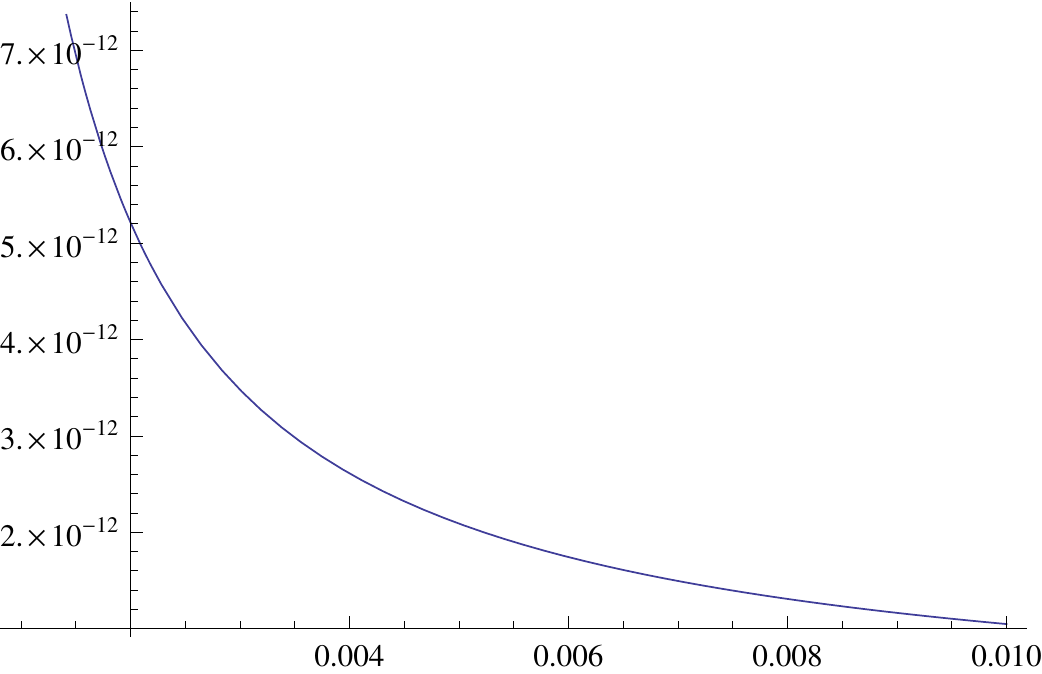}\hfil
\ing[width=0.45\textwidth]{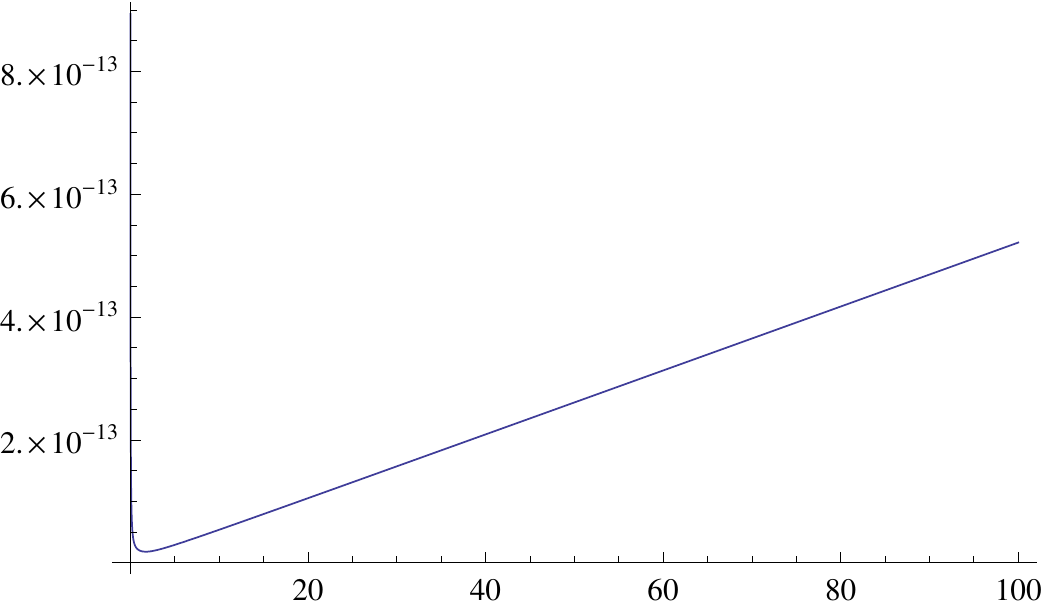}}
\hbox to \textwidth{\hbox to 0.45\textwidth{\hfil(M)\hfil}\hfil
\hbox to 0.45\textwidth{\hfil(N)\hfil}}
\hbox to \textwidth{\ing[width=0.45\textwidth]{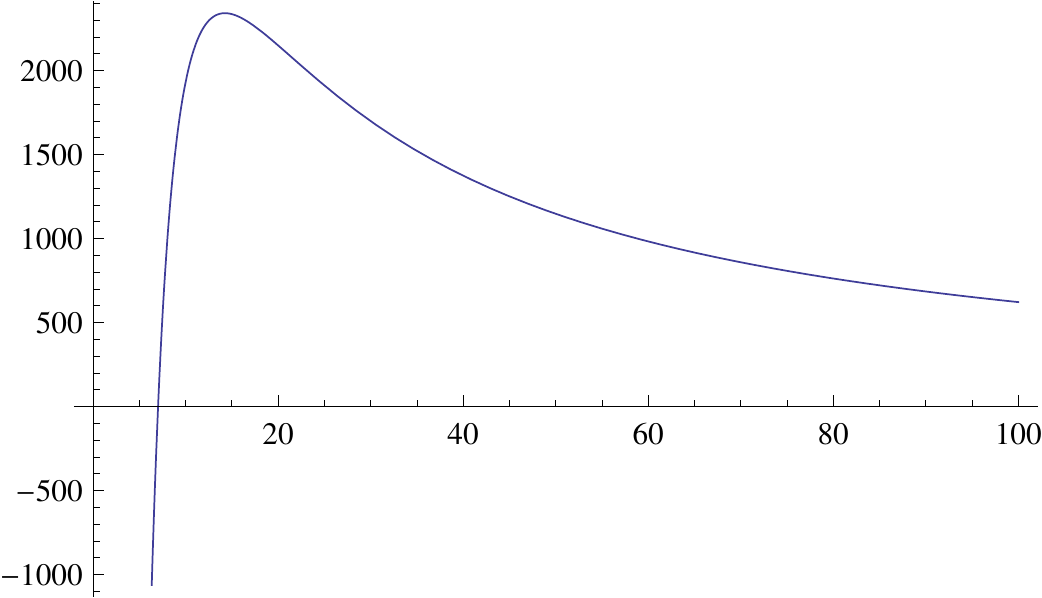}\hfil
\ing[width=0.45\textwidth]{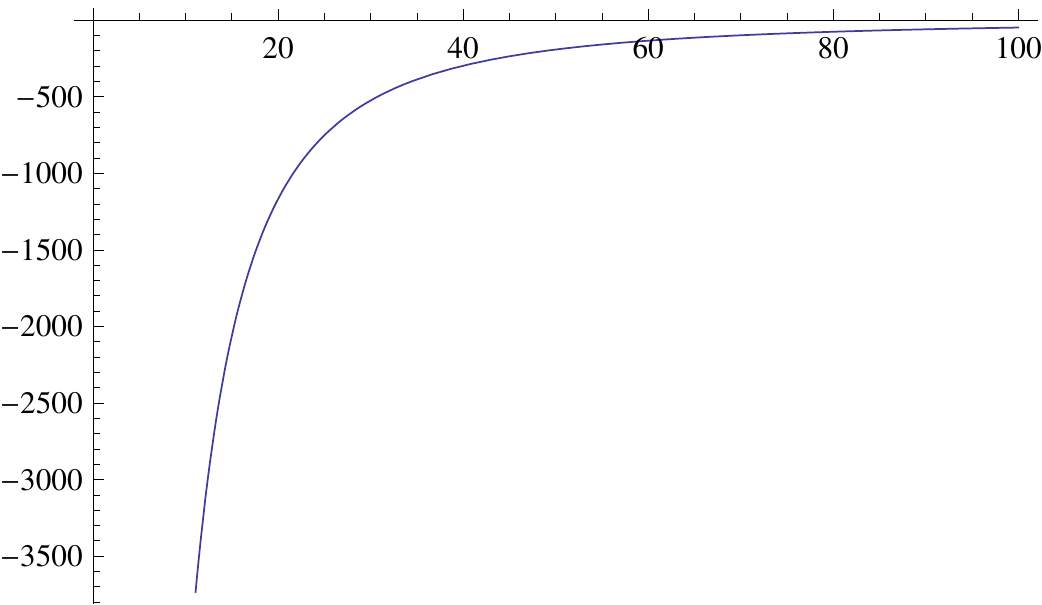}}
\hbox to \textwidth{\hbox to 0.45\textwidth{\hfil(O)\hfil}\hfil
\hbox to 0.45\textwidth{\hfil(P)\hfil}}
\medskip
{\small \noindent Figure \thefigure\ (cont.):
(I)---a plot of $\ov V_2(r)$ for $10^{-3}<r<100$,
(J)---a plot of $b_1(r)$ for $10^{-3}<r<100$,
(K)---a plot of $b_2(r)$ for $10^{-3}<r<100$,
(L)---a plot of $b_2(r)$ for $10^{-3}<r<10^{-2}$,
(M)---a plot of $-\ov U(r)$ for $10^{-3}<r<10^{-2}$,
(N)---a plot of $-\ov U(r)$ for $10^{-3}<r<100$,
(O)---a plot of $\eta_1(r)$ for $10^{-3}<r<100$,
(P)---a plot of $\eta_2(r)$ for $10^{-3}<r<100$.

}

\hbox to \textwidth{\ing[width=0.45\textwidth]{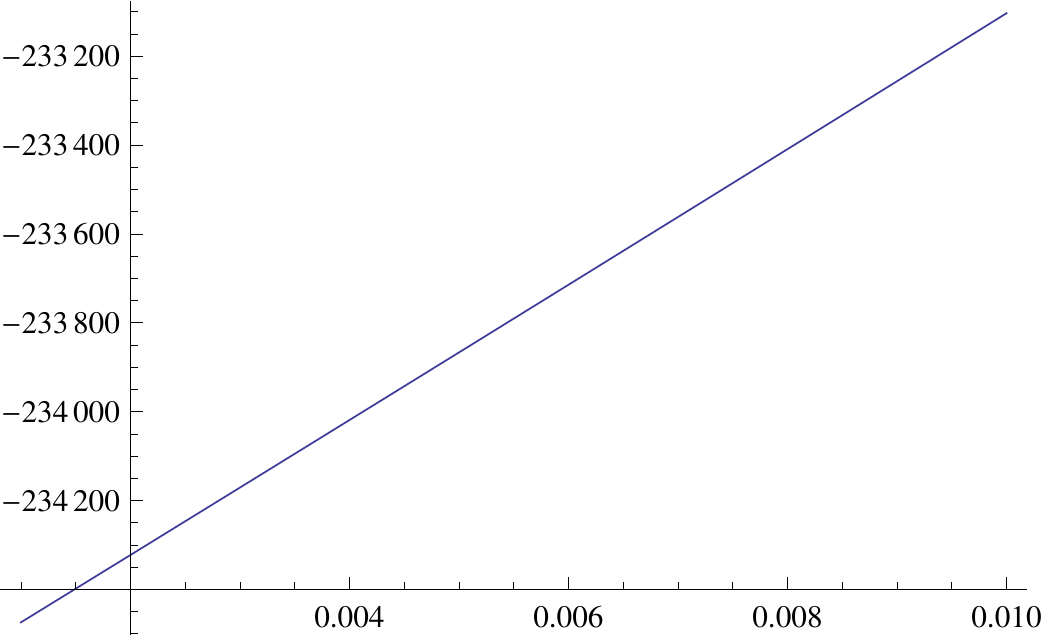}\hfil
\ing[width=0.45\textwidth]{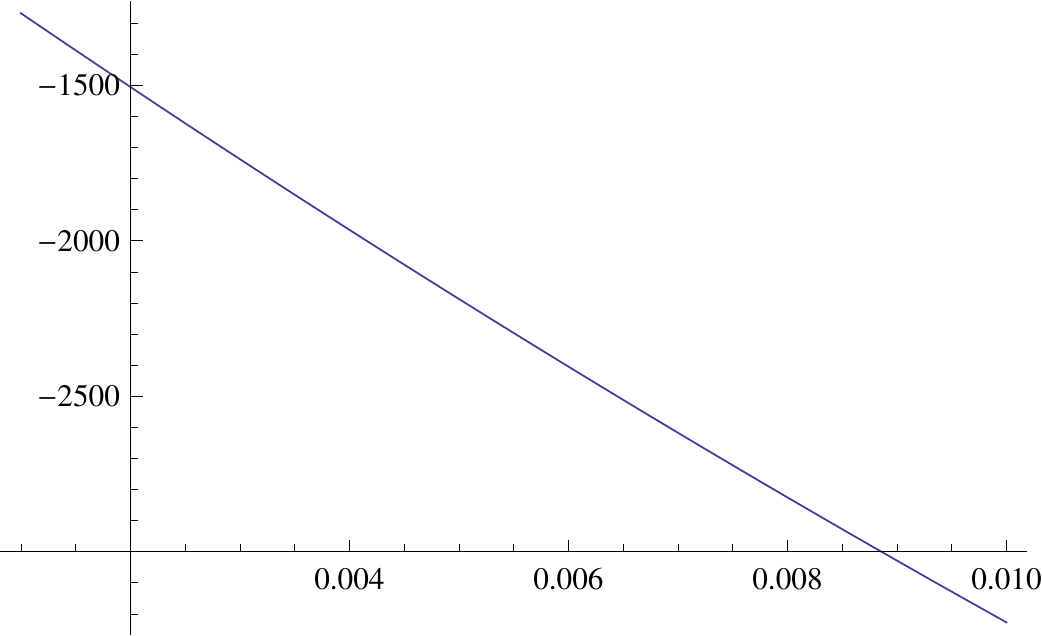}}
\hbox to \textwidth{\hbox to 0.45\textwidth{\hfil(Q)\hfil}\hfil
\hbox to 0.45\textwidth{\hfil(R)\hfil}}
\hbox to \textwidth{\ing[width=0.45\textwidth]{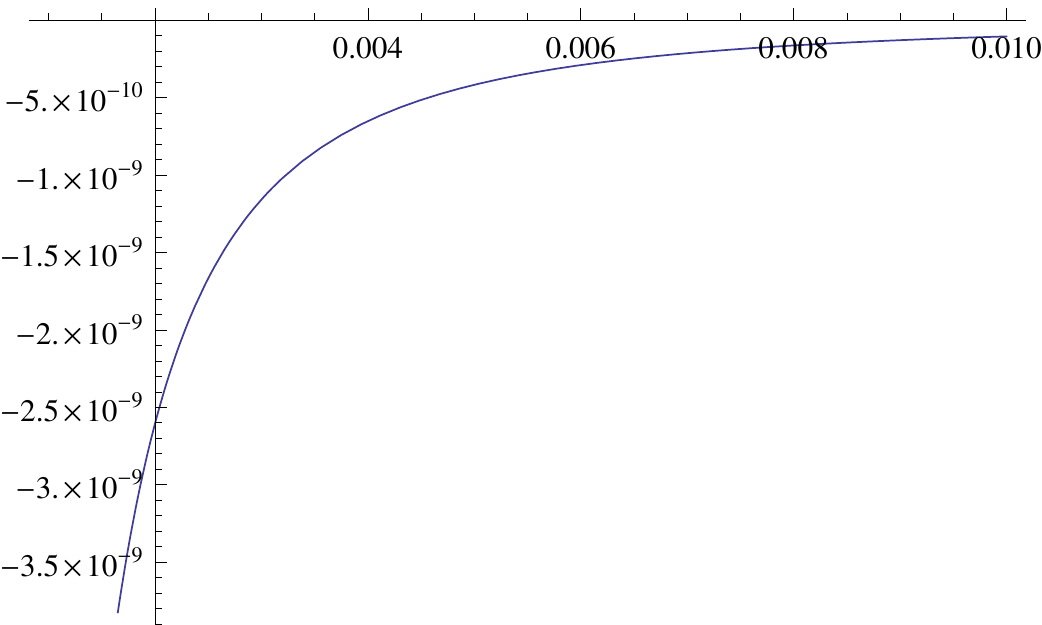}\hfil
\ing[width=0.45\textwidth]{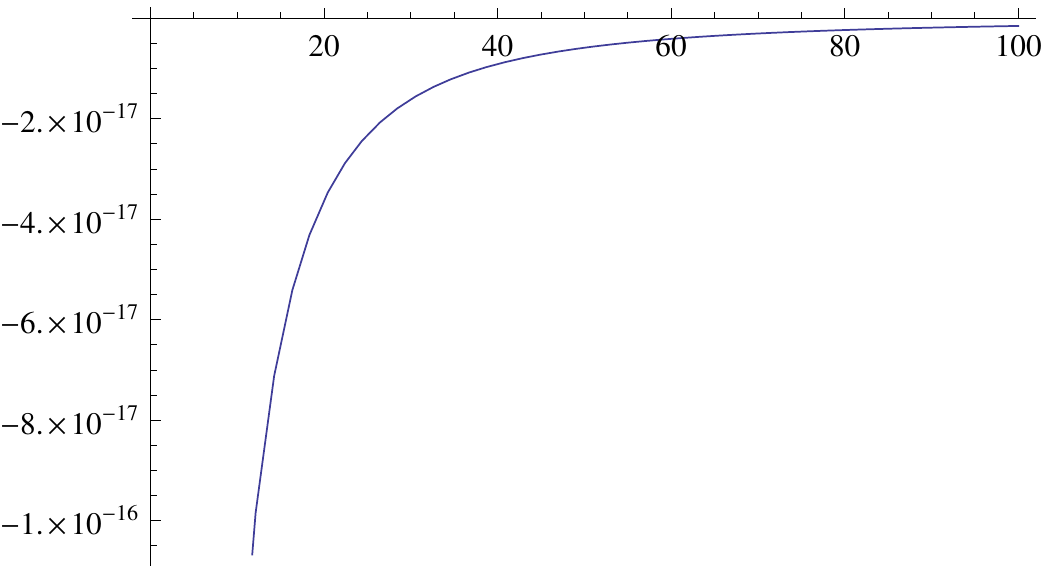}}
\hbox to \textwidth{\hbox to 0.45\textwidth{\hfil(S)\hfil}\hfil
\hbox to 0.45\textwidth{\hfil(T)\hfil}}
\hbox to \textwidth{\ing[width=0.45\textwidth]{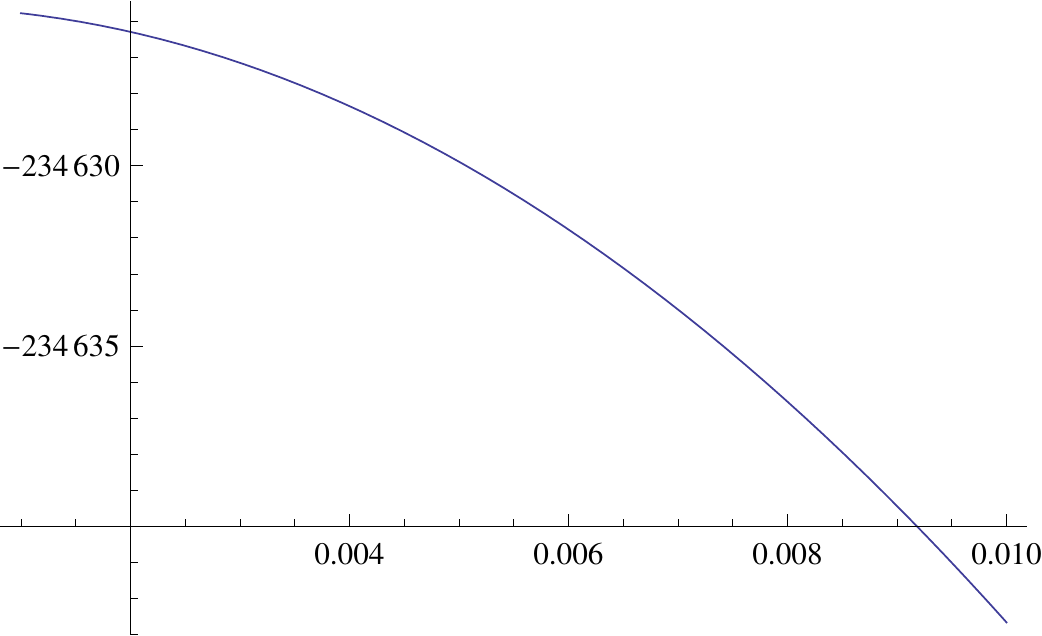}\hfil
\ing[width=0.45\textwidth]{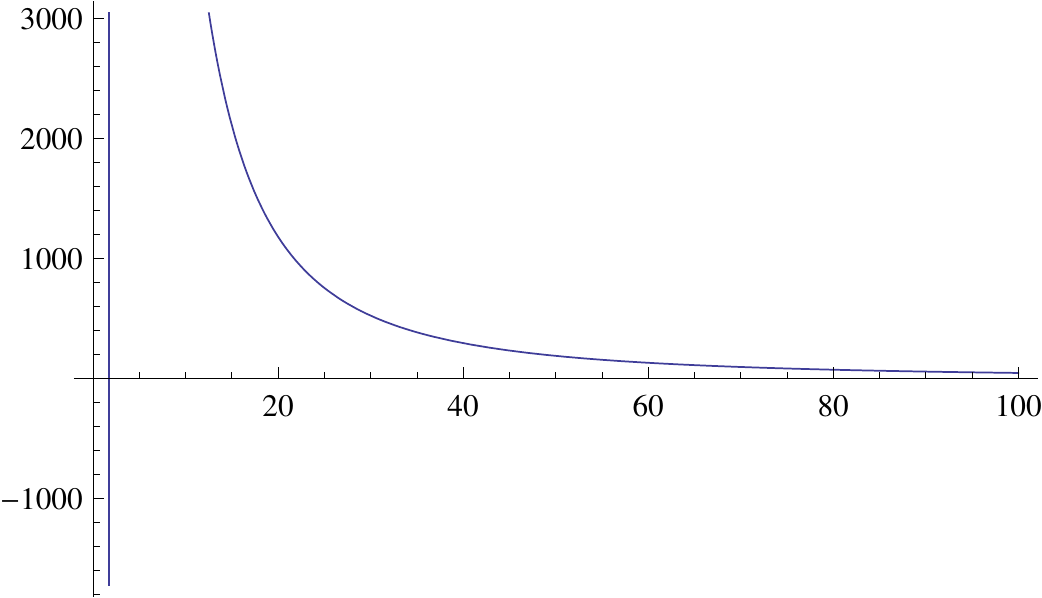}}
\hbox to \textwidth{\hbox to 0.45\textwidth{\hfil(U)\hfil}\hfil
\hbox to 0.45\textwidth{\hfil(V)\hfil}}
\hbox to \textwidth{\ing[width=0.45\textwidth]{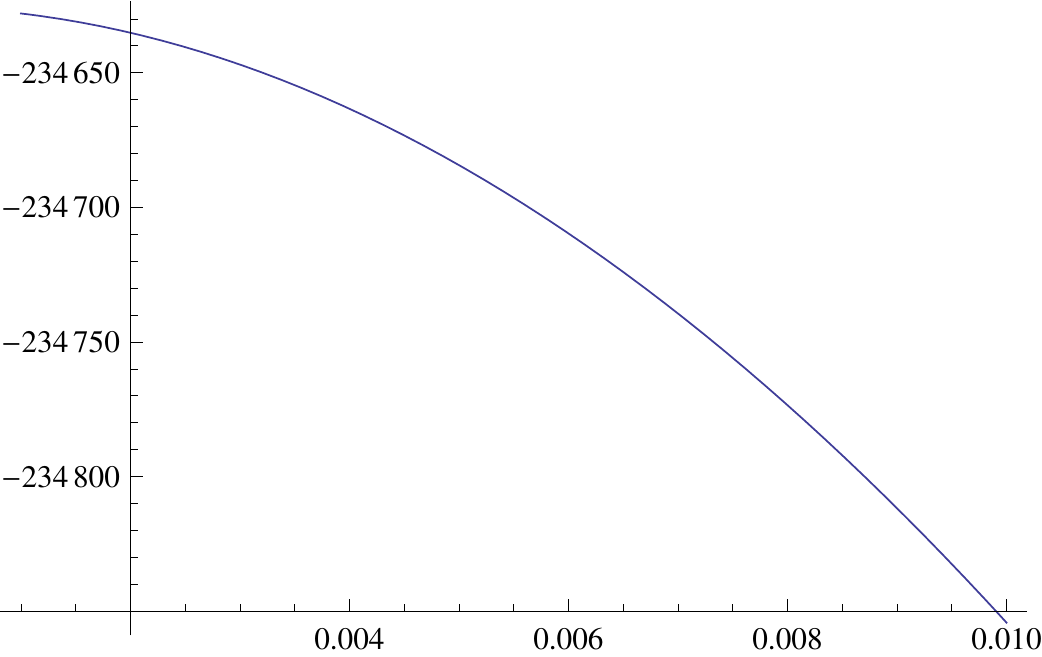}\hfil
\ing[width=0.45\textwidth]{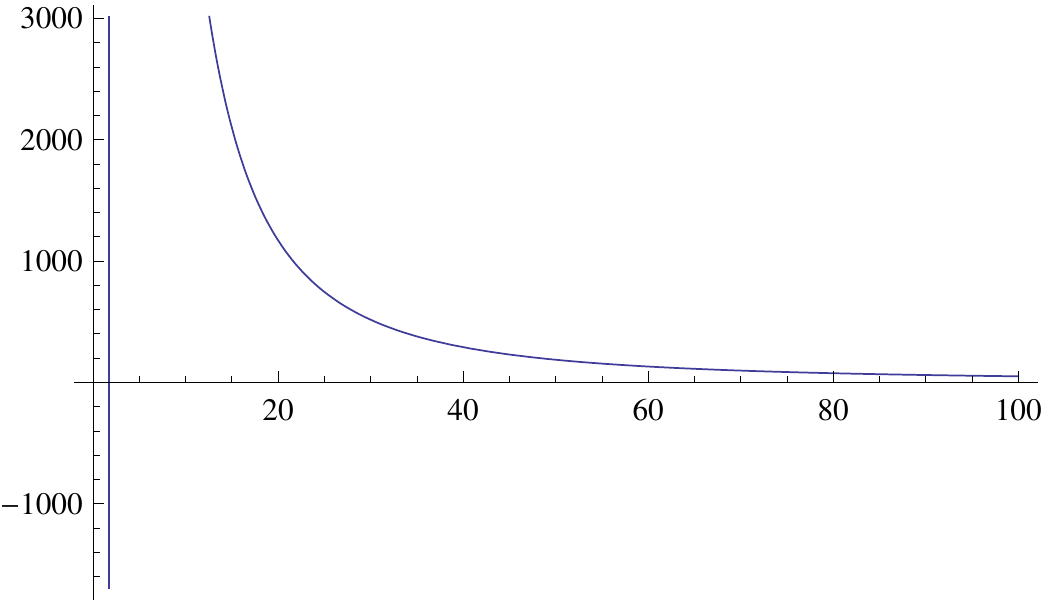}}
\hbox to \textwidth{\hbox to 0.45\textwidth{\hfil(X)\hfil}\hfil
\hbox to 0.45\textwidth{\hfil(Y)\hfil}}
\medskip
{\small \noindent Figure \thefigure\ (cont.):
(Q)---a plot of $\eta_1(r)$ for $10^{-3}<r<10^{-2}$,
(R)---a plot of $\eta_2(r)+233600$ for $10^{-3}<r<10^{-2}$,
(S)---a plot of $-\pz{\ov U}r(r)$ for $10^{-3}<r<10^{-2}$,
(T)---a plot of $-\pz{\ov U}r-5.21136\t10^{-15}(r)$ for $10^{-3}<r<100$,
(U)---a plot of $-\frac{b_1(r)}{\pz{\ov U}r(r)}$ for $10^{-3}<r<10^{-2}$,
(V)---a plot of $-\frac{b_1(r)}{\pz{\ov U}r(r)}$ for $10^{-2}<r<100$,
(X)---a plot of $-\frac{b_2(r)}{\pz{\ov U}r(r)}$ for $10^{-3}<r<10^{-2}$,
(Y)---a plot of $-\frac{b_2(r)}{\pz{\ov U}r(r)}$ for $10^{-2}<r<100$.

}

\bigskip

On Fig. \ref{q} we plot our results for $\wt V_1(r)$, $\wt V_2(r)$, $\wt
b_1(r)$, $\wt b_2(r)$ for several regions of~$r$.

\bigskip
\refstepcounter{figure}\label{q}
\hbox to \textwidth{\ing[width=0.43\textwidth]{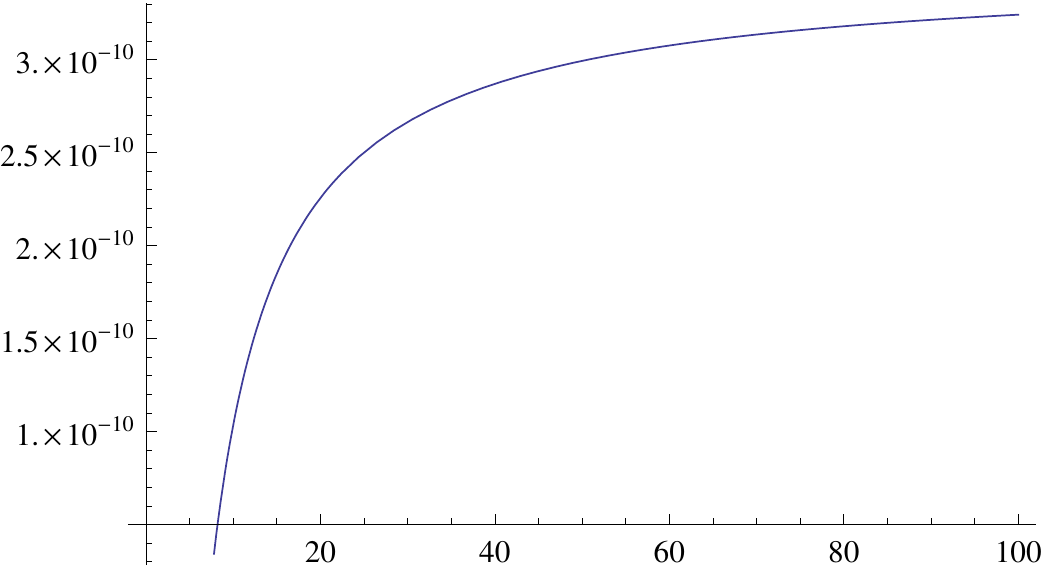}\hfil
\ing[width=0.43\textwidth]{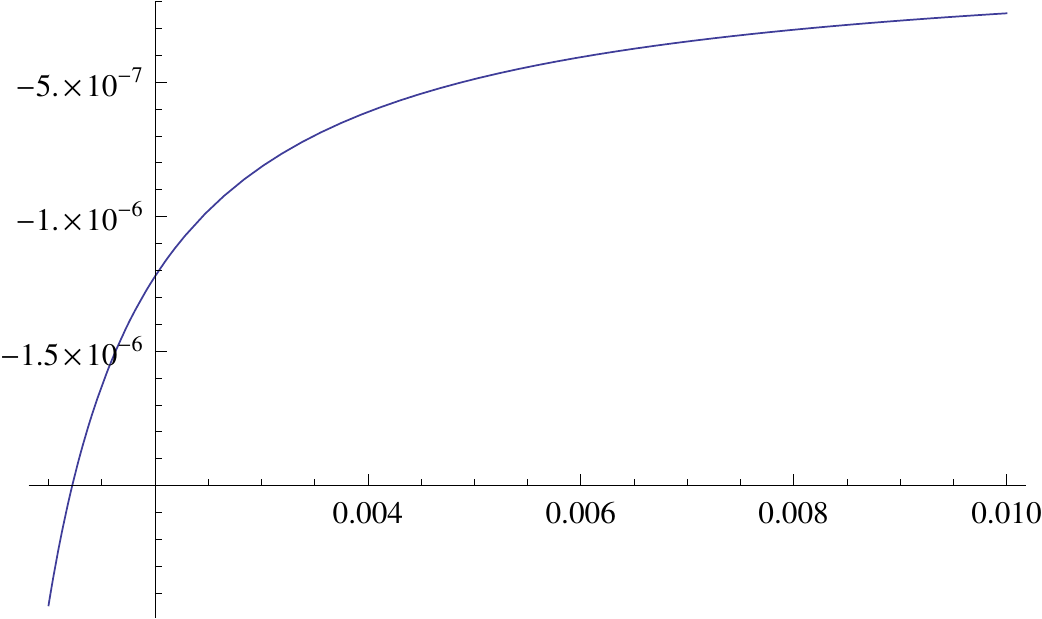}}
\hbox to \textwidth{\hbox to 0.43\textwidth{\hfil(A)\hfil}\hfil
\hbox to 0.43\textwidth{\hfil(B)\hfil}}
\hbox to \textwidth{\ing[width=0.43\textwidth]{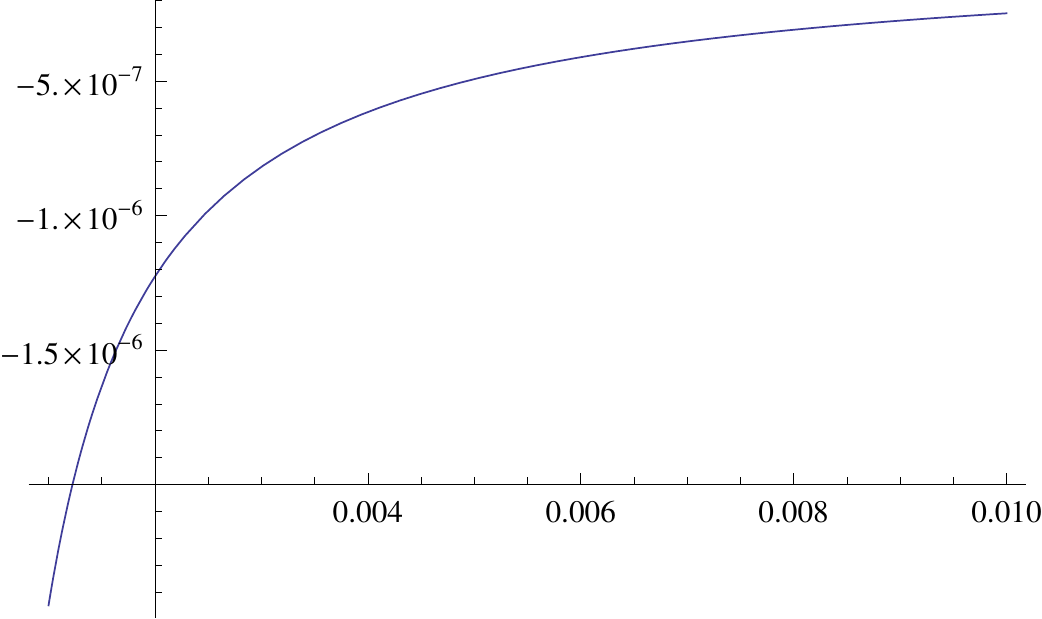}\hfil
\ing[width=0.43\textwidth]{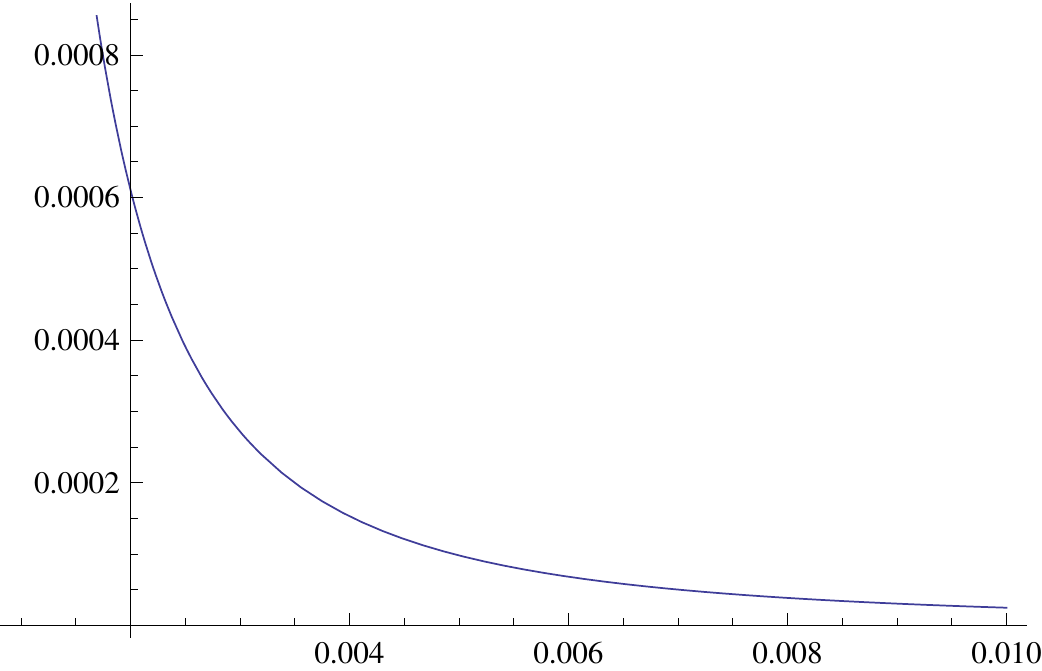}}
\hbox to \textwidth{\hbox to 0.43\textwidth{\hfil(C)\hfil}\hfil
\hbox to 0.43\textwidth{\hfil(D)\hfil}}
\hbox to \textwidth{\ing[width=0.43\textwidth]{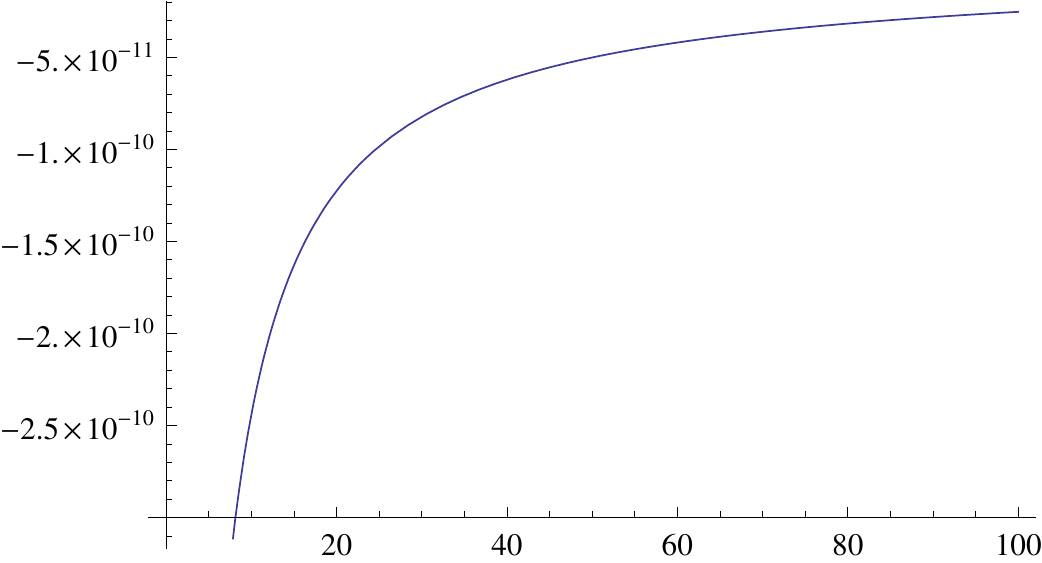}\hfil
\ing[width=0.43\textwidth]{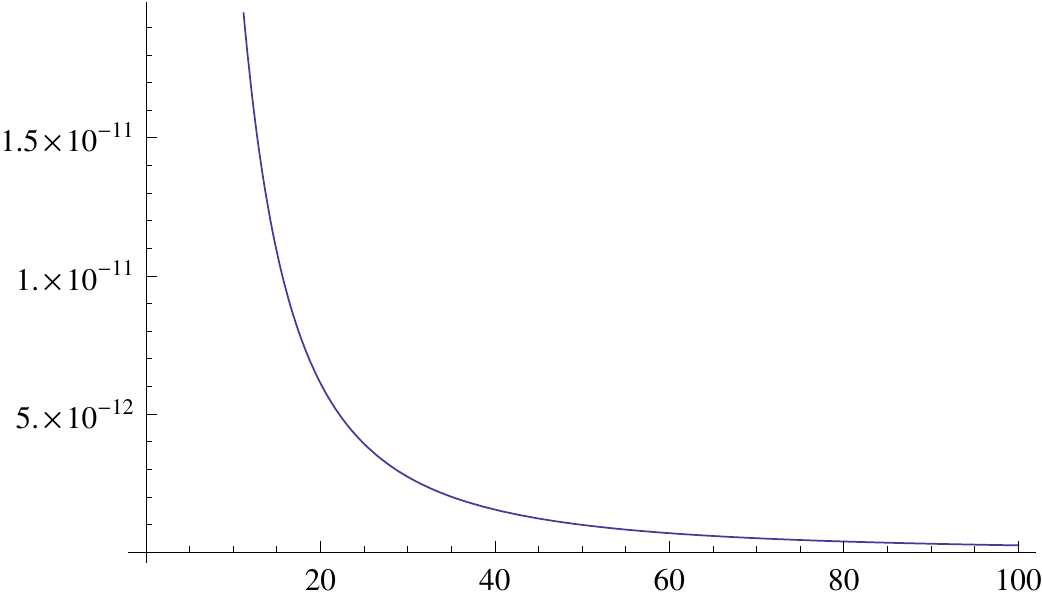}}
\hbox to \textwidth{\hbox to 0.43\textwidth{\hfil(E)\hfil}\hfil
\hbox to 0.43\textwidth{\hfil(F)\hfil}}
\hbox to \textwidth{\ing[width=0.43\textwidth]{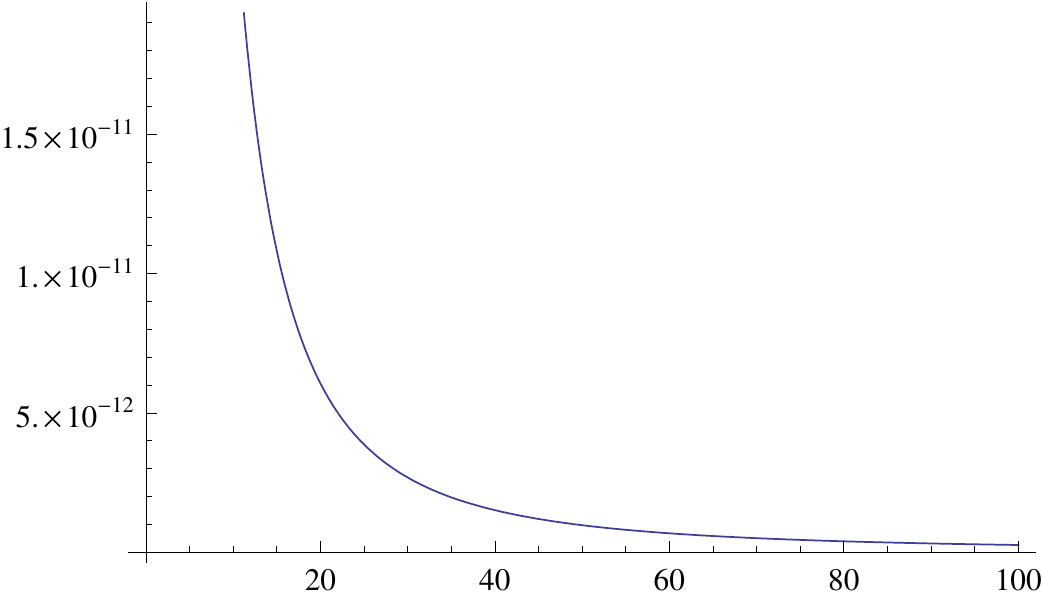}\hfil
\ing[width=0.43\textwidth]{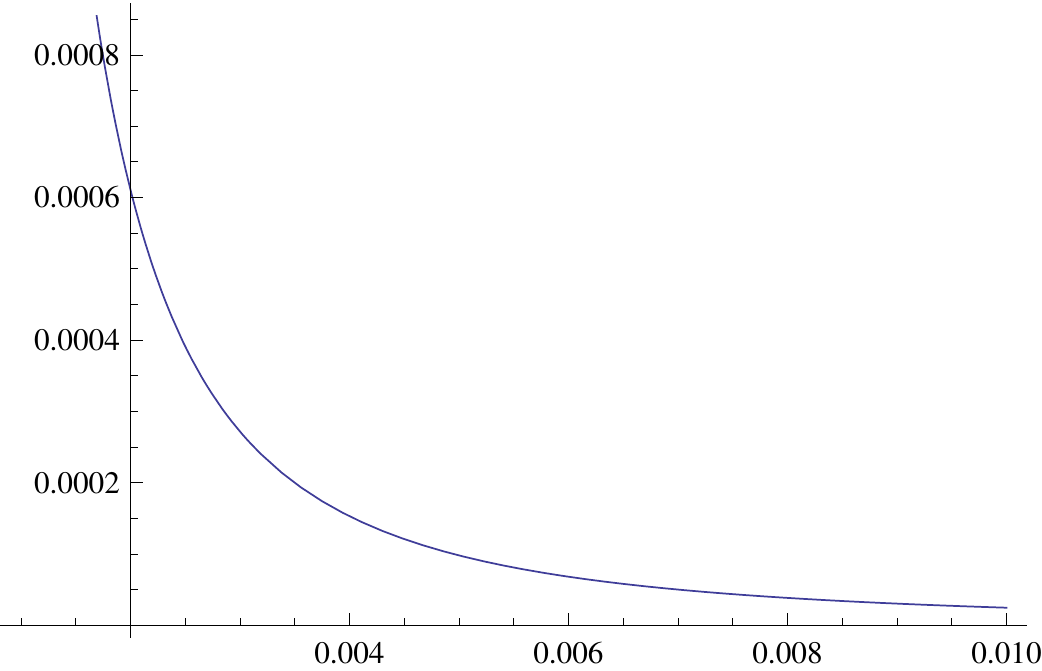}}
\hbox to \textwidth{\hbox to 0.43\textwidth{\hfil(G)\hfil}\hfil
\hbox to 0.43\textwidth{\hfil(H)\hfil}}
\smallskip \noindent
{\small Figure \thefigure:
(A)---a plot of $\wt V_1(r)$ for $10^{-3}<r<100$,
(B)---a plot of $\wt V_1(r)$ for $10^{-3}<r<10^{-2}$,
(C)---a plot of $\wt V_2(r)$ for $10^{-3}<r<10^{-2}$,
(D)---a plot of $\wt b_1(r)$ for $10^{-3}<r<10^{-2}$,
(E)---a plot of $\wt V_2(r)$ for $10^{-3}<r<100$,
(F)---a plot of $\wt b_1(r)$ for $10^{-3}<r<100$,
(G)---a plot of $\wt b_2(r)$ for $10^{-3}<r<100$,
(H)---a plot of $\wt b_2(r)$ for $10^{-3}<r<10^{-2}$.}

\hbox to \textwidth{\ing[width=0.45\textwidth]{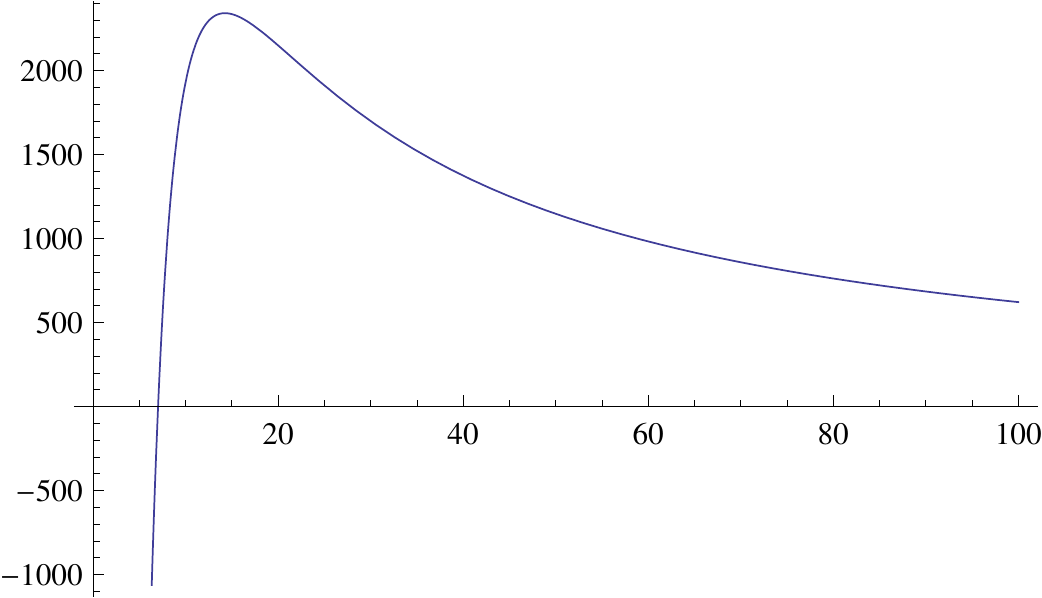}\hfil
\ing[width=0.45\textwidth]{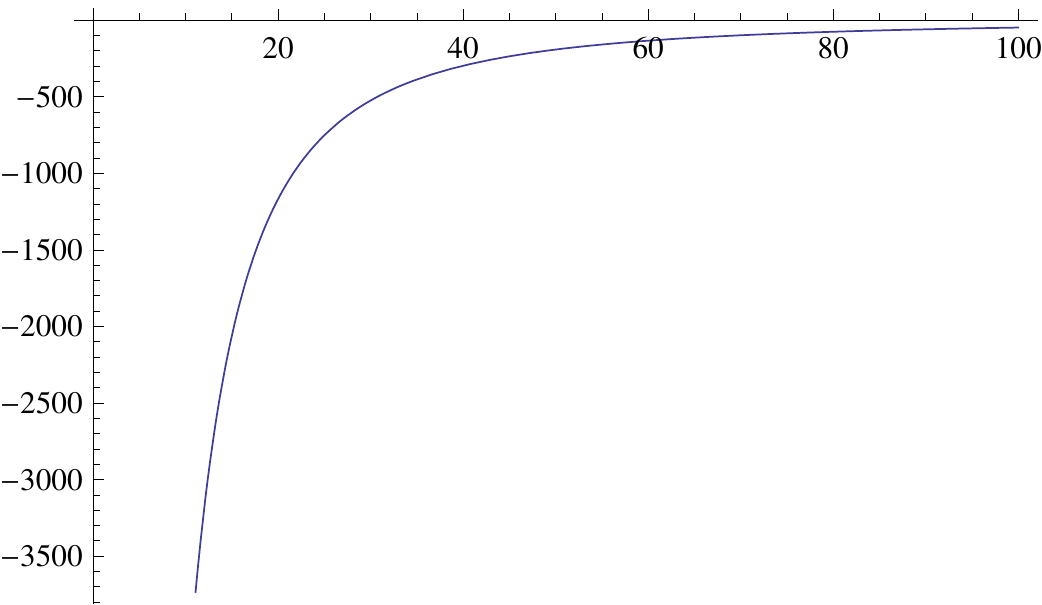}}
\hbox to \textwidth{\hbox to 0.45\textwidth{\hfil(I)\hfil}\hfil
\hbox to 0.45\textwidth{\hfil(J)\hfil}}
\hbox to \textwidth{\ing[width=0.45\textwidth]{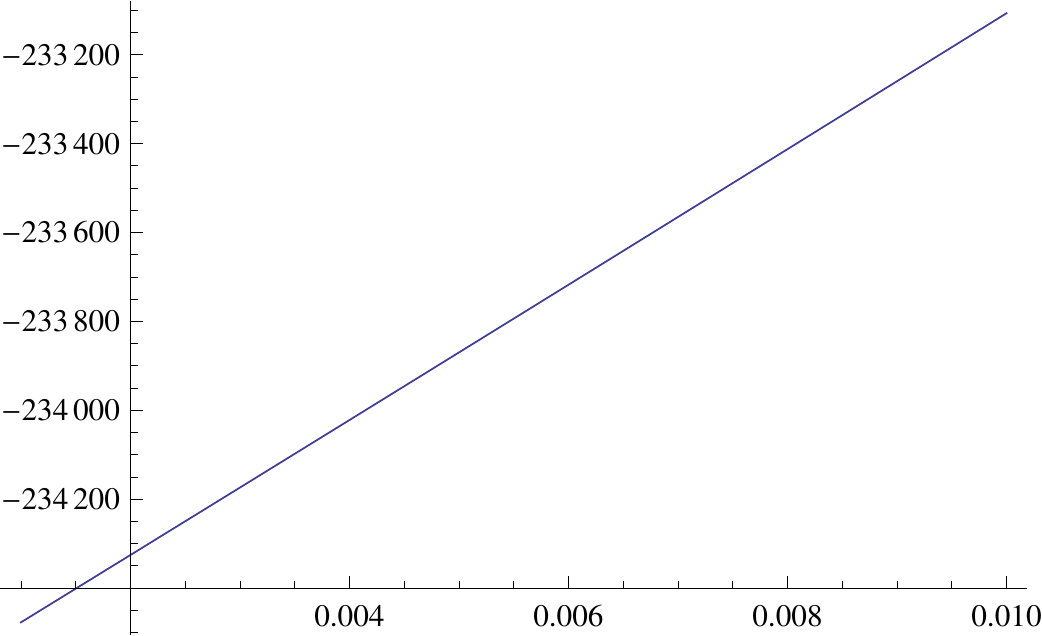}\hfil
\ing[width=0.45\textwidth]{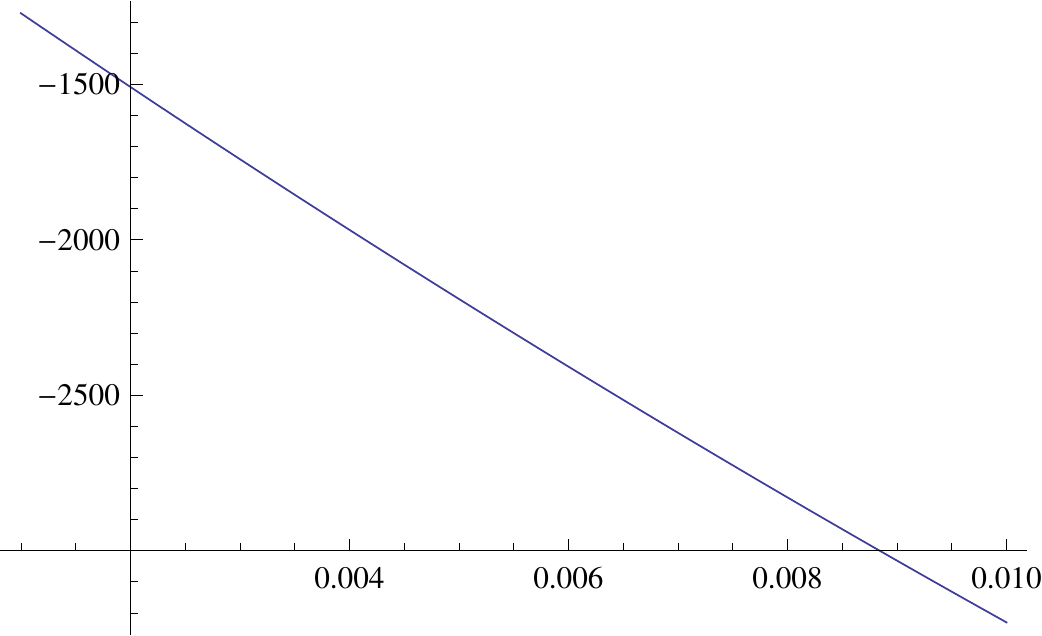}}
\hbox to \textwidth{\hbox to 0.45\textwidth{\hfil(K)\hfil}\hfil
\hbox to 0.45\textwidth{\hfil(L)\hfil}}
\hbox to \textwidth{\ing[width=0.45\textwidth]{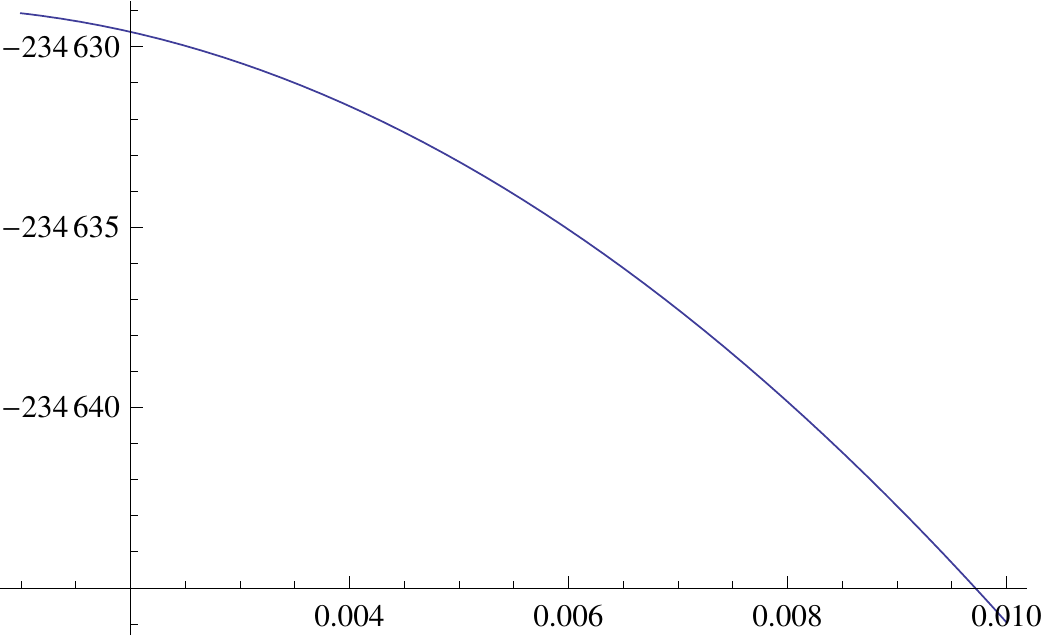}\hfil
\ing[width=0.45\textwidth]{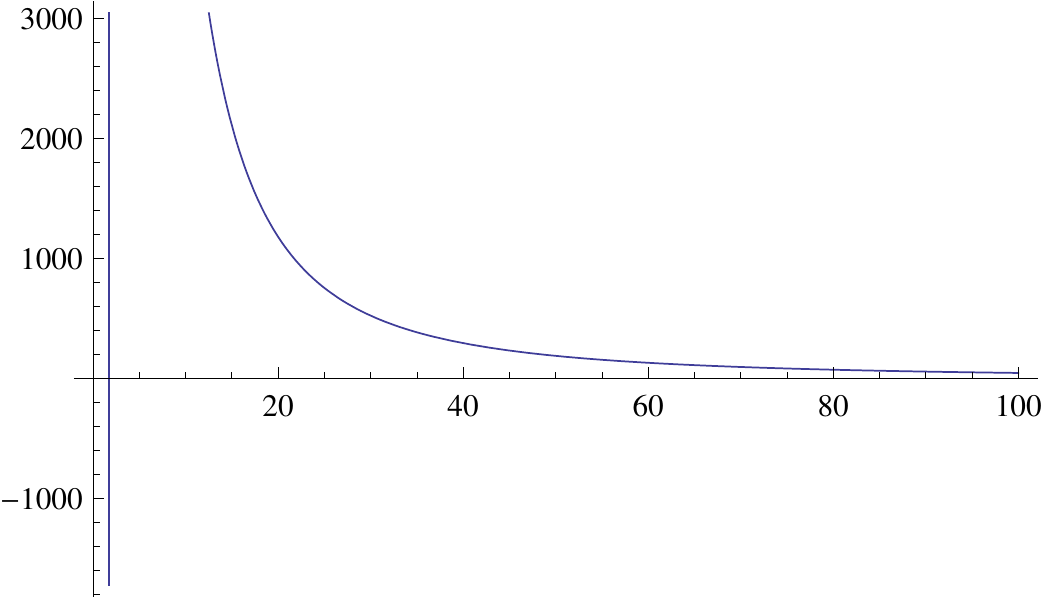}}
\hbox to \textwidth{\hbox to 0.45\textwidth{\hfil(M)\hfil}\hfil
\hbox to 0.45\textwidth{\hfil(N)\hfil}}
\hbox to \textwidth{\ing[width=0.45\textwidth]{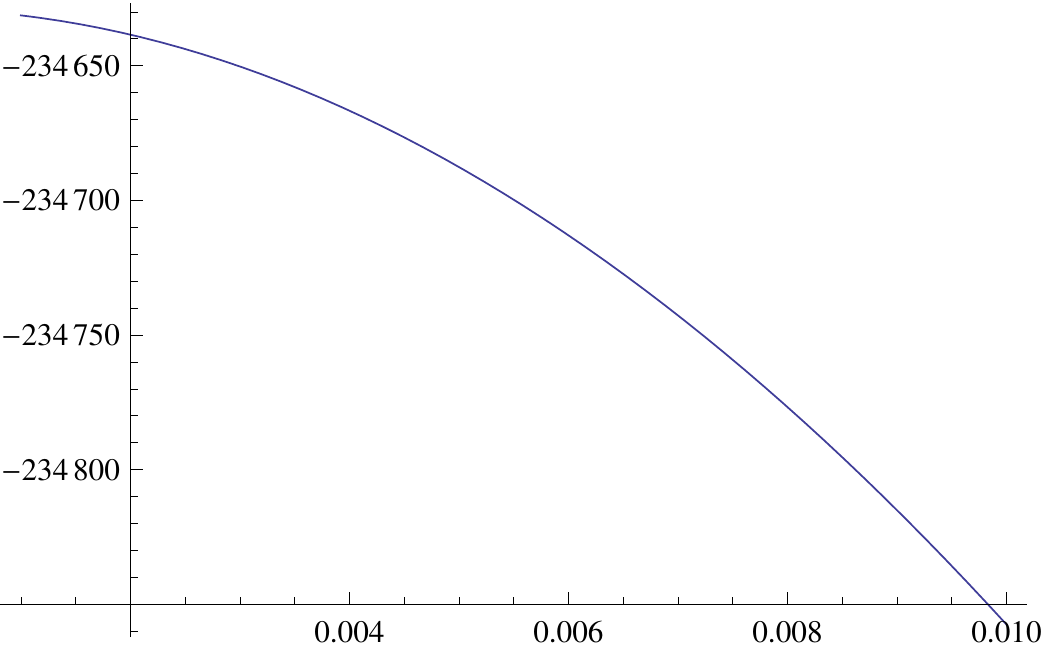}\hfil
\ing[width=0.45\textwidth]{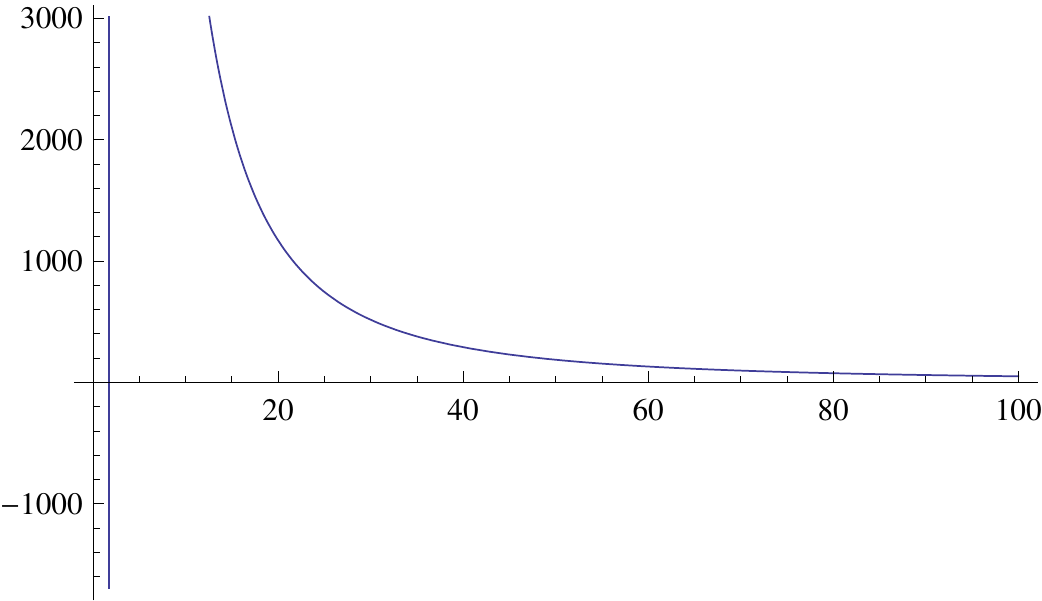}}
\hbox to \textwidth{\hbox to 0.45\textwidth{\hfil(O)\hfil}\hfil
\hbox to 0.45\textwidth{\hfil(P)\hfil}}
\medskip
{\small \noindent Figure \thefigure\ (cont.):
(I)---a plot of $\wt \eta_1(r)$ for $10^{-3}<r<100$,
(J)---a plot of $\wt \eta_2(r)$ for $10^{-3}<r<100$,
(K)---a plot of $\wt \eta_1(r)$ for $10^{-3}<r<10^{-2}$,
(L)---a plot of $\wt \eta_2(r)+233600$ for $10^{-3}<r<10^{-2}$,
(M)---a plot of $-\frac{\wt b_1(r)}{\pz{\ov U}r(r)}$ for $10^{-3}<r<10^{-2}$,
(N)---a plot of $-\frac{\wt b_1(r)}{\pz{\ov U}r(r)}$ for $10^{-3}<r<100$,
(O)---a plot of $-\frac{\wt b_2(r)}{\pz{\ov U}r(r)}$ for $10^{-3}<r<10^{-2}$,
(P)---a plot of $-\frac{\wt b_2(r)}{\pz{\ov U}r(r)}$ for $10^{-3}<r<100$.

}
\bigskip \goodbreak
We use definitions for $\ov V_1(r)$, $\ov V_2(r)$, $\wt V_1(r)$, $\wt
V_2(r)$, $b_1(r)$, $b_2(r)$, $\wt b_1(r)$, $\wt b_2(r)$ according to
\er{11.8}--\er{11.8a}, \er{11.9}--\er{11.10}, \er{11.11}--\er{11.12}.
$\eta_1(r)$, $\eta_2(r)$ are defined by Eqs \er{Da270}.
All the programmes written in Mathematica~7 are quoted in Appendix~E.

Let us notice that the difference between $\ov V_i(r)$ and $\wt V_i(r)$,
$i=1,2$, is insignificant. The same we have in the case of $b_i(r)$ and $\wt
b_i(r)$, $i=1,2$. This is reasonable for in our model $G\eff(r)$ changes very
slowly and $\La$---a \co ical \ct---is very small. Moreover, for large
distances the situation can change. For large distances we mean more than
$10^6$\,AU.

Let us notice that $e^{2(A(r)+B(r))}\ne1$ for our \so\ is not a Schwarzschild
\so. Moreover, $e^{2(A(r)+B(r))}$ is quite close to~1. The \rl\ model of an
\an\ \ac\ is close to our non\rl\ model from Section~3 on large distances
from the Sun up to 100\,AU. Moreover, it is different for small distances. On
very large distances it is significantly different. $\ov U(r)$ is defined by
the formula \er{Da269}
\bea{11.28}
\wt\eta_1(r)&=&-\frac{\wt V_1(r)}{\ov U(r)}\\
\wt\eta_2(r)&=&-\frac{\wt V_2(r)}{\ov U(r)}.\label{11.29}
\end{eqnarray}
$\ov\rho(r)$ is smaller than a density of an interplanetary dust (see
Appendix~D for details).

Our model of an \an\ \ac\ can be considered seriously only if it passes all
general \rl\ tests in the \SS. It means it should give predictions for \ph\
movement of Mercury, Venus, Earth and Icarus close to General Relativity
predictions. It should also give predictions for a bending of light and
Shapiro effect consistent with observations.

Let us consider a \ph\ movement. According to our considerations from
Appendix~D an additional \ph\ advance of a planet due to an \an\ \ac\ per one
revolution reads (see \er{Db301}):
\beq{11.30}
\D \ov\a=\pi \ov b_1
\end{equation}
where
\beq{11.31}
\ov b_1=-4.24764 \t 10^{-8} \t \frac1\pi \int_0^\pi \frac{
a_e^2(1-e^2)^2}{(1+e\cos\vf)^2} \cdot b_1\X2(\frac{0.243a_e(1-e^2)}{1+e\cos\vf}\Y2)
\cos\vf\,d\vf
\end{equation}
in appropriate systems of units \st $a_e$ is measured in AU.

Moreover, we need $\D\ov\a$ in arc seconds per century. Thus one gets
\beq{11.32}
\D\a''/{\rm century}=\frac{648\t10^4}{\pi T}\,\D\a
\end{equation}
where $T$ is a revolution time in years.

Calculating $\D\a''$ for Mercury, Venus, Earth and Icarus one gets in
comparison to GR results.

\bigskip
\centerline{\begin{tabular}{|l|c|c|}
\hline
&\vbox{\hsize=0.36\textwidth \parindent0pt \small
\begin{center} \vskip-8pt
General Relativity \ph\ advance in arc seconds per century
\end{center}\vskip-8pt }
&\vbox{\hsize=0.36\textwidth \parindent0pt \small
\begin{center} \vskip-8pt
Additional \ph\ movement due to an \an\ \ac\ in arc seconds per century
\end{center}\vskip-8pt }\\
\hline
Mercury&$43.1\pm0.5$&$2.09$\\
\hline
Venus&$8\pm5$&$0.17$\\
\hline
Earth&$5\pm1$&$0.47$\\
\hline
Icarus&$10\pm1$&$10.63$\\
\hline
\end{tabular}}

\medskip
It is easy to see that our model does not contradict GR results
for the Earth and Venus and due to this passes the first test. However, it is
in disagreement for Icarus and Mercury. This means that we should tune a
model (via initial conditions) on small distances from the Sun.

Let us consider the Shapiro effect in our case. This is the Shapiro effect
for a round trip of a radar signal from the Earth to Venus and back from
Venus to the Earth during a superior conjunction of Venus.
According to Appendix~D we get
\beq{11.33}
\D t=\D t_{GR}+2\X1(\d t(r_E,\ov r_0)+\d t(\ov r_0,r_V)\Y1)
\end{equation}
where $\D t_{GR}$ is a GR result equal $220\,\mu\rm s$ and
\beq{11.34}
\d t(\ov r_0,r)=\frac1c\int_{\ov r_0}^r \X2(1-\frac{r_s}r\Y2)^{-1}
\X1(\ov V_1(r)+\ov V_2(r)\Y1)\X2(1-\frac{b^2}{r^2}\X2(1-\frac{r_s}r\Y2)
\Y2)^{-1/2}dr.
\end{equation}
Let us estimate an additional correction.
Thus one gets from numerical calculations
$$
2\X1(\d t(r_E,\ov r_0)+\d t(\ov r_0,r_V)\Y1) \simeq 7\,\mu{\rm s}.
$$

Thus we are in agreement with Shapiro experiment. We should also consider a
bending of light. Moreover, in Appendix~D we consider this problem (see
\er{Db315}) getting consistency with observations.

\def\ea#1{e^{#1A(r)}}
Let us consider a motion of massive point bodies. We have the following \e s
\bea{11.39}
\X2(\pz r\tau\Y2)^2&=&\frac{H^2}{c^2\ea2\eb2}-\frac{h^2}{r^2\eb2}
-\frac{c^2}{\eb2}\\
\X2(\pz rt\Y2)^2&=&\frac{c^2\ea2}{\eb2}\X3(1-\frac{h^2c^2}{H^2}\,
\frac{\ea2}{r^2}-\frac{c^4\ea2}{H^2}\Y3)\label{11.40}\\
\X2(\pz r\vf\Y2)^2&=&\frac{H^2r^4}{h^2c^2\ea2\eb2}-\frac{r^2}{\eb2}
-\frac{r^4c^2}{h^2\eb2}\label{11.41}\\
\pz \vf t&=&\frac{c^2h\ea2}{Hr^2}\label{11.42}\\
\pz \vf \tau&=&\frac{h}{r^2(\tau)}.\label{11.43}
\end{eqnarray}
$t$ means a \cd\ time, $\tau$ means a proper time. A motion is considered in
the equatorial plane.

According to our parametrization,
\bea{11.44}
H&=&c^2+\frac{k^2}{2a_h}\\
\hbox{or} \q H&=&c^2-\frac{k^2}{2a_e}\label{11.45}\\
h&=&k\sqrt{a_h(e^2-1)}, \q e>1, \label{11.46}\\
\hbox{or} \q h&=&k\sqrt{a_e(1-e^2)}, \q e<1. \label{11.47}
\end{eqnarray}
Using $r_0$ unit as a unit of length and supposing that $a_h$ is measured in
AU, one gets
\bea{11.48}
\pz r\tau&=& \pm \frac c{\eb{}}\sqrt{\frac1{\ell_1}\,\ea{-2}-
\frac1{\ell_3r^2}-1}\\
\pz rt&=&\pm\frac{c\ea{}}{\eb{}}\sqrt{1-\frac{\ell_2\ea2}{r^2}-\ell_1\ea2}
\label{11.49}\\
\pz r\vf&=&\pm\frac r{\eb{}}\sqrt{\frac1{\ell_2}\,\ea{-2}-1-\ell_3r^2}
\label{11.50}\\
\pz \vf\tau &=&\frac h{r^2}\label{11.51}\\
\pz \vf t&=&\frac{\ell_4}{r^2}\,\ea2\label{11.52}
\end{eqnarray}
where
\bea{11.53}
c&=&4.889\t 10^{-4}\,[\tfrac{r_0}{\rm s}]\\
h&=&0.3764\sqrt{a_h(e^2-1)}\,\rm[\tfrac1s] \label{11.54}\\
\ell_1&=&\frac{c^2}{H^2}=\frac{a_h^2}{(a_h+4.831\t10^{-9})^2} \label{11.55}\\
\ell_2&=&\frac{h^2c^2}{H^2}=\frac{9.662\t 10^{-9}a_h^3(e^2-1)}
{(a_h+4.831\t10^{-9})^2}\label{11.56}\\
\ell_3&=&\frac{c^2}{h^2}=\frac{1.035\t 10^8}{a_h(e^2-1)} \label{11.57}\\
\ell_4&=&\frac{1.243 a_h^{3/2}(e^2-1)^{1/2}}{10^6a_h+4.831\t10^{-3}}
\,\rm[\tfrac1s] \label{11.57a}\\
\hbox{or \ }
\ell_4&=&\frac{3.919 \t 10^7 a_h^{3/2}(e^2-1)^{1/2}}{10^6a_h+4.831\t
10^{-3}}\,\rm[\tfrac1{yr}]\label{11.58}\\
h&=&11.87\t 10^6 \sqrt{a_h(e^2-1)}\,\rm[\tfrac1{yr}]\label{11.59}\\
c&=&1.542\t10^4\,[\tfrac{r_0}{\rm yr}].\label{11.60}
\end{eqnarray}

We are interested in unbounded orbits, i.e.\ hyperbolic-like. Moreover, it is
quite easy to generalize the above formulae to the elliptic-like case.

In a non\rl\ case $H\simeq c^2$ one can simplify the above formulae
\bea{11.61}
\pz r\tau&=& \pm \frac c{\eb{}}\sqrt{\ea{-2}-1-\frac1{\ell_3r^2}}\\
\pz rt&=&\pm\frac{c\ea{}}{\eb{}}\sqrt{1-\ea2-\frac{\ea2}{r^2\ell^3}}
\label{11.62}\\
\pz r\vf&=&\pm\frac r{\eb{}}\sqrt{\frac{r^2}{\ell_3}(\ea{-2}-1)-1}
\label{11.63}\\
\pz \vf\tau &=&\frac h{r^2}\label{11.64}\\
\pz \vf t&=&\frac{h}{r^2}\,\ea2\,.\label{11.65}
\end{eqnarray}

\bigskip
\refstepcounter{figure}\label{aq}
\hbox to \textwidth{\ing[width=0.45\textwidth]{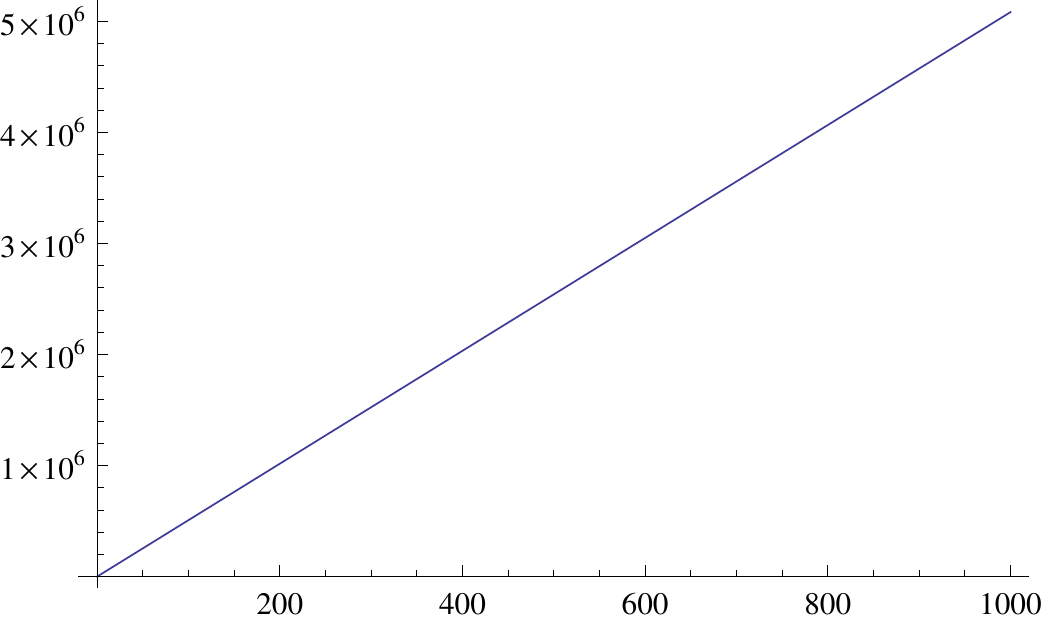}\hfil
\ing[width=0.45\textwidth]{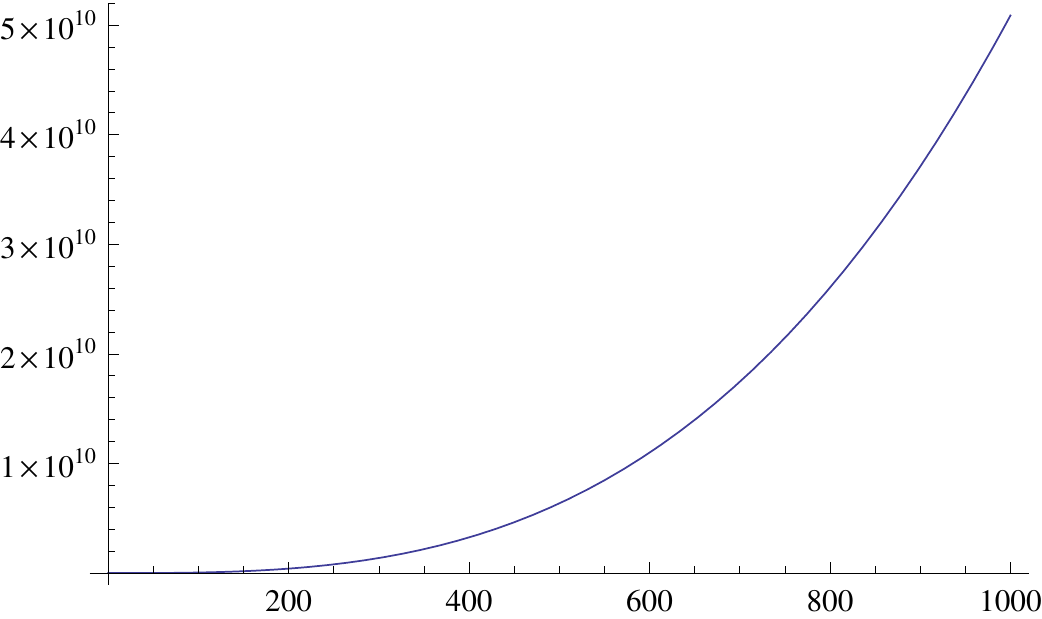}}
\hbox to \textwidth{\hbox to 0.45\textwidth{\hfil(A)\hfil}\hfil
\hbox to 0.45\textwidth{\hfil(B)\hfil}}
\hbox to \textwidth{\hfil \ing[width=0.67\textwidth]{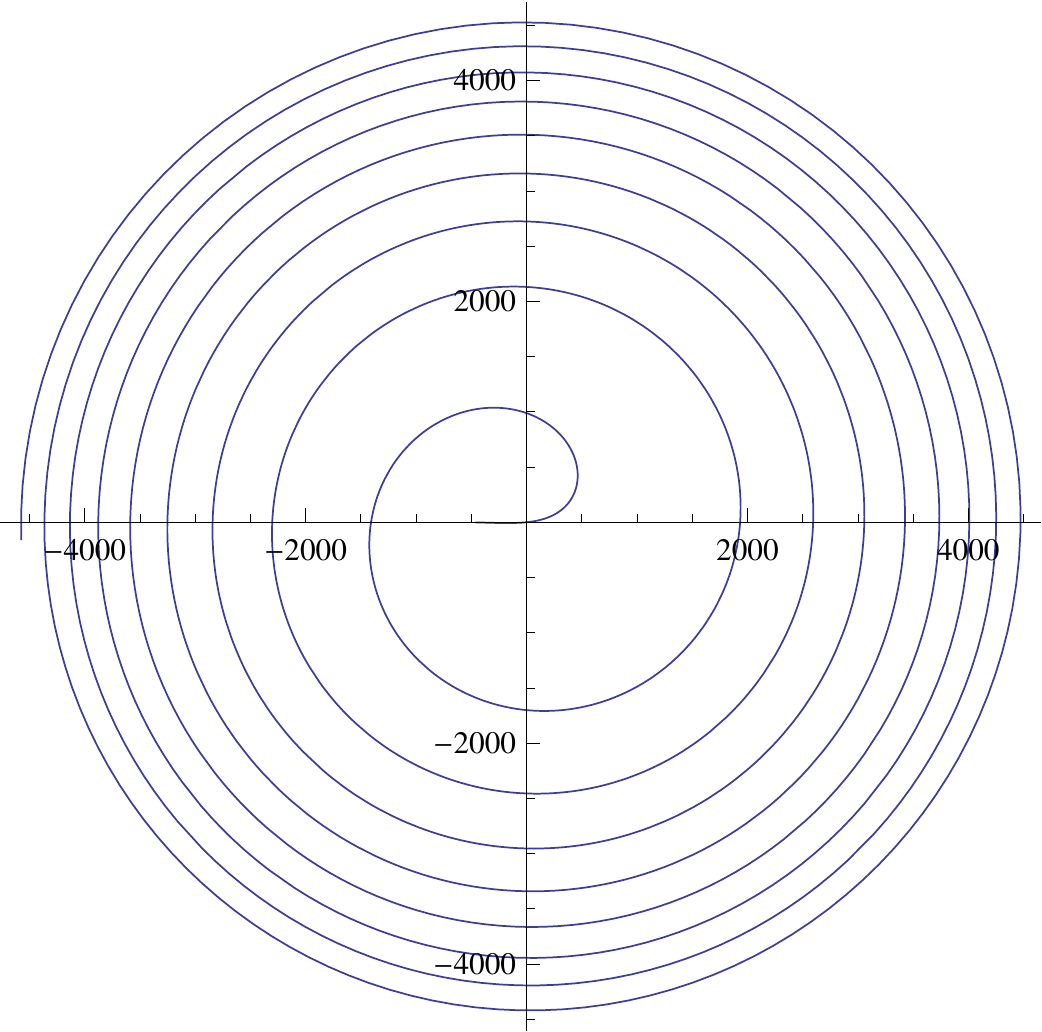}\hfil}
\hbox to \textwidth{\hfil (C)\hfil}
\smallskip \noindent
{\small Figure \thefigure:
(A)---a plot of $r(\tau)$ for a massive point body with $a_h=1.5\t
10^{-6}$\,AU, $e=5$,
(B)---a plot of $\vf(\tau)$ for a massive point body with $a_h=1.5\t
10^{-6}$\,AU, $e=5$.
(C)---a polar plot $r(\vf)$ for a massive point body with $a_h=1.5\t
10^{-6}$\,AU, $e=5$, $r<5000r_0\simeq 2\t10^4$\,AU.}

\bigskip

Using our parametrization one writes
\bea{11.66}
\pz rt&=&\pm\frac{c\ea{}}{\eb{}}\sqrt{2V_1(r)-\frac{\ea2}{r^2\ell_3}}\\
\pz r\vf&=&\frac r{\eb{}}\sqrt{\frac{2r^2}{\ell_3}\,V_1(r)-1}\,.\label{11.67}
\end{eqnarray}
In the above formulae $r$ is measured in $r_0=4.103\div4.104$\,AU. If we want
to get results in AU we should multiply by the factor $4.103\div 4.104$.

\bigskip
\hbox to \textwidth{\hfil \ing[width=0.65\textwidth]{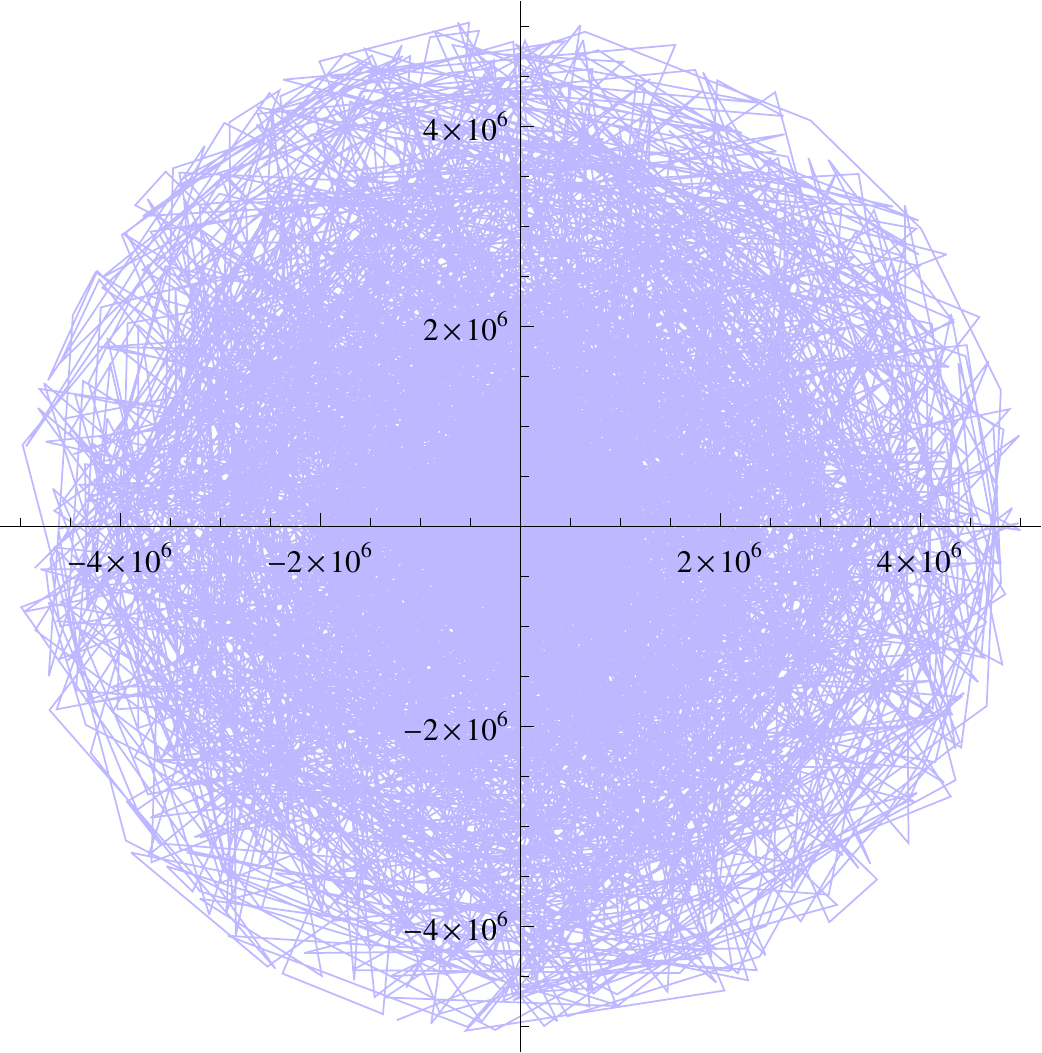}\hfil}
\hbox to \textwidth{\hfil (D)\hfil}
\smallskip \noindent
{\small Figure \thefigure\ (cont.):
(D)---a polar plot $r(\vf)$ for a massive point body with $a_h=1.5\t
10^{-6}$\,AU, $e=5$, $r<6\t10^6r_0\simeq 2.5\t10^7$\,AU.}

\bigskip

Let us consider two examples of motion for $a_h=1.5\t10^{-6}$\,AU, $e=5$ and
$a_h=1.5\t10^{-8}$\,AU, $e=5$. The first example corresponds to an asymptotic
\hy\ velocity of $0.1c$ ($c$---velocity of light). The second example is
highly \rl\ (\rl\ \sp). On Fig.~\ref{aq} we plot $r(\tau)$ and $\vf(\tau)$
for the first example. We plot also an orbit $r=r(\vf)$ in polar \cd s in two
cases for a scale of~$r_0$.

On Fig.~\ref{aqr} we plot $r(\tau)$ and $\vf(\tau)$ for the second example.
$\tau$ means a proper time. We plot also an orbit $r(\vf)$ in polar \cd s.

\medskip
\obraz{aqr}
\hbox to \textwidth{\ing[width=0.44\textwidth]{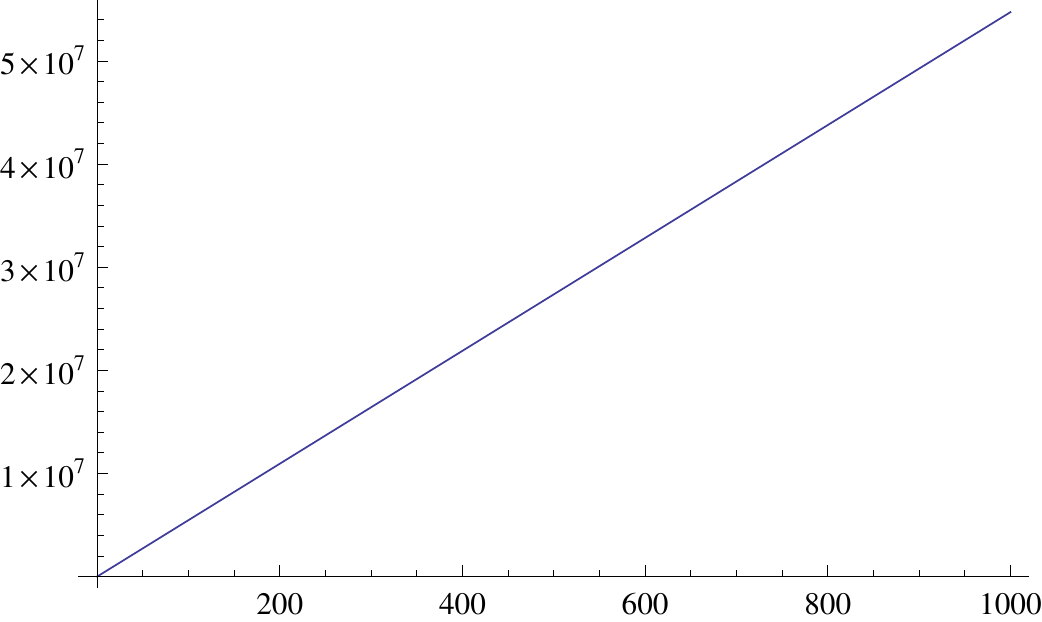}\hfil
\ing[width=0.44\textwidth]{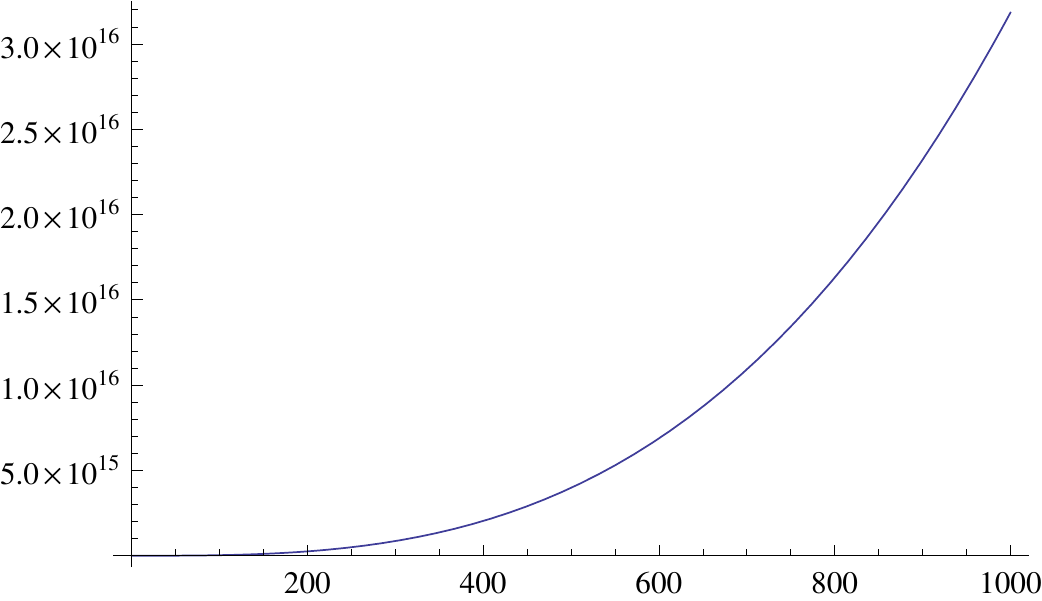}}
\hbox to \textwidth{\hbox to 0.44\textwidth{\hfil(A)\hfil}\hfil
\hbox to 0.44\textwidth{\hfil(B)\hfil}}
\medskip
\podpis{:
(A)---a plot of $r(\tau)$ for a massive point body with $a_h=1.5\t
10^{-8}$\,AU, $e=5$,
(B)---a plot of $\vf(\tau)$ for a massive point body with $a_h=1.5\t
10^{-8}$\,AU, $e=5$.}

\bigskip
\hbox to \textwidth{\hfil \ing[width=0.67\textwidth]{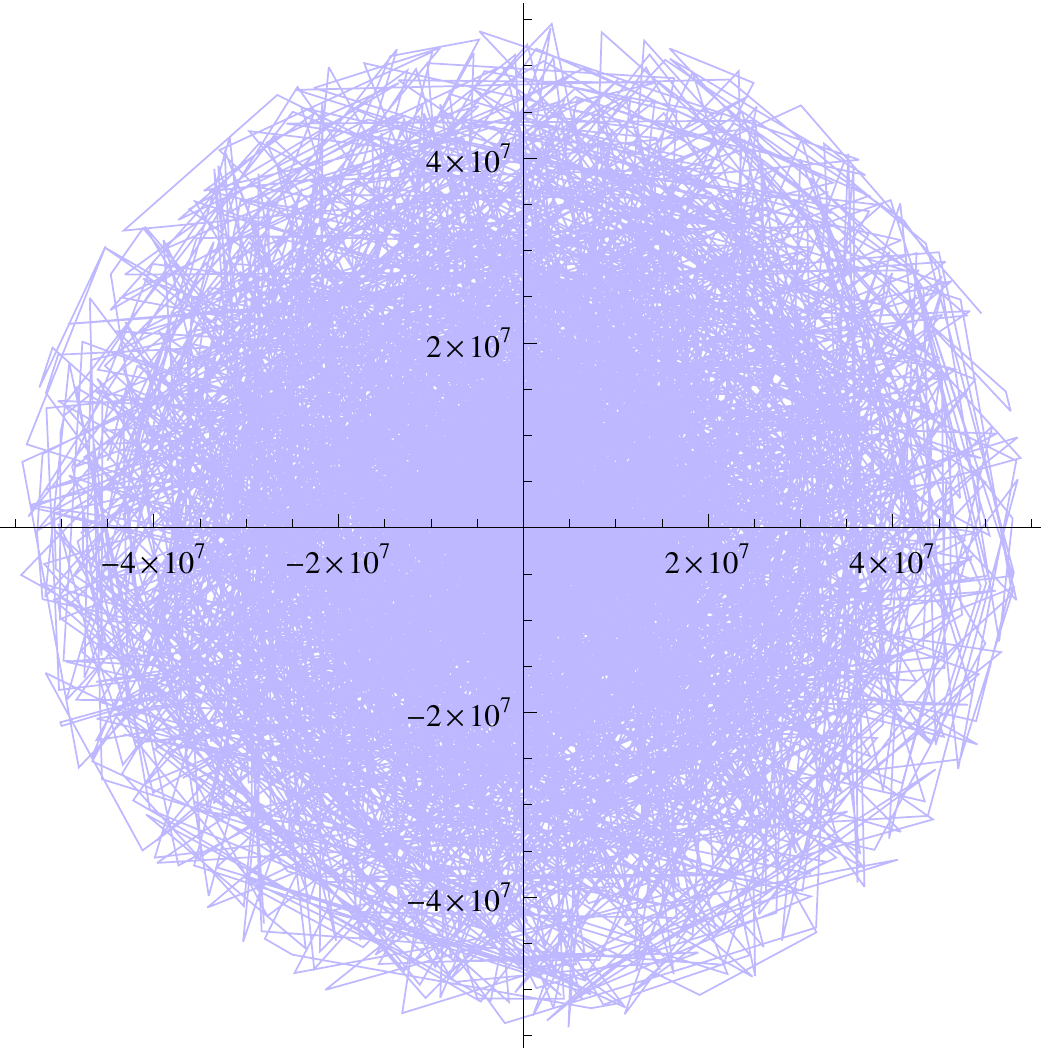}\hfil}
\hbox to \textwidth{\hfil (C)\hfil}
\smallskip \noindent
{\small Figure \thefigure\ (cont.):
(C)---a polar plot $r(\vf)$ for a massive point body with $a_h=1.5\t
10^{-8}$\,AU, $e=5$ for $r<6\t10^7r_0\simeq 2.5\t10^8$\,AU.}

\bigskip

\begin{figure}[h]
\hbox to \textwidth{\ing[width=0.45\textwidth]{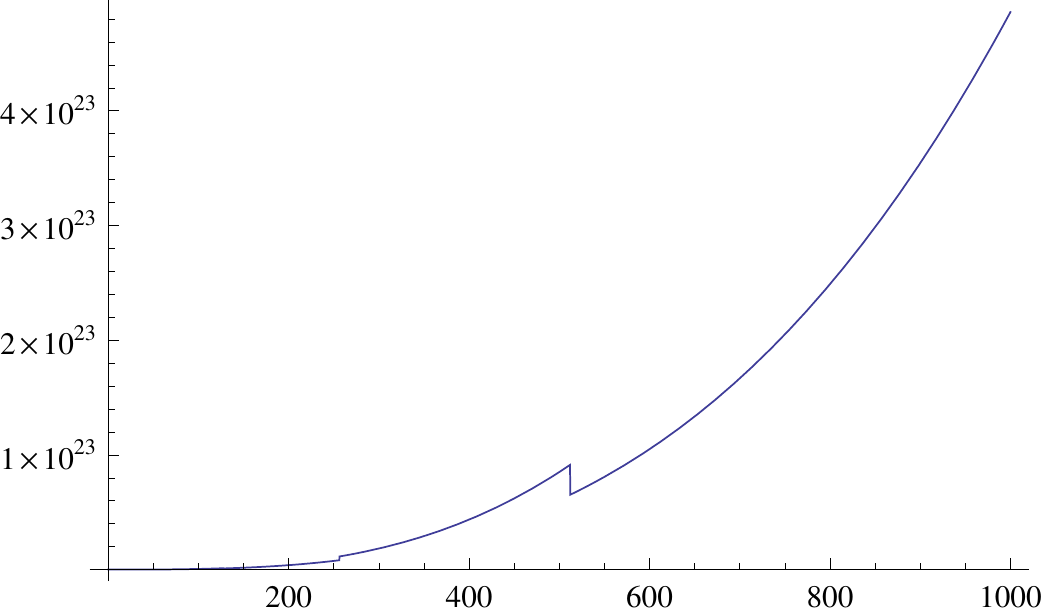}\hfil
\ing[width=0.45\textwidth]{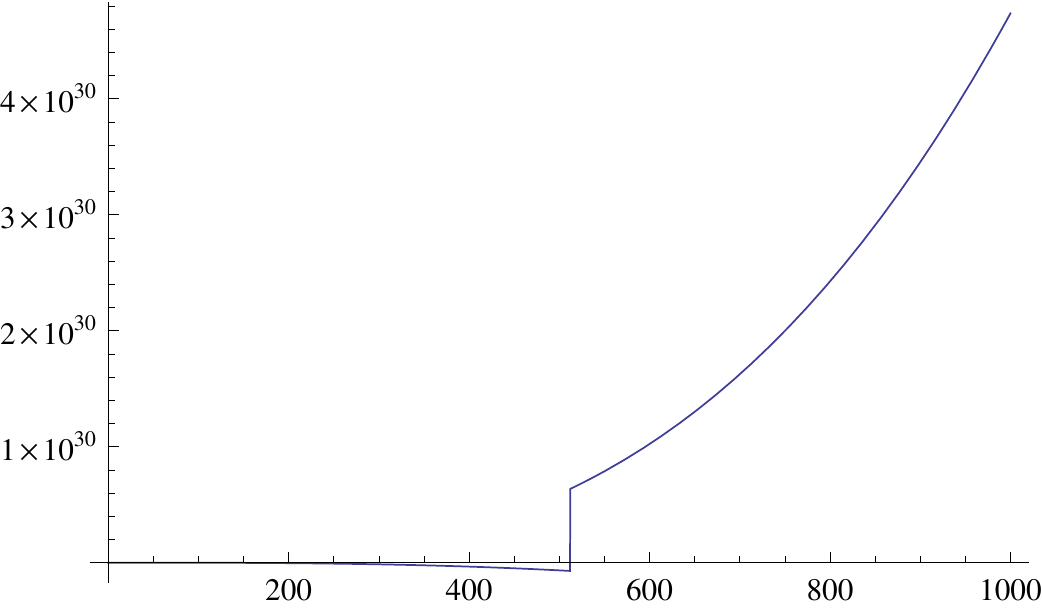}}
\hbox to \textwidth{\hbox to 0.45\textwidth{\hfil(A)\hfil}\hfil
\hbox to 0.45\textwidth{\hfil(B)\hfil}}
\caption{%
(A)---a plot of $\vf(t)$ ($t$ is a \cd\ time) for a massive point body,
$a_h=1.5\t10^{-6}$\,AU, $e=5$, $t$ measured in years;
(B)---a plot of $\vf(t)$ ($t$ is a \cd\ time) for a massive point body,
$a_h=1.5\t10^{-8}$\,AU, $e=5$, $t$ measured in years.}\label{aq5}
\end{figure}

On Fig.\ \ref{aq5} we give two plots of $\vf$ versus $t$ (a \cd\ time) for
$a_h=1.5\t 10^{-6}$\,AU, $e=5$ and $a_h=1.5\t10^{-8}$\,AU, $e=5$. It is easy
to see that in the second case (a highly \rl\ case) $\vf$ is almost \ct\
(equal to zero) during a time about 500~years. What does it mean? It means
that during this time an orbit is a straight line. However, we cannot
visualize it on Fig.~\ref{aqr}C and it is really absent if we use a proper
time~$\tau$. The \cd\ time $t$ in the case of Schwarzschild \so\ means a time
of infinitely distant observer. For our \so\ is not asymptotically flat and
it cannot be extended too far from the center of the \SS, this interpretation
fails. Thus we keep an interpretation of $t$ as a time of an observer at a
distance about $10^6$\,AU from the center of the \SS.

It is hard to visualize properties of the orbits in polar \cd s
\beq{11.77}
x(\tau)=r(\tau)\cos\vf(\tau), \q
y(\tau)=r(\tau)\sin\vf(\tau)
\end{equation}
for we get spirals with very many revolutions. In the first case, $a_h=1.5\t
10^{-6}$\,AU, about $10^{10}$; in the second case about $10^{17}$. It means
both orbits are unbounded and really chaotic. For $\tau$ is measured in years
and $r$ in $r_0$, we have to do with a spiral chaotic motion during $10^3$
years. Spirals make $10^{10}$--$10^{17}$ revolutions during $10^3$ years,
i.e.\ $10^7$--$10^{14}$ per year.

Let us consider a motion on \hy\ orbit
\beq{11.68}
\pz{^2u}{\vf^2}+u=\frac1{a_h(e^2-1)}+3\,\frac{k^2}{c^2}\,u^2-
\frac{c^2}{2k^2a_h(e^2-1)u^2}\,b_1\X2(\frac1u\Y2).
\end{equation}
Substituting numbers we get
\beq{11.69}
\pz{^2u}{\vf^2}+u=\frac{4.104}{a_h(e^2-1)}+2.898\t10^{-8}u^2
-\frac{21.24\t10^7}{a_h(e^2-1)u^2}\,b_1\X2(\frac1u\Y2)
\end{equation}
in such a way that $a_h$ is measured in AU.

We should also define initial conditions. We take them as follows:
\bg{11.70}
u(0)=\frac{4.104}{a_h(e-1)}\\
\pz u\vf(0)=0. \label{11.71}
\end{gather}
The orbit should be considered for $0\le\vf<\arccos(-\frac1e)$.

On Fig.\ \ref{pion} we give plots of \so s of Eq.\ \er{11.69} for
$r(\vf)=\frac 1{u(\vf)}$ in two cases $a_h=20$, $e=3$ and $a_h=20$, $e=5$.

\bigskip
\refstepcounter{figure}\label{pion}
\hbox to \textwidth{\ing[width=0.45\textwidth]{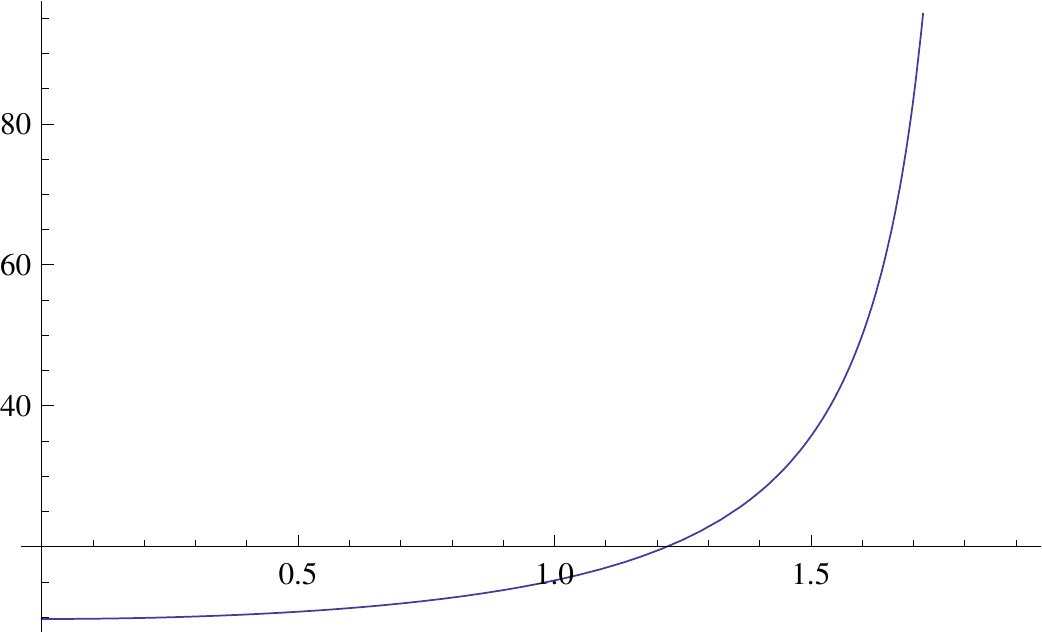}\hfil
\ing[width=0.45\textwidth]{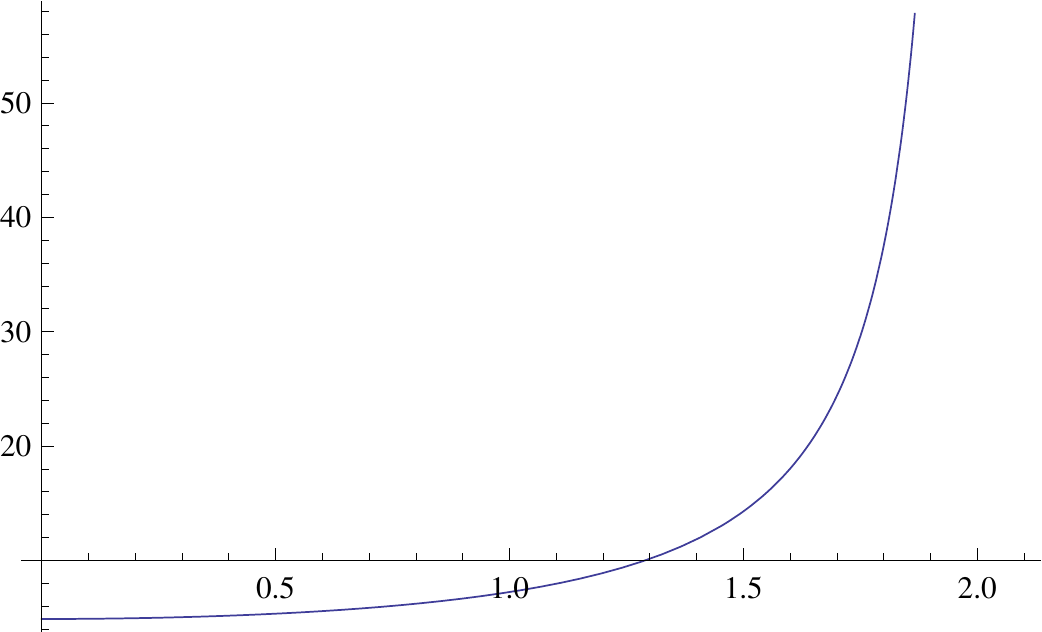}}
\hbox to \textwidth{\hbox to 0.45\textwidth{\hfil(A)\hfil}\hfil
\hbox to 0.45\textwidth{\hfil(B)\hfil}}
\smallskip \noindent
{\small Figure \thefigure:
(A)---a plot of $r(\vf)$ for $a_h=20$, $e=3$,
(B)---a plot of $r(\vf)$ for $a_h=20$, $e=2$.
}

\bigskip
\hbox to \textwidth{\ing[width=0.45\textwidth]{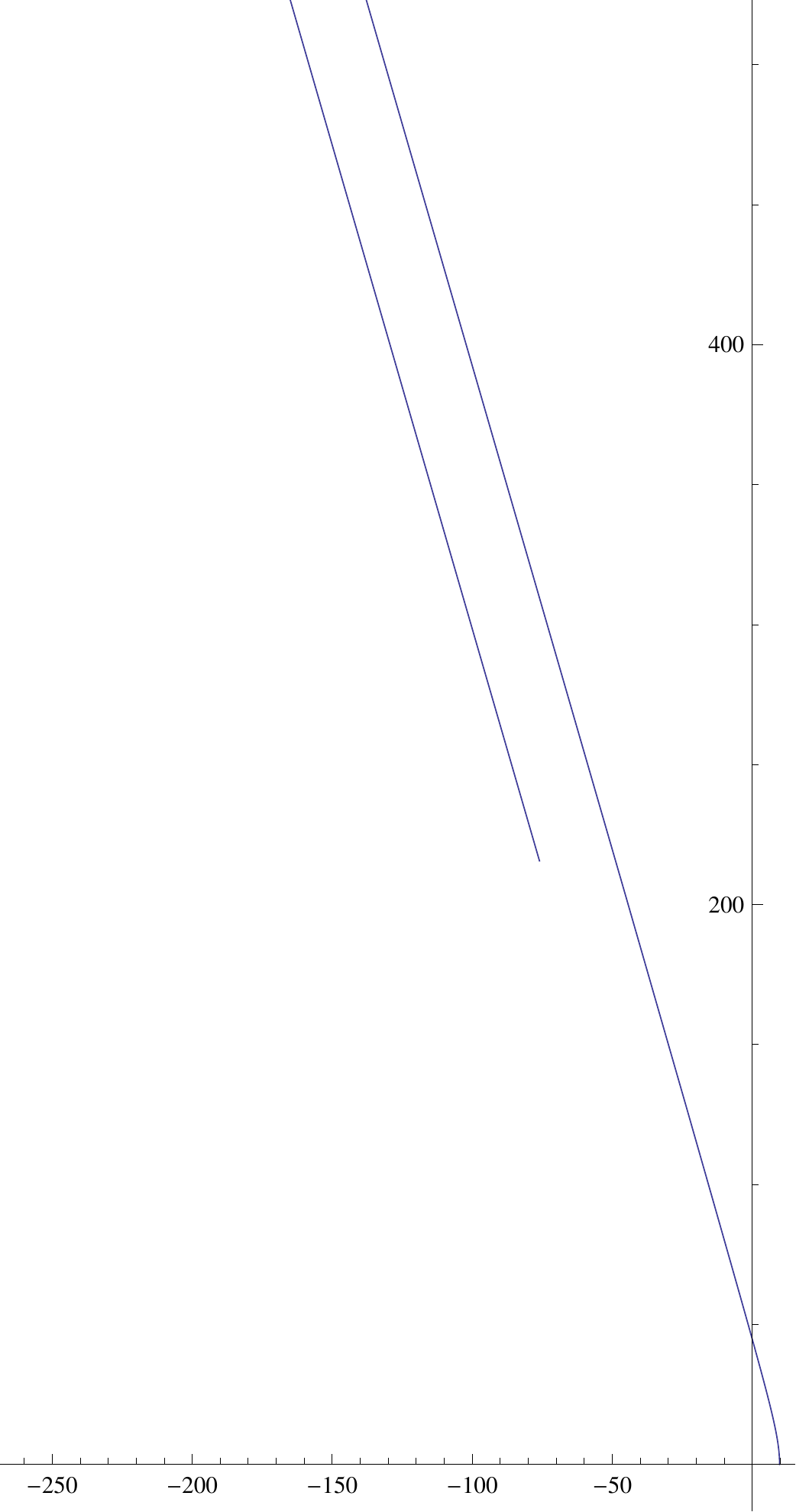}\hfil
\ing[width=0.45\textwidth]{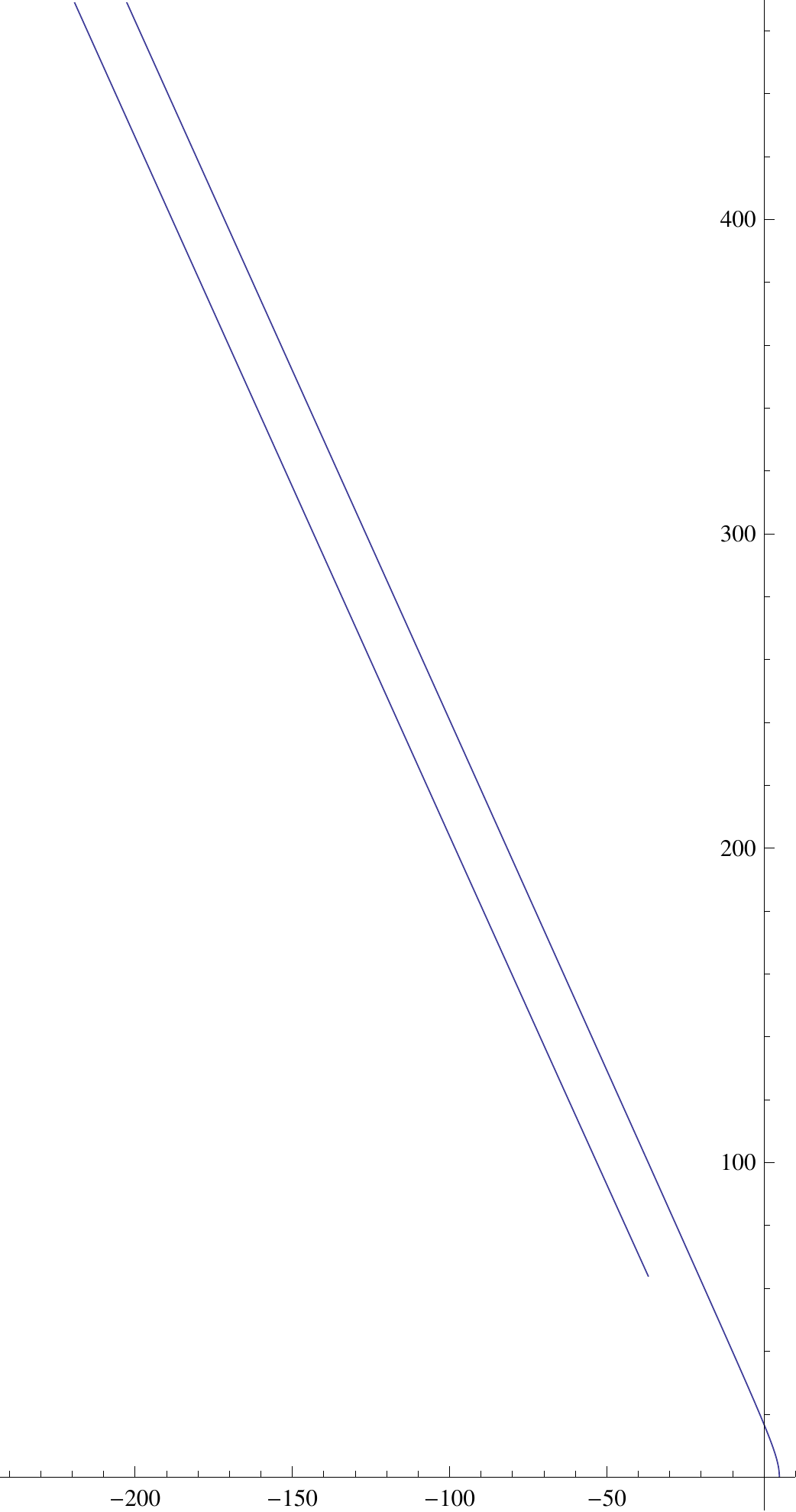}}
\hbox to \textwidth{\hbox to 0.45\textwidth{\hfil(C)\hfil}\hfil
\hbox to 0.45\textwidth{\hfil(D)\hfil}}
\smallskip \noindent
{\small Figure \thefigure\ (cont.):
(C)---a polar plot of an orbit for a massive point body with $a_h=20$, $e=3$,
(D)---a polar plot of an orbit for a massive point body with $a_h=20$, $e=2$.
}

\bigskip
\refstepcounter{figure}\label{piona}
\hbox to \textwidth{\ing[width=0.45\textwidth]{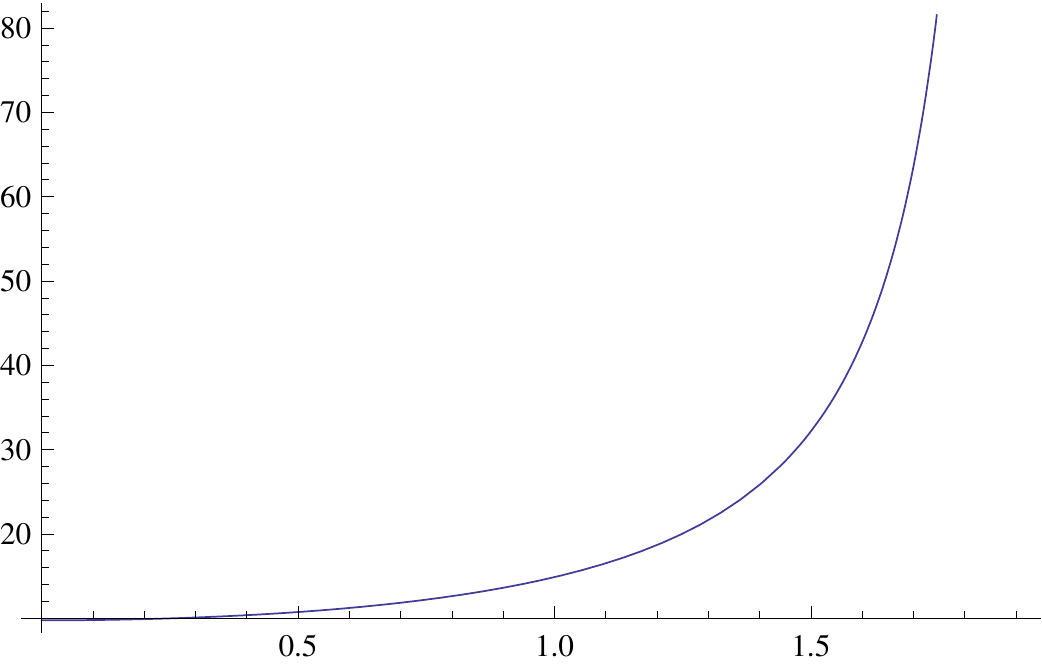}\hfil
\ing[width=0.45\textwidth]{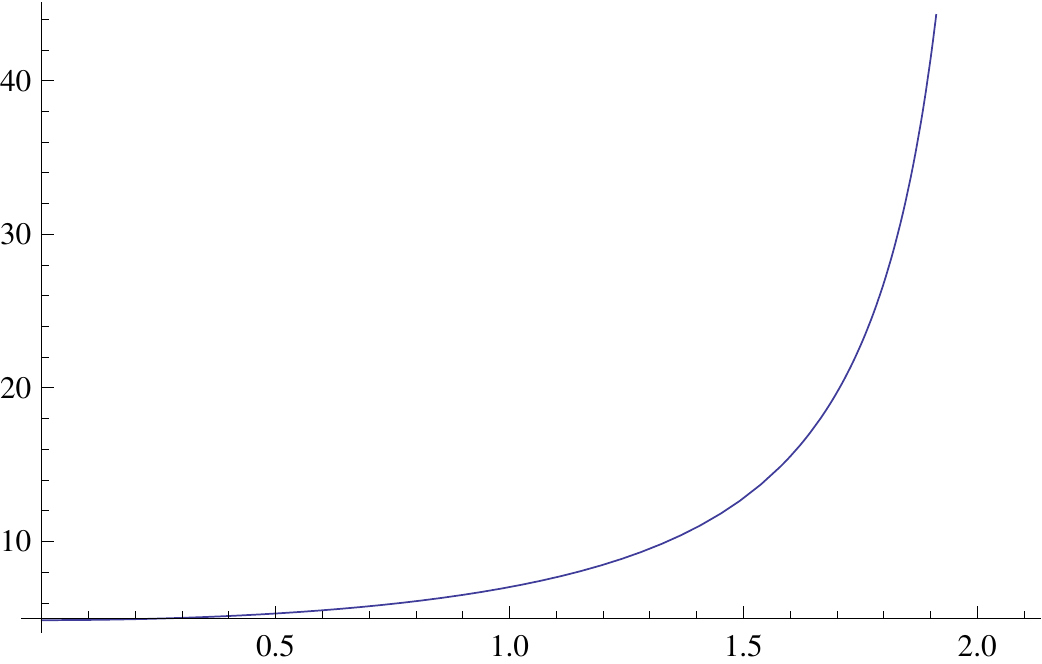}}
\hbox to \textwidth{\hbox to 0.45\textwidth{\hfil(A)\hfil}\hfil
\hbox to 0.45\textwidth{\hfil(B)\hfil}}
\smallskip \noindent
{\small Figure \thefigure:
(A)---a plot of $r(\vf)$ (without an \an\ \ac\ term) for $a_h=20$, $e=3$,
(B)---a plot of $r(\vf)$ (without an \an\ \ac\ term) for $a_h=20$, $e=2$.
}

\bigskip
\hbox to \textwidth{\ing[width=0.45\textwidth]{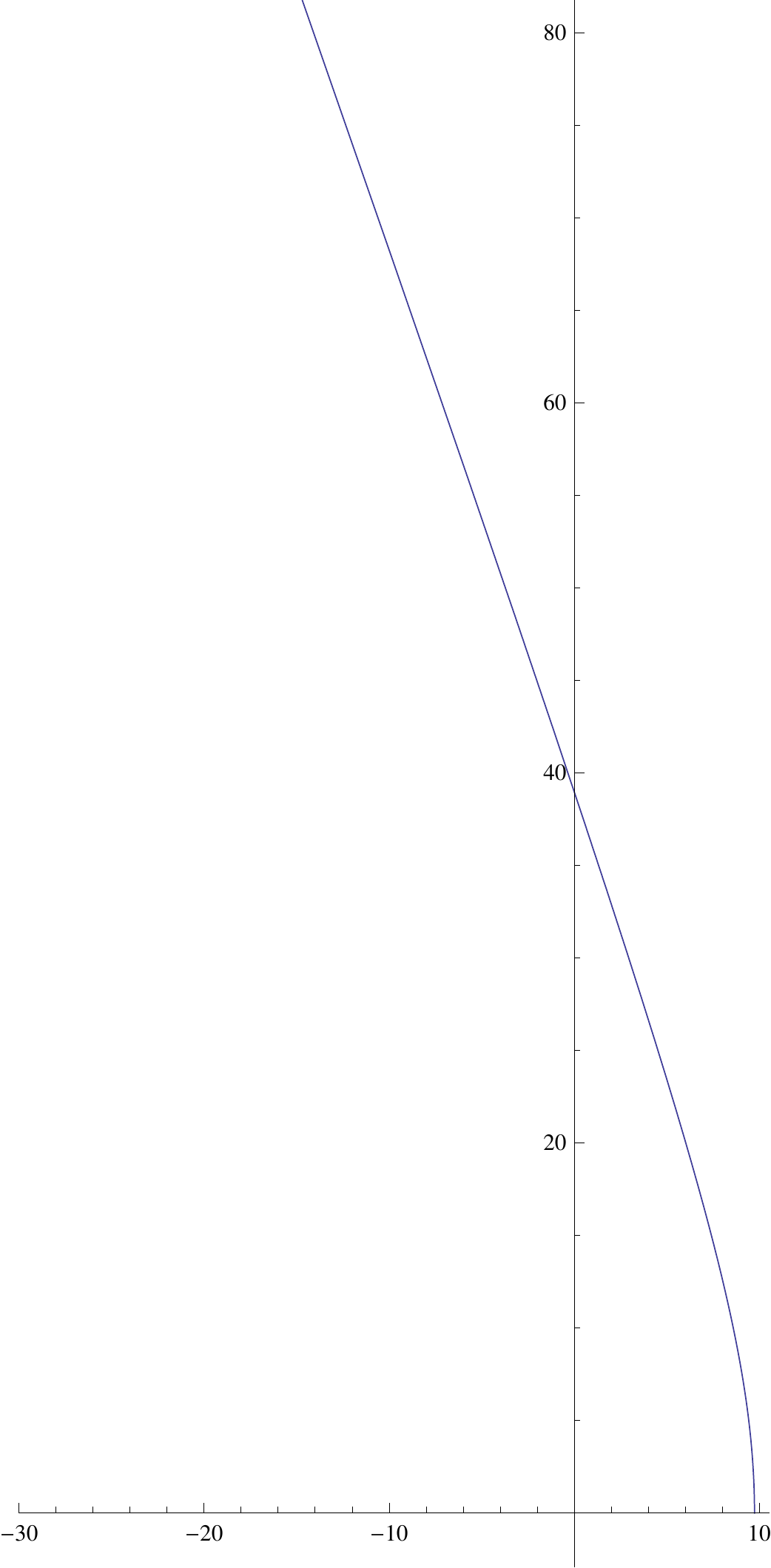}\hfil
\ing[width=0.45\textwidth]{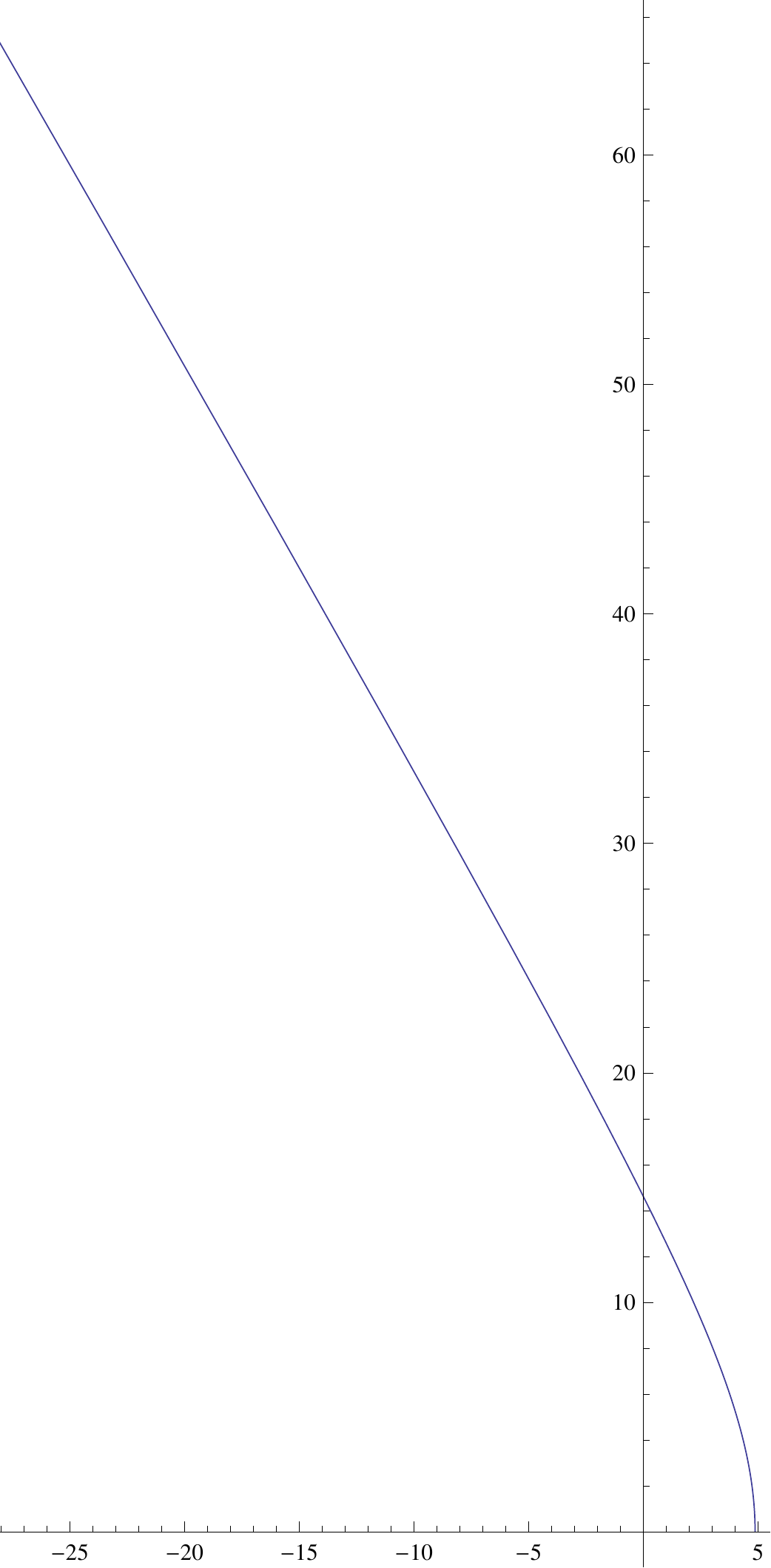}}
\hbox to \textwidth{\hbox to 0.45\textwidth{\hfil(C)\hfil}\hfil
\hbox to 0.45\textwidth{\hfil(D)\hfil}}
\smallskip \noindent
{\small Figure \thefigure\ (cont.):
(C)---a polar plot of an orbit for a massive point body
(\e s of motion without an \an\ \ac\ term) with $a_h=20$, $e=3$,
(D)---a polar plot of an orbit for a massive point body
(\e s of motion without an \an\ \ac\ term) with $a_h=20$, $e=2$.
}

On Fig.\ \ref{piona} we give plots of \so s of Eq.\ \er{11.69} without an
\an\ \ac\ term, i.e.\ only with GR-term. It means, we solve the \e
\beq{11.78}
\pz{^2u}{\vf^2}+u=\frac1{a_h(e^2-1)}+3\,\frac{k^2}{c^2}\,u^2
\end{equation}
with initial conditions
\beq{11.79}
u(0)=\frac1{a_h(e-1)}, \q \pz u\vf(0)=0.
\end{equation}
It is easy to see that GR orbits seem to be unbounded. Moreover, in the case
with an \an\ \ac\ they seem to be bounded.

From this point of view \P0/11 cannot escape from the \SS\ as we concluded in
Appendix~A ($a_h=20$ corresponds to the \P0/11 case). Moreover, in the case
of \rl\ \sp s  the escape is possible, i.e.\ in 1000 years a \sp\ with
$a_h=1.5\t 10^{-6}$ can reach a distance ${}\simeq2\t10^7$\,AU, a \sp\ with
$a_h=1.5\t10^{-8}$ a distance ${}\simeq2\t10^8$\,AU.

Let us consider our bounded orbits in the \SS\ for $I=3^\circ$, which
corresponds to the Pioneer case.

\atw0.2
\bigskip
\refstepcounter{figure}\label{pion3}
\hbox to \textwidth{\ing[width=0.46\textwidth]{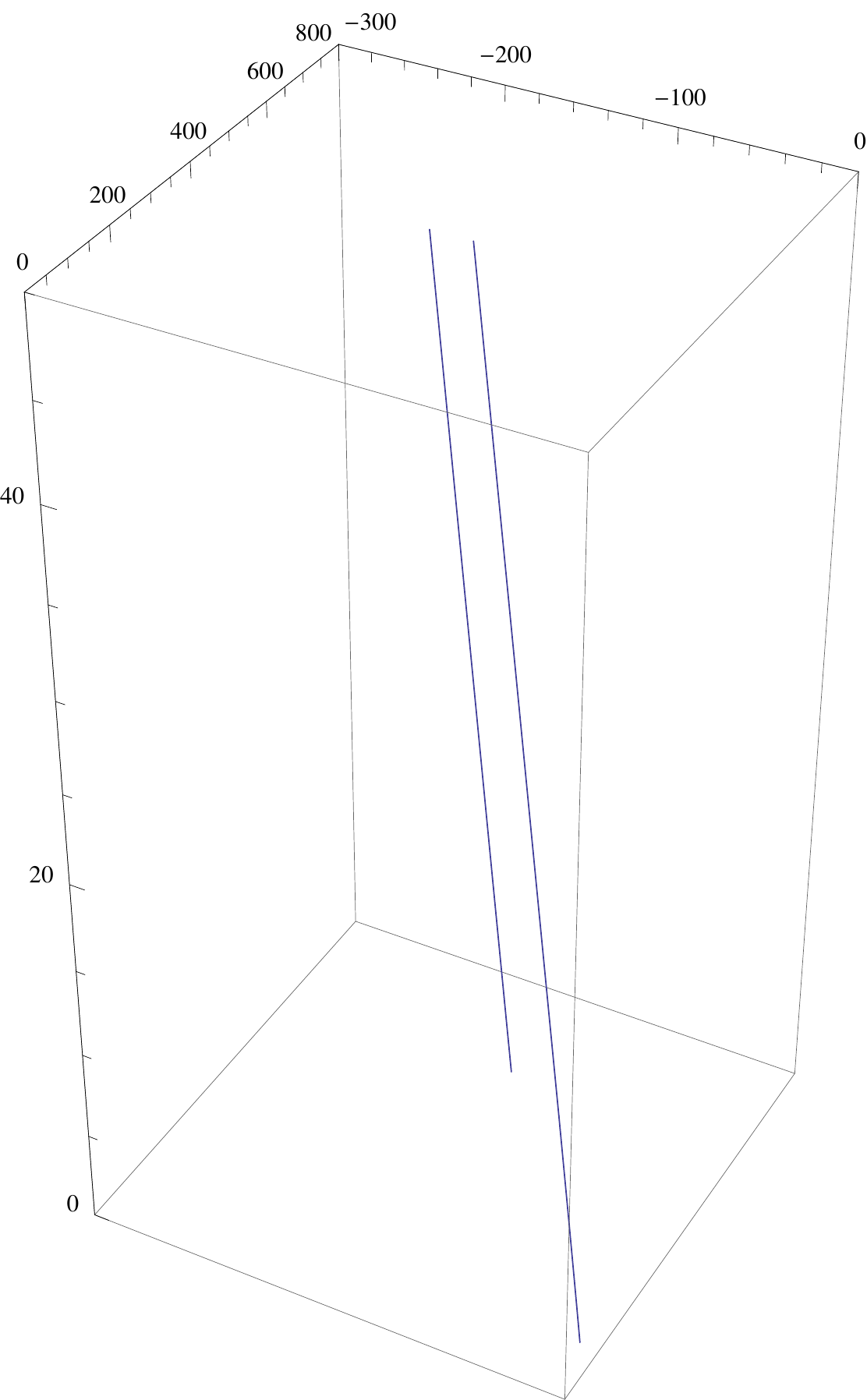}\hfil
\ing[width=0.46\textwidth]{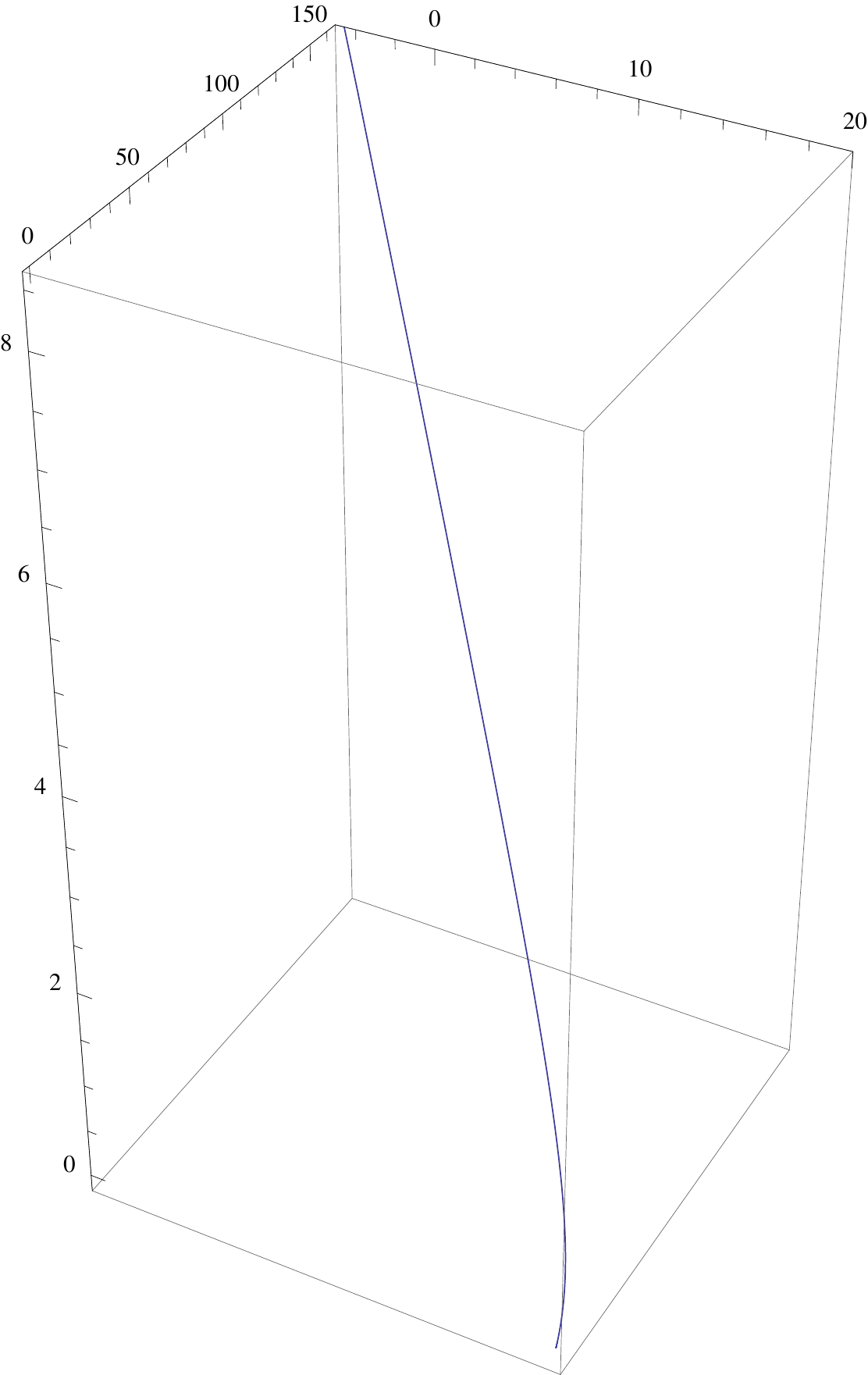}}
\hbox to \textwidth{\hbox to 0.46\textwidth{\hfil(A)\hfil}\hfil
\hbox to 0.46\textwidth{\hfil(B)\hfil}}
\smallskip \noindent
{\small Figure \thefigure:
(A)---a 3D polar plot of an orbit of a massive point body for $a_h=20$,
$e=3$, $I=3^\circ$,
(B)---a 3D polar plot of an orbit of a massive point body for $a_h=20$,
$e=5$, $I=3^\circ$.}

On Fig.\ \ref{pion3} we give a 3D plot of both orbits,
\beq{11.80}
\bal
x(\vf)&= r(\vf)\cos\vf\\
y(\vf)&= r(\vf)\sin\vf \cos I\\
z(\vf)&= r(\vf)\sin\vf \sin I.
\eal
\end{equation}
We suppose $\o=\O=0$ (see Eq.\ \er{7.2}).

On Fig.\ \ref{pion4} we give a full range plot in the equatorial plane and a
3D plot for $a_h=20$, $e=3$. On Fig.~\ref{pion2} we give plots for $a_h=20$,
$e=5$.

It is easy to see that these orbits are bounded. The range in both cases is
about $36496.6r_0\simeq 149782\,{\rm AU}\simeq 1.5\t10^5$\,AU and the time to
come back to the center of the \SS\ is about $10^6$ years. This range is
about the size of an Oort cloud.

Let us come back to the problem of Shapiro effect. We do not consider two
important cases of this effect, i.e.\ the effect measured by Viking and the
effect which includes the planet Mars as a passive mirror. Both problems
contain additional derivations concerning the movement of the \sp\ Viking and
the planet Mars during an experiment (see Ref.\ \cite{30}, there is a mirror
installed on the planet Mars devoted to this measurement).

Probably in order
to get an accuracy of these experiments we should fine tune initial
conditions. This will be done elsewhere. For a distance of $r>40$\,AU we
should redefine $k^2$ to
\beq{11.81}
k^2=G_N(M_\odot +\D M)
\end{equation}
where $\D M$ is a mass of all the planets and asteroids inside a sphere of
radius 40\,AU.

Let us come back to our model \er{11.1}--\er{11.5}. This model can be applied
to different systems than the \SS, using initial conditions different from
\er{11.6}. In these initial conditions we should include our knowledge of
this system similarly as we have used our knowledge of the \SS\ and Anderson
et al.\ data.

A practical model of an \an\ \ac\ can be obtained from our model using
perturbation formula for \hy\ orbits (see \er{7.18}--\er{7.22}).

One can also consider a little different initial conditions than \er{11.6},
i.e.
\beq{11.82}
\bal
A(643)&=-7.47\t10^{-12},\qquad
B(643)=7.47\t10^{-12}
\eal
\end{equation}
and the same for $\pz Br(643)$, $\wt\vF(643)$, $\pz{\wt\vF}r(643)$. At a
distance $r=7.03$ we get
\beq{11.83}
\bal
A(7.03)&=-9.18992549\t10^{-12}\\
B(7.03)&=3.22169883\t10^{-13}\\
\pz Br(7.03)&=-3.99454073\t10^{-14}\\
\wt\vF(7.03)&=-3.38827390\t10^{-12}\\
\pz{\wt\vF}r(7.03)&=5.12592617\t10^{-12}
\eal
\end{equation}

One can get the above initial conditions supposing that at $r=\ov R=643r_0$ a
total \gr al \pt\ $U(\ov R)=2b\ov R$ (Eq.\ \er{11.82}). Afterwards solving
Eqs \er{11.2}--\er{11.5} with \er{11.82} as initial conditions one gets
\er{11.83}. Eqs \er{11.83} can serve as new initial conditions and probably they
are more convenient to tune a model on small distances from the Sun.

According to the suggestions of IAU we should use an isotropic \cd\ system in
the \SS\ rather than Schwarzschild-like (see Refs \cite{22},
\cite{25}--\cite{29}, \cite{31}). We
introduce this system in Appendix~D (see Eqs \er{Db325}--\er{Db334}) with
transformations from one system to another one. The problem of fine tuning of
initial conditions using values of $A(r)$ and $B(r)$ close to $r=0$ (in $r_0$
unit) seems to be hard from the numerical point of view, for the left hand side of Eqs
\er{11.2}--\er{11.4} is singular at $r=0$. This demands very high precision
of numerical calculations.

In this model of an \an\ \ac\ the \an\ \ac\ is not \ct\ even asymptotically
and it decreases in a distance from the Sun (from a barycenter of the \SS).
Due to this an orbit for a \hy\ velocity $v_p=1.2\t10^4\,{\rm\frac ms}$ is
bounded. Moreover, an orbit with $v_p=0.1c$ is unbounded.

Let us notice the following fact. There is a distinction among such notions
as a reference system, a reference frame and a \cd\ system. A reference frame
is a practical realization of a reference system (e.g.\ a catalogue of radio
sources) and a \cd\ system gives \cd s in this realization. From our point of
view the most important is a \cd\ system (e,g.\ Schwarzschild-like or an
isotropic \cd\ system, see Section~7).

\bigskip
\hbox to \textwidth{\vbox{\hsize=0.45\textwidth
\refstepcounter{figure}\label{pion4}
\hbox to\hsize{\hfil \ing[width=0.95\hsize]{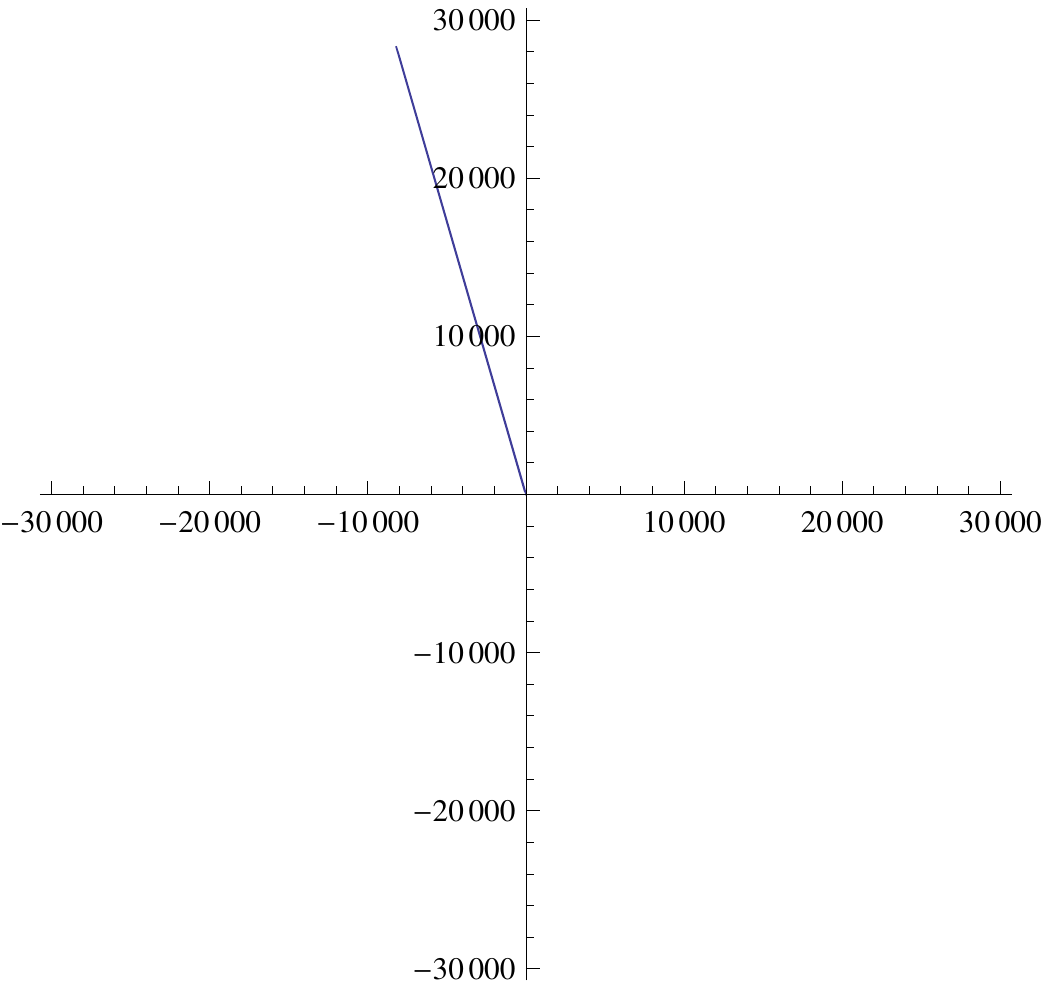}\hfil}
\hbox to\hsize{\hfil(A)\hfil}
\hbox to\hsize{\hfil \ing[width=0.95\hsize]{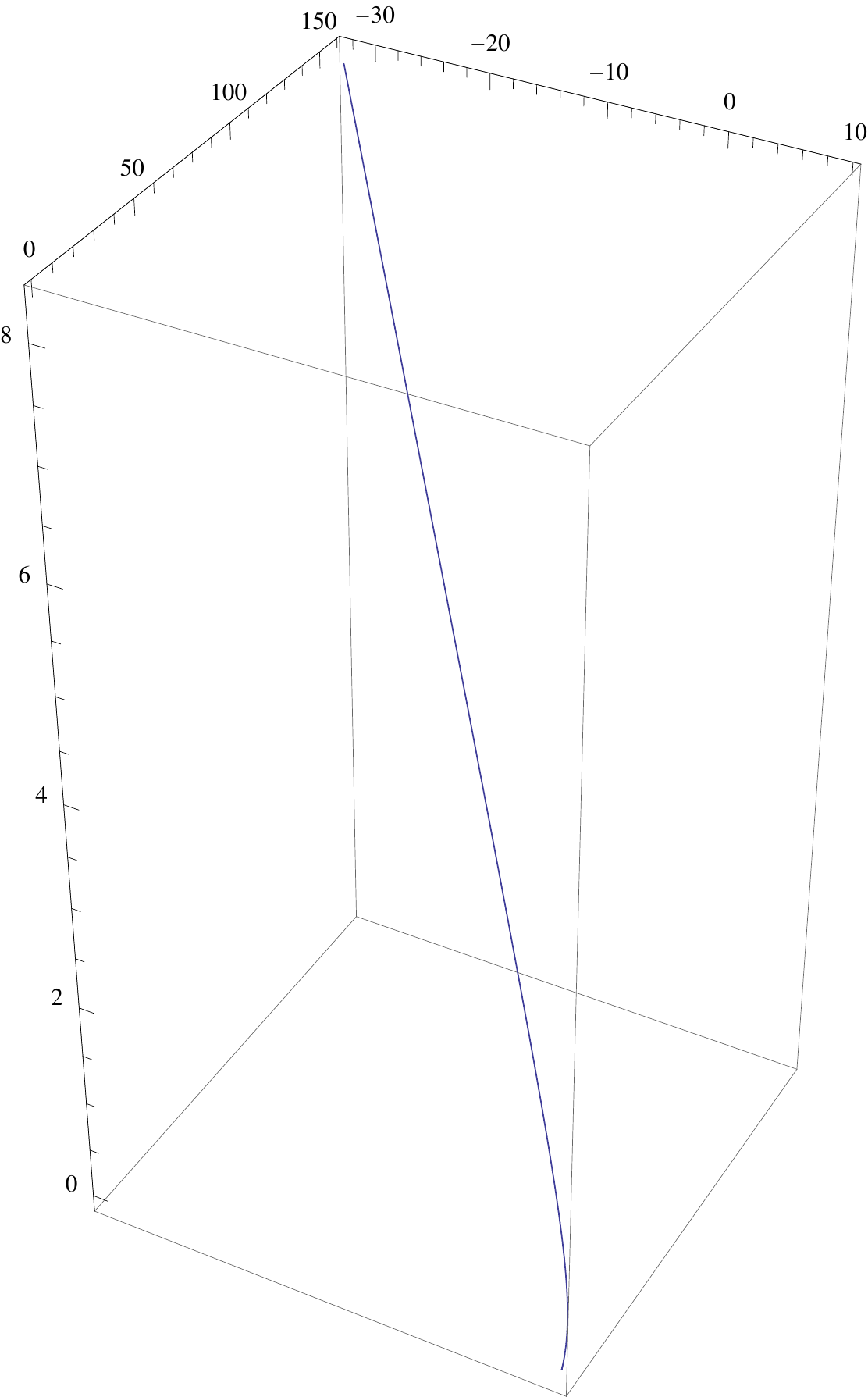}\hfil}
\hbox to\hsize{\hfil(B)\hfil}
\smallskip \noindent
{\small Figure \thefigure:
(A)---a polar plot of an orbit of a massive point body for a full range for
$a_h=20$, $e=3$ (in equatorial plane),
(B)---a 3D polar plot of an orbit of a massive point body for a full range
for $a_h=20$, $e=3$, $I=3^\circ$ ($r<150r_0$).}}\hfil
\vbox{\hsize=0.45\textwidth
\refstepcounter{figure}\label{pion2}
\hbox to\hsize{\hfil \ing[width=0.95\hsize]{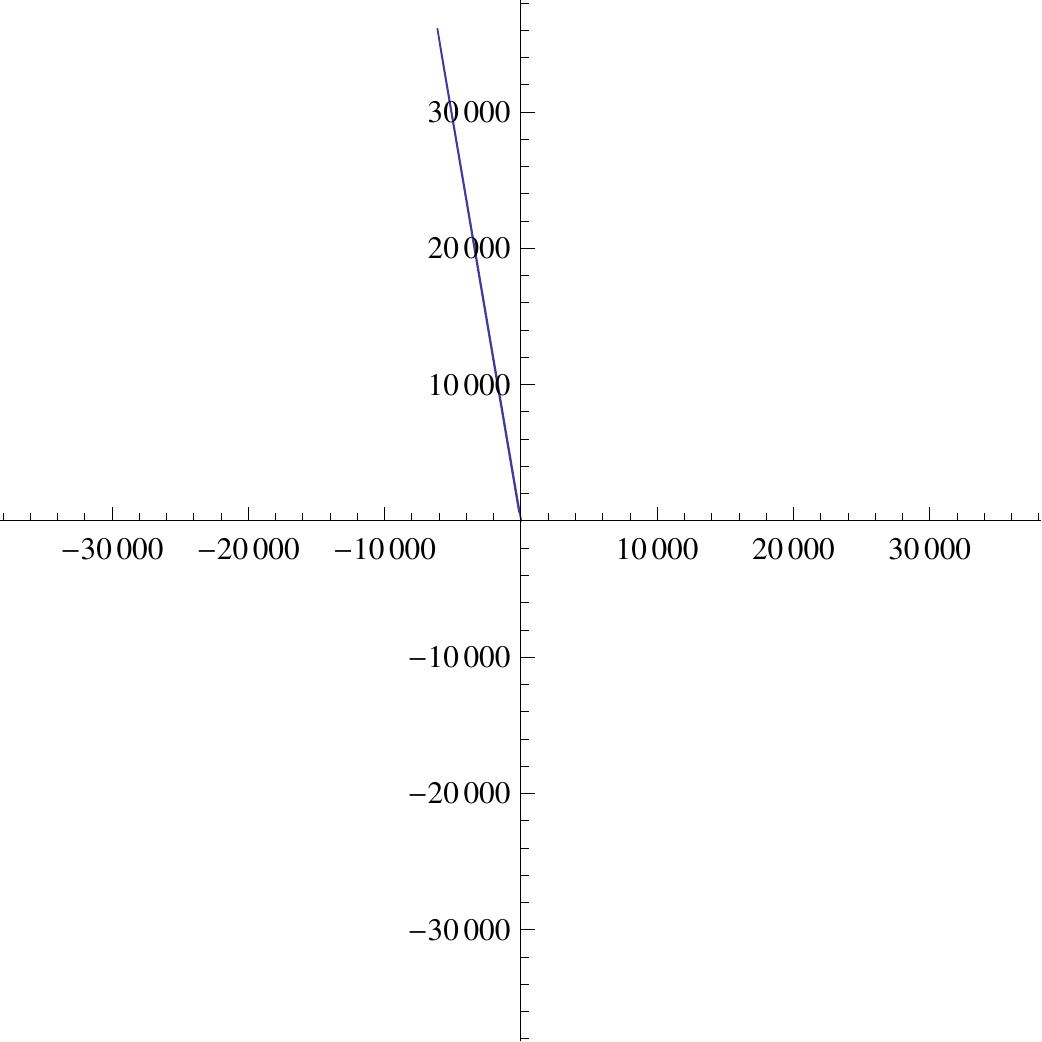}\hfil}
\hbox to\hsize{\hfil(A)\hfil}
\hbox to\hsize{\hfil\ing[width=0.95\hsize]{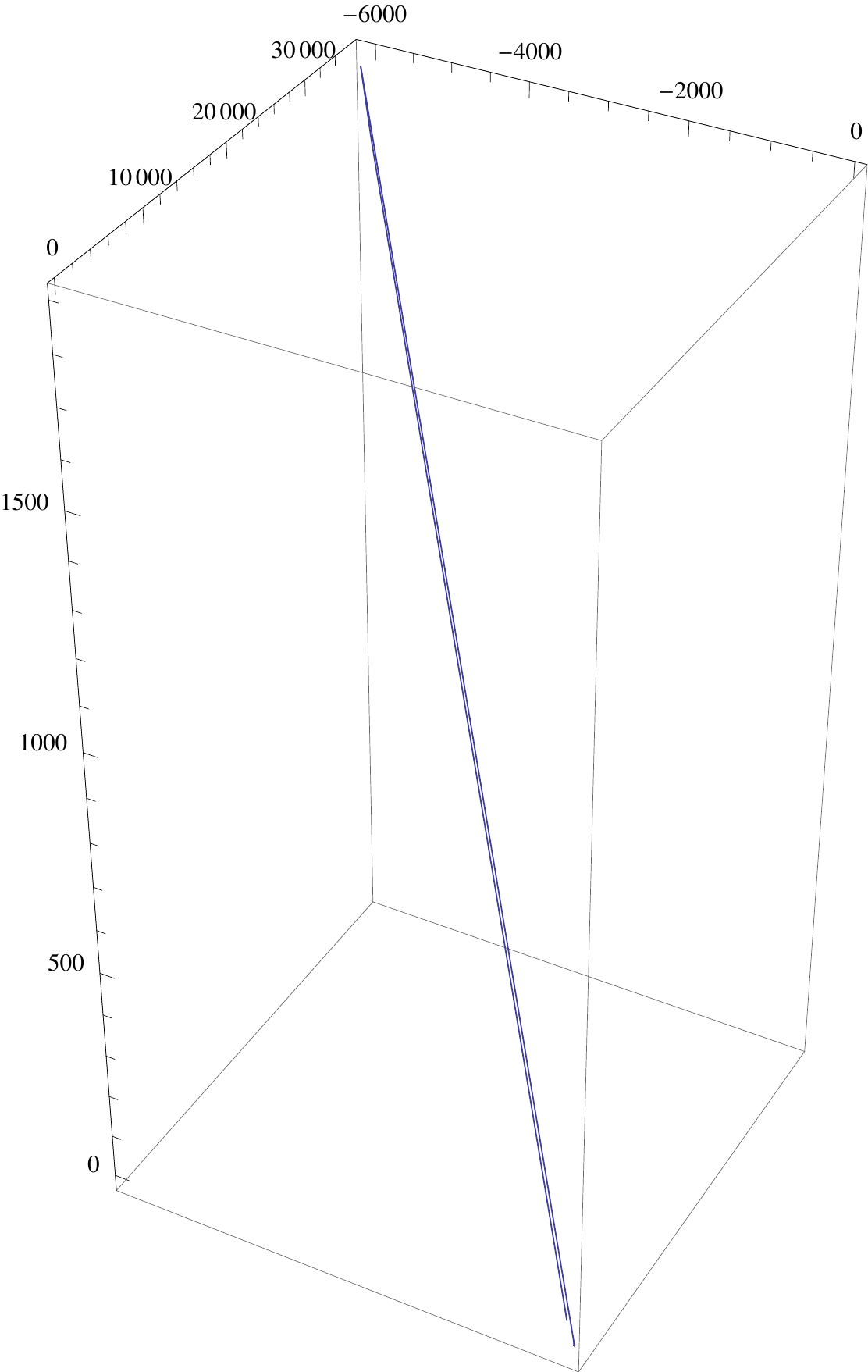}\hfil}
\hbox to\hsize{\hfil(B)\hfil}
\smallskip \noindent
{\small Figure \thefigure:
(A)---a polar plot of an orbit of a massive point body for a full range for
$a_h=20$, $e=5$ (in equatorial plane),
(B)---a 3D polar plot of an orbit of a massive point body for a full range
for $a_h=20$, $e=5$, $I=3^\circ$.}}}

\bigskip

\def\theequation{A.\arabic{equation}}
\def\thefigure{A.\arabic{figure}}
\setcounter{figure}0
\setcounter{equation}0
\section*{Appendix A}
In this appendix we consider an orbit in the second region, i.e.\ $r>20\,{\rm
AU}$ exactly. In order to do this let us examine Eqs (\ref{5.1}--\ref{5.2})
in details.

We consider the orbits exactly without \ap ions at distances comparable to
$10^3$\,AU or even more. We give a detail analysis of the properties of the
orbits dividing them into two classes: periodic and chaotic. Both orbits are
bounded. We consider also an influence of a \co ical \ct\ on the orbits at
distances more than $10^6$\,AU.

From this point of view we consider the orbits in classical Newtonian
mechanics. This approach can be used for a \hy-like orbit of an asymptotic
motion of \P0/11. Moreover, we consider also a relativistic or almost
relativistic motion of \sp s. We give some numerical examples for \P0/11
case. We consider also the full theoretical treatment of the \eu\nos\
Jordan--Thiry Theory, giving \e s for the scalar field $\vF$ and a metric
tensor $g_{\mu\nu}$ in a spherically \s, stationary case. In order to avoid a
confusion with notation we use a capital $\vF$ for a scalar field.

Let us notice that we change a notation and in place of~$r$ we
use~$\rho$ and
\bea{A.1}
\pz \rho\vf &=& \pm \frac{\rho^2}A \sqrt{2U(\rho)+2H -\frac{A^2}{\rho^2}}\\
\pz \rho t &=& \pm \sqrt{2U(\rho)+2H -\frac{A^2}{\rho^2}}\label{A.2}
\end{eqnarray}
where $A=\rho^2 \dot\vf$, $U(\rho)=-\frac {k^2}\rho -b\rho$, or
{\def\theequation{A.\arabic{equation}a}
\setcounter{equation}0
\bea{A.1a}
\pz \rho\vf &=& \pm \frac{\rho^2}A \sqrt{2H -\frac{2k^2}\rho -2b\rho -\frac{A^2}{\rho^2}}\\
\pz \rho t &=& \pm \sqrt{2H -\frac{2k^2}\rho -2b\rho -\frac{A^2}{\rho^2}}\label{A.2a}
\end{eqnarray}}

From Eq.\ (A.1a) one gets
\beq{A.3}
d\vf=\pm\frac{d\rho}{\rho
\sqrt{-\frac{2b}{A^2}\,\rho^3+\frac{2H}{A^2}\,\rho^2-\frac{2k^2}{A^2}\,\rho-1}},
\end{equation}
changing variables to $x=\root3\of{\frac{2b}{A^2}}\,\rho$, $\rho=
\root3\of{\frac{A^2}{2b}}\,x$ one gets
\beq{A.4}
\pm d\vf=\frac{\sqrt{|A|}\,dx}{x\sqrt{-x^3+\ov e x^2-fx-1}}
\end{equation}
where
$$
\ov e=\frac{2H}{\root3\of{4b^2A^2}}\,, \q f=\frac{2k^2}{A\root3\of{2bA}}\,.
$$
From Eq.\ \er{A.2a} one gets
{\def\theequation{A.\arabic{equation}a}
\setcounter{equation}3
\beq{A.4a}
\frac{x\,dx}{\sqrt{-x^3+\ov e x^2-fx-1}}=\pm\sqrt{|A|}
\root3\of{\frac{b^2}{A^2}}\,dt.
\end{equation}}

Let us consider a cubic \e
\beq{A.5}
x^3-\ov ex^2+fx+1=0.
\end{equation}
We define a discriminant $D$ of the \e. One gets
\beq{A.6}
f(H)=108A^4b^2D
\end{equation}
where
\beq{A.7}
f(H)=4H^2\X2(\root3\of4\,A^2(bA)^{2/3}-k^4\Y2)+36k^2A^2bH - b(27A^4b+4k^6).
\end{equation}
This is a quadratic \f\ \wrt $H$. In order to solve a cubic \e\ it is
necessary to know a sign of $D$ (or~$f(H)$). Thus we examine $f(H)$. Let us
calculate a discriminant of $f(H)$. One gets
\beq{A.8}
\D=16\X2(108k^4A^4b^2-4\root3\of4\,k^6A^{8/3}b^{5/3}+27\sqrt2\,
A^{20/3}b^{8/3}+4b^2k^{10}\Y2)=16\D_1.
\end{equation}

We make the following substitution
\beq{A.9}
A=\ve\,\frac{k^{3/2}}{b^{1/4}}\,p^{3/4}, \q \ve^2=1.
\end{equation}
One gets
\beq{A.10}
\D_1=k^{10}b\X2(27\sqrt2\,p^5+108p^3-4\root3\of4\,p^2+4\Y2).
\end{equation}
$\D_1(p)$ has only one real root
$$
p_0\simeq -0.31531.
$$
For this it is nonnegative for $p\ge p_0$. In our case $p\ge0$. Thus $\D$ is
always nonnegative. In this way we have two possibilities:
\refstepcounter{equation}\label{A.11}
$$
\dsl{
1^\circ \hfill 4\root3\of4 A^2(bA)^{2/3}-k^4\le 0 \hfill (\rm\theequation)\cr
\hbox{or}\hfill
A\le \frac{k^{3/2}}{\root4\of{4b}}\simeq 0.707\,
\frac{k^{3/2}}{b^{1/4}} \hfill (\rm\theequation a)\cr
\refstepcounter{equation}\label{A.12}
2^\circ \hfill 4\root3\of4 A^2(bA)^{2/3}-k^4\ge 0 \hfill (\rm\theequation)\cr
\hbox{or}\hfill
A\ge \frac{k^{3/2}}{\root4\of{4b}}\,.
\hfill (\rm\theequation a)}
$$

$f(H)$ has two real roots
\beq{A.13}
H_{1,2}=\frac{k\sqrt b\X2(9p^{3/2}\mp
\sqrt{27\sqrt2\,p^5+108p^3-4\root3\of4\,p^2+4}\Y2)}
{2\X1(\root3\of 4\,p^2-1\Y1)}\,.
\end{equation}
In this way we have
\refstepcounter{equation}\label{A.14}
$$
\dsl{
1^\circ \hskip40pt D\ge0, \hfill\cr
\hfill A\ge \frac{k^{3/2}}{\root4\of{4b}} \qh{and}
H\ge H_2 \hbox{ or }H\le H_1 \hfill (\rm\theequation)\cr
\refstepcounter{equation}\label{A.15}
\hbox{or}\hfill A\le \frac{k^{3/2}}{\root4\of{4b}} \qh{and}
H_1\le H\le H_2 \hfill (\rm\theequation)\cr
\refstepcounter{equation}\label{A.16}
2^\circ \hskip40pt D<0, \hfill\cr
\hfill A\ge \frac{k^{3/2}}{\root4\of{4b}} \qh{and}
H_1< H< H_2 \hfill (\rm\theequation)\cr
\refstepcounter{equation}\label{A.17}
\hbox{or}\hfill A\le \frac{k^{3/2}}{\root4\of{4b}} \qh{and}
H> H_2 \hbox{ or }H< H_1. \hfill (\rm\theequation)}
$$

In case $1^\circ$ we have only one real root. In case $2^\circ$ we have three
real roots. The case $2^\circ$ is more interesting from the physical point of
view (it corresponds to the case of \sp s).

In case $2^\circ$, $D<0$, we have three real roots $x_1,x_2,x_3$ given by Cardano formulae.
Let us define
\beq{A.18}
\cos\a=\frac{4H^3-18Hk^2b+27b^2A}{4(H^2A-3k^2bA)^{3/2}}
\end{equation}
and
\beq{A.19}
r=\frac23\,\frac{\sqrt{2H^2-3k^2}}{\root3\of{4b^2A}}>0.
\end{equation}
One gets
\beq{A.20}
\bal
x_1&=\frac{\,\ov e\,}3-2r\cos\frac\a3=\frac{\,\ov e\,}3+\ov e_1\cr
x_2&=\frac{\,\ov e\,}3+2r\cos\X2(\frac\pi3+\frac\a3\Y2)=\frac{\,\ov e\,}3+\ov e_2\cr
x_3&=\frac{\,\ov e\,}3+2r\cos\X2(\frac\pi3-\frac\a3\Y2)=\frac{\,\ov e\,}3+\ov e_3\,.
\eal
\end{equation}
In this case we have
\beq{A.21}
x_1<x_2<x_3.
\end{equation}
One gets
\bea{A.22}
x_1-x_3&=&-2\sqrt3\,r\cos\X2(\frac\a3-\frac\pi6\Y2)\\
x_2-x_3&=&-2\sqrt3\,r\sin\frac\a3 \label{A.23}\\
x_1-x_2&=&-2\sqrt3\,r\cos\X2(\frac\a3+\frac\pi6\Y2). \label{A.24}
\end{eqnarray}
In this case we integrate Eq.\ \er{A.4} getting
\bml{A.25}
\sq\X3(\arcsin\sqrt{\frac{x-x_1}{2\sqrt3\,r\cos(\frac\a3+\frac\pi6)}}\,,
-\frac{2\sqrt3\,r\cos(\frac\a3+\frac\pi6)}{x_1}\,,
\sqrt{\frac{\cos(\frac\a3+\frac\pi6)}{\cos(\frac\a3-\frac\pi6)}}\Y3)\\
{}=\frac{\root3\of4\,x_1\sqrt{r\cos(\frac\a3-\frac\pi6)}}{\sqrt2}\,(\vf-\vf_0)
\end{multline}
where $x=\root3\of{\frac{2b}{A^2}}\,\rho$.

It is easy to see that $x>x_1$ and
$$
0<\frac{x-x_1}{2\sqrt3\,r\cos(\frac\a3+\frac\pi6)}<1.
$$
Thus we get
\beq{A.26}
x_1<x<x_1+2\sqrt3\,r\cos\X2(\frac\a3+\frac\pi6\Y2)
\end{equation}
or in terms of $\rho$
\bg{A.27}
\frac{\root3\of4}{3A\root3\of{Ab}}\X2(2\sqrt{2H^2-3k^2b}-H\Y2)<\rho
<\frac{\root3\of4}{3A\root3\of{Ab}}\X2(H+2\sqrt{2H^2-3k^2b}\sin\X2(\frac\pi6
-\frac\a3\Y2)\Y2)\\
\D\rho=\rho_{\max}-\rho_{\min}=
\frac{2\root3\of4}{3A\root3\of{Ab}}\sqrt{2H^2-3k^2b}\cos\X2(\frac\pi6+
\frac\a3\Y2). \label{A.28}
\end{gather}
It is easy to see that for $x=x_3$ we have $\vf=\vf_0$. It means that for
$$
\rho=\frac{\root3\of4}{3A\root3\of{Ab}}\X2(2\sqrt{2H^2-3k^2b}-H\Y2),
\q \vf=\vf_0,
$$
$\sq$ is an elliptic integral of third kind,
\beq{A.29}
\sq(\psi,h,k)=\int_0^\psi
\frac{d\vf}{(1+h\sin^2\vf)(1-k^2\sin^2\vf)}\,.
\end{equation}

At the same time the second integral Eq.\ \er{A.2} can be integrated and we
get
\bml{A.30}
\frac{\sqrt2\,x_3}{\sqrt3 \sqrt{r\cos(\frac\a3-\frac\pi6)}}
F\X3(\arcsin\sqrt{\frac{x-x_1}{2\sqrt3\,r\cos(\frac\pi6+\frac\a3)}}\,,
\sqrt{\frac{\cos(\frac\pi6+\frac\a3)}{\cos(\frac\a3-\frac\pi6)}}\Y3)\\
{}-2\sqrt2\root4\of3\sqrt{r\cos\X2(\frac\a3-\frac\pi6\Y2)}
E\X3(\arcsin\sqrt{\frac{x-x_1}{2\sqrt3\,r\cos(\frac\pi6+\frac\a3)}}\,,
\sqrt{\frac{\cos(\frac\pi6+\frac\a3)}{\cos(\frac\a3-\frac\pi6)}}\Y3)\\
{}=\sqrt A \root3\of{\frac{4b^2}{A^2}}(t-t_0).
\end{multline}

It is easy to see that $t=t_0$ for $x=x_1$. Moreover, we have for $\rho
=\rho_{\max}$
\bg{A.31}
\sq\X3(\frac\pi2\,,-\frac{2\sqrt2\,r\cos(\frac\pi6+\frac\a3)}{x_1}\,,
\sqrt{\frac{\cos(\frac\pi6+\frac\a3)}{\cos(\frac\a3-\frac\pi6)}}\Y3)
=\frac{\root4\of3\,x_1\sqrt{r\cos(\frac\a3-\frac\pi6)}}{\sqrt2}
\X1(\vf_{\max}-\vf_0\Y1)\\
\hskip-230pt \frac{\sqrt2\,x_3}{\sqrt3\sqrt{r\cos(\frac\a3-\frac\pi6)}}
F\X3(\frac\pi2\,,
\sqrt{\frac{\cos(\frac\pi6+\frac\a3)}{\cos(\frac\a3-\frac\pi6)}}\Y3)\nn\\
\hskip30pt{}-2\sqrt2\root4\of3\sqrt{r\cos\X2(\frac\a3-\frac\pi6\Y2)}E\X3(\frac\pi2\,,
\sqrt{\frac{\cos(\frac\pi6+\frac\a3)}{\cos(\frac\a3-\frac\pi6)}}\Y3)
=\sqrt A\root3\of{\frac{4b^2}{A^2}}(t_{\max}-t_0)\label{A.32}
\end{gather}
where
\beq{A.33}
F(\psi,k)=\int_0^\psi \frac{d\vf}{\sqrt{1-k^2\sin^2\vf}}
\end{equation}
is an elliptic integral of the first kind,
\beq{A.34}
E(\psi,k)=\int_0^\psi \sqrt{1-k^2\sin^2\vf}\,d\vf
\end{equation}
is an elliptic integral of the second kind.

In the case of $D\ge0$ we have only one real root of Eq.\ \er{A.5} and we have
$x<x_1$ for the polynomial under a square root must be positive. One gets
\beq{A.40}
x_1=\frac{\,\ov e\,}3-2r\cosh \frac\a3\,,
\end{equation}
where
\beq{A.41}
\cosh\a=\frac{4H^3-18Hk^2b+27b^2A^2}{4(H^2A-3k^2bA)^{3/2}}\,.
\end{equation}

It is interesting to notice that we can express some interesting parameters
obtained from exact analysis of an orbit with $a_h$ and $e$ measured for
early time of movement. In this way one gets
\bg{A.42}
\D\rho=\frac{\sqrt2\root3\of4\sqrt{-6ba_h^2+k^2}}
{3a_h^{5/3}(e^2-1)^{2/3}\root3\of{kb}}
\cos\X2(\frac\pi6+\frac\a3\Y2)\\
\cos\a=\frac{k^6-18k^4ba_h^2+54b^2a_h^4(e^2-1)}
{k^{7/2}a_h^{3/4}(e^2-1)^{3/4}(k^2-12a_h^2b)^{3/2}}
\label{A.43}\\
r=\frac{\sqrt{2k^2-12a_h^2b}\root3\of k}
{3a_h^{4/3}\root3\of{4b^2(e^2-1)}}\label{A.44}\\
\ov e=\frac{k^{4/3}}{2a_h^{4/3}\root3\of{4b^2(e^2-1)}}\,.\label{A.45}
\end{gather}

In these formulae we put
\beq{A.*}
H=\frac{k^2}{2a_h} \qh{and} A=k\sqrt{a_h(e^2-1)}\,.
\end{equation}
Thus we obtain only bounded orbits.In order to get \e s of motion $\rho(t)$, $\vf(t)$ it is
necessary to invert \e s $\vf(\rho)$, $t(\rho)$, getting $\rho(\vf)$, $\rho(t)$
and using $A=\rho^2 \dot\vf$ to get $\vf(t)$.

Let us consider a deep space region. It means that we are far away from the
Sun, i.e., $\rho>10^5$\,AU. For such a region a newtonian \ac\ of the Sun is
smaller than an \an\ \ac. Thus we can put $k^2=0$ in the formulae. We get
$f=0$ and
\beq{A.47}
f(H)=4H^2\root3\of4\,A^2(bA)^{2/3}-27A^4b^2
\end{equation}
and
\beq{A.48}
H_{1,2}=\mp\frac{3\sqrt3}{2\root3\of2}\,A^{2/3}b^{1/6} \simeq \mp2.061
A^{2/3}b^{1/6}.
\end{equation}
In this case $D\ge0$ if $H<H_1$ or $H>H_2$, and $D<0$ if $H_1<H<H_2$.

For $D<0$ one gets
\bg{A.49}
r=\frac{2\sqrt2}{3\root3\of4}\,\frac H{\sqrt{b^2A^2}}\\
\cos\a=\frac{4H^3-27b^2A}{4H^3A^{3/2}} \label{A.50}\\
\frac{\root3\of4\,H}{3A\root3\of{Ab}}\X1(2\sqrt2-1\Y1)<\rho<
\frac{\root3\of4\,H}{3A\root3\of{Ab}}\X2(1+6\sqrt3\sin\X2(\frac\pi6
-\frac\a3\Y2)\Y2)\label{A.51}\\
\D\rho=\frac{2\sqrt2\root3\of4}{3A\root3\of{Ab}}\,\cos\X2(\frac\pi6
+\frac\a3\Y2).\label{A.52}
\end{gather}
For $D\ge0$ one finds
\beq{A.53}
\rho<\frac{2H}{3\root3\of{2Ab}}\X2(1+2\sqrt2\cos\X2(\frac\pi3-\frac\a3\Y2)\Y2).
\end{equation}

Let us consider two definite integrals ($D<0$)
\bea{A.49a}
I_1&=&\int _{x_1}^{x} \frac{dy}{y\sqrt{-y^3+\ov ey-fy-1}}\\
I_2&=&\int _{x}^{x_3} \frac{y\,dy}{\sqrt{-y^3+\ov ey-fy-1}}\label{A.50a}
\end{eqnarray}
and let us define a $\cP$-Weierstrass \f\ $\cP(z,\o_1,\o_2)$ \st
\beq{A.51a}
\cP'{}^2=4\cP^3-g_2\cP-g_3=4(\cP-\ov e_1)(\cP-\ov e_2)(\cP-\ov e_3)
\end{equation}
(see Eq.\ \er{A.20}).

One gets
\beq{A.52a}
\bal
g_3&=-4\ov e_1\ov e_2\ov e_3\\
g_2&=4\root3 \of4 \,\X1(\ov e_1\ov e_2+\ov e_2\ov e_3+\ov e_3\ov e_1\Y1)
\eal
\end{equation}
and
\bea{A.53a}
\o_1&=&2\int_0^\iy (4t^3-g_2t-g_3)^{-1/2}\,dt\\
\o_2&=&-2i\int_{-\iy}^{\bar e_3}(g_3+g_2t-4t^3)^{-1/2}.\label{A.54a}
\end{eqnarray}
Then $\cP$ has two periods $\o_1$ and $\o_2$. In the case of $D<0$ one is
real and the second pure imaginary.

One can easily calculate
\bea{A.55a}
g_2&=&-\frac{4\root3\of4\,(2H^2-3k^2b)}{3bA\root3\of{2bA}}\\
g_3&=&-\frac{27b^2A^2+54Hk^2-4H^3}{27b^2A^2} \label{A.56a}
\end{eqnarray}
and
\beq{A.57a}
\o_1=W K(k)
\end{equation}
where
\beq{A.58a}
K(k)=\int_0^{\pi/2}\frac{d\vf}{\sqrt{1-k^2\sin^2\vf}}
\end{equation}
is an elliptic integral of the first kind
\bg{A.59a}
W=\frac{2\sqrt2\,(2\ov e{}_1^2+\ov e_2\ov e_3)^{1/4}}
{\X1(3\ov e_1(2\ov e{}_1^2+\ov e_2\ov e_3)+(2(9\ov e{}_1^4+18g_3\ov e{}_1^2
-4))^{1/2}\Y1)^{1/2}}\\
k^2=1-\frac2
{(2\ov e{}_1^2+\ov e{}_2\ov e{}_3)^{3/2}
\X1(3\ov e_1(2\ov e{}_1^2+\ov e_2\ov e_3)^{1/2}
+(2(9\ov e{}_1^4+18g_3\ov e{}_1^2-4))^{1/2}\Y1)}\,.\label{A.60a}
\end{gather}

The integrals $I_1$ and $I_2$ can be transformed in the following way. First
we translate $x\to \ov x=x-\frac{\ov e}3$ and next transform $z=-\root3\of4
\,\cP(\ov x,\o_1,\o_2)$. The transformations transform real axis into real
axis. In this way one gets
\bea{A.61a}
I_1&=&\frac12 \int_{\g_1}\frac{dz}{\cP-\frac{\bar e}3}\\
I_2&=&\frac12 \int_{\g_2}\X2(\cP-\frac{\,\ov e\,}3\Y2)\,dz \label{A.62a}
\end{eqnarray}
where $\g_i$, $i=1,2$, are paths on real axis. They are images of $\langle
x_1,x \rangle$ and $\langle x,x_3\rangle$.

This integration cannot be easily done. Moreover, we have done it above in a
different way. $\cP$~is a periodic \f\ on a real axis with a period $\o_1$
(see Eq.~\er{A.60a}). Thus $x$ (or~$\rho$) should be a periodic \f\ of~$t$
and~$\vf$ and the periods can be calculated:
\bg{A.63a}
T=\o_1\,\frac2{\sqrt{|A|}}\sqrt{\frac{A^2}{b^2}}, \q
\varPhi=\frac{2\o_1}{\sqrt{|A|}},\\
\rho(t+T)=\rho(t), \q  \rho(\vf+\varPhi)=\rho(\vf). \label{A.64a}
\end{gather}
In this way orbits are closed. This is the only case of closed orbits. In the
case of $D>0$ orbits are only bounded (an appropriate $\cP$-Weierstrass \f\
has complex periods). They behave in a chaotic way. Eqs\ \er{A.63a},
\er{A.64a} work also for a deep space region.

Let us consider the case $D\ge0$ with more details. One gets from Eq.\
\er{A.4a}
\beq{A.65a}
\bal
\lefteqn{\sqrt{|A|}\root3\of{\frac b{A^2}}\,(t-t_0)}\\
&=-\frac{4\sqrt m (l-m)}
{(m(5-l)+l)\sqrt{3\X1(\sqrt{5u^2+5v^2+7uv}-(u+v)\Y1)}\,
\X1(\frac{\ov e}{3}+u+v+\sqrt{5u^2+5v^2+7uv}\Y1)}\\
&\t\X3(F(\psi,k)+\wt A\sq(\psi,h_1,k)+\wt B\sq(\psi,h_2,k)\\
&\q{}-\frac1{4D_1}\log\frac
{D_1(z-C_1)}{z+2D_1\sqrt{(m+z)(1+mz)}+m(2+mz)+C_1(1+m^2+2mz)}\\
&\q{}-\frac1{4D_2}\log\frac
{D_2(z-C_2)}{z+2D_2\sqrt{(m+z)(1+mz)}+m(2+mz)+C_2(1+m^2+2mz)}\Y3)
\eal
\end{equation}
where
\bea{A.66a}
u&=&\root3\of{\sqrt D-\ov q}, \q  v=\root3\of{\sqrt D+\ov q}, \q
\ov q=\frac H{27A^2b^3\sqrt b}\X2(\root3\of4\, HA^{2/3}+27k^2\root3\of b\Y2)-\frac12
\\
m&=&\frac{2\root4\of{5u^2+5v^2+7uv}-\sqrt{3(u+v)+2\sqrt{5u^2+5v^2+7uv}}}
{2\root4\of{5u^2+5v^2+7uv}+\sqrt{3(u+v)+2\sqrt{5u^2+5v^2+7uv}}} \label{A.67a}\\
l&=&\frac{\ov e/3+u+v-\sqrt{5u^2+5v^2+7uv}}{\ov e/3+u+v+\sqrt{5u^2+5v^2+7uv}}
\label{A.68a}\\
k^2&=&1-m^2 \label{A.69a}\\
h_{1,2}&=&\frac{m(l-5)-l}{l-m(l-2)\pm
\sqrt{m(m(l-2)^2-l(l+1))}}=-\frac1{u_{1,2}} \label{A.70a}\\
\wt A&=&\frac1{u_1}\X3(\frac{2(l-m(l-2))}{m(5-l)+l}+\frac{2l-m}{l-m}\Y3)
\frac{m(5-l)+l}{2u_1(l-m(l-2))}\nn\\
&&\q{}-\frac{l(m(5-l)+l)}{2(l-m)u_1(l-m(l-2))}-\frac{u_2}{u_1}\X2(u_2+\frac
{2l-m}{l-m}\Y2) \label{A.71a}\\
\wt B&=&\frac1{2u_2}\X2(\frac l{l-m}+u_2^2+\frac{2l-m}{l-m}\Y2)
\frac{m(5-l)+l}{l-m(l-2)} \label{A.72a}\\
C_{1,2}&=&2-l\pm\sqrt{1-l}, \q D_i=\sqrt{(C_i+m)(1+C_im)}, \ i=1,2. \label{A.73a}
\end{eqnarray}
$F$ and $\sq$ are elliptic \f s of the first and the third kind (see Eqs
\er{A.33}, \er{A.29}),
\bg{A.74a}
\psi=\arcsin\frac{Y-Z}{\sqrt{m^2(Y+Z)^2+(Y-Z)^2}}=\arcsin z_1,\\
z=\X3(\frac{Y-Z}{Y+Z}\Y3)^2,\label{A.75a}\\
Y=\root4\of{5u^2+5v^2+7uv}, \q
Z=\sqrt{\frac{\bar e}3+u+v-\root3\of{\frac{2b}{A^2}}\rho}\,,\nn
\end{gather}
$k$ and $h_i$, $i=1,2$ are usual parameters of elliptic \f s. $D$ is given by
Eq.\ \er{A.7}.

From Eq.\ \er{A.4a} one finds
\bml{A.76a}
\frac1{\sqrt{|A|}}(\vf-\vf_0)=
-\frac{4\X1(\frac{\bar e}3+u+v+\sqrt{5u^2+5v^2+7uv}\Y1)}
{\sqrt{3\X1(\sqrt{3(5u^2+5v^2+7uv)}-(u+v)\Y1)}}\\
{}\t\X4[\X3(\frac{2l(m+z)\sqrt{1+mz}}{(1-z)^3}
\X2(z^2(8m+3l-2ml+3m^2l)+z(2-12m-6m^2-8l)\\
{}+(-6-4m-6m^2+9l+10ml+9m^2)\Y2)\\
{}-3(1+m)(l(m-1)^2+4m)\sqrt{m+z} \log
\frac{(l(m-1)^2+4m)(1-z)}{2(1+m)^2\X1(\sqrt{m+z}+\sqrt{1+mz}\Y1)^2}\Y3)\\
{}+\frac{\sqrt m}{(m+1)^4}\X2[(l-m^3)F(\psi,k)
-\wt A_1\sq(\psi,-\tfrac1a,k)+\wt A_2H_2+\wt A_3H_3+\wt A_4H_4\Y2]\Y4],
\end{multline}
where
\bg{A.77a}
a=\frac1{m+1}=\frac{2\root4\of{5u^2+5v^2+7uv}
-\sqrt{3(u+v)+2\sqrt{5u^2+5v^2+7uv}}}
{4\root4\of{5u^2+5v^2+7uv}}\\
\bal
\wt A_1&=a_3\\
\wt A_2&=a_2+3a_3a\\
\wt A_3&=a_1+a(2a_2-3a_3+6a_3a)\\
\wt A_4&=a_0+a_2a(1-a)+a^2a_3(3-2a).
\eal \label{A.78a}
\end{gather}
The \cf s $a_i$, $i=0,1,2,3$, are given by the formulae
\bea{A.79a}
a_3&=&\frac{m^4-3m^3+(6l-7)m^2-(2l-7)m+4(l+1)}{m+1}\\
a_2&=&\frac{m^4(l-2)+m^3(8l-17)+m^2(13l-16)+m(6l-7)+6l}{(m+1)^2} \label{A.80a}\\
a_1&=&\frac{m^5(l-2)+m^4(12l-11)+28m^3(l-1)+m^2(31l-23)+m(18l-7)+8l}
{(m+1)^3} \label{A.81a}\\
a_0&=&\frac{m\X1(m^5(l{-}2)+5(2l{-}3)m^4+(31l{-}40)m^3+(44l{-}49)m^2
+(7l{-}4)m+(10l{-}7)\Y1)}{(m+1)^4}\q \label{A.82a}\\
H_n&=&\int \frac{dz_1}{(z_1^2-a)^n\sqrt{(1-z_1^2)(1-k^2z_1^2)}}\,, \q
n=2,3,4. \label{A.83a}
\end{eqnarray}
Let us notice that
\beq{A.84a}
H_1=\sq(\psi,-\tfrac1a,k), \q \psi=\arcsin z_1.
\end{equation}

For $H_2$ one gets
\bml{A.85a}
H_2=\frac{z_1\sqrt{(1-z_1^2)(1-k^2z_1^2)}}{2(z_1^2-a)a((k^2+1)a-k^2a^2-1)}
+\frac{1-2a(k^2+1)+3k^2a}{2a((k^2+1)a-k^2a^2-1)}\,H_1\\
{}-\frac{k^2}{2a((k^2+1)a-k^2a^2-1)}\,I_1+\frac{k^2}{2((k^2+1)a-k^2a^2-1)}
\,I_0
\end{multline}
where
\bea{A.86a}
I_0&=&\int\frac{dz_1}{\sqrt{(1-z_1^2)(1-k^2z_1^2)}}=F(\psi,k)\\
I_1&=&\int\frac{z_1^2\,dz_1}{\sqrt{(1-z_1^2)(1-k^2z_1^2)}}=
\frac1{k^2}\X1(F(\psi,k)-E(\psi,k)\Y1). \label{A.87a}
\end{eqnarray}
$z_1=\sin\psi$ is given by the formula \er{A.74a}.
\bml{A.88a}
H_3=\frac{3+7k^4a^4+3k^2a^2+(k^2+1)^2[6a^2-2a^3]
+(k^2+1)[6k^2a^4-4a-16a^3k^2]}{4a^2((k^2+1)a-k^2a^2-1)^2}\\
{}-\frac{3k^2(1-2a(k^2+1)+3k^2a^2)}{8a^2((k^2+1)a-k^2a^2-1)}\,I_1
+\frac{k^2(1-4a(k^2+1)+7k^2a^2)}{8a((k^2+1)a-k^2a^2-1)^2}\,I_0\\
{}+\frac{z_1\sqrt{(1-z_1^2)(1-k^2z_1^2)}}{4a(z_1^2-a)((k^2+1)a-k^2a^2-1)}
\X3[\frac1{z_1-a}+\frac{3(1-2a(k^2+1)+3k^2a^2)}{2a((k^2+1)a-k^2a^2-1)}\Y3]
\end{multline}
\bml{A.89a}
H_4=\frac{z_1\sqrt{(1-z_1^2)(1-k^2z_1^2)}}{6(z_1^2-a)^3a((k^2+1)a-k^2a^2-1)}
+\frac{5(1-2a(k^2+1)+3k^2a^2)}{6a((k^2+1)a-k^2a^2-1)}\,H_3\\
{}-\frac{2((k^2+1)a-3k^2a^2)}{3a((k^2+1)a-k^2a^2-1)}\,H_2
-\frac{k^2}{2a((k^2+1)a-k^2a^2-1)}\,H_1
\end{multline}

Using Eq.\ \er{A.89a} and Eqs \er{A.84a}, \er{A.88a} we find
\beq{A.89}
\bal
\kern-11pt
H_4&=H_1\,\frac{1-2a(k^2+1)+3k^2a^2}{24a^3((k^2+1)a-k^2a^2-1)}\,
\X1(6a(k^2+1)-24k^4a^3+22a^2(k^2+1)^2\\
&\qquad{}+35k^4a^4-39k^2a^2-48a^3(k^2+1)k^2
-10a^3(k^2+1)^2+30k^2a^4(k^2+1)\Y1)\\
&+I_1\,\frac{26(k^2+1)a-3k^2a^2-9}{24a^2((k^2+1)a-k^2a^2-1)^2}
+I_0\,\frac{3+21k^2a^2-12(k^2{+}1)a+80(k^2{+}1)+24ak^2}{12a((k^2+1)a-k^2a^2-1)^2}\\
&+\frac{z\sqrt{(1-z_1^2)(1-k^2z_1^2)}}{2a(z_1^2-a)
((k^2+1)a-k^2a^2-1)}
\X3(\frac1{3(z_1^2-a)}+\frac1{2(z_1^2-a)}\\
&\qquad{}+
\frac{45a^3k^2(k^2+1)-18a^2(k^2+1)^2-10+27a(k^2+1)-27k^4a^4
-33k^2a^2-k^2}{12a((k^2+1)a-k^2a^2-1)}\Y3)
\eal
\end{equation}

In this way we can rewrite Eq.\ \er{A.76a} as
\bml{A.90}
\frac1{\sqrt{|A|}}(\vf-\vf_0)=
-\frac{4\X1(\frac{\bar e}3+u+v+\sqrt{5u^2+5v^2+7uv}\Y1)}
{\sqrt{3\X1(\sqrt{3(5u^2+5v^2+7uv)}-(u+v)\Y1)}}
\X4[\X3(\frac{2l(m+z)\sqrt{1+mz}}{(1-z)^3}\\
{}\t\X2(z^2(8m+3l-2ml+3m^2l)+z(2-12m-6m^2-8l)
+(-6-4m-6m^2+9l+10ml+9m^2)\Y2)\\
{}-3(1+m)(l(m-1)^2+4m)\sqrt{m+z} \log
\frac{(l(m-1)^2+4m)(1-z)}{2(1+m)^2\X1(\sqrt{m+z}+\sqrt{1+mz}\Y1)^2}\Y3)\\
{}+\frac{\sqrt m}{(m+1)^4}\X4[F(\psi,k)
\X3[(l-m^3)+\frac{k^2\wt A_2}{2(k^2+1)a-k^2a^2-1}+
\frac{k^2(1-4a(k^2+1)+7a^2k^2)}{8a((k^2+1)a-k^2a^2-1)^2}\,\wt A_3\\
{}+\wt A_4\,\frac{3+21k^2a^2-12(k^2+1)a+80(k^2+1)+24ak^2}
{12a((k^2+1)a-k^2a^2-1)^2} - \frac{\wt A_2}{2a((k^2+1)a-k^2a^2-1)}\\
{}-\frac{3(1-2a(k^2+1)+3k^2a^2)\wt A_3}{8a^2((k^2+1)a-k^2a^2-1)^2}
+\frac{26(k^2+1)a-3k^2a^2-9}{24a^2k^2((k^2+1)a-k^2a^2-1)}\,\wt A_4\Y3]\\
{}+\sq\X1(\psi,-\tfrac1a,k\Y1)\X3[-\frac{\wt A_1}a
+\frac{1-2a(k^2+1)+3k^2a^2}{2a((k^2+1)-k^2a^2-1)}\,\wt A_2
+\wt A_3\,\frac{3+6a^2(k^2+1)^2}{4a^2((k^2+1)-k^2a^2-1)^2}\\
+\wt A_3\,\frac{7k^4a^4-4a(k^2+1)+3k^2a^2-16a^3(k^2+1)k^2
-2a^3(k^2+1)^2+6k^2a^2(k^2+1)}{4a^2((k^2+1)-k^2a^2-1)^2}\\
{}+\wt A_4\X2(6a(k^2+1)-24k^4a^3+15+22a^2(k^2+1)^2+35k^4a^4-39k^2a^2
\\
{}-48a^3(k^2+1)k^2-10a^3(k^2+1)^2+30k^2a^4(k^2+1)\Y2)
\frac{1-2a(k^2+1)+3k^2a^2}{24a^3((k^2+1)-k^2a^2-1)}\Y3]\\
{}+\frac{z_1\sqrt{(1-z_1)(1-k^2z_1^2)}}{2a(z_1^2-a)((k^2+1)-k^2a^2-1)}
\X3[\wt A_2+\frac{\wt A_3+\wt A_4}{2(z_1^2-a)}+\frac{\wt A_4}{3(z_1^2-a)}
+\frac{3\wt A_3(1-2a(k^2+1)+3k^2a^2)}{4a((k^2+1)-k^2a^2-1)}\\
{}+\X1(45a^3k^2(k^2+1)-18a^2(k^2+1)^2-10+27a(k^2+1)-27k^4a^4-33k^2a^2-k^2\Y1)\Y3]\\
\noalign{\eject}
{}+E(\psi,k)\X3[\frac{3(1-2a(k^2+1)+3k^2a^2)\wt A_3}{8a^2((k^2+1)-k^2a^2-1)^2}
-\frac{(26(k^2+1)a-3k^2a^2-9)\wt A_4}{24k^2a^2((k^2+1)-k^2a^2-1)^2}\\
{}+\frac{\wt A_2}{2a((k^2+1)-k^2a^2-1)}\Y3]\Y4]\Y4]
\end{multline}
$E(\psi,k)$ is an elliptic integral of the second kind (see Eq.\ \er{A.34}).

One can find that if $\rho\to\root3\of{\frac{A^2}{2b}}\X1(\frac{\bar e}3
+v+u\Y1)$ the right hand side of Eq.\ \er{A.65a} and Eq.\ \er{A.76a} is going
to infinity. It simply means that a point mass approaches a border of a
movement in an infinite time and $\vf$ is also going to infinity. The last
means it makes infinite $2\pi$ periods. One can find any details of elliptic
integrals calculations in my favourite book by {\ros G.~M. Fihtengolp1c}
{\rit Kurs differencialp1nogo i integralp1nogo isqisleniya} (see
Ref.~\cite{Fi}).

In this way we get $t=t(\rho)$ and $\vf=\vf(\rho)$ for the orbit in the case
$D\ge0$. This orbit is bounded, however, it never closes (see our argument
above). In this way the orbit is chaotic. It looks as a rosette.
The conditions to be closed or
chaotic orbit are given above in terms of an energy per a unit mass.

All the above formulae are applicable for a deep space region. Moreover in
this case we put
\beq{A.91}
\ov q=\frac{H^2}{\root3\of{2}\,b^{4/3}A^{4/3}}-\frac12
\end{equation}
and
\beq{A.92}
D=\frac{\root3\of4\,H^2}{27A^{4/3}b^{4/3}}-\frac14
\end{equation}
with the conditions $H<H_1$ or $H>H_2$ where $H_{1,2}$ are given by the
formulae \er{A.48}.

Let us consider a very deep region of space. In this case we should take
under consideration an influence of the \co ical \ct. Thus we have ($\lambda
=\frac{\La c^2}{6}$, where $\La$ is a \co ical \ct)
\beq{A.54}
U(\rho)=-b\rho-\la\rho^2,
\end{equation}
we neglected an ordinary \gr al interaction of the Sun.

Using Eq.~\er{A.1} one gets
\beq{A.55}
\vf-\vf_0=\pm\int\frac{d\rho}{\rho^2\sqrt{-2b\rho-2\la\rho^2+2H-\frac{A^2}{\rho^2}}}\,.
\end{equation}
Moreover we neglect the term $\frac{A^2}{\rho^2}$ ($\rho\gg1\,\rm AU$) and
one finds
\bml{A.56}
\frac1A(\vf-\vf_0)=-\frac{\sqrt{H-\rho(b+\la\rho)}}{\sqrt2\,H\rho}\\
{}+\frac b{2\sqrt2\,H^{3/2}}\log\frac{b\rho}
{2\sqrt2\,H^{3/2}\bigl(2H-b\rho+2\sqrt{H(H-\rho(b+\la\rho))}\bigr)}\,.
\end{multline}
We have
\beq{A.57}
0<\rho<\rho_{\max}=\frac1{2\la}\bigl(\sqrt{b^2+4\la b}-b\bigr).
\end{equation}
Thus we found an \e\ of an orbit.

From Eq.\ \er{A.2} one gets neglecting terms $\frac A{\rho^2}$ and
$\frac{k^2}{\rho^2}$:
\beq{A.58}
\sqrt{\la}(t-t_0)=\pm\int\frac{d\rho}{\sqrt{2H-2b\rho-2\la\rho^2}}\,.
\end{equation}
One easily integrates and gets
\beq{A.59}
\rho(t)=\frac1{\sqrt{2\la}}\bigl(-b+\sqrt{b^2+4\la H}\cdot\sin\sqrt{2\la}(t-t_0)\bigr).
\end{equation}
The orbit is bounded, $0<\rho(t)<\rho_{\max}$ or
$\arcsin(\frac b{\sqrt{b^2+4\la H}})\le \sqrt{2\la}(t-t_0)\le
\pi-\arcsin(\frac b{\sqrt{b^2+4\la H}})$.

Eventually we conclude. All the exact orbits are bounded.  Thus the \sp\
cannot escape from the \SS. It is a confinement (see Ref.~\cite8). It will
come back.

In all of these investigations we do not consider an influence of the Galaxy
background \gr al field and perturbations of different stars. Additional
influences can change some pesymistic conclusions.

Let us give some numerical examples. In order to do this we write a
polynomial under square root (see Eq.~(A.1a)). Using Eqs~\er{A.*} and a value
of~$b$ from Section~3 one gets
\beq{A.99}
g(\rho)=-\frac{2.9\times 10^{-7}}{a_h(e^2-1)}\,\rho^3 + \frac{\rho^2}
{a_h^2(e^2-1)} - \frac{2\rho}{a_h(e^2-1)}-1.
\end{equation}
$\rho$ and $a_h$ are measured in AU. In this way we parametrize the
polynomial by elements of a hyperbola. This is quite artificial, moreover
very useful. $a_h$ and~$e$ can be considered as elements of a hyperbola which
is a good approximation of the orbit not to far from the Sun, i.e.\ on
distances ${}<200\,\rm AU$. We consider case $1^\circ$~$D\le 0$ (three real
roots).

Let us take $a_h=6\,\rm AU$. This choice corresponds to the case of the
Pioneer orbit. In this case we have for $e=3, 7, 10$ the following roots of
$g(\rho)$:
\beq{A.100}
\bal
e=3,&\quad\rho_1=-11.9999,&\quad&\rho_2=24.077,&\quad &\rho_3=574701.0\\
e=7,&\quad\rho_1=-35.999,&\quad&\rho_2=48.0023,&\quad &\rho_3=574701.0\\
e=10,&\quad\rho_1=-53.9997,&\quad&\rho_2=66.0042,&\quad &\rho_3=574701.0.
\eal
\end{equation}
It is easy to see that the orbit  is bounded by
\beq{A.101}
\rho_{\min}=\rho_2<\rho<\rho_3=\rho_{\max}
\end{equation}
according to general theory. The roots are measured in AU.
Taking $a_h=100$, $e=20$ one gets
\beq{A.102}
\rho_1=-1853.3, \quad \rho_2=2173.06, \quad \rho_3=34163.0;
\end{equation}
if $a_h=1$, $e=3$, then
\beq{A.103}
\rho_1=-2, \quad \rho_2=4, \quad \rho_3=3.44827 \times 10^6.
\end{equation}

\begin{figure}
\centerline{\ing[height=0.26\textheight]{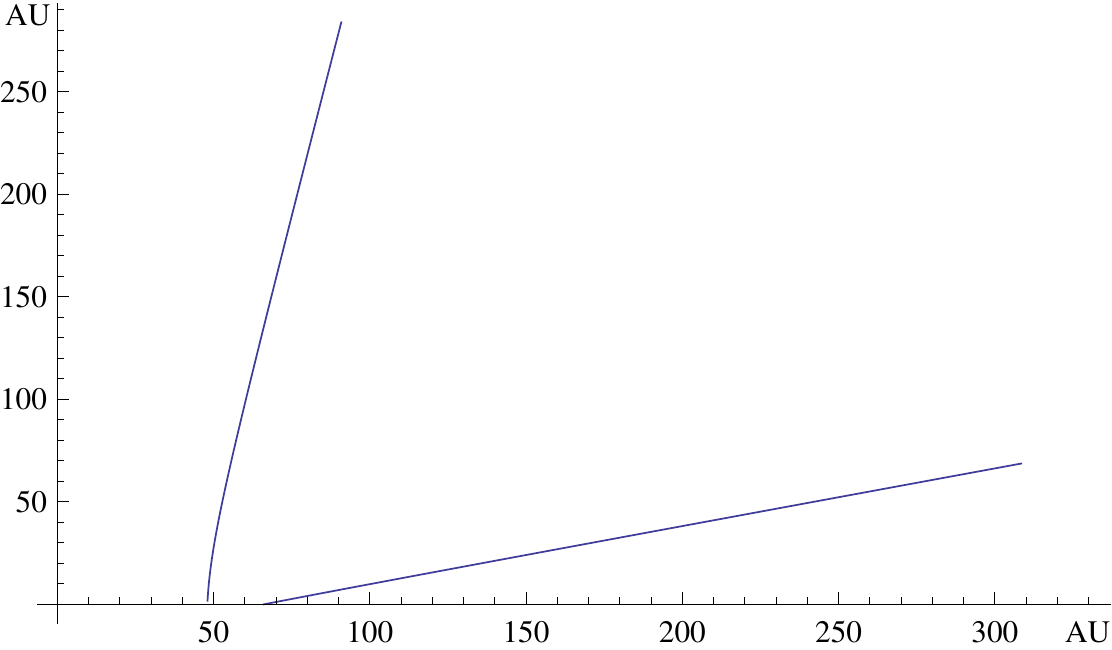}}
\centerline{(A)}
\centerline{\ing[height=0.26\textheight]{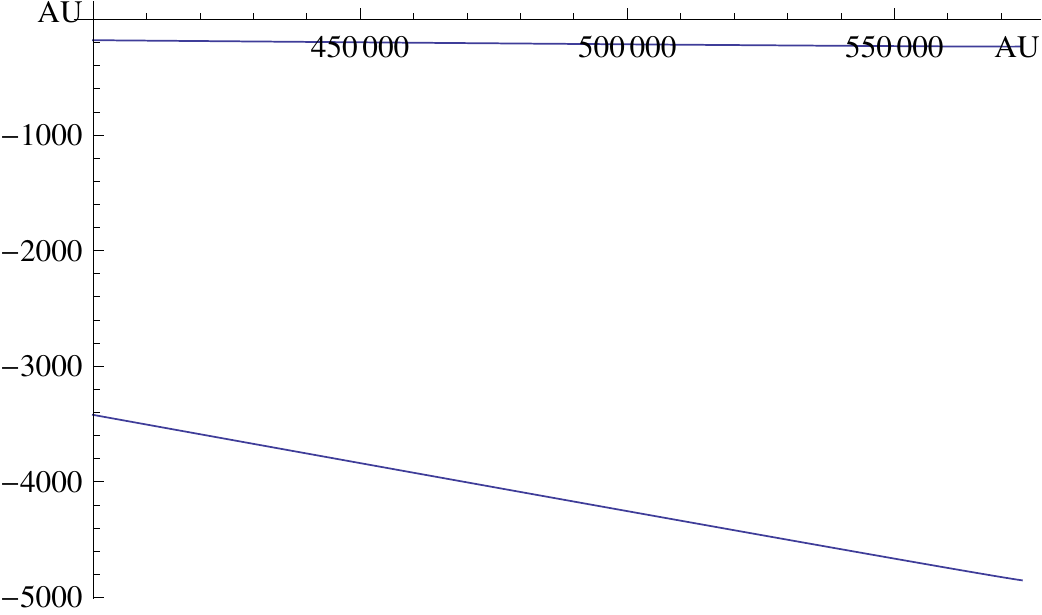}}
\centerline{(B)}
\centerline{\ing[height=0.26\textheight]{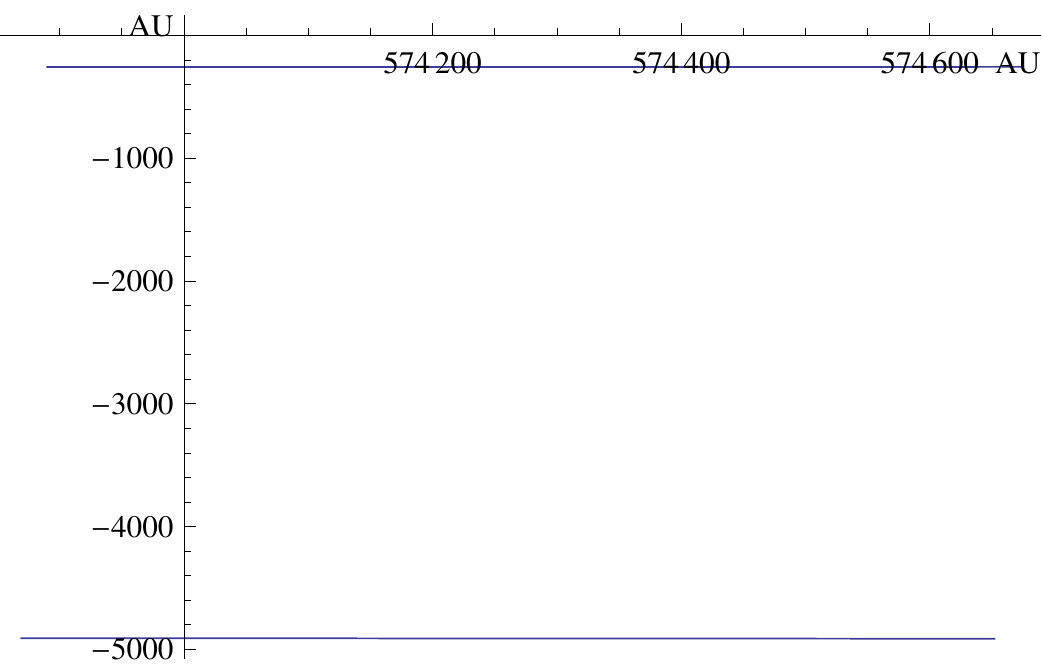}}
\centerline{(C)}
\caption{(A) orbits for $a_h=6$, $e=7,10$ and $\rho<400\,\rm AU$,
(B)~orbits for $a_h=6$, $e=7,10$ and $4\times10^5\,{\rm AU} < \rho <
573889\,\rm AU$,
(C)~orbits for $a_h=6$, $e=7,10$ and $573890\,{\rm AU}<\rho
<574673\,{\rm AU}$}\label{W}
\end{figure}

On Fig.~\ref{W} we give orbits for $a_h=6$, $e=7,10$ in cartesian \cd s
with an origin at the Sun for several regions of~$\rho$.
On Fig.~A for $\rho<400\,\rm AU$, $e=7,10$. On Fig.~B
for  $4\times10^5\,{\rm AU} < \rho < 573889\,\rm AU$, $e=7,10$.
On Fig.~C for  $573890\,{\rm AU}<\rho <574673\,{\rm AU}$, $e=7,10$.

On Fig.~A it looks similar to a part of hyperbolas, except $e=10$ which looks
like a straight line. This is reasonable---for $e=10$ a hyperbola is close to
its asymptote in this region. On Fig.~B and Fig.~C they look as straight
lines.

Let us notice that our model works well for $a_h=6$\,AU for $\rho>20\,\rm AU$ and
that $\rho_{\max}=57470.1\,{\rm AU}$ is of an order of an Oort cloud size.
The orbits considered here are very large in size from 40\,AU up to
$0.5\times 10^6\,\rm AU$. Moreover, they are very slim. In comparison to its
extent they are as straight lines.

Let us consider a different example: $a_h=1.5\times 10^{-6}\,\rm AU$ for
$e=3,7,10$. One gets
\beq{A.104}
\bal
{}&e=3,&\quad&\rho_1=-3\times 10^{-6},&\quad&\rho_2=6\times10^{-6},&\quad&
\rho_3=2.29885\times10^{12}\\
{}&e=7,&\quad&\rho_1=-9\times 10^{-6},&\quad&\rho_2=0.000012,&\quad&
\rho_3=2.29885\times10^{12}\\
{}&e=10,&\quad&\rho_1=-0.000135,&\quad&\rho_2=0.0000165,&\quad&
\rho_3=2.29885\times10^{12}
\eal
\end{equation}
Let us notice that
\beq{A.105}
\rho_3=2.29885\times10^{12}\,{\rm AU}=11.3467\,\rm Mpc
\end{equation}
is of order of the scale in our model (see Eq.~\er{2.23}). Thus we reach a
limit of our model. Moreover, $a_h=1.5\t 10^{-6}\,\rm AU$ corresponds to the
asymptotic \hy\ velocity of an order $0.1c$, where $c$ is a velocity of
light. This is also a border of our model for we consider here
nonrelativistic mechanics. We are still confined in the \SS. A~\sp\ with such a
velocity cannot escape the \SS. Moreover, we can plan a travel to distant
stars or even galaxies if we construct a \sp\ reaching $0.1c$ velocity. In
comparison an analogous velocity of the Pioneer is only $1.4\t 10^4 \rm\frac
ms$.

In order to be familiar with these new notions we consider also asymptotic
\hy\ velocities of order $0.05c$ and $0.01c$. One gets in the case of $0.05c$
$a_h=15\t10^{-5}\,\rm AU$ and in the case of $0.01c$ $a_h=6\t10^{-4}\,\rm
AU$. Thus we get for $a_h=15\t10^{-5}\,\rm AU$
\beq{A.107}
\bal
{}&e=3, &\q& \rho_1=-0.0003, &\q& \rho_2=0.0006, &\q& \rho_3=2.29885\t 10^{10}\\
{}&e=7, &\q& \rho_1=-0.0009, &\q& \rho_2=0.0012, &\q& \rho_3=2.29885\t 10^{10}\\
{}&e=10, &\q& \rho_1=-0.00135, &\q& \rho_2=0.00165, &\q& \rho_3=2.29885\t 10^{10}
\eal
\end{equation}
and for $a_h=6\t10^{-4}\,\rm AU$
\beq{A.108}
\bal
{}&e=3, &\q& \rho_1=-0.0012, &\q& \rho_2=0.0024, &\q& \rho_3=5.74713\t 10^9\\
{}&e=7, &\q& \rho_1=-0.0036, &\q& \rho_2=0.0048, &\q& \rho_3=5.74713\t 10^9\\
{}&e=10, &\q& \rho_1=-0.0054, &\q& \rho_2=0.0066, &\q& \rho_3=5.74713\t 10^9
\eal
\end{equation}

According to Section 3 we consider the model of an \an\ \ac\ up $10^9$\,AU.
This means that a velocity of $0.01c$ approaches this limit (as for a
velocity of $0.05c$). Moreover, velocities $0.01c$ and $0.05c$ are still
nonrelativistic. The velocity $0.1c$ is close to relativistic mechanics,
i.e., to Special Relativity. The \gr al field of the Sun is very small
starting from 20\,AU and any General Relativistic corrections on a \hy-like
orbit are not important. The \an\ \ac\ is an effect of non-relativistic
nature (however is non-newtonian). Thus we can use Special Relativity Theory
to describe a motion of a material point in this case. Using a special
relativistic energy conservation law we write:
\beq{A.110}
\frac{c^2}{\sqrt{1-\frac{v^2}{c^2}}}+U(\rho)=H,
\end{equation}
where $H=H'+c^2$ is a total energy with a rest energy per a unit mass, and
$H'$ is an energy without a rest energy per a unit mass, $c$~is a velocity of
light in a vacuum. One gets
\beq{A.111}
v^2=\frac{A^2}{\rho^2}\X2(1+\frac1{\rho^2}\X3(\pz \rho \vf\Y2)^2\Y3)
\end{equation}
(as in the nonrelativistic case). Thus we get finally
\bml{A.112}
\vf-\vf_0=\pm \int {A(H\rho+k^2+b\rho^2)}
\rho^{-1}\X2(b^2c^2\rho^6\\
{}+ 2Hbc^2\rho^5 + (H^2c^2+2k^2bc^2-A^2b^2)\rho^4
+(2Hk^2c^2-2HbA^2-c^6)\rho^3\\
{}+ (k^2c^2-A^2H^2-2k^2b^2A^2)\rho^2
-2A^2Hk^2\rho-k^2A^2\Y2)^{-1/2}\,d\rho
\end{multline}
\bml{A.113}
t-t_0=\pm \int \rho (H\rho +k^2+b\rho^2)
\X2(b^2c^2\rho^6\\
{}+ 2Hbc^2\rho^5 + (H^2c^2+2k^2bc^2-A^2b^2)\rho^4
+(2Hk^2c^2-2HbA^2-c^6)\rho^3\\
{}+ (k^2c^2-A^2H^2-2k^2b^2A^2)\rho^2
-2A^2Hk^2\rho-k^2A^2\Y2)^{-1/2}\,d\rho
\end{multline}

The above integrals are hyperelliptic integrals. They cannot be integrated as
in the nonrelativistic case ($c\to\iy$). They can be investigated only
numerically. Moreover, if we are planning a travel to distant stars or even
galaxies using relativistic \sp s we do not use these formulae in full. First
of all we should not launch a \sp\ with a relativistic velocity from the
Earth or even from the Moon. We should launch the \sp\ with reasonably low
velocity to a parking orbit distant from the Sun up to $10^3$\,AU. Why?
Travelling with so high velocity we will have troubles to maneuvre in the
asteroid belt and also in the Kuiper belt. Thus we should start with a
relativistic velocity from the orbit of $10^3$\,AU size. In this case one can
use an \ap ion with $k^2\to0$ (as in the nonrelativistic case). Thus one gets
\beq{A.114}
\vf-\vf_0=\pm \int \frac{A(H+b\rho)\,d\rho}
{\rho\sqrt{b^2c^2\rho^4 + 2Hbc^2\rho^3 + (H^2c^2-A^2b^2)\rho^2
-(2HbA^2+c^6)\rho -A^2H^2}}
\end{equation}
and
\beq{A.115}
t-t_0=\pm \int \frac{\rho(H+b\rho)\,d\rho}
{\sqrt{b^2c^2\rho^4 + 2Hbc^2\rho^3 + (H^2c^2-A^2b^2)\rho^2
-(2HbA^2+c^6)\rho -A^2H^2}}\,.
\end{equation}
Integrals \er{A.114} and \er{A.115} are still elliptic integrals and can be
calculated using methods from Ref.~\cite{Fi}. Let us notice the following
fact. There are some methods \ti{ab initio} to calculate orbits using
numerical integrations of \e s of motion. However in our case the orbits are
very large in size up $10^{10}$\,AU or even more. In this case numerical
integrations lead to large propagating round errors and it is better to have exact
formulae. These formulae can be treated numerically for various values of
parameters and the results are more precise than \ti{ab initio} methods.

Moreover, a further development of the theory of relativistic \sp s is beyond
the scope of this work. For this we will not proceed it here. Some of very
interesting issues in the light of an \an\ \ac\ will be done elsewhere.

Let us notice that according to Special Relativity Theory for a velocity
$v=0.01c$ a fraction of a kinetic energy measured in terms of a rest energy is
$5\t10^{-5}$, for $v=0.05c$ is $0.125\%$, for $v=0.1c$ is $0.5\%$. It means
that Newtonian \e s of motion can be used here with a caution. From the other
side a range of an orbit is close to the length scale of an application of
the model of an \an\ \ac. It means that in order to consider a full
relativistic motion of a \sp\ according to Special Relativity Theory (it is
enough to use Special Relativity Theory for a \gr al field is weak) it is
necessary to find a more refined model of nonnewtonian \ac. This can be
achieved by using a full formalism of the \eu\nos\ Jordan--Thiry Theory. The
\eu\nos\ Jordan--Thiry Theory has the following \lg
\beq{A.115a}
L_{\rm tot}=R-\frac{8\pi G_N}{c^4}\X1[\exp(-(n+2)\vF)L_{\rm matter}+L(\vF)\Y1],
\end{equation}
where $R$ is a scalar curvature. For we are working in a limit
$g_\m=g_{\nu\mu}$ this is a scalar curvature for a Riemannian \cn\ induced by
$g_\m$ on a \spt, $L_{\rm matter}$ is
a~\lg\ of a matter which is considered here in a hydrodynamic limit
\beq{A.116}
L(\vF)=\ov M g^\m \pa_\mu \vF\pa_\nu \vF+p
\X3[\frac{e^{(n+2)\vF}}{n+2} - \frac{e^{n\vF}}n
\Y3]+\ov\rho \exp(-(n+2)\vF).
\end{equation}

The field \e s read
\beq{A.117}
R_\m-\frac12 Rg_\m=\frac{8\pi G_N}{c^4}\X1[\exp(-(n+2)\vF)T_\m+T_\m(\vF)\Y1]
=\ov T_\m
\end{equation}
where
\bg{A.118}
T_\m=\ov\rho u_\mu u_\nu\\
T_\m(\vF)=2\ov M \pa_\mu \vF \pa_\nu\vF -g_\m L(\vF) \label{A.119}
\end{gather}
and
\beq{A.120}
-2\ov M\wt\nabla_\a \X1(g^{\a\b}\pa_\b \vF\Y1)+ p\exp(n\vF)(\exp(2\vF)-1)
-\frac{G_N}{8\pi c^4}\,(n+2)\exp(-(n+2)\vF)\ov\rho=0.
\end{equation}
$T_\m$ is an energy-momentum tensor of a pressureless matter (a~dust),
$\ov\rho$~is an energy density of a matter and $u^\mu$ its four-velocity,
\beq{A.121}
p=\la_{\rm co} \,\frac{n(n+2)}{2}.
\end{equation}
$\wt\nabla_\b$ means a covariant \dv\ \wrt the Riemannian \cn\ induced by a
metric $g_\m$.

In order to proceed investigations it is necessary to consider a stationary
and spherically \s\ cases, i.e.\ to consider metric in a form
\beq{A.122}
ds^2=e^{2A(r)}\,dt^2-e^{2B(r)}\,dr^2 -r^2 \,(d\theta^2+\sin^2\theta \,d\ov\vf{}^2).
\end{equation}
There is no possibility to confuse a radial \cd\ $r$ with $r$ considered above.
The programme can be done effectively for due to Bianchi identity and a symmetry
of the problem we have effectively four \f s: $A(r), B(r), \vF(r), \ov\rho(r)$ and
four \e s. In this way we can get an effective \gr al \ct
\beq{A.123}
G\eff(r)=G_N \exp(-(n+2)\vF(r)).
\end{equation}

Due to the same trick as in Section 8 we can find a dependence of $G\eff$
(and of course of an \an\ \ac) on a Hubble \ct, i.e.\ on a \co ical time.
Moreover this programme is beyond the scope of this work and will be done
elsewhere.

Let us remind that $\Ps=\Ps_0+\vF$ (we change the notation from $\vf$ to
$\vF$) and (see Section~2) $G_0=G_N\exp(-(n+2)\Ps_0)$.

In order to facilitate a future research we write down \e s in the case of
stationary and spherically \s\ case:
\begin{gather}\label{A.124}
 L(\vF(r))=\ov M e^{-2B(r)}\X2(\pz\vF r\Y2)^2
+pe^{n\vF(r)}\X2(\frac{e^{2\vF(r)}}{n+2}-\frac1n\Y2)
+\ov\rho(r) e^{-(n+2)\vF(r)},\\
\hskip-10pt \bal
\ov T_{44}&=\frac{8\pi G_N}{c^4}\X2(L(\vF(r))+\ov\rho(r) e^{-(n+2)\vF(r)}\Y2)
e^{2A(r)}\\
&=\frac{8\pi G_N}{c^4}\X3(\ov Me^{-2B(r)}\X2(\pz\vF r\Y2)^2
+p e^{n\vF(r)}\X2(\frac{e^{2\vF(r)}}{n+2}-\frac1n\Y2)+
2\ov\rho(r)e^{-(n+2)\vF(r)}\Y3)e^{2A(r)}\\
\ov T_{11}&=\frac{8\pi G_N}{c^4}\X3(\ov M \X2(\pz \vF r\Y2)^2
-pe^{n\vF(r)}\X2(\frac{e^{2\vF(r)}}{n+2}-\frac1n\Y2) e^{2B(r)}
-e^{2B(r)}\ov\rho(r) e^{-(n+2)\vF(r)}\Y3)\\
\ov T_{22}&=-\frac{8\pi G_N}{c^4}\,r^2 \X3(\ov M e^{-2B(r)}\X2(\pz \vF r\Y2)^2
+pe^{n\vF(r)}\X2(\frac{e^{2\vF(r)}}{n+2}-\frac1n\Y2)
+\ov\rho(r) e^{-(n+2)\vF(r)}\Y3)\\
\ov T_{33}&=\sin^2\theta \ov T_{22}
\eal \label{A.124a}
\end{gather}
and
\bml{A.125}
-\frac{2\ov M}{r^2}\, e^{-(A(r)+B(r))}\pz{}r \X2(r^2e^{(A(r)-B(r))}\,\pz\vF r\Y2)
+pe^{n\vF(r)}(e^{2\vF(r)}-1)\\
{}-(n+2)e^{-(n+2)\vF(r)}\ov\rho(r) =0.
\end{multline}
In the case of $G_\m=R_\m-\frac12Rg_\m$ one gets according to Refs~\cite{20},
\cite{46}, \cite{Sch}
\beq{A.126}
\bal
G_{44}&=\frac1{r^2}\,e^{2A(r)}\,\pz{}r \X1(r(1-e^{-2B(r)})\Y1)\\
G_{11}&=-\frac1{r^2}\,e^{2B(r)}(1-e^{-2B(r)})+\frac2r\,\pz Ar\\
G_{22}&=r^2 e^{-2B(r)}\X3(\pz{^2A}{r^2}+\X2(\pz Ar\Y2)^2+\frac 1r\,\pz Ar
-\pz Ar\,\pz Br-\frac1r\,\pz Br\Y3)\\
G_{33}&=\sin^2\theta G_{22}
\eal
\end{equation}

It is a real challenge to find a \so\ of the system of \e s
\beq{A.136}
R_\m-\frac12\,g_\m R=\frac{8\pi G_N}{c^4}\X2(T_\m(\vF)+
\exp(-(n+2)\vF)T_\m\Y2)=\ov T_\m
\end{equation}
in such a way that
\beq{A.127}
\wt\nabla_\mu \ov T{}^\mu{}_\nu=0, \q
\ov T{}^\mu{}_\nu=g^{\mu\a}\ov T{}_{\a\nu}.
\end{equation}

In order to proceed the programme we should consider Bianchi identities,
i.e.\ \er{A.127}. In this way we get from \er{A.127} Eq.~\er{A.120} and
\bg{A.130}
u^\mu\,\pp\vF{x^\mu}=0 \\
\wt\nabla_\mu(\ov \rho u^\mu)=0. \label{A.131}
\end{gather}
In the case of spherical symmetry and stationarity we get from
\er{A.130}--\er{A.131}
\bea{A.132}
\pp\vF t&=&0\\
\pp{\ov\rho}t&=&0. \label{A.133}
\end{eqnarray}
Thus $\vF$ and $\ov\rho$ are \f s of $r$ only according to our assumptions.
In this way \e s are consistent.

\def\theequation{B.\arabic{equation}}
\setcounter{equation}0
\def\thefigure{B.\arabic{figure}}
\setcounter{figure}0
\section*{Appendix B}
In this appendix we give formulae for $F(\vf)$ and $F_p(\vf)$ \f s without
simplifications from Sections 5~and~6. We also write down \e s for $\vf(t)$
in both cases (distorted hyperbola and parabola).

\eu\e s from Sections 5 and 6 are easier to handle from practical point of
view. Moreover, for an unusual precision of a measurement of positions
($\mu$arcs), obtained due to VLBI (Very Large Base Interferometry)
techniques, it is rational to derive these more precise
\e s. Thus for a \hy\ motion we get
\beq{B.1}
\frac {k(t-T)}{a_h^{3/2}(e^2-1)^{3/2}}=e\tan q-\log\tan\X2(\frac\pi4+\frac q2\Y2)
-\frac{ba_h^2(e^2-1)^2}{2ek}\,\ov F(\vf)
\end{equation}
where $\ov F(\vf)$ is given by the formula
\def\es{\sqrt{e^2-1}}
\bml{B.3}
\ov F(\vf)=K(e)\log\tan\X2(\frac\pi4+\frac q2\Y2)+L(e)\tan q\\
-\frac{e^2}{(e^2-1)^3}\cdot
\frac{4+4s^2-2e^4s^2+es(11+5s^2)+e^2(7s^2-1)-2e^3s(1+s^2)}
{(1+2es+s^2)^2}\\
-\frac1{(e+1)^3(e^2-1)^{1/2}}\X3(\log
\frac{\X1(\sqrt{e^2-1}-s-e\Y1)^2\X1(e+\sqrt{e^2-1}\Y1)}
{\X1(\sqrt{e^2-1}+s+e\Y1)^2\X1(e-\sqrt{e^2-1}\Y1)}
\log\X2(\sqrt2\cos\X2(\frac\pi4+\frac\vf2\Y2)\Y2)\\
+2\X2(\Li_2\X2(\frac{s-1}{\sqrt{1-e^2}-1-e}\Y2)-
\Li_2\X2(\frac{1-s}{1+e+\sqrt{1-e^2}}\Y2)\Y2)\Y3)\\
-\frac1{(e-1)^3(e^2-1)^{1/2}}\X3[
\log\frac{\cos^2\frac\vf2}{\cos^2\X1(\frac\pi4+\frac\vf2\Y1)}\cdot
\log\frac{\X1(1-e+\sqrt{e^2-1}\Y1)\X1(\sqrt{e^2-1}-e-s\Y1)}
{\X1(e-1+\sqrt{e^2-1}\Y1)\X1(\sqrt{e^2-1}+s+e\Y1)}\\
+2\X2(\Li_2\X2(\frac{1+s}{1-e+\es}\Y2)-\Li_2\X2(\frac{-(1+s)}{e-1+\es}\Y2)\Y2)\\
-\log\cos\X2(\frac\vf2+\frac\pi4\Y2)\cdot\log
\frac{\X1(e+s+\es\Y1)^2\X1(e-\es\Y1)}
{\X1(\es-e-s\Y1)^2\Y1(e+\es\Y1)}\Y3]\\
+\frac{2e(e^2+3)}{(e^2-1)^{5/2}}\X3[
\log\X1(e+s+\es\Y1)\cdot\log\frac{\X1(\es-e-s\Y1)^2}{4E^2(e^2-1)}\\
-2\log\cos\X2(\frac\vf2+\frac\pi4\Y2)\cdot
\log\frac{\X1(e+s+\es\Y1)^2\sqrt{e-\es}}
{\X1(\es-e-s\Y1)\X1(e+\es\Y1)\sqrt{2e}}\\
+\log\X2(e-\es+s\Y2)\log\frac{4\X1(e-\es\Y1)(e^2-1)}
{\X1(e+s+\es\Y1)^2}\\
+2\X3(\Li_2\X2(\frac{e-\es+s}{2\es}\Y2)+\Li_2\X2(\frac{e+\es+s}{2\es}\Y2)\Y3)\Y3]\\
+N(e)G(e,\vf)+M(e)\X2(\frac\vf2+\frac\pi4\Y2),
\end{multline}
where
\bea{*}
G(e,\vf)&=&\Re\Li_2\X2(\frac{1-es-s\es+i\X1(e-s+\es\Y1)}
{2e\X1(e+\es\Y1)}\Y2)\nn\\
&&\qquad-\Re\Li_2\X2(\frac{1-es+s\es+i\X1(\es-s-e\Y1)}
{2e\X1(e-\es\Y1)}\Y2)\\
K(e)&=&\frac{e\X1[(2+e^2)(e-1)^{1/2}+4e(e^2+3)(e-1)(e+1)^{5/2}\log
e+4e(e+1)^{3/2}\Y1]}
{2(e^2-1)^3(e-1)^{1/2}} \label{B.4}\\
L(e)&=&-\frac{4e^3}{(e-1)^{7/2}(e+1)^{3/2}}\,, \label{B.5}
\end{eqnarray}
$i=\sqrt{-1}$, $E=\lim\limits_{n\to\infty}\X1(1+\frac1n\Y1)^n$ (in order to
avoid misunderstanding with~$e$),
\bea{B.6}
M(e)&=&-\frac{4e \arctan\es}{(e^2-1)^{7/2}}\,\X2(e^4+e^2-6+(e^2+3)(e^2-1)
\arccot\X1(e+\es\Y1)\Y2)\\
\noalign{\allowbreak}
N(e)&=&\frac{4e(e^4+e^2-6)}{(e^2-1)^{7/2}} \label{B.7}\\
\noalign{\allowbreak}
s&=&\tan\X2(\frac\pi4+\frac\vf2\Y2) \label{B.8}\\
\noalign{\allowbreak}
q&=&2\arctan\X3(\sqrt{\frac{1+e}{1-e}}\tan\frac\vf2\Y3), \label{B.9}
\end{eqnarray}
$\Li_2(z)$ is a dilogarithm \f\ defined by
\beq{B.10}
\Li_2(z)=\sum_{n=1}^\infty \frac{z^n}{n^2} \qh{for}|z|<1.
\end{equation}
This \f \ can be defined in a different way,
\bml{B.11}
\Li_2(z)=z\int_0^\iy \frac t{e^t-z}\,dt=z \int_1^\iy
\frac{\log t}{t(t-z)}\,dt\\ = \frac{\pi^2}6 - \int_1^z \frac{\log(1-t)}t\,dt
=-\int_0^z\frac{\log(1-t)}t\,dt.
\end{multline}
Formula \er{B.10} can be used only for $|z|\le1$. Moreover, $\Li_2(z)$ is an
analytic \f\ in complex domain (an analytic extension to~$\C$) known since
eighteenth century (L.~Euler formula, 1768). It is quite easy to find $\Li_2(z)$
for $|z|>1$ using the following formula
\beq{B.12}
\Li_2(z)=-\Li_2\X2(\frac1z\Y2)-\frac1z\log^2(-z)-\frac{\pi^2}6
\end{equation}
(J. Landen formula, 1780).
The last formula is very important in order to calculate $\Re\Li_2(z)$ (the
real part of $\Li_2(z)$).

Moreover, for a calculation in a real domain outside the interval $(-1,1)$ it
is better to use an integral representation. A dilogarithm \f\ has very
important applications in High Energy Physics. In order to solve a motion it
is necessary to invert \e~\er{B.1} finding a \f\ $\vf=\vf(t)$. We substitute
this \f\ to Eq.~\er{5.17} getting $r(\vf(t))$ (without linear \ap ion as in
Section~5). In this way we have $r(\vf(t))$, $\vf(t)$ and we have a \so.
Eq.~\er{5.17} still gives us a shape of an orbit (a~distorted hyperbola). Let
us notice that our formula is valid up to $10^3$--$10^4$\,AU and for
distances greater than 20\,AU. Thus one gets
\beq{B.13}
20<r(\vf)<10^4.
\end{equation}

In this way one gets
\beq{B.14}
\arccos\frac{a_h(e^2-1)-20}{20e}<\vf<\frac\pi2+\arcsin
\frac{10^4-a_h(e^2-1)}{10^4e}\simeq\frac\pi2+\arcsin\frac1e\,,
\end{equation}
$a_h$ is measured in AU.

We can also consider the orbit in the \SS\ (as in Section~7). In this way we
can perturb a motion using Eqs \er{7.2}, \er{7.18}--\er{7.22}. Let us
consider a parabolic case. One gets
\bml{B.15}
\tan\frac\vf2+\frac13\tan^3\frac\vf3-\frac{bp^2}{2k^2}
\X3(\frac{(2+\cos\vf)\sin\vf}{3(1+\cos\vf)^2}\\{}+\frac\pi2+\vf+\tan\frac\vf2
\log\left|\sin\vf\right|+\frac{\sqrt2\cos(\frac\pi4+\frac\vf2)}{\cos\frac\vf2}\Y3)
=\frac{2k}{p^{3/2}}(t-T).
\end{multline}
Solving Eq.\ \er{B.15} we get $\vf(t)$ and afterwards we put it to
Eq.~\er{6.4} getting $r(\vf(t))$. This gives us a \so\ to the problem.

The shape of an orbit (a distorted parabola) is the same as in Section~6
(Eq.~\er{6.4}). We can also write the condition \er{B.13} for a distorted
parabola. One gets
\beq{B.16}
\arccos\frac{p-20}{20}<\vf<\frac\pi2+\arcsin\frac{10^4-p}{10^4}\simeq\pi,
\end{equation}
$p$ is measured in AU.

Parabolic orbits in general are more important in celestial mechanics than
\hy\ orbits. Moreover, in our case (Pioneer 10/11) a \hy\ (distorted) orbit
is in a real case of study.

Let us consider $\Re\Li_2(z)$. One gets for $|z|<1$
\beq{B.16a}
\Re\Li_2(z)=\sum_{n=1}^\iy \frac{T_n(\cos\arg z)}{n^2}\,|z|^n.
\end{equation}
In the case of $|z|>1$ using J.~Landen formula \er{B.12} we find
\beq{B.17}
\Re\Li_2(z)=\sum_{n=1}^\iy \frac{T_n(\cos\arg z)}{n^2|z|^n}-\frac12
\X1(\log^2|z|-(\arg z)^2\Y1)-\frac{\pi^2}6
\end{equation}
where $T_n(x)$ are Chebyshev polynomials
\beq{B.18}
T_n(x)=\frac{(x+\sqrt{x^2-1})^n+(x-\sqrt{x^2-1})^n}{2^n}, \q n=1,2,\dots
\end{equation}
The series in \er{B.16a} and \er{B.17} converge quickly. This is evident if
we consider a formula
\beq{B.19}
\frac{1-y^2}{1-2yx+y^2}=1+2\sum_{n=1}^\iy T_n(x)y^n
\end{equation}
where $|x|<1$, $|y|<1$.

The formula \er{B.19} gives a different definition of $T_n(x)$. Let us apply
\er{B.16a}--\er{B.17} to Eq.~\er{B.3}. In this \e\ we have $\Re\Li_2(z_1)
-\Re\Li_2(z_2)$, where
\bea{B.20}
z_1&=&\frac{(1-es-s\es)+i(e-s+\es)}{2e(e+\es)}\nn\\
&=&\frac{(es^2+e-2s)^{1/2}}{\sqrt2\,e(e+\es)^{1/2}}
\exp\X3(i\arctan\frac{e-s+\es}{1-es-s\es}\Y3)=r_1\ex{i\a_1}\\
z_2&=&\frac{(1-es+s\es)+i(\es-s-e)}{2e(e-\es)}\nn\\
&=&\frac{\X1(e(s^2+1)-\es(s+1)\Y1)^{1/2}}{\sqrt{2e}(e-\es)}
\exp\X3(i\arctan\frac{\es-s-e}{1-es+s\es}\Y3)=r_2\ex{i\a_2}\,.\qquad \label{B.21}
\end{eqnarray}
One gets for $|z_1|>1$
\beq{B.22}
2\arctan s_2-\frac\pi2 <\vf<2\arctan s_1-\frac\pi2
\end{equation}
and for $|z_1|<1$
\beq{B.23}
\vf>2\arctan s_1-\frac\pi2 \qh{or} \vf<2\arctan s_2-\frac\pi2
\end{equation}
where
\beq{B.24}
s_{1,2}=-1\pm\sqrt{1+2e^3(e+\es)}\,.
\end{equation}

For $|z_2|<1$
\beq{B.25}
2\arctan s_2'-\frac\pi2 < \vf < 2\arctan s_1'-\frac\pi2
\end{equation}
and for $|z_2|>1$
\beq{B.26}
\vf>2\arctan s_1'-\frac\pi2 \qh{or} \vf<2\arctan s_2'-\frac\pi2
\end{equation}
where
\beq{B.27}
s_{1,2}'=\frac{\es\pm\sqrt{e^2+4e(2e-1)(e-\es)}}{2e}\,.
\end{equation}

One can easily check that $z_1,z_2\ne0,1$. In this way we define four \f s
$G_i(\vf)$, $i=1,2,3,4$,
$$
\dsl{
\refstepcounter{equation}\label{B.28}
\indent G_1(\vf)=\sum_{n=1}^\iy \frac{r_1^nT_n(q_1)-r_2^nT_n(q_2)}{n^2},
\hfill |z_1|<1,\ |z_2|<1;\indent (\rm\theequation)\cr
\refstepcounter{equation}\label{B.30}
\indent G_2(\vf)=\sum_{n=1}^\iy \frac1{n^2}\X2(r_1^n T_n(q_1)
-\frac1{r_2^n}\,T_n(q_2)\Y2)
+\frac12\X1(\log^2r_2-(\arccos q_2)^2\Y1)+\frac{\pi^2}6,\hfill\cr
\hfill |z_1|<1,\ |z_2|>1;\indent (\rm\theequation)}
$$
$$
\dsl{
\refstepcounter{equation}\label{B.32}
\indent G_3(\vf)=\sum_{n=1}^\iy \frac1{n^2}\X2(\frac1{r_1^n}\,T_n(q_1)
-r_2^nT_n(q_2)\Y2)-\frac12\X1(\log^2r_1-(\arccos q_1)^2\Y1)-\frac{\pi^2}6,
\hfill\cr
\hfill |z_1|>1,\ |z_2|<1;\indent (\rm\theequation)\cr
\refstepcounter{equation}\label{B.34}
\indent G_4(\vf)=\sum_{n=1}^\iy
\frac1{n^2}\X2(\frac1{r_1^n}\,T_n(q_1)-\frac1{r_2^n}\,T_n(q_2)\Y2)
-\frac12\X1(\log^2r_1-\log^2r_2-(\arccos q_1)^2+(\arccos q_2)^2\Y1)\hfill\cr
\hfill |z_1|>1,\ |z_2|>1.\indent (\rm\theequation)}
$$
where
\bea{B.36}
r_1&=&\frac{(es^2+e-2s)^{1/2}}{\sqrt2\,e(e+\es)^{1/2}}\\
r_2&=&\frac{\X1(e(s^2+1)-\es(s+1)\Y1)^{1/2}}{\sqrt{2e}(e-\es)} \label{B.37}\\
q_1&=&\frac{1-es-s\es}{\sqrt2\X1(e^2-2es-2\es+e^2s^2+e\es+es^2\es\Y1)^{1/2}}
\label{B.38}\\
q_2&=&\frac{1-es+s\es}{\sqrt2\X1(e^2-e\es+e^2s^2-es^2\es\Y1)^{1/2}}.
\label{B.39}
\end{eqnarray}
for $e>1$, $\frac\pi2<\vf<\arccos(-\frac1e)<\pi$.

Let us notice that the \f\ series defined above are strictly and uniformly
convergent. Thus they define continuous \f s under assumptions given above.

\begin{figure}[ht]
\centerline{\ing{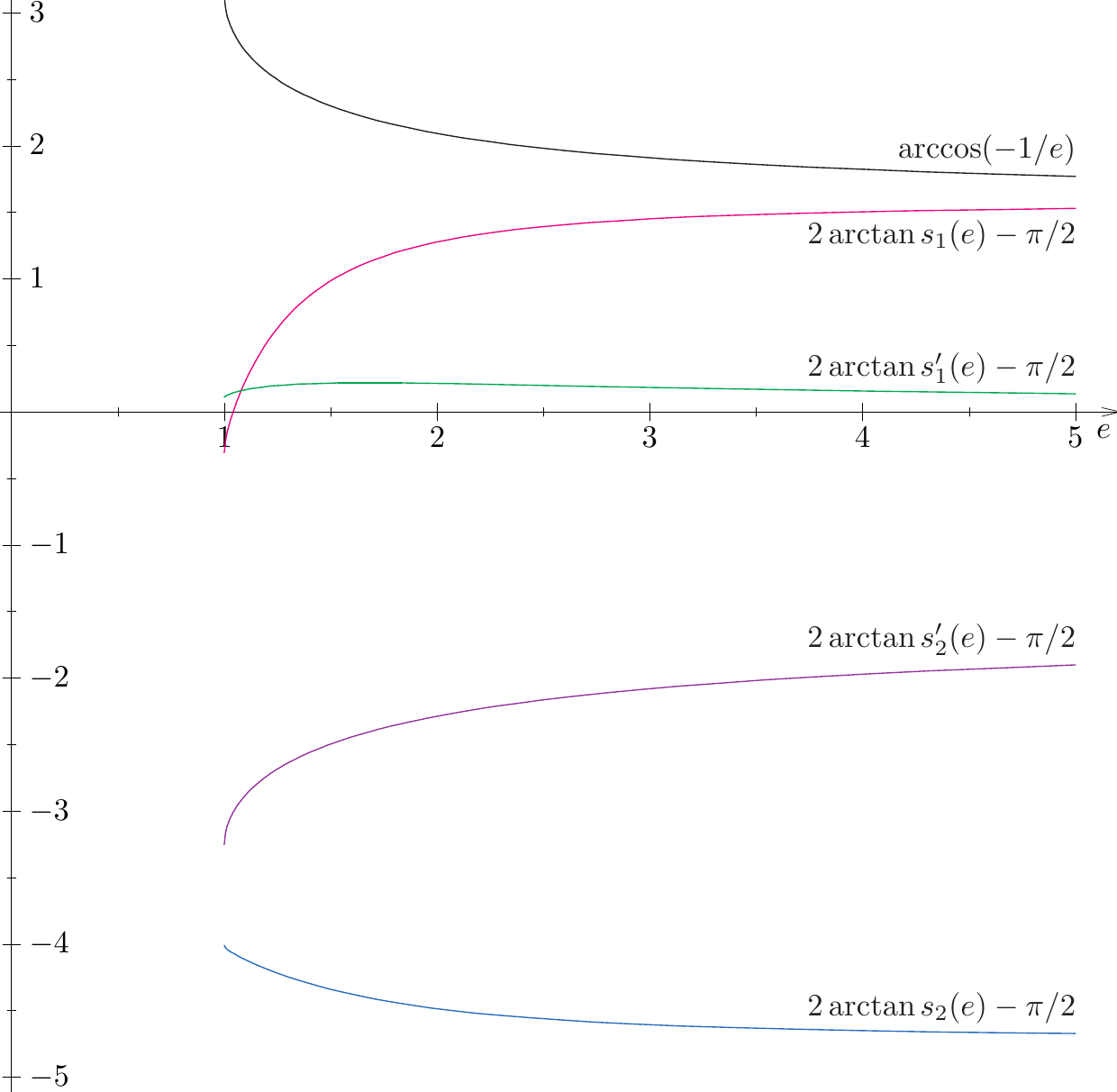}}
\caption{A plot of \f s $\arccos(-1/e)$, $2\arctan s_1(e)-\pi/2$,
$2\arctan s_1'(e)-\pi/2$,
$2\arctan s_2(e)-\pi/2$ and $2\arctan s_2'(e)-\pi/2$ (see a text).}\label{f5}
\end{figure}

\eu\f s $2\arctan s_i(e)-\pi/2$, $2\arctan s_i'(e)-\pi/2$, $i=1,2$, are
plotted on Fig.~\ref{f5} together with $\arccos(-1/e)$. From Fig.~\ref{f5} we easily get
that $G_i$, $i=1,2,3$ have empty domains.
In this way we get
\beq{B.40}
G(e,\vf)=G_4(\vf), \q \vf\in(\pi/2,\arccos(-1/e)).
\end{equation}

This can be obtained in a more formal way. Let us define four sets
$D_i(e)\subset \R$, $i=1,2,3,4$, $e>1$.
\begin{eqnarray*}
D_1(e)&=&\X3\{\vf:
\max\X2[2\arctan s_2'(e)-\frac\pi2,2\arctan s_2(e)-\frac\pi2\Y2]<\vf\\
&&\qquad {}<
\min\X2[2\arctan s_1'(e)-\frac\pi2,2\arctan s_1(e)-\frac\pi2\Y2]\Y3\}\\
D_2(e)&=&\X2\{\vf:2\arctan s_2(e)-\frac\pi2<\vf<2\arctan s_1(e)-\frac\pi2\Y2\}\\
\noalign{\allowbreak}
&&\qquad{}\cap \X3[\X2\{\vf: \vf>2\arctan s_1'(e)-\frac\pi2\Y2\}\cup
\X2\{\vf: \vf<2\arctan s_2'(e)-\frac\pi2\Y2\}\Y3]\\
\noalign{\allowbreak}
D_3(e)&=&\X2\{\vf:2\arctan s_2'(e)-\frac\pi2<\vf<
2\arctan s_1'(e)-\frac\pi2\Y2\}\\
&&\qquad {}\cap
\X3[\X2\{\vf:\vf>2\arctan s_1(e)-\frac\pi2\Y2\}\cup
\X2\{\vf: \vf<2\arctan s_2(e)-\frac\pi2\Y2\}\Y3]\\
\noalign{\allowbreak}
D_4(e)&=&\X3\{\vf: \vf<\min\X2[2\arctan s_2(e)-\frac\pi2,
2\arctan s_2'(e)-\frac\pi2\Y2]\Y3\}\\
&&\qquad{}\cup
\X3\{\vf>\max\X2[2\arctan s_1(e)-\frac\pi2,2\arctan s_1'(e)-\frac\pi2\Y2]\Y3\}
\end{eqnarray*}
These sets are domains of the \f s $G_i(\vf)$, $i=1,2,3,4$. Moreover, in our
case $\frac\pi2<\vf<\arccos(-\frac1e)$ and one gets
$$
\gathered
\wt D_1(e)=\wt D_2(e)=\wt D_3(e)=\varnothing\\
\wt D_4(e)=\X2\{\vf: \frac\pi2<\vf<\arccos\X2(-\frac1e\Y2)\Y2\}
\endgathered
$$
where $\wt D_i(e)=D_i(e)\cap \X1\{\vf: \frac\pi2<\vf<\arccos(-\frac1e)\Y1\}$.

Let
\beq{B.41}
\Li_2(\z)=U(x,y)+iV(x,y)
\end{equation}
where $\z=x+iy$ and $U$ and $V$ are harmonic \f s on $\R^2$. Let us define a
\f
\beq{B.42}
\ov H(e,z)=U(x_1,y_1)-U(x_2,y_2) \qh{for} 0\le z<1,
\end{equation}
where
\bea{B.43}
\Li_2(z_j)&=&U(x_j,y_j)+iV(x_j,y_j), \q z_j=x_j+iy_j,\ j=1,2,\\
x_1&=&\frac{1-e\ov s(z)-\ov s(z)\es}{2e(e+\es)} \label{B.44}\\
y_1&=&\frac{e-\ov s(z)+\es}{2e(e+\es)} \label{B.45}\\
x_2&=&\frac{1-e\ov s(z)+\ov s(z)\es}{2e(e-\es)} \label{B.46}\\
\noalign{\eject}
y_2&=&\frac{\es-\ov s(z)-e}{2e(e-\es)} \label{B.47}\\
\ov s(z)&=&\cot\X2(\frac\pi4(1-z)\Y2) \label{B.48}
\end{eqnarray}

\begin{figure}[ht]
\hbox to \textwidth{\ing[width=0.45\textwidth]{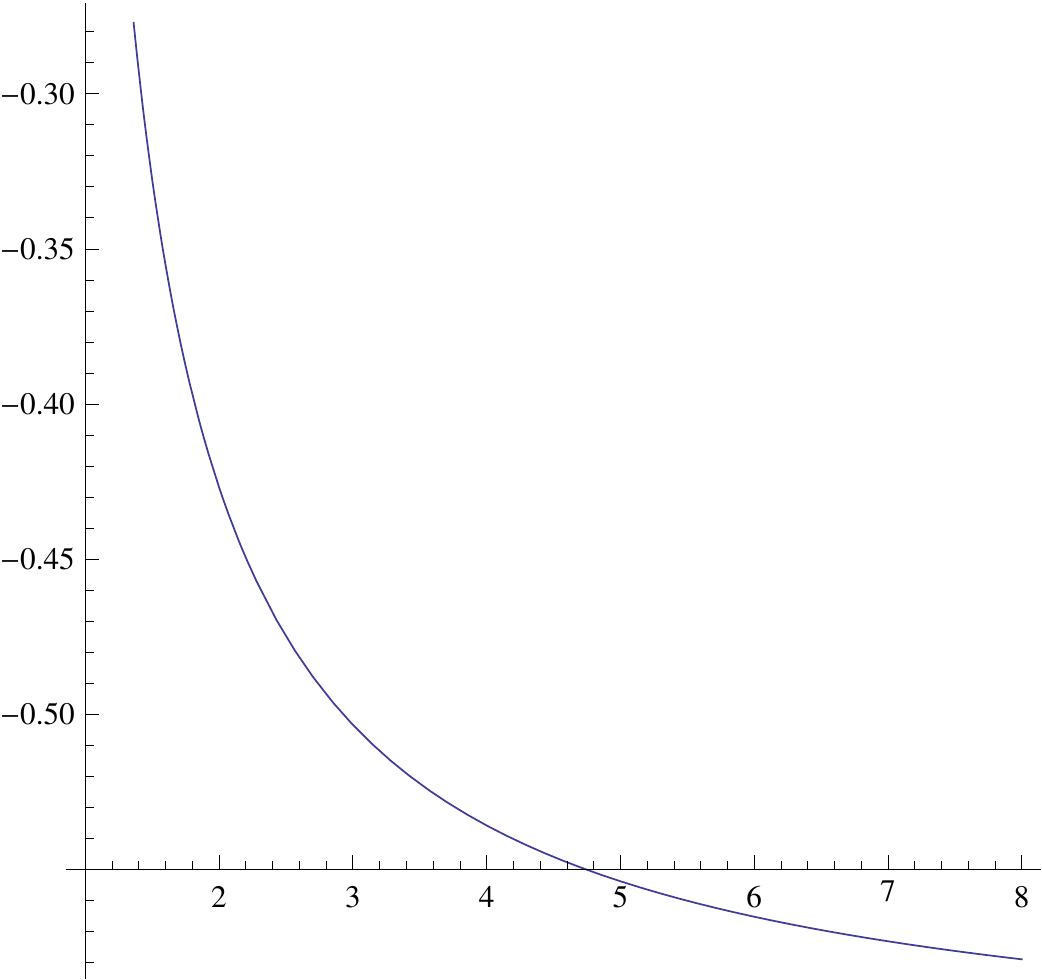}\hfil\ing[width=0.45\textwidth]{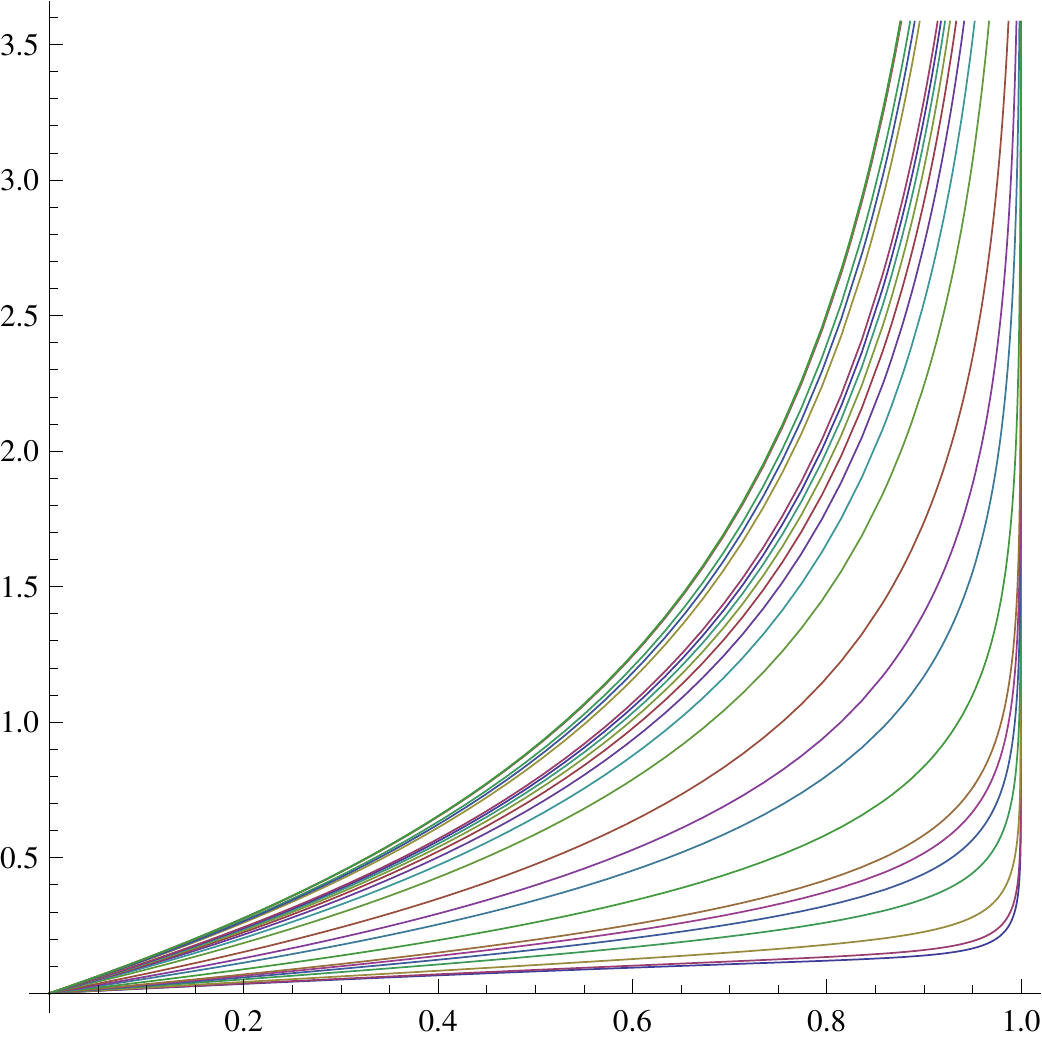}}
\hbox to \textwidth{\hbox to 0.45\textwidth{\hfil(A)\hfil}\hfil
\hbox to 0.45\textwidth{\hfil(B)\hfil}}
\hbox to \textwidth{\ing[width=0.45\textwidth]{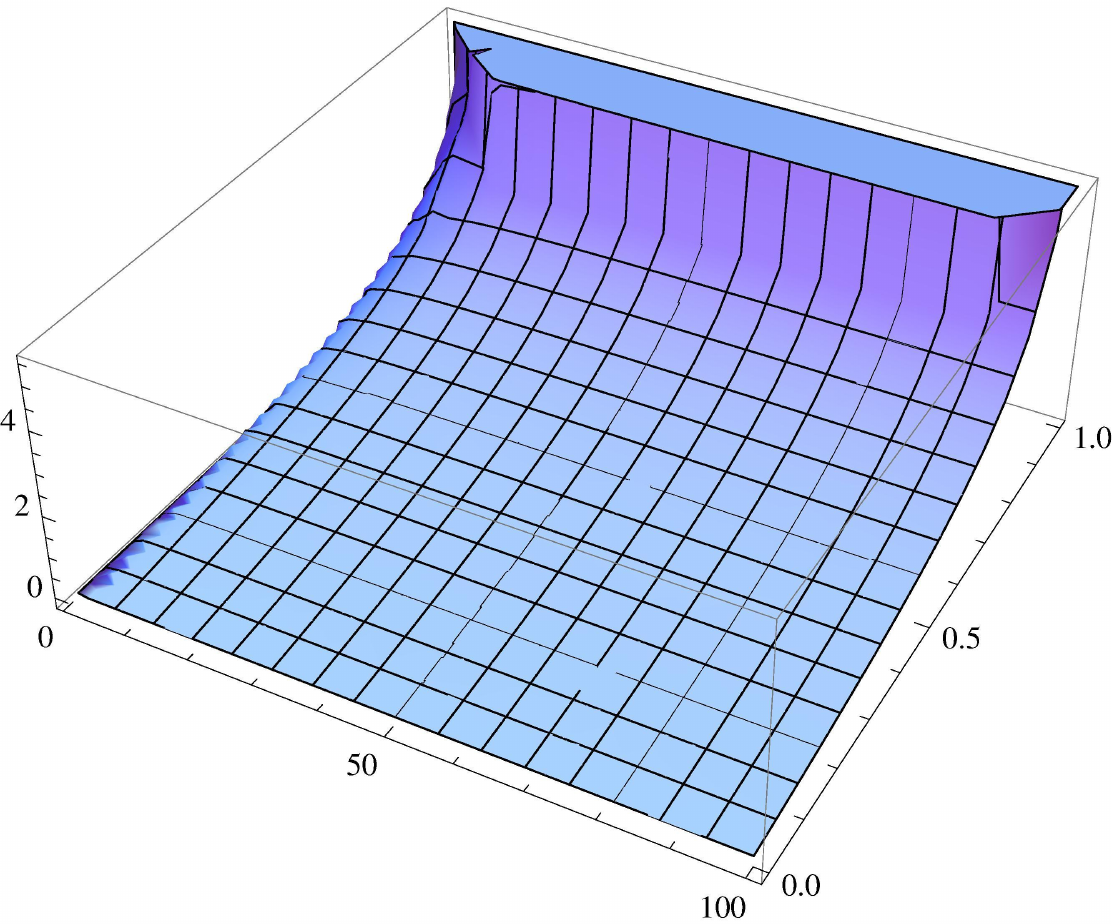}\hfil \ing[width=0.45\textwidth]{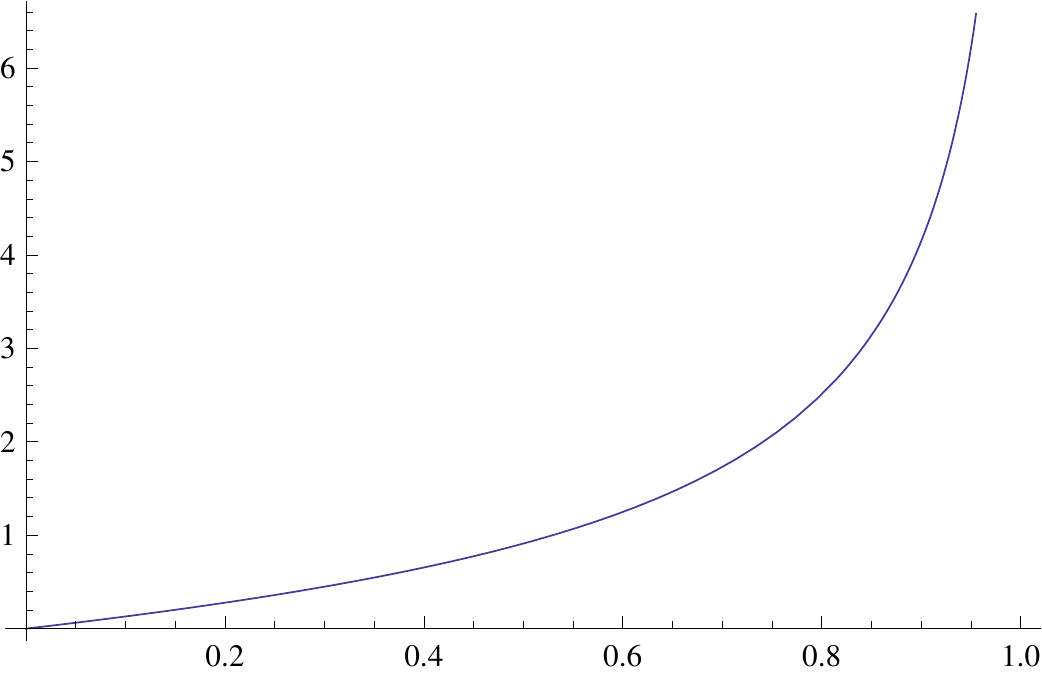}}
\hbox to \textwidth{\hbox to 0.45\textwidth{\hfil(C)\hfil}\hfil
\hbox to 0.45\textwidth{\hfil(D)\hfil}}
\caption{(A) is a plot of $H(e)$,
(B) is a plot of the \f\ $H(e,z)$ for several values of~$e$,
(C)~is a 3D plot of the \f\ $H(e,z)$,
(D) is a plot of the \f\ $H(\iy,z)=g(z)$.}\label{f6}
\end{figure}

Using $\ov H(e,z)$ we define a different \f
\beq{B.49}
H(e,z)=\ov H(e,z)-H(e), \q H(e)=\ov H(e,0).
\end{equation}
The \f\ $H(e,z)$ is plotted on Fig.~\ref{f6} for several values of a parameter
$e>1$. On Fig.~\ref{f6} we plot a 3D plot of $H(e,z)$, a \f\ $H(e)$ and a \f\
$g(z)=\lim\limits _{e\to\infty}H(e,z)=-U(1,-\ov s(z))$. It is easy to see that
\bg{B.50}
H(e_1,z)>H(e_2,z), \q 0<z<1, \ e_1>e_2,\\
H(e,0)=0, \q \lim_{e\to1}H(e,z)= U\X3(\frac{1-\ov s(z)}2\,,\frac{1-\ov s(z)}2
\Y3)-U\X3(\frac{1-\ov s(z)}2\,,-\frac{1+\ov s(z)}2\Y3)
\label{B.51}\\
\lim_{z\to\iy}H(e,z)=+\iy. \label{B.52}
\end{gather}
Let us notice that we can extend $H(e,z)$ to a unit complex disc $|z|<1$.
Such a \f\ is a continuous \f\ of a complex variable~$z$, $|z|<1$. This \f\
has a first \dv\ \wrt $z$, $|z|<1$.

One gets
\bml{B.53}
\pz Hz=\frac \pi{8e\sin^2(\frac\pi4(1-z))}\X3[\pp Ux(x_1,y_1)-\pp Ux(x_2,y_2)\\
{}+\pp Uy(x_1,y_1)\,\frac1{e+\es}-\pp Uy(x_2,y_2)\,\frac1{e-\es}\Y3].
\end{multline}
$\pz Hz$ is continuous \wrt $z$, $|z|<1$. Thus $H(e,z)$ is a holomorphic \f\
for $|z|<1$ for every $e>1$. In this way we can expand $H(e,z)$ in a power
series
\beq{B.54}
H(e,z)=\sum_{n=1}^\iy a_n(e)z^n, \q |z|<1.
\end{equation}
One can easily find $a_1(e)$:
\bml{B.55}
a_1(e)=\frac\pi{4e}\X4[\pp Ux\X3(\frac{1+e+\es}{2e(e+\es)},
\frac{e-1+\es}{2e(e+\es)}\Y3)\\
{}-\pp Ux\X3(\frac{1+e-\es}{2e(e-\es)},\frac{\es+1-e}{2e(e-\es)}\Y3)\\
{}+\pp Uy\X3(\frac{1+e+\es}{2e(e+\es)},
\frac{e-1+\es}{2e(e+\es)}\Y3) \frac1{e+\es}\\
{}-\pp Uy\X3(\frac{1+e+\es}{2e(e+\es)},\frac{e-1+\es}{2e(e+\es)}\Y3)
\frac1{e-\es}\Y4].
\end{multline}

The \f\ $H(e,z)$ can be analytically extended to the whole complex plane and
we get a one parameter family of analytic \f s. $H(e,z)$ has a pole at $z=1$.
This is a new special \f. Using this \f\ we easily write
\beq{B.56}
G(e,\vf)=H(e,2(1-\tfrac\vf\pi))+H(e), \q
\tfrac\pi2 <\vf<\arccos(-\tfrac1e).
\end{equation}
Let us notice the following properties of the analytic \f\ $H(e,z)$:
\beq{B.57}
H(e,z)=H(e,z+2n), \q n=\pm1,\pm2,\pm3.
\end{equation}
Thus it is a periodic \f\ with a period 2. The \f\ has an infinite number of
zeros on a real axis at
\beq{B.58}
\z_k=2k, \q k=0,\pm1,\pm2,\dots.
\end{equation}
It has also an infinite number of singularities (even poles of infinite order)
on the real axis at
\beq{B.59}
\eta_k=2k+1, \q k=0,\pm1,\pm2,\dots.
\end{equation}
The fundamental region of this \f\ is
\beq{B.60}
0<\Re z<2.
\end{equation}

We give a programme written in Mathematica 7 to calculate and plot
$H(e,z)$ for $0\le z<1$. This is the listing
\begin{verbatim}
s[z_] := Cot(Pi*(1-z)/4)
z1[x_,z_] := (1-x*s[z]-s[z]*(x^2-1)^(1/2)+I*(x-s[z]+(x^2-1)^(1/2))) *
    (2*x*(x-(x^2-1)^(1/2)))^(-1)
z2[x_,z_] := (1-x*s[z]+s[z]*(x^2-1)^(1/2)+I*((x^2-1)^(1/2)-s[z]-x)) *
    (2*x*(x-(x^2-1)^(1/2)))^(-1)
l[x_,z_] := PolyLog[2,z1[x,z]]-PolyLog[2,z2[x,z]]
g1[x_,z_] := Re[l[x,z]]
h[x_,z_] := g1[x,z]-g1[x,0]
\end{verbatim}

In the listing {\tt x} means $e$, {\tt g1}---$\ov H$ and {\tt
h} means $H$. The
\f\ $H(e,z)$ can be written as $H(e,z)=h(e,\cot(\frac\pi4(1-z)))$.

\begin{figure}[ht]
\hbox to \textwidth{\ing[width=0.45\textwidth]{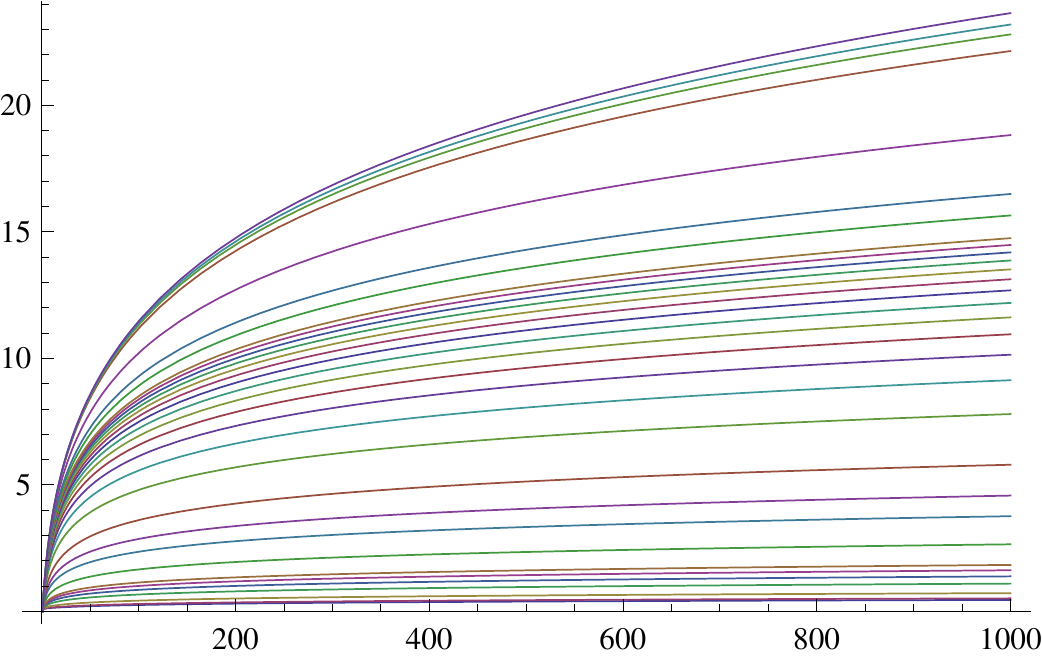}\hfil \ing[width=0.45\textwidth]{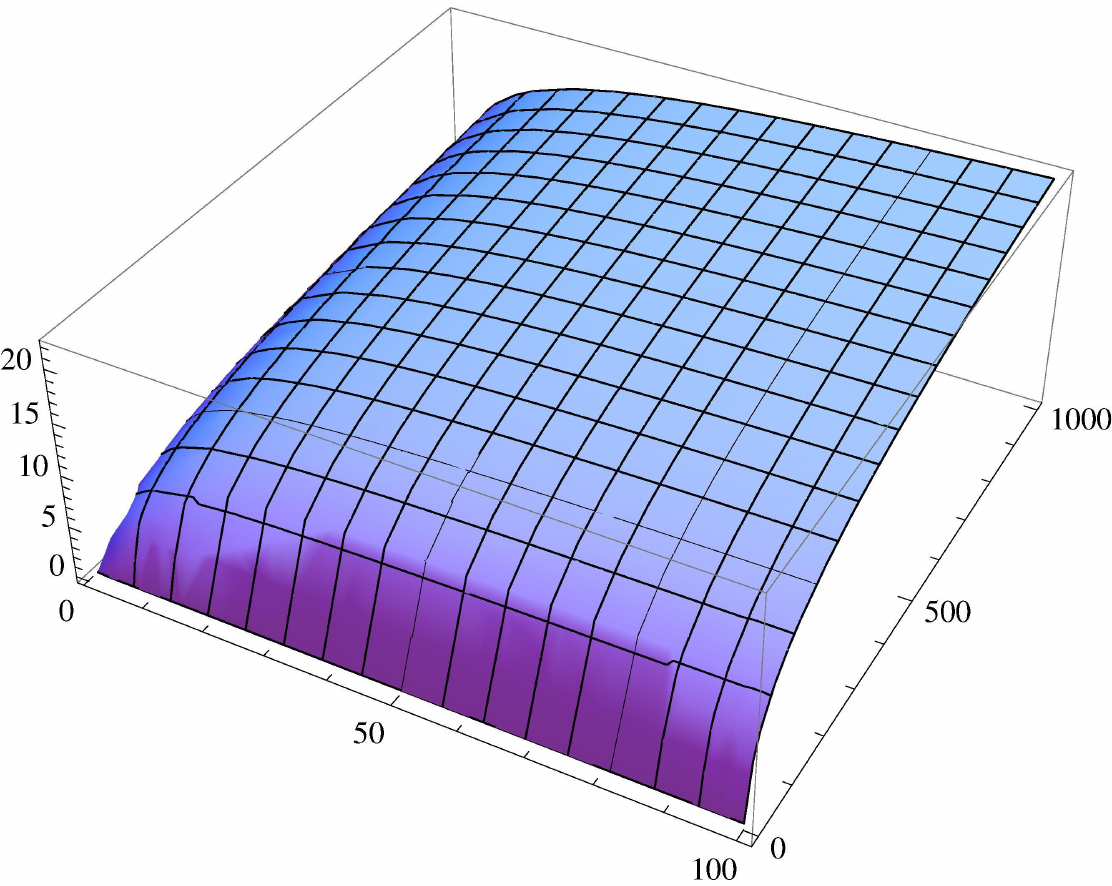}}
\hbox to \textwidth{\hbox to 0.45\textwidth{\hfil(A)\hfil}\hfil
\hbox to 0.45\textwidth{\hfil(B)\hfil}}
\hbox to \textwidth{\ing[width=0.45\textwidth]{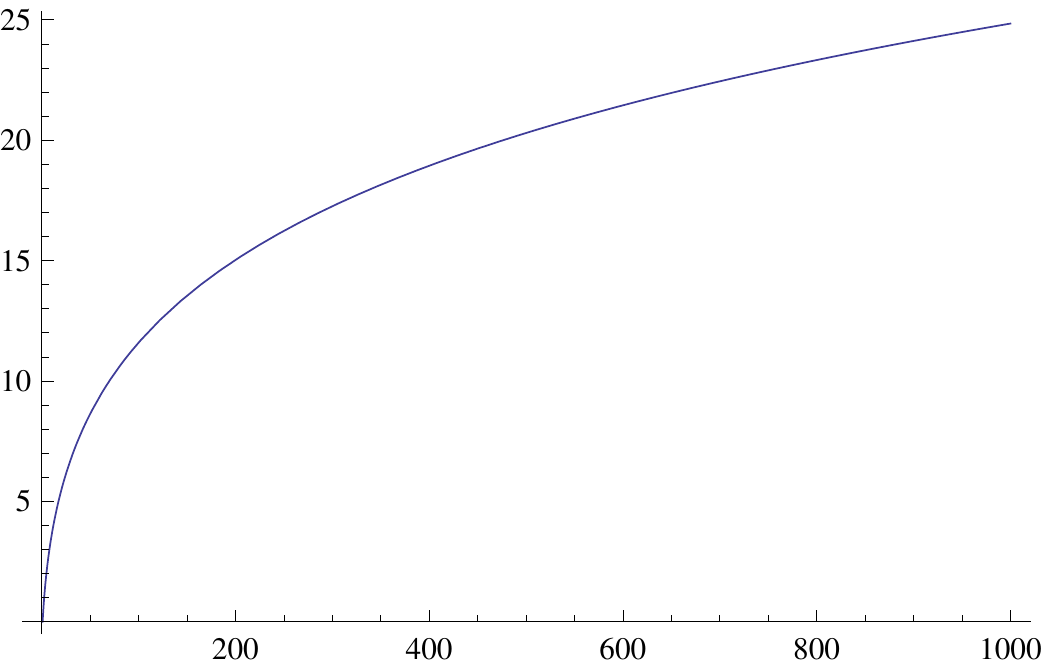}\hfil \ing[width=0.45\textwidth]{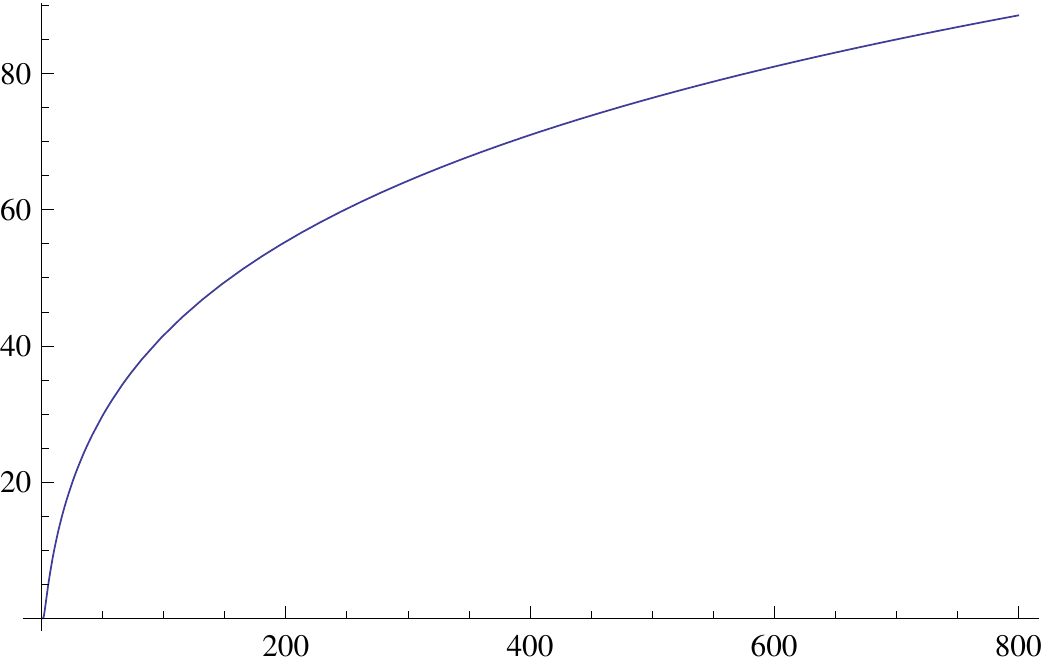}}
\hbox to \textwidth{\hbox to 0.45\textwidth{\hfil(C)\hfil}\hfil
\hbox to 0.45\textwidth{\hfil(D)\hfil}}
\caption{(A)---a plot of a \f\ $h(e,x)$ for several values of a parameter $e$,
(B)---3D plot of a \f\ $h(e,x)$,
(C)---a plot of a \f\ $h(e,x)$ for a large $e=10^5$,
(D)---a plot of a \f\ $h(\iy,x)$.}\label{f7}
\end{figure}

We plot $h(e,x)$ for several values of the parameter $e>1$ and we give a 3D
plot. We plot also $h(e,x)$ for a large $e$ on Fig.~\ref{f7}.
This is a listing of a programme
written in Mathematica~7.
\begin{verbatim}
z1[x_,s_] := (1-x*s-s*(x^2-1)^(1/2)+I*(x-s+(x^2-1)^(1/2))) *
    (2*(x+(x^2-1)^(1/2)))^(-1)
z2[x_,s_] := (1-x*s-s*(x^2-1)^(1/2)+I*((x^2-1)^(1/2)-s-x)) *
    (2*x*(x-(x^2-1)^(1/2)))^(-1)
l[x_,s_] := PolyLog[2,z1[x,s]] - PolyLog[2,z2[x,s]]
g1[x_,s_] := Re[l[x,s]]
h[x_,s_] := g1[x,s]-g1[x,1]
\end{verbatim}
Here $x$ means $e$ and $s$ means $x$. We plot $h$ for $e=10^7$ with range
$10^3$ and $h(\iy,s)$ (see Fig.~\ref{f7}).

Let us consider a \f
\bg{B.68}
H_1(e,z)=\frac1{a_1(e)}\,\ov H(e,z), \q |z|<1,\ e>1; \\
H_1(e,z)=z+\sum_{n=2}^\iy \wt a_n(e)z^n, \label{B.69}\\
\wt a_n(e)=\frac{a_n(e)}{a_1(e)}\,.\label{B.70}
\end{gather}
$H_1(e,z)$ is a univalent \f\ in $K_1=\{z: |z|<1\}$ and $H_1(e,K_1)=\C$.

In this way we can apply a Bieberbach conjecture (1916) proved by de~Branges
(1985), known also as the de Branges theorem. It means that
\beq{B.71}
|\wt a_n(e)|\le n,\q n\ge2
\end{equation}
for $e>1$.

Let us take a compact set $E\subset K_1$. One gets
\beq{B.72}
|H_1(e,z)|\le 1+\sum_{n=2}^\iy |\wt a_n(e)|\,|z|^n
\le 1+\sum_{n=2}^\iy n|z|^n=\frac{|z|}{(1-|z|)^2}\,.
\end{equation}
Thus there is a \ct\ $M(E)$ \st
\beq{B.73}
|H_1(e,z)|\le M(E)
\end{equation}
for $z\in E$, $e>1$.

This means that a family of holomorphic \f s
$$
\F_1=\{H_1(e,z): e>1\},\q z\in K_1,
$$
satisfies a Montel criterion. It means that $\F_1$ is a normal family of
holomorphic \f s in~$K_1$.

Simultaneously a family of holomorphic \f s ($a_1(e)$ is bounded as a \f\ of
$e>1$)
$$
\F=\{\ov H(e,z): e>1,\ |z|<1\}
$$
is a normal family of holomorphic \f s in~$K_1$. Thus from any infinite
sequence $\ov H(e_n,z)$, $e_n>1$, we can choose a subsequence
$\ov H(e_{n_k},z)$ \st it is uniformly convergent to a holomorphic
\f\ $\ov H_g(z)$ or it is divergent to infinity. $\ov H_g(z)$ need not belong
to the family~$\F$.

The \f\ $h(e,\cot(\frac\pi4(1-z)))$ is a periodic \f\ for $z\in\C$.
Simultaneously it has poles and zeros at a real axis. According to a general
theorem concerning a double periodic \f\ in a complex domain (an elliptic
\f), it should be a meromorphic \f\ of two variables of appropriate
$\cP$--Weierstrass \f\ and its \dv. In this case this is
$\cot(\frac\pi4(1-z))$. Moreover, in this case it is only
$\cot(\frac\pi4(1-z))$. In this way $h(e,\z)$ is more than a meromorphic \f. It is
an entire \f\ for every $e>1$ and $\lim\limits_{\z\to\iy}h(e,\z)=\iy$,
$h(e,1)=0$, only one zero at $\z=1$.

It is easy to see that $h(e,\ov s)$ is an entire \f\ if we remind that
$$
h(e,\ov s)=U\X1(x_1(\ov s),y_1(\ov s)\Y1)-U\X1(x_2(\ov s),y_2(\ov s)\Y1)
-U\X1(x_1(1),y_1(1)\Y1)+U\X1(x_2(1),y_2(1)\Y1)
$$
(see Eqs\ \er{B.44}--\er{B.47}).

Let us consider the following \e\ (see Eq.\ \er{B.11}):
\beq{B.74}
\pz{\Li_2(z)}{z}=-\frac{\log (1-z)}z
\end{equation}
and let us define \f s
\bg{B.75}
f_i(s)=\Li_2(z_i(s)), \q i=1,2,\\
f(s)=\Re(f_1(s)-f_2(s)). \label{B.76}
\end{gather}
One gets
\beq{B.77}
\pz{f_i}{s}=-\frac{\log(1-z_i(s))}{z_i(s)}\X2(\pz {z_i(s)}s\Y2)
\end{equation}
and finally
\beq{B.78}
\pz fs=\frac1{2e}\X3[n_1(s)-n_2(s)-\X2(\frac{m_1(s)}{e+\es}-\frac{m_2(s)}
{e-\es}\Y2)\Y3]
\end{equation}
where
\beq{B.79}
n_i(s)=\Re\X2(\frac{\log(1-z_i(s))}{z_i(s)}\Y2), \q m_i(s)=\Im
\X2(\frac{\log(1-z_i(s))}{z_i(s)}\Y2),\q i=1,2.
\end{equation}
It is easy to see that $f(s)=h(e,s)$ for $f(1)=0$. In this way we can write
$G(e,\vf)=H(e)+h(e,\tan (\frac\pi4+\frac\pi2))$.

Let us consider Eq.~\er{B.78} with an initial condition $f(1)=0$ for
$e\to\iy$. One gets
\bg{B.80}
\pz{h(\iy,s)}s=\frac{s\log s}{(1+s^2)}-\frac\pi{2(1+s^2)}\\
h(\iy,1)=0 \nn\\
h(\iy,s)=\int_1^{s^2}\frac{\log y}{1+y}\,dy-\frac\pi2\,\X2(\arctan
s-\frac\pi4\Y2).
\end{gather}
We plot this \f\ on Fig.~\ref{f8}D.

Let us consider Eq.~\er{B.78} in more details. Using Eq.~\er{B.79}, Eqs
\er{B.75}--\er{B.76}, Eqs \er{B.20}--\er{B.21}, Eqs
\er{B.36}--\er{B.37} one gets
\beq{B.82}
h(e,s)=\int_1^s K(e,y)\,dy,\q e>1,
\end{equation}
where
\beq{B.83}
\bal
K(e,s)&=\frac12 \X3[\frac12(1-es-s\es)\log\frac
{k_+(e,s)+(e-s+\es)^2 }{4e^2(e+\es)}\\
&-(e-s+\es)\arctan\frac{e-s+\es}
{2e(e+\es)-1+es+s\es}\\
&-\frac{(e-\es)(1-es+s\es)}{2e\X1(e(s^2+1)-\es(s+1)\Y1)}
\log\frac{k_-(e,s)+(\es-s-e)^2}{2e(e-\es)}\\
&-\frac{(e-\es)(\es-s-e)}{e\X1(e(s^2+1)-\es(s+1)\Y1)}\arctan
\frac{\es-s-e}{2e(e-\es)-1+es-s\es}\\
&-\frac{1-es-s\es}{e+\es} \arctan\frac
{e-s+\es}{2e(e+\es)-1+es+s\es}\\
&+\frac{e-s+\es}{2(e+\es)} \log \frac
{k_+(e,s)+(e-s+\es)^2 }{4e^2(e+\es)}\\
&+\frac{1-es+s\es}{e\X1(e(s^2+1)-\es(s+1)\Y1)} \arctan\frac
{\es-s-e}{2e(e-\es)-1+es-s\es}\\
&+\frac{\es-s-e}{2\X1(e(s^2+1)-\es(s+1)\Y1)} \log \frac
{k_-(e,s)+(\es-s-e)^2}{2e(e-\es)}\Y3],\\
k_+(e,s)&=\X1(2e(e+\es)-1+es+s\es\Y1)^2,\\
k_-(e,s)&=\X1(2e(e-\es)-1+es-s\es\Y1)^2.
\eal
\end{equation}

\begin{figure}[ht]
\hbox to \textwidth{\ing[width=0.45\textwidth]{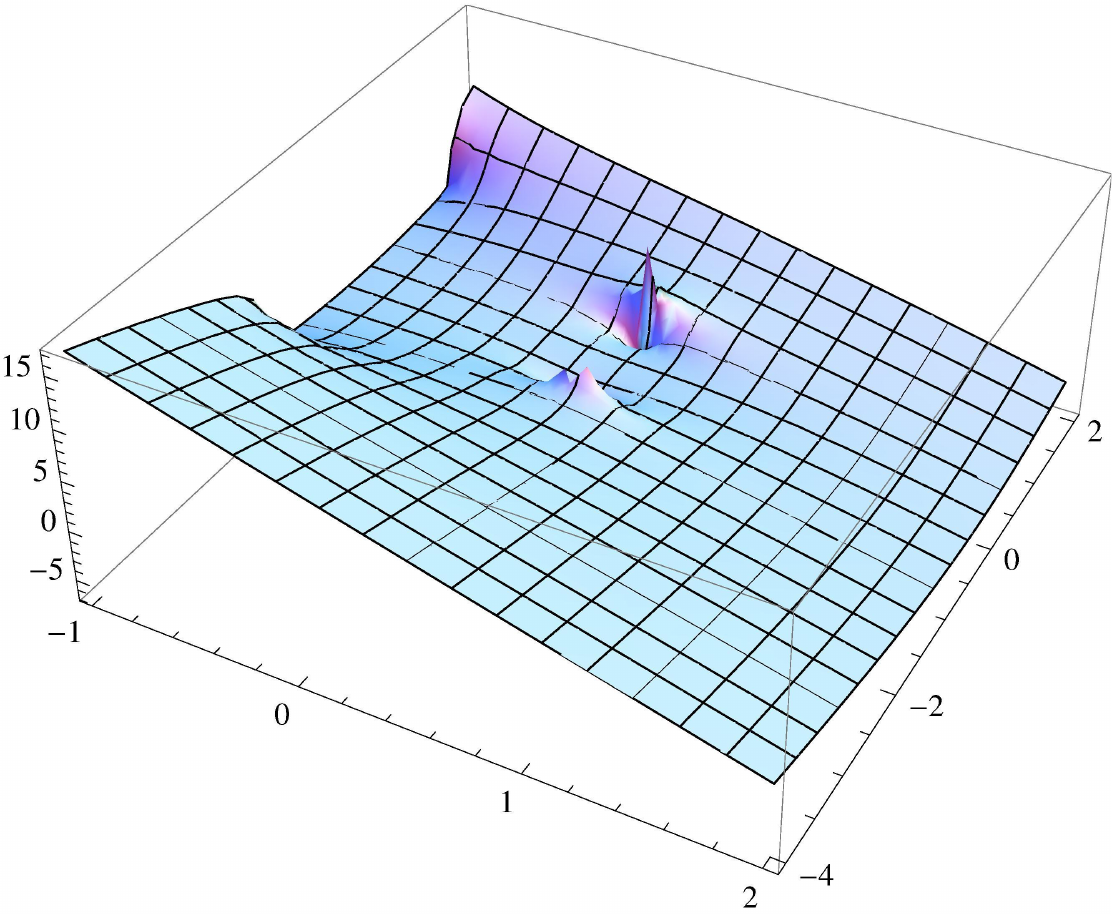}\hfil \ing[width=0.45\textwidth]{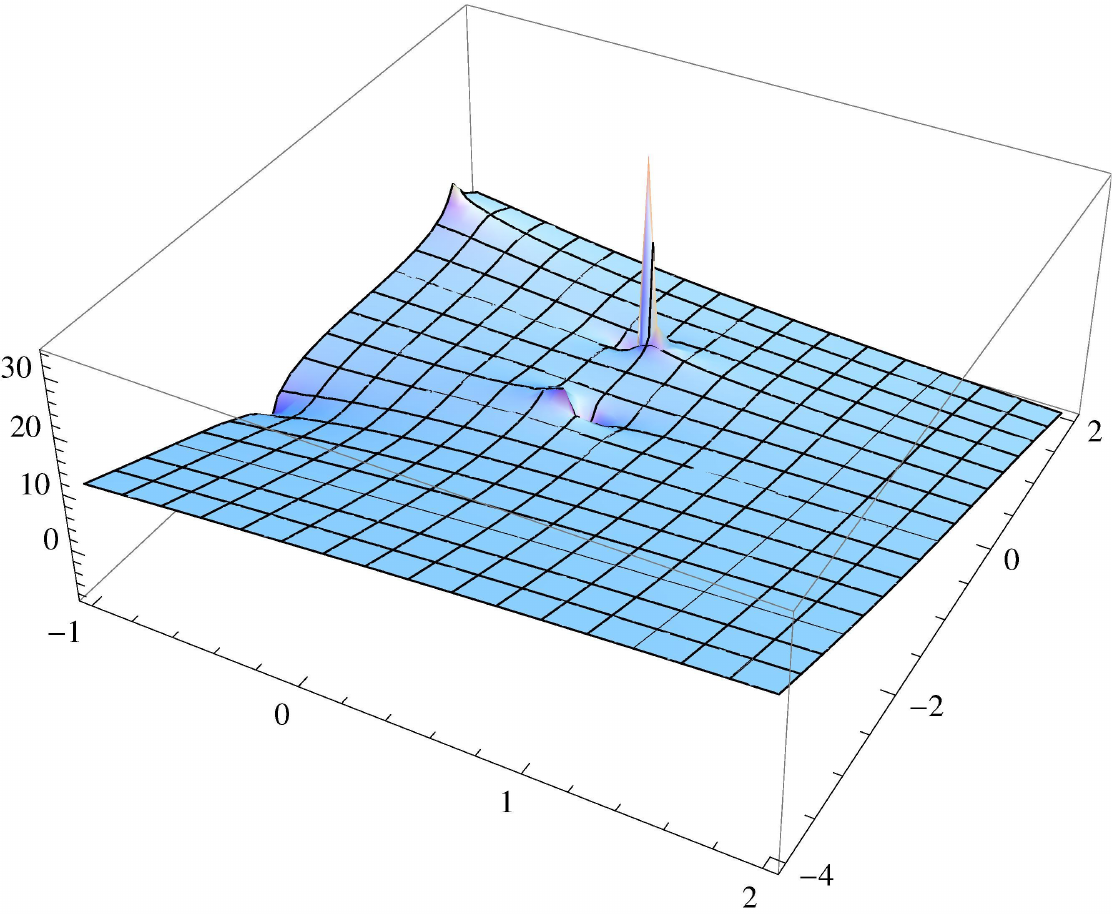}}
\hbox to \textwidth{\hbox to 0.45\textwidth{\hfil(A)\hfil}\hfil
\hbox to 0.45\textwidth{\hfil(B)\hfil}}
\hbox to \textwidth{\ing[width=0.45\textwidth]{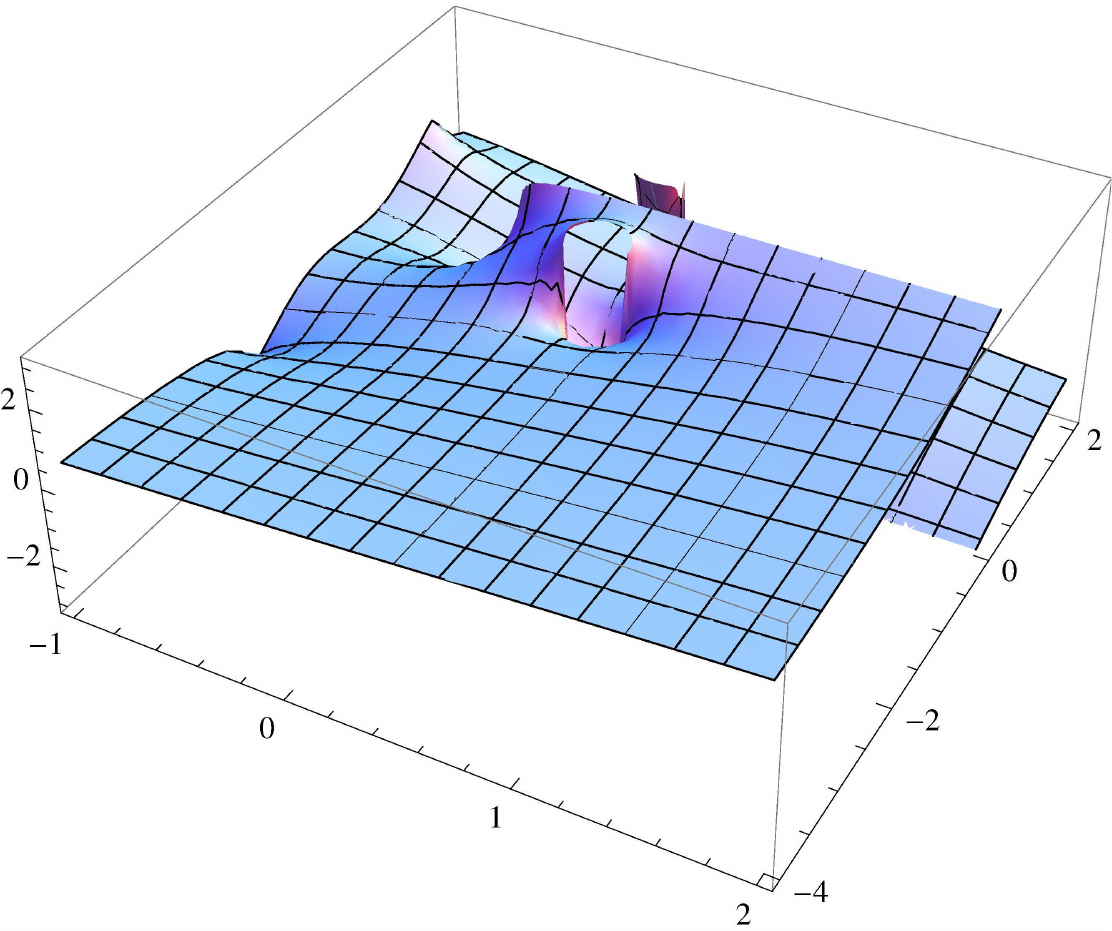}\hfil \ing[width=0.45\textwidth]{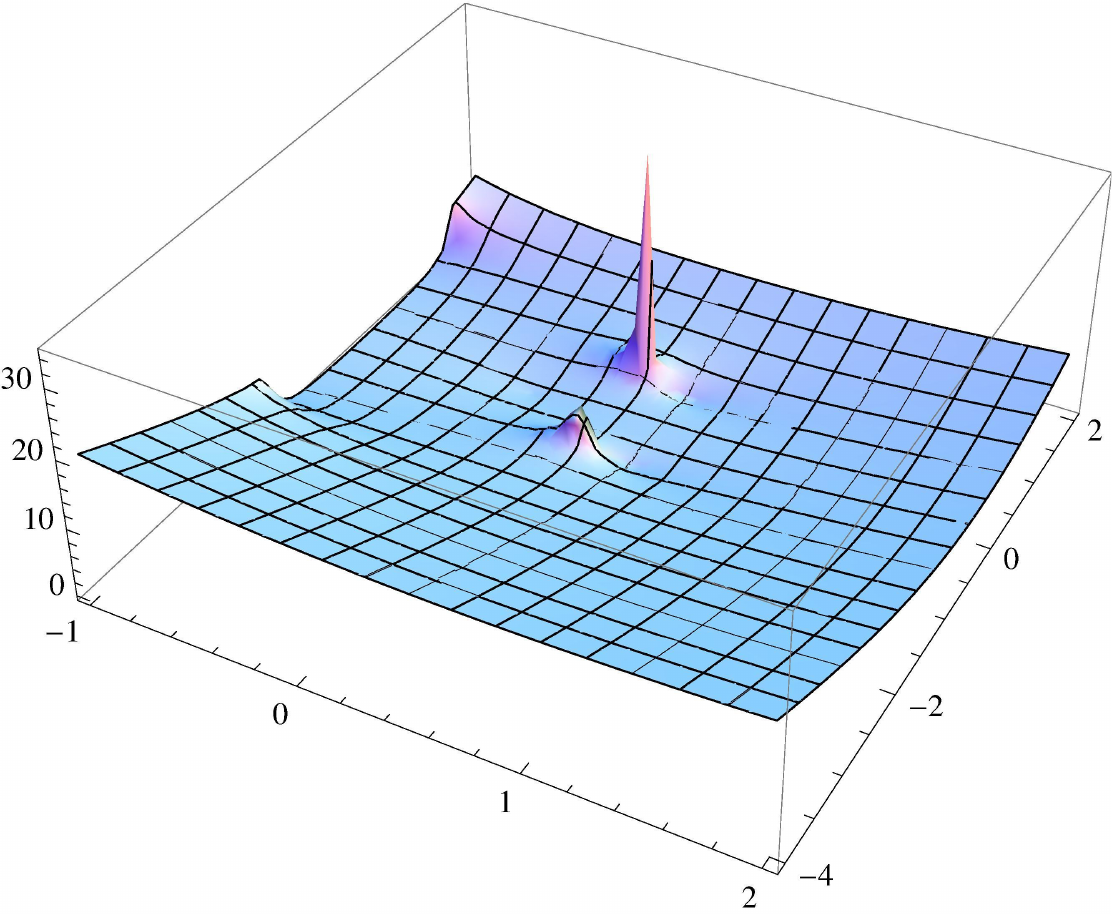}}
\hbox to \textwidth{\hbox to 0.45\textwidth{\hfil(C)\hfil}\hfil
\hbox to 0.45\textwidth{\hfil(D)\hfil}}
\caption{(A)---a 3D plot of a \f\ $\Re K(1.2,s)$,
(B)---a 3D plot of a \f\ $\Im K(1.2,s)$,
(C)---a 3D plot of a \f\ $|K(1.2,s)|$,
(D)---a 3D plot of a \f\ $\Arg K(1.2,s)$.}\label{WYK1}
\end{figure}

\begin{figure}[ht]
\hbox to \textwidth{\ing[width=0.45\textwidth]{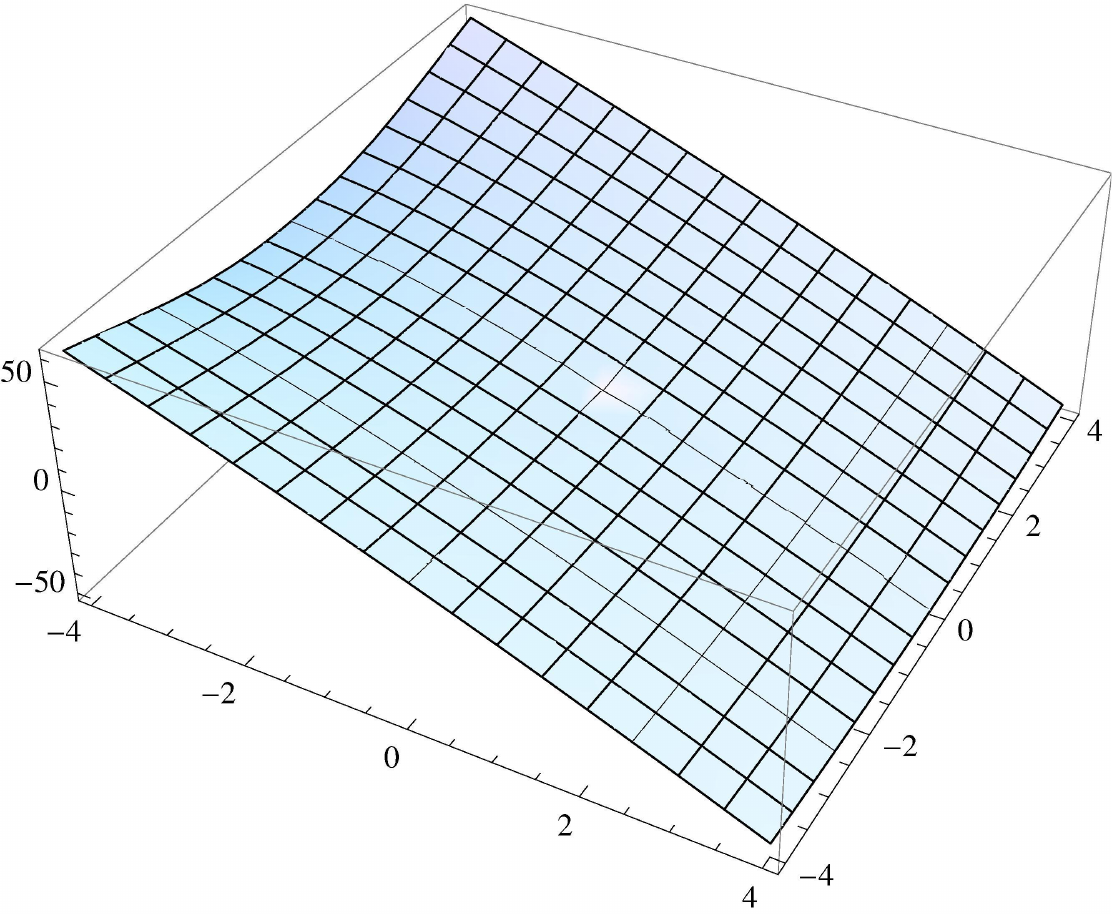}\hfil \ing[width=0.45\textwidth]{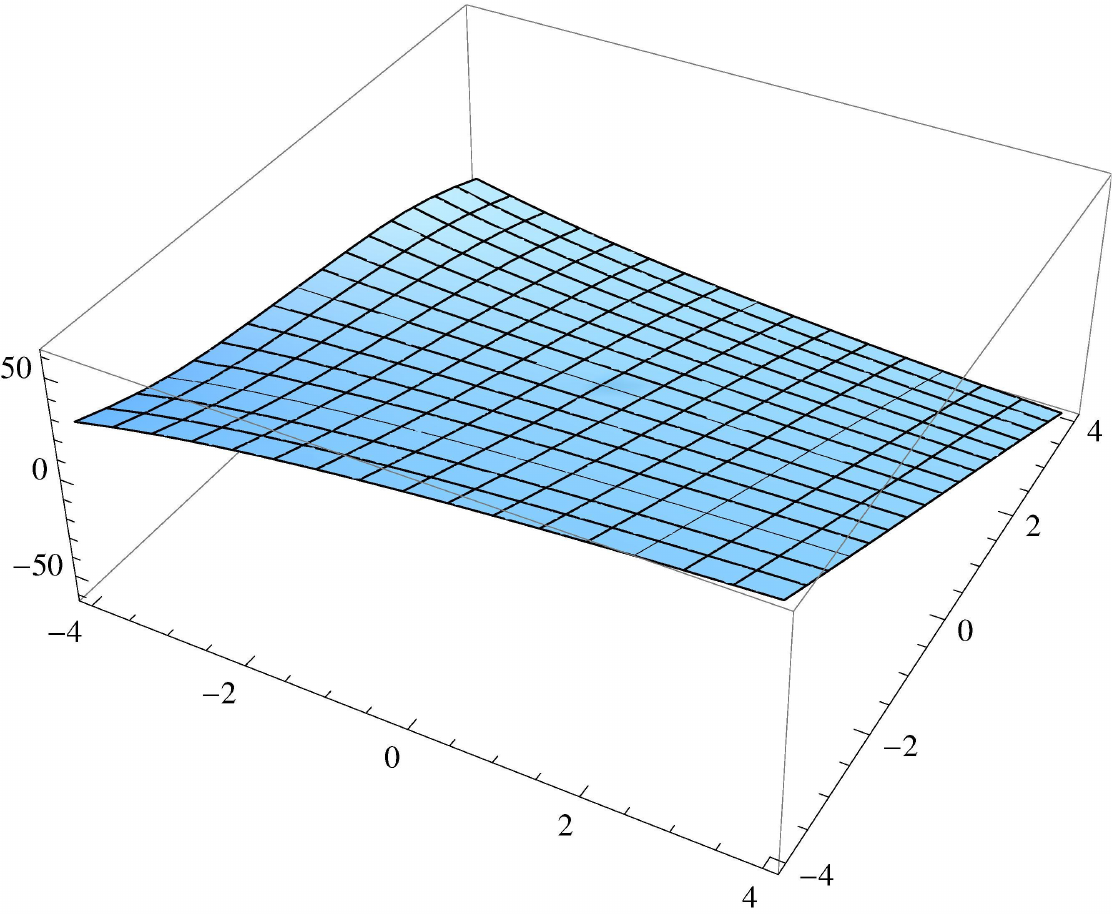}}
\hbox to \textwidth{\hbox to 0.45\textwidth{\hfil(A)\hfil}\hfil
\hbox to 0.45\textwidth{\hfil(B)\hfil}}
\hbox to \textwidth{\ing[width=0.45\textwidth]{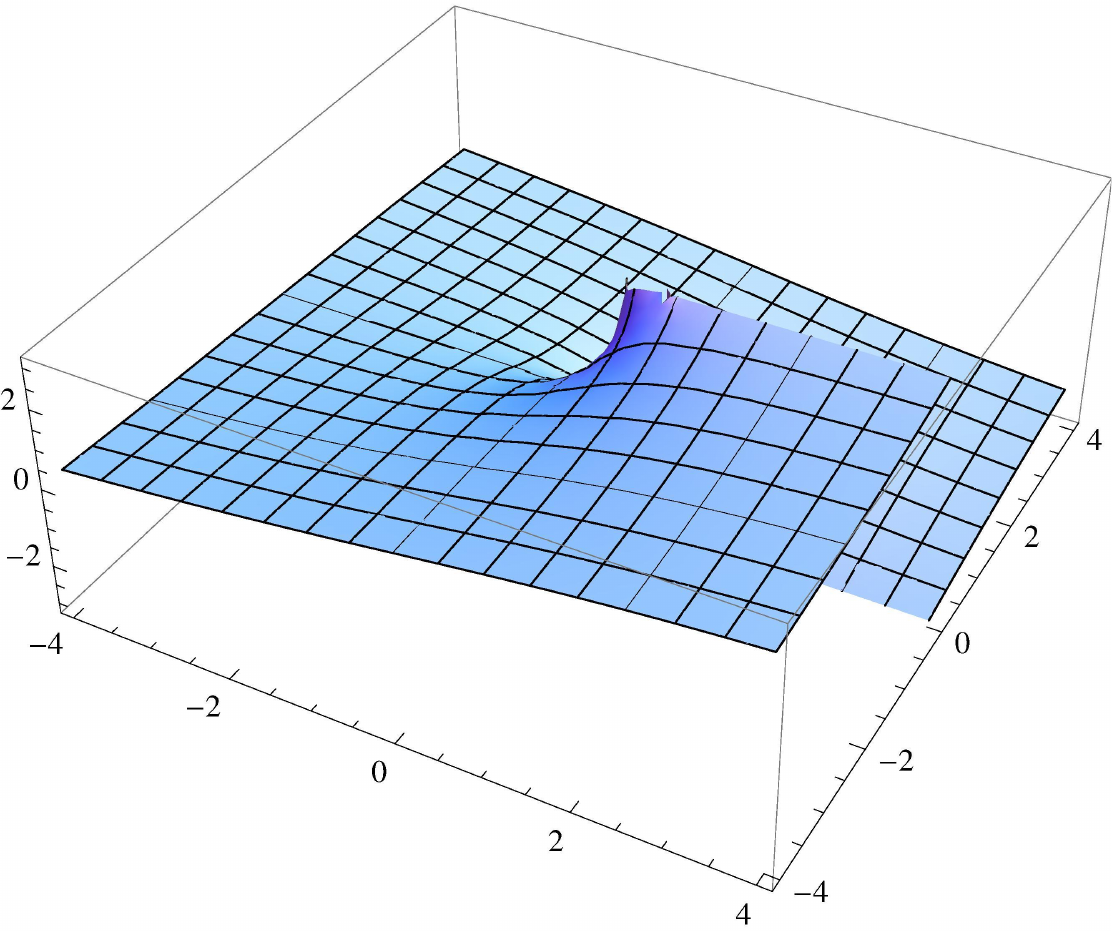}\hfil \ing[width=0.45\textwidth]{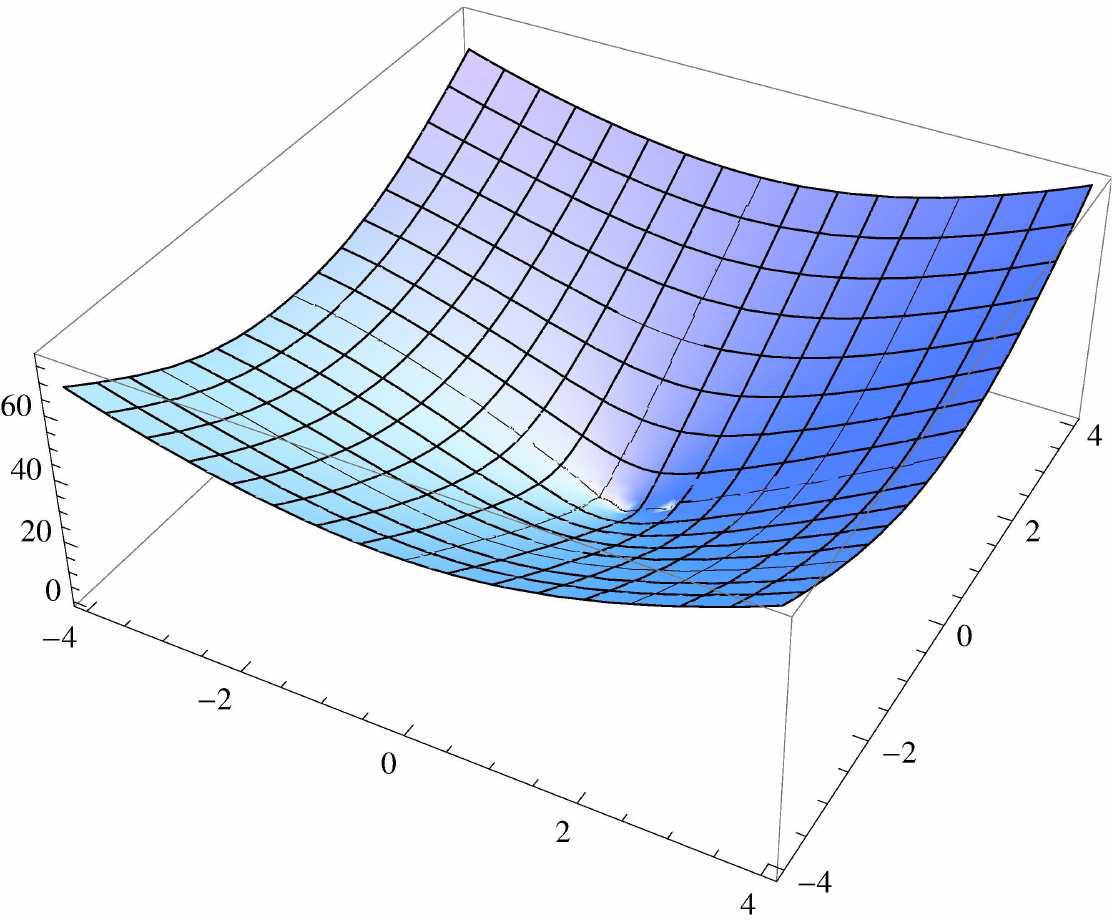}}
\hbox to \textwidth{\hbox to 0.45\textwidth{\hfil(C)\hfil}\hfil
\hbox to 0.45\textwidth{\hfil(D)\hfil}}
\caption{(A)---a 3D plot of a \f\ $\Re K(3.2,s)$,
(B)---a 3D plot of a \f\ $\Im K(3.2,s)$,
(C)---a 3D plot of a \f\ $|K(3.2,s)|$,
(D)---a 3D plot of a \f\ $\Arg K(3.2,s)$.}\label{WYK2}
\end{figure}

It is easy to see that $K(e,s)$ can be extended to a complex plane, i.e.\
$s\in \C$. In this way $h(e,s)$ has a \dv\ in  a complex domain. It means
$h(e,s)$ is an analytic \f\ for $s\in\C$. Thus we can define $h(e,s)$ as a
curve integral on a complex plane.
\beq{B.84}
h(e,s)=\int_C K(e,y)\,dy,
\end{equation}
where
$$
C=\X1\{y(t)\in\C: y(t_1)=1,\ y(t_2)=s, \ t_1<t<t_2\Y1\}.
$$

The \f\ $K(e,s)$ has two complex poles for $1<e<\sqrt{\frac{11+8\sqrt2}7}
\simeq1.783$ at
\beq{B.85}
s_{1,2}=\frac{\es}{2e}\pm i\, \frac{\sqrt{3e^2+1-4e\es}}{2e}\,.
\end{equation}
In the case of $e>\frac{11+8\sqrt2}7=e_1$ it has two real poles at
\beq{B.86}
s_{1,2}=\frac{\es\pm \sqrt{4e\es-3e^2-1}}{2e}\,.
\end{equation}
The complex poles have equal modules smaller than one. In the case of real
poles they are smaller than one with a limit for $e\to\infty$ equal one for a
bigger one, $\lim\limits_{s\to\iy}K(e,s)=\iy$. The \f\ $K(e,s)$ has four
logarithmic singularities in the complex domain, given by the formula
\beq{B.87}
\ov s_{1,2,3,4}=\frac{2e(e+\es)-1+\eta(\ve e+\es)i}
{(e+\ve\es)+\eta i}
\end{equation}
where $\ve^2=\eta^2=1$.

One finds residua for $K(e,s)$ at $s=s_k$, $k=1,2$.
\bml{B.nn}
\res K(e,s)|_{s=s_k}\\ = -\frac{\eta_k s_k}{e\sqrt{4e\es-3e^2-1}}
\X2[e\X1(\es-e\Y1)\log a(s_k)-\es\arctan b(s_k)\Y2],
\end{multline}
where
\bea{B.88}
a(s_k)&=&\frac{\X2(\X1(2e(e-\es)-1+es_k-s_k\es\Y1)^2
+\X1(\es-s_k-e\Y1)^2\Y2)^{1/2}}
{\sqrt{2e}(e-\es)^{1/2}}\,,\\
b(s_k)&=&\frac{\es-s_k-e}{2e(e-\es)-1+es_k-s_k\es}\,, \label{B.89}
\end{eqnarray}
$\eta_1=1$, $\eta_2=-1$, $\sqrt{4e\es-3e^2-1}=i\sqrt{3e^2+1-4e\es}$ for
$1<e<e_1$.

\begin{figure}[ht]
\hbox to \textwidth{\ing[width=0.45\textwidth]{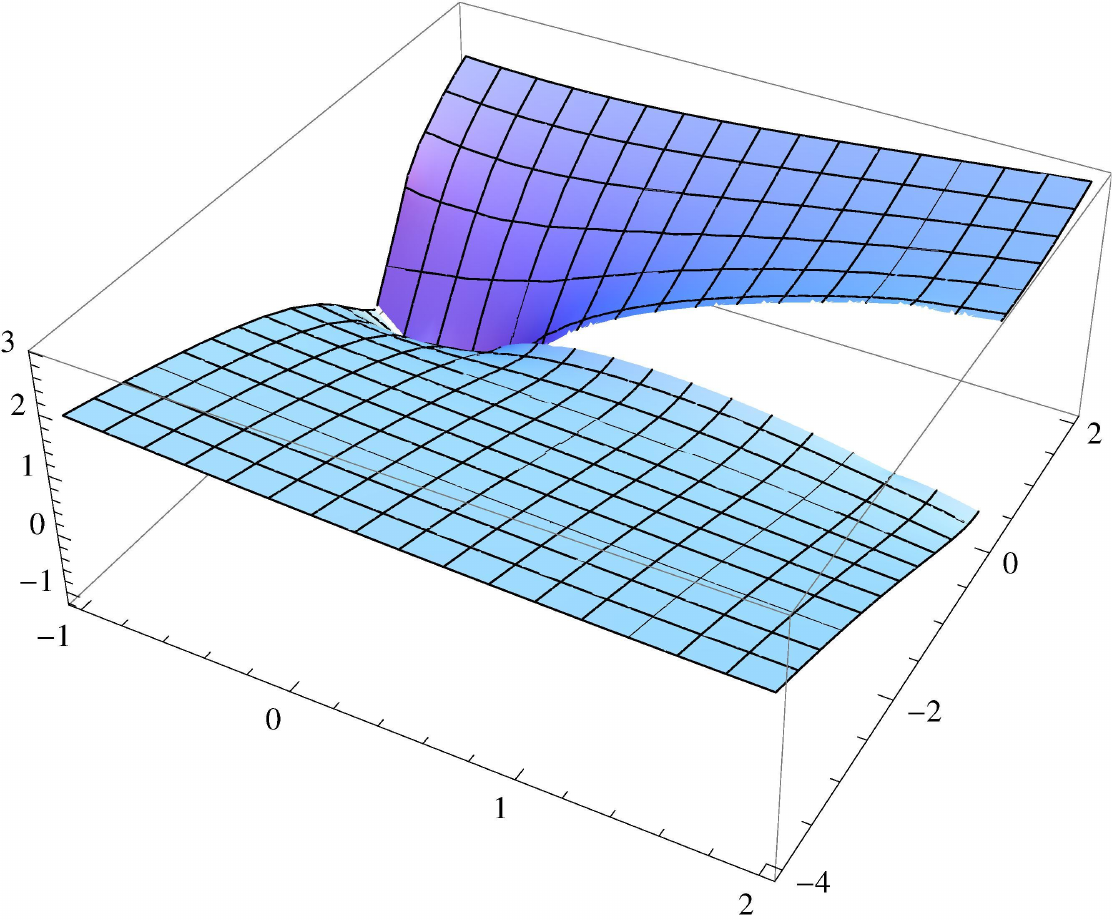}\hfil \ing[width=0.45\textwidth]{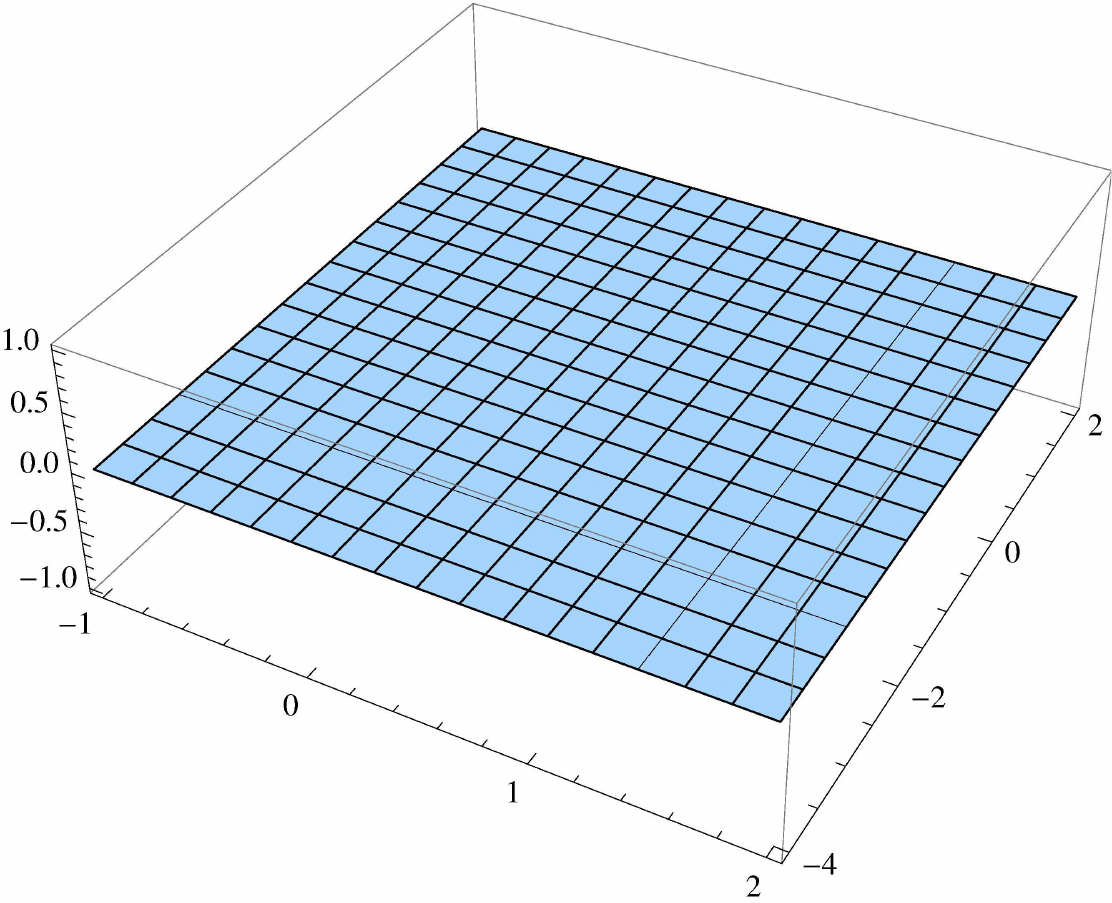}}
\hbox to \textwidth{\hbox to 0.45\textwidth{\hfil(A)\hfil}\hfil
\hbox to 0.45\textwidth{\hfil(B)\hfil}}
\hbox to \textwidth{\ing[width=0.45\textwidth]{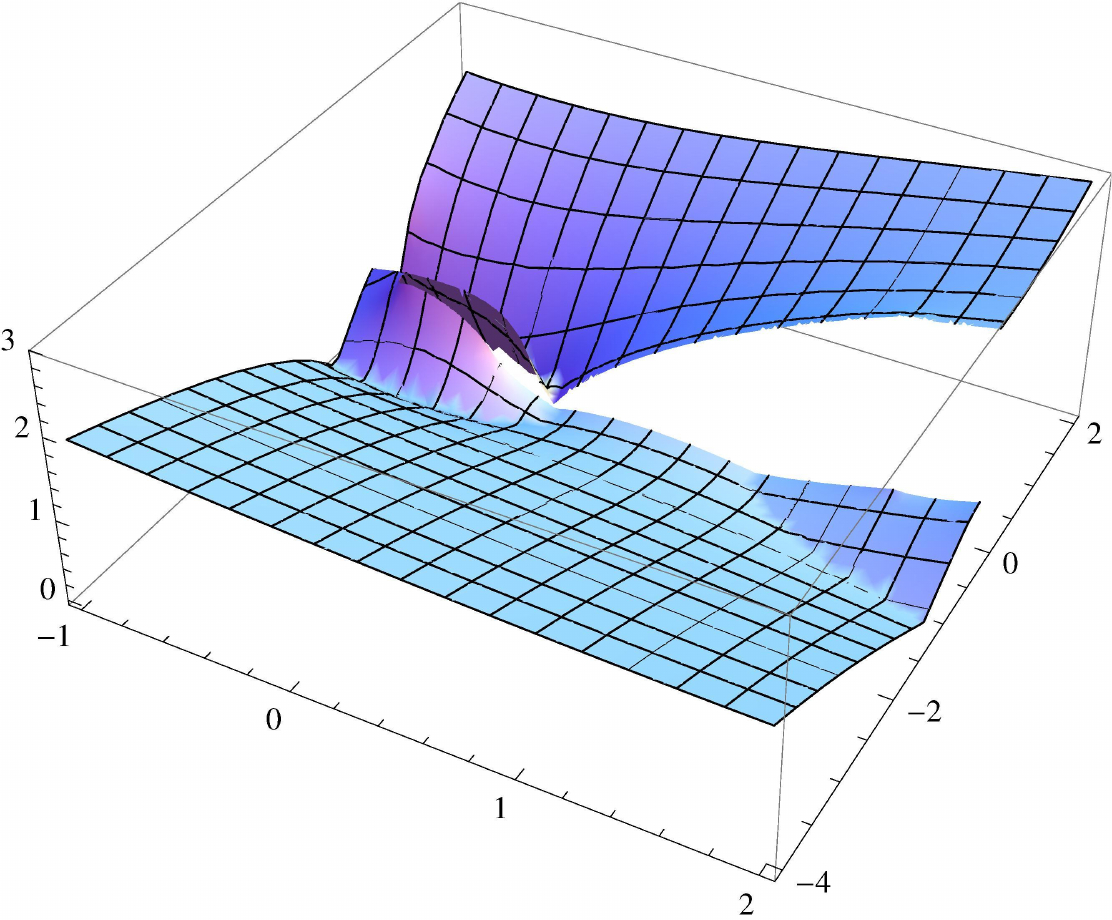}\hfil \ing[width=0.45\textwidth]{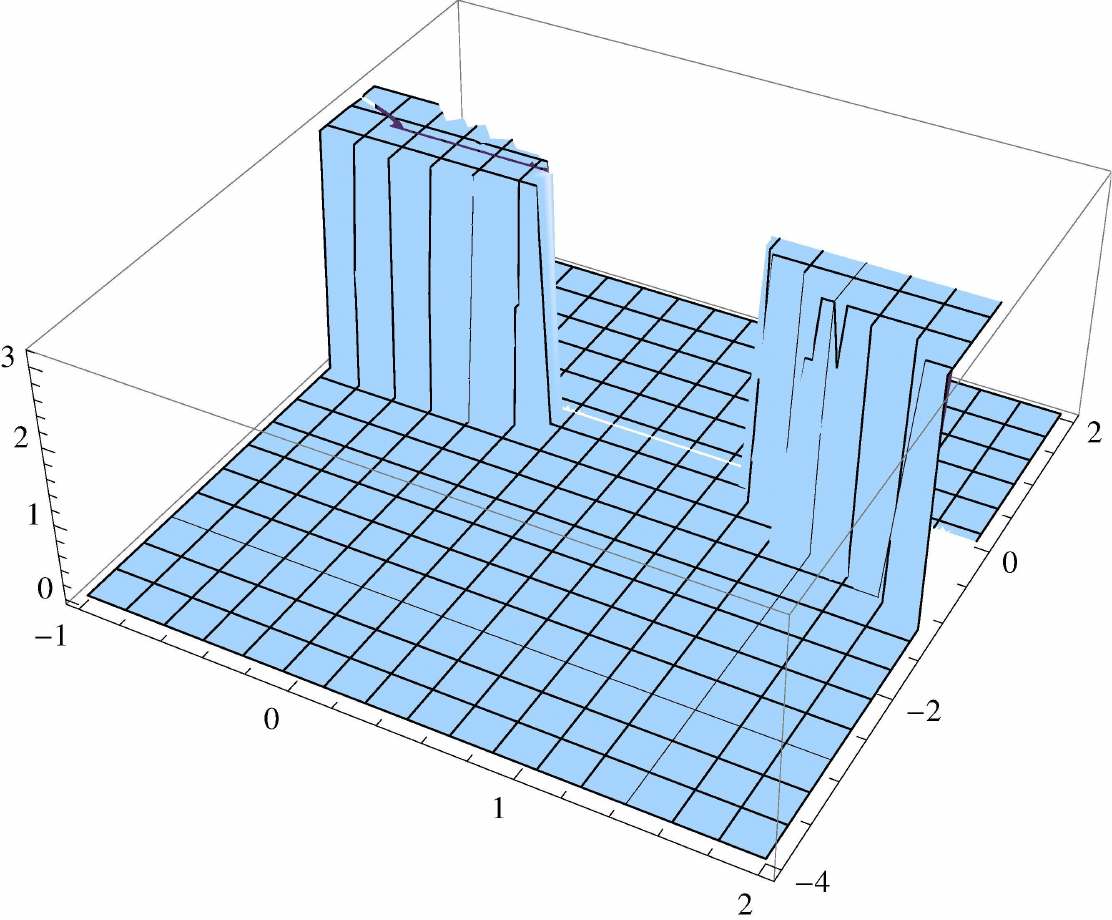}}
\hbox to \textwidth{\hbox to 0.45\textwidth{\hfil(C)\hfil}\hfil
\hbox to 0.45\textwidth{\hfil(D)\hfil}}
\caption{(A)---a plot of a \f\ $\Re \ov G(1.2,s)$,
(B)---a plot of a \f\ $\Im \ov G(1.2,s)$,
(C)---a plot of a \f\ $|\ov G(1.2,s)|$,
(D)---a plot of a \f\ $\Arg \ov G(1.2,s)$.}\label{WYKL1}
\end{figure}

\begin{figure}[ht]
\hbox to \textwidth{\ing[width=0.45\textwidth]{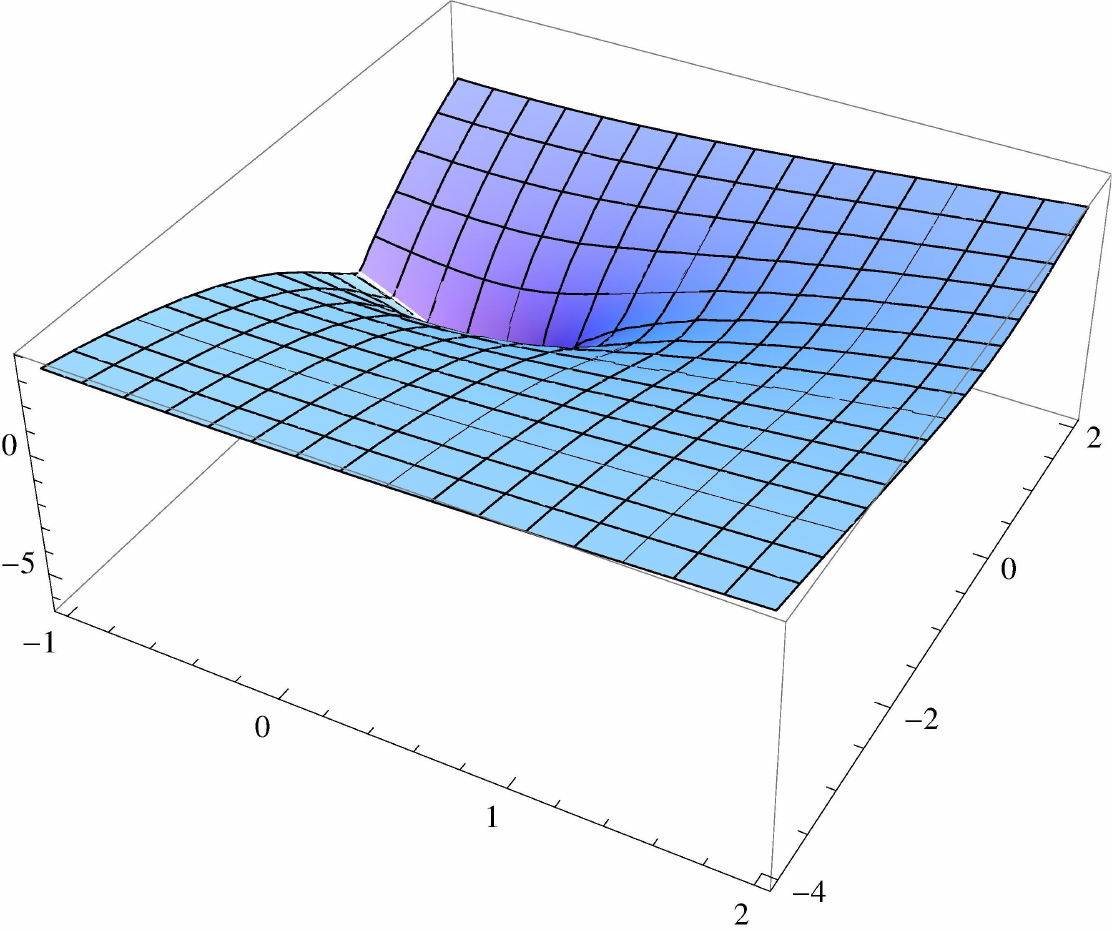}\hfil \ing[width=0.45\textwidth]{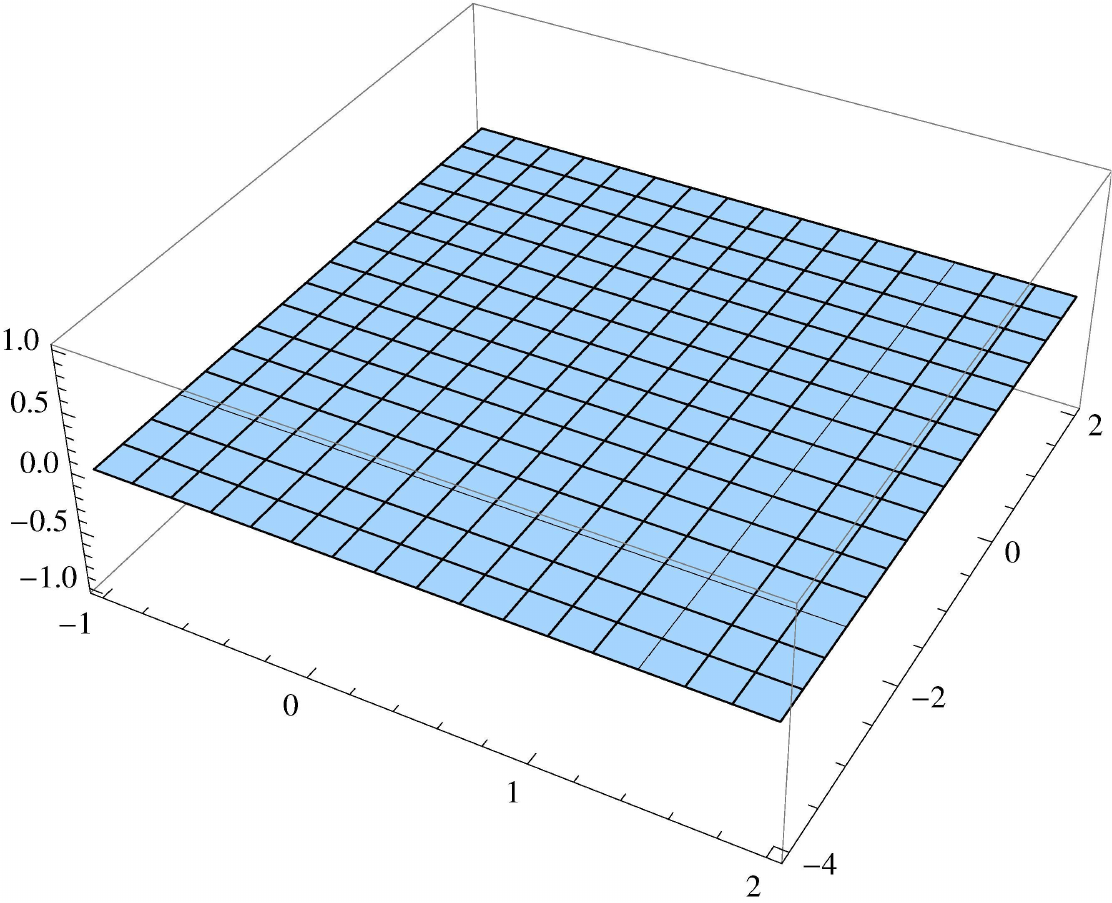}}
\hbox to \textwidth{\hbox to 0.45\textwidth{\hfil(A)\hfil}\hfil
\hbox to 0.45\textwidth{\hfil(B)\hfil}}
\hbox to \textwidth{\ing[width=0.45\textwidth]{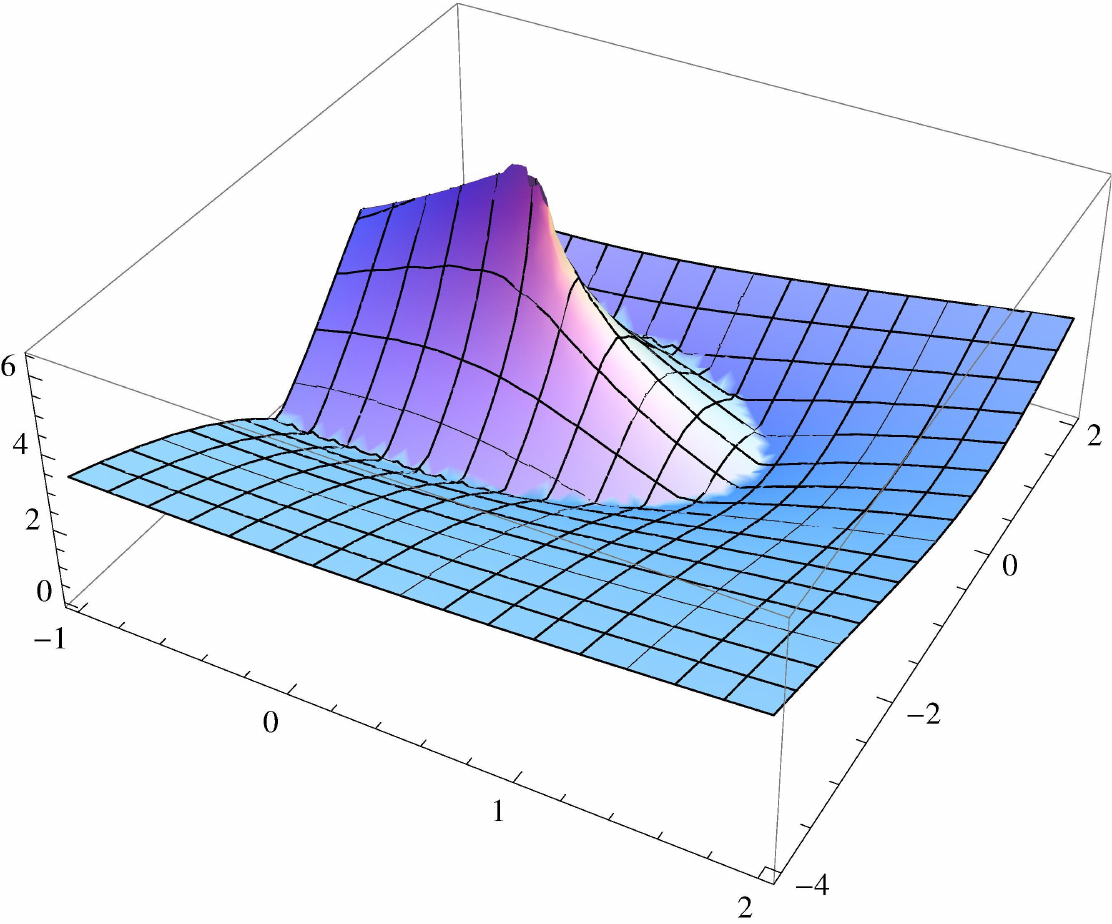}\hfil \ing[width=0.45\textwidth]{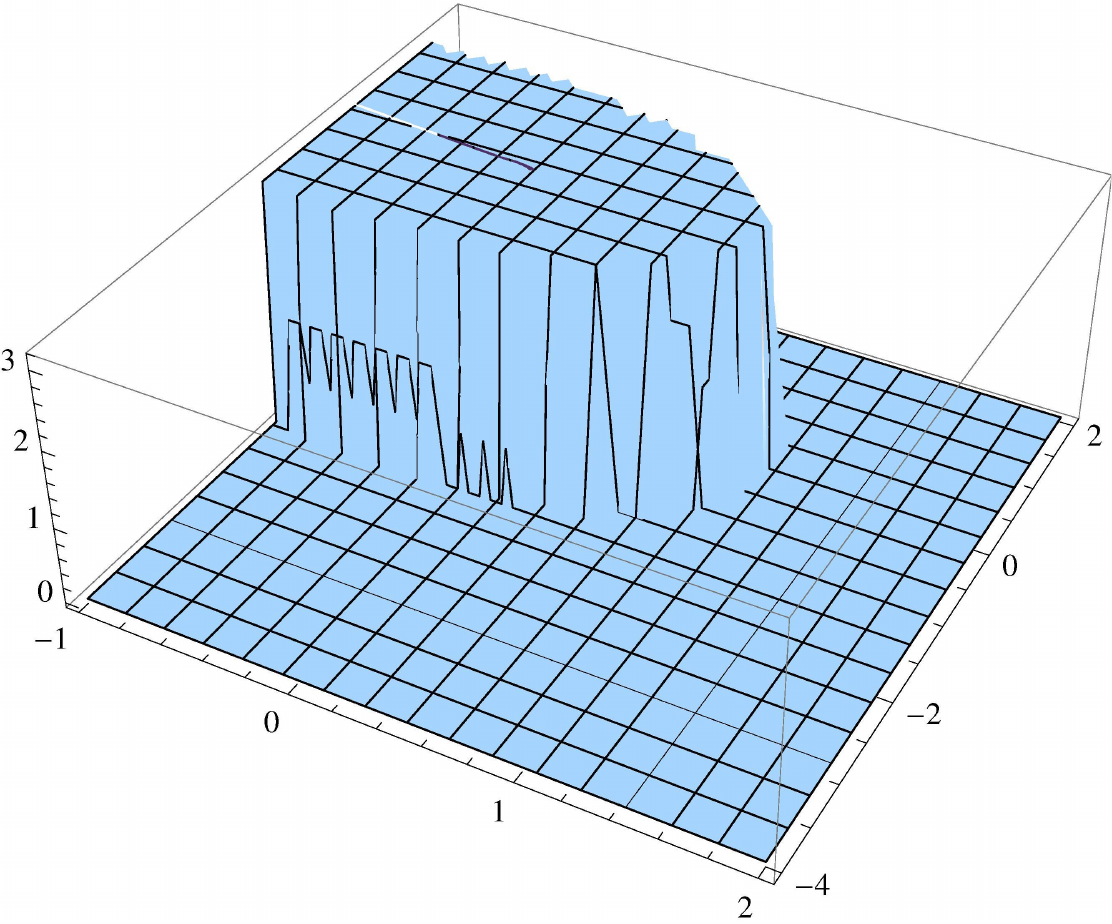}}
\hbox to \textwidth{\hbox to 0.45\textwidth{\hfil(C)\hfil}\hfil
\hbox to 0.45\textwidth{\hfil(D)\hfil}}
\caption{(A)---a plot of a \f\ $\Re \ov G(3.2,s)$,
(B)---a plot of a \f\ $\Im \ov G(3.2,s)$,
(C)---a plot of a \f\ $|\ov G(3.2,s)|$,
(D)---a plot of a \f\ $\Arg \ov G(3.2,s)$.}\label{WYKL2}
\end{figure}

Eq.~\er{B.nn} can be rewritten in a more convenient form
\beq{B.90}
\ov s_{1,2,3,4}=\frac{2e(e+\es)-1+\ve}{2e}
-i\eta\,\frac{(1-\ve)(e+\es)}{e+\ve\es}\,.
\end{equation}
For $\ve=1$ we get
\beq{B.91}
|\ov s_{1,2}|=\frac{\X1(2e^2+2e\es-1\Y1)^{1/2}}{\sqrt{2e(e+\es)}}\,.
\end{equation}
For $\ve=-1$ we get
\beq{B.92}
|\ov s_{3,4}|=\frac{\X1(30e^2+26e\es-13\Y1)^{1/2}}{2e(e-\es)}\,.
\end{equation}
The modulus for $s_{1,2}$ equals
\beq{B.93}
|s_{1,2}|=\sqrt{1-\frac{\es}e}\,.
\end{equation}

In order to define $h(e,s)$ as a curve integral we should choose $C$ very
carefully to avoid singularities. For poles and for first two ($\ve=1$)
logarithmic singularities it is enough to choose $|y(t)|>1$. However, for the
second two logarithmic singularities ($\ve=-1$) it is not enough. It is
necessary to suppose $\left|\Im y(t)\right|<\frac{2(e+\es)}{e-\es}$. Thus a
safe $C$ is defined as
\beq{B.94}
|y(t)|>1, \qquad \left|\Im y(t)\right|<\frac{2e(e+\es)}{e-\es}\,.
\end{equation}
There are different possibilities, i.e. $\Im y(t)>\frac{2e(e+\es)}{e-\es}$
or $\Im y(t)<\frac{2e(e+\es)}{e-\es}$\,. Due to this $h(e,s)$ has several
cuts on a complex plane. In order to define $h(e,s)$ for $|s|<1$ it is
necessary to use residuum theorem (for poles) and to work very carefully with
logarithmic singularities inside a unit circle.

On Fig.~\ref{WYK1} we give 3D plots of $\Re K(e,s)$, $\Im K(e,s)$, $|K(e,s)|$
and $\Arg K(e,s)$ for $e=1.2<e_1$.
On Fig.~\ref{WYK2} we give 3D plots of $\Re K(e,s)$, $\Im K(e,s)$, $|K(e,s)|$
and $\Arg K(e,s)$ for $e=3.2>e_1$, $s=x+iy$. On Fig.~\ref{WYK2}C $|K(3.2,s)|$
shows a cusp-like singularity.

Let us consider a \f
\bml{B.96}
\ov G(e,s)=\Re\X3[\Li_2
\X2(\frac{1-es-s\es+i(e-s+\es)}{2e(e+\es)}\Y2)\\
{}-\Li_2\X2(\frac{1-es+s\es+i(\es-s-e)}{2e(e-\es)}\Y2)\Y3].
\end{multline}
It is easy to see that
\beq{B.97}
\ov G(e,s)=h(e,s)+H(e).
\end{equation}
For our purposes it is even more convenient to consider $\ov G(e,s)$ than
$h(e,s)$. Thus we plot on Fig.~\ref{WYKL1} and on Fig.~\ref{WYKL2} $\Re(\ov
G(e,s))$, $\Im(\ov G(e,s))$, $|\ov G(e,s)|$, $\Arg(\ov G(e,s))$, $s=x+iy$, in
such a way that on Fig.~\ref{WYKL1} we give plots for $e=1.2<e_1$ and on
Fig.~\ref{WYKL2} for $e=3.2>e_1$. Plots for $e=1.2$ and for $e=3.2$ look
qualitatively different as someone can suspect.

Further properties of this special \f\ will
be examined elsewhere.

In order to use our results for $\ov F(\vf)$ and $F_p(\vf)$ including properties
of our new special \f\ it is enough to find from Eqs \er{B.1} and \er{B.15}
$\vf(t)$ and $\vf_p(t)$ using some numerical procedure.
These \so s in the regions Eqs \er{B.14} and \er{B.16}
describe a motion under an \an\ \ac. They give us together with an \e\ of an
orbit a full \e\ of motion under an \an\ \ac\ for distances ${}>20$\,AU up
$10^3$\,AU with higher precision than in Sections 5~and~6. The results can
give more precise positions and a time of a position to be comparable with
VLBI measurements. Positions and times should be measured in modern reference
frames.

\def\thefigure{C.\arabic{figure}}
\setcounter{figure}0
\section*{Appendix C}
In this appendix we give a programme to calculate the function $f(x,y)$. It
is a listing written in Mathematica 7. {\tt xx}, {\tt yy}, {\tt ss} define
tables with the data of Anderson et al. (see Refs~\cite1, \cite2, \cite3).

{\advance\baselineskip by-0.15pt
\begin{verbatim}
xx=Table[0,{j,1,19}];
xx[[1]]=5.80;
xx[[2]]=9.39;
xx[[3]]=12.16;
xx[[4]]=14.00;
xx[[5]]=16.83;
xx[[6]]=18.90;
xx[[7]]=22.25;
xx[[8]]=23.30;
xx[[9]]=26.60;
xx[[10]]=29.50;
xx[[11]]=26.36;
xx[[12]]=28.88;
xx[[13]]=31.64;
xx[[14]]=34.34;
xx[[15]]=35.58;
xx[[16]]=37.33;
xx[[17]]=40.59;
xx[[18]]=43.20;
xx[[19]]=45.70;

yy=Table[0,{j,1,19}];
yy[[1]]=0.69;
yy[[2]]=1.56;
yy[[3]]=6.28;
yy[[4]]=8.05;
yy[[5]]=8.15;
yy[[6]]=9.03;
yy[[7]]=8.13;
yy[[8]]=8.98;
yy[[9]]=8.56;
yy[[10]]=8.33;
yy[[11]]=8.68;
yy[[12]]=8.88;
yy[[13]]=8.59;
yy[[14]]=8.43;
yy[[15]]=7.67;
yy[[16]]=8.43;
yy[[17]]=7.45;
yy[[18]]=8.09;
yy[[19]]=8.24;

ss=Table[0,{j,1,19}];
ss[[1]]=1.48;
ss[[2]]=6.85;
ss[[3]]=1.77;
ss[[4]]=2.16;
ss[[5]]=0.75;
ss[[6]]=0.41;
ss[[7]]=0.69;
ss[[8]]=0.30;
ss[[9]]=0.15;
ss[[10]]=0.30;
ss[[11]]=0.50;
ss[[12]]=0.27;
ss[[13]]=0.32;
ss[[14]]=0.55;
ss[[15]]=0.23;
ss[[16]]=0.37;
ss[[17]]=0.46;
ss[[18]]=0.20;
ss[[19]]=0.20;
g[z_]:=1/z*(Exp[z]-1-z)
f[x_,a_,b_]:=b*g[z]/.z->a*x^2
mm[a_,b_]:=FullSimplify[1/2*Sum[(f[xx[[j]],a,b]
           -yy[[j]])^2*ss[[j]]^(-2),{j,1,19}]]
nn1[x_,y_]:=FullSimplify[mm[a,b]/.{a->x-0.05938468430992681`,
            b->y-8.521513551131733`}]
nn[x_,y_]:=FullSimplify[2*(nn1[x,y]-nn1[0,0])]
nn1[0,0]
2*nn1[0,0]
nn[x,y]
\end{verbatim}}

$$
\bal
{}& f(x,y)=-46.3557\\
&\kern-5.3pt+25.0\X3(8.24-\frac{6.56281\cdot 10^{-58} (8.9757\cdot 10^{55}+E^{2088.49 x}-1.52374\cdot 10^{57} x) (-8.52151+y)}{-0.0593847+x}\Y3)^2\\
&\kern-5.3pt+25.0\X3(8.09-\frac{3.96169\cdot 10^{-52} (1.48545\cdot 10^{50}+E^{1866.24 x}-2.52418\cdot 10^{51} x) (-8.52151+y)}{-0.0593847+x}\Y3)^2\\
&\kern-5.3pt+4.7259 \X3(7.45-\frac{1.95961\cdot 10^{-46} (2.99946\cdot 10^{44}+E^{1647.55 x}-5.10305\cdot 10^{45} x) (-8.52151+y)}{-0.0593847+x}\Y3)^2\\
&\kern-5.3pt+7.3046 \X3(8.43-\frac{8.24441\cdot 10^{-40} (7.11598\cdot 10^{37}+E^{1393.53 x}-1.21294\cdot 10^{39} x) (-8.52151+y)}{-0.0593847+x}\Y3)^2\\
&\kern-5.3pt+18.9036 \X3(7.67-\frac{1.77227\cdot 10^{-36} (3.30619\cdot 10^{34}+E^{1265.94 x}-5.64247\cdot 10^{35} x) (-8.52151+y)}{-0.0593847+x}\Y3)^2\\
&\kern-5.3pt+3.30579 \X3(8.43-\frac{3.27637\cdot 10^{-34} (1.78663\cdot 10^{32}+E^{1179.24 x}-3.05215\cdot 10^{33} x) (-8.52151+y)}{-0.0593847+x}\Y3)^2\\
&\kern-5.3pt+9.76563 \X3(8.59-\frac{1.517\cdot 10^{-29} (3.84877\cdot 10^{27}+E^{1001.09 x}-6.59197\cdot 10^{28} x) (-8.52151+y)}{-0.0593847+x}\Y3)^2\\
&\kern-5.3pt+11.1111 \X3(8.33-\frac{4.13261\cdot 10^{-26} (1.40917\cdot 10^{24}+E^{870.25 x}-2.41978\cdot 10^{25} x) (-8.52151+y)}{-0.0593847+x}\Y3)^2\\
&\kern-5.3pt+13.7174 \X3(8.88-\frac{3.69977\cdot 10^{-25} (1.57268\cdot 10^{23}+E^{834.054 x}-2.70287\cdot 10^{24} x) (-8.52151+y)}{-0.0593847+x}\Y3)^2\\
&\kern-5.3pt+44.4444 \X3(8.56-\frac{7.97907\cdot 10^{-22} (7.26543\cdot 10^{19}+E^{707.56 x}-1.25328\cdot 10^{21} x) (-8.52151+y)}{-0.0593847+x}\Y3)^2\\
&\kern-5.3pt+4.0\X3(8.68-\frac{1.72835\cdot 10^{-21} (3.35265\cdot 10^{19}+E^{694.85 x}-5.78586\cdot 10^{20} x) (-8.52151+y)}{-0.0593847+x}\Y3)^2\\
&\kern-5.3pt+11.1111 \X3(8.98-\frac{1.83618\cdot 10^{-17} (3.13382\cdot 10^{15}+E^{542.89 x}-5.44608\cdot 10^{16} x) (-8.52151+y)}{-0.0593847+x}\Y3)^2\\
&\kern-5.3pt+2.1004 \X3(8.13-\frac{3.44715\cdot 10^{-16} (1.66412\cdot 10^{14}+E^{495.063 x}-2.90095\cdot 10^{15} x) (-8.52151+y)}{-0.0593847+x}\Y3)^2\\
&\kern-5.3pt+5.94884 \X3(9.03-\frac{1.71583\cdot 10^{-12} (3.29784\cdot 10^{10}+E^{357.21 x}-5.8281\cdot 10^{11} x) (-8.52151+y)}{-0.0593847+x}\Y3)^2\\
&\kern-5.3pt+1.77778 \X3(8.15-\frac{1.74871\cdot 10^{-10} (3.19402\cdot 10^{8}+E^{283.249 x}-5.7185\cdot 10^{9} x) (-8.52151+y)}{-0.0593847+x}\Y3)^2\\
&\kern-5.3pt+0.214335 \X3(8.05-\frac{4.49591\cdot 10^{-8} (1.20738\cdot 10^{6}+E^{196.0x}-2.22424\cdot 10^{7} x) (-8.52151+y)}{-0.0593847+x}\Y3)^2
\eal
$$
$$
\bal
&\kern-5.3pt+0.319193 \X3(6.28-\frac{1.039\cdot 10^{-6} (50646.8+E^{147.866 x}-962468.0x) (-8.52151+y)}{-0.0593847+x}\Y3)^2\\
&\kern-5.3pt+0.0213117 \X3(1.56-\frac{0.0000603492 (796.087+E^{88.1721 x}-16570.2 x) (-8.52151+y)}{-0.0593847+x}\Y3)^2\\
&\kern-5.3pt+0.456538 \X3(0.69-\frac{0.00403231 (7.35514+E^{33.64 x}-247.997 x) (-8.52151+y)}{-0.0593847+x}\Y3)^2
\eal
$$

\def\ebf#1#2{e^{#1B(r)+#2\vF(r)}}
\def\nd{(n+2)}
\def\eb#1{e^{#1B(r)}}
\def\ef#1{e^{#1\vF(r)}}
\def\<#1>{\left\langle #1\right\rangle}
\def\qe{quintessence}
\def\ep{\X3(\exp\X2(\frac{1+\sqrt5}4\,r^{(1+\sqrt5)/2}\Y2)-1\Y3)}
\def\epm{\X2(\exp\X1(\tfrac{1+\sqrt5}4\,r^{(1+\sqrt5)/2}\Y1)-1\Y2)}
\def\cosec{\mathop{\rm cosec}\nolimits}

\def\theequation{D.\arabic{equation}}
\setcounter{equation}0
\def\thefigure{D.\arabic{figure}}
\setcounter{figure}0
\section*{Appendix D}
In this appendix we consider in more details spherically \s\ and stationary
case of the \eu\nos\ Jordan--Thiry \e s announced in Appendix~A. We consider
Eqs \er{A.124}--\er{A.136} getting
\bg{D.1}
\frac1{r^2}\,\pz{}r\X1(r\X1(1-e^{-2B(r)}\Y1)\Y1)=\k\X3(2\ov\rho(r)e^{-(n+2)\vF(r)}
+\ov M\X2(\pz\vF r\Y2)^2 +pe^{n\vF(r)}\X2(\frac{e^{2\vF(r)}}{n+2}-\frac1n
\Y2)\Y3)\\
-\frac1{r^2}\,e^{2B(r)}+\frac1{r^2}+\frac2r\,\pz{A(r)}r \hskip200pt \nonumber\\
\hskip40pt {}=\k\X3(\ov M\X2(\pz\vF r\Y2)^2
-pe^{2B(r)}e^{n\vF(r)}\X2(\frac{e^{2\vF(r)}}{n+2}
-\frac1n \Y2)-\ov\rho(r)\eb2 e^{-(n+2)\vF(r)}\Y3)\label{D.2}\\
\pz{^2A(r)}{r^2}+\X2(\pz{A(r)}r\Y2)^2+\frac1r\X2(\pz{A(r)}r\Y2)
-\pz{A(r)}r\,\pz{B(r)}r-\frac1r\,\pz{B(r)}r\hskip100pt \nonumber \\
\hskip40pt{}=\k\X3(-\ov M\X2(\pz{\vF(r)} r\Y2)^2
-pe^{2B(r)}e^{n\vF(r)}\X2(\frac{e^{2\vF(r)}}{n+2}-\frac1n\Y2)
-\ov\rho e^{2B(r)}e^{-(n+2)\vF(r)}\Y3)\label{D.3}\\
\frac{2\ov M}{r^2}\,e^{-(A(r)+B(r))}\pz{}r\X2(r^2e^{A(r)-B(r)}\pz{\vF(r)}r\Y2)
\hskip100pt\nonumber\\
\hskip100pt{}-pe^{n\vF(r)}(e^{2\vF(r)}-1)-(n+2)e^{-(n+2)\vF(r)}
\ov \rho(r)=0, \label{D.4}
\end{gather}
where $\k=\frac{8\pi G_N}{c^4}$, $p=\frac{\la_{\rm co}n(n+2)}2$\,.
This is a system of \f-differential \e s.

From Eq.\ \er{D.1} we can obtain
\bml{D.5}
\ov\rho(r)=\frac12 \X3[-\ov M\X2(\pz{\vF(r)}r\Y2)^2\eb{-2}-pe^{n\vF(r)}
\X2(\frac{e^{2\vF(r)}}{n+2}-\frac1n\Y2)\\
{}+\frac1\k \X2[\frac1{r^2}\X1(1-e^{-2B(r)}\Y1)+\frac2{r}\,e^{-2B(r)}
\pz {B(r)}r\Y2]\Y3]
e^{(n+2)\vF(r)}.
\end{multline}
We put this result to Eqs\ \er{D.2}--\er{D.4}. Simultaneously we use such system of
units that $p=1$ (this unit of length is $\xi\times10$\,Mpc, where $\xi$ is
of order one and contains some uncertainty in a value of a \co ical \ct).
Moreover, we change also a unit of time \st $c=1$ and eventually a unit of
mass in such a way that $\k=\frac{8\pi G_N}{c^4}=1$.

In this way one gets
\bg{D.6}
\frac1r\,\pz{B(r)}r-\frac1{2r^2}\,e^{2B(r)}+\frac1{2r^2}+\frac2r\,\pz{A(r)}r=
\frac{3\ov M}2\X2(\pz{\vF(r)} r\Y2)^2 -\frac12\,e^{2B(r)}e^{n\vF(r)}\X2(\frac{e^{2\vF(r)}}{n+2}
-\frac1n \Y2)\\
\pz{^2A(r)}{r^2}+\X2(\pz{A(r)}r\Y2)^2+\frac1r\,\pz{A(r)}r
-\pz{A(r)}r\,\pz{B(r)}r\hskip120pt\nonumber \\
\hskip40pt{}=-\frac{\ov M}2\X2(\pz{\vF(r)} r\Y2)^2
-\frac12\, e^{2B(r)}e^{n\vF(r)}\X2(\frac{e^{2\vF(r)}}{n+2}-\frac1n\Y2)
-\frac1{2r^2}(e^{2B(r)}-1)\label{D.7}\\
-2\ov M e^{-2B(r)}\X3(\pz{^2\vF(r)}{r^2}+\frac2r\,\pz{\vF(r)}r
+\pz{\vF(r)}r\X2(\pz{A(r)}r-\pz{B(r)}r\Y2)\Y3)+\ov M
\X2(\pz{\vF(r)} r\Y2)^2 (n+2)e^{-2B(r)}\hskip10pt\nonumber\\
\hskip20pt{}+\frac12\,e^{n\vF(r)}\X2(3e^{2\vF(r)}-\frac{3n+2}n\Y2)
-(n+2)\X3(\frac1{2r^2}+e^{-2B(r)}
\X2(\frac1r\,\pz{B(r)}r-\frac1{2r^2}\Y2)\Y3)=0.
\label{D.8}
\end{gather}

The system of ordinary differential Eqs \er{D.6}--\er{D.8} seems to be of
fifth order (it means we need five integration \ct s). Moreover, this is
the case as we see below. First of all we calculate
\beq{D.9}
\pz{A(r)}r = \frac r2\X3(\frac{3\ov M}2\X2(\pz{\vF(r)}r\Y2)^2-\frac12\,e^{2B(r)}e^{n\vF(r)}
\X2(\frac{e^{2\vF(r)}}{n+2}-\frac1n\Y2)+\frac1{r^2}\,e^{2B(r)}-\frac1{r^2}\Y3)
-\frac12\,\pz{B(r)}r
\end{equation}
and
\bml{D.10}
\pz{^2B(r)}{r^2}=\frac14\X3(e^{2B(r)}\X2(-\frac{e^{n\vF(r)}}{n}
+\frac{\ef{(n+2)}}{n+2}-\frac{1}{r^2}\Y2)+\frac{1}{r^2}\\
{}+\frac{2\eb2\X1(\ef{(n+2)}nr^2+(n+2)(n-\ef nr^2)\Y1)\pz{B(r)}r}{n(n+2)r}\\
{}-2\,\pz{^2B(r)}{r^2}+\pz{\vF(r)}r \, \X2(\ebf2n(-1+\ef2)r
+3\ov M\,\pz{\vf(r)}r + 6\ov Mr\,\pz{\vF(r)}r\Y2)\Y3).
\end{multline}
Let us note that $\pz{^2\vF(r)}{r^2}(r)$ and $\pz{^2B(r)}{r^2}$
appear in \er{D.10}.

Now we substitute \er{D.9} into \er{D.7} and \er{D.9}--\er{D.10} into
\er{D.8}. After this substitution, $\pz{^2\vF(r)}{r^2}(r)$ appears in the \e\
derived from~\er{D.7}. Using the \e\ derived from \er{D.8} after a
substitution we get the following result (calculating $\pz{^2B(r)}{r^2}$)
\bml{D.11}
\pz{^2B(r)}{r^2}=\frac12 \X4(
3\X2(\pz{B(r)}r\Y2)^2 + 3\,\pz{B(r)}r\,\pz{\vF(r)}r
\X2(-\eb2\nd+(-2+3\eb2)\ov M\,\pz{\vF(r)}r\Y2)\\
{}+\frac1{4n^2\nd^2r^2}\X4(2\ebf2{\nd}(7+\eb2)n^2\nd r^2+\ebf4{2\nd}n^2r^4\\
{}+\nd\X2((9-10\eb2+\eb4)n^2\nd-2\ebf2n(7+\eb2)n\nd r^2\\
{}-\ebf4{2n}\X1(-2+(-1+2\ef2)n\Y1)r^4\Y2)\\
{}-n\nd r\,\pz{\vF(r)}r\X4(2\eb2\nd\X2(3(-1+\eb2)n^2+6\ebf2n r^2\\
{}+n\X1(-6+6\eb2+2\ef nr^2+9\ebf2n r^2-\ef{\nd}(2+9\eb2)r^2\Y1)\Y2)\\
{}+\ov Mr\,\pz{\vF(r)}r\X3(6\ebf2{\nd}(-1+\eb2)nr^2\\
{}+\nd\X3(-6\ebf2n(-1+\eb2)r^2+n\X2(-26+6\eb4 - 9\ov Mr^2
\X2(\pz{\vF(r)}r\Y2)^2 \\
{}+6\eb2\X2(6+r\,\pz{\vF(r)}r\X2(-2-n+3\ov Mr\,\pz{\vF(r)}r\Y2)\Y2)\Y2)
\Y3)\Y3)\Y4)\Y4)\Y4)
\end{multline}
Substituting Eq.\ \er{D.11} to Eq.\ \er{D.8} one gets
\bml{D.12}
\pz{^2\vF(r)}{r^2}=\frac1{4\ov Mr^2}\,\eb2 \X4(
2+n+\frac{\eb2\X1(-n\nd +\ef n(3(\ef2-1)n-2)r^2\Y1)}n\\
{}+\frac1{n\nd}\X3(-\ebf2n\ov M\X1(-2+(-1+\ef2)n\Y1)r^3\,\pz{\vF(r)}r\\
{}+n\nd r\X3(-(7+\eb2)\ov M\,\pz{\vF(r)}r\\
{}+\X2(n+2-3\ov M\,\pz{\vF(r)}r\Y2)\X2(-2\,\pz{B(r)}r + \ov Mr
\X2(\pz{\vF(r)}r\Y2)^2\Y2)\Y3)\Y3)\Y4)
\end{multline}

Eqs \er{D.9}, \er{D.11}--\er{D.12} are equivalent to Eqs \er{D.1}--\er{D.4}.
Eqs \er{D.9}, \er{D.11}--\er{D.12} form a system of ordinary differential \e
s for \f s $A(r)$, $B(r)$ and $\vF(r)$. After solving this system we
substitute the results into Eq.\ \er{D.5} getting a \so\ for $\ov\rho(r)$.

Let us introduce the \f s
$$
\chi(r)=\pz\vF r(r), \q \vt(r)=\pz Br(r).
$$
Substituting these \f s into Eqs \er{D.9}, \er{D.11}--\er{D.12} we get the following
ordinary differential \e
\beq{D.14}
\pz {V(r)}r = F(V(r),r)
\end{equation}
where $V(r)=\X1(A(r),B(r),\vt(r),\vF(r),\chi(r)\Y1)\in \R^5$,
$F(V,r)\in \R^5$.

Thus our system is of fifth order and we need five integration \ct s. $F(V,r)$
is a $C^\iy$-\f\ except $r=0$. Thus we can pose a well defined
initial Cauchy problem for the system \er{D.14} in an interval
$I=\langle a,b\rangle\subset \R$, $0\notin I$, \st $V(a)=V_0$, getting a
unique \so\ according to well known theorems. Let us consider our system for
small $0<r\ll 1$. In this case the system simplifies and we get
\bea{D.15}
\pz{A(r)}{r}&=&\frac{\eb2-1}{4r}\\
\pz{^2\vF(r)}{r^2}&=&-\frac{\nd(\eb2-1)}{4r^2\ov M}\,\eb2 \label{D.16}\\
\pz{^2B(r)}{r^2}&=&\frac{9-10\eb2+\eb4}{8r^2}\label{D.17}
\end{eqnarray}

In the case of large $r\gg1$ one gets
\bea{xxx}
\pz{^2B(r)}{r^2}&=&\frac{r^2}{8n^2\nd^2}\X4[
4\ebf4{2n}(1+n-n\ef2)+\ebf4{2n}n^2(1-\ef2)^2\nn\\
&&{}-24\ebf2n \ov Mn(1-2\eb2)(1+n)\X2(\pz{\vF(r)}r\Y2)^2\nn\\
&&{}+12\ebf2{\nd}\ov Mn^2(1-2\eb2)\X2(\pz{\vF(r)}r\Y2)^2\nn\\
&&{}-6\ebf2n \ov Mn^3 (1-\eb2)(1-\ef2)\X2(\pz{\vF(r)}r\Y2)^2\nn\\
&&{}+9\ov M{}^2n^2\nd^2(1-\eb2)\X2(\pz{\vF(r)}r\Y2)^4\Y4]
\label{D.18}\\
\pz{^2\vF(r)}{r^2}&=&-\frac{\eb2 \,\pz{\vF(r)}r \,r\X1(\ebf2{\nd}n
-\ebf2n \nd+3\ov Mn\nd \X1(\pz{\vF(r)}r\Y1)^2\Y1)}{4n\nd}\hskip30pt
\label{D.19}\\
\pz{A(r)}r&=&\frac{\X1(\ebf2{\nd}n-\ebf2n \nd+3\ov Mn\nd \X1(\pz{\vF(r)}r\Y1)^2
\Y1)r}{4n\nd}
\label{D.20}
\end{eqnarray}

The system of units which we are using here can be expressed in SI as
$$
\bal
L&=\xi\cdot 3.09 \times 10^{23} \rm m\\
T&=\xi \times 10^{15} \rm s\\
M_0&=\xi \cdot 1.57 \times 10^{49} \rm kg.
\eal
$$
An \ac\ unit $\frac L{T^2}=\xi\cdot 3.09 \times 10^{23} \rm \frac m{s^2}$ and
a density unit $\frac{M_0}{L^3}=\frac1{\xi^2}\cdot 5.3\t
10^{-22}\rm\frac{kg}{m^3}$. If we solve the system \er{D.9}, \er{D.11}--\er{D.12} we
get a \spt\ with a metric
\beq{D.21}
ds^2=\ov A(r)\,dt^2 - \ov B(r)\,dr^2 - r^2(d\theta^2+\sin^2\theta\,d\vf^2)
\end{equation}
where $\ov A(r)=e^{2A(r)}$, $\ov B(r)=\eb2$. In the \spt\ we can consider a
geodetic \e\ for massive and massless particles (photons)
\beq{D.22}
\pz{^2x^\mu}{\la^2}+\gd\wt\G,\mu,\a\b,\,\pz{x^\a}\tau\,\pz{x^\b}\tau=0
\end{equation}
where $\la$ is an affine parameter for a geodesic. In the case of a massive
particle it is a proper time ($\tau$). $\gd\wt\G,\mu,\a\b,$ are Christoffel's
symbols for a metric \er{D.21}, $x^\mu=(r,\theta,\vf,t)$.

One can easily find an \e\ for a massive particle (in the equatorial plane
$\theta=\frac\pi2$)
\beq{D.23}
\pz{^2r}{\tau^2}+\frac1{2\ov B(r)}\,\pz{\ov B(r)}r \X2(\frac{H^2}{\ov A(r)}
-1-\frac h{r^2}\Y2)+\frac12\,\pz{\ov A(r)}r\,\frac{H^2}{\ov A{}^2(r)\ov B(r)}
-\frac h{\ov B(r)r^3}=0,
\end{equation}
where $H$ is a total energy per a unit mass and $h$ is an angular momentum
per a unit mass (in Appendix~A it is denoted by~$A$); $\tau$ is a proper
time. In the case of a photon we get
\beq{D.24}
\pz{^2r}{\la^2}+\frac1{2\ov B(r)}\,\pz{\ov B(r)}r\X2(
\frac{H^2}{\ov A(r)}-\frac h{r^2}\Y2)+\frac12\,\pz{\ov A(r)}r\,\frac{H^2}
{\ov A(r)\ov B(r)}-\frac h{\ov B(r)r^3}=0.
\end{equation}
$\la$ is an affine parameter along a photon path in such a way that
\beq{D.25}
ds^2=\ov A(r)\X2(\pz t\la\Y2)^2 - \ov B(r)\X2(\pz r\la\Y2)-r^2
\X2(\pz\vf\la\Y2)^2=0.
\end{equation}

Eq.\ \er{D.23} gives us an \an\ \ac\ (if we subtract an analogous \ac\ due to
Schwarzschild metric). Moreover, we can write down an \e\ for an orbit
\refstepcounter{equation}\label{D.26}
$$
\dsl{
\hfill \X2(\pz r\tau\Y2)^2 = \frac{H^2}{c^2\ov A(r)\ov B(r)}
-\frac{h^2}{r^2\ov B(r)} -\frac{c^2}{\ov B(r)}\hfill\rm(\theequation)\cr
\hfill\X2(\pz rt\Y2)^2=\frac{c^2\ov A(r)}{\ov B(r)} \X2(1-\frac{h^2c^2}{H^2}\,
\frac{\ov A(r)}{r^2}- \frac{c^4\ov A(r)}{H^2}\Y2)\hfill\rm(\theequation a)}
$$
and
\beq{D.27}
\X2(\pz r\vf\Y2)^2=\frac{H^2r^4}{h^2c^2\ov A(r)\ov B(r)}-\frac{r^2}{\ov B(r)}
-\frac{r^4c^2}{h^2\ov B(r)}.
\end{equation}
An \e\ for a photon can be also written as
\beq{D.28}
\pz{k^\mu}\la +\gd\wt\G,\mu,\a\b,k^\a k^\b=0
\end{equation}
where $k^\mu$ is a four-vector of a photon $g_\m k^\mu k^\nu=0$. It is easy
to find non-zero Christoffel symbols for \er{D.21} (see Ref.~\cite{20}).
In the formulae \er{D.26}, (\ref{D.26}a), \er{D.27} we keep $c$ (a velocity
of light) for a future convenience, even in our system of units $c=1$. In
formula (\ref{D.26}a) $t$~is a \cd\ time.
We came back to this problem later in this Appendix using Mathematica~7 and
getting preliminary orbits.

Moreover, we can consider \e s for $0<r\ll1$, i.e.\ \er{D.18}--\er{D.20}. It
is easy to notice that these \e s are scale invariant. In this way we can
consider a transformation of an independent variable
\beq{D.29}
r\to x=\frac r{r_0}
\end{equation}
where $r_0$ is a scale of length. One gets
\bea{D.30}
\pz{^2\wt B(x)}{x^2}&=&\frac{9-10e^{2\tilde B(x)}+e^{4\tilde B(x)}}{8x^2}\\
\pz{^2\wt\vF(x)}{x^2}&=&-\frac{-1+e^{2\tilde B(x)}}{4x^2}\,e^{2\tilde B(x)}
\label{D.31}\\
\pz{\wt A(x)}x&=&\frac{-1+e^{2\tilde B(x)}}{4x} \label{D.32}
\end{eqnarray}
where
\bea{D.33}
B(r)&=&\wt B\X2(\frac r{r_0}\Y2)\\
A(r)&=&\wt A\X2(\frac r{r_0}\Y2) \label{D.34}\\
\vF(r)&=&\frac{n+2}{\ov M}\,\wt \vF\X2(\frac r{r_0}\Y2). \label{D.35}
\end{eqnarray}

Let us consider a Cauchy initial value problem for Eqs \er{D.18}--\er{D.20}, i.e.
\beq{D.36}
\bga
B(R)=B_0, \q \pz{B}r(R)=B_1, \q A(R)=A_0,\\
\vF(R')=\vF_0, \q \pz\vF r(R')=\vF_1
\ega
\end{equation}
for established $R$ and $R'$. Eqs \er{D.36} can be rewritten in terms of $\wt
A(x)$, $\wt B(x)$, $\wt\vF(x)$ in the following way:
\beq{D.37}
\bga
\wt B(x_0)=B_0, \q \pz{\wt B}x(x_0)=r_0B_1, \q \wt A(x_0)=A_0,\\
\wt\vF(x_0')=0, \q \pz{\wt\vF} x(x_0')=0,\\
x_0=\frac R{r_0}, \q x_0'=\frac{R'}{r_0}.
\ega
\end{equation}
In this way
\beq{D.38}
\vF(r)=\frac{n+2}{\ov M}\,\wt\vF\X2(\frac r{r_0}\Y2)+\ov \vF_0r+\ov \vF_1
\end{equation}
where $\ov\vF_0$ and $\ov\vF_1$ are \ct s.

In order to simplify an initial problem we suppose that $B_1=\frac1{r_0}$. In
this way
\beq{D.39}
\pz{\wt B}x(x_0)=1.
\end{equation}
The question is how to define $r_0$ and $R$. For we want to consider a \spt\
with an \an\ \ac\ it is natural to use some knowledge from our model. It
means we take for $r_0$ and $R\simeq 0.6\ov R$ two time scales from Section~3 (see Eqs
\er{3.35}--\er{3.36}) and we get
\beq{D.40}
x_0=\frac R{r_0}\simeq 3.7 \t 10^2.
\end{equation}
Now it is necessary to establish $B_0$ and $A_0$. We proceed in the following
way. Let $U(\ov R)$ be a total \gr al \pt\ including an \an\ \ac\ at $r=\ov R$.
According to Section~3 we get
\beq{D.41}
U(\ov R)=2b\ov R.
\end{equation}

Let us consider the following \ap ion for $e^{2A(\ov R)}$, $e^{2B(\ov R)}$
\bea{D.42}
e^{2A(\ov R)}&=&1-\frac{2U(\ov R)}{c^2}\simeq 1-\frac{4b\ov R}{c^2}\\
e^{2B(\ov R)}&=&\frac1{1-\frac{2U(\ov R)}{c^2}}\simeq 1+\frac{4b\ov R}{c^2}.\label{D.43}
\end{eqnarray}
For $A(\ov R)$ and $B(\ov R)$ are small we get (we have for $R\simeq 0.6\ov R)$
\bea{D.44}
A(R)&=&-\frac{2bR}{c^2}=-4\t 10^{-12}\\
B(R)&=&\frac{2bR}{c^2}=4\t 10^{-12}.\label{D.45}
\end{eqnarray}
In this way we get the following initial Cauchy problem
\beq{D.46}
\bga
\wt A(3.7\t10^2)=-4\t10^{-12}, \q \wt B(3.7\t10^2)=4\t10^{-12}, \q
\pz{\wt B}x(3.7\t10^2)=1,\\
\wt\vF(5)=0, \q \pz{\wt\vF}x(5)=0.
\ega
\end{equation}

The system \er{D.30}--\er{D.32} with initial conditions \er{D.46} can be
easily solved numerically and the results are plotted on Fig.~\ref{yk}.
We plot here $A(r)$, $B(r)$, $\wt A(r)$, $\wt B(r)$ and $\wt\vF(r)$. The unit
of length is chosen to be $r_0=4.103$\,AU (see Section~3). It is easy to
conclude that the \spt\ derived here is not asymptotically flat, even $\ov
A(r)$ seems to go to one, $\ov B(r)$ seems to be not bounded by~1 in the
region considered. Further development of the problem demands detailed
numerical studies of the system \er{D.30}--\er{D.32} or even \er{D.9},
\er{D.11}, \er{D.12}.

\bigskip
\refstepcounter{figure}\label{yk}
\hbox to \textwidth{\ing[width=0.45\textwidth]{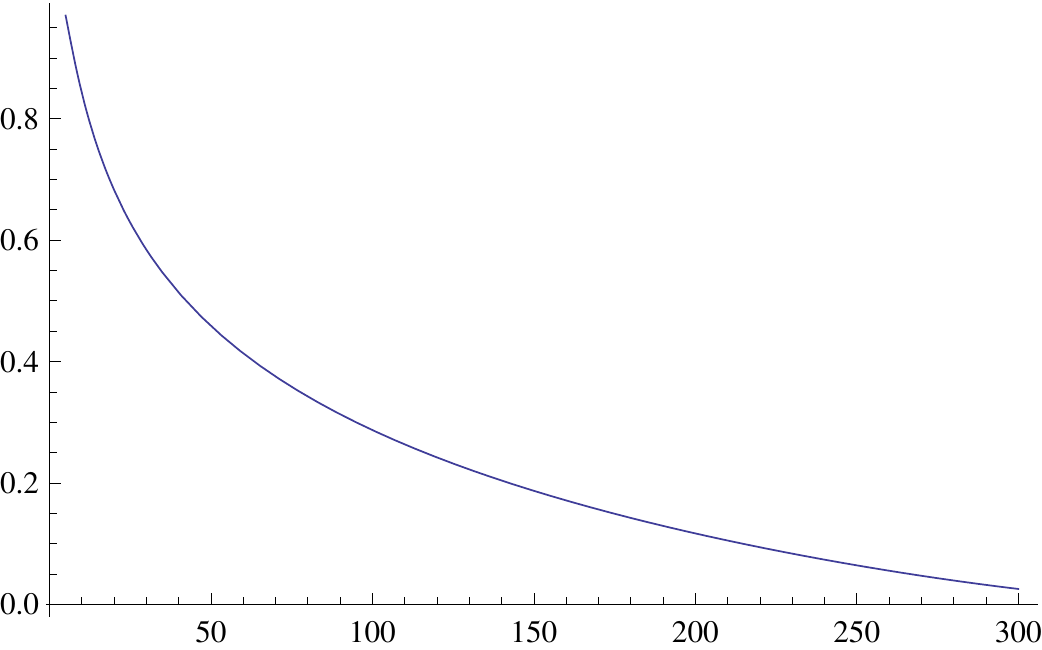}\hfil
\ing[width=0.45\textwidth]{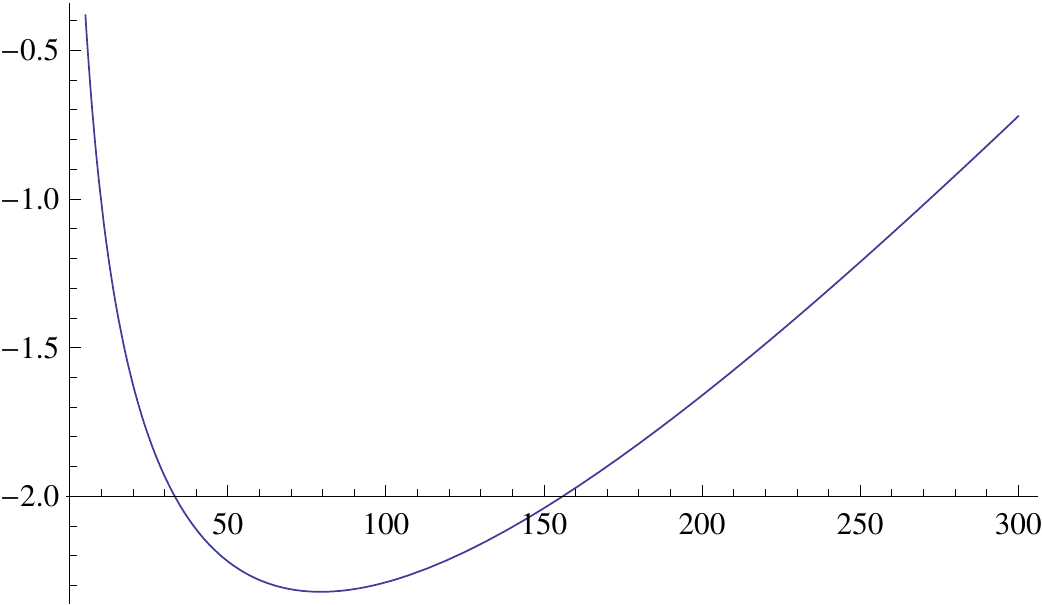}}
\hbox to \textwidth{\hbox to 0.45\textwidth{\hfil(A)\hfil}\hfil
\hbox to 0.45\textwidth{\hfil(B)\hfil}}
\hbox to \textwidth{\ing[width=0.45\textwidth]{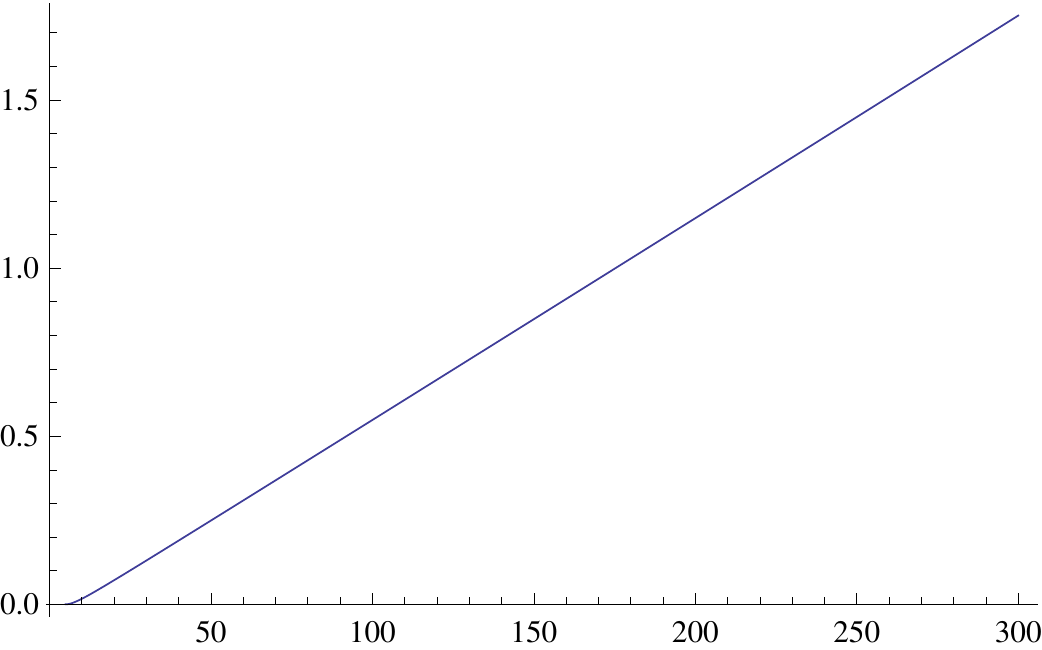}\hfil
\ing[width=0.45\textwidth]{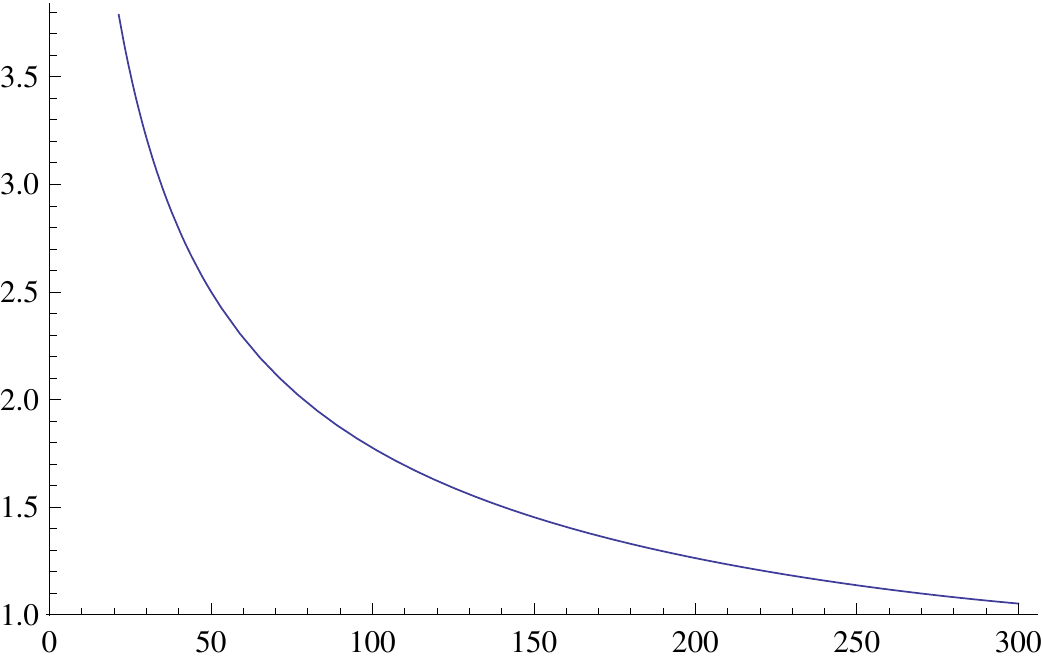}}
\hbox to \textwidth{\hbox to 0.45\textwidth{\hfil(C)\hfil}\hfil
\hbox to 0.45\textwidth{\hfil(D)\hfil}}
\hbox to \textwidth{\ing[width=0.45\textwidth]{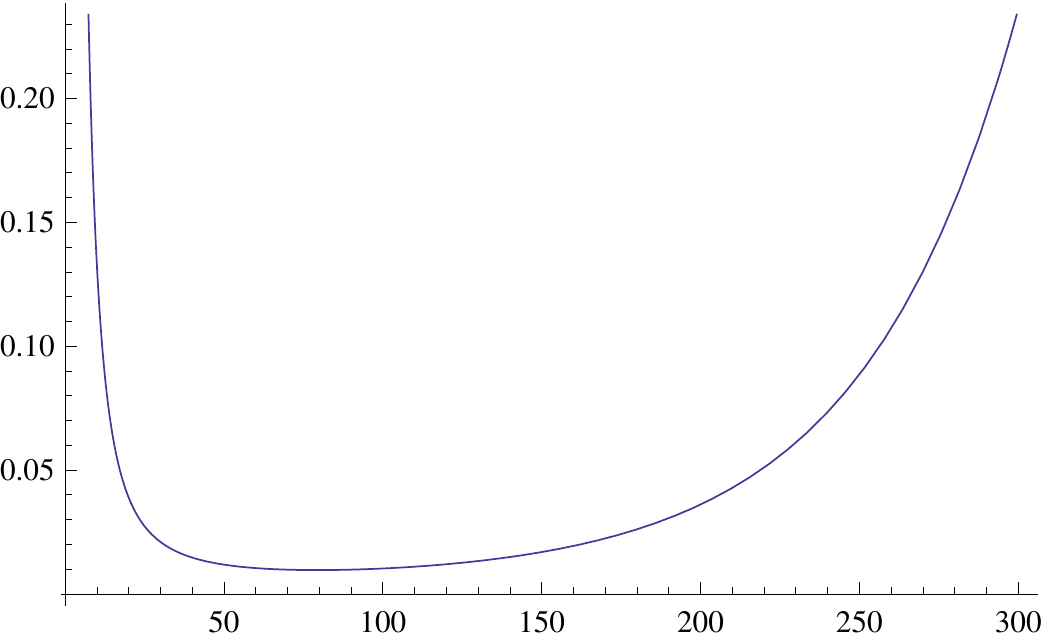}\hfil
\ing[width=0.45\textwidth]{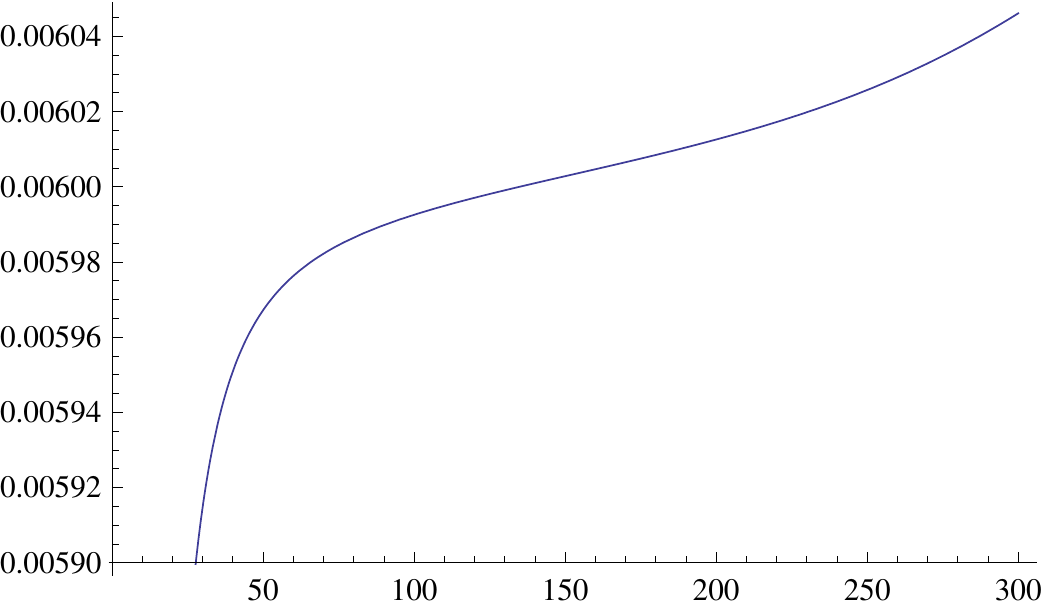}}
\hbox to \textwidth{\hbox to 0.45\textwidth{\hfil(E)\hfil}\hfil
\hbox to 0.45\textwidth{\hfil(F)\hfil}}
\smallskip \noindent
{\small Figure \thefigure: (A)---a plot of the \f\ $A(r)$,
(B)---a plot of the \f\ $B(r)$, (C)---a plot of the \f\ $\wt\vF(r)$,
(D)---a plot of the \f\ $\ov A(r)$, (E)---a plot of the \f\ $\ov B(r)$,
(F)---a plot of the \f\ $\pz{\wt\vF}r(r)$.

}\bigskip

Moreover, there is a problem of a dependence on a Hubble \ct. This can be
posed even on the level of this simple model. Let us notice that
$b(t)=1.557\cdot c\cdot H(t)$ (see \er{8.16} and \er{8.20}). In this way
\bea{D.46a}
A(R)=-4\t10^{-12}\X2(\frac{H(t)}{H(t_0)}\Y2)\\
B(R)=\eta\cdot4\t10^{-12}\X2(\frac{H(t)}{H(t_0)}\Y2)\label{D.47}
\end{eqnarray}
where $\eta$ is a \ct\ of order~1 which can give us a better estimation of
the \so.

Let us consider a density of matter $\ov\rho(r)$. This \f\ should be
multiplied by a factor $\frac1{\xi^2}\cdot5.3\t10^{-22}$ in order to be
in~SI.

If we want to be in an agreement with observational data we should have
\beq{D.48}
\ov\rho(r)<\xi^2\cdot 0.96\t10^{-4}
\end{equation}
density  lower than a density of an interplanetary matter
\beq{D.49}
\ov\rho(r)<\xi^2\cdot 0.18\t10^{-4}
\end{equation}
density  lower than a density of an interstellar matter
\beq{D.50}
\ov\rho(r)<\xi^2\cdot 0.18\t10^{-10}
\end{equation}
density  lower than a density of an intergalactic matter. $r$ is measured in
$r_0=4.103$\,AU units. Let us consider Eq.\ \er{D.5}. In this equation we
have a factor $\ef{\nd}$. According to Eq.\ \er{D.38} $\vF(r)$ contains a
linear term $\vF_0r$ where $\vF_0$ is an arbitrary \ct. If we take a
sufficient negative $\vF_0$ we can satisfy all these conditions even on the
level of the Sun radius.

Finally, let us write a photon trajectory in our \spt. One gets
$$\dsl{
\refstepcounter{equation}\label{D.51}
\hskip100pt \X2(\pz r\la\Y2)^2= \frac{H^2}{c^2\ov A(r)\ov B(r)}-\frac{h^2}{\ov
B(r)r^2} \hfill \rm(\theequation)\cr
\hskip100pt \X2(\pz rt\Y2)^2 = \frac{c^2\ov A(r)}{\ov B(r)} \X2(1-\frac{h^2c^2}
{H^2}\,\frac{\ov A(r)}{r^2}\Y2) \hfill \rm(\theequation a)\cr
\refstepcounter{equation}\label{D.52}
\hskip100 pt \X2(\pz r\vf\Y2)^2= \frac{H^2r^4}{c^2\ov A(r)\ov B(r)h^2}
-\frac{r^2}{\ov B(r)}\,.\hfill \rm(\theequation)}
$$
In this way according to a general equivalence principle we get an influence
of an \an\ \ac\ on photon movement. $t$ is a \cd\ time. We keep $c$ in the
formulae.

Let us consider a case with $g_\m\ne g_{\nu\mu}$ in more details. Thus we
write down \e s of the \eu\nos\ Jordan--Thiry Theory (see Refs \cite{4},
\cite{5}, \cite7).

\def\ef#1{e^{#1\vF}}
One gets
\bg{D.53}
\ov R_{\a\b}(\ov W)-\frac12\,g_{\a\b}\ov R(\ov W)=\nad{\rm tot}T_{\a\b}\\
\falg^{[\m]}{}_{,\nu}=0 \label{D.54}\\
g_{\m,\a}-g_{\z\nu}\ov \G_{\mu\si}-g_{\mu\z}\gd\ov\G{},\z,\si\nu,=0 \label{D.55}\\
\gd\ov W{},\a,\b,=\gd\ov W{},\a,\b\g,\ov\th{}^\g \label{D.56}\\
\gd\ov \o{},\a,\b,=\gd\ov \G{},\a,\b\g,\ov\th{}^\g \label{D.57}\\
\noalign{\eject}
\gd\ov W{},\a,\b,=\gd\ov\o{},\a,\b,-\frac23\,\gd\d,\a,\b,\ov W \label{D.58}\\
\ov W=\ov W_\g \ov\th{}^\g=\frac12\X1(\gd\ov W{},\si,\g\si,
-\gd\ov W{},\si,\si\g,\Y1)\ov\th{}^\g. \label{D.59}
\end{gather}
We remind to the reader that in the \nos\ field theory we have to do with two
\cn s $\gd \ov W{},\a,\b,$ and $\gd \ov \o{},\a,\b,$. The first is
unconstrained and the second is constrained, i.e.
\bg{D.60}
\ov \G_\g=\ov\G_{[\g\a]}^\a=0\\
g^{\a\b}g_{\d\b}=g^{\b\a}g_{\b\d}=\gd\d,\a,\d,, \label{D.61}
\end{gather}
where the order of indices is important. In Eq.~\er{D.53} we use our system
of units $p=1$, $c=1$, $\k=1$.
\beq{D.61a}
\nad{\rm tot}T_{\a\b}=\ef{-(n+2)}\ov\rho u_\a u_\b+T_{\a\b}(\vF)
\end{equation}
is a total energy momentum tensor
\beq{D.62}
T_{\a\b}(\vF)=2\ov M\pa_\a\vF\pa_\b\vF - g_{\a\b}L(\vF).
\end{equation}
Notice that $T_{\a\b}(\vF)$ is now \nos\ for $g_{\a\b}$ is \nos
\beq{D.63}
L(\vF)=\ov Mg^\m\pa_\mu \vF\pa_\nu\vF+\X2(\frac{\ef{(n+2)}}{n+2}
-\frac{\ef n}{n}\Y2)+ \ov\rho \ef{(n+2)}.
\end{equation}
Now $g^\m$ is also \nos\ and $g^{(\m)}$ is not an inverse tensor of
$g_{(\m)}$.
\beq{D.64}
\ov R(\ov W)=g^{\mu\a}\ov R_{\mu\a}(\ov \G)+\frac23\,g^\[\mu\a]
\ov W_\[\mu \a]
\end{equation}
is a Moffat--Ricci scalar of a curvature for a \cn\ $\gd\ov W{},\a,\b,$ and
$\ov R_{\mu\a}(\ov\G)$ is a Moffat--Ricci tensor for a \cn\
$\gd\ov\o{},\a,\b,$. In particular
\beq{D.65}
\ov R_{\m}(\ov\G)=\gd\ov R{},\a,\m\a,(\ov \G)+\frac12\,\gd\ov R,\a,\a\m,
(\ov\G),
\end{equation}
where $\gd\ov R,\a,\m\b,(\ov \G)$ is a curvature tensor for a \cn\ $\gd\ov\o,
\a,\b,$.

From the Bianchi identity we get an \e\ for the scalar field $\vF$,
\beq{D.66}
-2\ov M\ov\nabla_\a(g^{\a\b}\pa_\b\vF)+\ef n(e^{2\vF}-1)
-(n+2)\ef{-(n+2)}\ov\rho=0.
\end{equation}
$\ov\nabla_\a$ means a covariant \dv\ \wrt a \cn\ $\gd\ov\o,\a,\b,$ (this
\cn\ is not a Levi-Civit\`a \cn). Eq.~\er{D.66} can be rewritten in a more
convenient form
\bg{D.67}
-\frac{2\ov M}{\sqrt{-g}}\,\pa_\a\X1(\falg^{\a\b}\pa_\b\vF\Y1)+
\ef n(e^{2\vF}-1)-(n+2)\ef{-(n+2)}\ov\rho=0\\
g=\det g_{\a\b} \label{D.68}\\
\falg^{\a\b}=\sqrt{-g}\,g^{\a\b}. \label{D.69}
\end{gather}
$\ov\rho$ means a density of a dust, $u^\mu=\pz{x^\mu}\tau$ is a
four-velocity (as usual).

Let us calculate a trace for a $\nad{\rm tot}T_{\a\b}$, i.e.
\beq{D.70}
\nad{\rm tot}T=g^{\a\b}\nad{\rm tot}T_{\a\b}=-\X2(
2\ov Mg^{\a\b}\pa_\a \vF\pa_\b\vF+4\X2(\frac{\ef{(n+2)}}{n+2}
-\frac{\ef n}n\Y2)+3\ov \rho\ef{-(n+2)}\Y2).
\end{equation}
Let us define a tensor
\bg{D.71}
\wt T_{\a\b}=\nad{\rm tot}T_{\a\b}-\frac12\,g_{\a\b}\nad{\rm tot}T\\
\wt T_{\a\b}=\ef{-(n+2)}\ov\rho u_\a u_\b +2\ov M\pa_\a \vF\pa_\b\vF
+g_{\a\b}\X2(\ef n\X2(\frac{\ef2}{n+2}-\frac1n\Y2)+
\frac{\ov\rho}2 \,\ef{-(n+2)}\Y2). \label{D.72}
\end{gather}
In terms of $\wt T_{\a\b}$ Eq.~\er{D.53} can be rewritten in a more
convenient form (using Eqs \er{D.56}--\er{D.60}, Eq.~\er{D.64}):
\bg{D.73}
\ov R_{(\a\b)}(\ov\G)=\wt T_{(\a\b)}\\
\ov R_{[[\a\b],\g]}(\ov\G)-\wt T_{[[\a\b],\g]}=0. \label{D.74}
\end{gather}
Eq.~\er{D.54} can be rewritten as
\beq{D.75}
\ov\G_\g=0.
\end{equation}
Thus the full field \e s consist of \er{D.73}, \er{D.74}, \er{D.75},
\er{D.55}, \er{D.67}. Due to the Bianchi identity they are consistent modulo
some constraints imposed on $\ov\rho,u^\mu,\vF$.

Let us consider a stationary, spherically \s\ case of those \e s. In this
case $\ov\rho,\vF$ are \f s of $r$ only and $u^\mu$ has only one temporal
component (the dust does not move). The most general \nos\ tensor in
stationary and spherically \s\ case reads
\beq{D.76}
g_\m=\mt{-\a & 0 & 0 & \o \\
0 & -\b & f\sin\th & 0 \\
0 & -f\sin\th & -\b\sin^2\th & 0 \\
-\o & 0 & 0 & \g}
\end{equation}
where $\a,\b,\g,f$ and $\o$ are real \f s of $r$ and $t$. For we are really
working in a static case they are \f s of $r$ only. The additional
constraints for $\ov\rho$ and~$\vF$ are trivially satisfied, $\a,\g>0$. For
$g^\m$, the only nonvanishing components  are
\bea{D.77}
g^{11}&=&\frac\g {\o^2-\a\g}\\
g^{22}&=&g^{23}\sin^2\th=-\frac\b{\b^2+f^2} \label{D.78}\\
g^{44}&=&-\frac\a{\o^2-\a\g} \label{D.79}\\
g^{[14]}&=&\frac\o{\o^2-\a\g} \label{D.80} \\
g^{[23]}\sin\th&=&\frac f{\b^2+f^2} \label{D.81}
\end{eqnarray}
We suppose $\o^2-\a\g\ne0$, and $\b^2+f^2\ne0$. From Eq.~\er{D.54} we get
\beq{D.82}
\frac{\o^2}{\a\g-\o^2}=\frac{\ell^4}{\b^2+f^2}
\end{equation}
where $\ell^2$ is an integration \ct. This \ct\ has an interpretation as a fermion
charge. From Eq.~\er{D.74} we get
\beq{D.83}
\ov R_\[23](\ov G)-\wt T_\[23]=C_1\sin\th
\end{equation}
where $C_1$ is an integration \ct\ and
\beq{D.84}
\frac{\wt T_\[23]}{\sin\th}=-f\X2(\ef n\X2(\frac{\ef2}{n+2}-\frac1n\Y2)
+\frac{\ov\rho}2\,\ef{-(n+2)}\Y2).
\end{equation}

Note that the Moffat--Ricci tensor is a linear combination of the ordinary
Ricci tensor and the second contraction of the curvature tensor. Moreover, we
have $\ov\G_\g=0$ and
\beq{D.85}
\gd\ov\G,\b,\nu\b,=\X1[\log(-g)^{1/2}\Y1]_{,\nu}\,.
\end{equation}
Consequently, the Moffat--Ricci tensor in this case is identically equal to
the ordinary Ricci tensor for the second contraction
\beq{D.86}
\gd\ov R,\a,\a\m,=\frac12\X2(\gd\ov\G,\b,{(\mu\b),\nu},-\gd\ov\G,\b,
{(\nu\b),\mu},\Y2)=0
\end{equation}
equals zero. We denote an ordinary Ricci tensor by $A_\m(\G)$. Thus we get \e
s
\bg{D.87}
A_\m(\ov\G)=\wt T_{(\m)} \\
A_\[23](\ov\G)-\wt T_\[23]=C_1\sin\th \label{D.88}
\end{gather}
and Eq.~\er{D.67} which can be rewritten in the form
\beq{D.89}
\frac{2\ov M \ell^4}{\o(\b^2+f^2)}\,\pz{}r\X2(\frac\g\o\,\pz\vF r\Y2)
+\ef n(\ef2-1)-(n+2)\ov\rho \ef{-(n+2)}=0.
\end{equation}
We use Eq.\ \er{D.82} and we put
\beq{D.90}
\sqrt{-g}=\sin\th\X1[(\a\g-\o^2)(\b^2+f^2)\Y1]^{1/2}.
\end{equation}

One can easily notice that
\bea{D.91}
\wt T_{11}&=&2\ov M\X2(\pz \vF r\Y2)^2 - \a\X2(\ef n\X2(\frac{\ef2}{n+2}
-\frac1n\Y2)+\frac{\ov\rho}2\,\ef{-(n+2)}\Y2)\\
\wt T_{44}&=&\X2(\frac32\ef{-(n+2)}\ov\rho + \ef n\X2(\frac{\ef2}{n+2}-\frac
1n\Y2)\Y2)\g \label{D.92}\\
\wt T_{22}&=&-\b\X2(\ef n\X2(\frac{\ef2}{n+2}-\frac 1n\Y2)+\frac{\ov \rho}2
\ef{-(n+2)}\Y2) \label{D.93}\\
\wt T_{33}&=&\sin^2\th \wt T_{22}. \label{D.94}
\end{eqnarray}
Using Refs \cite4, \cite{D78}, \cite{D79} we also get
\beq{D.95}
A_{33}(\ov\G)=\sin^2\th A_{22}.
\end{equation}
Thus we get the \e s
\bea{D.96}
A_{44}(\ov\G)&=&\wt T_{44}\\
A_{11}(\ov\G)&=&\wt T_{11} \label{D.97}\\
A_{22}(\ov\G)&=&\wt T_{22} \label{D.98}\\
A_{(14)}(\ov\G)&=&0. \label{D.99}\\
\noalign{\eject}
A_{11}(\ov\G)&=&-\frac12\,\pz{^2\Ps}{r^2}-\frac18\X2(\X2(\pz\Ps r\Y2)^2+4C^2\Y2)
+\frac1{4\a}\,\pz \a r\,\pz\Ps r \nonumber \\
&&{}+\X2(\frac{\o^2}{2\a\g}\,\pz\Ps r+
\frac1{2\g}\,\pz\g r\Y2)\X2(\frac1{2\a}\,\pz\a r-\frac{\o^2}{2\a\g}\,\pz
\Ps r-\frac1{2\g}\,\pz\g r\Y2)+\pz{}r\X2(\frac{\o^2}{2\a\g}+\frac1{2\g}
\,\pz\g r\Y2)\Y3)\hskip30pt \label{D.100}\\
A_{44}(\ov\G)&=&\frac{\o^2}{8\a^2}\X2(3\X2(\pz\Ps r\Y2)^2+4C^2\Y2)
\label{D.101} \\
A_{22}(\ov \G)&=&\frac1{4\a}\X2(2fC-\b\,\pz\Ps r\Y2)
+\frac1{8\a}\X2(2fC-\b\,\pz\Ps r\Y2)\cdot \pz{}r\log(\o^2(\b^2+f^2))\nonumber\\
&&{}+\frac C{4\a}\X2(f\,\pz\Ps r+2\b C\Y2) \label{D.102}\\
A_\(14)&=&0 \label{D.103}
\end{eqnarray}
where
\beq{D.104}
\Ps=\log(\b^2+f^2), \q C=\frac{f\,\pz\b r - \b\,\pz fr}{\b^2+f^2}.
\end{equation}
\bg{D.105}
A_\[14](\ov\G)=\frac \o{8\a}\X2(\X2(\pz\Ps r\Y2)^2+4C^2\Y2)
+\frac{\o^2}{4\a}\X2(\pz\Ps r\Y2)^2\\
\frac{A_\[23](\ov\G)}{\sin\th}=\pz{}r \X2(\frac1{4\a}\X2(f\,\pz\Ps r\Y2)\Y2)
-\frac C{4\a}\X2(2fC-\b\pz\Ps r\Y2)\hskip60pt \nonumber\\
\hskip60pt {}+\frac1{8\a}
\X2(f\,\pz\Ps r+2\b C\Y2)\X2(\frac1\a\,\pz\a r+\frac{\o^2}{\a\g}
+\frac1\g\,\pz\g r\Y2). \label{D.106}
\end{gather}

\def\edn{\X2(\frac{\ef2}{n+2}-\frac1n\Y2)}
\def\pzd#1{\X2(\pz{#1}r\Y2)^2}
From Eq.\ \er{D.96} one can find a \f\ $\ov\rho$
\beq{D.107}
\ov\rho(r)=\frac2{3\g}\,\ef{(n+2)}\X3(\frac{\o^2}{8\a^2}
\X2(3\X2(\pz\Ps r\Y2)^2+4C^2\Y2)-\g\ef n\edn\Y3).
\end{equation}
We substitute \er{D.106} to \er{D.91}, \er{D.93} and \er{D.84} getting
\bea{D.108}
\wt T_{11}&=&2\ov M\pzd\vF-\a\X3(\frac23\,\ef n\edn+
\frac{\o^2}{12\a^2\g}\X2(3\pzd\Ps+4C^2\Y2)\Y3)\\
\wt T_{12}&=&-\b\X3(\frac23\,\ef n\edn+\frac{\o^2}{12\a^2\g}\X2(3\pzd\Ps
+4C^2\Y2)\Y3) \label{D.109}\\
\frac{\wt T_\[23]}{\sin\th}&=&f\X3(\frac23\,\ef n\edn+\frac{\o^2}{12\a^2\g}\X2(3\pzd\Ps
+4C^2\Y2)\Y3) \label{D.110}
\end{eqnarray}

Substituting \er{D.107} into \er{D.89} one gets
\beq{D.110a}
\frac{2\ov M\ell^4}{\o(\b^2+f^2)}\,\pz{}r\X2(\frac\g\o\,\pz\vF r\Y2)
-\frac{(n+2)\o^2}{12\a^2\g}\X2(3\pzd\Ps+4C^2\Y2)
+\ef n\X2(\frac53\,\ef2-\frac{5n+4}{3n}\Y2)=0.
\end{equation}

Thus we get the following \e s from \er{D.95}--\er{D.99}
\bg{nic}
-\frac12\,\pz{^2\Ps}{r^2}-\frac18\X2(\pzd\Ps+4C^2\Y2)+\frac1{4\a}\,\pz\a
r\,\pz\Ps r\hskip220pt \nn\\
{}+\X2(\frac{\o^2}{2\a\g}\,\pz\Ps r+\frac1{2\g}\,\pz\g r\Y2)
\X2(\frac1{2\a}\,\pz\a r-\frac{\o^2}{2\a\g}\,\pz\Ps r-\frac1{2\g}\,\pz\g r\Y2)
+\pz{}r\X2(\frac{\o^2}{2\a\g}+\frac1{2\g}\,\pz\g r\Y2)\nonumber\\
\hskip80pt {}=2\ov M\pzd \vF-\a\X2(\frac23\,\ef n\edn + \frac{\o^2}{12\a^2\g}\,
\X2(3\pzd\Ps+4C^2\Y2)\Y2) \label{D.111}\\
\frac1{4\a}\X2(2fC-\b\,\pz \Ps r\Y2)+\frac1{8\a}\X2(2fC-\b\,\pz\Ps r\Y2)
\cdot \pz{}r\log(\o^2(\b^2+f^2))+\frac C{4\a}\X2(f\,\pz\Ps r+2\b C\Y2)
\hskip60pt \nn\\
\hskip80pt {}=-\b\X3(\frac23\,\ef n\edn+\frac{\o^2}{12\a^2\g}\,\X2(3\pzd\Ps
+4C^2\Y2)\Y3) \label{D.112}\\
\pz{}r\X2(\frac1{4\a}\X2(f\,\pz\Ps r-2\b C\Y2)\Y2)-\frac C{4\a}
\X2(2fC-\b\,\pz\Ps r\Y2)+\frac1{8\a}\X2(f\,\pz\Ps r+2\b C\Y2)
\X2(\frac1\a\,\pz\a r+\frac{\o^2}{\a\g}\,\pz\Ps r+\frac1\g\,\pz\g r\Y2)\nn\\
\hskip100pt {}-f\X3(\frac23\,\ef n\edn+\frac{\o^2}{12\a^2\g}\X2(3\pzd\Ps+4C^2\Y2)\Y3)=C_1
\label{D.113}
\end{gather}

Thus we get Eq.\ \er{D.110a} and Eqs \er{D.111}--\er{D.113} and an algebraic
\e\ \er{D.82} for $\Phi(r)$, $\g(r)$, $\a(r)$, $\o(r)$, $f(r)$, $\b(r)$. It
means we have five \e s for six \f s. One can eliminate $\g$ using
Eq.~\er{D.82}:
\beq{D.114}
\g=\frac{\o^2(f^2+\b^2+\ell^4)}{\ell^4\a}\,.
\end{equation}
We have also two integration \ct s $C_1$ and $\ell^2$.

\def\ef#1{e^{#1\vF(r)}}
\def\pzd#1{\X2(\pz{#1}r(r)\Y2)^2}
In this way we have more \f s than \e s. Thus we have one degree of freedom.
One can use this degree of freedom supposing that
\beq{D.115}
\b=r^2.
\end{equation}
Thus Eqs \er{D.110a}--\er{D.113}, condition \er{D.115} and
\beq{D.116}
\g=\frac{\o^2(f^2+r^4+\ell^4)}{\ell^4\a}
\end{equation}
give us four \e s for $\vF(r)$, $\o(r)$, $\a(r)$, $f(r)$. Eqs
\er{D.110a}--\er{D.113} are reduced to
\bml{D.117}
\frac18\X4(\frac{16\ef n(-2+(\ef2-1)n)\a(r)}{3n\nd}\\
{}+\X3(4\,\pz\a r(r)\X2(\o(r)\X2(r^3\X1(8r^4+\ell^4(r+4)\Y1)
+f(r)\X2((\ell^4+8r^3)f(r)+2(\ell^4+2r^4+2f^2(r))\pz fr(r)\Y2)\Y2)\\
{}+3(r^4+f^2(r))(\ell^4+r^4+f^2(r))\pz\o r(r)\Y2)
-4(r^4+f^2(r))(\ell^4+r^4+f^2(r))\o(r)\pz{^2\a}{r^2}(r)\Y3)\\
\hskip100pt {}\times\X2(\a(r)(r^4+f^2(r))(\ell^4+r^4+f^2(r))\o(r)\Y2)^{-1}\\
\noalign{\eject}
+\frac4{3(r^4+f^2(r))^2(\ell^4+r^4+f^2(r))^2\o^2(r)}\hskip180pt\\
\X4(-12(r^4+f^2(r))^2(\ell^4+r^4+f^2(r))^2\pzd\o
-\o^2(r)\X3(6r^6\X1(\ell^8(r-4)+2\ell^4(r-3)r^4+6r^8\Y1)\\
{}+12\ov Mf(r)^8\pzd\vF + r^4(\ell^4+r^4)\X2((7\ell^4+3r^4)\pzd f+12\ov M
r^4(\ell^4+r^4)\pzd\vF\Y2)\\
{}+3f(r)^6\X2(4r^2+3\pzd f+8\ov M(\ell^4+2r^4)\pzd\vF\Y2)\\
{}+f^2(r)\X3(2r^2\X1(42r^8+4\ell^4r^4(2+3n)+\ell^8(20+3r)\Y1)\\
{}+(-9\ell^8-2\ell^4r^4+15r^8)\pzd f+24\ov Mr^4(\ell^4+r^4)(\ell^4+2r^4)\pzd\vF\Y3)\\
{}+f(r)^4\X2(4r^2(15r^4+\ell^4(13+3r))+3(-4\ell^4+7r^4)\pzd f
+12\ov M(\ell^8+6\ell^4r^4+6r^8)\pzd\vF\Y2)\\
{}+6f(r)^5\X2((\ell^4+4r^3)\pz fr(r)+\ell^4\pz{^2f}{r^2}(r)\Y2)\\
{}+r^3f(r)\X2(\X1(24r^8+\ell^8(3r-64)+2\ell^4r^4(3r-44)\Y1)
+6\ell^4r(\ell^4+r^4)\pz {^2f}{r^2}(r)\Y2)\\
{}+f(r)^3\X2(\X1(3\ell^8+48r^7+4\ell^4r^3(3r-22)\Y1)\pz fr(r)
+6\ell^4(\ell^4+2r^4)\pz{^2f}{r^2}(r)\Y2)\Y3)\\
{}-3(r^4+f^2(r))(\ell^4+r^4+f^2(r))\o(r)\X3(\X2(r^3(8r^4+\ell^4(r+4))\\
{}+f(r)\X2((\ell^4+8r^3)f(r)+2(\ell^4+2r^4+2f^2(r))\pz fr(r)\Y2)\Y2)\pz \o r(r)\\
{}-2(r^4+f^2(r))(\ell^4+r^4+f^2(r))\pz{^2\o}{r^2}(r)\Y3)\Y4)\Y4)=0.
\end{multline}

\bml{D.118}
\frac1{12}\,r\X4(-\frac{12\X1(r^4-f^2(r)+rf(r)\pz fr(r)\Y1)}
{\a(r)(r^4+f^2(r))}
+\frac{6(2f(r)-rf(r))\X1(4r^3f(r)+(-r^4+f^2(r))\pz fr(r)\Y1)}
{\a(r)(r^4+f^2(r))^2}\\
{}+r\X3(8\ef n\edn+\frac{4\ell^4\X1(4r^2(3r^4+f^2(r))+8r^3f(r)\pz fr(r)
+(r^4+3f^2(r))\pzd f\Y1)}
{\a(r)(r^4+f^2(r))^2(\ell^4+r^4+f^2(r))}\Y3)\\
{}-\frac{12(r^4-f^2(r)+rf(r))\X1(\o(r)\X1(2r^3+f(r)\pz fr(r)\Y1)
+(r^4+f^2(r))\pz \o r(r)\Y1)}
{\a(r)(r^4+f^2(r))^2 \o(r)}\Y4)=0.
\end{multline}

\bml{D.119}
\frac18\X4(-8C_1-\frac{4\pz\a r(r)\pz fr(r)}{\a^2(r)}
-\frac{8r^2\X1(-2f(r)+r\pz fr(r)\Y1)\X1(r^4-f^2(r)+rf(r)\pz fr(r)\Y1)}
{\a(r)(r^4+f^2(r))^2}\\
{}-\frac23\,f(r)\X3(8\ef n\edn+
\frac{4\ell^4\X1(4r^2(3r^4+f^2(r))+8r^3f(r)\pz fr(r)+(r^4+3f^2(r))\pz fr(r)\Y1)}
{\a(r)(r^4+f^2(r))^2(\ell^4+r^4+f^2(r))}\Y3)\\
+\frac{4\pz fr(r)\X2(\frac{2r^3+f(r)\pz fr(r)}{r^4+f^2(r)}+\frac{\pz\o r(r)}
{\o(r)}\Y2)}{\a(r)}+\frac{4\pz{^2f}{r^2}(r)}{\a(r)}\Y4)=0.
\end{multline}

\bml{D.120}
\frac{\ef n(-4+5(\ef2-1)n)}{n}\\
{}-\frac{\ell^4\nd\X1(4r^2(3r^4+f^2(r))+8r^3f(r)\pz fr(r)+(r^4+3f^2(r))
\pzd f\Y1)}{\a(r)(r^4+f^2(r))^2(\ell^4+r^4+f^2(r))}\\
{}+\frac{6\ov M}{\a^2(r)(r^4+f^2(r))\o(r)}
\X3(\pz\vF r(r)\X3(-(\ell^4+r^4+f^2(r))\o(r)\pz\a r(r)\\
{}+\a(r)\X2(2\o(r)\X2(2r^3+f(r)\pz fr(r)\Y2)+(\ell^4+r^4+f^2(r))\pz \o
r(r)\Y2)\Y3)\\
+\a(r)(\ell^4+r^4+f^2(r))\o(r)\pz{^2\vF}{r^2}(r)\Y3).
\end{multline}

By using Eqs \er{D.117}--\er{D.120}, $\g(r)$ and $\ov\rho(r)$ are given in
terms of $\vF(r)$, $f(r)$, $\o(r)$, $\a(r)$, i.e.
\bml{D.121}
\ov\rho(r)=\frac{\ell^4\ef\nd}{(\ell^4+r^4+f^2(r))(r^4+f^2(r))}
\X3(3\X2(2r+f(r)\,\pz fr(r)\Y2)^2-4\X2(2f(r)r-r\,\pz fr(r)\Y2)^2\Y3)\\
{}-\frac23\,\ef{2(n+1)}\edn
\end{multline}
and Eq.\ \er{D.116} for $\g(r)$.

Moreover, we are interested in \e s for small $r$, i.e.\ $r\ll 1$. Using Eqs
\er{D.117}--\er{D.120} we derive the following \e s
\bea{D.122}
\pz{\o(r)}r&=&-\frac{f(r)+2\pz fr(r)}{f(r)}\,\o(r)\\
\pz{^2\a(r)}{r^2}&=&\frac1{3n\nd f^2(r)(\ell^4+f^2(r))^2} \nn\\
&&\X4(-3n\nd f^2(r)(\ell^4+f^2(r))^2\pz\a r(r)\X2(3f^2(r)\X2(f(r)-\pz fr(r)\Y2)\nn\\
&&\ {}+2\ell^4\X2(f(r)+5\pz fr(r)\Y2)\Y2)
+\a(r)\X3(-16\ef\nd n\a(r)f^6(r)+2\nd f^4(r)\nn\\
&&\ \ \X3(8\ef n\a(r)f^2(r)-3n\X2(4f(r)\pz fr(r)+2\pzd f \nn\\
&&\ \ \ {}+f(r)\X2(4C_1\a(r)+f(r)+2\ov Mf(r)\pzd\vF\Y2)\Y2)\Y3)\nn\\
&&\ {}+\ell^4f^2(r)\X3(-40\ef\nd n\a(r)f^2(r)\nn\\
&&\ \ {}+\nd\X2(40\ef n\a(r)f^2(r)-3n\X2(14\pzd f \nn\\
&&\ \ \ {}+20f(r)\X2(C_1\a(r)+\pz fr(r)\Y2)
+f^2(r)\X2(3+8\ov M\pzd\vF\Y2)\Y2)\Y2)\Y3)\nn\\
&&\ {}+3\ell^8\X3(-8\ef\nd n\a(r)f^2(r)+\nd \X2(8\ef n\a(r)f^2(r)-n\X2(11f(r)\pz fr(r)\nn\\
\noalign{\eject}
&&\ \ {}+18\pzd f+f(r)\X2(12C_1\a(r)
+f(r)-4\ov Mf(r)\pzd\vF\Y2)\Y2)\Y2)\Y3)\Y3)\Y4) \label{D.123}\\
\pz{^2f(r)}{r^2}&=&\frac1{3n\nd \a(r)f(r)(\ell^4+f^2(r))^2} \nn\\
&&\X4(f^2(r)\X3(-8\ef n\a^2(r)f^2(r) \nn\\
&&{}+3n^2\X3(2C_1\a^2(r)f(r)
+\pz fr(r)\X2(f(r)\pz\a r(r)+\a(r)\X2(f(r)+\pz fr(r)\Y2)\Y2)\Y3) \nn\\
&&\ {}+2n\X3(6C_1\a^2(r)-2\ef n\a^2(r)f^2(r)+2\ef\nd \a^2(r)f^2(r) \nn\\
&&\ \ {}+3\pz fr(r)\X2(f(r)\pz\a r(r)+\a(r)\X2(f(r)+\pz fr(r)\Y2)\Y2)\Y3)\Y3) \nn\\
&&\ {}+\ell^4\X3(-8\ef n\a^2(r)f^2(r)+2n\X2(6C_1\a^2(r)f(r)-2\ef n
\a^2(r)f^2(r)\nn\\
&&\ \ {}+2\ef\nd\a^2(r)f^2(r)+3\pz fr(r)\X2(\a(r)f(r)+f(r)\pz\a r(r)+3\a(r)\pz fr(r)\Y2)\Y2) \nn\\
&&\ {}+3n^2\X2(2C_1\a^2(r)f(r)+\pz fr(r)\X2(f(r)\pz\a r(r)+\a(r)
\X2(f(r)+3\pz fr(r)\Y2)\Y2)\Y2)\Y3)\Y4) \label{D.124}\\
\pz{^2\vF(r)}{r^2}&=&\frac1{\a(r)(\ell^4+f^2(r))\o(r)} \nn\\
&&\X3(\frac{\ef n\X1(5(\ef2-1)-4\Y1)}n
-\frac{3\ell^4\nd f^2(r)\pzd f}{\a(r)f^2(r)(\ell^4+f^2(r))}\nn \\
&&+\frac{6\ov M\,\pz\vF r(r)\X2((\ell^4+f^2(r))\X1(\pz \o r(r)-
\pz\a r(r)\,\o(r)\Y1)+2\a(r)\o(r)f(r)\pz fr(r)\Y2)}
{\a^2(r)f^2(r)\o(r)}\Y3)\q\q \label{D.125}
\end{eqnarray}

Eq.\ \er{D.122} can easily be integrated
\beq{D.126}
\o(r)=f^2(r)e^{\ov c-r}
\end{equation}
where $\ov c$ is an integration \ct. Moreover, the remaining \e s are hard to
handle. The interesting point in Eqs \er{D.123}--\er{D.125} is that we can
pose an initial Cauchy problem for $r=0$. This means a \so\ can be
nonsingular at $r=0$. This is of course beyond the scope of this work.

We can consider the \cn\ $\gd\ov\G,\a,\b\g,$ and the Christoffel symbols for
a \nos\ metric considered here. The Christoffel symbols are formed for a \s\
part of $g_\m$
\beq{D.127}
g_\(\m)=\mt{-\a & 0 & 0 & 0\\
0 & -r^2 & 0 & 0 \\
0 & 0 & -r^2\sin^2\th & 0\\
0 & 0& 0& \g }.
\end{equation}
An \e\ of motion for a particle reads
\beq{D.128}
\pz{^2x^\a}{\tau^2}+\gd\ov\G,\a,(\b\g),\,\pz{x^\b}\tau\,\pz{x^\g}\tau=0
\end{equation}
which is a geodetic \e\ for a \cn\ $\gd\ov\G,\a,\b\g,$. They are called
nonextremal geodetics. In some sense Eq.~\er{D.128} are very closed to the
Galileo principle.

Moreover, we can also consider
\beq{D.129}
\pz{^2x^\a}{\tau^2}+\genfrac\{\}{0pt}0 \a{\b\g}_g \pz{x^\b}\tau
\cdot \pz{x^\g}\tau=0
\end{equation}
extremal geodetics. They can be reduced as in the Riemannian case to Eqs
\er{D.24}--\er{D.27}, \er{D.51}--\er{D.52} by making a replacement $\ov A(r)
\to \a(r)$, $\ov B(r)\to \g(r)$, $\gd\wt\G,\a,\b\g, \to \genfrac\{\}{0pt}1
\a{\b\g}_g$. These \e s can be considered as \e s of motion for massive point
particles and photons under an \an\ \ac.

The Lagrangian \er{D.63} is in some sense  a toy model in comparison to the
full formalism of the \eu\nos\ Jordan--Thiry Theory developed in Refs \cite4,
\cite5, \cite7. In this theory the lagrangian for a scalar field reads
\beq{D.130}
\wh L(\vF)=\X2(\ov M\wt g^\(\g\nu)+n^2g^\[\m]g_{\d\mu}\wt g^\(\d\g)
\Y2)\pa_\nu\vF \cdot \pa_\g\vF +
\X2(\frac{e^{\nd \vF}}{n+2}-\frac{e^{n\vF}}n\Y2)+
\wh{\ov \rho}e^{-\nd\vF},
\end{equation}
$\wh{\ov\rho}$ is a density of a matter (a dust).

An energy momentum tensor for a field $\vF$ is as follows:
\bml{D.131}
\wh T_{\a\b}(\vF)=(g_{\k\a}g_{\o\b}+g_{\o\a}g_{\k\b})\wt g^\(\g\k)\wt
g^\(\nu\o) \X2(\frac{n^2}2 (g^{\z\mu}g_{\nu\z}-\gd\d,\mu,\nu,)\pa_\mu\vF
+\ov M\pa_\nu\vF\Y2)\pa_\g\vF\\ {}-g_{\a\b}\X2(\X1(\ov M\wt g^\(\g\nu)
+n^2g^\[\m]g_{\d\mu} \wt g^\(\g\d)\Y1) \pa_\nu\vF \cdot \pa_\g\vF
+\X2(\frac{e^{\nd \vF}}{n+2}-\frac{e^{n\vF}}n\Y2)+\wh{\ov \rho} \ef{-\nd}\Y2)
\end{multline}
and the trace of $\wh T_{\a\b}$
\beq{D.132}
\wt T=g^{\a\b}\wh T_{\a\b}(\vF)=-2\ov M\wt g^\(\g\nu)\pa_\nu\vF\cdot \pa_\g\vF
-2n^2 g^{\z\nu}g_\[\mu\z]\wt g^\(\g\mu)\pa_\nu\vF \cdot\pa_\g\vF - 4U(\vF)
\end{equation}
where
\beq{D.133}
\wt g^\(\g\nu)g_\(\g\a)=\gd\d,\nu,\a,.
\end{equation}
$\wt g^\(\g\nu)$ is an inverse tensor for a \s\ part of $g_{\g\a}, g_\(\g\a)$,
\beq{D.134}
U(\vF)=\X2(\frac{e^{\nd \vF}}{n+2}-\frac{e^{n\vF}}n\Y2)+
\wh{\ov\rho}\ef{-\nd}.
\end{equation}

\def\ef#1{e^{#1\vF}}
Now we proceed all the mathematical manipulations starting from \er{D.53} up
to \er{D.129}, shifting $L(\vF)$ to $\wh L(\vF)$ and $T_{\a\b}$ to $\wh
T_{\a\b}$. In this way we get
\bea{D.135}
\nad{\rm tot}{\wh T}_{\a\b}&=&\ef{-\nd} \wh{\ov \rho}u_\a u_\b \\
\nad{\rm tot}{\wh T}&=&\nad{\rm tot}{\wh T}_{\a\b}g^{\a\b}
=-2\ov M\wt g^\(\g\nu)\pa_\nu\vF \cdot \pa_\g\vF \nn\\
&&\q{}-2n^2g^{\z\nu}g_\[\mu\z]\wt g^\(\g\nu)\cdot \pa_\nu\vF\cdot \pa_\g\vF
-4\ef n\edn-3\wh{\ov\rho}\ef{-\nd} \label{D.136}\\
\wt{\wh T}_{\a\b}&=&\nad{\rm tot}{\wh T}_{\a\b}
-\frac12\,g_{\a\b}\nad{\rm tot}{\wh T}=
\ef{-\nd}\wh{\ov\rho} u_\a u_\b \nn\\
&&\q {}+(g_{\k\a}g_{\o\b}+g_{\o\a}g_{\k\b})\wt g^\(\g\k)\wt g^\(\nu\o)
\X2(\frac{n^2}2(g^{\z\mu}g_{\nu\z}-\gd\d,\mu,\nu,)\pa_\mu\vF
+\ov M\pa_\nu\vF\Y2)\pa_\g\vF \nn\\
&&\q {}+g_{\a\b}\X2(\ef n\edn +\frac12\,\ov \rho \ef{-\nd}\Y2). \label{D.137}
\end{eqnarray}
An \e\ for a scalar field $\vF$ reads
\beq{D.138}
-2\ov\nabla_\g\X1(\ov M\wt g^\(\g\nu)+n^2g^\[\m]g_{\d\mu}\wt g^\(\d\g)\pa_\nu
\vF\Y1) + \ef n\edn -\nd\wh{\ov\rho}\ef{-\nd}=0
\end{equation}
or
\bml{D.139}
-\frac2{\sqrt{-g}}\pa_\g\X2(\sqrt{-g}\X1(\ov M\wt g^\(\g\nu)+n^2g^\[\m]
g_{\d\mu}\wt g^\(\d\g)\Y1)\pa_\nu\vF\Y2)\\
{}+\ef n\edn-\nd\wh{\ov\rho}\ef{-\nd}=0.
\end{multline}

\def\pzd#1{\X2(\pz{#1}r\Y2)^2}
Considering spherically \s\ and static case we get
\bea{D.140}
\wt{\wh T}_{11}&=& 2\ov M\pzd\vF - \a\X2(\ef n\edn + \frac{\wh{\ov\rho}}2
\ef{-\nd}-\frac{2\o^2n^2}{\o^2-\a\g}\pzd\vF\Y2) \\
\wt{\wh T}_{44}&=& \frac{\o^2}{\a^2}\pzd\vF \X2(2\ov M-\frac{n^2\o^2}{\o^2-\a\g}
\Y2)+\g\X2(\ef n\edn + \frac{3\wh{\ov \rho}}2 \ef{-\nd}\Y2) \label{D.141}\\
\wt{\wh T}_{22}&=& -\b\X2(\ef n\edn +\frac{\ov \rho}2 \ef{-\nd}\Y2) \label{D.142}\\
\wt{\wh T}_{33}&=& \sin^2\th \wt{\wh T}_{22} \label{D.143}\\
\frac{\wt{\wh T}_{[23]}}{\sin\th}&=& f\X2(\ef n \edn +\frac{\wh{\ov\rho}}2
\ef{-\nd}\Y2)
\label{D.144}
\end{eqnarray}

Eq.\ \er{D.139} can be rewritten as
\bml{D.145}
\frac2{\b^2+f^2}\,\pz{}r \X3(\frac{(\b^2+f^2)\o}\a
\X2(-\ov M+\frac{n^2\ell^4}{\b^2+f^2}\,\pz\vF r\Y2)\Y3)\\
{}+\ef n(\ef2-1)-\nd \wh{\ov\rho} \ef{-\nd}=0.
\end{multline}

Now we use Eqs \er{D.87}--\er{D.88} and Eqs \er{D.100}--\er{D.106} in the
case of $\wt{\wh T}_{\a\b}$ and we get
\bml{D.146}
\wh{\ov\rho}(r)=\frac2{3\g}\X3[\frac{\o^2}{8\a^2}\X2(3\pzd\Ps+4C^2\Y2)
-\frac{\o^2}{\a^2}\pzd\vF \X2(2\ov M-\frac{n^2\o^2}{\o^2-\a\g}\Y2)\Y3]
\ef\nd\\{} - \frac23\,\ef n\edn \ef\nd
\end{multline}
which differs from Eq.\ \er{D.107}.

In a similar way as before we get
\bml{D.147}
-\frac12\,\pz{^2\Ps}{r^2}-\frac18\X2(\pzd\Ps+4C^2\Y2)
+\frac1{4\a}\,\pz\a r\,\pz\Ps r\\
{}+\X2(\frac{\o^2}{2\a\g}\,\pz\Ps r
+\frac1{2\g}\,\pz\g r\Y2)\X2(\frac1{2\a}\,\pz\a r-\frac{\o^2}{2\a\g}\,\pz\Ps r
-\frac1{2\g}\,\pz\g r\Y2)+\pz{}r\X2(\frac{\o^2}{2\a\g}+\frac1{2\g}\,\pz\g r\Y2)\\
{}=2\ov M\X2(1+\frac{\o^2\a}{3\g}\Y2)-\frac{n^2\o^4\a}{3\g(\o^2-\a\g)}
\pzd\vF\\ {}- \frac{\o^2}{12\a\g}\X2(3\pzd\Ps+4C^2\Y2)
-\frac{2\a}3 \,\ef n\edn
\end{multline}
\bml{D.148}
\hskip-10pt
\frac1{4\a}\X2(2fC-\b\,\pz\Ps r\Y2)+\frac1{8\a}\X2(2fC-\b\,\pz\Ps r\Y2)
\pz{}r\log\X1(\o^2(\b^2+f^2)\Y1)+\frac C{4\a}\X2(f\,\pz\Ps r+2\b C\Y2)\\
\hskip-20pt
{}=-\b\X3(\frac23\,\ef n\edn +\frac{\o^2}{12\a^2\g}\X2(3\pzd\Ps+4C^2\Y2)
-\frac{\o^2}{3\g\a^2}\pzd\vF \X2(2\ov M-\frac{n^2\o^2}{\o^2-\a\g}\Y2)\Y3)
\mskip-15mu
\end{multline}

\bml{D.149}
\pz{}r\X3(\frac1{4\a}\X2(f\pz\Ps r-2\b C\Y2)\Y3)-\frac C{4\a}\X2(2fC-\b\pz\Ps r\Y2)
+\frac1{8\a}\X2(f\pz\Ps r+2\b C\Y2)\X2(\frac1\a\,\pz\a r+\frac{\o^2}{\a\g}
\,\pz\Ps r+\frac1\g\,\pz\g r\Y2)\\
{}-f\X3(\frac23\,\ef n\edn-\frac{\o^2}{12\a^2\g}\X2(3\pzd\Psi+4C^2\Y2)
-\frac{\o^2}{3\g\a^2}\pzd\vF\X2(2\ov M-\frac{n^2\o^2}{\o^2-\a\g}\Y2)\Y3)\\
{}=C_1
\end{multline}

\bml{D.150}
-\frac2{(\b^2+f^2)\o}\,\pz{}r \X3(\frac{(\b^2+f^2)\o}\a\X2(-\ov M
+\frac{n^2\ell^4}{\b^2+f^2}\Y2)\pz\vF r\Y3)+\ef
n\X2(\frac53\,\ef2-\frac{5n+4}{3n}\Y2)\\
{}-\frac{2\nd}{3\g}\X3(\frac{\o^2}{2\a^2}\X2(3\pzd\Ps+4C^2\Y2)-\frac{\o^2}{\a^2}
\pzd\vF\X2(2\ov M-\frac{n^2\o^2}{\o^2-\a\g}\Y2)\Y3)
\end{multline}
The derived \e s differ from Eqs \er{D.110a}--\er{D.113}.

\def\ef#1{e^{#1\vF(r)}}
\def\pzd#1{\X2(\pz{#1}r(r)\Y2)^2}
We suppose as before that $\b=r^2$, getting four \e s for four \f s $\a(r)$,
$f(r)$, $\o(r)$ and $\vF(r)$. $\Ps,C,\g$ have the same meaning as before,
$C_1$ is an integration \ct\ (as before). Using definitions of $\Ps,C$ and
Eq.~\er{D.116} for $\g$, one gets after some simplifications
\bg{pusty}
\frac16\X4(\frac{4\ef n\X1(-2+(\ef2-1)n\Y1)\a(r)}{n\nd} \hskip200pt\nn\\
{}+\frac{1}{\a(r)(r^4+f^2(r))(\ell^4+r^4+f^2(r))\o(r)}\X3(3\pz\a r(r)
\X2(\o(r)\X2(r^3(8r^4+\ell^4(r+4))\nn\\
{}+f(r)\X2((\ell^4+8r^3)f(r)+2(\ell^4+2r^4+2f^2(r))\pz fr(r)\Y2)\Y2)\nn\\
{}+3(r^4+f^2(r))(\ell^4+r^4+f^2(r))\pz\o r(r)\Y2)\nn\\
{}-3(r^4+f^2(r))(\ell^4+r^4+f^2(r))\o(r)\pz{^2\a}{r^2}(r)\Y3)\nn\\
{}-\frac1{(r^4+f^2(r))^2(\ell^4+r^4+f^2(r))^2\o^2(r)}\X3(12(r^4+f^2(r))^2
(\ell^4+r^4+f^2(r))^2\pzd\o\nn\\
{}+\o^2(r)\X3(6r^6\X1(\ell^8(r-4)+2\ell^4(r-3)r^4+6r^8\Y1)
+r^4(\ell^4+r^4)(7\ell^4+3r^4)\pzd f\nn\\
{}+3f^6(r)\X2(4r^2+3\pzd f\Y2)+f^4(r)\X2(4r^2\X1(15r^4+\ell^4(3r+13)\Y1)
\label{D.151}\\
{}+3(-4\ell^4+7r^4)\pzd f\Y2)
+f^2(r)\X2(2r^2\X1(42r^8+4\ell^4r^4(3r+2)+\ell^8(3r+20)\Y1)\nn\\
{}+(-9\ell^8-2\ell^4r^4+15r^8)\pzd f\Y2)
+6f^5(r)\X2((\ell^4+4r^3)\pz fr(r)+\ell^4\pz{^2f}{r^2}(r)\Y2)\nn\\
{}+r^3f(r)\X2(\X1(24r^8+\ell^8(3r-64)+2\ell^4r^4(3r-44)\Y1)\pz fr(r)
+6\ell^4r(\ell^4+r^4)\pz{^2f}{r^2}(r)\Y2)\nn\\
{}+f^3(r)\X2(\X1(3\ell^8+48r^7+4\ell^4r^3(3r-22)\Y1)\pz fr(r)
+6\ell^4(\ell^4+2r^4)\pz{^2f}{r^2}(r)\Y2)\Y3)\nn\\
\noalign{\eject}
{}+3(r^4+f^2(r))(\ell^4+r^4+f^2(r))\o(r)\X3(\X2(r^3(8r^4+\ell^4(r+4))\nn\\
{}+f(r)\X2((\ell^4+8r^3)f(r)+2(\ell^4+2r^4+2f^2(r))\pz fr(r)\Y2)\Y2)\pz\o r(r)\nn\\
{}-2(r^4+f^2(r))(\ell^4+r^4+f^2(r))\pz{^2\o}{r^2}\Y3)\Y3)\Y4)
=\frac23\,\ov M\X3(3+\frac{\ell^4\a^2(r)}{\ell^4+r^4+f^2(r)}\Y3)
\pzd\vF,\nn
\end{gather}

\bml{D.152}
-\frac{12\X1(r^4-f^2(r)+rf(r)\pz fr(r)\Y1)}{\a(r)(r^4+f^2(r))}\\
+\frac{6r\X1(2f(r)-r\pz fr(r)\Y1)\X1(4r^3f(r)+(-r^4+f^2(r))\pz fr(r)\Y1)}
{\a(r)(r^4+f^2(r))^2}
\X3(\frac{8\ef n((\ef2-1)-2)}{n\nd}\\
{}+\frac{4\ell^4\X1(12r^6+4r^2f^2(r)+8r^3f(r)\pz fr(r)+r^4\pzd f+3f^2(r)
+3f^2(r)\pzd f}
{\a(r)(r^4+f^2(r)^2)(\ell^4+r^4+f^2(r))}\\
{}-\frac{(r^4+f^2(r))\X1(\ell^4n^2+2\ov M(r^4+f^2(r))\Y1)\pzd \vF\Y1)}
{\a(r)(r^4+f^2(r)^2)(\ell^4+r^4+f^2(r))}\Y3)\\
{}-\frac{12\X1(r^4-f^2(r)+rf(r)\pz fr(r)\Y1)\X1(\o(r)\X1(2r^3
+f(r)\pz fr(r)\Y1)+(r^4+f^2(r))\pz\o r(r)\Y1)}
{\a(r)(r^4+f^2(r))^2\o(r)}=0,
\end{multline}

\bml{D.153}
-8C_1-\frac{4\pz\a r(r)\pz fr(r)}{\a^2(r)}
-\frac{8r^2\X1(-2f(r)+r\pz fr(r)\Y1)\X1(r^4-f^2(r)+rf(r)\pz fr(r)\Y1)}
{\a(r)(r^4+f^2(r))^2}\\
{}-\frac23\,f(r)\X3(\frac{8\ef n (-2+(\ef2-1)n)}{n\nd}\\
{}+\frac1{\a(r)(r^4+f^2(r))^2(\ell^4+r^4+f^2(r))}\X3(
4\ell^4\X2(12r^6+4r^2f^2(r)+8r^3f(r)\pz fr(r)\\
{}+r^4\pzd f+3f^2(r)\pzd f-(r^4+f^2(r))\X1(\ell^4n^2+2\ov M(r^4+f^2(r))\Y1)
\pzd\vF\Y2)\Y3)\Y3)\\
{}+\frac{4\dfrac{df}{dr}(r)\X2(\dfrac{2r^3+f(r)\pz fr(r)}{r^4+f^2(r)}
+\dfrac{\pz\o r(r)}{\o(r)}\Y2)+4\dfrac{d^2f}{dr^2}(r)}{\a(r)}=0,
\end{multline}

\bml{D.154}
\frac{\ef n(5(\ef2-1)n-4)}n
+\frac{2\ell^4\nd (\ell^4n^2+2\ov M(r^4+f^2(r)))\pzd \vF}
{\a(r)(r^4+f^2(r))(\ell^4+r^4+f^2(r))}\\
{}-\frac{\ell^4\nd \X2(4r^2(3r^4+f^2(r))+8r^3f(r)\pz fr(r)+(r^4+3f^2(r))
\pzd f\Y2)}{\a(r)(r^4+f^2(r))(\ell^4+r^4+f^2(r))}\\
{}+\frac1{\a^2(r)(r^4+f^2(r))\o(r)}
\X3(6\,\pz \vF r(r)\X2((\ell^4n^2-\ov M(r^4+f^2(r)))\o(r)\pz \a r(r)\\
\noalign{\eject}
{}+\a(r)\X2(2\ov M\o(r)\X2(2r^3+f(r)\pz fr(r)\Y2)+
(-\ell^4n^2+\ov M(r^4+f^2(r)))\pz\o r(r)\Y2)\Y2)\\
{}+6\a(r)(-\ell^4n^2+\ov M(r^4+f^2(r)))\o(r)\pz{^2\vF}{r^2}(r)\Y3)=0.
\end{multline}

Moreover, we are still interested in the limit for $r\to0$, i.e.\ $r\ll 1$.
After some algebra one gets the following \e s from Eqs
\er{D.151}--\er{D.154}.

\def\lf{(\ell^4+f^2(r))}
\beq{D.155}
\pz\o r(r)=-\frac{\o(r)\X1(f(r)+2\pz fr(r)\Y1)}{f(r)}
\end{equation}

\bml{D.156}
\pz{^2\a}{r^2}(r)=\frac1{3n\nd f^2(r)(\ell^4+f^2(r))^2}\\
\X4(-3n\nd f(r)\lf\pz\a r(r)\X2(3f^2(r)\X2(f(r)+2\,\pz fr(r)\Y2)
+2\ell^4\X2(f(r)+5\,\pz fr(r)\Y2)\Y2)\\
{}+\a(r)\X3(-16\ef\nd n\a(r)f^6(r)+2\nd f^4(r)\\
\X2(8\ef n\a(r)f^2(r)-3n\X2(4f(r)\pz fr(r)+2\pzd f+f(r)\X2(4C_1\a(r)
+f(r)+2\ov Mf(r)\pzd\vF\Y2)\Y2)\Y2)\\
{}+\ell^4f^2(r)\X3(-40\ef \nd n\a(r)f^2(r)+\nd\X2(40\ef n\a(r)f^2(r)\\
{}-3n\X2(14\pzd f+20f(r)\X2(C_1\a(r)+\pz fr(r)\Y2)+f^2(r)\X2(3+8\ov M\pzd \vF
\Y2)\Y2)\Y2)\Y3)\\
{}+3\ell^8\X3(-8\ef\nd n\a(r)f^2(r)+\nd \X2(8\ef n\a(r)f^2(r)
-n\X2(11f(r)\pz fr(r)\\
{}+18\pzd f+f(r)\X2(12C_1\a(r)+f(r)+4\ov Mf(r)\pzd\vF\Y2)\Y2)\Y2)\Y3)\Y3)\Y4)
\end{multline}

\bml{D.157}
\pz{^2f}{r^2}(r)=\frac1{3n\nd \a(r)f(r)(\ell^4+f^2(r))^2}\\
\X4(f^2(r)\X3(-8\ef n\a^2(r)f^2(r)+3n^2\X2(2C_1\a^2(r)f(r)+\pz fr(r)
\X2(f(r)\pz\a r(r)+\a(r)\X2(f(r)+\pz fr(r)\Y2)\Y2)\Y2)\\
{}+2n\X2(6C_1\a^2(r)f(r)-2\ef n\a^2(r)f^2(r)+2\ef\nd\a^2(r)f^2(r)\\
{}+3\pz fr(r)\X2(f(r)\pz\a r(r)+\a(r)\X2(f(r)+\pz fr(r)\Y2)\Y2)\Y2)\Y3)\\
{}+\ell^4\X3(-8\ef n\a^2(r)f^2(r)+2n\X2(6C_1\a^2(r)f(r)-2\ef n\a^2(r)f^2(r)
+2\ef\nd\a^2(r)f^2(r)\\
{}+3\pz fr(r)\X2(\a(r)f(r)+f(r)\pz\a r(r)+3\a(r)\pz fr(r)\Y2)\Y2)\\
+3n\X2(2C_1\a^2(r)f(r)+\pz fr\X2(f(r)\pz\a r(r)+\a(r)\X2(f(r)+3\pz fr(r)
\Y2)\Y2)\Y2)\Y3)\Y4)
\end{multline}

\bml{D.158}
\pz{^2\vF}{r^2}(r)=\frac1{6\ov Mn\a(r)f(r)\lf^2}
\X3(6\ov Mnf(r)\lf^2\pz\a r(r)\pz\vF r(r)\\
{}+\a(r)\X2(f(r)\X2(-5\ef \nd n\a(r)f^2(r)
\lf\\
{}+\ef n(5n+4)\a(r)f^2(r)\lf+3\ell^4 n\nd \pzd f\Y2)\\
+6\ov Mn\lf\X2(f^3(r)+\ell^4\X2(f(r)+2\pz fr(r)\Y2)\Y2)\pz \vF r\Y2)\Y3)
\end{multline}

Eq.\ \er{D.155} can be easily integrated getting
\beq{D.159}
\o(r)=f^2(r)e^{\ov c-r}
\end{equation}
where $\ov c$ is an integration \ct. Let us notice that Eq.~\er{D.155} is
exactly the same as Eq.~\er{D.122} and \er{D.159} as \er{D.126}.

Let us consider $\wh{\ov\rho}(r)$. One gets
\bml{D.160}
\wh{\ov\rho}(r)=\frac{\ell^4\ef\nd}{(\ell^4+r^4+f^2(r))(r^4+f^2(r))}
\X3[3\X2(2r+f(r)\pz fr(r)\Y2)^2-4r^2\X2(2f(r)-\pz fr(r)\Y2)^2\\
{}-\frac{\o^2}\a\pzd\vF\X2(2\ov M+\frac{\ell^4 n^2}{r^4+f^2(r)}\Y2)\Y3]
-\frac23\,\ef{2\nd}\edn.
\end{multline}

In the case of $r\ll 1$ we have $\o^2=f^4e^{2(\ov c-r)}$.

All the considerations concerning the possibility of nonsigular \so s can be
repeated here. We can also repeat formulae \er{D.127}--\er{D.129} and remarks
below. We give here only some examples.

Further development of this approach including \so s of field \e s and their
applications to the problem of motion of massive particles and photons is beyond
a scope of this work. We consider only some transformation of those equations
which will be useful in further work.

Moreover, let us consider a problem to embed a \spt\ with a metric
\beq{D.161}
ds_1^2=e^{2A(r)}\,dt^2-e^{2B(r)}\,dr^2-r^2(d\th^2+\sin^2\th\,d\vf)
\end{equation}
to a \co ical \so\ with a metric
\beq{D.162}
ds_2^2=dt^2 - R^2(t)\X1(dr^2+r^2(d\th^2+\sin^2\th\,d\vf^2)\Y1).
\end{equation}
In order to do it we should write \er{D.161} in isotropic \cd s (see Refs
\cite{D80}, \cite{D81}, \cite{D82}):
\beq{D.163}
ds_1^2=d\ov t^2-e^{\ov\mu}\X1(d\ov r{}^2+\ov
r{}^2(d\th^2+\sin^2\th\,d\vf^2)\Y1).
\end{equation}
One gets
\bg{D.164}
r \to \ov r=\ov f(r)=\exp\X3(\int\frac{e^{B(r)}}r\,dr\Y3)\\
e^{\mu(r)}=Cr^2\X3(\exp\X3(2\int\frac{e^{B(r)}}r\,dr\Y3)\Y3)^{-1}=
e^{\ov\mu(\ov r)}
\label{D.165} \\
t\to\ov t=te^{A(r)} \label{D.166}
\end{gather}
$C$ is an integration \ct.

The matching condition is as follows:
\bg{D.166a}
e^{\ov\mu(\ov r)}=R^2=e^{\mu(r)}\\
\pz{\mu(r)}{r}=\pz{R^2(\ov t)}r \label{Dn169}
\end{gather}
for $r=r_1$, where $r_1$ means a radius of a sphere on which the match
takes place. Inside this sphere the spherically \s\ and static metric
\er{D.161} describes a \spt. If we use a metric satisfying a GR model
\er{D.15}--\er{D.17} we  can try in some sense to connect an \an\ \ac\ to \co
y.

From \er{D.164}--\er{Dn169} one gets
\bg{Dn170}
\frac{r^2}{C^2}\exp\X2(-2\int\frac{\eb{}}r\,dr\Y2)=R^2(e^{A(r)}t)\\
4\X3(\frac{1-\frac{\eb{}}r}{\eb2-1}\Y3)=te^{A(r)}H(e^{A(r)}t)\label{Dn171}
\end{gather}
where $H=\frac{\dot R}R$  is a Hubble \ct\ for a \co ical model considered
here. \er{Dn170} can be rewritten as
\beq{Dn172}
\frac{r_1}C\exp\X3(-\int_{r_0}^{r_1}\frac{\eb{}}r\,dr\Y3)=R(e^{A(r_1)}t).
\end{equation}
It is easy to see that from the point of view of Eq.\ \er{Dn172} $r_1$ is
arbitrary for $R$ is a scale factor and has not a physical meaning.
However a Hubble \ct\ $H$ has a physical meaning and $r_1$ is a \so\ of the
\e
\beq{Dn173}
4\X3(\frac{1-\frac{e^{B(r_1)}}{r_1}}{e^{2B(r_1)}-1}\Y3)=te^{A(r_1)}H(e^{A(r_1)}t)
.
\end{equation}
In this way $r_1=r_1(t)$. This is different from a Schwarzschild \so. Let us
take a \co ical model which is a paradigm of modern \co y (CDM$\La$ model,
see Section~8). One gets
\beq{Dn174}
\frac{r_1-e^{B(r_1)}}{r_1e^{A(r_1)}(e^{2B(r_1)}-1)}=\frac14
\sqrt{\frac\La3} \,t\coth\X3(\frac32\,e^{A(r_1)}\sqrt{\frac\La3}\,t\Y3)
\end{equation}
(we put 1 for a velocity of light).

$r_1=r_1(t)$ depends on a \co ical time. Moreover, this dependence is weak.
If we take a \co ical model filled with a dust we get
\beq{Dn175}
r_1e^{2B(r_1)}+6e^{B(r_1)}-7r_1=0
\end{equation}
and no dependence of time. In this case $H(t)=\frac2{3t}$.

Let us remind to the reader that $r_1$ is measured in our $L$ unit
($\sim10$\,Mpc). It is interesting to apply our simplified \so\ to
\er{Dn175}. It happens that the left-hand side of Eq.~\er{Dn175} equals zero up
a machine accuracy for all $r$ considered.

\begin{figure}[h]
\hbox to \textwidth{\ing[width=0.45\textwidth]{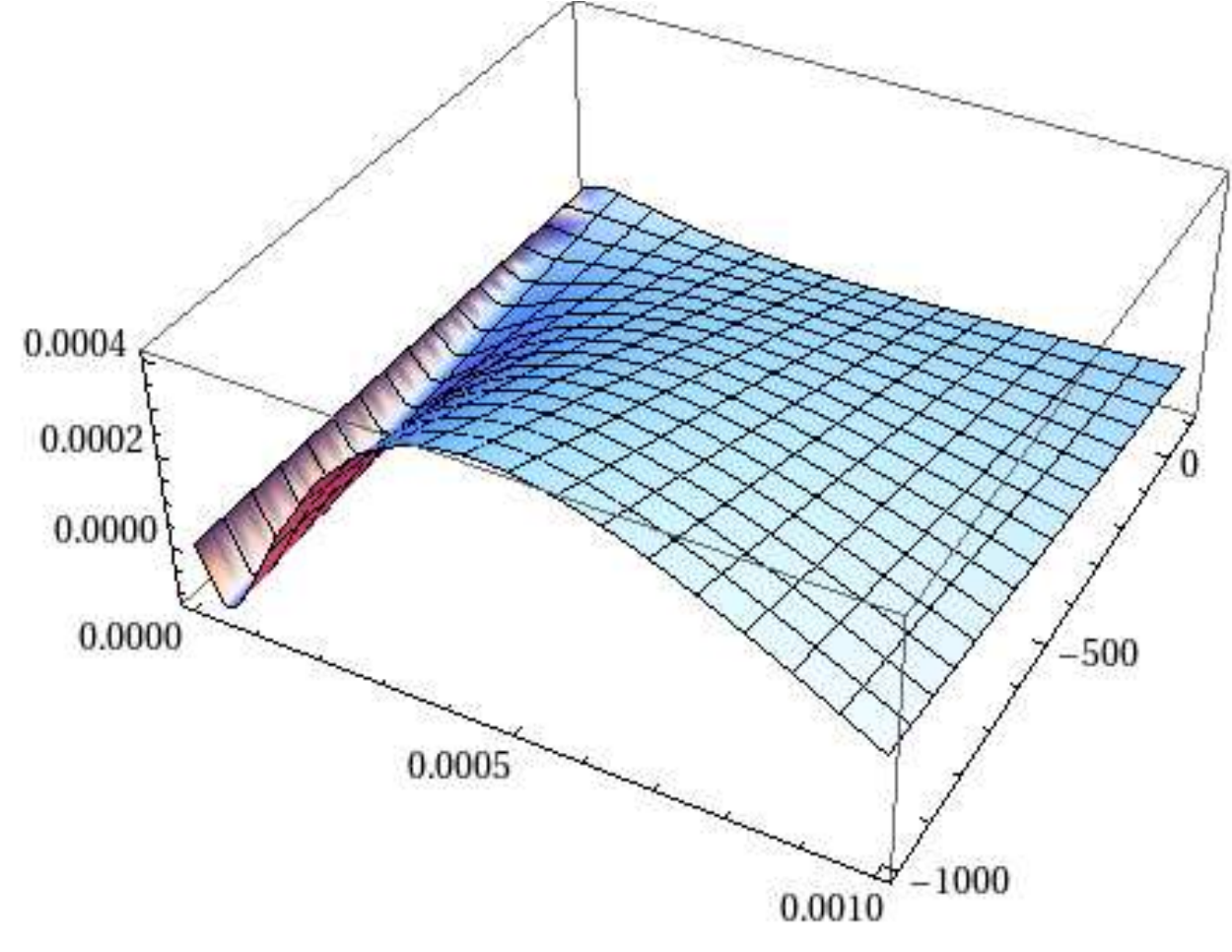}\hfil
\ing[width=0.45\textwidth]{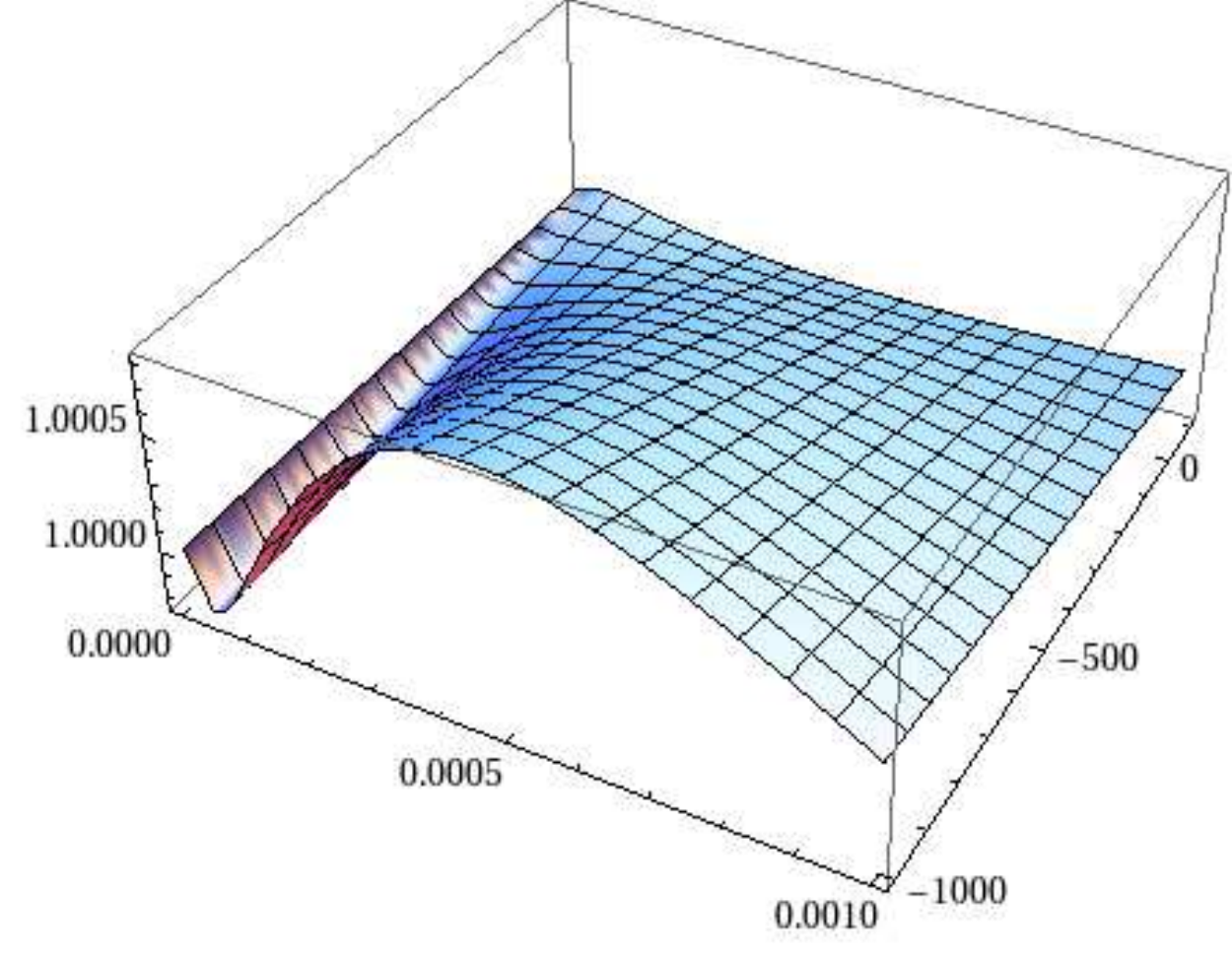}}
\hbox to \textwidth{\hbox to 0.45\textwidth{\hfil(A)\hfil}\hfil
\hbox to 0.45\textwidth{\hfil(B)\hfil}}
\hbox to \textwidth{\ing[width=0.45\textwidth]{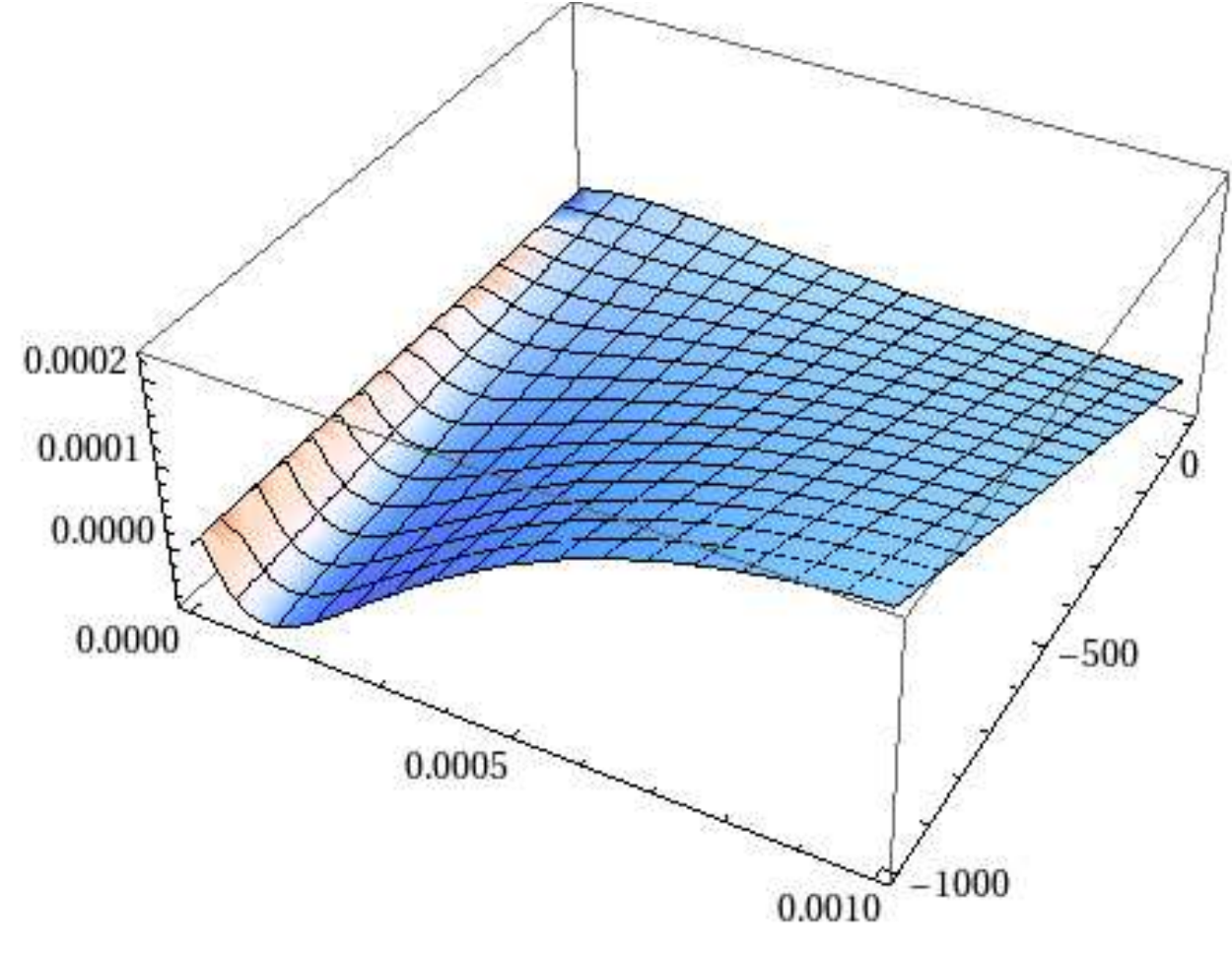}\hfil
\ing[width=0.45\textwidth]{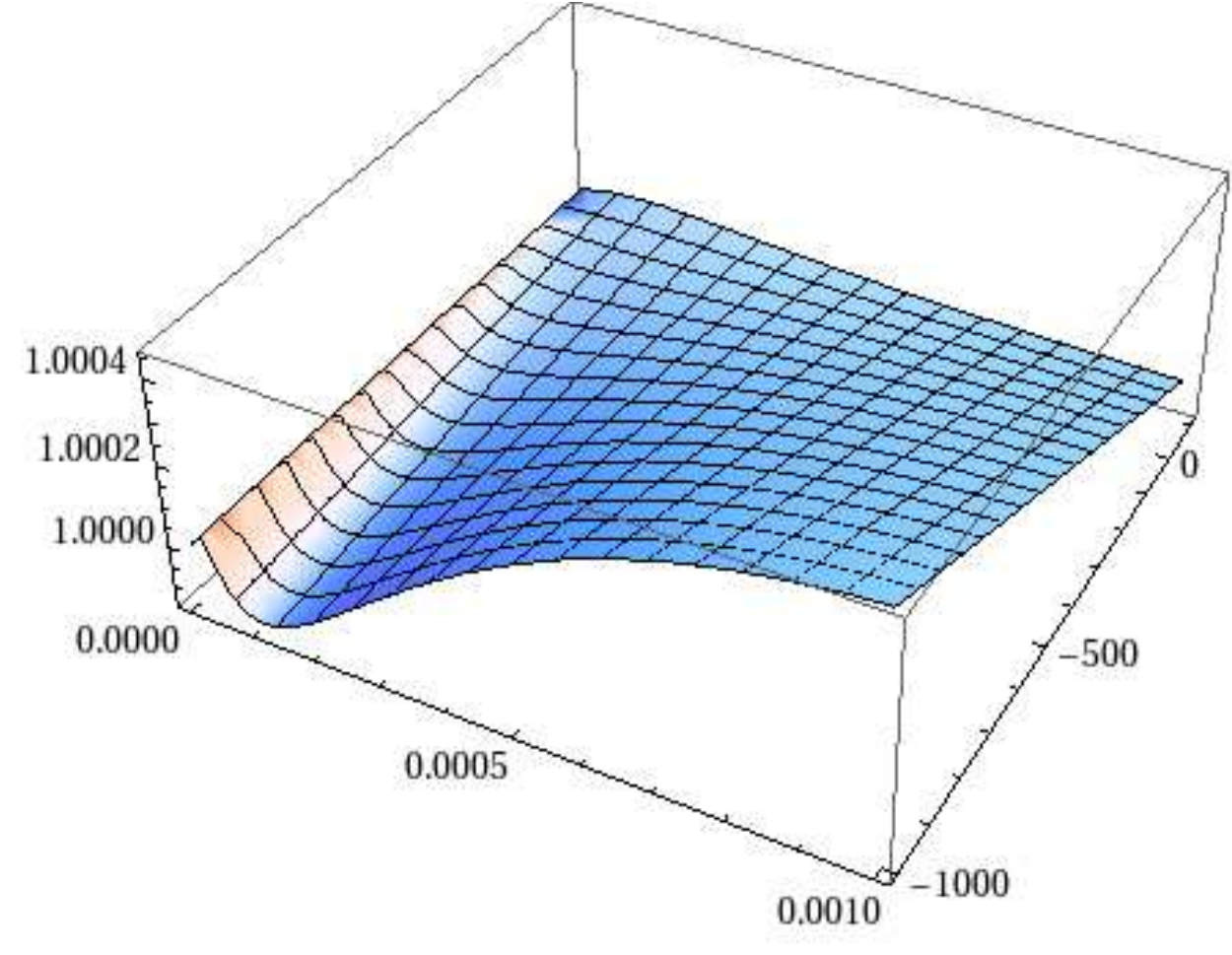}}
\hbox to \textwidth{\hbox to 0.45\textwidth{\hfil(C)\hfil}\hfil
\hbox to 0.45\textwidth{\hfil(D)\hfil}}
\caption{(A)---3D plot of $B(r,B_1)$, (B)---3D plot of $e^{2B(r,B_1)}$,
(C)---3D plot of $A(r,B_1)$,\break (D)---3D plot of $e^{2A(r,B_1)}$ (see text)}
\label{ykll}
\end{figure}

Let us consider the following Cauchy initial problem
\beq{Dn176}
\bal
{}&\pz{^2B(r)}{r^2}=\frac{9-10\eb2+\eb4}{8r^2}\\
&\pz{A(r)}r=\frac{-1+\eb2}{4r}\\
&A(7.36\t10^{-10})=-4\t10^{-12}\\
&B(7.36\t10^{-10})=4\t10^{-12}\\
&\pz Br(7.36\t10^{-10})=B_1
\eal
\end{equation}
In this way $A,B$ can be considered as \f s of $r$ and $B_1$:
\beq{Dn177}
A=A(r,B_1), \q B=B(r,B_1).
\end{equation}
The value $7.36\t10^{-10}$ is measured in $L$ unit and it corresponds to
$3.7\t10^2$ in $r_0$ unit. The \f s \er{Dn177} are differentiable \wrt both
arguments. On Fig.~\ref{ykll} we give 3D plots of $B(r,B_1)$, $e^{2B(r,B_1)}$,
$A(r,B_1)$, $e^{2A(r,B_1)}$. It is easy to find that $B_1=5\t10^{11}$
corresponds to $\wt B_1=1$ from the problem \er{D.37}.

Let us consider the following problem connected with the embedding of the
\so\ in a \co ical \spt. If we remove a ball of radius $r_2$ from the \co
ical space (in our case this is $E^3$) and put here space with a geometry
described by metric \er{D.161}, we should remove also a matter (a dust).

Moreover, we should take such a radius $r_2$ that a total mass of the model
\er{D.161} is equal to the mass of a dust. According to \cite{D82} a mass
inside a ball of radius $r_2$ equals
\beq{Dn178}
m_1(r_2)=\frac12\,r_2(1-e^{-2B(r_2)}).
\end{equation}
Moreover, a mass of a dust inside a ball of radius $r_2$ equals
\beq{Dn179}
m_2(r_2)=\frac43\,\pi r_2^3\rho_{\rm tot}\,.
\end{equation}
Both masses must be equal
\beq{Dn180}
m_1(r_2)=m_2(r_2).
\end{equation}
$\rho_{\rm tot}$ is a density of a dust in a \co ical model
\beq{Dn181}
\rho_{\rm tot}=1.878\t 10^{-26}h^2{\rm \frac{kg}{m^3}}
\end{equation}
where $h=0.7$ (see Section 8).

Moreover, we are using our system of units $L,T,M_0$ and we should transform
\er{Dn181} to our density unit. In this way we get the following \e\ from
\er{Dn178}--\er{Dn181}
\beq{Dn182}
-1+e^{-2B(r_2)}+0.182\t 10^{-4}r_2^2=0.
\end{equation}
Moreover, it could be reasonable to have $r_1=r_2$. Is it possible? The
answer is \ti{yes}. Let us consider the system of \e s
\beq{Dn183}
\bga
-1+e^{-2B(r_1,B_1)}+0.146\t10^{-4}r_1^2=0\\
r_1e^{2B(r_1,B_1)}+6e^{B(r_1,B_1)}-7r_1=0
\ega
\end{equation}
where we consider $B=B(r,B_1)$ as a \f\ of two variables $r$ and $B_1$ (see
Eqs \er{Dn176}--\er{Dn177}. The first \e\ gives us a match of masses, the
second a match of geometries.

The \so\ reads
\beq{Dn184}
\bal r_1&=0.0036536\\
B_1&=14.311691. \eal
\end{equation}
$r_1\ll 1$ and the results are consistent with our assumptions.
\beq{Dn185}
r_1\simeq 36536.9\,{\rm pc}=119167\,{\rm ly}=7.5361\t10^9\,{\rm AU}
=1.12742\t10^{18}\,{\rm km}.
\end{equation}

However, this value is quite big and in a ball of radius $r_1$ we find many
stars, not only the \SS. Moreover, it can be considered as a preliminary
result. Let us consider an initial value problem with $B_1=14.311691$ for
\er{Dn176} with an \e\ for $\wt\vF(r)$
\beq{Dn186}
\bal
{}&\pz{^2\wt\vF}{r^2}(r)=\frac{-1+\eb2}{4r^2}\\
&\wt\vF(7.36\t10^{-10})=0\\
&\pz{\wt\vF}r(7.36\t10^{-10})=0
\eal
\end{equation}

\begin{figure}[h]
\hbox to \textwidth{\ing[width=0.45\textwidth]{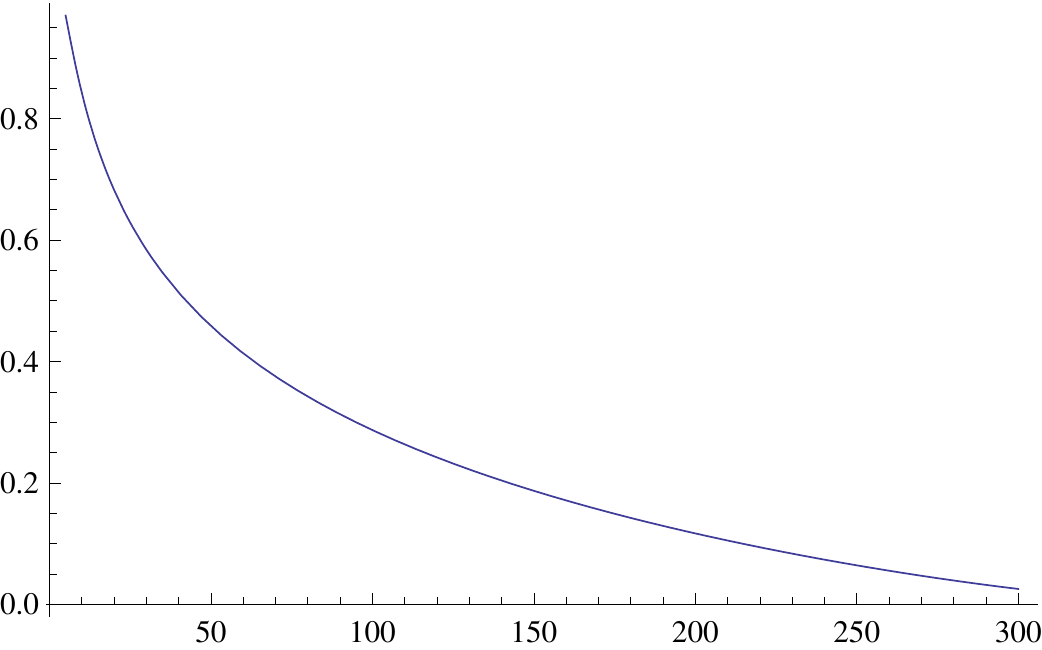}\hfil
\ing[width=0.45\textwidth]{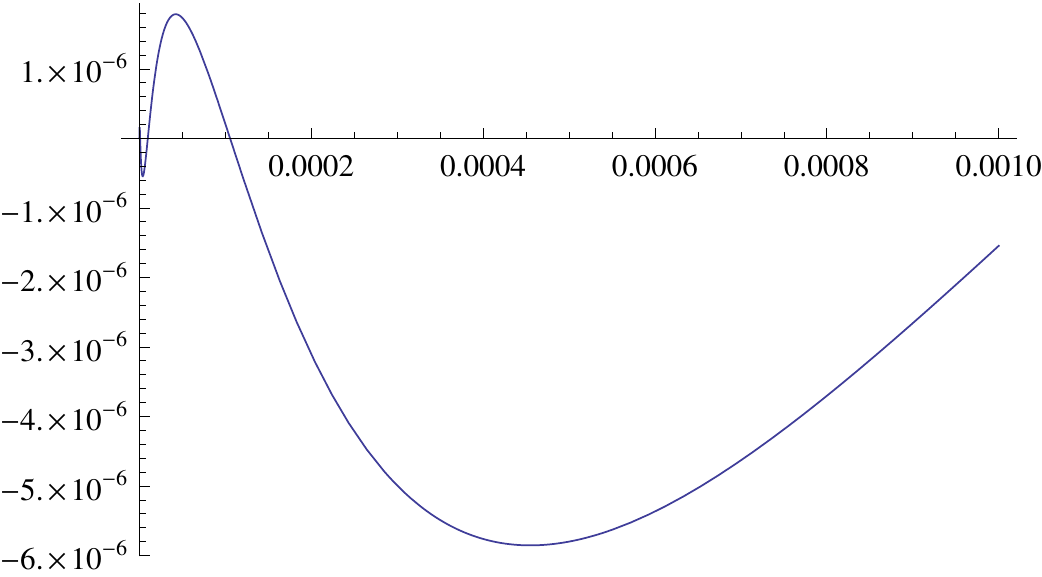}}
\hbox to \textwidth{\hbox to 0.45\textwidth{\hfil(A)\hfil}\hfil
\hbox to 0.45\textwidth{\hfil(B)\hfil}}
\hbox to \textwidth{\ing[width=0.45\textwidth]{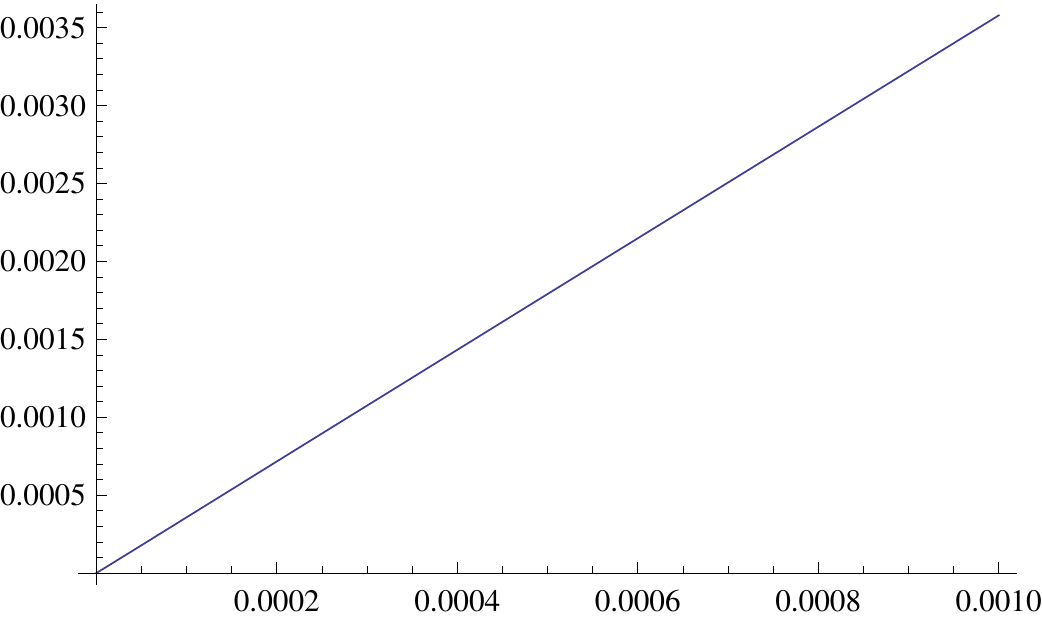}\hfil
\ing[width=0.45\textwidth]{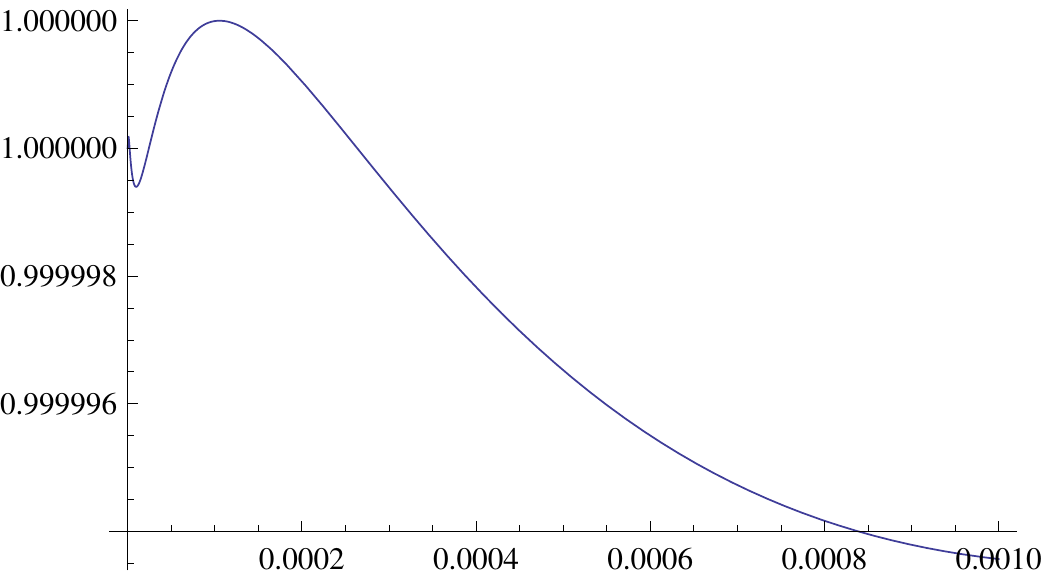}}
\hbox to \textwidth{\hbox to 0.45\textwidth{\hfil(C)\hfil}\hfil
\hbox to 0.45\textwidth{\hfil(D)\hfil}}
\caption{(A)---a plot of $A(r)$, (B)---a plot of $B(r)$,
(C)---a plot of $\vF(r)$, (D)---a plot of $\pz{\wt\vF}r(r)$ (see text)}
\label{ykl1}
\end{figure}

\begin{figure}[h]
\hbox to \textwidth{\ing[width=0.45\textwidth]{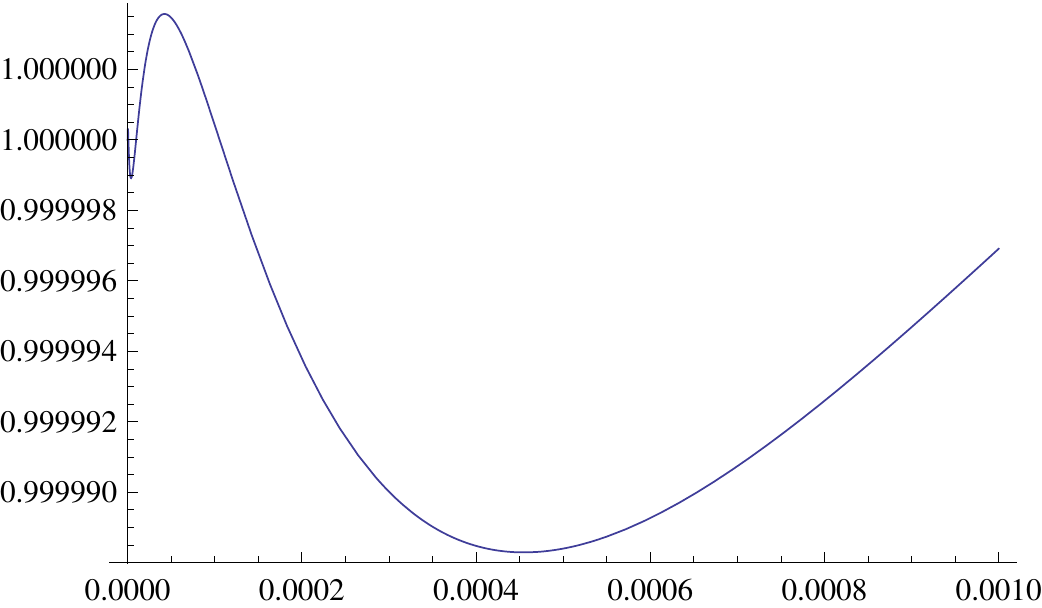}\hfil
\ing[width=0.45\textwidth]{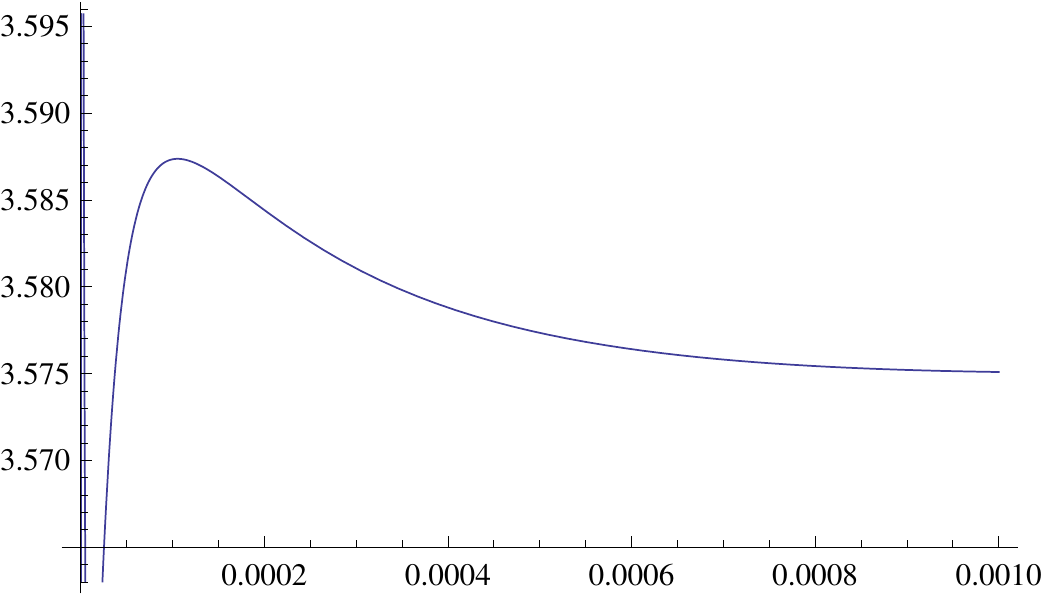}}
\hbox to \textwidth{\hbox to 0.45\textwidth{\hfil(A)\hfil}\hfil
\hbox to 0.45\textwidth{\hfil(B)\hfil}}
\hbox to \textwidth{\hfil \ing[width=0.45\textwidth]{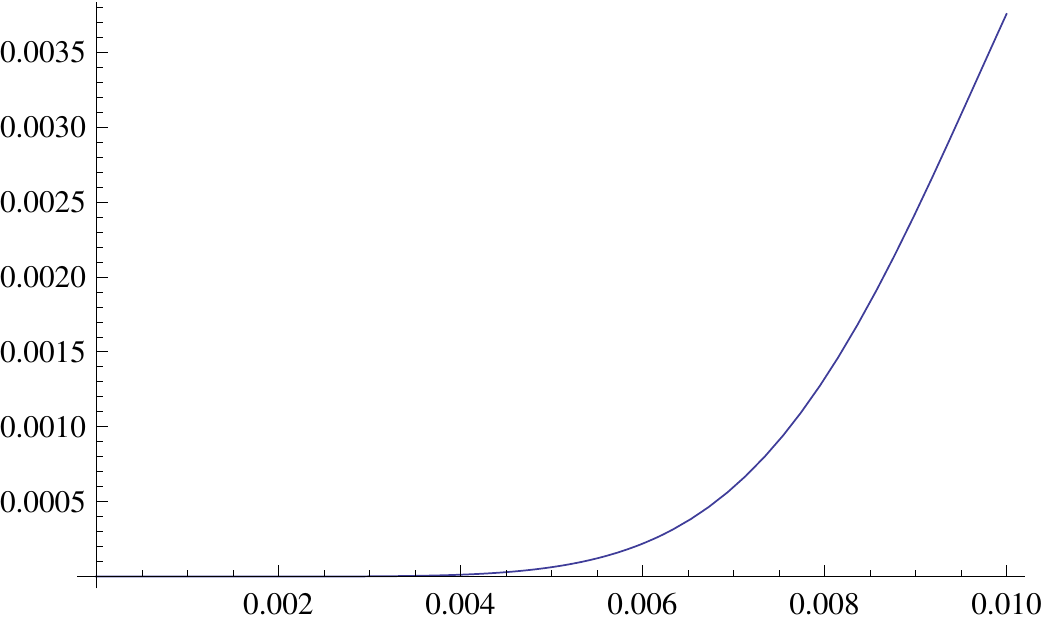}\hfil}
\hbox to \textwidth{\hfil\hbox to 0.45\textwidth{\hfil(C)\hfil}\hfil}
\caption{(A)---a plot of $e^{2A(r)}$, (B)---a plot of $e^{2B(r)}$,
(C)---a plot of $m_1(r)$}
\label{ykl2}
\end{figure}

On Figs \ref{ykl1}, \ref{ykl2} we give plots of $A(r)$, $B(r)$, $\wt\vF(r)$,
$e^{2A(r)}$, $\eb2$, $\pz{\wt\vF}r(r)$ and $m_1(r)=\frac r2(1-\eb{-2})$ which
represents a mass for radius~$r$. The \f\ $\vF(r)$ reads
\beq{Dn187}
\vF(r)=\frac{n+2}{\ov M}\,\wt\vF(r)+\vF_0r+\vF_1
\end{equation}
as before and the argument for a negligible $\ov\rho(r)$ still works.

Now we have only a problem to match $\vF(r)$  for $r=r_1$. $\vF$ satisfies an
\e\ of the second order. Thus we should match $\vF(r_1)$, $\pz\vF r(r_1)$,
$\pz{^2\vF}{r^2}(r_1)$. For $r>r_1$, $\vF$ satisfies a partial differential
\e\ of second order, i.e.,
\bml{Dn188}
-\frac{2\ov M}{r^2} \pp{}r\X2(r^2\,\pp{\vF(r,t)}r\Y2)
+e^{n\vF(r,t)}(e^{2\vF(r,t)}-1)\\ {}-(n+2)\rho_{\rm tot}e^{-\nd \vF(r,t)}
+\frac1{R^3(t)}\,\pp{}t \X2(R^3(t)\,\pp{\vF(r,t)}t\Y2)=0
\end{multline}
where $\rho_{\rm tot}$ is a density of a \co ical matter.

Thus we have
\beq{Dn189}
\bal
\vF(r_1,t)=\frac{n+2}{\ov M}\,\wt\vF(r_1)+\vF_0 r_1+\vF_1\\
\pp\vF r(r_1,t)=\frac{n+2}{\ov M}\,\pz{\wt\vF(r_1)}r+\vF_0 \\
\pp{^2\vF}{r^2}(r_1,t)=\frac{n+2}{\ov M}\,\pz{^2\wt\vF(r_1)}{r^2}
\eal
\end{equation}

Moreover, for Eq.\ \er{Dn188} we have the following initial condition
\beq{Dn190}
\bal
\vF(r,t_0)&=a(r)\\
\pp\vF t(r,t_0)&=b(r), \ t_0>0.
\eal
\end{equation}
Simultaneously we should suppose that $\vF_0$ and $\vF_1$ are \f s of time
\beq{Dn191}
\vF_0=\vF_0(t), \q \vF_1=\vF_1(t).
\end{equation}
Eqs \er{Dn189} and \er{Dn190}--\er{Dn191} should be consistent for $r=r_1$
and $t=t_0$. This gives
\bg{Dn192}
\frac{n+2}{\ov M}\,\wt\vF(r_1)+r_1\vF_0(t_0)+\vF_1(t_0)=a(r_1)\\
r_1\,\pz{\vF_0}{t}(t_0)+\pz{\vF_1}{t}(t_0)=b(r_1) \label{Dn193}\\
\pz ar(r_1)=\frac{n+2}{\ov M}\,\pz{\wt\vF(r_1)}{r}+\vF_0(t_0) \label{Dn194}\\
\pz {^2a}{r^2}(r_1)=\frac{n+2}{\ov M}\,\pz{^2\wt\vF(r_1)}{r^2}\label{Dn195}\\
\pz{^2b}{r^2}(r_1)=0. \label{Dn196}
\end{gather}

In this way we can match an initial problem for a hyperbolic \e\ \er{Dn188}
to the conditions \er{Dn189} getting a time dependence of a scalar
field~$\vF$.

Let us come back to Eq.\ \er{Dn174}. From this \e\ we get $r_1=r_1(t)$. If we
use this \e\ in place of the second \e\ in Eq.~\er{Dn183}, we find also a
time dependence of $B_1$, $B_1=B_1(t)$ and of course $r_2=r_2(t)=r_1(t)$. If
we come back to Eqs \er{D.46a}--\er{D.47}, we can conclude the time dependence
of our \so\ is due to an initial problem which parametrically depends on time
via a Hubble \ct. This can influence an \an\ \ac\ of massive particles and
photons due to time dependence of a \cn\ and geodesics.

Let us consider Eqs \er{D.46a}--\er{D.47} for two types of \co ical models.
One gets
\bea{Dn197}
A(7.36\t10^{-10})&=&-3.44\t 10^{-12}\coth(0.003t)\\
B(7.36\t10^{-10})&=&\eta\cdot3.44\t 10^{-12}\coth(0.003t) \label{Dn198}
\end{eqnarray}
for a CDM$\La$ model and
\bea{Dn199}
A(7.36\t10^{-10})&=&-1.72\t 10^{-9}\,\tfrac 1t\\
B(7.36\t10^{-10})&=&\eta\cdot1.72\t 10^{-9}\,\tfrac 1t\label{Dn200}
\end{eqnarray}
for a dust filled model.

Eq.\ \er{Dn173} can be written for a CDM$\La$ model.
\beq{Dn201}
\frac{r-e^{B(r,B_1(t))}}
{re^{A(r,B_1(t))}(e^{2B(r)B_1(t))}-1)}
=0.5\t10^{-3}t\coth(0.003t).
\end{equation}

\begin{figure}[h]
\hbox to \textwidth{\ing[width=0.45\textwidth]{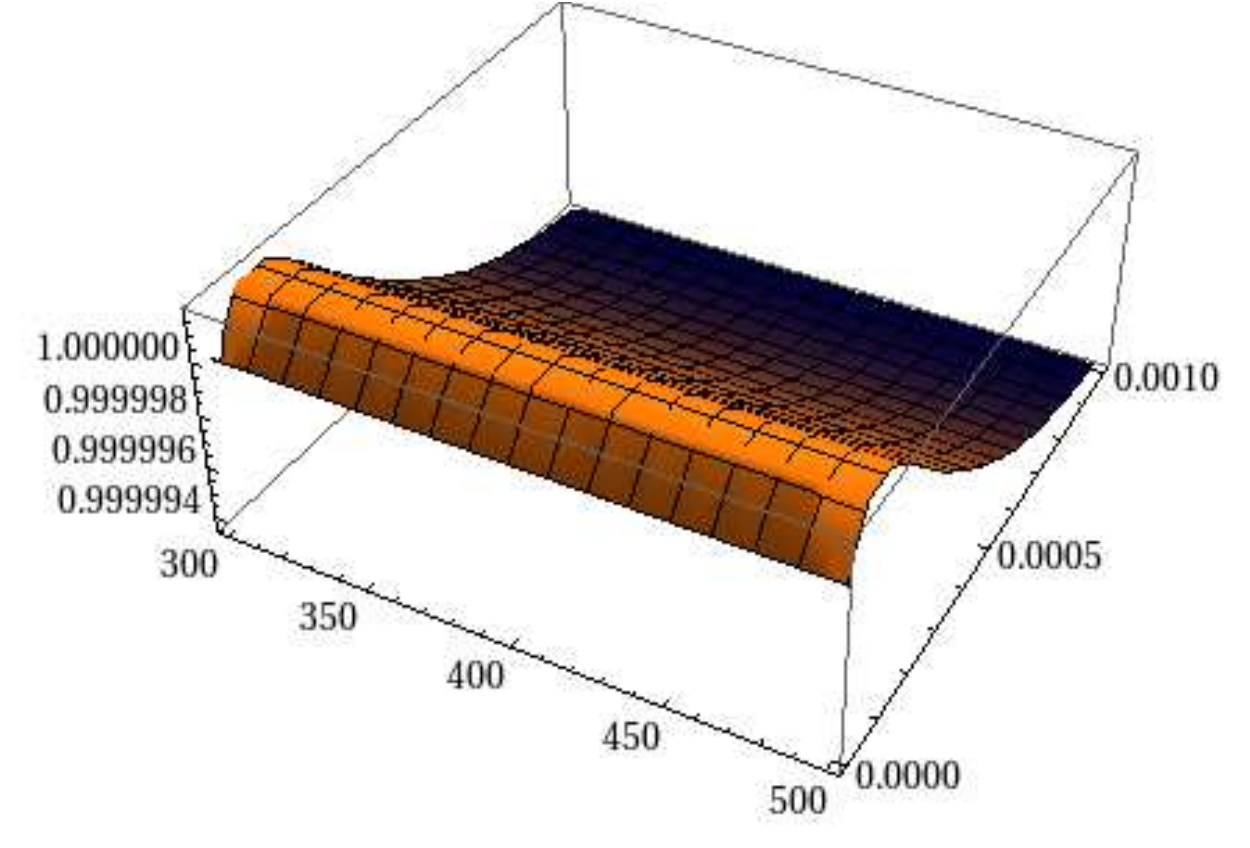}\hfil
\ing[width=0.45\textwidth]{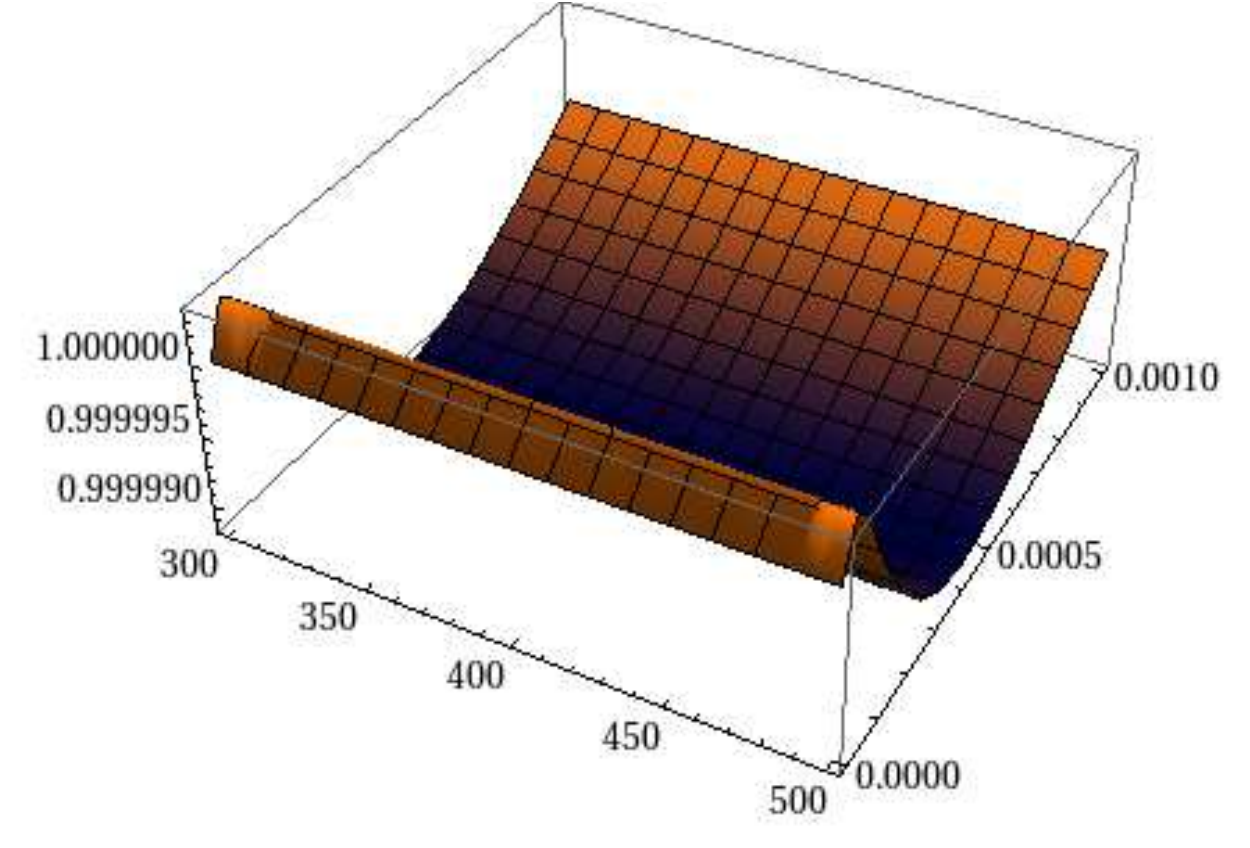}}
\hbox to \textwidth{\hbox to 0.45\textwidth{\hfil(A)\hfil}\hfil
\hbox to 0.45\textwidth{\hfil(B)\hfil}}
\hbox to \textwidth{\ing[width=0.45\textwidth]{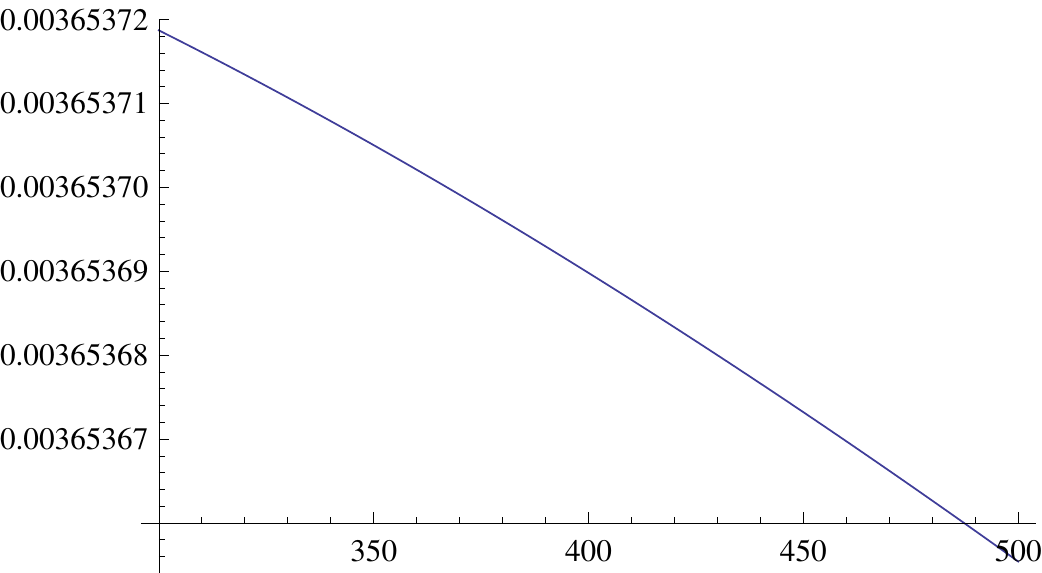}\hfil
\ing[width=0.45\textwidth]{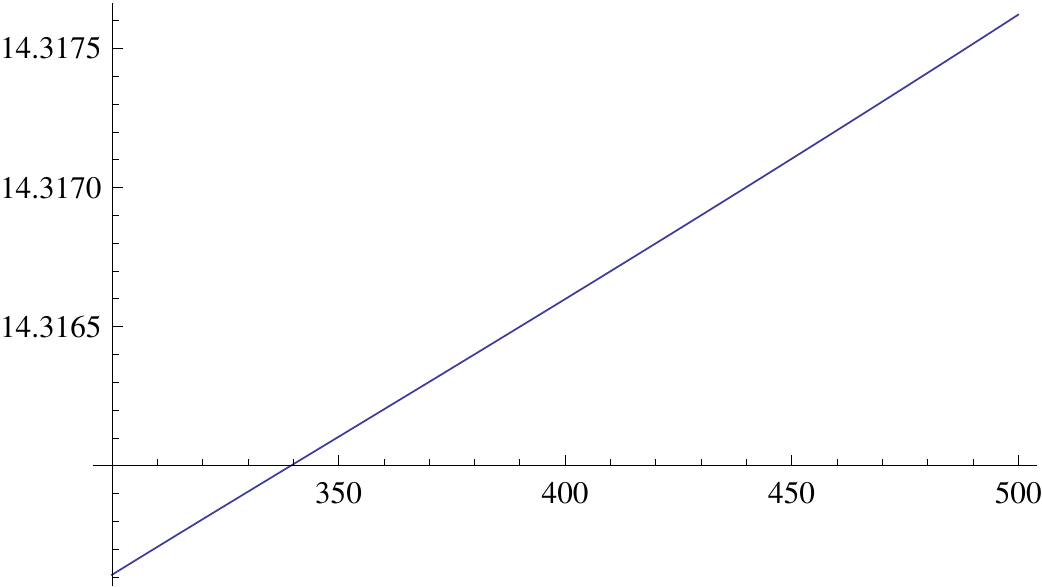}}
\hbox to \textwidth{\hbox to 0.45\textwidth{\hfil(C)\hfil}\hfil
\hbox to 0.45\textwidth{\hfil(D)\hfil}}
\caption{(A)---a 3D plot of $e^{2A(r,t)}$ in the case of CDM$\La$ model,
(B)---a 3D plot of $e^{2B(r,t)}$ in the case of CDM$\La$ model,
(C)---a plot of $r_1(t)$, (D)---a plot of $B_1(t)$ (see text)}
\label{qp}
\end{figure}

In this case a time $t$ is measured in our unit ($T$) and $r$ is measured in
our unit ($L$). First two \e s of Eqs \er{Dn176} and \er{Dn197}--\er{Dn198}
have been solved together with the first \e\ of \er{Dn183} and \er{Dn201}.
The results have been plotted on Fig.~\ref{qp}. We plot here $e^{2A(r,t)}$ and
$e^{2B(r,t)}$ as 3D plot and $r_1(t)$ and $B_1(t)$. It is easy to see that
during a period from $300\,T$ (9.5\,Gyr) to $500\,T$ (16\,Gyr) $e^{2A(r,t)}$ and
$e^{2B(r,t)}$ do not change significantly. In the case of $r_1(t)$ (during the
same period) it changes from 36537.2\,pc$=$7536345558\,AU to
36536.65\,pc$=$7536232112\,AU. Thus a difference reads
$$
\D r=r_1(300\,T)-r_1(500\,T)=113446\,AU=
16960177\t10^3\,{\rm km}, \q \frac{\D r}{r_1(t)}\simeq 10^{-5},
$$
$r_1(t)$ is of course a decreasing \f\ of time. We put here $\eta=\xi=1$.
Further research on this subject will be done elsewhere.

\begin{figure}[h]
\hbox to \textwidth{\ing[width=0.45\textwidth]{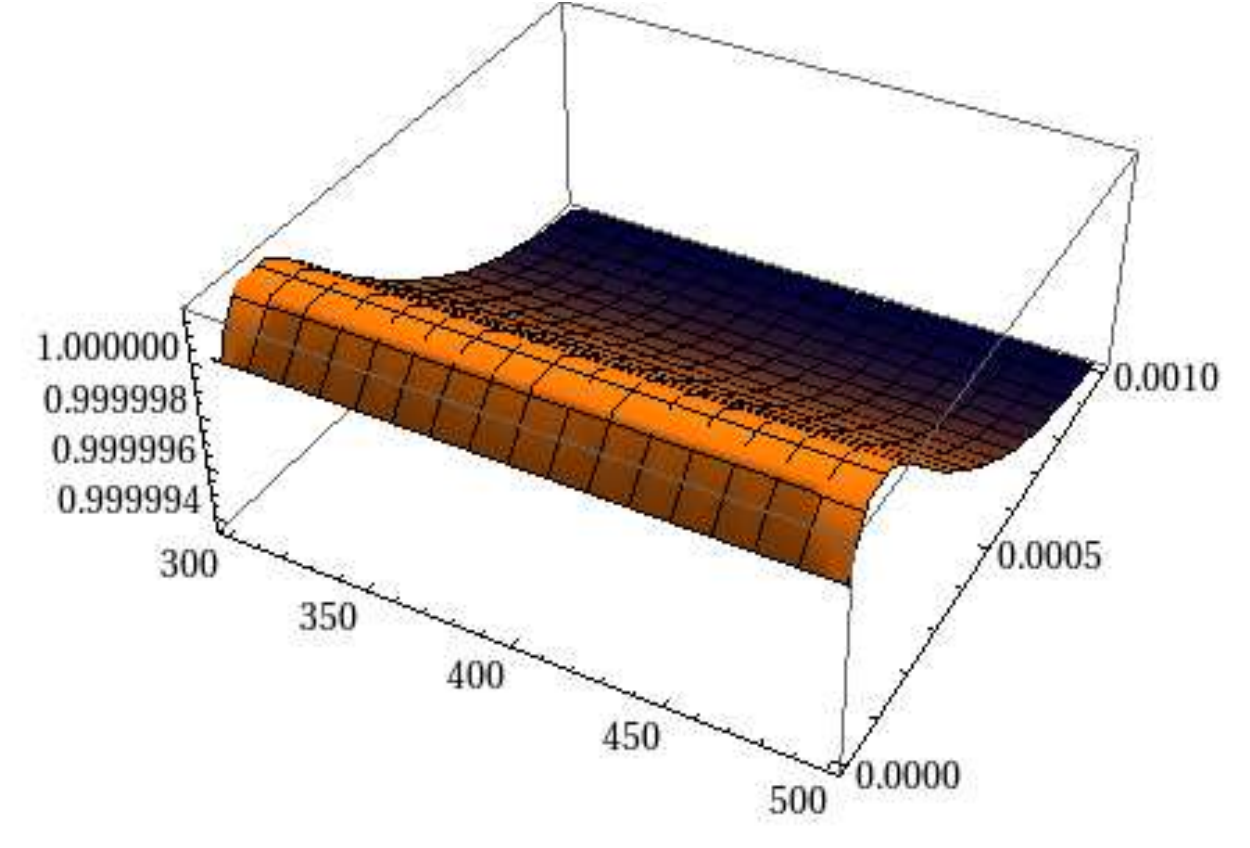}\hfil
\ing[width=0.45\textwidth]{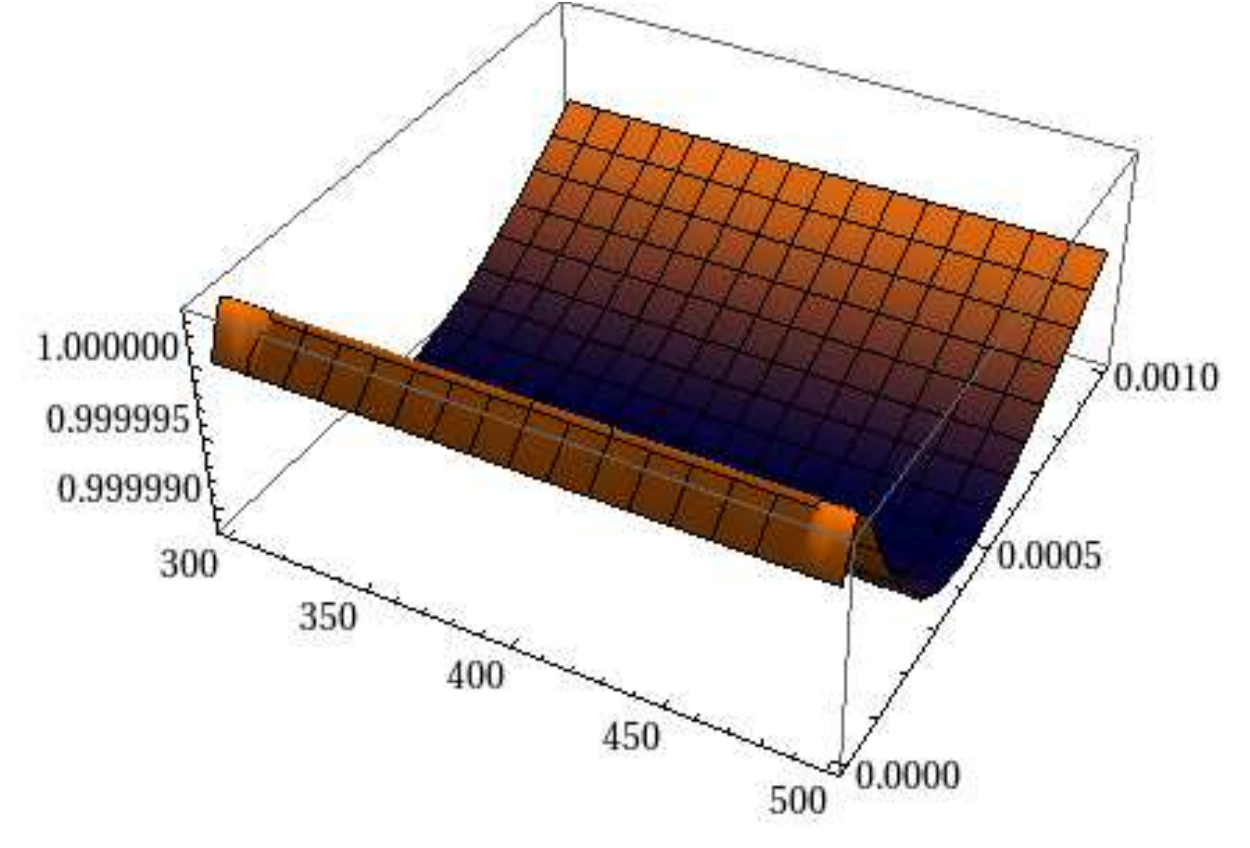}}
\hbox to \textwidth{\hbox to 0.45\textwidth{\hfil(A)\hfil}\hfil
\hbox to 0.45\textwidth{\hfil(B)\hfil}}
\hbox to \textwidth{\ing[width=0.45\textwidth]{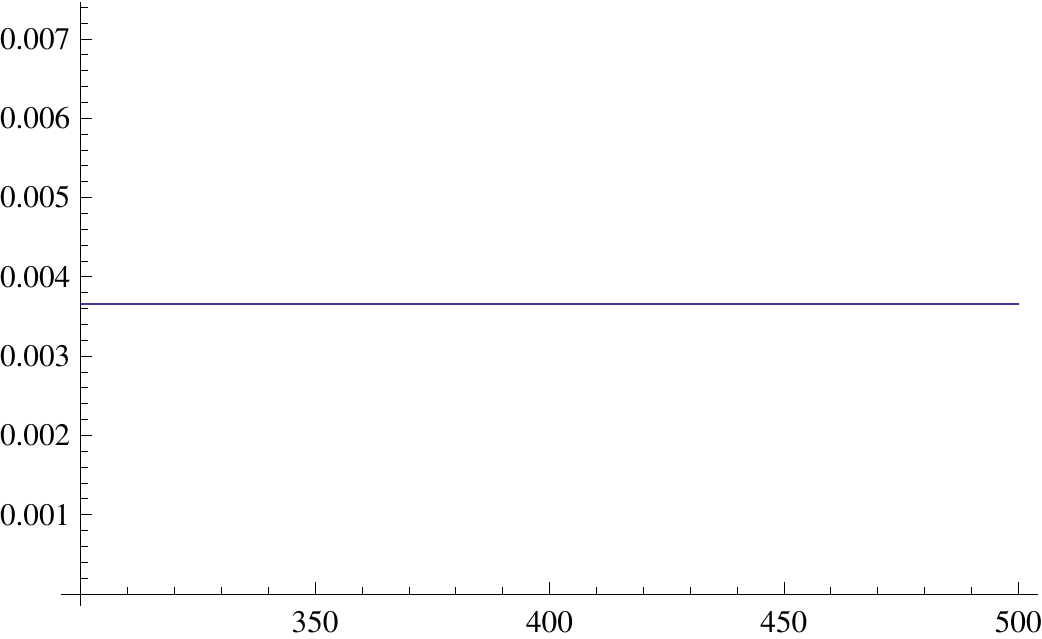}\hfil
\ing[width=0.45\textwidth]{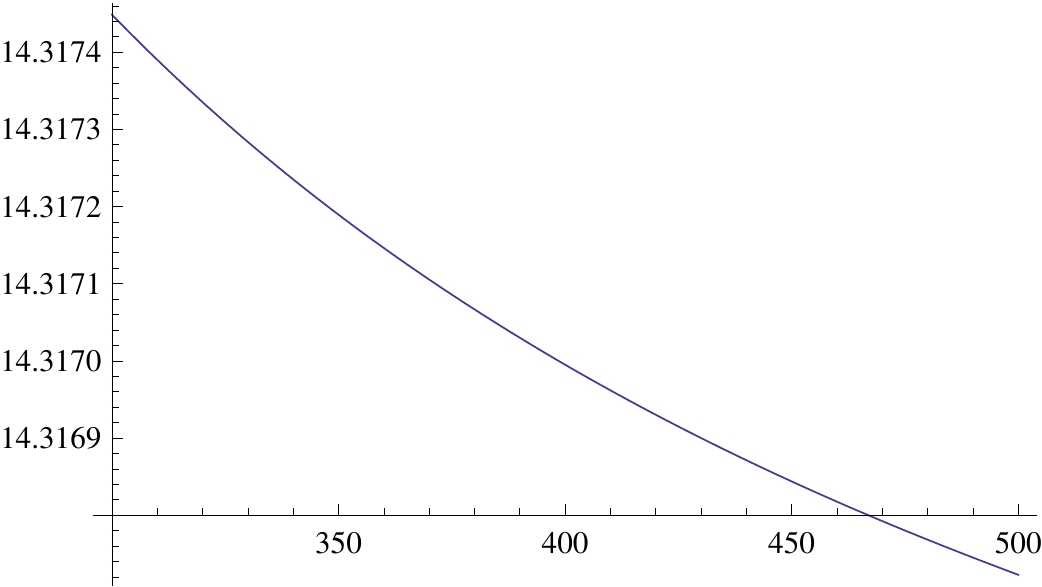}}
\hbox to \textwidth{\hbox to 0.45\textwidth{\hfil(C)\hfil}\hfil
\hbox to 0.45\textwidth{\hfil(D)\hfil}}
\caption{(A)---a 3D plot of $e^{2A(r,t)}$ in the case of a dust filled \co
ical model,
(B)---a 3D plot of $e^{2B(r,t)}$ in the case of a dust filled \co
ical model,
(C)---a plot of $r_1(t)$ in a dust filled \co ical model,
(D)---a plot of $B_1(t)$ in a dust filled \co ical model (see text)}
\label{rc}
\end{figure}

Moreover, we can do similar calculations using a different \co ical model
(matter filled \Un), i.e.\ the first two \e s of Eq.\ \er{Dn176} and Eqs
\er{Dn199}--\er{Dn200}. The results have been plotted on Fig.~\ref{rc} as
3D plots of $e^{2A(r,t)}$, $e^{2B(r,t)}$, $r_1(t)$ and $B_1(t)$. If we
compare Fig.~\ref{qp} A,~B and Fig.~\ref{rc} A,~B, we conclude that they look
almost the same, but $r_1(t)$ and $B_1(t)$ look differently, $r_1(t)=\rm
const.$

Let us notice that $m_1$ should also depend on $t$ via $e^{2A}$. Moreover
(see Fig.\ \ref{qp}A and Fig.~\ref{rc}A), this dependence is very weak.

Below we give a listing of programmes calculating in Mathematica~7 both cases.
For CDM$\La$ model

\medskip
{\parindent 0pt
\tt eqn1 = D[D[b[x, b1, t], x], x] == $\dfrac{\tt 9 - 10 * e^{2\,b[x,b1,t]}
+e^{4\,b[x,b1,t]}}{\tt 8 * x^2}$

\vskip2pt
eqn2 =  D[D[b[x, b1, t], x], x] == $\dfrac{\tt -1 + e^{2\,b[x,b1,t]}}
{\tt 4 * x^2}$

\vskip2pt
eqn3 = D[a[x, b1, t], x] == $\dfrac{\tt -1 + e^{2\,b[x,b1,t]}}{\tt 4 * x}$

\vskip2pt
eqn4 = a[7.36 * 10\^{ }-10, b1, t] == -3.44 * 10\^{ }-12 * Coth[0.003 * t]

eqn5 = b[7.36 * 10\^{ }-10, b1, t] == 3.44 * 10\^{ }-12 * Coth[0.003 * t]

eqn6 = (D[b[x, b1, t], x] /.\ x $\to$ 7.36 * 10\^{ }-10) == b1

ss = NDSolve[\char123eqn1, eqn3, eqn4, eqn5, eqn6\char125, \char123a, b\char125,

\ \ \ \char123x, 10\^{ }-11, 10\^{ }-3\char125, \char123b1, -100, 100\char125, \char123t, 300, 500\char125, MaxSteps $\to$ 50 000]

}
\medskip

For dust (matter) filled model

\medskip
{\parindent 0pt
\tt eqn1 = D[D[b[x, b1, t], x], x] == $\dfrac{\tt 9 - 10 * e^{2\,b[x,b1,t]}
+e^{4\,b[x,b1,t]}}{\tt 8 * x^2}$

\vskip2pt
eqn2 =  D[D[b[x, b1, t], x], x] == $\dfrac{\tt -1 + e^{2\,b[x,b1,t]}}
{\tt 4 * x^2}$

\vskip4pt
eqn3 = D[a[x, b1, t], x] == $\dfrac{\tt -1 + e^{2\,b[x,b1,t]}}{\tt 4 * x}$

\vskip2pt
eqn4 = a[7.36 * 10\^{ }-10, b1, t] == -1.72 * 10\^{ }-9 * 1 / t

eqn5 = b[7.36 * 10\^{ }-10, b1, t] == 1.72 * 10\^{ }-9 * 1 / t

eqn6 = (D[b[x, b1, t], x] /. x $\to$ 7.36 * 10\^{ }-10) == b1

ss1 = NDSolve[\char123eqn1, eqn3, eqn4, eqn5, eqn6\char125, \char123a, b\char125,

\ \ \ \char123x, 10\^{ }-11, 10\^{ }-3\char125, \char123b1, -100, 100\char125, \char123t, 300, 500\char125, MaxSteps $\to$ 50 000]

}
\medskip

Here {\tt x} corresponds to $r$, {\tt a} to $A$, {\tt b} to $B$, {\tt b1} to
$B_1$ (see Appendix~E). Plots are given in both cases for $B_1=14.31$.

\begin{figure}[h]
\hbox to \textwidth{\ing[width=0.45\textwidth]{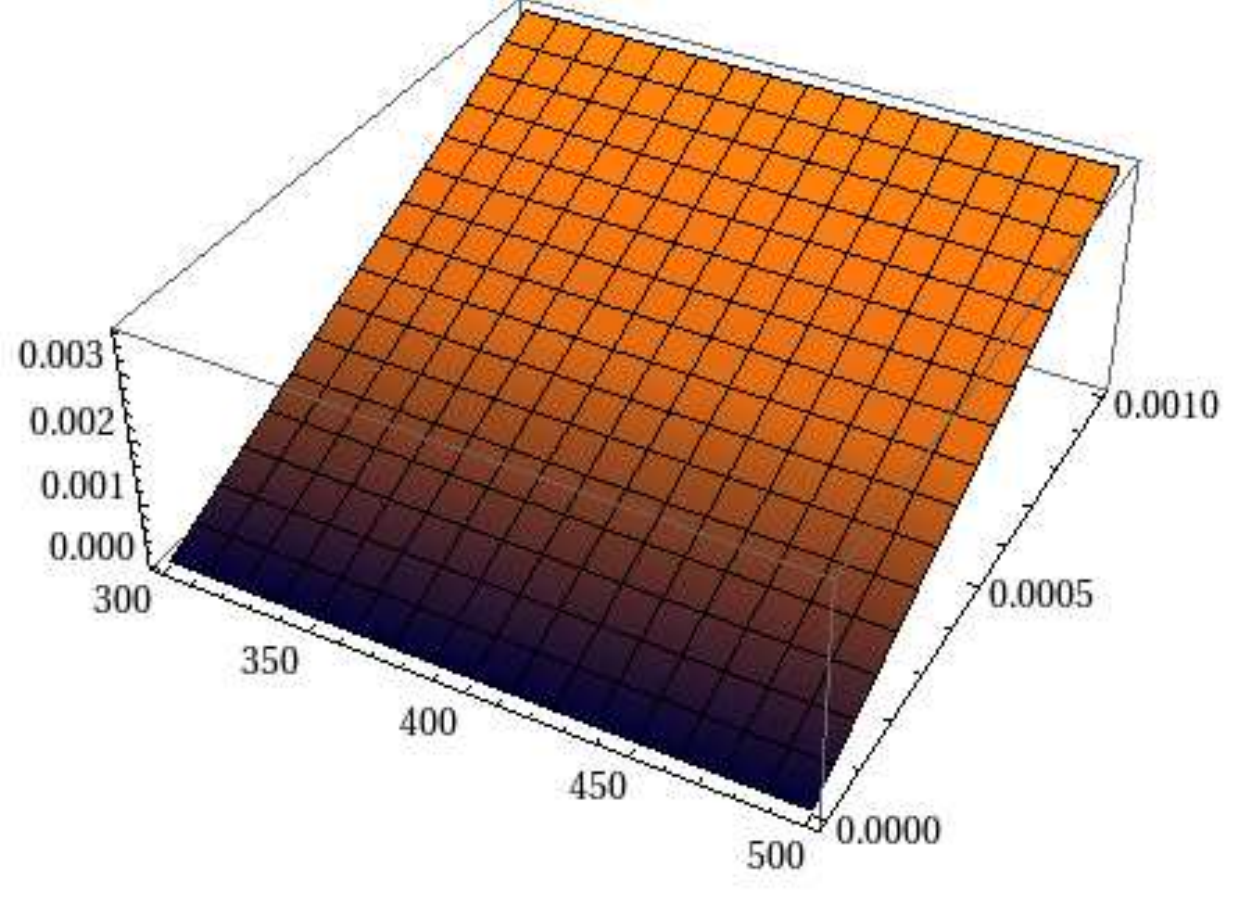}\hfil
\ing[width=0.45\textwidth]{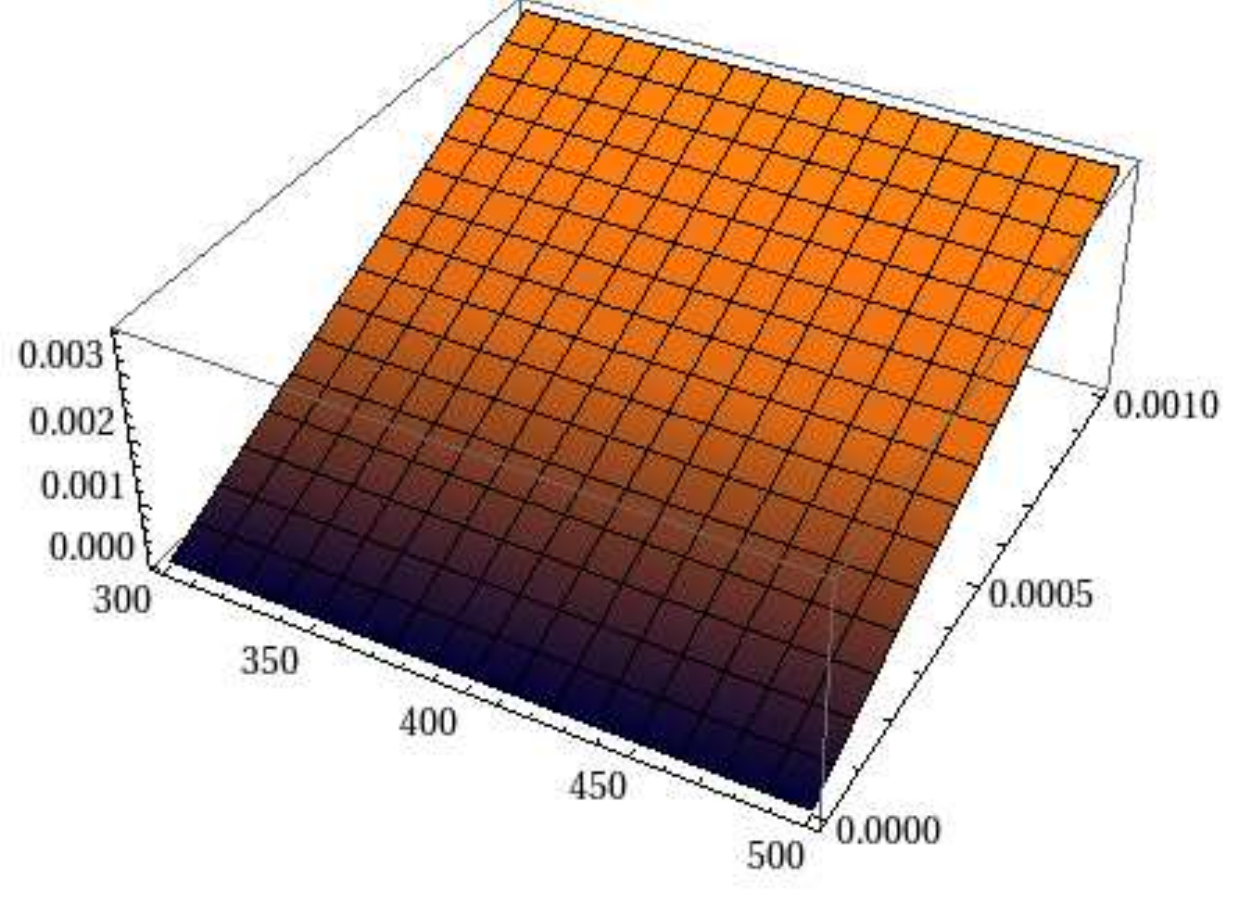}}
\hbox to \textwidth{\hbox to 0.45\textwidth{\hfil(A)\hfil}\hfil
\hbox to 0.45\textwidth{\hfil(B)\hfil}}
\hbox to \textwidth{\ing[width=0.45\textwidth]{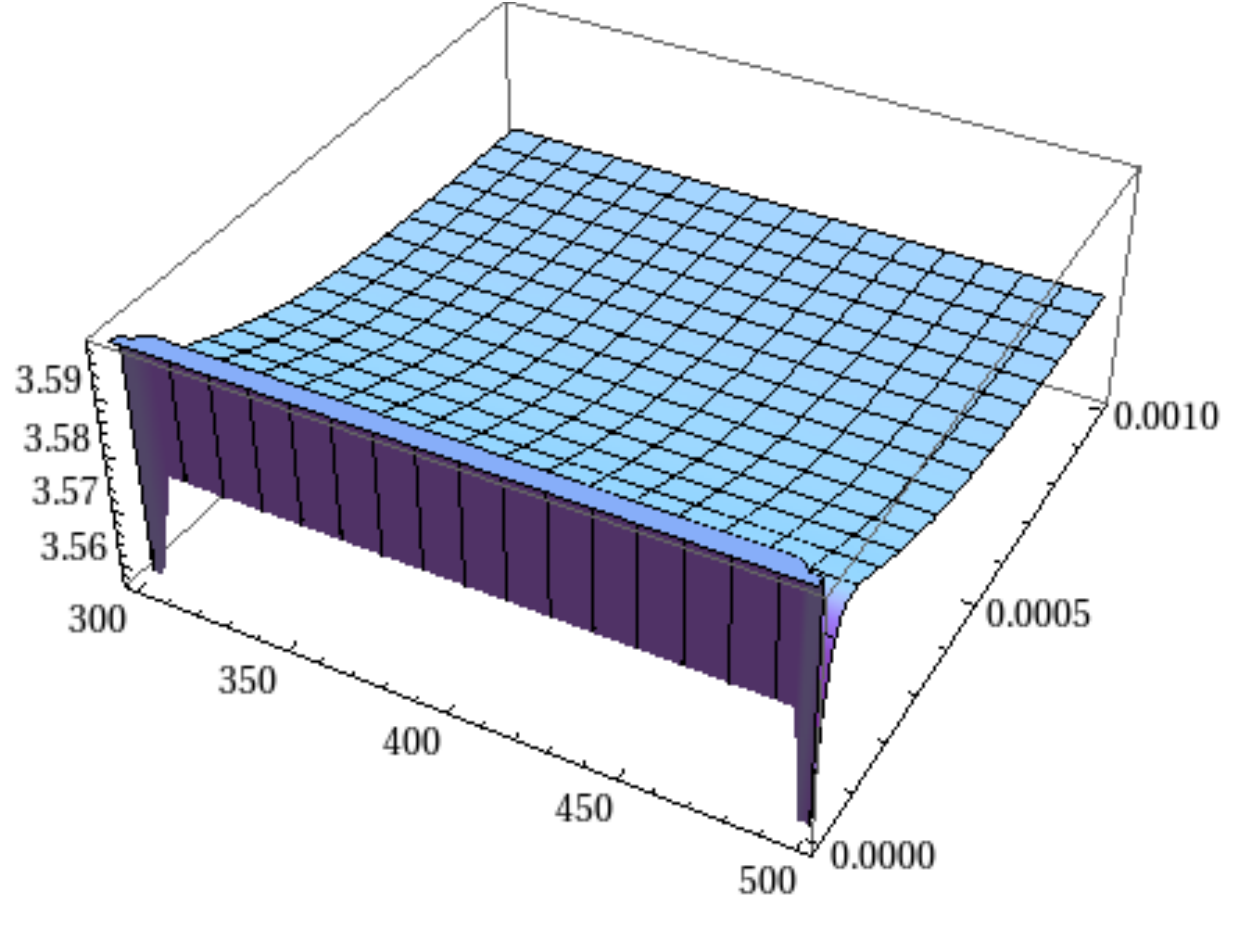}\hfil
\ing[width=0.45\textwidth]{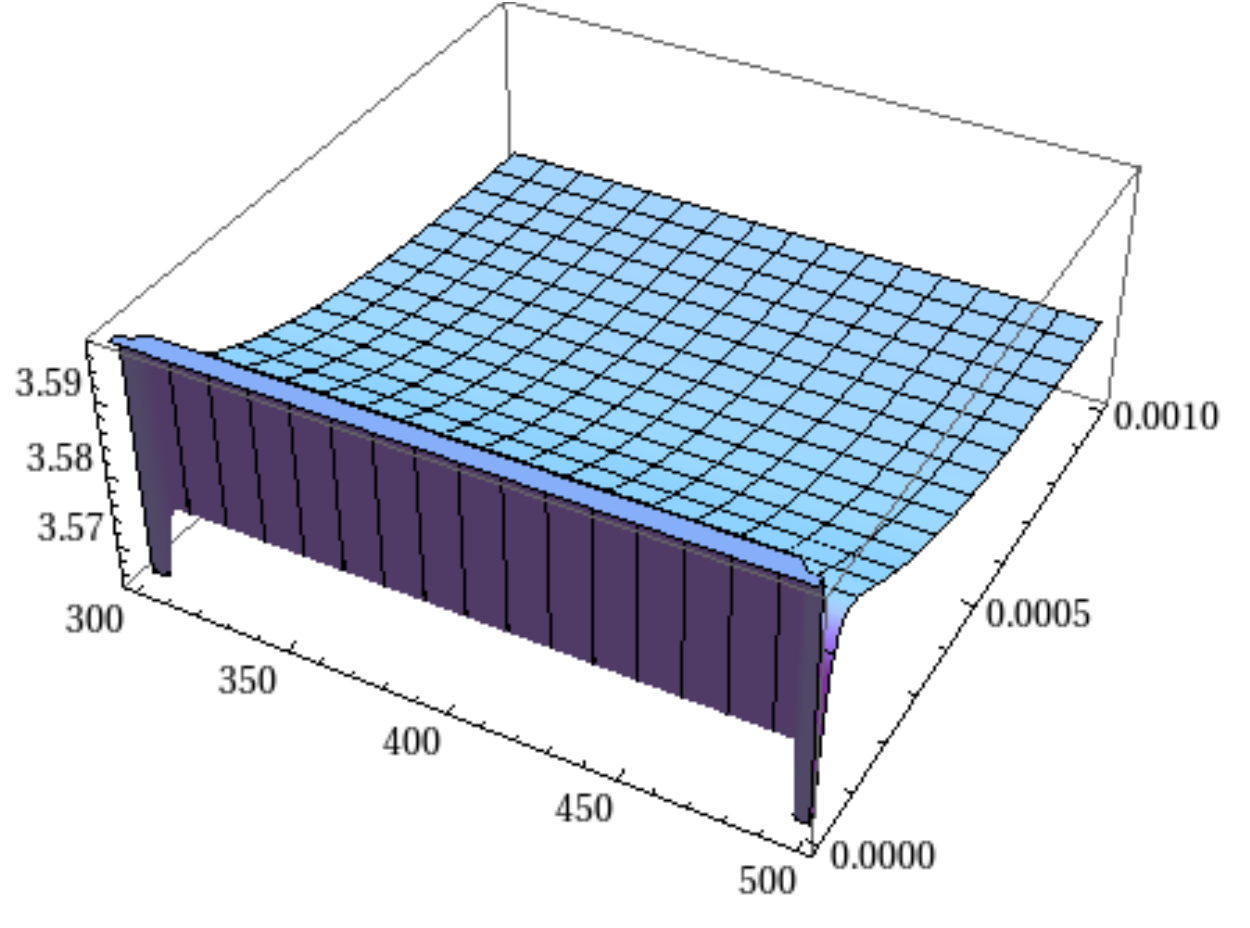}}
\hbox to \textwidth{\hbox to 0.45\textwidth{\hfil(C)\hfil}\hfil
\hbox to 0.45\textwidth{\hfil(D)\hfil}}
\caption{(A)---a 3D plot of $\wt\vF(r,t)$ in a CDM$\La$ model,
(B)---a 3D plot of $\pp{\wt\vF}r(r,t)$ in a CDM$\La$ model,
(C)---a 3D plot of $\wt\vF(r,t)$ in a dust filled model,
(D)---a 3D plot of $\pp{\wt\vF}r(r,t)$ in a dust filled model (see text)}
\label{nm}
\end{figure}

Let us notice that a value $r_1(t)$ can be considered as a range of the model
being of order $0.003\,L$, i.e. $r_1(t)\ll 1$. On Fig.~\ref{nm} we give
3D plots of $\wt\vF(r,t)$ and $\pp{\wt\vF}r(r,t)$ in both cases. They look
almost the same. In the case of the dust filled model $r_1(t)=\rm const.$
However, in the CDM$\La$ model $r_1(t)$ really depends on time. Thus
in this case we should consider moving matching conditions for the field
$\vF(r,t)$. In order to do this let us consider Eq.~\er{Dn188} in the case of
a \ct\ \so. One gets
\beq{Dn202}
y_{(n)}^{n+2}-y_{(n)}^{n+1}-0.17\t10^{-4}\nd=0
\end{equation}
where $y_{(n)}=e^{2\ov\vF(n)}$, $\ov\vF(n)$ being a \ct\ \so. A \ct\ \so\
of Eq.~\er{Dn188} in our \Un\ is dictated by the fact that a \co ical ``\ct''
is really \ct. One gets
\beq{Dn203}
\bal
\ov\vF(100)&=0.00074527\\
\ov\vF(120)&=0.00084459\\
\ov\vF(133)&=0.000900619\\
\ov\vF(248)&=0.001179582.
\eal
\end{equation}
We suppose temporally that $\ov\vF=0$. In more precise calculations one can
use a non-zero value for an established $n>14$. Thus we should match
$\vF(r,t)$ on $r_1(t)$ to zero.

Using
\beq{Dn204}
\vF(r,t)=\frac{n+2}{\ov M}\wt\vF(r,t)+\vF_0(t)r+\vF_1(t)
\end{equation}
one gets for $r=r_1(t)$ from
\beq{Dn205}
\bal
\vF(r_1(t),t)&=0\\
\pp\vF r(r_1(t),t)&=0
\eal
\end{equation}
the equality
\beq{Dn206}
\vF(r,t)=\frac{n+2}{\ov M}\X2[\X1(\wt\vF(r,t)-\wt\vF(r_1(t),t)\Y1)
+\pp{\wt\vF}r(r_1(t),t)(r_1(t)-r)\Y2].
\end{equation}
Eq.\ \er{Dn206} gives us a match of the \so\ inside a ball $r<r_1(t)$ and a
zero \so\ outside the ball governed by the Eq.~\er{Dn188}.

\def\Ch#1#2#3{\genfrac\{\}{0pt}0 #1{#2#3}_g}
Moreover, to facilitate a future research we can do some formalities. In both
non-Riemannian cases for $r\ll 1$ one gets
\beq{D.167}
\g=\frac{f^4(f^2+r^4+\ell^4)}{\ell^4\a}\,e^{2(\ov c-r)}.
\end{equation}
\goodbreak
\noindent For in the \e\ of motion appears a Levi-Civit\`a \cn\ induced by a \s\ part
of a \nos\ tensor
$g_\(\a\b)$ it is really important to quote Christoffel symbols for $g_\(\a\b)$.
\beq{D.168}
\bal
\Ch111&=\frac1{2\a}\,\pz\a  r\\
\Ch122&=\frac r\a\\
\Ch133&=\frac r\a \sin^2\th\\
\Ch233&=-\frac12\sin2\th\\
\Ch332&=\cot\th\\
\Ch144&=\frac1{2\a}\,\pz\g r\\
\Ch221&=\Ch331=\frac 1r\\
\Ch441&=-\frac1{2\g}\,\pz\g r
\eal
\end{equation}
where
\bml{D.169}
\pz\g r=\frac{e^{2\ov c-r}f^3}{\a\ell^4}\X2[2\pz fr(3f^2+2r^4+2\ell^4f)\\
{}-f\,\pz{}r\log\a \cdot(f^2+r^4+\ell^4)+2f(2r^3-f^2-r^4-\ell^4)\Y2].
\end{multline}
The remaining Christoffel symbols are zero.
These Christoffel symbols are applicable for both cases (moreover with
different $\a$ and~$f$).

In order to get GR case it is enough to shift
$\genfrac\{\}{0pt}1 \a{\b\g}_g \to \gd\wt\G,\a,\b\g,$, $\a\to \ov A$, $\g \to
\ov B$ in Eq.\ \er{D.168}.

It is interesting to consider Eqs \er{D.151}--\er{D.154} for large $r$, $r\gg
1$. One gets
\bml{D.170}
\pz{^2\a}{r^2}(r)=\frac{8r\ef{(n+1)}\a(r)}{3n\nd \o(r)}\X1(-\cosh
(\vF(r))+(n+1)\sinh(\vF(r))\Y1)\pz\a r(r)\\
{}+n\nd \a(r)\sinh(\vF(r))\pz\vF r(r)+\a(r)\X2((n+1)\sinh(\vF(r))-\cosh
(\vF(r))\pz\o r(r)\Y2)
\end{multline}
\bea{D.171}
\pz{^2\vF}{r^2}(r)&=&-\frac{2r\ef n(n+2-\ef2 n)\a(r)\pz \vF r(r)}{3n\nd}\\
\pz{^2f}{r^2}(r)&=&-\frac{2r\ef n(n+2-\ef2 n)\a(r)\pz fr(r)}{3n\nd}\label{D.172}\\
\pz{\o}{r}(r)&=&-\frac{2r\ef n(n+2-\ef2 n)\a(r)\o(r)}{3n\nd}\label{D.173}
\end{eqnarray}
One easily gets from Eqs \er{D.171}--\er{D.173}
\bea{D.174}
\pz\vF r(r)&=&\ov{\ov c}\o(r)\\
\pz fr(r)&=&\ov{\ov c}_1\o(r)\label{D.175}
\end{eqnarray}
and
\beq{D.176}
\vF(r)=\wt cf(r)+a_1
\end{equation}
where $\wt c=\ov{\ov c}/\ov{\ov c}_1$, $\ov{\ov c},\ov{\ov c}_1$
and $a_1$ are integration \ct s.
\beq{D.177}
f(r)=\ov{\ov c}_1 \int \o(r)\,dr+d_1
\end{equation}
where $d_1$ is an integration \ct.

These results will be considered in further development of the \eu\nos\
Jordan--Thiry Theory.

Let us notice the following fact. Even if for a geodetic motion we use
Christoffel symbols (a~\cn\ for a \s\ part of $g_\m$, $g_{(\m)}$), skew-\s\
part of $g_\m$, i.e.\ $f(r)$, enters those symbols. It means the fact that in a
theory which uses a \nos\ metric, a skew-\s\ part $g_\[\mu\nu]$ has an influence even on its Riemannian part.
Finally we consider a match of a \nos\ theory \so\ to a \co ical \so\ which
we suppose as a \s\ one. The match of a \s\ part of $g_\m$, $g_\(\m)$ is
exactly the same as in the GR case (of course in terms of $\a$ and~$\g$).
Moreover, we have to do also with skew-\s\ part $g_\[\m]$. In this case we
suppose that on the sphere of the match $r=r_1$ we have
\beq{Dn218}
\bal
\o(r_1)&=0\\
f(r_1)&=0\\
\pz \o r(r_1)&=0\\
\pz fr(r_1)&=0
\eal
\end{equation}

Using Eq.\ \er{D.126} or Eq.\ \er{D.159} one easily gets that from
$\o(r_1)=0$, $f(r_1)=0$ and from $\pz\o r(r_1)=0$
\beq{Dn219}
\pz fr(r_1)=0.
\end{equation}
It means we can add initial conditions
\beq{Dn220}
\bal
\o(r_1)&=0\\
\pz\o r(r_1)&=0
\eal
\end{equation}
on a surface of the match.

These problems are of course beyond the scope of this work and will be
considered elsewhere.

Let us consider full relativistic orbits in an equatorial plane for massive
particles and photons, i.e.\ \er{D.27} and \er{D.52},
\bea{Dn222}
\pz r\vf&=&\pm\sqrt{e^{-2(A(r)+B(r))}\,\frac{H^2}{h^2}\,r^4-\frac{r^2}{e^{2B(r)}}
-\frac{r^4}{h^2e^{2B(r)}}}\\
\pz r\vf&=&\pm\sqrt{e^{-2(A(r)+B(r))}\,\frac{H^2}{h^2}\,r^4-\frac{r^2}{e^{2B(r)}}
}\label{Dn223}
\end{eqnarray}
in a background \gr al field described by a metric \er{D.21}.
We consider also \e s \er{D.26} and \er{D.51}
\bea{Dn224a}
\pz r\tau&=&\pm\sqrt{e^{-2(A(r)+B(r))}H^2-\frac{h^2}{r^2}\,e^{-2B(r)}
-\frac{1}{e^{2B(r)}}}\\
\pz r\si&=&\pm\sqrt{e^{-2(A(r)+B(r))}H^2-\frac{h^2}{r^2}\,e^{-2B(r)}}
\label{Dn225a}
\end{eqnarray}
where $\tau$ is a proper time for massive test particle and $\si$ is an
affine parameter along a photon orbit.

\begin{figure}
\hbox to \textwidth{\ing[width=0.43\textwidth]{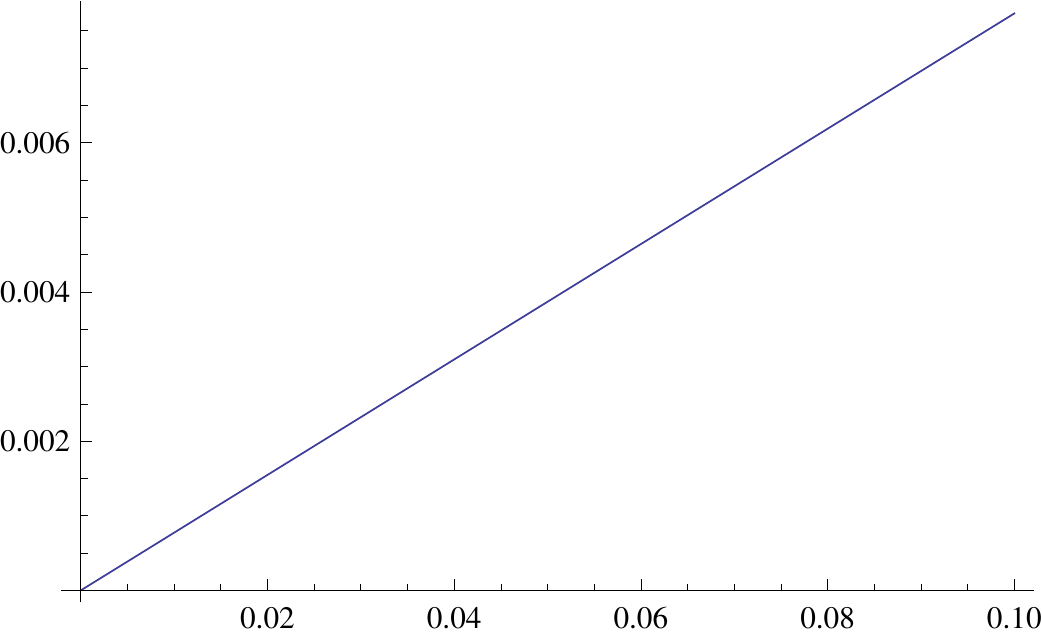}\hfil
\ing[width=0.43\textwidth]{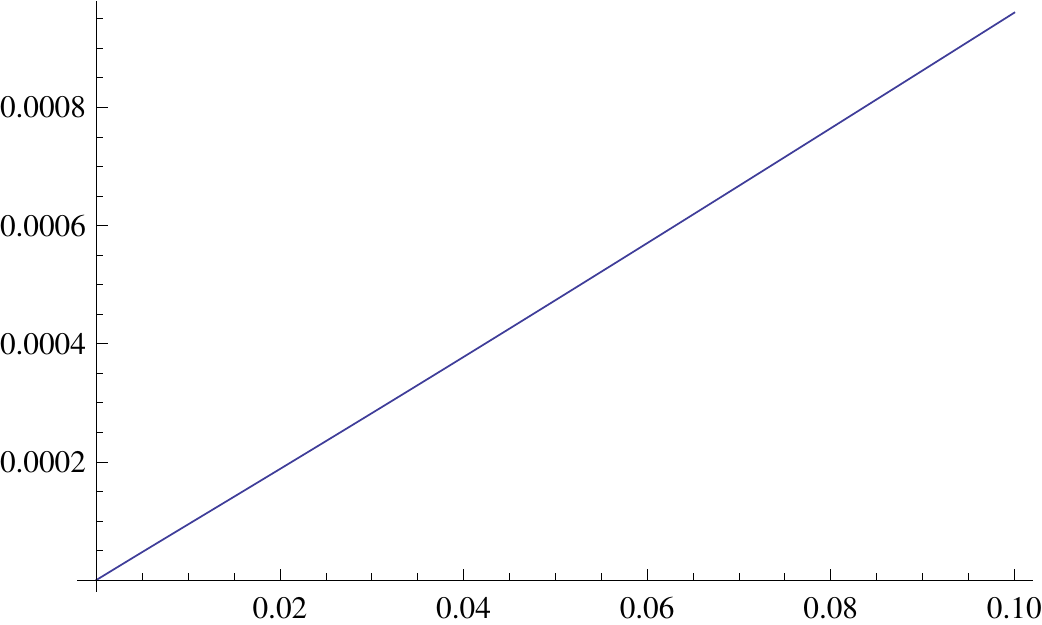}}
\hbox to \textwidth{\hbox to 0.43\textwidth{\hfil(A)\hfil}\hfil
\hbox to 0.43\textwidth{\hfil(C)\hfil}}
\hbox to \textwidth{\ing[width=0.43\textwidth]{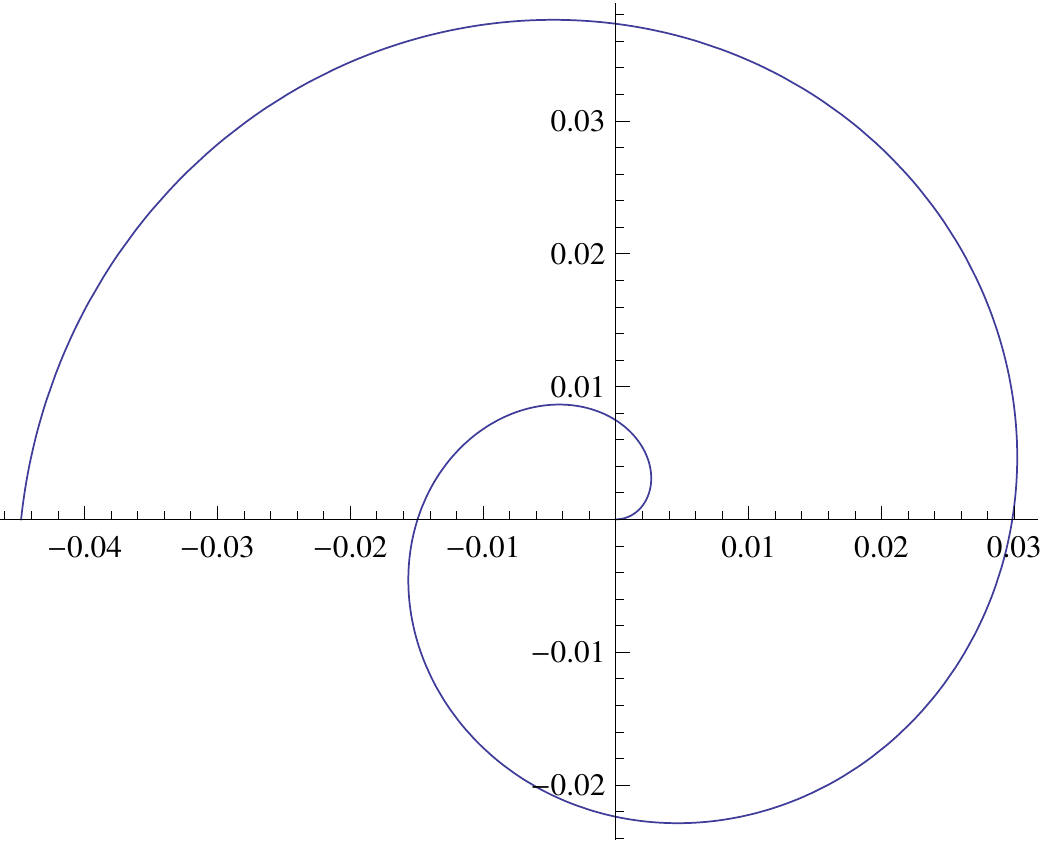}\hfil
\ing[width=0.43\textwidth]{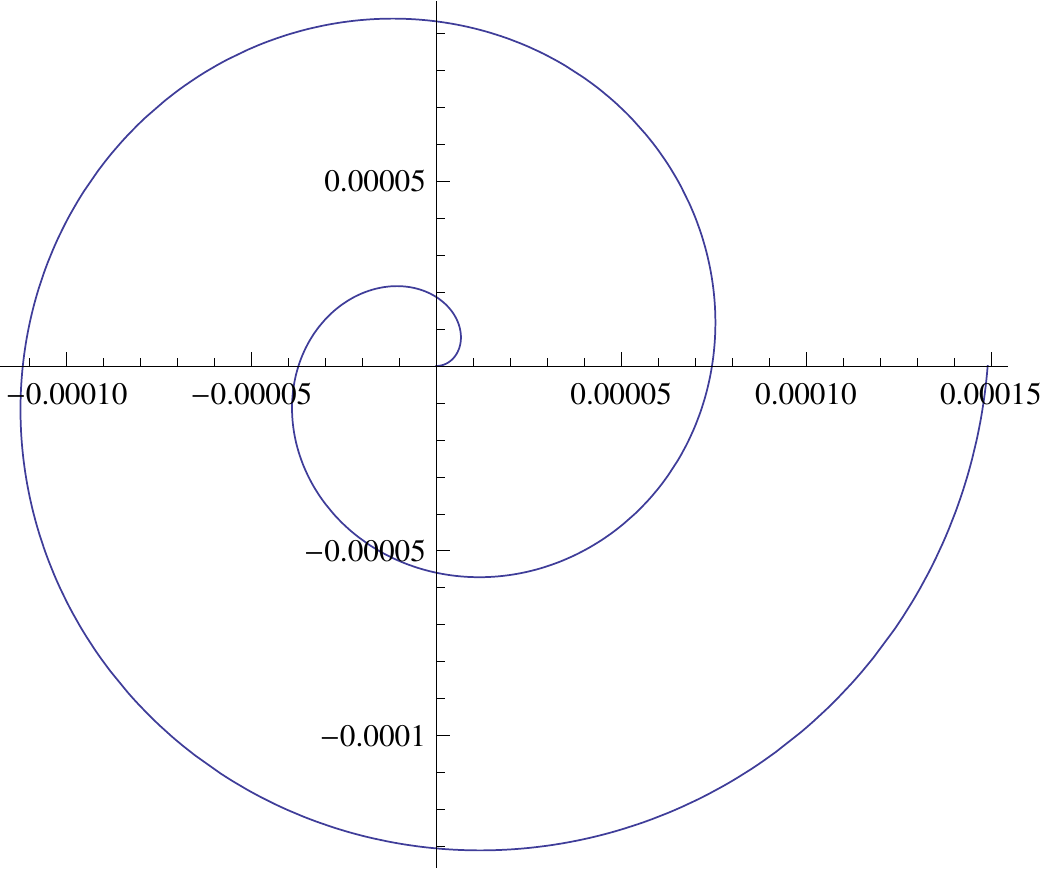}}
\hbox to \textwidth{\hbox to 0.43\textwidth{\hfil(B)\hfil}\hfil
\hbox to 0.43\textwidth{\hfil(D)\hfil}}
\caption{(A)---a plot of $r(\tau)$ for a massive point body (a test particle)
for parameters $a_h=1.5\t10^{-6}$, $e=3$, $r(0)=10^{-10}$;
(B)---a polar plot of an orbit for a massive point body (a test particle)
for parameters $a_h=1.5\t10^{-6}$, $e=3$, $r(0)=10^{-10}$;
(C)---a plot of $r(\tau)$ for a massive point body (a test particle)
for parameters $a_h=10^{-4}$, $e=17$, $r(0)=10^{-10}$;
(D)---a polar plot of an orbit for a massive point body (a test particle)
for parameters $a_h=10^{-4}$, $e=17$, $r(0)=10^{-10}$.
\label{lrr-a}}
\end{figure}

The field \e s
for $A(r)$ and $B(r)$ are given by
\beq{Dn224}
\bal
{}&\pz{A(r)}r=\frac{\eb2-1}{4r}\q\q &
&\pz{^2B(r)}{r^2}=\frac{9-10\eb2+\eb4}{8r^2}\\
&A(7.36\t10^{-10})=-4\t10^{-12}\\
&B(7.36\t10^{-10})=4\t10^{-12}\q\q &
&\pz Br(7.36\t10^{-10})=14.31
\eal
\end{equation}
The last \e s are really \e s \er{Dn176} with $B_1=14.13$. This value is from
Eq.~\er{Dn184}. In order to consider the problem we should parametrize $H$
and~$h$.

\begin{figure}
\hbox to \textwidth{\vbox{%
\hbox{\ing[width=0.45\textwidth]{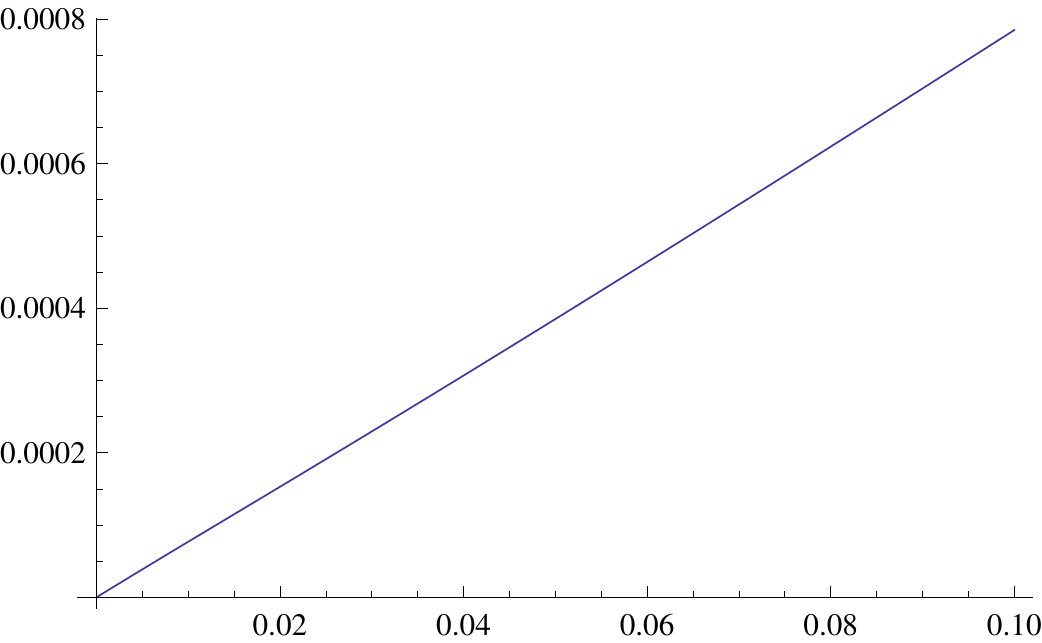}}
\hbox to0.45\textwidth{\hfil(A)\hfil}
\hbox{\ing[width=0.45\textwidth]{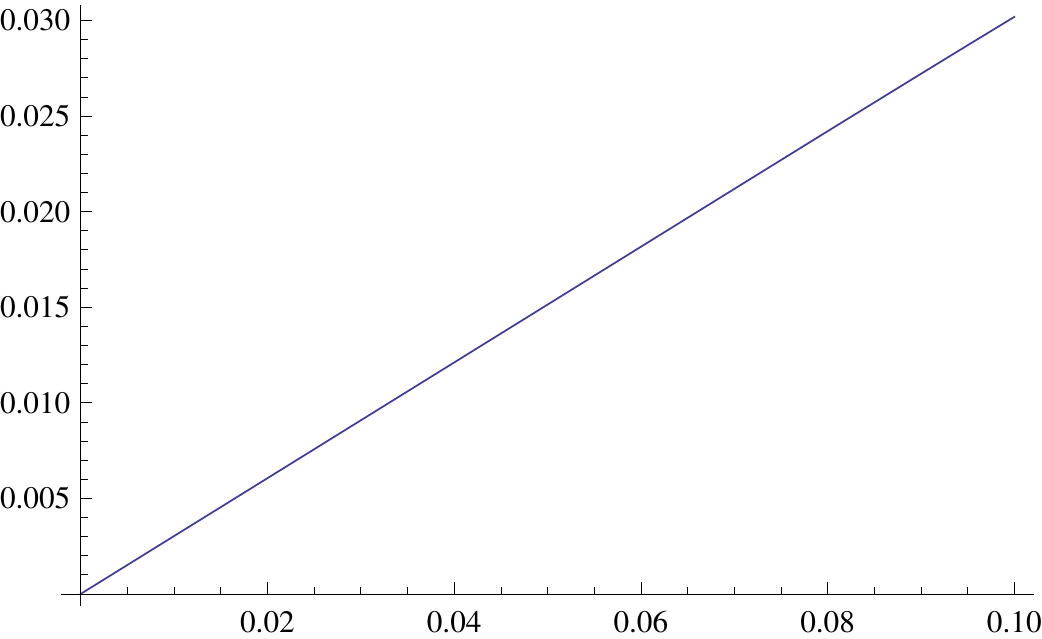}}
\hbox to0.45\textwidth{\hfil(C)\hfil}
\hbox{\ing[width=0.45\textwidth]{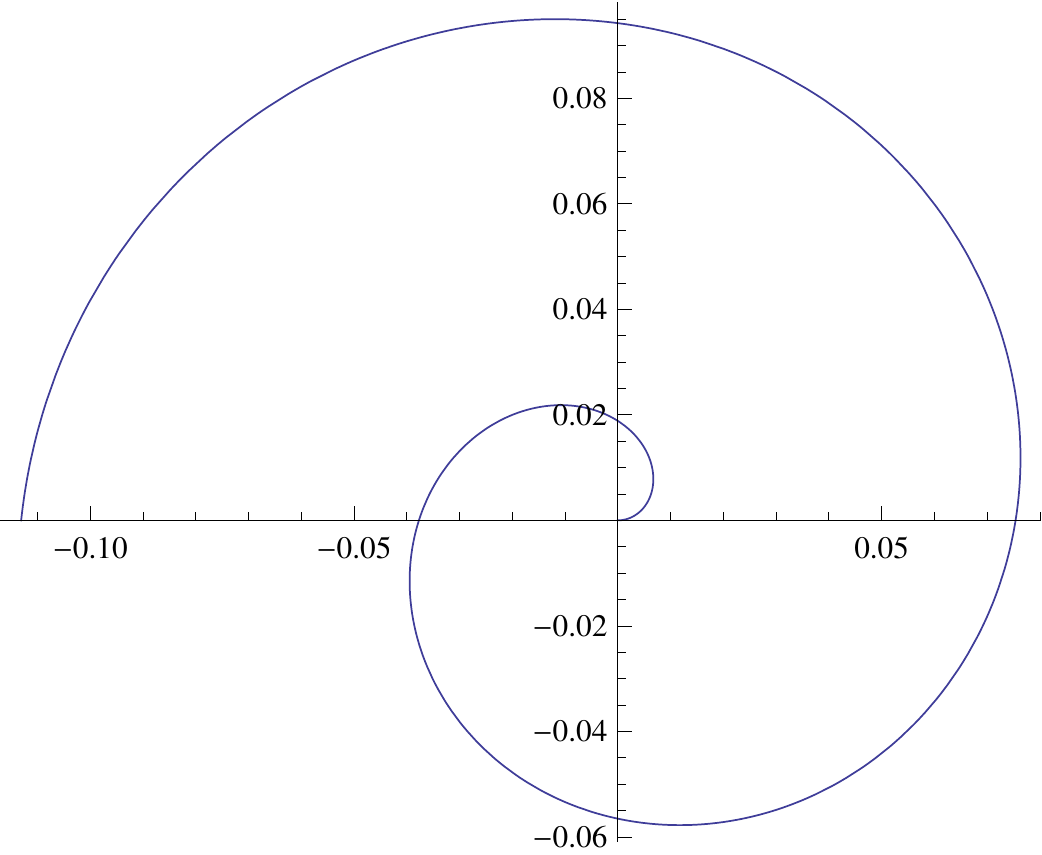}}}\hfil
\ing[width=0.45\textwidth]{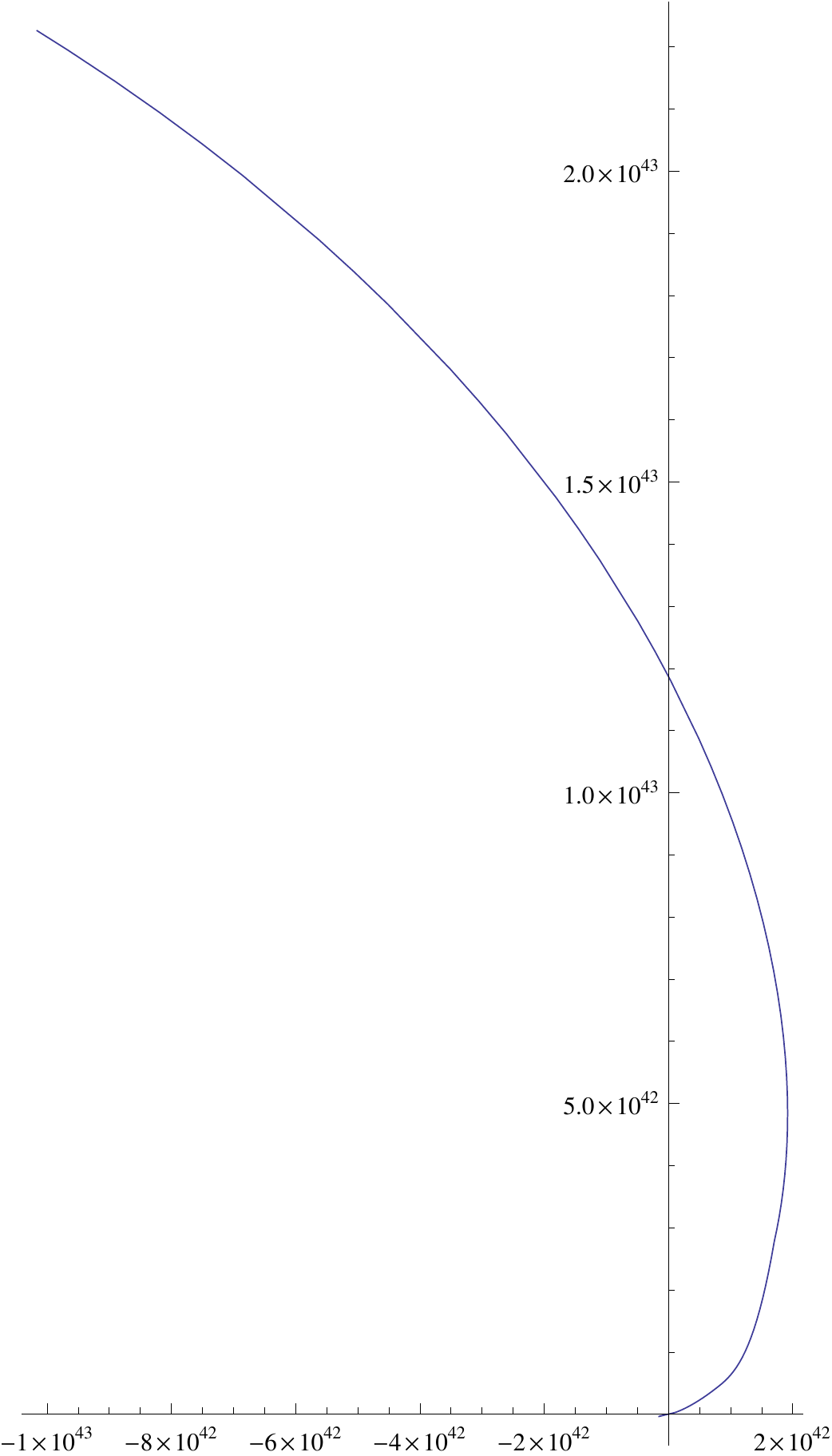}}
\hbox to \textwidth{\hbox to 0.45\textwidth{\hfil(D)\hfil}\hfil
\hbox to 0.45\textwidth{\hfil(B)\hfil}}
\caption{(A)---a plot of $r(\tau)$ for a massive point body (a test particle)
for parameters $a_h=15\t10^{-5}$, $e=3$, $r(0)=10^{-10}$;
(B)---a polar plot of an orbit for a massive point body (a test particle)
for parameters $a_h=15\t10^{-5}$, $e=3$, $r(0)=10^{-10}$;
(C)---a plot of $r(\tau)$ for a massive point body (a test particle)
for parameters $a_h=10^{-7}$, $e=17$, $r(0)=10^{-10}$;
(D)---a polar plot of an orbit for a massive point body (a test particle)
for parameters $a_h=10^{-7}$, $e=17$, $r(0)=10^{-10}$.
\label{lrr-b}}
\end{figure}

We use the following formulae for massive particles:
\bea{Dn225}
H&=&1+0.448\t10^{-8}\,\frac1{a_h}\\
h&=&0.47\t10^{-16}\sqrt{(e^2-1)a_h}, \q e>1 \label{Dn226}
\end{eqnarray}
where $a_h$ is semi-major axis for a hyperbola measured in AU and $e$ is its
excentricity.

For a photon we get
\beq{Dn227}
H=0.448\t10^{-8}\,\frac1{a_h}
\end{equation}
and for $h$ the same formula \er{Dn226}. All of these formulae are coming
from the known formulae (in the case of massive particles)
\bea{Dn228}
H&=&c^2+\frac{k^2}{2a_h}\\
h&=&k\sqrt{(e^2-1)a_h} \label{Dn229}\\
k^2&=&G_NM_\odot, \label{Dn230}
\end{eqnarray}
where $M_\odot$ is a mass of the Sun, $G_N$ a \gr al \ct, $c$ a velocity of
light written in our system of units in such a way that $a_h$ is measured in
AU.

\begin{figure}[h]
\hbox to \textwidth{\ing[width=0.4\textwidth]{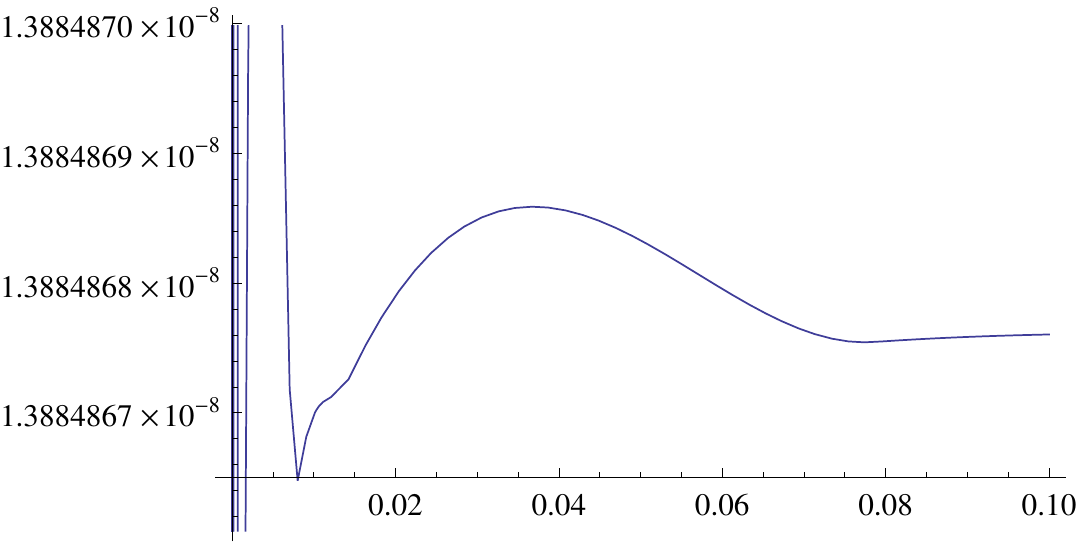}\hfil
\ing[width=0.44\textwidth]{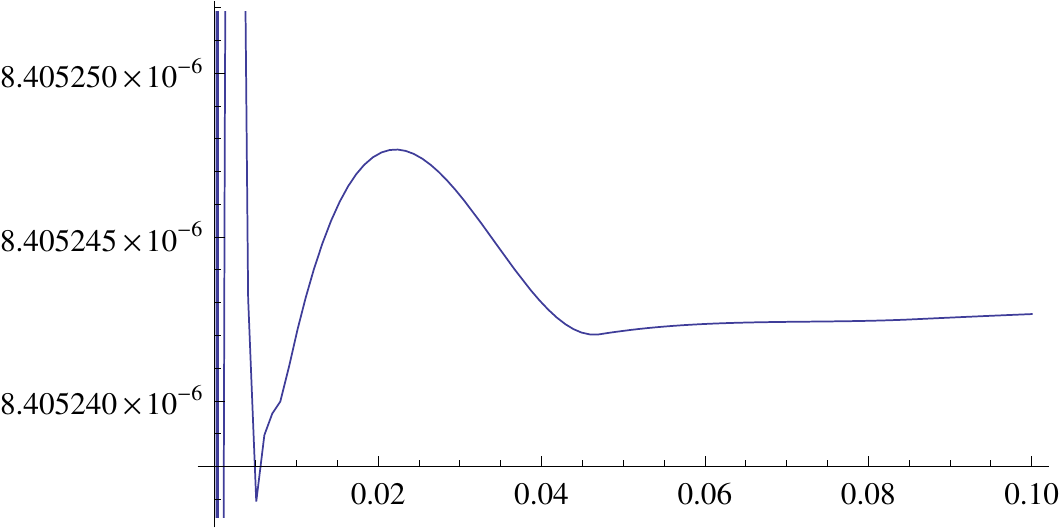}}
\hbox to \textwidth{\hbox to 0.44\textwidth{\hfil(A)\hfil}\hfil
\hbox to 0.44\textwidth{\hfil(B)\hfil}}
\hbox to \textwidth{\ing[width=0.44\textwidth]{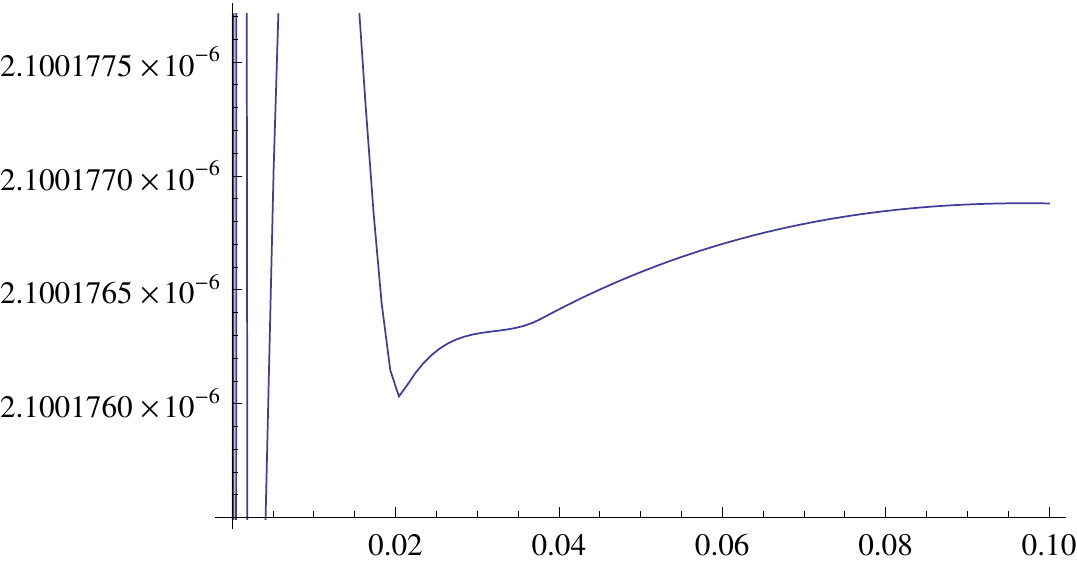}\hfil
\ing[width=0.44\textwidth]{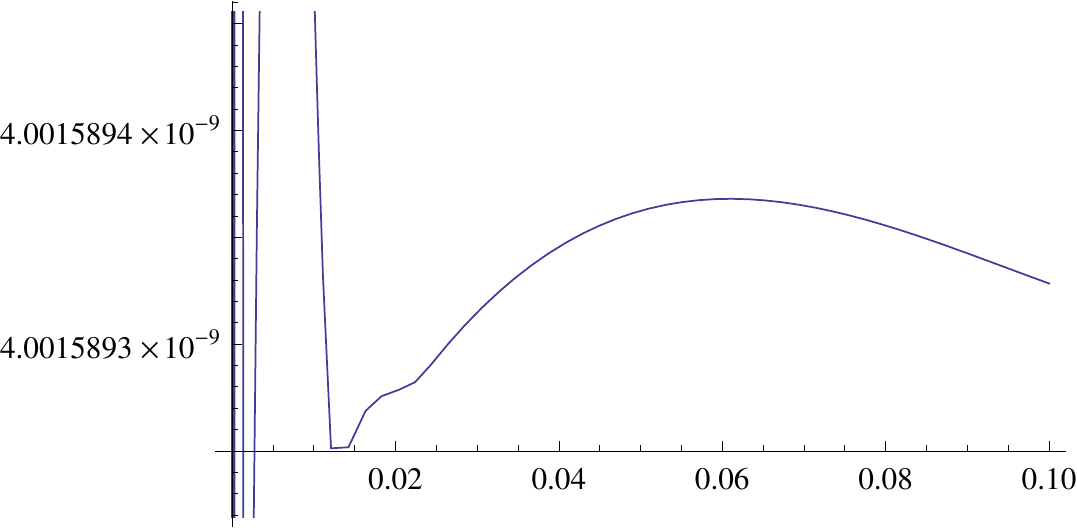}}
\hbox to \textwidth{\hbox to 0.44\textwidth{\hfil(C)\hfil}\hfil
\hbox to 0.44\textwidth{\hfil(D)\hfil}}
\caption{(A)---a plot of $\vf(\tau)$ for a massive point body (a test particle)
for parameters $a_h=1.5\t10^{-6}$, $e=3$, $\vf(0)=0$;
(B)---a plot of $\vf(\tau)$ for a massive point body (a test particle)
for parameters $a_h=10^{-4}$, $e=17$, $\vf(0)=0$;
(C)---a plot of $\vf(\tau)$ for a massive point body (a test particle)
for parameters $a_h=15\t10^{-5}$, $e=3$, $\vf(0)=0$;
(D)---a plot of $\vf(\tau)$ for a massive point body (a test particle)
for parameters $a_h=10^{-7}$, $e=17$, $\vf(0)=0$.
\label{aa}}
\end{figure}

In the case of a photon
\beq{Dn231}
H=\frac{k^2}{2a_h}\,.
\end{equation}
We suppose here that an energy per unit mass of a massive particle can be
divided into two parts ($c^2$) and a remaining total energy parametrized by a
value $a_h$ which has a sound meaning for a nonrelativistic case. The same
ideology has been applied for an angular momentum for a unit mass. This has
been extrapolated to a photon case. In this case $a_h$ can be also considered
as a semi-major axis of an ellipse, i.e.,
$$
H=-0.448\t10^{-8}\,\frac1{a_e}, \q h=0.47\t10^{-16}\sqrt{(1-e^2)a_e}, \q
0\le e<1.
$$
Thus we consider Eqs \er{Dn222},
\er{Dn223} and \er{Dn224a} for several values of $a_h$ and $e$ and initial
value $r(0)=10^{-10}$. We suppose sign $+$ in Eqs \er{Dn222}--\er{Dn225a}.

Using Mathematica 7 we get the following results. In the case of massive
point bodies (point particles) we plot $r(\tau)$ and orbits in polar \cd s
(see Fig.~\ref{lrr-b}).

Let us consider the following integral of motion for a massive test particle:
\refstepcounter{equation}\label{Dn233}
$$
\dsl{
\hfill h=r^2(\tau)\pz \vf \tau(\tau),\hfill \rm(\theequation)\cr
\hbox{or }\hfill \pz\vf t=\frac{c^2he^{2A(r)}}{Hr^2}\hfill \rm(\theequation a)}
$$
where $\tau$ is a proper time and $t$ is a \cd\ time. Using a parametrization
\er{Dn226} and a dependence of~$r(\tau)$ given by \er{Dn224a} we plot
$\vf(\tau)$ for several values of $a_h$ and $e$ on Fig.~\ref{aa}.

In the case of a photon we plot orbits in polar \cd s
(see Fig.~\ref{plrr-a}).

On both axes of an orbit plot the length is measured in our unit $L$
($\sim10$\,Mpc), as before for massive particles. $\tau$---proper time is
measured in our unit of time ($\sim10^{19}$\,s). The plotted orbits in both
cases give us a taste of an influence of the \gr al field with an \an\ \ac\
taken into account on movement of massive point bodies and photons. Further
investigations consist in changing initial conditions for a \gr al field and
for test particles. In Appendix~E we write down listings of programmes
written in Mathematica~7 to calculate orbits described above. Let us notice
that $r(0)=10^{-10}\,L$ corresponds to $r(0)=206.3\,$AU.

\refstepcounter{figure}\label{plrr-a}
\hbox to \textwidth{\ing[width=0.44\textwidth]{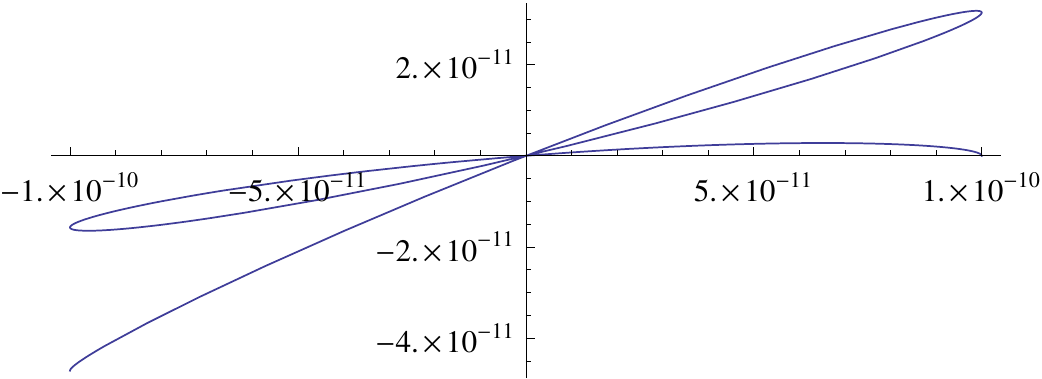}\hfil
\ing[width=0.44\textwidth]{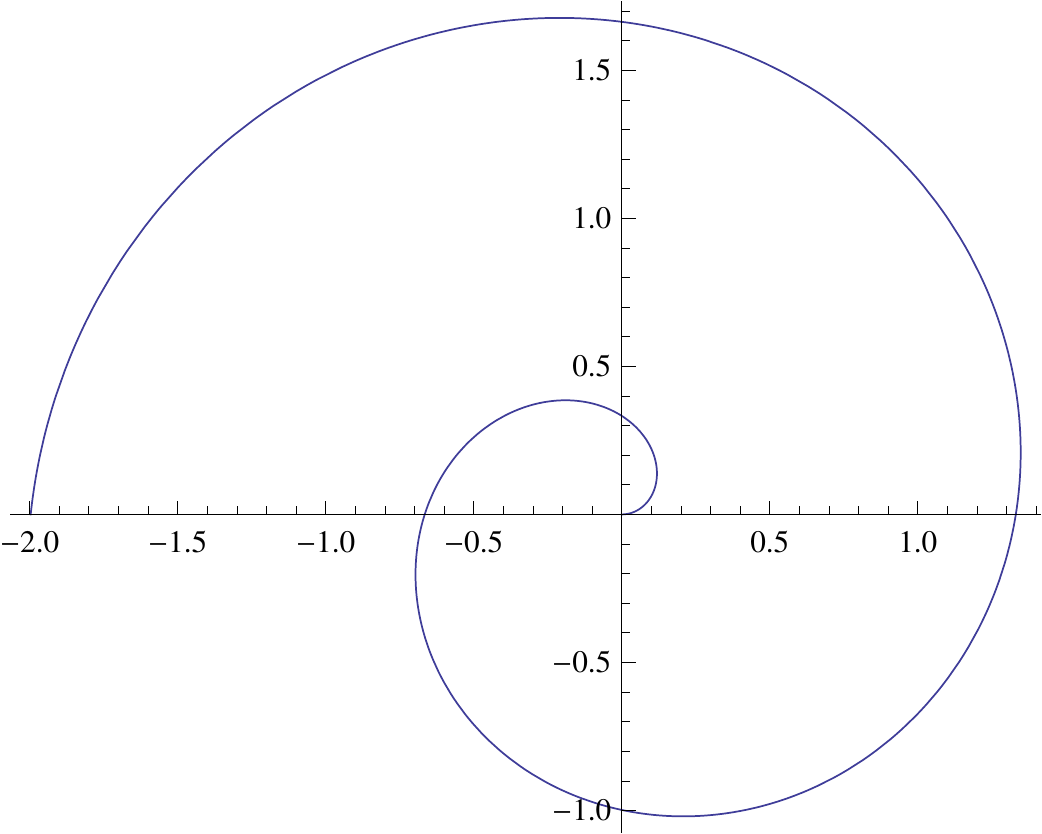}}
\hbox to \textwidth{\hbox to0.44\textwidth{\hfil(A)\hfil}\hfil
\hbox to0.44\textwidth{\hfil(B)\hfil}}
\noindent
{\small Figure \thefigure: (A)---a polar plot for a photon orbit
for parameters $a_h=20$, $e=10$, $r(0)=10^{-10}$;
(B)---a polar plot for a photon orbit
for parameters $a_h=1.5\t10^{-6}$, $e=3$, $r(0)=10^{-10}$.}

\hbox to \textwidth{\hfil
\ing[width=0.17\textwidth]{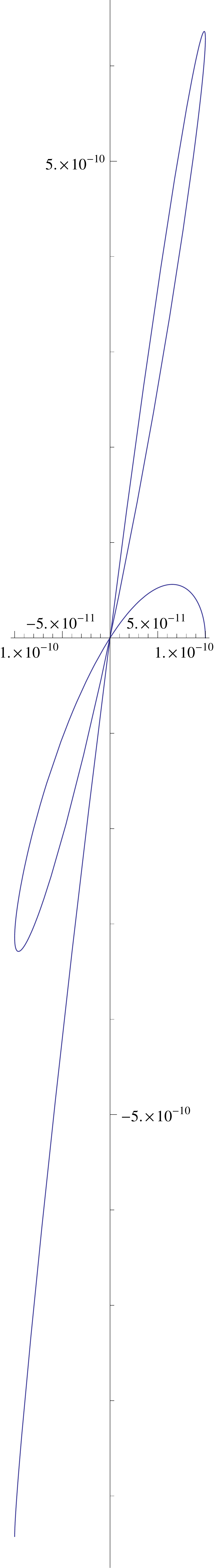}\hfil
\vbox{\hsize=0.48\textwidth
\centerline{\ing[width=0.45\textwidth]{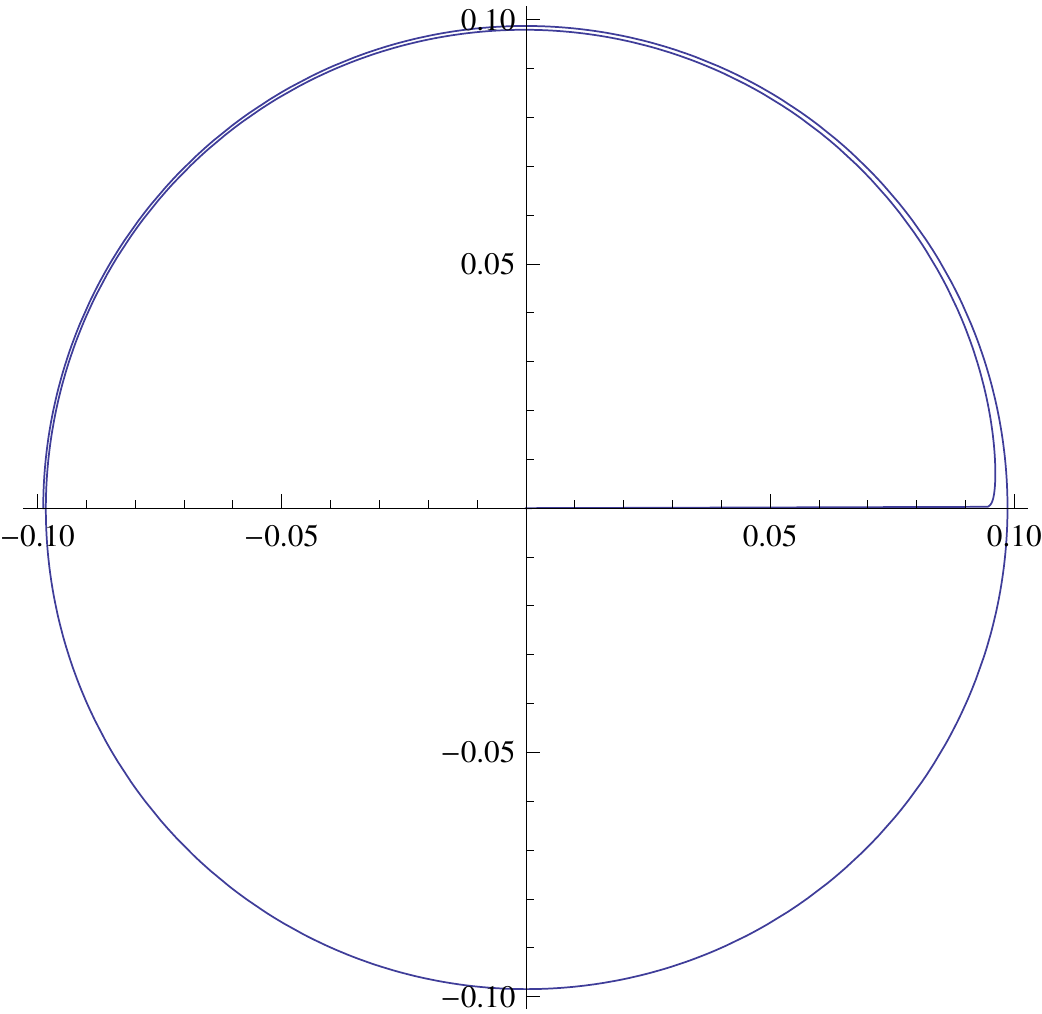}}
\centerline{(D)}

\vskip30pt
\centerline{\ing[width=0.45\textwidth]{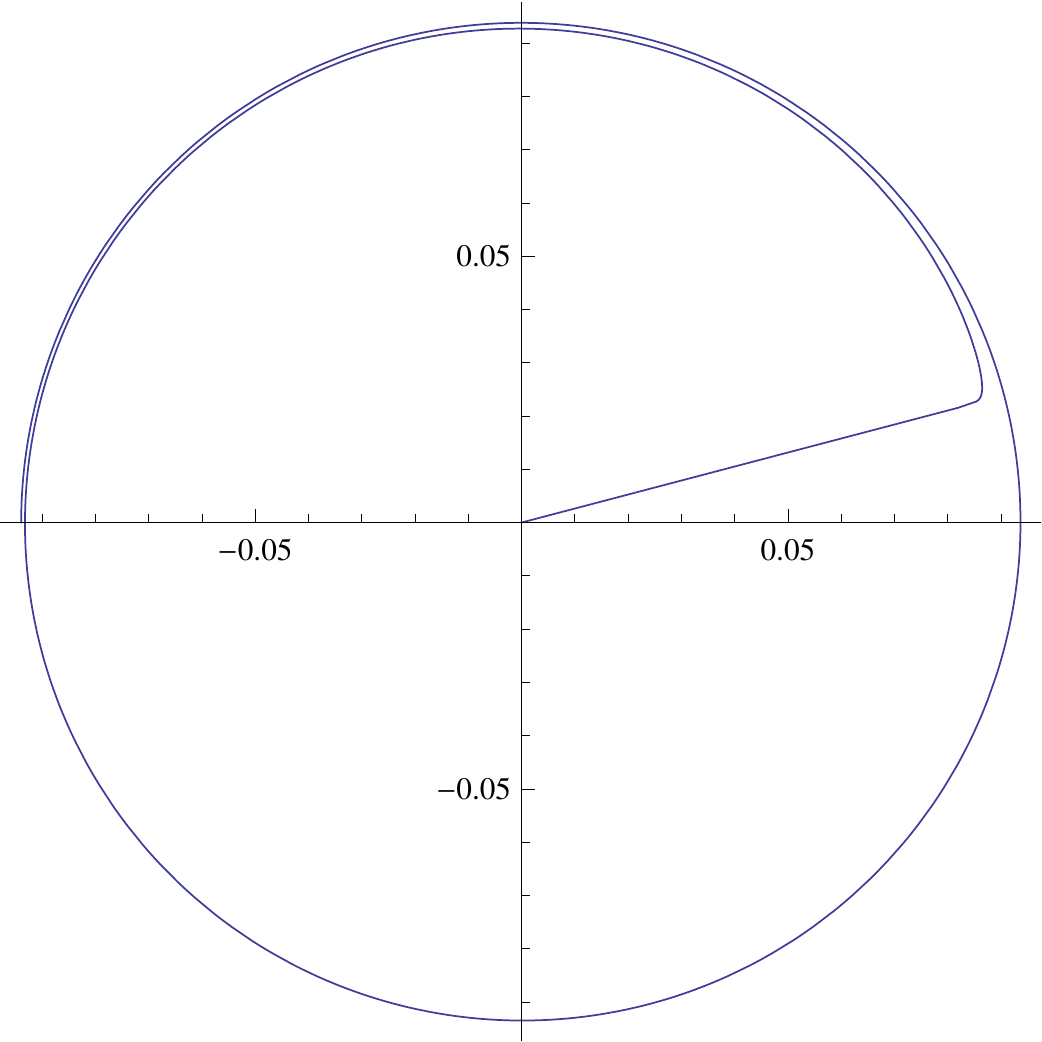}}}\hfil}
\hbox to \textwidth{\hfil\hbox to 0.17\textwidth{\hfil(C)\hfil}\hfil
\hbox to 0.45\textwidth{\hfil(E)\hfil}\hfil}
\medskip
{\small \noindent Figure \thefigure\ (cont.): (C)---a polar plot for a photon orbit
for parameters $a_h=20$, $e=7$, $r(0)=10^{-10}$;
(D)---a polar plot for a photon orbit
for parameters $a_e=0.001$, $e=0$, $r(0)=10^{-10}$;
(E)---a polar plot for a photon orbit
for parameters $a_e=0.001$, $e=0.1$, $r(0)=10^{-10}$.}

\newpage
It is worth to notice that orbits for massive test particles are not bounded
for $H-c^2>0$. It means that relativistic extension of an \an\ \ac\ does not
confirm a pessimistic prediction on a \cfn\ in the \SS. In the case of photon
orbits are closed for some cases of~$a$ and~$e$ (as in GR), e.g.\ circular
orbits. They can also be unbounded showing a complicating behaviour.

Let us consider Eq.\ \er{D.28} for a metric \er{D.21} \st $\ov
A(r)=e^{2A(r)}$, $\ov B(r)=e^{2B(r)}$, $A(r)$ and $B(r)$ satisfying
\er{Dn224}. One gets for $k^\mu=(k_r,k_\th,k_\vf,\frac\o c)$, where $c$ is a
velocity of light.
\beq{Dn234}
\bga
\pz\o\si - \pz Br\,\o\cdot k_r=0\\
\pz{k_r}\si + \frac r{e^{2A(r)}}\,k_\th^2
+k_r^2\,\pz Ar+ \frac r{e^{2A(r)}}\,k_\vf^2 + e^{2(B(r)-A(r))}
\,\pz Br\,\o^2=0\\
\pz{k_\th}\si - \frac12\,k_\vf^2+\frac1r\,k_\th k_\vf=0\\
\pz{k_\vf}\si + \frac1r\,k_\vf k_r=0
\ega
\end{equation}
where we use Christoffel symbols for a metric \er{D.21}, supposing an
equatorial motion, $\th=\frac\pi2$, $\si$~is an affine parameter.

Simultaneously  we have
\beq{Dn235}
0=g_\m k^\mu k^\nu= e^{2A(r)}\,\frac{\o^2}{c^2} - e^{2B(r)}k_r^2-r^2k_\vf^2.
\end{equation}
In this way we can suppose $k_\th=0$. Moreover, to simplify \e s we put also
$k_\vf =0$. Thus one gets
\beq{Dn236}
e^{A(r)}\,\frac\o c=e^{B(r)}k_r.
\end{equation}
Using Eqs \er{Dn230} and \er{Dn234} one gets
\beq{Dn237}
\pz\o\si - \frac1c\,\pz Br\,\o^2 e^{(A(r)-B(r))}.
\end{equation}

A \so\ of Eq.\ \er{Dn237} can be written in the form
\beq{Dn238}
T=T_0+\D T
\end{equation}
where
\bea{Dn239}
T&=&\frac{2\pi}\o=\frac1\nu\\
T_0&=&\frac{2\pi}{\o_0}=\frac1{\nu_0} \label{Dn240}
\end{eqnarray}
and $\D T$ satisfies the \e
\beq{Dn241}
\bga
\pz{(\D T(r))}\si=-\frac{2\pi}c\,\pz Br\,e^{A(r)-B(r)}\\
\D T(0)=0.
\ega
\end{equation}
In this way we get
\beq{Dn242}
\D \nu=\nu_0^2\,\frac{\D T}{1+\nu_0\D T}.
\end{equation}
$T_0$ means an integration \ct\ for Eq.\ \er{Dn237}.

\begin{figure}[h]
\hbox to \textwidth{\ing[width=0.44\textwidth]{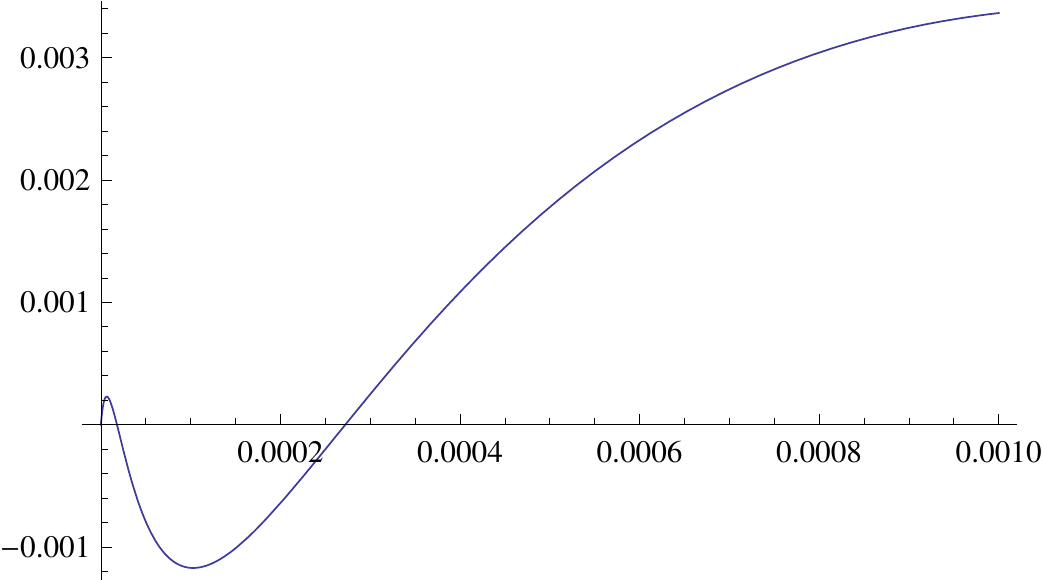}\hfil
\ing[width=0.44\textwidth]{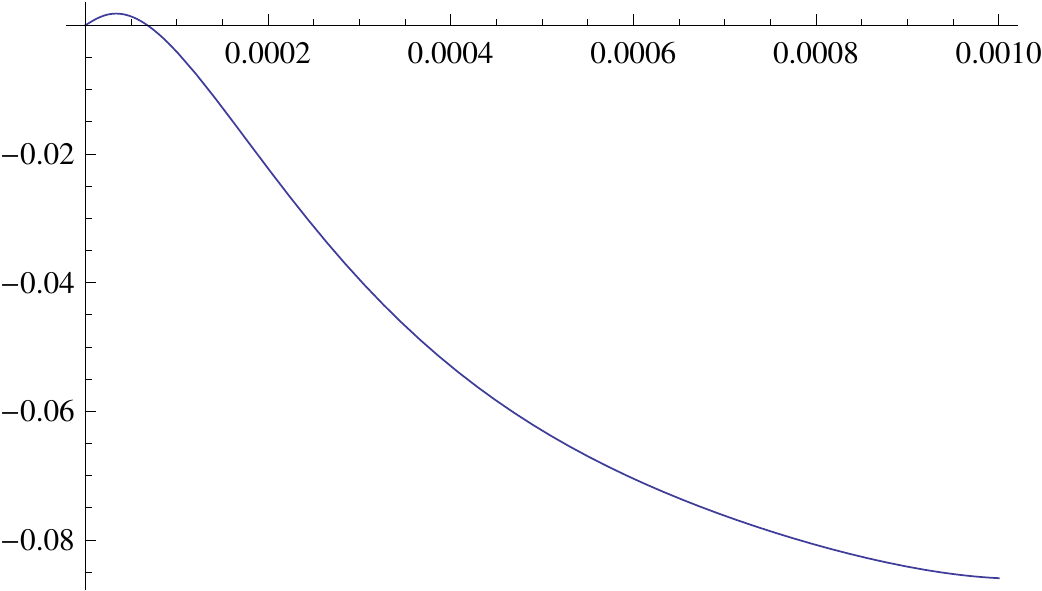}}
\hbox to \textwidth{\hbox to0.44\textwidth{\hfil(A)\hfil}\hfil
\hbox to0.44\textwidth{\hfil(B)\hfil}}
\hbox to \textwidth{\hfil \ing[width=0.44\textwidth]{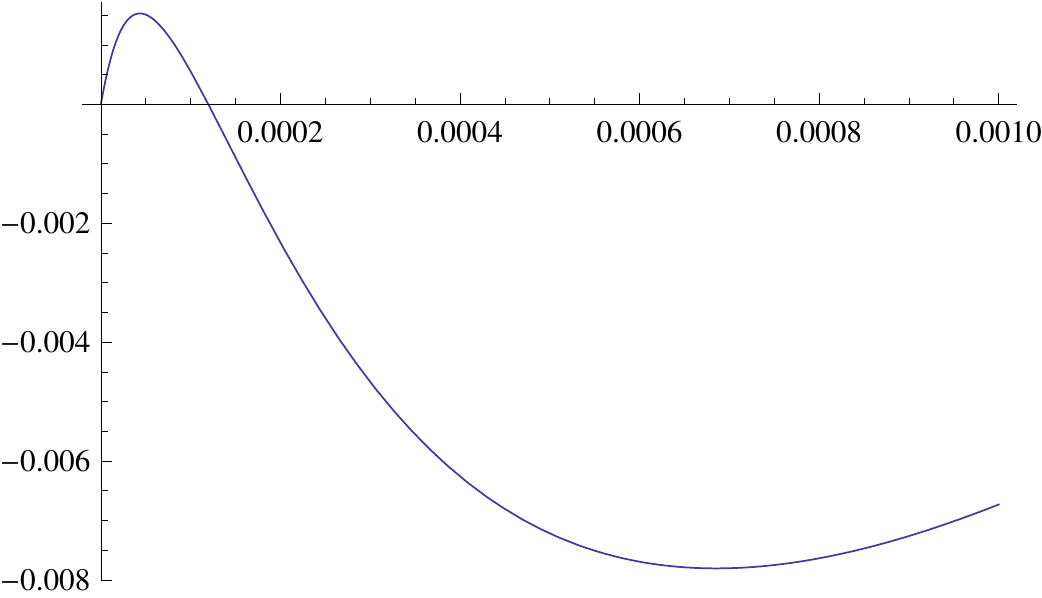}\hfil}
\hbox to \textwidth{\hfil\hbox to0.44\textwidth{\hfil(C)\hfil}\hfil}
\caption{(A)---a plot of $\D T$
for $a_h=1.5\t10^{-6}$, $e=3$, $r(0)=10^{-10}$;
(B)---a plot of $\D T$
for $a_h=20$, $e=3$, $r(0)=10^{-10}$;
(C)---a plot of $\D T$
for $a_e=10^{-3}$, $e=0.001$, $r(0)=10^{-10}$.
\label{pqr}}
\end{figure}

Moreover, $T$ means here period, $\nu$ a frequency of light shifting from $T_0$
and $\nu_0$ by an amount of $\D T$ due to \gr al effects on an orbit of a
photon. Supposing that $r=r(\si)$ for a photon orbit (see Eq.\ \er{Dn225a})
and using several values for $a$, $e$ and $r(0)=10^{-10}$, we plot a shift of
the period $\D T$ on Fig.\ \ref{pqr}. $\D T$~is measured here in ${\rm ns}=
10^{-9}$\,s and $\si$ in $L$ ($\sim10$\,Mpc). Moreover,
$\la=\frac{2\pi}{k_r}$ and we get
\beq{Dn243}
\la=(\la_0+c\D T(\si))e^{B(r(\si))-A(r(\si))}
\end{equation}
where $\la$ is a length of an \elm c wave with a frequency $\nu$ and a
period~$T$ on an orbit of a photon  in the \gr al field described by our
metric. $\la$ is measured in meters.

In this way we get an influence of an \an\ \ac\ on photon orbits, a frequency
and a wave length of an \elm c wave. A~listing of a programme written in
Mathematica~7 to calculate and plot $\D T$ is quoted in Appendix~E.

Let us consider Eqs \er{D.123}--\er{D.125}. They form a system of ordinary
differential \e s of the second order. In order to give a taste of the full
\NK{\JT\ } we solve these \e s for some special initial Cauchy conditions,
\beq{Da245}
\bal
{}&\a(0)=C_2=0.1 &\qquad&\pz\a r(0)=C_3=0\\
&\vF(0)=C_4=0.1 &\q&\pz\vF r(0)=C_5=0.01\\
&f(0)=C_6=0.1&\q &\pz fr(0)=C_7=0
\eal
\end{equation}

\begin{figure}
\hbox to \textwidth{\ing[width=0.48\textwidth]{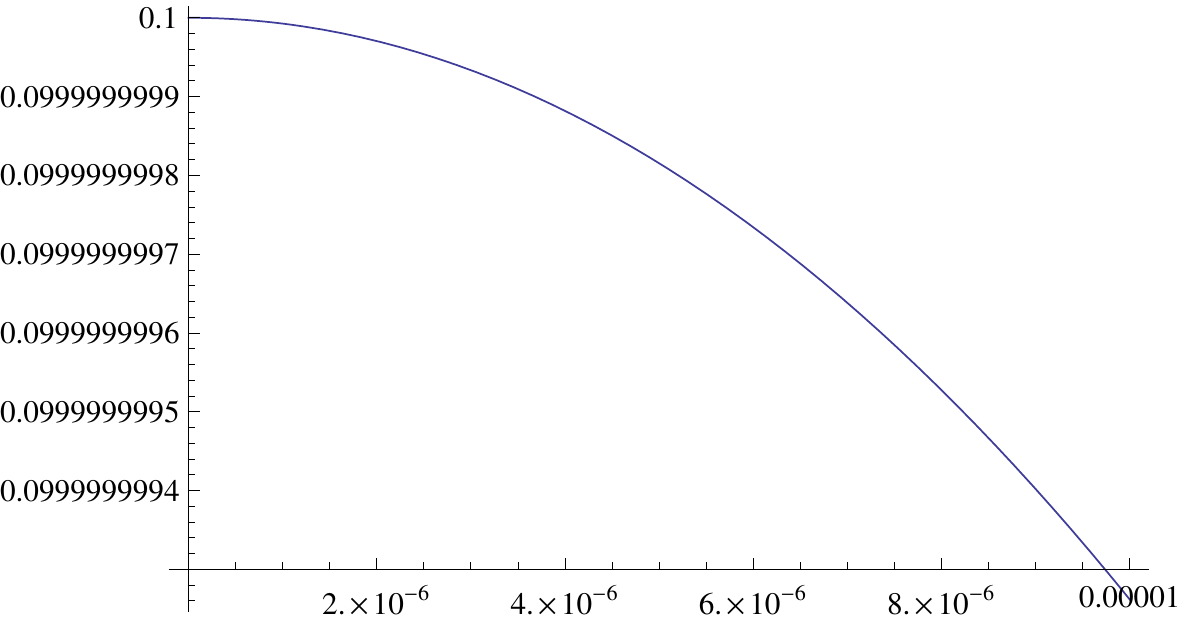}\hfil
\ing[width=0.48\textwidth]{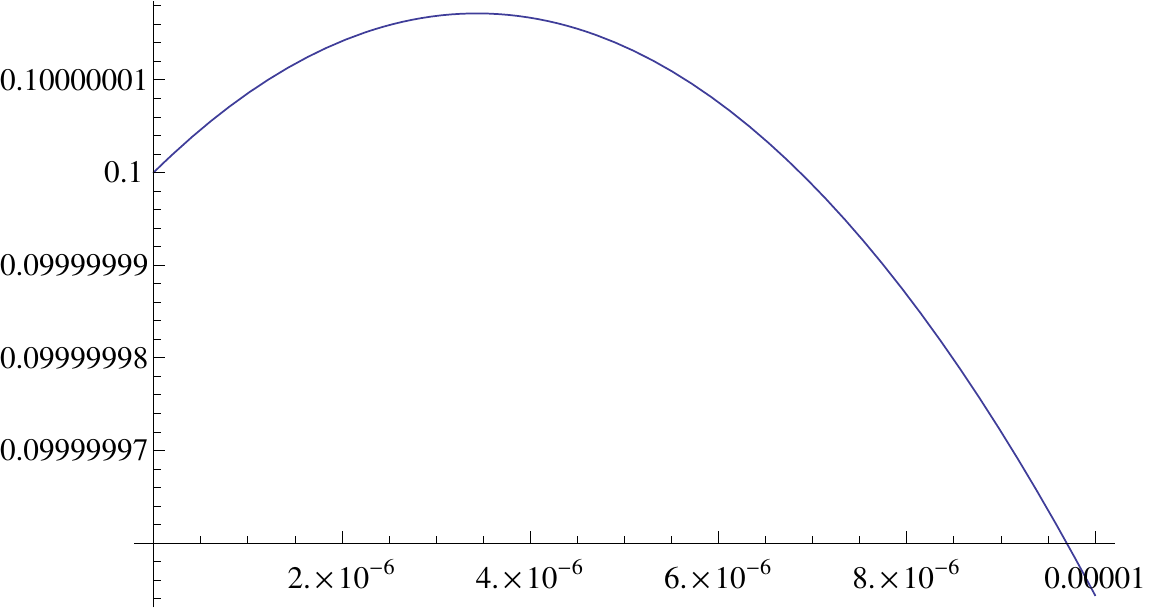}\hfil}
\hbox to \textwidth{\hbox to0.48\textwidth{\hfil(A)\hfil}\hfil
\hbox to0.48\textwidth{\hfil(B)\hfil}\hfil}
\hbox to \textwidth{\ing[width=0.48\textwidth]{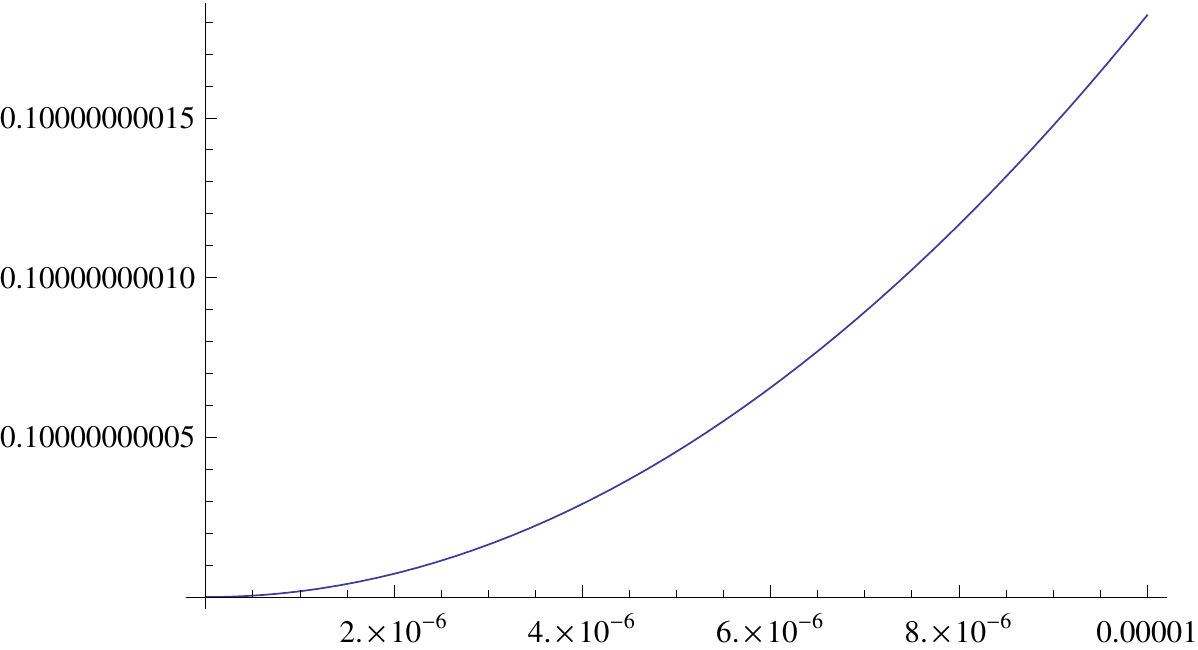}\hfil
\ing[width=0.45\textwidth]{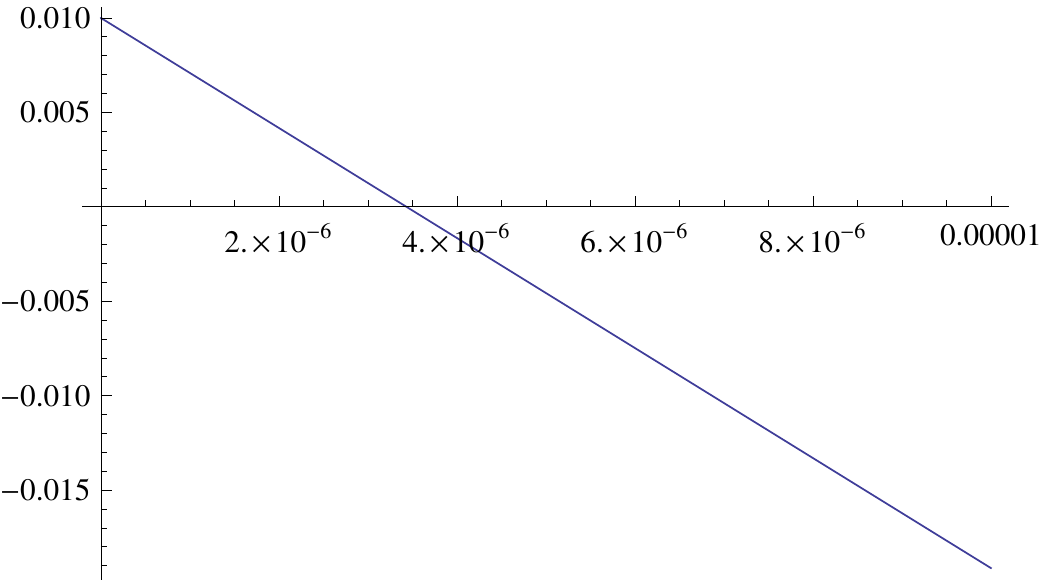}\hfil}
\hbox to \textwidth{\hbox to0.48\textwidth{\hfil(C)\hfil}\hfil
\hbox to0.45\textwidth{\hfil(D)\hfil}\hfil}
\hbox to \textwidth{\ing[width=0.45\textwidth]{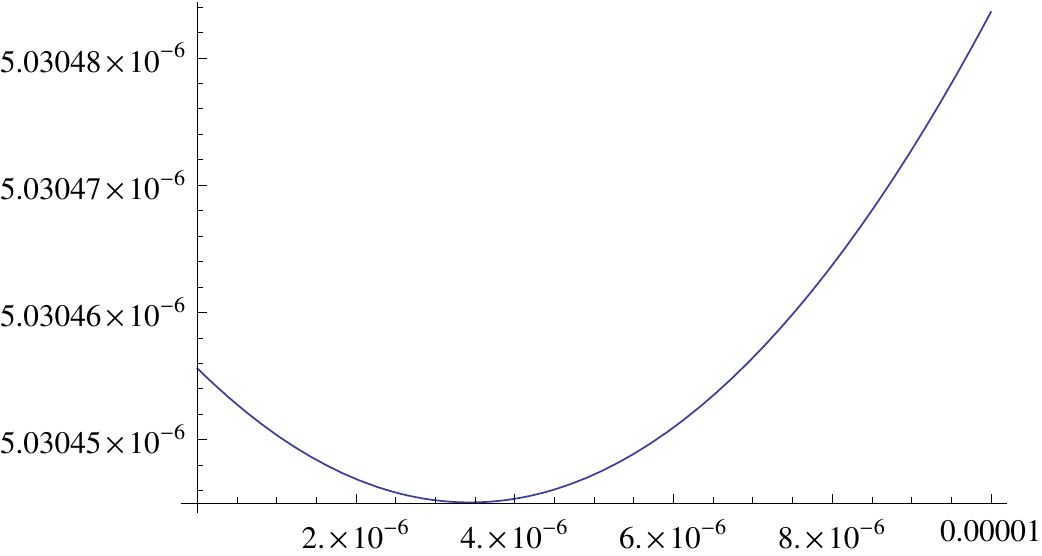}\hfil
\ing[width=0.45\textwidth]{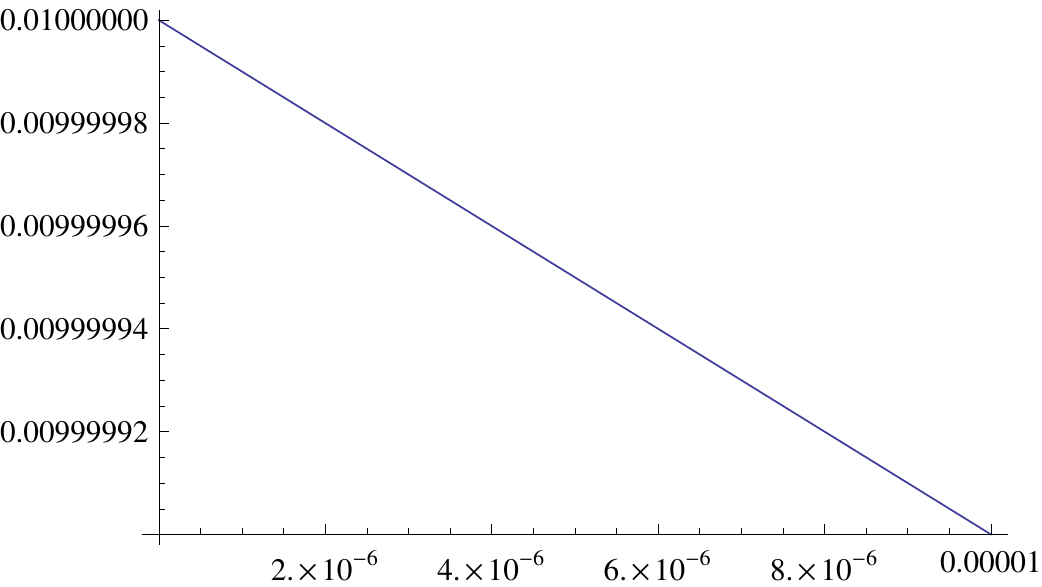}\hfil}
\hbox to \textwidth{\hbox to0.45\textwidth{\hfil(E)\hfil}\hfil
\hbox to0.45\textwidth{\hfil(F)\hfil}\hfil}
\hbox to \textwidth{\hfil \ing[width=0.45\textwidth]{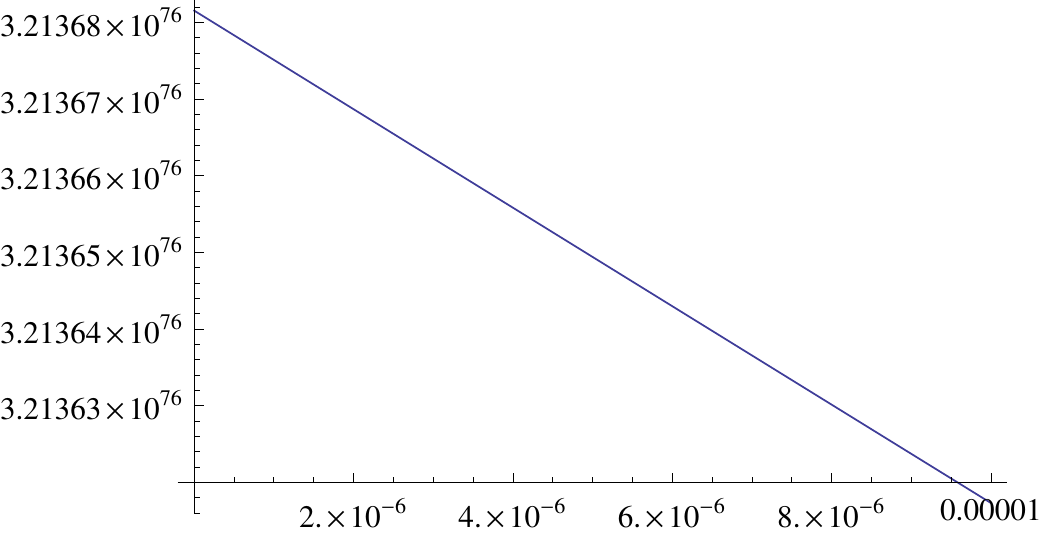}\hfil}
\hbox to \textwidth{\hfil\hbox to0.45\textwidth{\hfil(G)\hfil}\hfil}
\caption{(A)---a plot of $\a(r)$;
(B)---a plot of $\vF(r)$;
(C)---a plot of $f(r)$;
(D)---a plot of $\pz\vF r(r)$;
(E)---a plot of $\exp(-122\vF(r))=G_{\rm eff}(r)$;
(F)---a plot of $\o(r)$;
(G)---a plot of $\g(r)$.
\label{kk}}
\end{figure}

In general we have here six \ct s ($C_i$, $i=2,\dots,7$) in initial
conditions and three integration \ct s $C_1,\ell^2$ and $\ov c$. Together
nine \ct s. We take for simplicity $C_1=\ov c=0$. For an integration \ct\
$\ell$ we take $\ell=1.3\t10^3\,{\rm km}=0.42\t10^{-20}\,L$ which is smaller
than $\ell$ from Moffat's paper (see Ref.~\cite{37}).

The results have been plotted on Fig.~\ref{kk} and a programme has been
quoted in Appendix~E.

\begin{figure}
\hbox to \textwidth{\hfil \ing[width=0.75\textwidth]{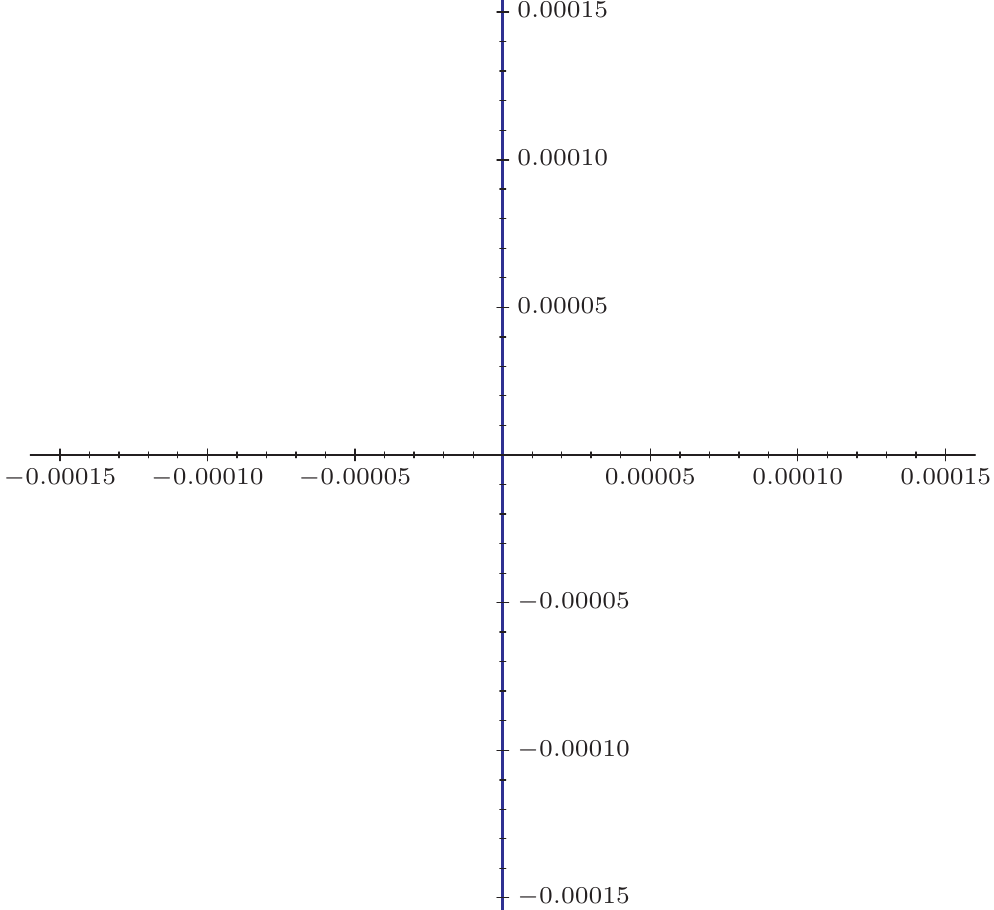}\hfil}
\caption{A polar plot for an orbit of a massive point body for parameters
$a_h=20$, $e=7$, $r(0)=10^{-4}$.
\label{uu}}
\end{figure}

We put $\ov M=1$ and $n=\dim\SO(16)=120$ ($H=SO(16)$). There is no
possibility to confuse a group $H$ with a \ct\ of integration $H$. It is
interesting to consider a motion of a massive test particle in the \gr al
field described by this \so. In order to do this we suppose that massive
point bodies move along geodesics in the Riemann geometry of $\Ch\a\b\g$ (see
\er{D.168}). Using Eqs \er{Dn223}--\er{Dn225a} one gets
$$
\dsl{
\refstepcounter{equation}\label{Da246}
\hskip50pt \pz r\vf=\pm r\sqrt{\frac{H^2r^2\ell^4}{h^2c^2f^6(r)}\,e^{2(r-\bar c)}
-\frac1{\a(r)}-\frac{c^2r^2}{h^2\a(r)}}\hfill\rm(\theequation)\cr
\refstepcounter{equation}\label{Da247}
\hskip50pt \pz r\tau=\pm\sqrt{\frac{H^2\ell^4}{c^2f^6(r)}\,e^{2(r-\bar c)} -
\frac{h^2}{r^2\a(r)} - \frac{c^2}{\a(r)}}\hfill\rm(\theequation)\cr
\hskip50pt \pz rt=\pm \frac{cf^3(r)}{\sqrt H\,\ell^2 \sqrt{\a(r)}}\,e^{-(r-\bar c)}
\sqrt{\frac{H^2\ell^4}{h^2f^6(r)c^2}\,e^{2(r-\bar c)} - \frac{h^2}{r^2\a(r)}
- \frac{c^2}{\a(r)}}\hfill\rm(\theequation a)\cr
\refstepcounter{equation}\label{Da248}
\hskip50pt \pz \vf\tau=\frac h{r^2(\tau)}\,. \hfill\rm(\theequation)}
$$
$\tau$ is a proper time and $t$ is a \cd\ time.

For photons one gets
$$
\dsl{
\refstepcounter{equation}\label{Da249}
\hskip120pt \pz r\vf=\pm\frac{re^{r-\ov c}\ell^2}{f^3(r)}
\sqrt{\frac{H^2}{c^2h^2}\,r^2-\a(r)}\hfill\rm(\theequation)\cr
\refstepcounter{equation}\label{Da250}
\hskip120pt \pz r\si=\pm\frac1c\, \frac{\ell^2e^{r-\bar c}}{f^3(r)}
\sqrt{H^2r^2-h^2\a(r)}\hfill\rm(\theequation)\cr
\hskip120pt \pz rt=\pm\frac{c\a(r)\ell^2e^{r-\bar c}}{f^3(r)}
\sqrt{1-\frac{h^2c^2}{H^2}\,\frac{\a(r)}{r}}\hfill\rm(\theequation a)\cr
\refstepcounter{equation}\label{Da251}
\hskip120pt \pz \vf\si=\frac h{r^2(\si)}\,.\hfill \rm(\theequation)
}
$$
$\si$ is an affine parameter and $t$ is a \cd\ time.

\begin{figure}
\hbox to \textwidth{\ing[width=0.45\textwidth]{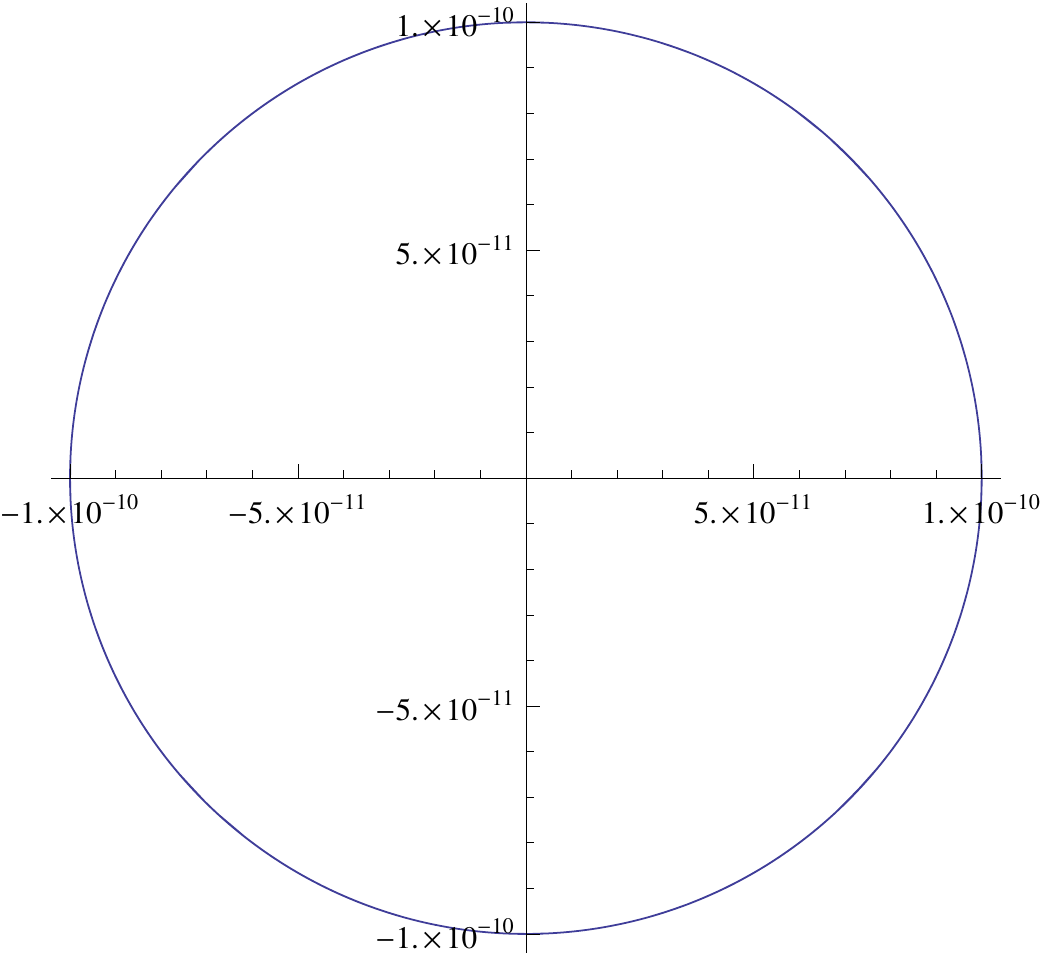}\hfil
\ing[width=0.45\textwidth]{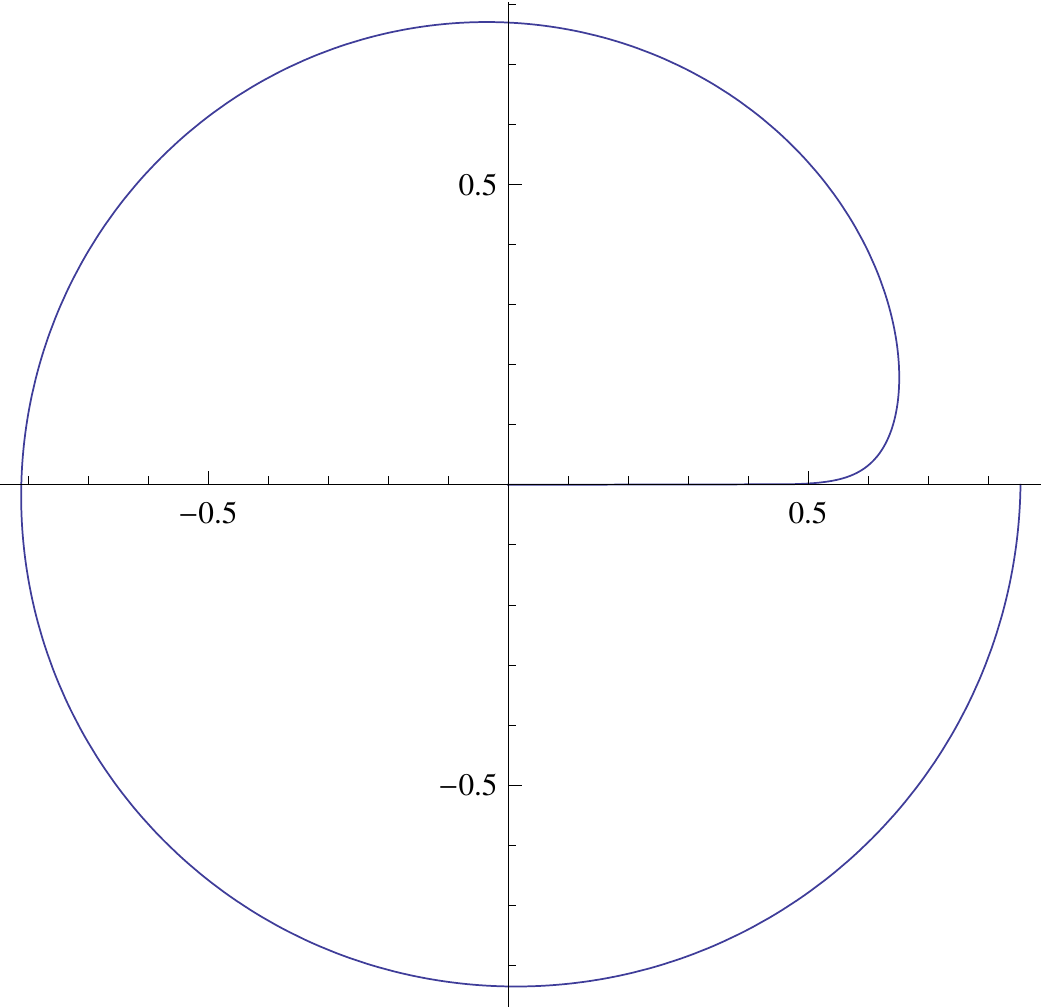}\hfil}
\hbox to \textwidth{\hbox to0.45\textwidth{\hfil(A)\hfil}\hfil
\hbox to0.45\textwidth{\hfil(B)\hfil}\hfil}
\caption{(A)---a polar plot $r(\vf)$ of an orbit for a photon for $a_h=20$,
$e=7$, $r(0)=10^{-4}$;
(B)---a polar plot $r(\vf)$ of an orbit for a photon for $a_h=10^{-30}$,
$e=7$, $r(0)=10^{-4}$;
\label{bb}}
\end{figure}

It is possible also to consider a \gr al Doppler effect in this case. One
gets in similar \ap ion as before (see \er{Dn241})
\bg{Da252}
\pz{(\D T(r(\si)))}{\si}=-\frac \pi c\,\pz\a r(r(\si))\frac{\ell^2e^{r(\si)
-\ov c}}{f^3(r(\si))}\\
\la=\X1(\la_0+c\D T(r(\si))\Y1)\cdot \frac{f^3(r(\si))e^{-r(\si)+\ov c}}
{\a(r(\si))\ell^2}\,.\label{Da253}
\end{gather}
In all of these formulae we use an \ap ion for $r\ll1$, i.e.
\bg{Da254}
\o(r)=f^2(r)e^{-r+\ov c}\\
\g(r)=\frac{\o^2(r)f^2(r)}{\ell^4\a(r)}=\frac{f^6(r)}{\ell^4\a}\,
e^{-2(r-\ov c)}.\label{Da255}
\end{gather}

In the case of massive point bodies we use a parametrization \er{Dn226}. The
results of numerical calculations for massive particle have been plotted on
Fig.~\ref{uu} for $a_h=10^{15}$, $e=7$. On Fig.~\ref{bb} we plotted an orbit
for a photon and an orbit for a massive particle in polar \cd s. In Eqs
\er{Da246}--\er{Da249} we put a sign $+$. The programmes have been quoted in
Appendix~E.

\medskip
\obraz{kkf}
\hbox to \textwidth{\ing[width=0.43\textwidth]{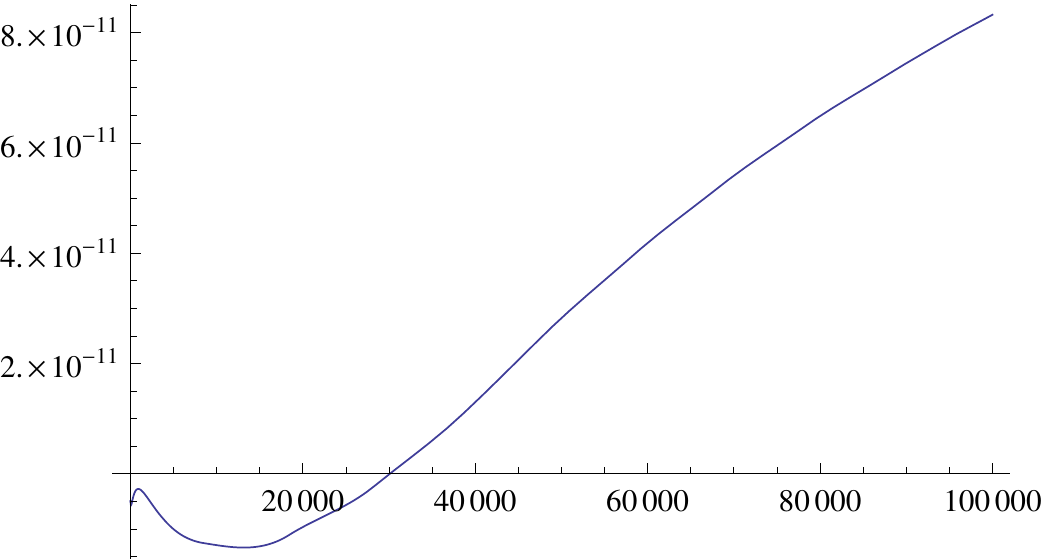}\hfil
\ing[width=0.43\textwidth]{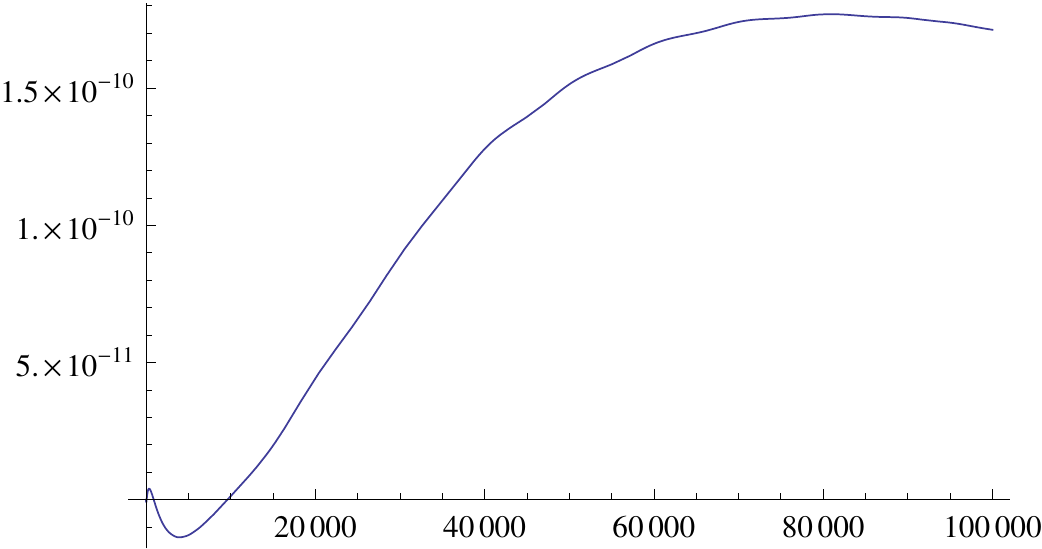}}
\hbox to \textwidth{\hbox to 0.43\textwidth{\hfil(A)\hfil}\hfil
\hbox to 0.43\textwidth{\hfil(B)\hfil}}
\hbox to \textwidth{\ing[width=0.43\textwidth]{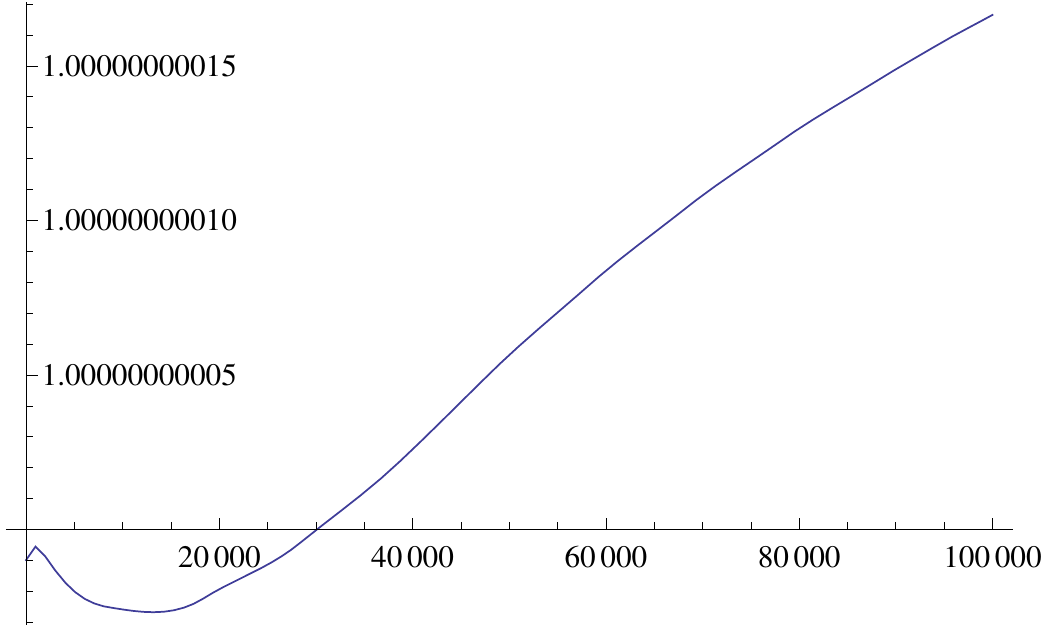}\hfil
\ing[width=0.43\textwidth]{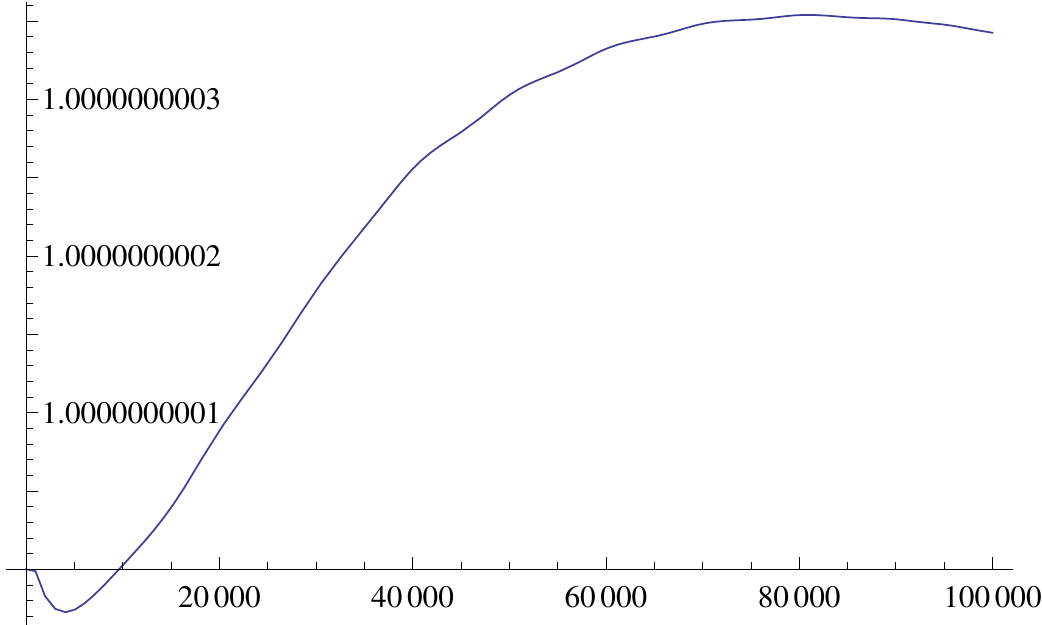}}
\hbox to \textwidth{\hbox to 0.43\textwidth{\hfil(C)\hfil}\hfil
\hbox to 0.43\textwidth{\hfil(D)\hfil}}
\hbox to \textwidth{\ing[width=0.43\textwidth]{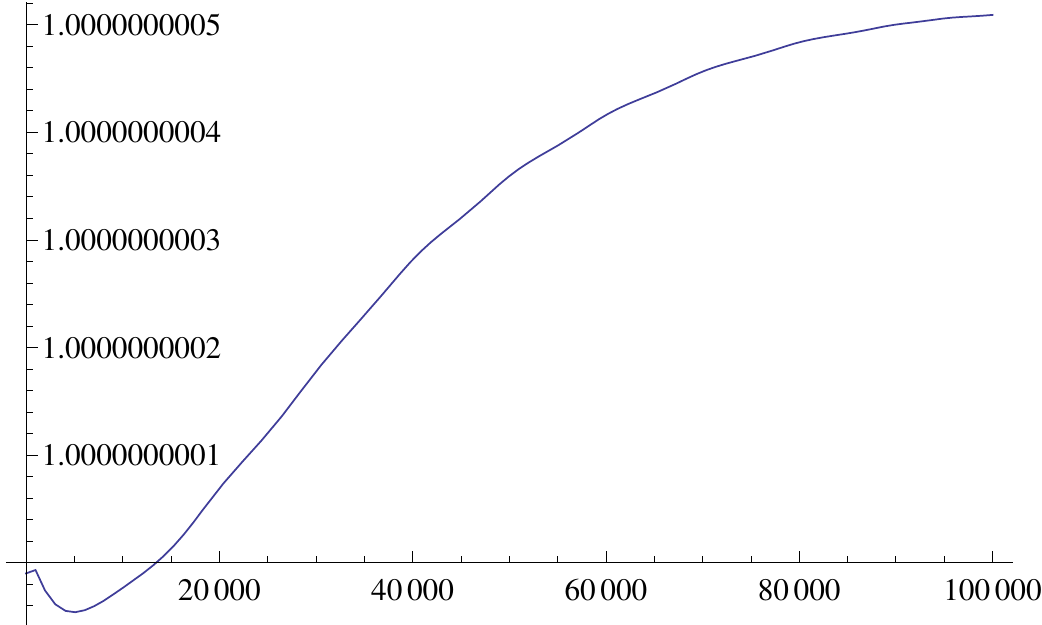}\hfil
\ing[width=0.43\textwidth]{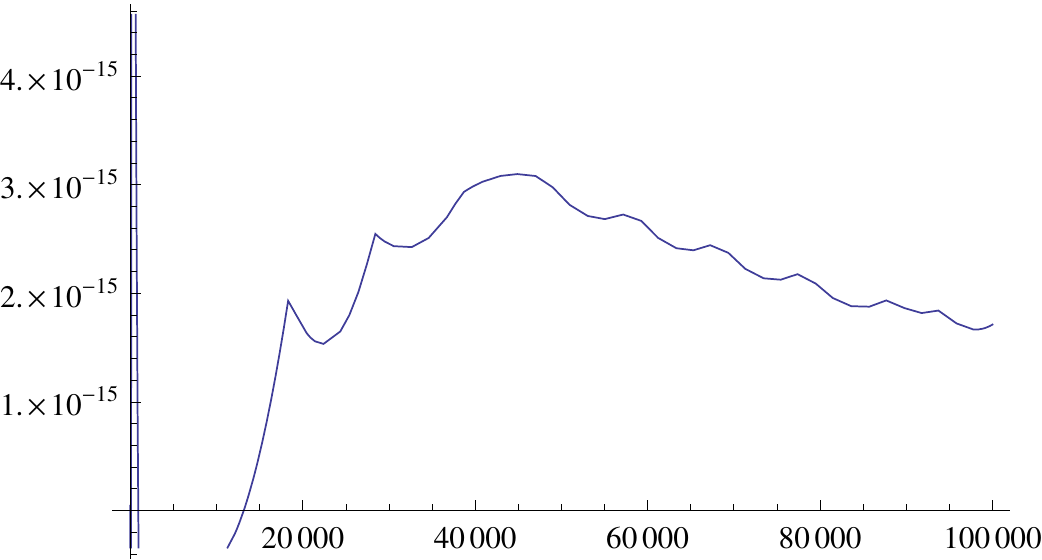}}
\hbox to \textwidth{\hbox to 0.43\textwidth{\hfil(E)\hfil}\hfil
\hbox to 0.43\textwidth{\hfil(F)\hfil}}
\hbox to \textwidth{\ing[width=0.43\textwidth]{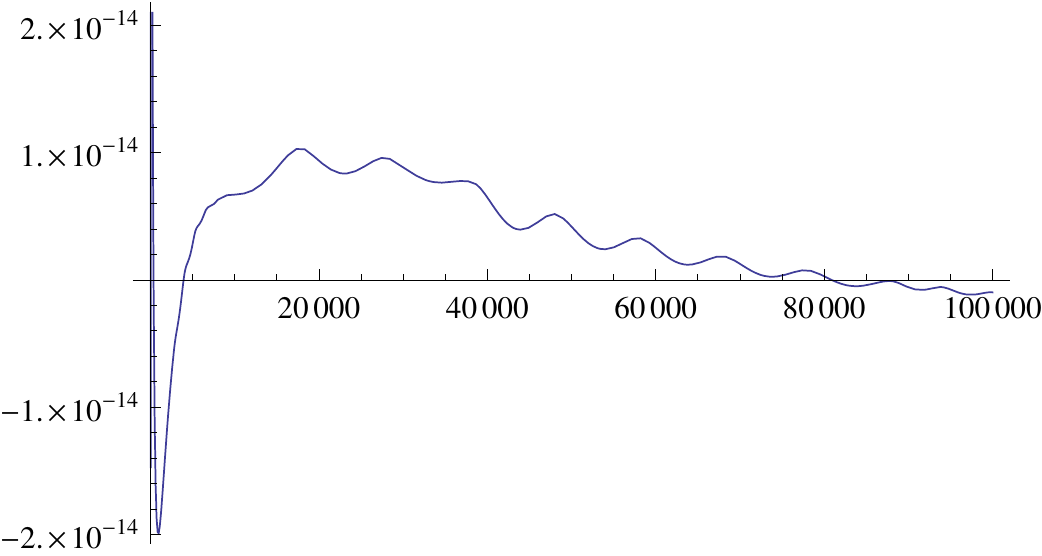}\hfil
\ing[width=0.43\textwidth]{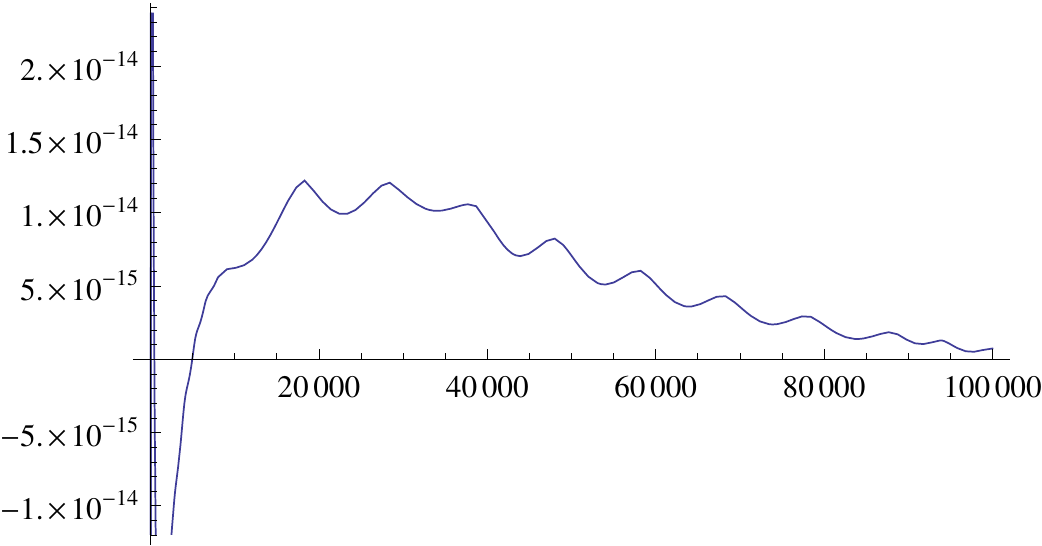}}
\hbox to \textwidth{\hbox to 0.43\textwidth{\hfil(G)\hfil}\hfil
\hbox to 0.43\textwidth{\hfil(H)\hfil}}
\podpis{: (A)---a plot of $A(r)$ for $10^{-3}<r<10^5$;
(B)---a plot of  $B(r)$ for $10^{-3}<r<10^5$;
(C)---a plot of  $e^{2A(r)}$ for $10^{-3}<r<10^5$;
(D)---a plot of  $e^{2B(r)}$ for $10^{-3}<r<10^5$;
(E)---a plot of  $e^{2(A(r)+B(r))}$ for $10^{-3}<r<10^5$;
(F)---a plot of  $\pz {}r e^{2A(r)}$ for $10^{-3}<r<10^5$;
(G)---a plot of $\pz{}re^{2B(r)}$ for $10^{-3}<r<10^5$;
(H)---a plot of $\pz{}r e^{2(A(r)+B(r))}$ for $10^{-3}<r<10^5$.
}

\hbox to \textwidth{\ing[width=0.45\textwidth]{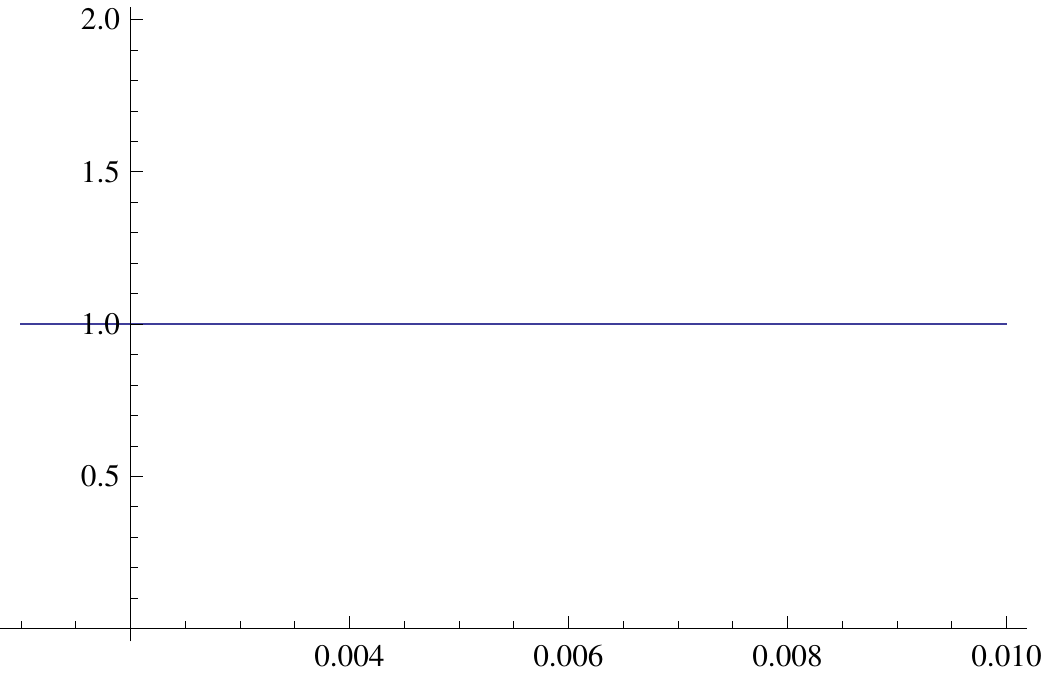}\hfil
\ing[width=0.45\textwidth]{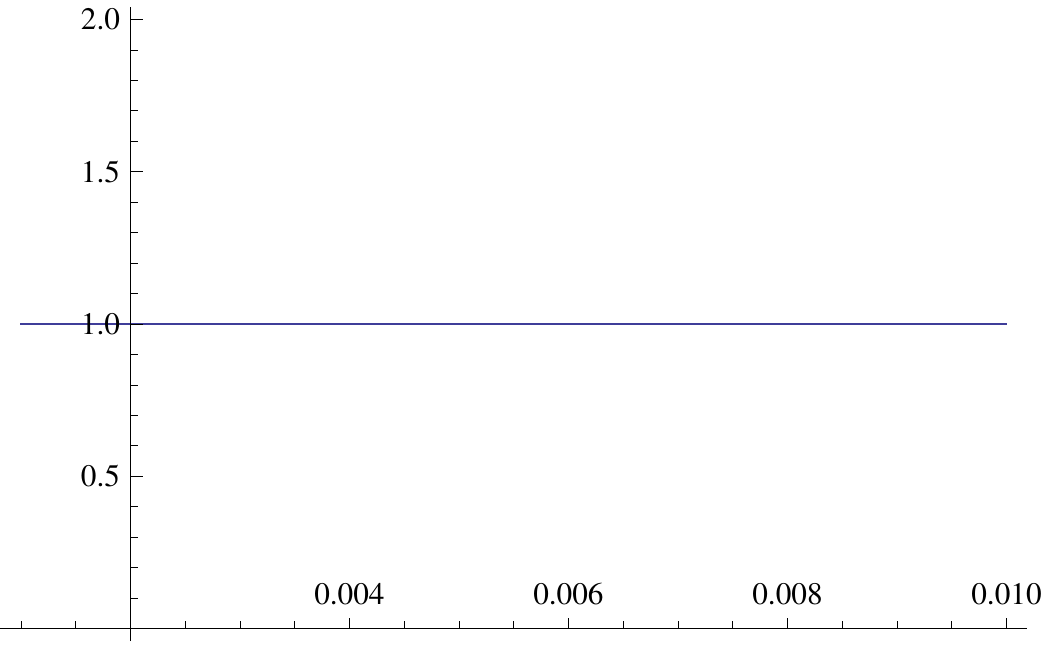}}
\hbox to \textwidth{\hbox to 0.45\textwidth{\hfil(I)\hfil}\hfil
\hbox to 0.45\textwidth{\hfil(J)\hfil}}
\hbox to \textwidth{\hfil \ing[width=0.45\textwidth]{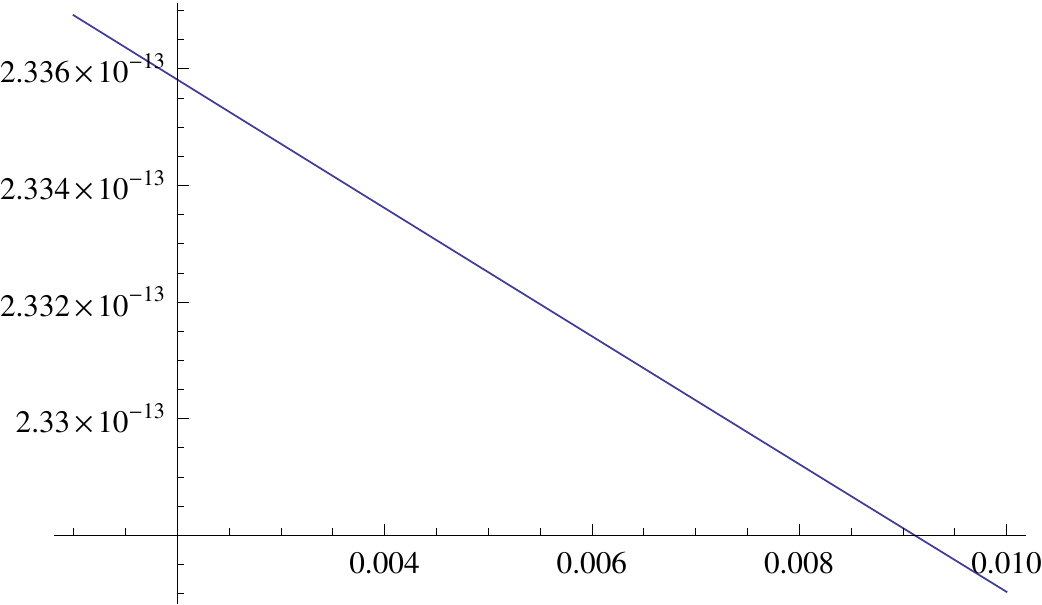}\hfil}
\hbox to \textwidth{\hfil\hbox to 0.45\textwidth{\hfil(K)\hfil}\hfil}
\smallskip
{\small \noindent Figure \thefigure\ (cont.):
(I)---a plot of $e^{2A(r)}$ for $10^{-3}<r<10^{-2}$;
(J)---a plot of $e^{2B(r)}$ for $10^{-3}<r<10^{-2}$;
(K)---a plot of $\pz{}r e^{2(A(r)+B(r))}$ for $10^{-3}<r<10^{-2}$.}

\bigskip
Let us come back to the system of Eqs \er{D.30}--\er{D.32}. We consider the
system with the following initial conditions
\beq{Da256}
\bal
B(3.7\t10^2)&=4\t10^{-12}\q\q &
\pz Br(3.7\t10^2)&=0\\
A(3.7\t10^2)&=-4\t10^{-12}\\
\wt\vF(3.7\t10^2)&=0\q\q &
\pz{\wt\vF}r(3.7\t10^2)&=0
\eal
\end{equation}
$r$ is measured in $r_0$ unit ($r\simeq4.103$\,AU). Using Mathematica~7
(NDSolve instruction) we solved the system \er{D.30}--\er{D.32} with
\er{Da256}. The results have been plotted on Fig.~\ref{kkf}. We plot here
$A(r)$, $B(r)$, $e^{2A(r)}$, $e^{2B(r)}$, $e^{2(A(r)+B(r))}$,
$\pz{}re^{2A(r)}$, $\pz{}re^{2B(r)}$, $\pz{}re^{2(A(r)+B(r))}$, for $10^{-3}
<r<10^5$, and $e^{2A(r)}$, $e^{2B(r)}$, $\pz{}r e^{2(A(r)+B(r))}$ for
$10^{-3} <r <10^{-2}$. $r$~is measured in $r_0$ unit.

We have the following conclusions: $A(r)\ne-B(r)$ which is different than
 for a Schwarzschild \so.
Moreover, a product $e^{2(A(r)+B(r))}$ is quite close to one as it should be
for we consider this \so\ as a disturbed Schwarzschild \so\ by an \an\ \ac.
On Fig.~\ref{kkkf} we consider various properties of $\wt\vF(r)$, and
connected to it an effective \gr al \ct.
\beq{Da257}
G\eff=G_N(-(n+2)\vF(r))=G_N\X2(-\frac{(n+2)^2}{\ov M}\,\wt\vF(r)\Y2).
\end{equation}
Taking $n=120$ and $\ov M=1$ we get
\beq{Da258}
G\eff=G_N(-14884\,\wt\vF(r)).
\end{equation}
\medskip \obraz{kkkf}
\hbox to \textwidth{\ing[width=0.45\textwidth]{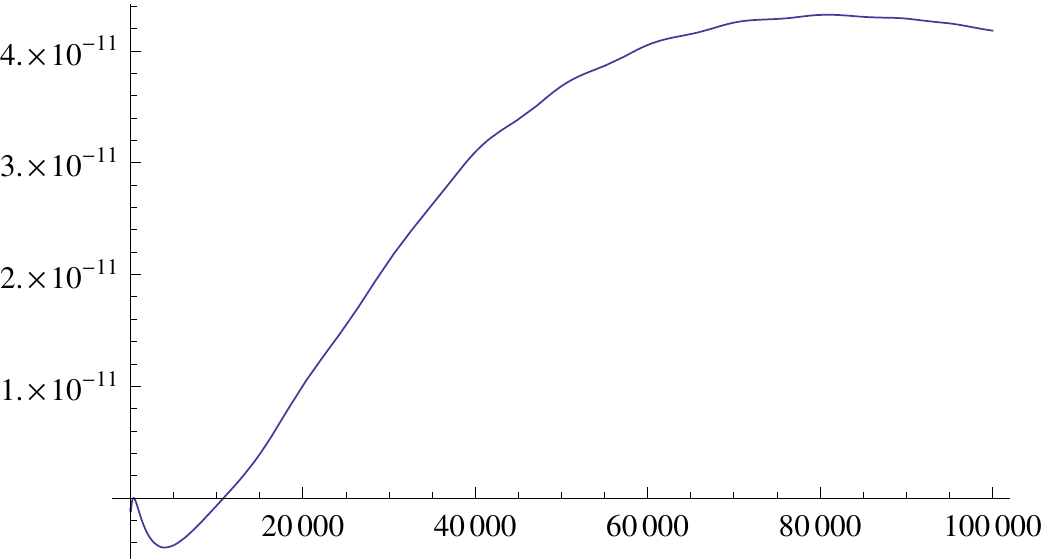}\hfil
\ing[width=0.45\textwidth]{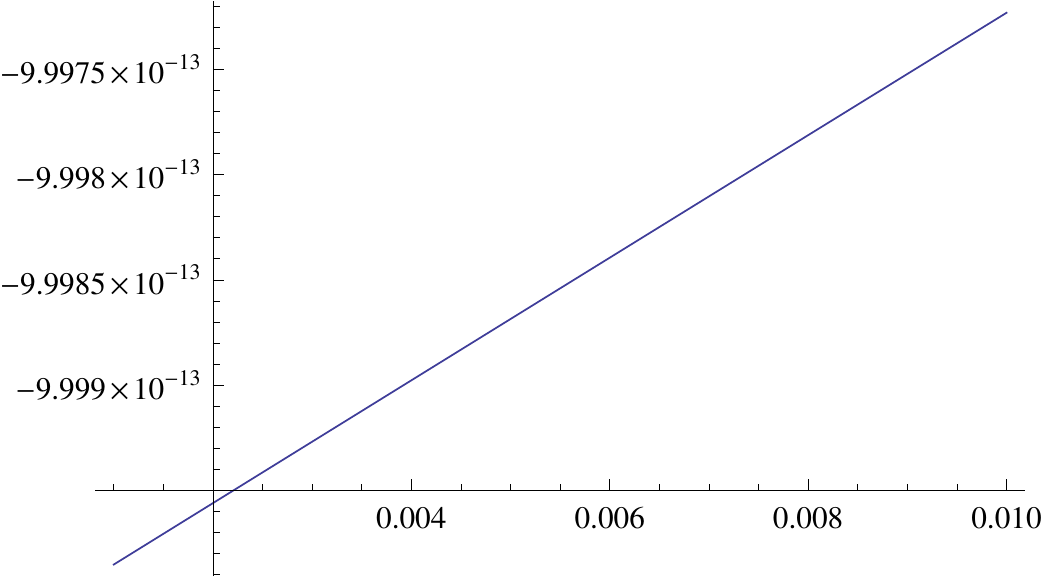}}
\hbox to \textwidth{\hbox to 0.45\textwidth{\hfil(A)\hfil}\hfil
\hbox to 0.45\textwidth{\hfil(B)\hfil}}
\hbox to \textwidth{\ing[width=0.45\textwidth]{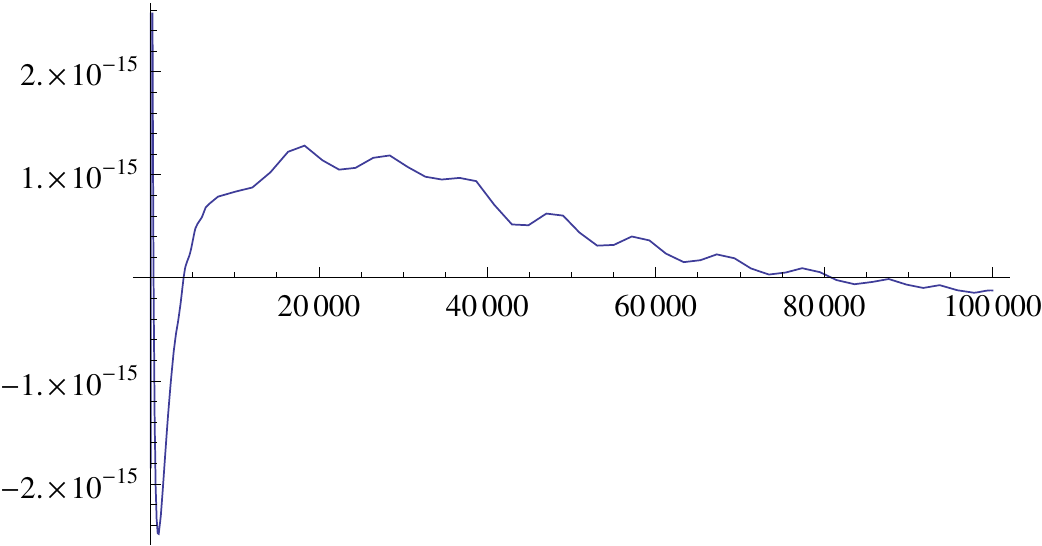}\hfil
\ing[width=0.45\textwidth]{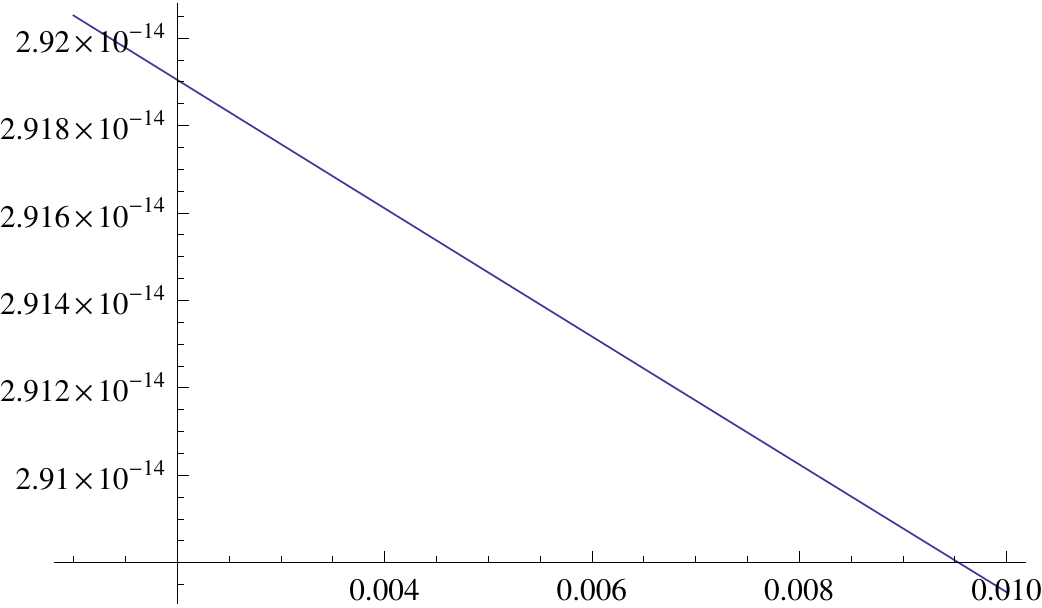}}
\hbox to \textwidth{\hbox to 0.45\textwidth{\hfil(C)\hfil}\hfil
\hbox to 0.45\textwidth{\hfil(D)\hfil}}
\hbox to \textwidth{\ing[width=0.45\textwidth]{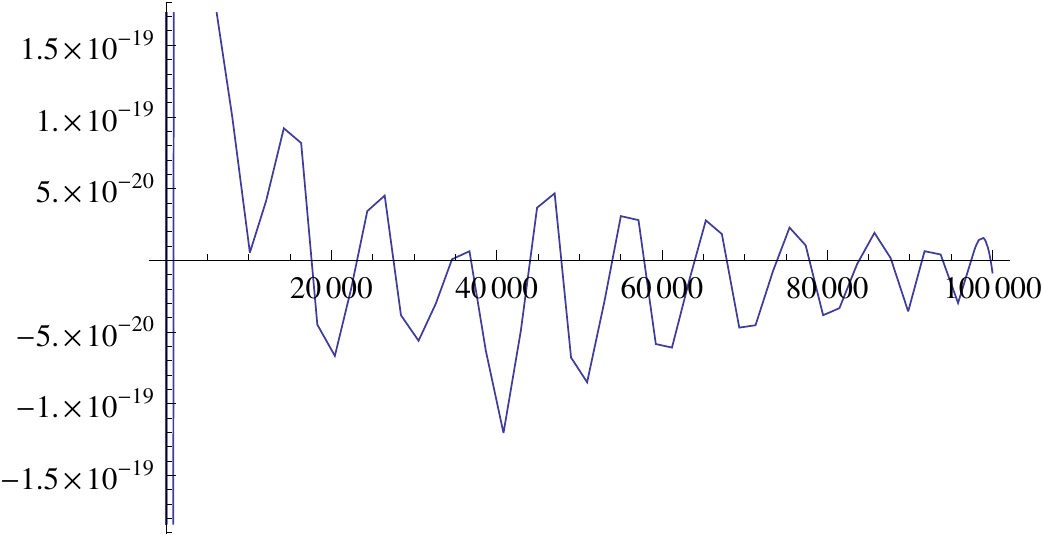}\hfil
\ing[width=0.45\textwidth]{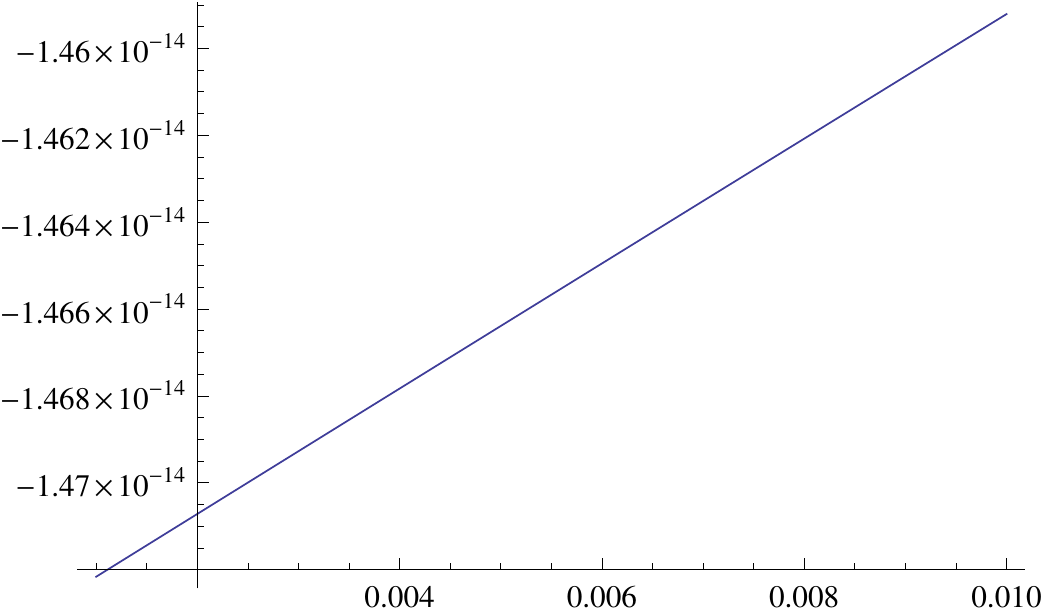}}
\hbox to \textwidth{\hbox to 0.45\textwidth{\hfil(E)\hfil}\hfil
\hbox to 0.45\textwidth{\hfil(F)\hfil}}
\hbox to \textwidth{\ing[width=0.45\textwidth]{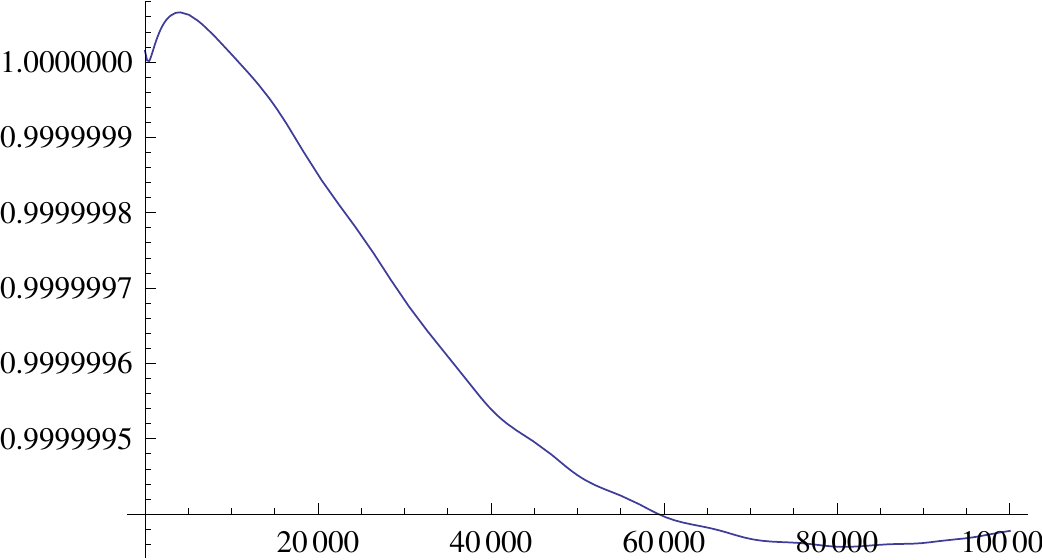}\hfil
\ing[width=0.45\textwidth]{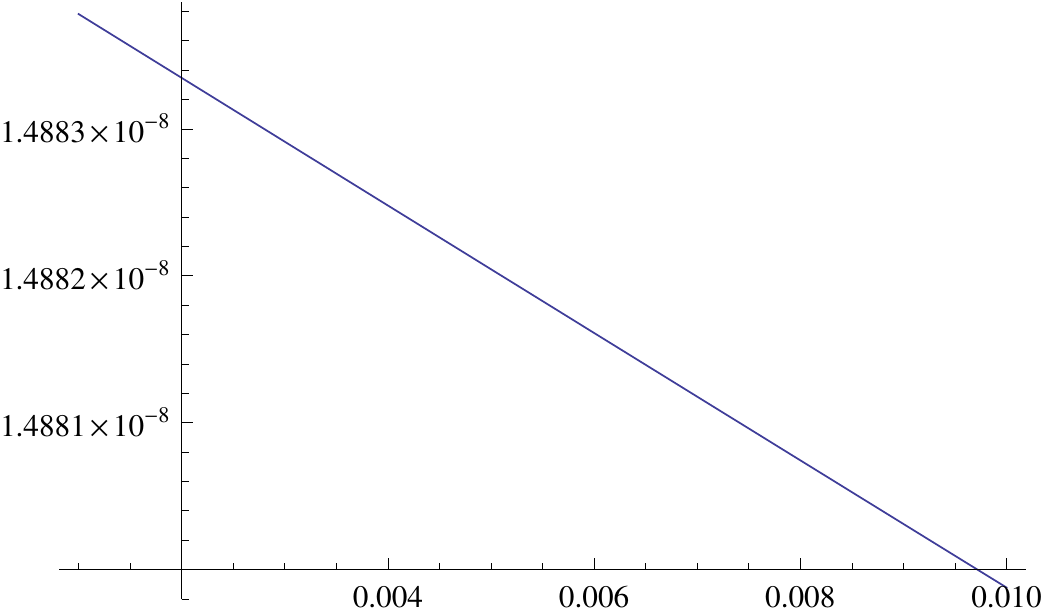}}
\hbox to \textwidth{\hbox to 0.45\textwidth{\hfil(G)\hfil}\hfil
\hbox to 0.45\textwidth{\hfil(H)\hfil}}
\podpis{:(A)---a plot of $\wt\vF(r)$ for $10^{-3}<r<10^5$;
(B)---a plot of  $\wt\vF(r)$ for $10^{-3}<r<10^{-2}$;
(C)---a plot of  $\pz{\wt\vF}r(r)$ for $10^{-3}<r<10^5$;
(D)---a plot of  $\pz{\wt\vF}r$ for $10^{-3}<r<10^{-2}$;
(E)---a plot of  $\pz{^2\wt\vF}{r^2}(r)$ for $10^{-3}<r<10^5$;
(F)---a plot of  $\pz {^2\wt\vF}{r^2}(r)$ for $10^{-3}<r<10^{-2}$;
(G)---a plot of $\frac1{G_N}G\eff(r)$ for $10^{-3}<r<10^5$;
(H)---a plot of $\frac1{G_N}G\eff(r)-1$ for $10^{-3}<r<10^{-2}$.}

\hbox to \textwidth{\ing[width=0.45\textwidth]{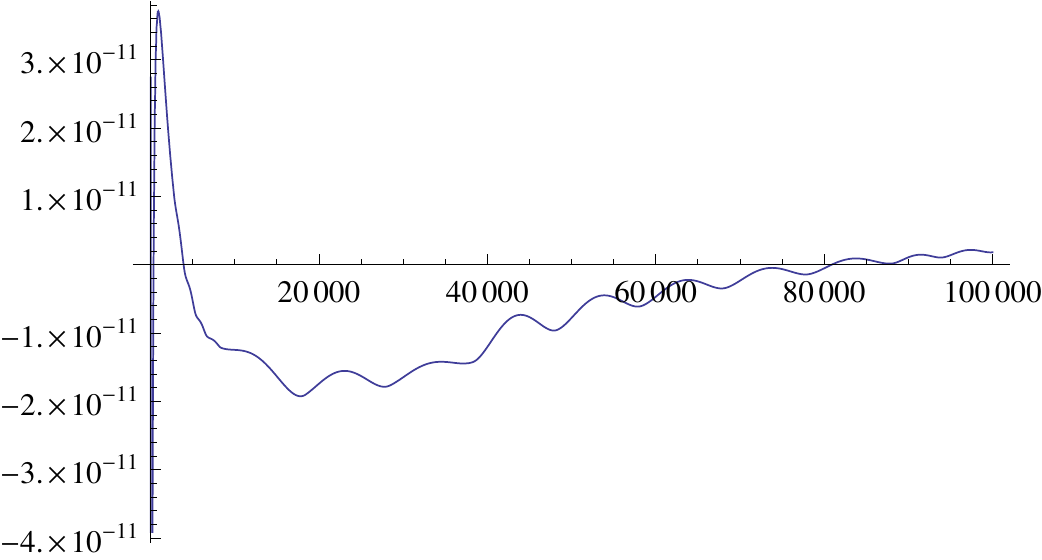}\hfil
\ing[width=0.45\textwidth]{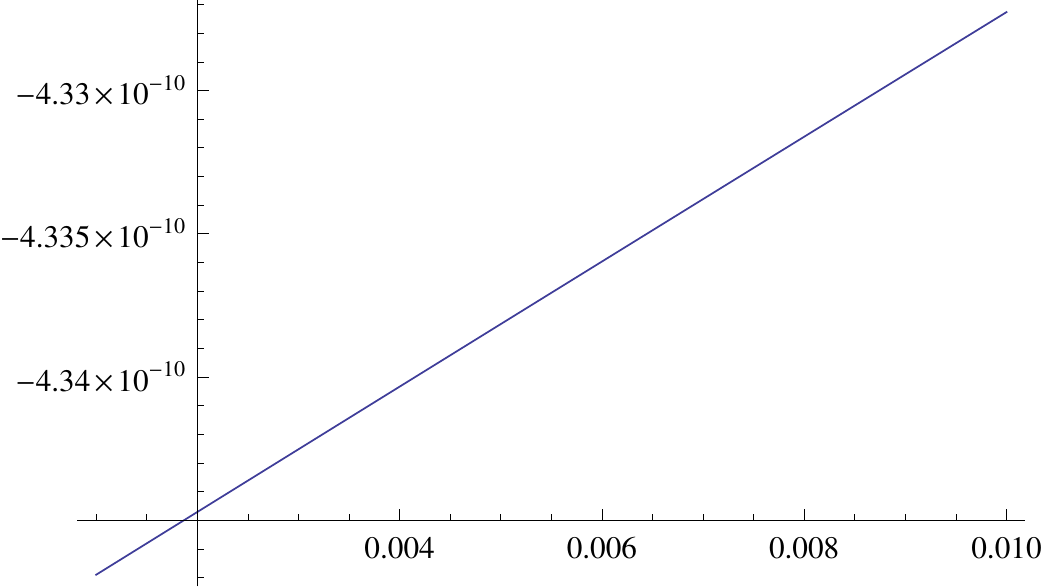}}
\hbox to \textwidth{\hbox to 0.45\textwidth{\hfil(I)\hfil}\hfil
\hbox to 0.45\textwidth{\hfil(J)\hfil}}
\smallskip
{\small \noindent Figure \thefigure\ (cont.):
(I)---a plot of $\frac1{G_N}\pz{G\eff(r)}r$ for $10^{-3}<r<10^{5}$;
(J)---a plot of $\frac1{G_N}\pz{G\eff(r)}r $ for $10^{-3}<r<10^{-2}$.}

\bigskip
We plot here $\wt\vF(r)$, $\pz{\wt\vF}r(r)$, $\pz{^2\wt\vF}{r^2}(r)$ for
$10^{-3}<r<10^5$ and for
$10^{-3} < r < 10^{-2}$, a normalized effective \gr al \ct\ $\frac1{G_N}
\,G\eff(r)$, $\frac1{G_N}\,\pz{G\eff}r(r)$ for $10^{-3}<r<10^{-2}$ and
$10^{-3} < r < 10^5$, $r$ is measured in $r_0$ (4.103\,AU). A~conclusion from
these plots is as follows: a \gr al \ct\ changes very slowly (as it should
be). There is a very important characterization of our \so. It is a density
of dust $\ov\rho(r)$ which we introduced to the field \e s (see
Eq.~\er{D.5}). Using Eq.~\er{D.5} and proper system of units (also for $r_0$)
we get
\bml{Da259}
\ov\rho(r)=1.785\t10^{-24}\X3(-14884\X2(\pz{\wt\vF}{r}\Y2)^2\eb{-2}
+\X2(\frac1{r^2}(1-\eb{-2})+\frac2r\,\eb{-2}\,\pz Br\Y2)\\
{}-0.28\t10^{24}\exp(14640\,\wt\vF(r))\X2(\frac{\exp(244\,\wt\vF(r))}{122}
-\frac1{120}\Y2)\Y3)\exp(14884\,\wt\vF(r)).
\end{multline}

\begin{figure}
\hbox to \textwidth{\ing[width=0.49\textwidth]{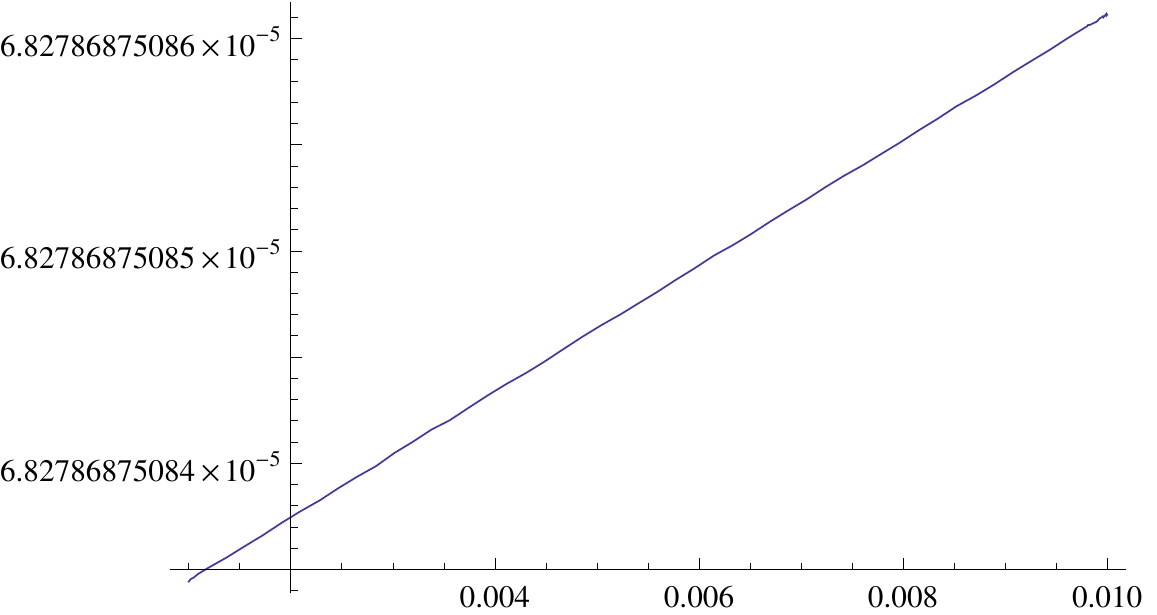}\hfil
\ing[width=0.455\textwidth]{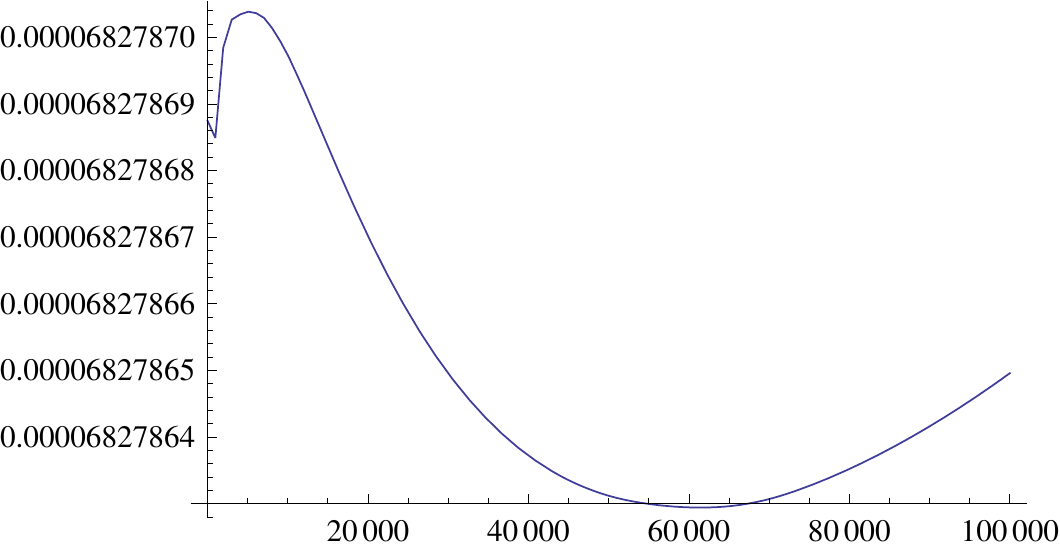}}
\hbox to \textwidth{\hbox to 0.49\textwidth{\hfil(A)\hfil}\hfil
\hbox to 0.455\textwidth{\hfil(B)\hfil}}
\hbox to \textwidth{\ing[width=0.495\textwidth]{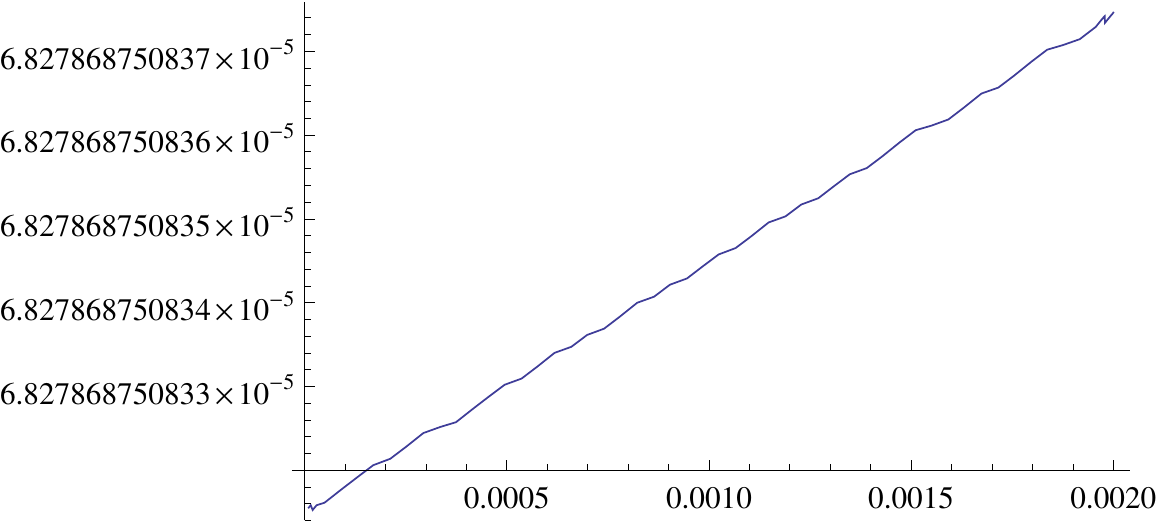}\hfil
\ing[width=0.45\textwidth]{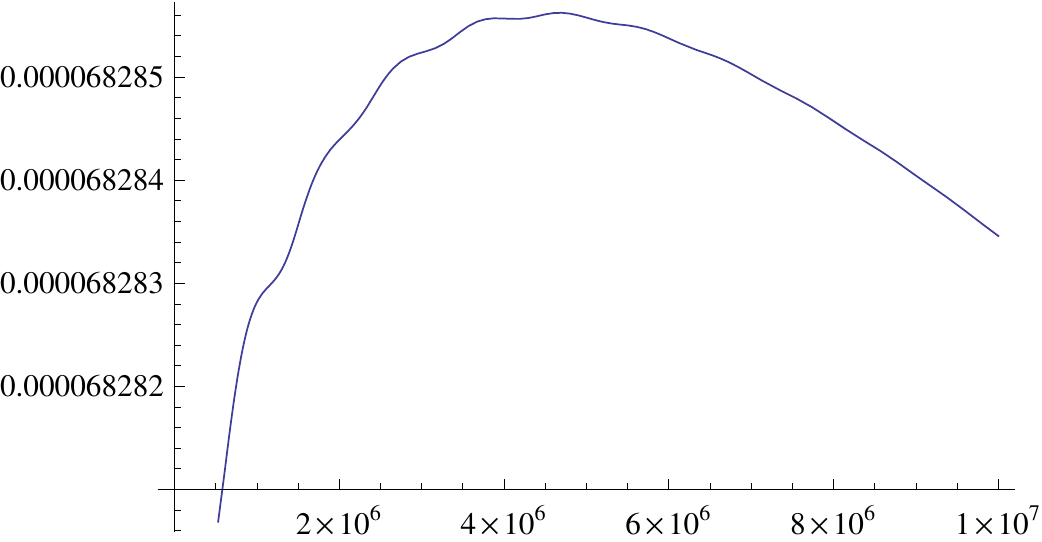}}
\hbox to \textwidth{\hbox to 0.495\textwidth{\hfil(C)\hfil}\hfil
\hbox to 0.45\textwidth{\hfil(D)\hfil}}
\caption{(A)---a plot of $\ov\rho(r)$ for $10^{-3}<r<10^{-2}$;
(B)---a plot of  $\ov\rho(r)$ for $10^{-3}<r<10^5$;
(C)---a plot of  $\ov\rho(r)$ for $10^{-5}<r<10^{-3}$;
(D)---a plot of  $\ov\rho(r)$ for $10^{5}<r<10^{7}$.
\label{kkkfx}}
\end{figure}

\atw0.2 \advance\abovedisplayskip by-1.5pt \advance\belowdisplayskip by-1.5pt
We plot $\ov\rho(r)$ on Fig.~\ref{kkkfx} for $10^{-3}<r<10^{-2}$,
for $10^{-3}<r<10^5$, for $10^{-5}<r<10^{-3}$, for $10^5<r<10^7$. It is easy
to see that
\beq{Da260a}
\ov\rho(r)<0.96\t10^{-4}.
\end{equation}
It means (see Eq.~\er{D.48}) this density is smaller than a density of an
interplanetary matter. Let us consider the following formulae
\bea{Da260}
e^{2A(r)}&=&1-2V_1(r)\\
\eb2&=&\frac1{1-2V_2(r)},\label{Da261}
\end{eqnarray}
\bea{Da262}
2V_1(r)&=&\frac{r_s}r+2\ov V_1(r)\\
2V_2(r)&=&\frac{r_s}r+2\ov V_2(r).\label{Da263}
\end{eqnarray}
In this way we get
\bea{Da264}
\ov V_1(r)&=&\frac12\X2(1-\frac{r_s}r-e^{2A(r)}\Y2)\\
\ov V_2(r)&=&\frac12\X2(1-\frac{r_s}r-e^{-2B(r)}\Y2),\label{Da265}
\end{eqnarray}
where $r_s$ is a Schwarzschild radius for the Sun
\beq{Da266}
r_s\simeq 3\,{\rm km}\simeq 4.8907 \t 10^{-9}r_0.
\end{equation}
$V_1(r)$ and $V_2(r)$ have an interpretation of \gr al \pt\ in our model and
$\ov V_1(r)$ and $\ov V_2(r)$ \an\ \gr al \pt s. Let us consider their \dv s,
\bea{Da267}
b_1(r)=\pz{\ov V_1}r\\
b_2(r)=\pz{\ov V_2}r\label{Da268}
\end{eqnarray}
$b_1(r)$ and $b_2(r)$ can be considered as \an\ \ac s.

Let us consider $\ov U(r)$ from Section 5 (see Eq.~\er{5.4}). In the system
of units considered here one gets
\beq{Da269}
\ov U(r)=-5.2112\t10^{-15}\X1(\exp(-r)+1+r^2+\sqrt\pi\,\operatorname{erf}(r)\Y1)
\end{equation}
where $r$ is measured in $r_0$. Moreover, to compare $\ov U(r)$ with $\ov
V_1(r)$ and $\ov V_2(r)$ we should use $-\ov U(r)$.

\begin{figure}
\hbox to \textwidth{\ing[width=0.44\textwidth]{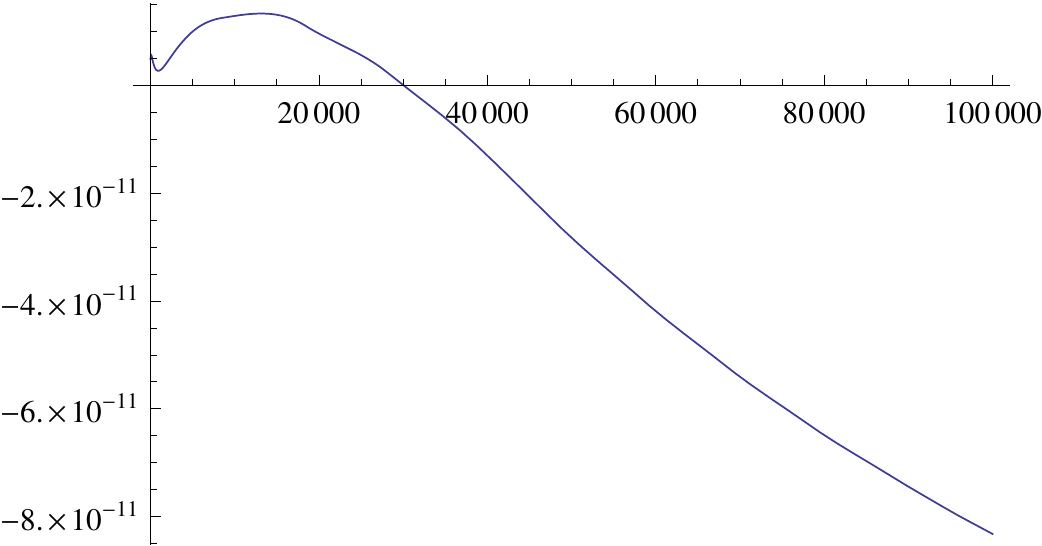}\hfil
\ing[width=0.44\textwidth]{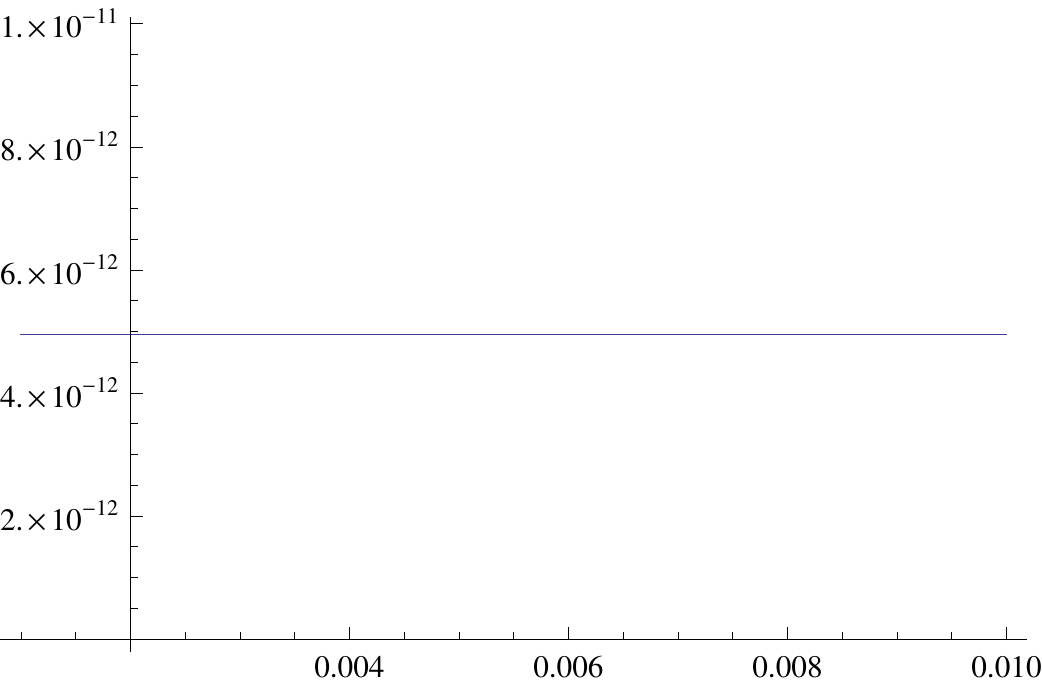}}
\hbox to \textwidth{\hbox to 0.44\textwidth{\hfil(A)\hfil}\hfil
\hbox to 0.44\textwidth{\hfil(B)\hfil}}
\hbox to \textwidth{\ing[width=0.44\textwidth]{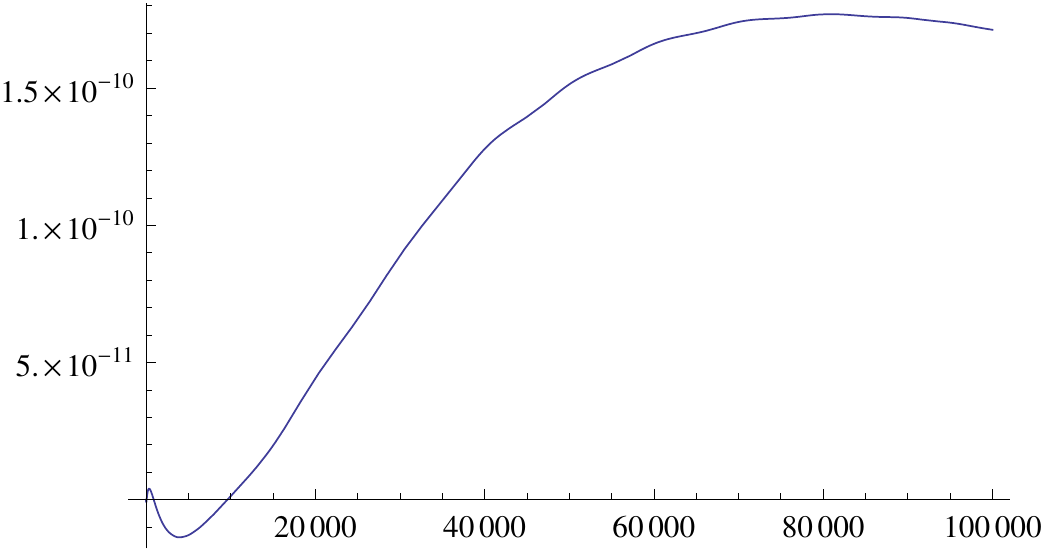}\hfil
\ing[width=0.44\textwidth]{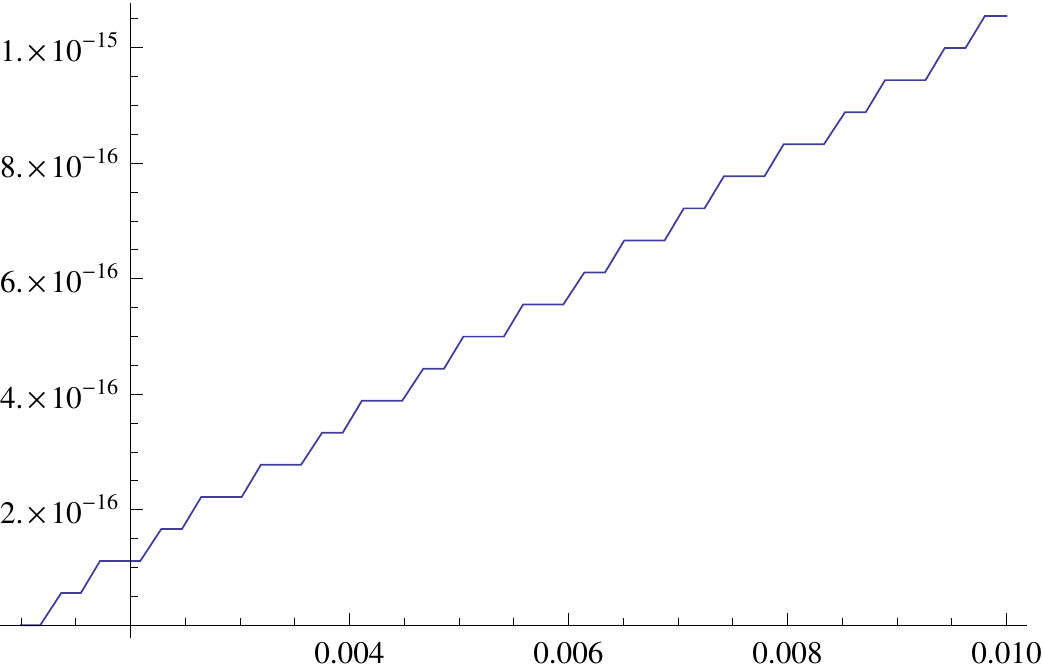}}
\hbox to \textwidth{\hbox to 0.44\textwidth{\hfil(C)\hfil}\hfil
\hbox to 0.44\textwidth{\hfil(D)\hfil}}
\hbox to \textwidth{\ing[width=0.44\textwidth]{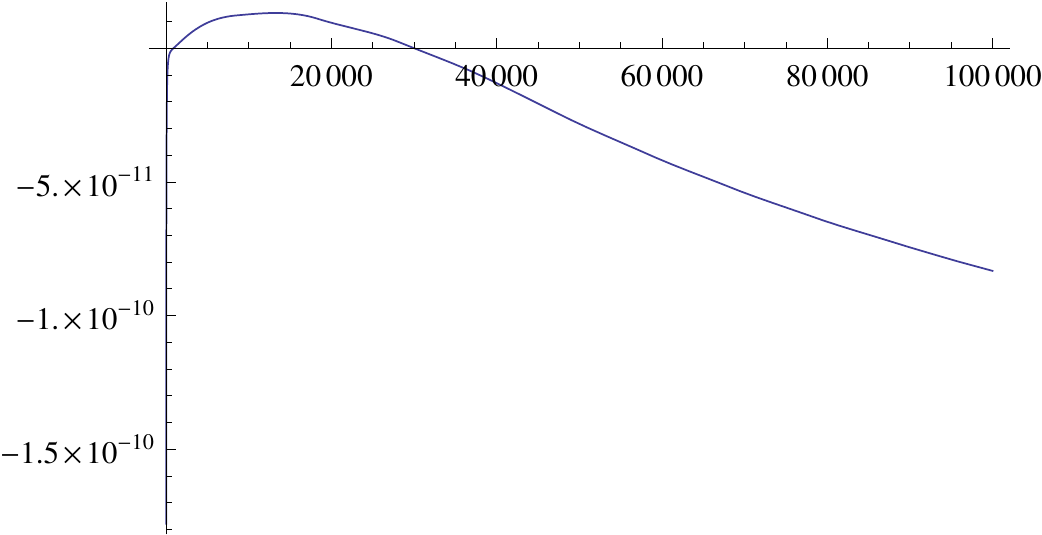}\hfil
\ing[width=0.44\textwidth]{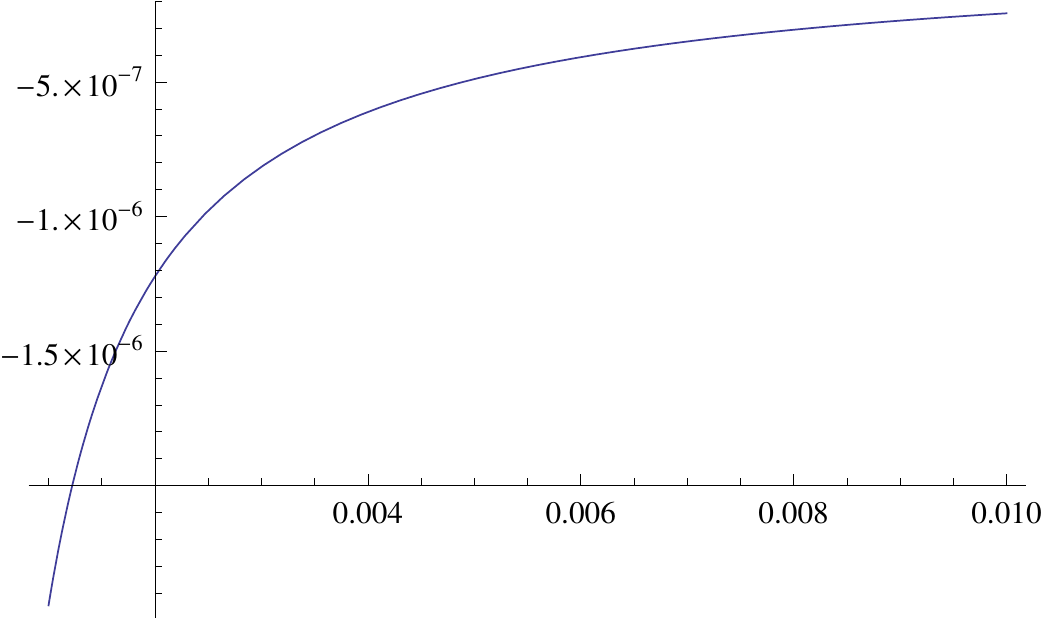}}
\hbox to \textwidth{\hbox to 0.44\textwidth{\hfil(E)\hfil}\hfil
\hbox to 0.44\textwidth{\hfil(F)\hfil}}
\hbox to \textwidth{\ing[width=0.44\textwidth]{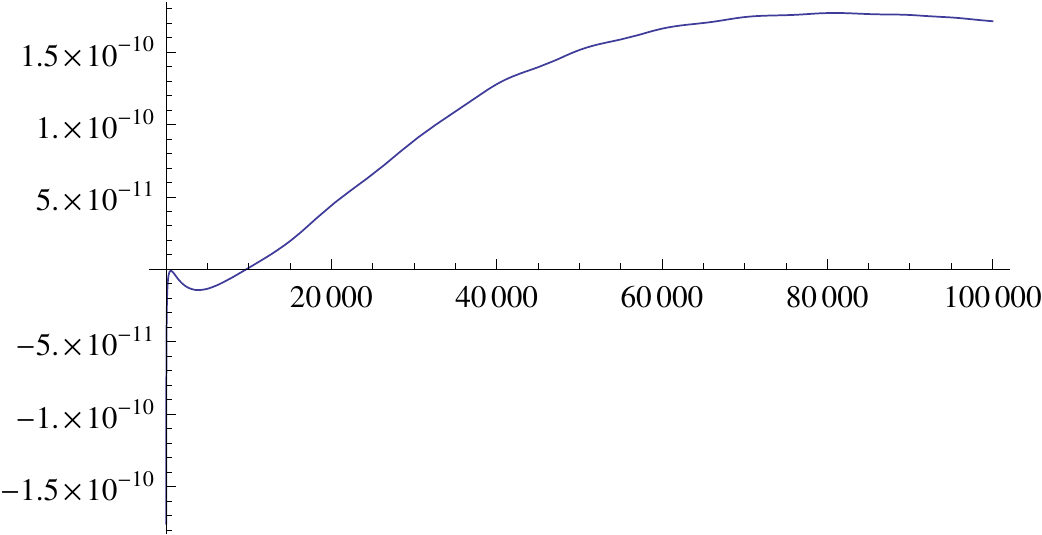}\hfil
\ing[width=0.44\textwidth]{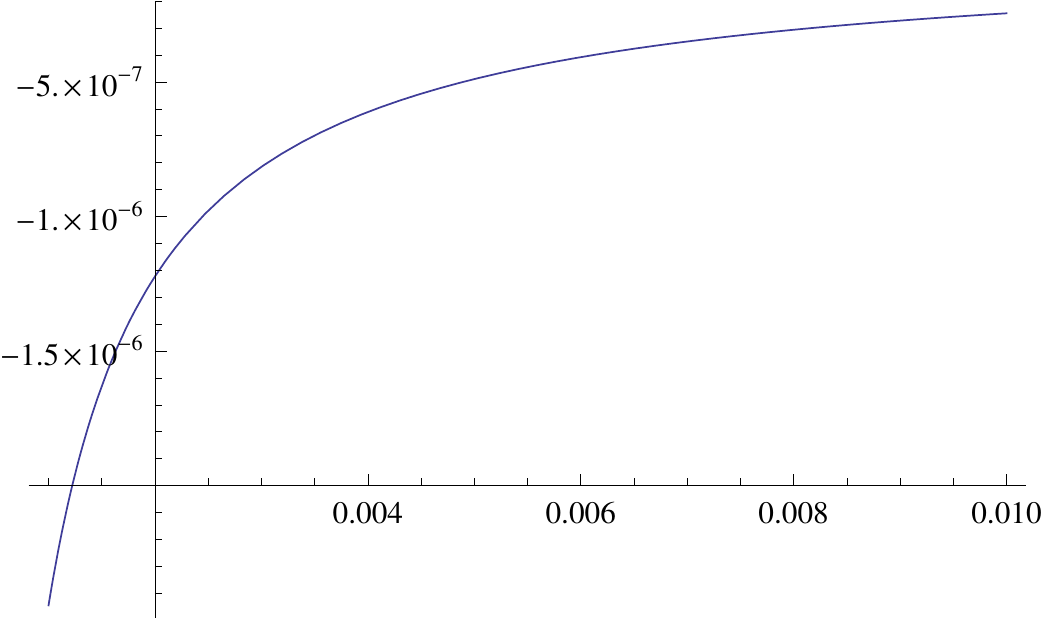}}
\hbox to \textwidth{\hbox to 0.44\textwidth{\hfil(G)\hfil}\hfil
\hbox to 0.44\textwidth{\hfil(H)\hfil}}
\caption{(A)---a plot of $V_1(r)$ for $10^{-3}<r<10^5$;
(B)---a plot of  $V_1(r)$ for $10^{-3}<r<10^{-2}$;
(C)---a plot of  $V_2(r)$ for $10^{-3}<r<10^5$;
(D)---a plot of  $V_2(r)$ for $10^{-3}<r<10^{-2}$;
(E)---a plot of  $\ov V_1(r)$ for $10^{-3}<r<10^5$;
(F)---a plot of  $\ov V_1(r)$ for $10^{-3}<r<10^{-2}$;
(G)---a plot of  $\ov V_2(r)$ for $10^{-3}<r<10^{5}$;
(H)---a plot of  $\ov V_2(r)$ for $10^{-3}<r<10^{-2}$.
\label{kkkb}}
\end{figure}

On Fig.\ \ref{kkkb} we plot $V_1(r)$, $V_2(r)$, $\ov V_1(r)$, $\ov V_2(r)$,
$b_1(r)$, $b_2(r)$, $-\ov U(r)$,
\beq{Da270}
\eta_1(r)=-\frac{\ov V_1(r)}{\ov U(r)}, \q
\eta_2(r)=-\frac{\ov V_2(r)}{\ov U(r)}
\end{equation}
for $10^{-3}<r <10^5$ and for $10^{-3}<r<10^{-2}$.

\hbox to \textwidth{\ing[width=0.45\textwidth]{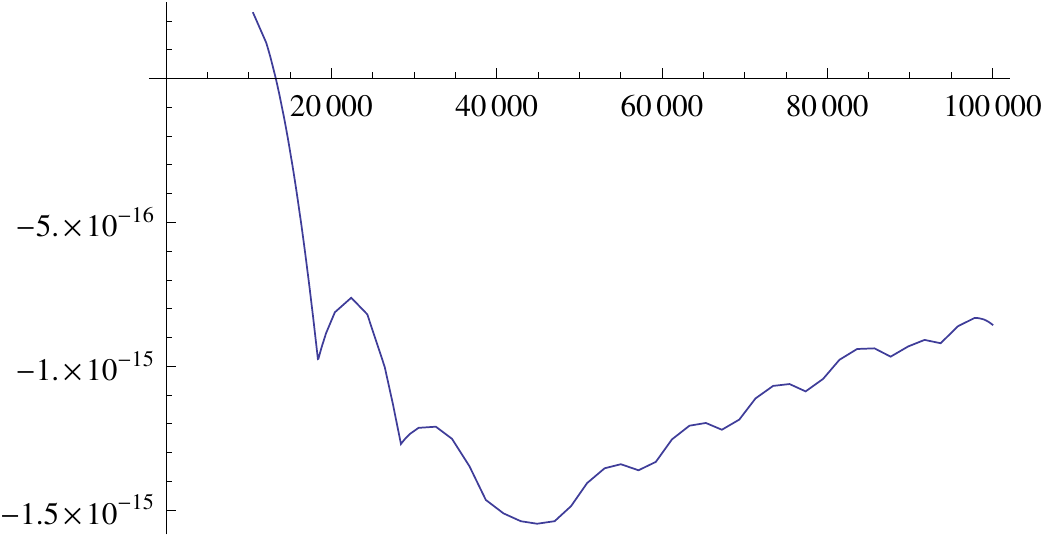}\hfil
\ing[width=0.45\textwidth]{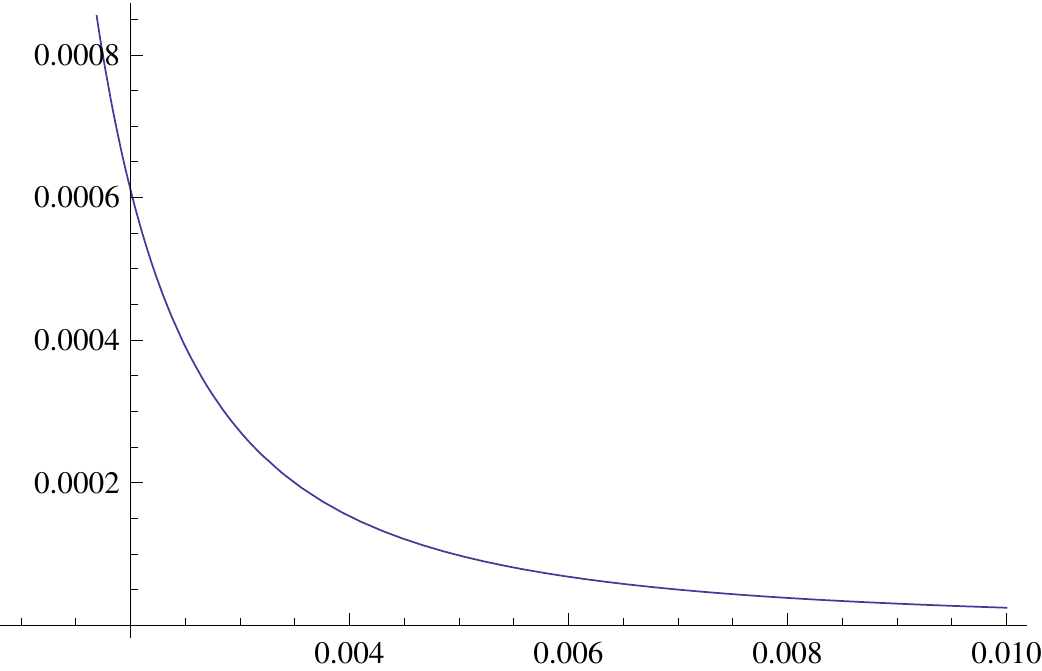}}
\hbox to \textwidth{\hbox to 0.45\textwidth{\hfil(I)\hfil}\hfil
\hbox to 0.45\textwidth{\hfil(J)\hfil}}
\hbox to \textwidth{\ing[width=0.45\textwidth]{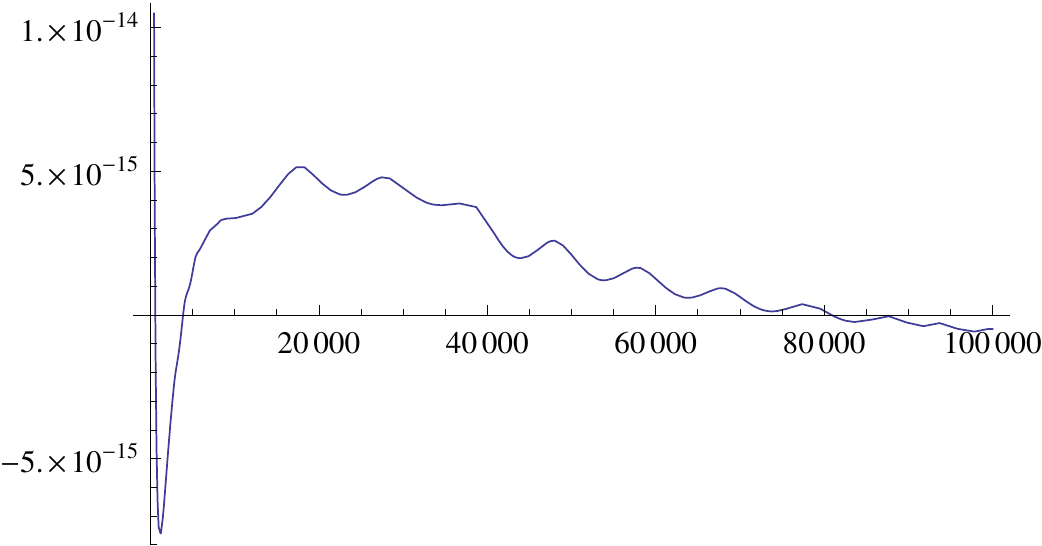}\hfil
\ing[width=0.45\textwidth]{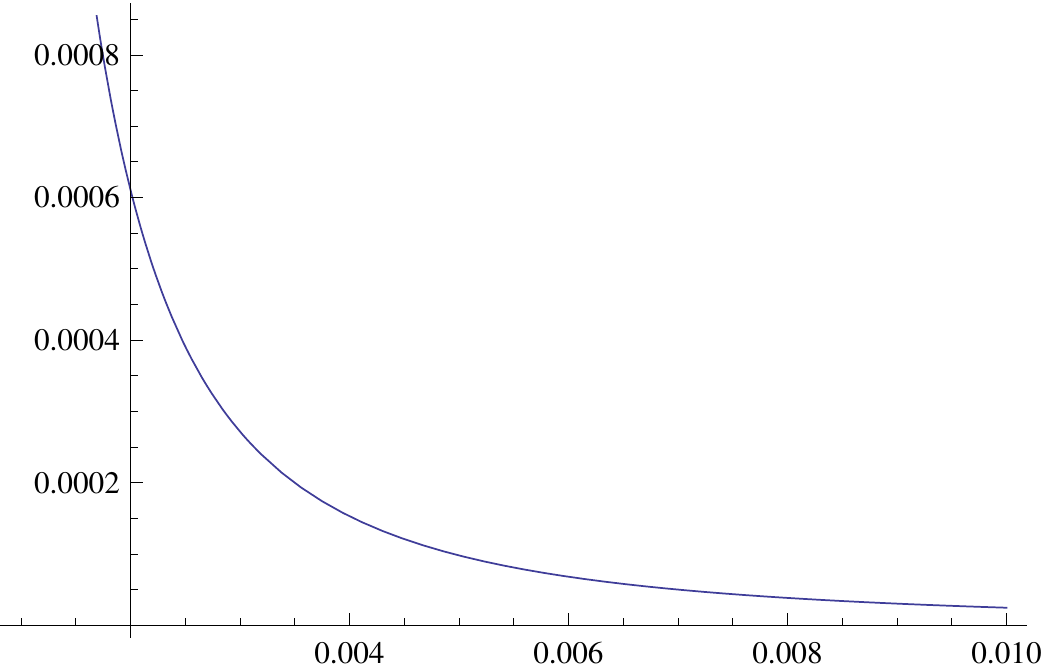}}
\hbox to \textwidth{\hbox to 0.45\textwidth{\hfil(K)\hfil}\hfil
\hbox to 0.45\textwidth{\hfil(L)\hfil}}
\hbox to \textwidth{\ing[width=0.45\textwidth]{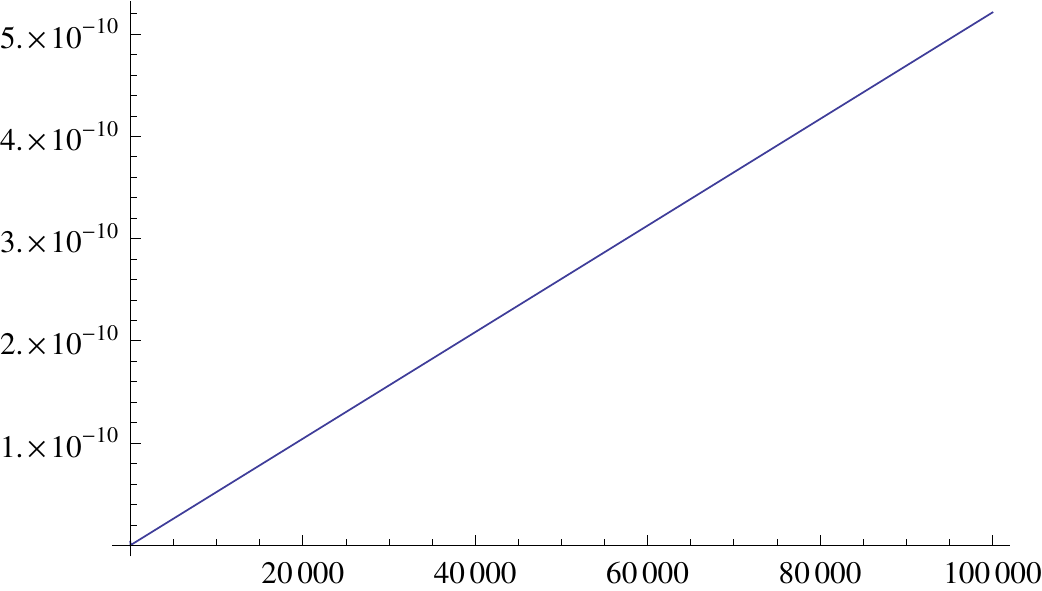}\hfil
\ing[width=0.45\textwidth]{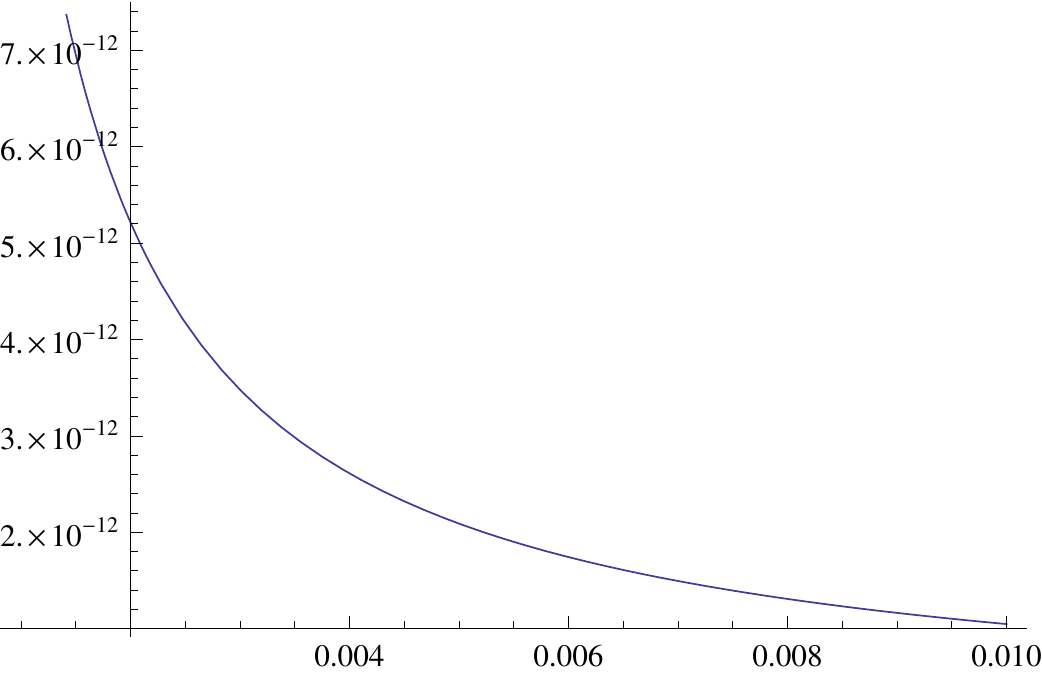}}
\hbox to \textwidth{\hbox to 0.45\textwidth{\hfil(M)\hfil}\hfil
\hbox to 0.45\textwidth{\hfil(N)\hfil}}
\hbox to \textwidth{\ing[width=0.45\textwidth]{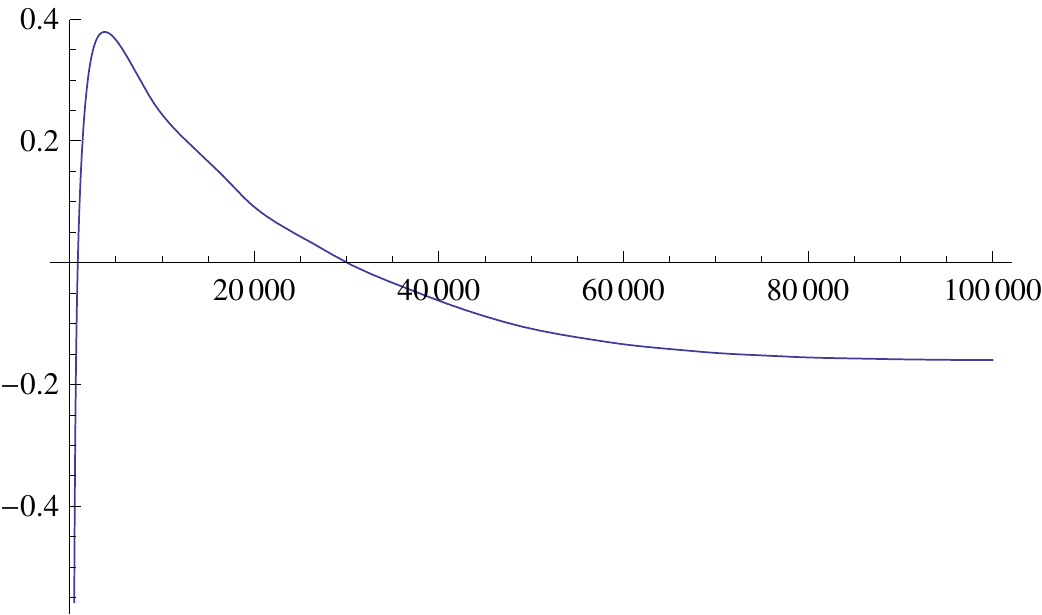}\hfil
\ing[width=0.45\textwidth]{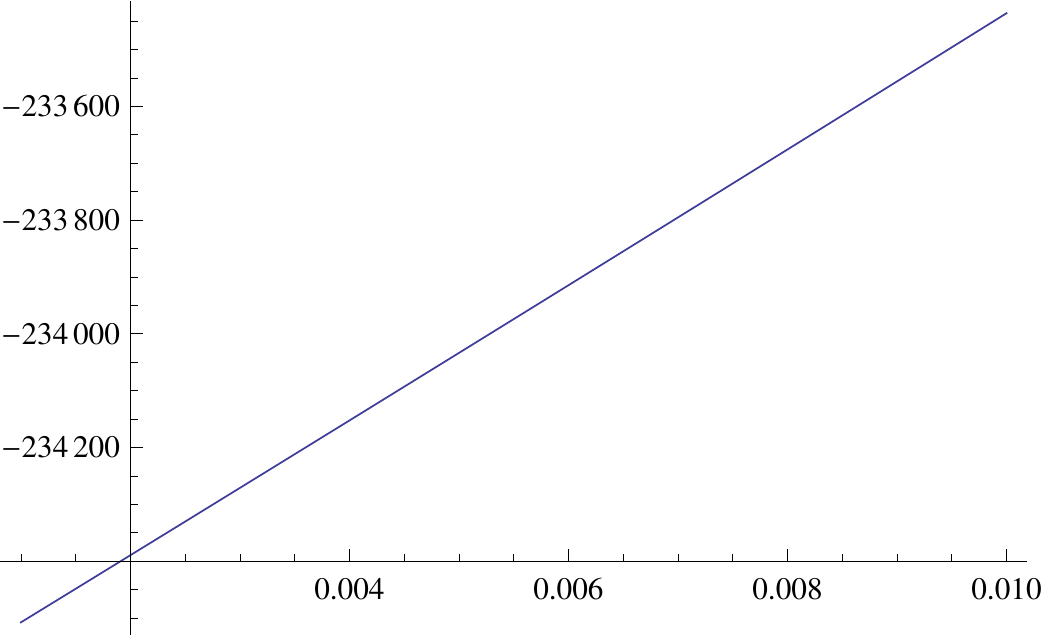}}
\hbox to \textwidth{\hbox to 0.45\textwidth{\hfil(O)\hfil}\hfil
\hbox to 0.45\textwidth{\hfil(P)\hfil}}
\smallskip
{\small \noindent Figure \thefigure\ (cont.):
(I)---a plot of  $b_1(r)$ for $10^{-3}<r<10^{5}$;
(J)---a plot of  $b_1(r)$ for $10^{-3}<r<10^{-2}$;
(K)---a plot of  $b_2(r)$ for $10^{-3}<r<10^{5}$;
(L)---a plot of  $b_2(r)$ for $10^{-3}<r<10^{-2}$;
(M)---a plot of  $-\ov U(r)$ for $10^{-3}<r<10^{5}$;
(N)---a plot of  $-\ov U(r)$ for $10^{-3}<r<10^{-2}$;
(O)---a plot of  $\eta_1(r)$ for $10^{-3}<r<10^{5}$;
(P)---a plot of  $\eta_1(r)$ for $10^{-3}<r<10^{-2}$.}

\hbox to \textwidth{\ing[width=0.45\textwidth]{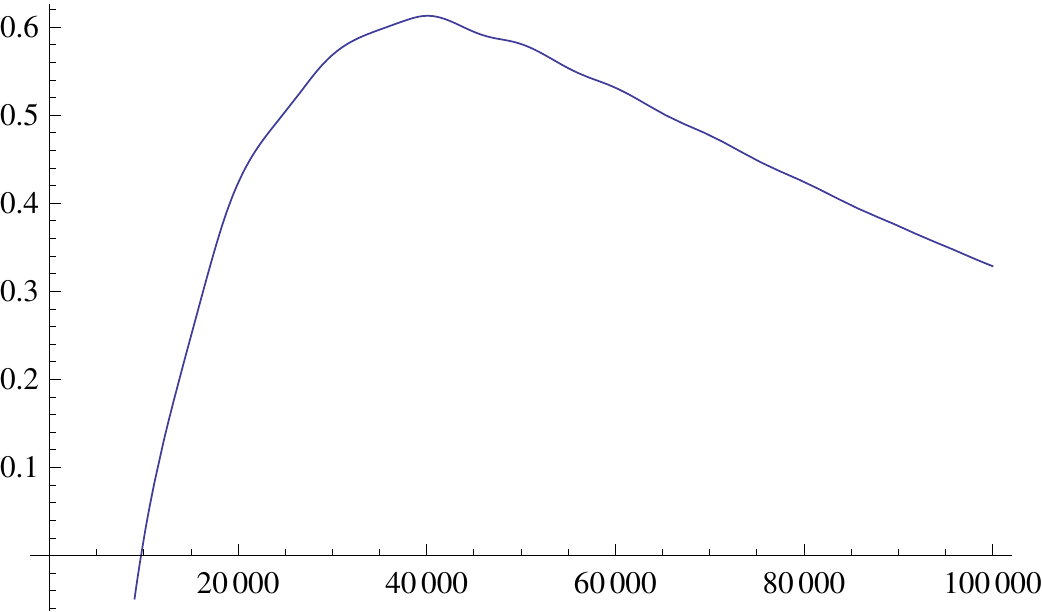}\hfil
\ing[width=0.45\textwidth]{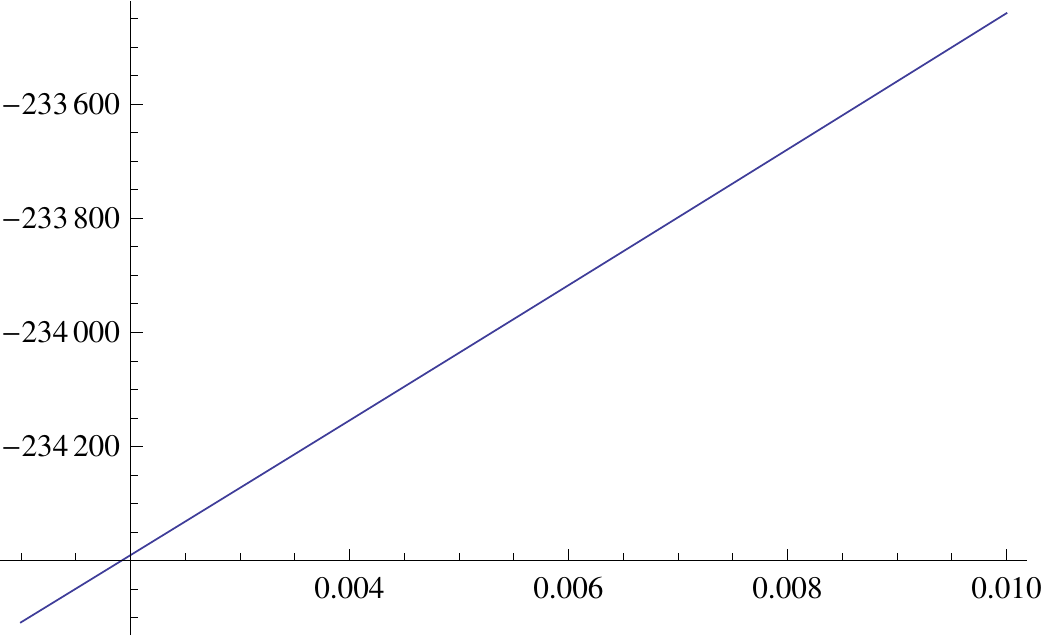}}
\hbox to \textwidth{\hbox to 0.45\textwidth{\hfil(Q)\hfil}\hfil
\hbox to 0.45\textwidth{\hfil(R)\hfil}}
\smallskip
{\small \noindent Figure \thefigure\ (cont.):
(Q)---a plot of  $\eta_2(r)$ for $10^{-3}<r<10^{5}$;
(R)---a plot of  $\eta_2(r)$ for $10^{-3}<r<10^{-2}$.}

\atw-0.2 \advance\abovedisplayskip by1.5pt \advance\belowdisplayskip by1.5pt
\bigskip
From these plots it is easy to see that $\ov V_1(r)$ and $\ov V_2(r)$ are
quite close to $-\ov U(r)$ on large distances. However, they are not close on
small distances. This is reasonable for on small distances Anderson et al.\
data have large fluctuations. We introduce to the model parameters of the Sun
($k^2=G_NM_\odot$) and Anderson et al.\ data ($r_0,b,R$) via our model from
Section~3 in the initial conditions.

All the programmes written in Mathematica~7 we quoted in Appendix E.

Let us consider a bending of light and Shapiro effect in our \spt. Using Eq.\
\er{Dn223} and Ref.~\cite{46} one gets
\beq{Da272}
\D\vf=2\int_{\ov r_1}^{\ov{\ov R}} \frac{e^{B(r)}}r \X2(\frac1{b^2}\,
r^2e^{-2A(r)}-1\Y2)^{-1/2}\,dr
\end{equation}
where $b=\frac{hc}{H}$ is an impact parameter, $\ov r_1$ is the closest
approach of a photon \st $\ov r_1>\ov r_0$ where $\ov r_0$ is a \so\ of an \e
\beq{Da273}
\ov r_0e^{-A(\ov r_0)}=b.
\end{equation}
In the case of GR, $\ov{\ov R}=+\iy$. However in our case for our \spt\ is
not asymptotically flat, $\ov{\ov R}<+\iy$ and $\ov{\ov R}$ should be chosen
carefully together with $\ov r_1$ and $b$.

The Shapiro effect (see Ref.\ \cite{46})---a time delay of a radar echo of
radar signal coming from Earth to Venus and back when Venus is in a superior
conjunction with the Sun---can be calculated
\beq{Da274}
\D t=2\X2(t(r_E,\ov r_0)+t(\ov r_0,r_V)-\frac{(r_E^2-\ov r{}_0^2)^{1/2}}c
-\frac{(r_V^2-\ov r{}_0^2)^{1/2}}c\Y2)
\end{equation}
where
\beq{Da275}
t(r,\ov r_0)=\frac1c \int_{\ov r_0}^r e^{B(r)-A(r)}\X2(1-\frac{b^2}{r^2}
e^{2A(r)}\Y2)^{-1/2}\,dr,
\end{equation}
$r_E$ and $r_V$ are distances of Earth and Venus from the Sun and $\ov r_0$
is given by the formula \er{Da273}. Moreover, the result depends on $b$,
$r_E$ and $r_V$ and should be fitted to the measurement. In the case of a \s\
part of a \nos\ metric \er{D.127} one gets similarly
\beq{Da276}
\D\vf=2\int_{\ov r_1}^{\ov{\ov R}} \frac{f^3(r)e^{-(r-\bar c)}}
{r\ell^2\sqrt{\a(r)}}\X2(\frac{r^2}{b^2\a(r)}-1\Y2)^{-1/2}\,dr,
\end{equation}
$\ov r_1>\ov r_0$ \st
\bg{Da277}
\ov r_0=\sqrt{\a(\ov r_0)}b \\
t(r,\ov r_0)=\frac{\ell^2e^{-\bar c}}c \int_{\ov r_0}^r \frac{\a(r)e^r}{f^3(r)}
\X2(1-\frac{b^2}r\,\a(r)\Y2)^{-1/2}\,dr \label{Da278}
\end{gather}
with the same comments as for GR.

Let us come back to the system of Eqs \er{D.30}--\er{D.32} and to Eqs
\er{Da256}. Now we change initial conditions for $B(r)$ and $A(r)$ in such a
way that
\beq{Da279}
\bal
B(3.7\t10^2)&=4\t10^{-12}\ve_1\\
\pz Br(3.7\t10^2)&=\ve_2\\
A(3.7\t10^2)&=\ve_3.
\eal
\end{equation}
$\ve_1,\ve_2$ and $\ve_3$ are free parameters \st $\ve_1$ and $\ve_3$ are close to
one and $\ve_2$ is close to zero. In this way
\beq{Da280}
\bal
B&=B(r,\ve_1,\ve_2)\\
A&=A(r,\ve_1,\ve_2,\ve_3)\\
\wt\vF&=\wt\vF(r,\ve_1,\ve_2).
\eal
\end{equation}
$A, B$ and $\wt\vF$ are differentiable \f s of their arguments.

In this way $\D\vf=\D\vf(\ve_1,\ve_2,\ve_3)$ and $\D t=\D
t(\ve_1,\ve_2,\ve_3)$ and we can try to fit data using $\ve_1,\ve_2,\ve_3$
via an inverse problem for Eqs \er{D.30}--\er{D.32} and \er{Da280}. Moreover,
we should also fit an \an\ \ph\ movement of Mercury. This can be achieved in
the following way. Let $r_M(\vf,\ve_1,\ve_2,\ve_3)$ be an \e\ of an orbit for
Mercury in polar \cd\ in an equatorial plane. Let us notice that $r_M$
depends also on $\ve_1,\ve_2,\ve_3$ via \er{Da280}. Let us suppose that
$r_M(0,\ve_1,\ve_2,\ve_3)=r_p$, where $r_p$ is a \ph\ distance for Mercury
and let $\D\a$ be a measured \an\ \ph\ advance for Mercury per one revolution
(in arc's measure). Thus we have
\beq{Da281}
r_p=r_M(2\pi+\D\a,\ve_1,\ve_2,\ve_3).
\end{equation}
This is the third \e\ to find $\ve_1,\ve_2,\ve_3$ with some freedom to choose
$b$, $\ov r_1$ and $\ov{\ov R}$. It seems that we can fit the data and get a
model of a \gr al field in the \SS\ with an \an\ \ac\ which pass general
relativistic tests. We can also apply a nonlinear statistical analysis to get
an estimation for $\ve_1,\ve_2$ and $\ve_3$ via Anderson et al.\ data and
data for the bending of light,  \ph\ advance of Mercury data, Shapiro test
data and some additional \SS\ data. In the case of the model from the full
\eu\nos\ Jordan--Thiry Theory with nine parameters (six from initial
conditions and three integration \ct s) we can do more fitting \ph\ advance
of Icarus, Venus and Earth. This is a real task for inverse problem in
numerical \so s of differential \e s and for a nonlinear statistical analysis
and will be considered elsewhere.

Let us consider Eq.\ \er{Dn222}, using a parametrization given by Eqs
\er{Da264} and \er{Da265}. One gets
\beq{Db282}
\X2(\frac h{r^2}\,\pz r\vf\Y2)^2+\frac{h^2}{r^2}=
\X3(\frac {H^2}{c^2}\,\frac{1-\frac{r_s}r-2\ov V_2(r)}{1-\frac{r_s}r-2\ov V_1(r)}
-c^2\Y3) + \frac{hr_s}{r^3}+\frac{r_sc^2}{r}+2\ov V_2(r)\X2(\frac{h^2}{r^2}
+c^2\Y2).
\end{equation}
Substituting $r=\frac1u$ and differentiating both sides \wrt $\vf$ one obtains
\beq{Db283}
\pz{^2u}{\vf^2}+u=T_0+T_{\rm GR}(u)+T(u)
\end{equation}
where
\beq{Db284}
T_0=\frac{r_sc^2}{2h^2}
\end{equation}
is a usual newtonian \ct\ term,
\beq{Db285}
T_{\rm GR}(u)=\frac{3r_su^2}2
\end{equation}
a term known from General Relativity,
\bml{Db286}
T(u)=-\frac {H^2}{h^2c^2(1-r_su-2\ov V_1(\frac 1u))} \cdot
\X3[-\frac{\D\ov V(\frac1u)(r_s+\frac2{u^2}\,\frac{d\ov V_1}{dr}(\frac1u))}
{1-r_su-2\ov V_1(\frac1u)}+\frac1{u^2}\,\D b\X2(\frac1u\Y2)\Y3]\\
{}+2\ov V_2\X2(\frac1u\Y2)u-b_2\X2(\frac1u\Y2)\X2(1+\frac{c^2}{h^2u^2}\Y2)
\end{multline}
is a new term, an \an\ \ac\ term. Let us notice that Eq.\ \er{Db283} is exact.
\bea{Db287}
\D\ov V(r)&=&\ov V_2(r)-\ov V_1(r)\\
\D b(r)&=& b_2(r)- b_1(r) \label{Db288}
\end{eqnarray}
(see Eqs \er{Da267}--\er{Da268}).

Eq.\ \er{Db283} can be applied for a perturbation of newtonian-elliptical
orbits in the following way. Let us consider an elliptic orbit
\beq{Db289}
u_0=\frac{1+e\cos\vf}{a_e(1-e^2)}, \q 0\le e<1,
\end{equation}
and a perturbation \st
\beq{Db290}
u=u_0+\D u_{\rm GR}+\D u
\end{equation}
where $\D u_{\rm GR}$ is a usual perturbation from GR and $\D u$ is a
perturbation due to an \an\ \ac. We consider both perturbations as very small
and we linearize \e\ \er{Db283} \wrt $\D u_{\rm GR}$ and $\D u$. One gets
\bea{Db291}
&&\pz{^2\D u_{\rm GR}}{\vf^2}+\D u_{\rm GR}=\frac{3k^2u_0^2}{c^2}\\
&&\pz{^2\D u}{\vf^2}+\D u=F(\vf) \label{Db292}
\end{eqnarray}
where
\beq{Db293}
F(\vf)=T\X2(\frac{1+e\cos\vf}{a_e(1-e^2)}\Y2).
\end{equation}
Eq.\ \er{Db291} has a classical \so\ known in GR going to \an\ \ph\ advance.

After some obvious simplifications one gets
\bml{Db294}
F(\vf)=-\frac{(c^2-\frac{k^2}{2a_e})^2}{c^2a_e(1-e^2)k^2}
\cdot \X3(\frac{a_e^2(1-e^2)^2}{(1+e\cos\vf)^2}\,\D b\X2(\frac{a_e(1-e^2)}
{1+e\cos\vf}\Y2)\Y3)\\
{}+2\ov V_2\X2(\frac{a_e(1-e^2)}{1+e\cos\vf}\Y2)\cdot\frac{1+e\cos\vf}
{a_e(1-e^2)} - b_2\X2(\frac{a_e(1-e^2)}{1+e\cos\vf}\Y2)\cdot
\X3(1+\frac{c^2(1-e^2)a_e}{(1+e\cos\vf)^2k^2}\Y3)
\end{multline}
or even simpler
\beq{Db294a}
F(\vf)=\frac{c^2a_e(1-e^2)}{k^2}\,b_1\X2(\frac{a_e(1-e^2)}{1+e\cos\vf}\Y2)
\frac1{(1+e\cos\vf)^2}\,.
\end{equation}
Now we proceed as in Section~4 (see Eqs \er{4.12}--\er{4.17}) and develop
$F(\vf)$ in a Fourier series:
\beq{Db295}
F(\vf)=\frac{b_0}2+\sum_{n=1}^\iy b_n\cos n\vf;
\end{equation}
we develop $\D u$ in a Fourier series too:
\beq{Db296}
\D u(\vf)=\frac{\,\ov u_0\,}2 + \sum_{n=1}^\iy u_n\cos n\vf,
\end{equation}
where
\bea{Db297}
u_n&=&\frac2\pi \int_0^\pi \D u(\vf)\cos n\vf\,d\vf\\
b_n&=&\frac2\pi \int_0^\pi F(\vf)\cos n\vf\,d\vf. \label{Db298}
\end{eqnarray}
We get similar \e s as in Section~4 getting \so s
\bea{Db299}
u_n(\vf)&=&\frac{b_n}{1-n^2}\cos n\vf, \q n\ge2\\
\ov u_0(\vf)&=&b_0 \label{Db300}\\
u_1(\vf)&=&\frac{b_1}2\,\vf \sin\vf. \label{Db301}
\end{eqnarray}
From \er{Db301} we get an \an\ movement of a \ph\ per one revolution
\beq{Db302}
\D \ov\a=\pi b_1 (1-e^2)a_e.
\end{equation}
From Eqs \er{Db296}, \er{Db299}, \er{Db300} we get an upper bound of a
distortion of an elliptic orbit (see Section~4 for details)
\beq{Db303}
|u-u_0-\D u_{\rm GR}|=\frac{h^2}{k^2}\X3|\frac{b_0}2 + \sum_{n=2}^\iy
\frac{b_n}{1-n^2}\cos n\vf\Y3| < \frac{b_0h^2}{2k^2} (\pi^2+1)=\ve.
\end{equation}
In this way we get a distortion of a major semi-axis of an ellipse
\bea{Db304}
\D a_e&=&a_e^2\,\frac{b_0}2\,(\pi^2+1)(1-e^2) \\
\frac{\D a_e}{a_e}&=&\frac{b_0}2\,(\pi^2+1)a_e(1-e^2). \label{Db305}
\end{eqnarray}
It is easy to see that we can use Eq.\ \er{Db302} in order to fit an \an\
movement of Mercury, in such a way that
\beq{Db306}
\D\a(\ve_1,\ve_2,\ve_3)=\D\a_{\rm GR}+\D\ov \a(\ve_1,\ve_2,\ve_3)
=\D\a_{\rm GR}+\pi b_1(\ve_1,\ve_2,\ve_3)(1-e^2)a_e
\end{equation}
and
\beq{Db307}
|\pi b_1(\ve_1,\ve_2,\ve_3)(1-e^2)a_e|<\d,
\end{equation}
where $\d$ is an error of measurement of $\D\a$.

In the case of a \hy\ orbit we get
\beq{Db308}
\pz{^2u}{\vf^2}+u=\frac1{a_h(e^2-1)}+3\,\frac{k^2}{c^2}\,u^2+T(u).
\end{equation}
Eq.\ \er{Db308} is a generalization of Eq.\ \er{9.8} without a term from NGT.

Let us consider Eq.\ \er{Da272} (a bending of light) and Eqs
\er{Da274}--\er{Da275} (Shapiro effect) using a parametrization given by Eqs
\er{Da264} and \er{Da265}. We get after some simplification
\beq{Db309}
\D\vf=2\int_{\ov r_0}^{\ov{\ov R}}\frac1{r^2}\X3(\frac1{b^2}-\frac{1-\frac
{r_s}r}{r^2}\Y3)^{-1/2}\,dr+2\int_{\ov r_0}^{\ov{\ov R}}
\frac{\ov V_2(r)+\ov V_1(r)}{r^2}\X3(\frac1{b^2}-\frac{1-\frac
{r_s}r}{r^2}\Y3)\,dr
\end{equation}
or
\bg{Db310}
\D\vf=\D\vf_{\rm GR}+2\X4(\int_{\ov r_0}^{\ov{\ov R}}\frac{\ov V_2(r)+\ov V_1(r)}{r^2}
\X3(\frac1{b^2}-\frac{1-\frac{r_s}r}{r^2}\Y3)^{-1/2}\,dr
-\int_{\ov{\ov R}}^\iy
\frac1{r^2}\X3(\frac1{b^2}-\frac{1-\frac{r_s}r}{r^2}\Y3)^{-1/2}\,dr \Y4)\\
\D\vf_{\rm GR}=2\int_{\ov r_0}^\iy
\frac1{r^2}\X3(\frac1{b^2}-\frac{1-\frac{r_s}r}{r^2}\Y3)\,dr. \label{Db311}
\end{gather}
For $\ov{\ov R}$ is large,
\beq{Db312}
\int_{\ov{\ov R}}^\iy
\frac1{r^2}\X3(\frac1{b^2}-\frac{1-\frac{r_s}r}{r^2}\Y3)^{-1/2}\,dr \simeq
\int_{\ov{\ov R}}^\iy \frac b{r^2}\,dr=\frac b{\,\ov{\ov R}\,}\,.
\end{equation}
In this way we can estimate a value of $\ov{\ov R}$ taking a term in brackets
in Eq.\ \er{Db310} equal to zero,
\beq{Db313}
\frac b{\,\ov{\ov R}\,}=\int_{\ov r_0}^{\ov{\ov R}}\frac{\ov V_2(r)+\ov V_1(r)}{r^2}
\X3(\frac1{b^2}-\frac{1-\frac{r_s}r}{r^2}\Y3)^{-1/2}\,dr.
\end{equation}
One can write the integral
\beq{Db314}
\int_{\ov r_0}^{\ov{\ov R}}\frac{\ov V_2(r)+\ov V_1(r)}{r^2}
\X3(\frac1{b^2}-\frac{1-\frac{r_s}r}{r^2}\Y3)^{-1/2}\,dr\simeq
\frac12\<\ov V_2+\ov V_1>\D\vf_{\rm GR}
\end{equation}
where $\<\ov V_2+\ov V_1>$ is an average value of $\ov V_2(r)+\ov V_1(r)$ for $\ov r_0<r<\ov
{\ov R}$. $\<\ov V_2+\ov V_1>$ is of order $1.5\t10^{-10}$. Taking $b$ equal to the
radius of the Sun, $b=R_\odot=6.96\t10^5$\,km, one gets
\beq{Db315}
\ov{\ov R}\simeq 1.094\t10^{21}\,{\rm km}\simeq 7.31\t10^{12}\,{\rm AU}
\simeq 35.48\,{\rm Mpc}.
\end{equation}
$\ov{\ov R}$ is of order of our unit $L$.

For $\D\vf_{\rm GR}$ we take $1.75''$ which is equal to $8.48\t10^{-6}$\,rd.
In more precise calculations we should take
\beq{Db316}
\ov{\ov R}=k_1\cdot 35.48\,{\rm Mpc}
\end{equation}
where $k_1$ is a \ct\ to be fitted.

Let us consider Eqs \er{Da274}--\er{Da275}. One gets
\bml{Db317}
t(\ov r_0,r)=\frac1c \int_{\ov r}^r \X2(1-\frac{r_s}r\Y2)^{-1}
\X3(1+\frac{\ov V_2(r)}{1-\frac{r_s}r}\Y3)
\X3(1+\frac{\ov V_1(r)}{1-\frac{r_s}r}\Y3)
\X3(1-\frac{b^2}{r^2}\X2(1-\frac{r_s}r\Y2)\Y3)^{-1/2}\,dr\\
{}\simeq t_{\rm GR}(\ov r_0,r)+\d t(\ov r_0,r)
\end{multline}
where
\beq{Db318}
t_{\rm GR}(\ov r_0,r)=\frac1c \int_{\ov r_0}^r \X2(1-\frac{r_s}r\Y2)^{-1}
\X3(1-\frac{b^2}{r^2}\X2(1-\frac{r_s}r\Y2)\Y3)^{-1/2}\,dr
\end{equation}
is $t(\ov r_0,r)$ for General Relativity and $\d t(\ov r_0,r)$ is an
additional correction due to an \an\ \ac
\beq{Db319}
\d t(\ov r_0,r)=\frac1c \int_{\ov r_0}^r \X2(1-\frac{r_s}r\Y2)^{-2}
\X1(\ov V_1(r)+\ov V_2(r)\Y1)\X2(1-\frac{b^2}{r^2}\X2(1-\frac{r_s}r\Y2)\Y2)
^{-1/2}\,dr.
\end{equation}
In this way
\bml{Db320}
\D t=2\X3(t(r_E,\ov r_0)+t(\ov r_0,r_V)-
\frac{(r_E^2-\ov r{}_0^2)^{1/2}}c - \frac{(r_V^2-\ov r{}_0^2)^{1/2}}c \Y3)\\
{}=\D t_{\rm GR}+2\X1(\d t(r_E,\ov r_0)+\d t(\ov r_0,r_V)\Y1).
\end{multline}

Let us come back to the problem of initial conditions of Eqs
\er{D.156}--\er{D.158} in order to apply it for the \SS\ \gr al field (we put
$n=120$, $\ov M=1$). We pose similarly as in the Riemannian geometry case
(i.e.\ $g_{\mu\nu}=g_{\nu\mu}$)
\beq{Db321}
\bal
\a(7.36\t10^{-10})&=1+8\t10^{-12}\\
\pz\a r(7.36\t10^{-10})&=0.
\eal
\end{equation}
In the case of the \f\ $f$ we pose (due to some heuristic considerations)
\beq{Db322}
\bal
f(7.36\t10^{-10})&=0\\
\pz fr(7.36\t10^{-10})&=0.
\eal
\end{equation}
In the case of the \f\ $\vF$
\beq{Db323}
\bal
\vF(7.36\t10^{-10})&=-5.681\t10^{-3}\\
\pz \vF r(7.36\t10^{-10})&=0.
\eal
\end{equation}
The first condition from \er{Db323} has some justification from our model of
an \an\ \ac\ of Section~3. The second one is posed in order to prevent a
significant changing of $G_{\rm eff}$ in the \SS. \eu\ct s $C_1,\ov c$ and
$\ell^2$ are still arbitrary. Using the same ideas as before we write initial
conditions
\beq{Db324}
\bal
\a(7.36\t10^{-10})&=1+8\ve_1\t10^{-12}\\
\pz\a r(7.36\t10^{-10})&=\ve_2\\
f(7.36\t10^{-10})&=\ve_3\\
\pz fr(7.36\t10^{-10})&=\ve_4\\
\vF(7.36\t10^{-10})&=-\ve_5\cdot 5.681\t10^{-3}\\
\pz \vF r(7.36\t10^{-10})&=\ve_6.
\eal
\end{equation}
$C_1$ and $\ov c$ are still completely arbitrary and
$$
\ell=\ve_7\cdot 3.1\t10^3\,{\rm km}=\ve_7\cdot 1.006\t10^{-14}L.
$$
\eu\ct s $\ve_1$, $\ve_5$ are of order one, $\ve_2,\ve_3,\ve_4,\ve_6$ are
around zero, $\ve_7$ is of order $10^{-2}$. The initial condition is posed at
the same point as before but a distance from the Sun is measured in our
system of units (${}\sim 10$\,Mpc).

In our approach we have to do with a dust matter $\ov \rho(r)$,
$\wh{\ov\rho}(r)$. The interesting point is to ask what is the nature of such a
matter. This is a dark matter (not barionic). Moreover, in the \eu\nos\
Jordan--Thiry Theory (see Ref.~\cite{10}) we have natural candidates for a
dark matter. It is a skewon and a \qe. Skewon is connected with $g_\[\m]$ in
a linear \ap ion of field \e s. The particle is massive due to a \co ical
\ct. The second particle is a \qe\ particle connected with the scalar field
in a linear \ap ion. This particle is also massive. Both particles are
interacting only \gr ally and very weak. In this theory we have three kinds
of \gr al particles: gravitons, skewons and \qe\ particles. Only the last two
are massive. Due to this fact they can be considered as a dark matter. Thus
we consider $\ov \rho(r)$, $\wh{\ov\rho}(r)$ as a dust of skewons and \qe\
particles. In the case of $\wh{\ov \rho}(r)$ it is obvious that it contains
skewons and \qe\ particles. In the case of
$\ov\rho(r)$ it is obvious that it contains \qe\
particles. The appearence of skewons is not so obvious. Moreover, we should
remember that in the full field \e s we have to do with a skew-\s\ part of
the metric. Even if we suppose that $g_\[\m]=0$ in the field \e s, any
perturbations of the \nos\ metric can produce in a linear \ap ion a skewon
field. Thus a dark matter in the \SS\ consists of skewons and \qe \
particles. Of course $\ov\rho(r)$ and $\wh{\ov\rho}(r)$ should be smaller
than a density of interplanetary matter which is of barionic origin.

Some researchers claim that changing a theory of \gr\ they can avoid
introducing a dark matter and dark energy. In our approach we try to explain
the appearence of dark matter and dark energy via \eu\nos\ Jordan--Thiry
Theory. We have candidates for dark matter particles. Simultaneously we
explain \co ical \ct\ via our \qe\ scenario.

There is an important problem to match our \so\ to \cd\ systems using in the
\SS\ according to the IAU suggestions (see Section~7). According to the
suggestions we should use an isotropic \cd\ system. It means, we should use a
metric
\beq{Db325}
ds^2=c^2\ov A{}'(\ov r)\,dt^2 - \ov B{}'(\ov r)\X1(d\ov r{}^2+\ov r{}^2(d\th^2
+\sin^2\th \,d\vf^2)\Y1)
\end{equation}
or
\beq{Db326}
ds^2=c^2\ov A{}'(\ov r)\,dt^2 - \ov B{}'(\ov r)\X1(dx^2+dy^2+dz^2\Y1)
\end{equation}
where $\ov r{}^2=x^2+y^2+z^2$.

Our \cd\ system is different
\beq{Db327}
ds^2=c^2\ov A(r)\,dt^2-\ov B(r)\,dr^2-r^2(d\th^2+\sin^2\th\,d\vf^2).
\end{equation}
Moreover, we can transform $r$ to $\ov r$ and vice-versa in the following way
using usual notation.
\bg{Db328}
\bal
\ov A{}'(\ov r)&=e^{2A'(\ov r)}\\
\ov B{}'(\ov r)&=e^{2B'(\ov r)}
\eal \\
\ov r=g(r)=\exp\X2(\int \frac1r \, e^{B(r)}\,dr\Y2) \label{Db329}\\
A'(g(r))=A(r) \label{Db330}\\
e^{B'(g(r))}=r\exp\X2(-\int\frac1r \,\eb{}\,dr\Y2) \label{Db331}
\end{gather}
or
\bg{Db332}
A'(\ov r)=A(\wt f(\ov r)) \\
e^{B'(\ov r)}=\wt f(\ov r)\exp\X3(-\int^{f(\ov r)}\frac1\tau \,e^{B(\tau)}\,
d\tau\Y3) \label{Db333}
\end{gather}
in such a way that
\beq{Db334}
\bal
\wt f(g(r))&=r\\
g(\wt f(\ov r))&=\ov r.
\eal
\end{equation}

This system of \cd s corresponds to isotropic system of \cd s for a
Schwarzschild \so. We consider this system of \cd s (see Eqs
\er{D.164}--\er{D.165}). Moreover, our time \cd\ is simply $t$ as before. In
the case of a \s\ part of a metric \er{D.127}, i.e.
\beq{Db335}
ds^2=\g(r)\,dt^2-\a(r)\,dr^2-r^2(d\th^2+\sin^2\th\,d\vf^2)
\end{equation}
one gets
\bg{Db336}
ds^2=\ov \g(\ov r)\,dt^2-\ov\a(\ov r)\X1(d\ov r{}^2+\ov r{}^2(d\th^2
+\sin^2\th\,d\vf^2)\Y1)\\
\ov r=g(r)=\exp\X2(\int\frac1r\sqrt{\a(r)}\,dr\Y2) \label{Db337}\\
\ov\g(g(r))=\g(\ov r) \label{Db338}\\
\ov \a (g(r))=r^2\exp\X2(-2\int\frac1r\sqrt{\a(r)}\,dr\Y2). \label{Db339}
\end{gather}
or
\bg{Db340}
\ov\a(\ov r)=\a\X1(\wt f(\ov r)\Y1)\\
\g(\ov r)=\wt f(\ov r)\exp\X2(-2\int^{\wt f(\ov r)}\frac1\tau \sqrt{\a(\tau)}
\,d\tau\Y2)\label{Db341}
\end{gather}
where $\wt f$ and $g$ satisfy Eq.\ \er{Db334}.

For the skew-\s\ part of the metric one gets
\bea{Db342}
\ov\o(g(r))&=&\frac r{\sqrt{\a(r)}}\exp\X2(-\int \frac1r \sqrt{\a(r)}\,dr\Y2)
\o(r)\\
\ov f(g(r))&=&f(r). \label{Db343}
\end{eqnarray}
Considering the movement of massive point bodies or photons we should use
rather \cd\ time~$t$ than a proper time $\tau$.

In order to facilitate future investigations let us parametrize $\a(r)$ and
$\g(r)$ in the following way:
\bea{Db344}
\a(r)&=&\frac1{1-2U_2(r)}=\frac1{1-\frac{r_s}r -2\ov U_2(r)}\\
\g(r)&=&\X2(1+\frac{\ell^4}{r^4}\Y2)(1-2U_1(r))=\X2(1+\frac{\ell^4}{r^4}\Y2)
\X2(1-\frac{r_s}r-2\ov U_1(r)\Y2). \label{Db345}
\end{eqnarray}
One gets
\bea{Db346}
U_2(r)&=&\frac12\X2(1-\frac1{\a(r)}\Y2)\\
\ov U_2(r)&=&\frac12\X2(1-\frac1{\a(r)}-\frac{r_s}{r}\Y2) \label{Db347}\\
U_1(r)&=&\frac12\X2(1-\frac{\g(r)}{1+\frac{\ell^4}{r^4}}\Y2) \label{Db348}\\
\ov U_1(r)&=&\frac12\X2(1-\frac{\g(r)}{1+\frac{\ell^4}{r^4}} -\frac{r_s}r \Y2).\label{Db349}
\end{eqnarray}
Using Eq.\ \er{D.167} one writes
\bea{Db350}
U_1(r)&=&\frac12\X3(1-\frac{f^6(r)e^{2(\ov c-r)}}{\ell^4\a(r)(1+\frac{\ell^4}{r^4})}\Y3)\\
\ov U_1(r)&=&\frac12\X3(1-\frac{f^6(r)e^{2(\ov c-r)}}{\ell^4\a(r)(1+\frac{\ell^4}{r^4})}
-\frac{r_s}r \Y3). \label{Db351}
\end{eqnarray}
The form of $\g(r)$ Eq.\ \er{Db345} has been obtained from a spherically-\s\
static \so\ in NGT (see Ref.~\cite9). In the case of $\o(r)$ we give the
following parametrization (see Refs \cite4,~\cite5)
\beq{Db352}
\o(r)=\frac \ell {r^2+h(r)}
\end{equation}
or (see Eq.\ \er{D.159})
\beq{Db353}
h(r)=\frac{\ell^2}{\o(r)}-r^2=\frac{\ell^2}{f^2(r)}\,e^{r-\ov c}-r^2.
\end{equation}

Let us consider an additional \an\ \ph\ movement, a bending of light and
Shapiro effect in a general treatment of the \eu\nos\ Jordan--Thiry Theory
using parametrization \er{Db344}--\er{Db349}. One gets similarly as before
for a \s\ metric case
\bml{Db355}
\X2(\frac h{r^2}\,\pz r\vf\Y2)^2+\frac{h^2}{r^2}\\{}=
\X3(\frac{H^2}{c^2}\,\frac{1-\frac{r_s}r-\ov U_2(r)}
{(1-\frac{r_s}r-\ov U_1(r))(1+\frac{\ell^4}{r^4})}-c^2\Y3)
+\frac{h^2r_s}r+\frac{r_sc^2}{r^2}+\frac{h^2\ov U_2(r)}{r^2}+c^2U_2(r).
\end{multline}
Taking $u=\frac1r$ one obtains
\bml{Db356}
\X2(\pz u\vf\Y2)^2+u^2=\frac1{h^2}
\X3(\frac{H^2}{c^2}\,\frac{1-\frac{r_s}r-\ov U_2(r)}
{(1-\frac{r_s}r-\ov U_1(r))(1+\frac{\ell^4}{r^4})}-c^2\Y3)\\
{}+\frac{r_sc^2u}{h^2}+r_su^3+\ov U_2\X2(\frac1u\Y2)\X2(u^2+\frac{c^2}{h^2}\Y2)
\end{multline}
or
\bml{Db357}
\X2(\pz u\vf\Y2)^2+u^2=\frac1{h^2}
\X2(\frac{H^2}{c^2}-c^2\Y2)+\frac{H^2}{h^2c^2}\,
\frac{\D\ov U(\frac1u)}{(1+\ell^4u^4)(1-r_su-\ov U_1(\frac1u))}\\
{}-\frac{H^2\ell^4u^4}{h^2c^2(1+u^4\ell^4)}+\frac{r_sc^2u}{h^2}+r_su^3
+\ov U_2\X2(\frac1u\Y2)\X2(u^2+\frac{c^2}{h^2}\Y2).
\end{multline}

We differentiate both sides of Eq.\ \er{Db357} \wrt $\vf$ getting
\bml{Db358}
\pz{^2u}{\vf^2}+u=\frac{r_sc^2}{2h^2}+\frac{3r_su^2}2
-\frac{2H^2\ell^4u^3}{h^2c^2(1+\ell^4u^4)^2}\\
{}+\frac{H^2}{2h^2c^2}\,\frac{-\frac1{u^2}\,\pz{\D\ov U}r(\frac1u)
(1+\ell^4u^4)(1-r_su-\ov U_1(\frac1u))-4\ell^4u^3\D\ov U(\frac1u)
(1-r_su-\ov U_1(\frac1u))}{(1-r_su-\ov U_1(\frac1u))^2(1+u^4\ell^4)^2}\\
{}+\D\ov U\X2(\frac1u\Y2)(1+u^4\ell^4)\X2(r_s-\frac1{u^2}\,
\pz{\ov U_1}{r}\X2(\frac1u\Y2)\Y2)+u\ov U_2\X2(\frac1u\Y2)
-\frac12\,\pz{\ov U_2}r\X2(\frac1u\Y2)\X2(1+\frac{c^2}{u^2h^2}\Y2).
\end{multline}
Taking $H\simeq c^2$ and making some simplifications one gets
\beq{Db359}
\pz{^2u}{\vf^2}+u=\frac{r_sc^2}{2h^2}+\frac{3r_su^2}2-\frac{2H^2\ell^2u^3}
{h^2c^2}-\frac{c^2}{2h^2u^2}\,\ov b_1\X2(\frac1u\Y2)
\end{equation}
where
\bg{Db360}
\ov U(r)=\ov U_1(r)-\ov U_2(r)\\
\ov b_1(r)=\pz{\ov U_1}r(r) \label{Db361}
\end{gather}

In the formula \er{Db358} we have a newtonian term, a General Relativity term
(as before) and a term from NGT and an \an\ \ac\ term. Supposing we are on an
elliptic orbit,
\beq{Db362}
u=u_0+\D u_{\rm GR}+\D u_{\rm NGT}+\D u, \q
u_0=\frac{1+e\cos\vf}{(1-e^2)a_e}\,.
\end{equation}
We get
\beq{Db363}
\pz{^2\D u}{\vf^2}+\D u=-\frac{c^2a_e(1-e^2)}{2k^2}\,\ov b_1\X2(
\frac{1+e\cos\vf}{a_e(1-e^2)}\Y2)\,\frac1{(1+e\cos\vf)^2}\,.
\end{equation}
In the case of a \hy\ orbit we get
\beq{Db364}
\pz{^2u}{\vf^2}+u=\frac{1}{a_h(e^2-1)}+3k^2u^2-\frac{2\ell^2u^3}{k^2a_h(e^2-1)}
-\frac{c^2}{2k^2a_h(e^2-1)u^2}\,\ov b_1\X2(\frac1u\Y2).
\end{equation}

The last \e\ is a generalization of Eq.\ \er{9.8}. Let us notice that we can
proceed all the considerations concerning an additional \an\ \ph\ movement
and a distortion of an elliptic orbit exactly as in the \s\ case (i.e.,
instead of $b_1$ we take $\ov b_1$). The formalism presented above is easily
applicated in the \nos\ case. Let us consider a bending of light. From Eqs
\er{Da272} and \er{Da276} one gets using parametrization \er{Db344}--\er{Db349}:
\beq{Db365}
\bal
\D\vf&=2\int_{\ov r_1}^{\ov{\ov R}}\frac1r\X2(1-\frac{r_s}r-2\ov U_2(r)\Y2)
^{1/2}\cdot \X2(\frac1{b^2}\,\frac{r^2}{(1+\frac{\ell^4}{r^4})(1-\frac{r_s}r
-2\ov U_1(r))}-1\Y2)^{-1/2}dr\\
&\simeq 2\int_{\ov r_1}^{\ov{\ov R}}\X2(1-\frac{r_s}r\Y2)^{1/2}
\X3(\frac1{b^2}\,\frac{r^2}{(1+\frac{\ell^4}{r^4})(1-\frac{r_s}r)}-1\Y3)
^{-1/2}dr\\
&{}-2\int_{\ov r_1}^{\ov{\ov R}}\frac{\ov U_1(r)+\ov U_2(r)}r
\X3(\frac1{b^2}\,\frac{r^2}{(1+\frac{\ell^4}{r^4})(1-\frac{r_s}r)}-1\Y3)
^{-1/2}dr\\
&= \D\vf_{\rm NGT}+2\X3(\frac b{\,\ov{\ov R}\,}+
\int_{\ov r_1}^{\ov{\ov R}}\frac{\ov U_1(r)+\ov U_2(r)}r\cdot
\X3(\frac1{b^2}\,\frac{r^2}{(1+\frac{\ell^4}{r^4})(1-\frac{r_s}r)}-1\Y3)
^{-1/2}dr\Y3)
\eal
\end{equation}
where
\beq{Db367}
\D\vf_{\rm NGT}=\int_{\ov r_1}^\iy \X2(1-\frac{r_s}r\Y2)^{1/2}
\X3(\frac1{b^2}\,\frac{r^2}{(1+\frac{\ell^4}{r^4})(1-\frac{r_s}r)}-1\Y3)
^{-1/2}dr
\end{equation}
is a bending of light obtained in NGT. If we take
\beq{Db368}
\frac b{\,\ov{\ov R}\,}+
\int_{\ov r_1}^{\ov{\ov R}}\frac{\ov U_1(r)+\ov U_2(r)}r\cdot
\X3(\frac1{b^2}\,\frac{r^2}{(1+\frac{\ell^4}{r^4})(1-\frac{r_s}r)}-1\Y3)
^{-1/2}dr=0
\end{equation}
we satisfy $\D\vf=\D\vf_{\rm NGT}$ (see Ref.~\cite9) getting a value of
$\ov{\ov R}<\iy$ (as before in \s\ case).

In the case of Shapiro effect we get
\beq{Db369}
\bal
t(r,\ov r_0)&=\frac1c \int_{\ov r_0}^r \X2(1+\frac{\ell^4}{r^4}\Y2)^{1/2}
\X2(1-\frac{r_s}r-2\ov U_1(r)\Y2)^{1/2}\\
&\qquad \X2(1-\frac{r_s}r-2\ov U_2(r)\Y2)^{1/2}\X2(1-\frac{b^2}{r^2}\X2(1+
\frac{\ell^4}{r^4}\Y2)\X2(1-\frac{r_s}r-2\ov U_1(r)\Y2)\Y2)^{-1/2}\\
&\simeq \frac1c \int_{\ov r_0}^r \X2(1+\frac{\ell^4}{r^4}\Y2)^{1/2}
\X2(1-\frac{r_s}r\Y2)
\X3(1-\frac{b^2}{r^2}\X2(1-\frac{r_s}r\Y2)\X2(1+\frac{\ell^4}{r^4}\Y2)
\Y3)^{-1/2}dr\\
&\qquad {}+\frac1c \int_{\ov r_0}^r \X2(1+\frac{\ell^4}{r^4}\Y2)^{1/2}
(\ov U_1(r)+\ov U_2(r))
\X3(1-\frac{b^2}{r^2}\X2(1-\frac{r_s}r\Y2)\X2(1+\frac{\ell^4}{r^4}\Y2)
\Y3)^{-1/2}dr
\eal
\end{equation}
In this way one gets
\bml{Db370}
\D t=2\X2(t(r_E,\ov r_0)+t(\ov r_0,r_V)
-\frac{(r_E^2-\ov r{}_0^2)^{1/2}}c-\frac{(r_V^2-\ov r{}_0^2)^{1/2}}c\Y2)\\
{}=\D t_{\rm NGT}+\d t(r_E,\ov r_0)+\d t(\ov r_0,r_V)
\end{multline}
where
\beq{Db371}
\D t_{\rm NGT}=2\X2(t_{\rm NGT}(r_E,\ov r_0)+t_{\rm NGT}(\ov r_0,r_V)
-\frac{(r_E^2-\ov r{}_0^2)^{1/2}}c-\frac{(r_V^2-\ov r{}_0^2)^{1/2}}c\Y2)
\end{equation}
is a Shapiro effect in NGT
\beq{Db372}
t_{\rm NGT}(r,\ov r_0)=\frac1c \int_{\ov r_0}^r \X2(1+\frac{\ell^4}{r^4}\Y2)^{1/2}
\X2(1-\frac{r_s}r\Y2)
\X3(1-\frac{b^2}{r^2}\X2(1-\frac{r_s}r\Y2)\X2(1+\frac{\ell^4}{r^4}\Y2)
\Y3)^{-1/2}dr
\end{equation}
(see Ref.\ \cite{37}) and
\beq{Db373}
\d t(r,\ov r_0)=\frac1c \int_{\ov r_0}^r \X2(1+\frac{\ell^4}{r^4}\Y2)^{1/2}
(\ov U_1(r)+\ov U_2(r))
\X3(1-\frac{b^2}{r^2}\X2(1-\frac{r_s}r\Y2)\X2(1+\frac{\ell^4}{r^4}\Y2)
\Y3)^{-1/2}dr.
\end{equation}
A bending of light and a Shapiro effect can be satisfied in NGT if $\ell \simeq
3.1\t 10^3$\,km (see Refs \cite9, \cite{37}). Thus the correction
$\d t(r_E,\ov r_0)+\d t(\ov r_0,r_V)$ must be smaller than an uncertainty
from the measurement.

Let us notice that an \an\ \ac\ can be written in the following way
\beq{Db374}
b=-\frac{c^2}2\,b_1(r)
\end{equation}
or in a proper system of units
\beq{Db375}
b=-7.33\t 10^4 b_1(r)
\end{equation}
in such a way that $b$ is measured in $\frac{\rm m}{\rm s^2}$. In the
formulae \er{Db374}--\er{Db375} we can put $b_1(r)=\pz{\ov V_1}{r}$,
$\ov b_1(r)=\pz{\ov U_1}r$.

\def\ef#1{e^{#1\vF(r)}}
Let us consider Eqs \er{D.156}--\er{D.158} and let us change a scale of
length in these \e s
\beq{Db376}
r \to \frac r{r_0}
\end{equation}
\eu\e s are transformed into
\beq{Db377}
\bal
{}&\pz{^2\a}{r^2}(r)=\frac1{3n\nd f(r)^2(\ell^4+f(r)^2)^2}\,r_0^2\\
&\ \q\X3(4\a(r)^2f(r)\X1(-3n\nd C_1+2\ef n(n+2-\ef2 n)f(r)\Y1)
(\ell^4+f(r)^2)(3\ell^4+2f(r)^2)\\[-1pt]
&\ {}-\frac{3n\nd f(r)(\ell^4+f(r)^2)\pz\a r(r)\X1(r_0f(r)(2\ell^4+3f(r)^2)
+2(5\ell^4+3f(r)^2)\pz fr(r)\Y1)}{r_0^2}\\[-1pt]
&\ {}+\frac{3n\nd \a(r)}{r_0^2}\X2(-r_0^2f(r)^2(\ell^8+3\ell^4f(r)^2+2f(r)^4)
-r_0f(r)(11\ell^8+20\ell^4f(r)^2+8f(r)^4)\pz fr(r)\\[-1pt]
&\ {}-2(9\ell^8+7\ell^4f(r)^2+2f(r)^4)\X2(\pz fr(r)\Y2)^2-4\ov Mf(r)^2(\ell^4
+f(r)^2)^2\X2(\pz\vF r(r)\Y2)^2\Y2)\Y3)\\[-1pt]
&\pz{^2f}{r^2}(r)=\X2(2r_0^2\a(r)^2f(r)\X1(3n\nd C_1+2\ef n(-2+(\ef 2-1)n)
f(r)\Y1)(\ell^4+f(r)^2)\\[-1pt]
&\ {}+3n\nd f(r)(\ell^4+f(r)^2)\pz \a r(r)\pz fr(r)+3n\nd \a(r)\pz fr(r)\\[-1pt]
&\ \q
\X2(r_0f(r)(\ell^4+f(r)^2)+(3\ell^4+f(r)^2)\pz fr(r)\Y2)\Y2)
\X1(3n\nd \a(r)f(r)(\ell^4+f(r)^2)\Y1)^{-1}\\[-1pt]
&\pz{^2\vF}{r^2}(r)=\X1(6\ov Mn\a(r)f(r)(\ell^4+f(r)^2)^2\Y1)^{-1}
\X3(\ef n(4-5(\ef2-1)n)r_0^2\a(r)^2f(r)^3(\ell^4+f(r)^2)\\[-1pt]
&\ {}+6\ov Mnf(r)(\ell^4+f(r)^2)^2\pz\a r(r)\pz\vF r(r)
+3n\a(r)\X2(\ell^4\nd f(r)\X2(\pz \vF r(r)\Y2)^2\\[-1pt]
&\ {}+2\ov M(\ell^4+f(r)^2)
\X2(r_0f(r)(\ell^4+f(r)^2)+2\ell^4\pz fr(r)\Y2)\pz\vF r(r)\Y2)\Y3)
\eal
\end{equation}
where we keep for a new \cd\ $r$ as before. The set of initial conditions
\er{Db321}--\er{Db323} changes into
\beq{Db378}
\bal
\a\X2(\frac{7.36\t10^{-10}}{r_0}\Y2)&=1+8\t10^{-12}\\
\hbox{or}\quad\a\X2(\frac{7.36\t10^{-10}}{r_0}\Y2)&=1+14.94\t10^{-12}
\qh{(see Eq.\ \er{11.82})}\\
\pz\a r\X2(\frac{7.36\t10^{-10}}{r_0}\Y2)&=0\\
f\X2(\frac{7.36\t10^{-10}}{r_0}\Y2)&=0\\
\pz fr\X2(\frac{7.36\t10^{-10}}{r_0}\Y2)&=0\\
\vF\X2(\frac{7.36\t10^{-10}}{r_0}\Y2)&=-5.681\t10^{-3}\\
\pz \vF r\X2(\frac{7.36\t10^{-10}}{r_0}\Y2)&=0.
\eal
\end{equation}

Let us notice that we define initial conditions at $R\simeq 0.6\ov R$. In
\er{Db378} $r_0$ is still arbitrary. If we take $r_0=4.103\,{\rm AU}
\simeq 1.9897\t10^{-12}\ov\xi L$ (our first length scale) we can use an uncertainty
in~$L$ definition (of order~1) and we have ($\ov\xi$ is of order 1):
\beq{Db379}
\bal
\a(643)&=1+8\t10^{-12}\\
\hbox{or}\q \a(643)&=1+14.94\t10^{-12}\\
\pz\a r(643)&=0\\
f(643)&=0\\
\pz fr(643)&=0\\
\vF(643)&=-5.681\t10^{-3}\\
\pz \vF r(643)&=0.
\eal
\end{equation}
Eqs \er{Db377} can be transformed by substituting for $r_0=1.9897\t10^{-12}$
and the result has been quoted in Appendix~E.

Moreover, now we take $L=10$\,Mpc and all uncertainty in $L$ definition has
been shifted to an uncertainty of $r_0$ definition.

Let us notice the following fact. The right hand sides of \e s \er{Db377} are
polynomials of three variables $r_0, C_1$ and $\ell^2$. Moreover, $\ell^2$
appears also in \cf s of these polynomials via $(\ell^4+f^2(r))^n$, $n=-1,
-2$. If $\ell^2$ is small in comparison to $f(r)$, it does not have any
meaning. According to this approximation the real meaning has a power of
$\ell^2, C_1$ and $r_0$ in an expansion of the right hand side of Eqs
\er{Db377}.

\def\Lf{(\ell^4+f^2(r))^2}
In order to facilitate a future research we expand Eqs \er{Db377} \wrt $r_0$,
$C_1$ and $\ell^2$. One gets
\beq{De379}
\bal
\pz{^2\a}{r^2}(r)&=
-\frac{12C_1\ell^8r_0^2\a^2(r)}{f(r)\Lf}
-\frac{20C_1\ell^4r_0^2\a^2(r)f(r)}{\Lf}
-\frac{8C_1r_0^2\a^2(r)f^3(r)}{\Lf}\\
&+\frac{\ell^8r_0^2\a(r)}{\Lf}
\X3(-1+8\a(r)\X2(\frac{\ef n}n-\frac{\ef\nd}{\nd}\Y2)\Y3) \\
&+\frac{\ell^4r_0^2\a(r)f^2(r)}{\Lf}\X3(
-3+\frac{40}3\,\a(r)\X2(\frac{\ef n}{n}
-\frac{\ef\nd}{\nd}\Y2)\Y3)\\
&+\frac{r_0^2\a(r)f^4(r)}{\Lf}\X3(-2+\frac{16}3\,\a(r)
\X2(\frac{\ef n}{n}-\frac{\ef\nd}{\nd}\Y2)\Y3)\\
&-\frac{2\ell^8r_0\pz\a r(r)}{\Lf}
-\frac{5\ell^4r_0f^2(r)\pz\a r(r)}{\Lf}
-\frac{3r_0f^4(r)\pz\a r(r)}{\Lf}\\
&-\frac{11\ell^8r_0\a(r)\pz fr(r)}{f(r)\Lf}
-\frac{20\ell^4r_0f(r)\a(r)\pz fr(r)}{\Lf}
-\frac{8r_0f^3(r)\a(r)\pz fr(r)}{\Lf}\\
&-\frac{10\ell^8\pz\a r(r)\pz fr(r)}{f(r)\Lf}
-\frac{16\ell^4f(r)\pz\a r(r)\pz fr(r)}{\Lf}
-\frac{6f^3(r)\pz\a r(r)\pz fr(r)}{\Lf}\\
&-\frac{18\ell^8\a(r)\X1(\pz fr(r)\Y1)^2}{f^2(r)\Lf}
-\frac{14\ell^4\a(r)\X1(\pz fr(r)\Y1)^2}{\Lf}
-\frac{4f^2(r)\a(r)\X1(\pz fr(r)\Y1)^2}{\Lf}\\
&-4\ov M\a(r)\X2(\pz\vF r(r)\Y2)^2
\eal
\end{equation}
\bea{nic2}
\pz{^2f}{r^2}(r)&=&2C_1r_0^2\a(r)+
\frac43\,r_0^2\a(r)f(r)\X2(\frac{\ef\nd}{n+2}-
\frac{\ef n}n\Y2) + r_0\,\pz fr(r)\nonumber\\
&+&\frac{\pz\a r(r)\pz fr(r)}{\a(r)}+\frac{(3\ell^4+f^2(r))\X1(\pz fr(r)\Y1)^2}
{(\ell^4+f^2(r))f(r)}\label{De380}\\
\pz{^2\vF}{r^2}(r)&=&\frac{r_0^2\a(r)f^2(r)}{\ov M(\ell^4+f^2(r))}
\X2(\frac{5n+4}{6n}\,\ef n-\frac56\,\ef\nd\Y2)\nonumber\\
&+&\frac{\nd\ell^4\X1(\pz fr(r)\Y1)^2}{2\ov M(\ell^4+f^2(r))^2}
+r_0\pz\vF r(r)+\frac{\pz \a r(r)\,\pz\vF r(r)}{\a(r)}
+\frac{2\ell^4\pz fr(r)\,\pz \vF r(r)}{f(r)(\ell^4+f^2(r))}.\label{De381}
\end{eqnarray}

In order to facilitate a future research we introduce $\a(r)=\ov\a(r)+1$. In
this way Eq.\ \er{Db377} changes to
$$
\bal
{}&\pz{^2\ov\a}{r^2}(r)=\frac1{3n\nd f(r)^2(\ell^4+f(r)^2)^2}\,r_0^2\\
&\ \q\X3(4(\ov\a(r)+1)^2f(r)\X1(-3n\nd C_1+2\ef n(n+2-\ef2 n)f(r)\Y1)
(\ell^4+f(r)^2)(3\ell^4+2f(r)^2)\\
&\ {}-\frac{3n\nd f(r)(\ell^4+f(r)^2)\pz{\ov \a} r(r)\X1(r_0f(r)(2\ell^4+3f(r)^2)
+2(5\ell^4+3f(r)^2)\pz fr(r)\Y1)}{r_0^2}\\
&\ {}+\frac{3n\nd (\ov\a(r)+1)}{r_0^2}\X2(-r_0^2f(r)^2(\ell^8+3\ell^4f(r)^2+2f(r)^4)
-r_0f(r)(11\ell^8+20\ell^4f(r)^2\\
&{}+8f(r)^4)\pz fr(r)
-2(9\ell^8+7\ell^4f(r)^2+2f(r)^4)\X2(\pz fr(r)\Y2)^2-4\ov Mf(r)^2(\ell^4
+f(r)^2)^2\X2(\pz\vF r(r)\Y2)^2\Y2)\Y3)
\eal
$$
\bg{Dd379}
\bal
{}&\pz{^2f}{r^2}(r)=\X2(2r_0^2(\ov\a(r)+1)^2f(r)\X1(3n\nd C_1+2\ef n(-2+(\ef 2-1)n)
f(r)\Y1)(\ell^4+f(r)^2)\\
&\ {}+3n\nd f(r)(\ell^4+f(r)^2)\pz {\ov \a} r(r)\pz fr(r)+3n\nd (\ov\a(r)+1)\pz fr(r)\\
&\ \q
\X2(r_0f(r)(\ell^4+f(r)^2)+(3\ell^4+f(r)^2)\pz fr(r)\Y2)\Y2)
\X1(3n\nd (\ov\a(r)+1)f(r)(\ell^4+f(r)^2)\Y1)^{-1}\\
&\pz{^2\vF}{r^2}(r)=\X1(6\ov Mn(\ov\a(r)+1)f(r)(\ell^4+f(r)^2)^2\Y1)^{-1}
\X3(\ef n(4-5(\ef2-1)n)r_0^2(\ov\a(r)+1)^2f(r)^3\\
&\ {}\cdot(\ell^4+f(r)^2)+6\ov Mnf(r)(\ell^4+f(r)^2)^2\pz{\ov \a} r(r)\pz\vF r(r)
+3n(\ov\a(r)+1)\X2(\ell^4\nd f(r)\X2(\pz \vF r(r)\Y2)^2\\
&\ {}+2\ov M(\ell^4+f(r)^2)
\X2(r_0f(r)(\ell^4+f(r)^2)+2\ell^4\pz fr(r)\Y2)\pz\vF r(r)\Y2)\Y3)
\eal
\end{gather}
The initial conditions \er{Db378} and \er{Db379} change into
\beq{Dd380}
\bal
\ov\a\X2(\frac{7.36\t10^{-10}}{r_0}\Y2)&=8\t10^{-12}\\
\hbox{or}\quad\ov\a\X2(\frac{7.36\t10^{-10}}{r_0}\Y2)&=14.94\t10^{-12}\\
\pz{\ov\a} r\X2(\frac{7.36\t10^{-10}}{r_0}\Y2)&=0\\
f\X2(\frac{7.36\t10^{-10}}{r_0}\Y2)&=0\\
\pz fr\X2(\frac{7.36\t10^{-10}}{r_0}\Y2)&=0\\
\vF\X2(\frac{7.36\t10^{-10}}{r_0}\Y2)&=-5.681\t10^{-3}\\
\pz \vF r\X2(\frac{7.36\t10^{-10}}{r_0}\Y2)&=0.
\eal
\end{equation}
and
\beq{Dd381}
\bal
\ov\a(643)&=8\t10^{-12}\\
\hbox{or}\q \ov\a(643)&=14.94\t10^{-12}\\
\pz{\ov\a} r(643)&=0\\
f(643)&=0\\
\pz fr(643)&=0\\
\vF(643)&=-5.681\t10^{-3}\\
\pz \vF r(643)&=0.
\eal
\end{equation}

$\ov\a(r)$ is better than $\a(r)$ from the numerical point of view.
Moreover, we still have arbitrary \ct s $C_1$, $\ov c$ and $\ell^2$. To tune
these \ct s we define two conditions for $\a$ and $\g$ at $r=R_\odot$ (a
radius of the Sun):
\bea{Dd382}
\ov\a(\ov R_\odot)&=&\frac{r_s}{R_\odot} \\
\g(R_\odot)&=&\X2(1+\frac{\ell^4}{R_\odot^4}\Y2)\X2(1-\frac{r_s}{R_\odot}\Y2)
\simeq 1+\frac{\ell^4}{R_\odot^4}-\frac{r_s}{R_\odot} \label{Dd383}
\end{eqnarray}
We suppose also that
\beq{Dd384}
\o(R_\odot)=\frac{\ell^2}{R_\odot^2}\,.
\end{equation}
Using Eq.\ \er{D.126} one gets
\beq{Dd385}
\frac{\ell^2}{R_\odot^2}\,e^{R_\odot}=f^2(R_\odot)e^{\ov c}.
\end{equation}
Moreover, from Eqs \er{Da255} and \er{Dd385} one easily gets using some
simplifications
\beq{Dd386}
f(R_\odot)=R_\odot^2+\frac{\ell^4}{2R_\odot^2}+r_sR_\odot\,.
\end{equation}
Thus one gets
\beq{Dd387}
\ov
c=\log\X3(\frac{\ell^2e^{R_\odot}}{R_\odot^6}\X2(1-\frac{\ell^4}{R_\odot^4}
+\frac{2r_s}{R_\odot}\Y2)\Y3) .
\end{equation}
In this way we can express $f(R_\odot)$ and $\ov c$ by $\ell^2$.
Moreover, we also have
$$
\ov\a(R_\odot)=\frac{r_s}{R_\odot}\,.
$$
The right hand side of Eq.\ \er{Dd379} depends on two arbitrary \ct s
$\ell^2$ and $C_1$. In this way we have not any arbitrary \ct s and the
initial value problem can be solved without any arbitrality (except a
possibility to fine tuning).

Taking
\beq{Dd388}
R_\odot=696\t10^3\,{\rm km}=0.1134\t10^{-2}r_0
\end{equation}
one gets
\bea{Dd389}
{}&&\ov \a(0.1134\t10^{-2})=4.310\t10^{-6}\\
&&\ov c=39.1+\log\X1(4.7078\ell^2(1-6.047\t10^{11}\ell^4-8.620\t10^{-6})\Y1)
\label{Dd390}\\
{}&&f(0.1134\t10^{-2})=0.0128\t10^{-4}+3.889\t10^5\ell^4+0.055\t10^{-10}
\label{Dd391}
\end{eqnarray}
or
\beq{Dd392}
f(0.1134\t10^{-2})=0.0128\t10^{-4}+3.889\t10^5\ell^4.
\end{equation}

All above calculations are of heuristic origin. They are based on the
following observation. An exact \so\ in NGT (also in Einstein Unified Field
Theory) has the shape (see Ref.~\cite9)
\beq{Dd393}
\mt{-\a(r) &\ & 0 &\ & 0 &\ & \frac{\ell^2}{r^2}\\
0 && -r^2 && 0 &&  0\\
0 && 0 && -r^2\sin^2\th && 0 \\
-\frac{\ell^2}{r^2} && 0 && 0 && \g(r)}
\end{equation}
in stationary and spherically \s\ case, where
\bea{Dd394}
{}&&\a(r)=\X2(1-\frac{r_s}r\Y2)^{-1}\\
&&\g(r)=\X2(1+\frac{\ell^4}{r^4}\Y2)\X2(1-\frac{r_s}r\Y2).\label{Dd395}
\end{eqnarray}
This \so\ for a sufficiently small $\ell^2$ describes very well a \gr al
field for the Sun. Thus it is natural to consider \er{Dd393} as a quite good
\ap ion for $r=R_\odot$. Due to \er{Dd393} at $r=R_\odot$ we can expect to
satisfy a Shapiro effect and perihelion movement of Mercury and Icarus. Any
fine tuning of initial conditions can also be applied.
We can substitute in Eq.\ \er{Dd379} $r_0=1.9897\t10^{-12}
$ and the result is
quoted in Appendix~E.

Let us consider Eqs \er{Dd379}. Supposing that $\ell^2$ is small and
$r_0\simeq 1.9897\t10^{-12}
$ we can simplify the right hand side of the \e s
getting
\bml{Dd396}
\pz{^2\ov \a}{r^2}(r)=-\frac2{\nd f^2(r)(\ell^4+f^2(r))^2}
\X3(2+\ov C_1(\ov\a(r)+1)^2f(r)\X2(3\ell^8\nd\\
{} +5\ell^4\nd f^2(r)+4f^4(r)\Y2)
+\nd f(r)(\ell^4+f^2(r))^2(5\ell^4+3f^2(r))\pz{\ov\a}r(r)\pz fr(r)\\
{}+\nd (\ov\a+1)\X2((9\ell^8+7\ell^4f^2(r)+2f^4(r))\X2(\pz fr(r)\Y2)^2\\
{}+2\ov M\X2(\pz fr(r)\Y2)^2(\ell^4+f^2(r))^2\X2(\pz\vF r(r)\Y2)^2\Y2)\Y3)
\end{multline}
\bea{Dd397}
\pz{^2f}{r^2}(r)&=&\frac{2\ov C_1(\ov \a(r)+1)(\ell^4\nd+2f^2(r))}
{\nd (\ell^4+f^2(r))}
+\frac{\pz{\ov\a}r(r)\cdot \pz fr(r)}{\ov\a(r)+1}
+\frac{(3\ell^4+f^2(r))\X1(\pz fr(r)\Y1)^2}{f(r)(\ell^4+f^2(r))}\hskip30pt\\
\pz{^2\vF}{r^2}(r)&=&\frac1{2\ov M(\ov\a(r)+1)f(r)(\ell^4+f^2(r))^2}
\X3(2\ov M f(r)(\ell^4+f^2(r))^2\pz{\ov \a}r(r)\,\pz{\vF}r(r)\nonumber\\
&+&\ell^4(\ov\a(r)+1)\pz fr(r)\X2(\nd f(r)\,\pz fr(r)+4\ov M(\ell^4+f^2(r))
\pz\vF r(r)\Y2)\Y3) \label{Dd398}
\end{eqnarray}
where
\beq{Dd399}
\ov C_1=C_1r_0^2
\end{equation}
is still an arbitrary \ct.

Now we suppose some additional supplementary conditions
\bea{Dd400}
\ov\a(R_\odot)&=&\frac{r_s}{R_\odot}\\
\g(R_\odot)&\simeq&1-\frac{r_s}{R_\odot}\label{Dd401}
\end{eqnarray}
($\ell$ is considered to be small).

From \er{Dd401} one gets
\beq{Dd402}
\exp(\ov c)=\frac{\ell^2 e^{R_\odot}}{f^3(R_\odot)}\,.
\end{equation}
We remind to the reader that $r$, $R_\odot$ and $r_s$ are measured in $r_0$
unit.

In this way we get field \e s \er{Dd396}--\er{Dd398}, initial conditions
\er{Dd381} and supplementary conditions
\bg{Dd403}
\ov\a(0.1134\t10^{-2})=4.310\t10^{-6}\\
\exp(\ov c)=\frac{1.00113\ell^2}{f^3(0.1134\t10^{-2})}\,.\label{Dd404}
\end{gather}
Thus the \ct\ $\ov c$ is expressed by $\ell^2$ and $f(R_\odot)$. Moreover,
the right hand sides of Eqs \er{Dd396}--\er{Dd398} depend on two \ct s
$\ell^2$  and $\ov C_1$. Conditions \er{Dd400} or \er{Dd403} can establish a
constraint on $\ell^2$ and~$C_1$. In this way one gets
\beq{Dd405}
\bal
\ov\a&=\ov\a(r,\ell^2,\ov C_1)\\
f&=f(r,\ell^2,\ov C_1)\\
\vF&=\vF(r,\ell^2,\ov C_1)
\eal
\end{equation}
in such a way that locally
\beq{Dd406}
\ov C_1=\ov C_1(\ell^2).
\end{equation}
Thus
\beq{Dd407}
\bal
\ov\a&=\ov\a(r,\ell^2)\\
f&=f(r,\ell^2)\\
\vF&=\vF(r,\ell^2).
\eal
\end{equation}
The \ct\ $\ell^2$ can be fixed  by some additional conditions. For a \ct\
$\ov c$ we have
\beq{Dd408}
\exp(\ov c)=\frac{1.00113\ell^2}{f^3(0.1134\t10^{-2},\ell^2)}\,.
\end{equation}

Let us consider \e s \er{Dd396}--\er{Dd398}. The integration \ct\ $\ell^2$ is
considered to be small. Let us neglect all terms with $\ell^2$ ($\ell^2$ is
small, but we do not consider it to be zero). One gets
\bea{niccc}
\pz{^2\ov\a}{r^2}(r)&=&-\frac
{2\X2(8\ov C_1(\ov\a(r)+1)^2f(r)+3\nd f(r) \pz{\ov\a}r(r)\pz fr(r)\Y2)}
{\nd f^2(r)}\nonumber\\
&-&\frac{2(\ov\a(r)+1)\X1(2(\pz fr(r))^2+2\ov Mf^2(r)(\pz\vF r(r))^2\Y1)}
{f^2(r)}\label{Dd409}\\
\pz{^2f}{r^2}(r)&=&\frac{4\ov C_1(\ov\a(r)+1)}{\nd}+\pz{\ov \a}r(r)
\pz fr(r)\,\frac1{(\ov\a(r)+1)}+\frac1{f(r)}\X2(\pz fr(r)\Y2)^2 \label{Dd410}\\
\pz{^2\vF}{r^2}(r)&=&\frac1{(\ov\a(r)+1)}\,\pz{\ov\a}r(r)\pz\vF r(r).
\label{Dd411}
\end{eqnarray}

Let us consider \er{Dd411}. One gets
\beq{Dd412}
\pz\vF r(r)=D(\ov\a(r)+1)
\end{equation}
where $D$ is an integration \ct.
In this way we get \e s
\bea{nicccc}
\pz{^2\ov\a}{r^2}(r)&=&-\frac
{2\X2(8\ov C_1(\ov\a(r)+1)^2f(r)+3\nd f(r) \pz{\ov\a}r(r)\pz fr(r)\Y2)}
{\nd f^2(r)}\nonumber\\
&-&\frac{2(\ov\a(r)+1)\X1(2(\pz fr(r))^2+2\ov Mf^2(r)D^2(\ov\a(r)+1)^2\Y1)}
{f^2(r)}\label{Dd413}\\
\pz{^2f}{r^2}(r)&=&\frac{4\ov C_1(\ov\a(r)+1)}{\nd}+\pz{\ov \a}r(r)
\pz fr(r)\,\frac1{(\ov\a(r)+1)}+\frac1{f(r)}\X2(\pz fr(r)\Y2)^2 \label{Dd414}\\
\vF(r)&=&D\int (\ov\a(r)+1)\,dr+\vF_0
\label{Dd415}
\end{eqnarray}
where $\vF_0$ is an integration \ct. If we suppose that
\beq{Dd416}
\vF(643)=\pz\vF r(643)\ne0
\end{equation}
one gets
\beq{Dd417}
D=0.
\end{equation}

Moreover, we are not forced to suppose initial conditions \er{Dd381} and we
can release
\beq{Dd418}
\pz\vF r(643)=0.
\end{equation}
Moreover, we can consider
\beq{Dd419}
D=0 \qh{and }\vF_0=-5.681\t10^{-3}.
\end{equation}
In this way Eqs \er{Dd413}--\er{Dd414} are simplified:
\bea{Dd420}
\pz{^2\ov\a}{r^2}(r)&=&-\frac
{2\X2(8\ov C_1(\ov\a(r)+1)^2f(r)+3\nd f(r) \pz{\ov\a}r(r)\pz fr(r)\Y2)}{\nd f^2(r)}
-4\frac{(\ov\a(r)+1)(\pz fr(r))^2}{f^2(r)}\hskip30pt\\
\pz{^2f}{r^2}(r)&=&\frac{4\ov C_1(\ov\a(r)+1)}{\nd}+\pz{\ov \a}r(r)
\pz fr(r)\,\frac1{(\ov\a(r)+1)}+\frac1{f(r)}\X2(\pz fr(r)\Y2)^2 \label{Dd421}
\end{eqnarray}
and
\beq{Dd422}
\vF(r)=\vF_0=-5.681\t10^{-3}.
\end{equation}

Let us consider $\wh{\ov \rho}(r)$. Using Eq.\ \er{D.160} one gets
\bml{Dd423}
\wh{\ov\rho}(r)=\frac{\ell^4 r_0^2\ef\nd}{f^4(r)}
\X3[3f^2(r)\X2(\pz fr(r)\Y2)^2-\frac{f^4(r)e^{2(\ov c-r)}}{\a(r)}
\X2(2\ov M+\frac{\ell^4 n^2}{f^2(r)}\Y2)\Y3]\\
-\frac23\,\ef{2\nd}\edn.
\end{multline}
Moreover, if we use the fact that $\vF(r)=0$ ($D=0$ and $\vF_0=0$) one finds
\beq{Dd424}
\wh{\ov \rho}(r)=\frac{r_0^2\ell^4}{f^2(r)}\X2(\pz fr(r)\Y2)^2
-\frac4{3n\nd}\,.
\end{equation}
For $n=120$ the last term in \er{Dd424} reads $9.107\t10^{-5}$.

Moreover, if we release an assumption $D=\vF_0=0$ one gets
\bml{Dd425}
\wh{\ov\rho}(r)=\frac{\ell^4 r_0^2\ef\nd}{f^2(r)}
\X3[3\X2(\pz fr(r)\Y2)^2-D^2\a(r)e^{2(\ov c-r)}\X1(\ell^4n^2
+2\ov Mf^2(r)\Y1)\Y3]\\
-\frac23\,\ef{2\nd}\edn
\end{multline}
where
$$
\vF(r)=D\int\a(r)+\vF_0.
$$

One can consider a system of differential \e s \er{Dd420}--\er{Dd421} with
initial conditions
\beq{De426}
\bal
\ov\a(643)&=14.94\t10^{-12}\\
\pz{\ov\a} r(643)&=0\\
f(643)&=10^{-7}\\
\pz fr(643)&=0
\eal
\end{equation}
with supplementary conditions
\bea{De427}
\ov\a(R_\odot)&=&\ov\a(0.1134\t10^{-2})=\frac{r_s}{R_\odot}=4.310\t10^{-6}\\
\g(R_\odot)&=&\g(0.1134\t10^{-2})=1-\frac{r_s}{R_\odot}=1-4.310\t10^{-6}.
\label{De428}
\end{eqnarray}
In this way we can satisfy Eqs \er{Dd420}--\er{Dd421}, \er{De426}--\er{De427}
in the following numerical programme written in Mathematica~7.

\medskip
{\parindent 0pt
\tt eq1 = D[a[x, t], x, x] == -(122 * f[x, t])\^{ }-1 * 2 * $\X1($
8 * t * (a[x, t] + 1)$\tt^2$

\ \ \ \ + 3 * 122 * D[a[x, t], x] * D[f[x, t], x]

\ \ \ \ + 2 * 122 * (a[x, t] + 1) * D[f[x, t], x]$\Y1)$

\vskip3pt
eq2 = D[f[x, t], x, x] == $\tt \dfrac{\hbox{4 * t * (a[x, t] + 1)}}{122}$

\ \ \ \ + $\tt\dfrac{\hbox{d[a[x, t], x] * D[f[x, t], x]}}{\hbox{(a[x, t] + 1)}}
\ +\ \dfrac{\hbox{D[f[x, t], x]}^2}{f[x,\ t]}$

\vskip3pt
eq3 = (a[x, t] /.\ x $\to$ 643.) == 14.94 * 10\^{ }-12

eq4 = (D[a[x, t], x] /.\ x $\to$ 643.) == 0

eq5 = (f[x, t] /.\ x $\to$ 643.) == 0.0000001

eq6 = (D[f[x, t], x] /.\ x $\to$ 643.) ==0.

so1 = NDSolve[\char123eq1, eq2, eq3, eq4, eq5, eq6\char125, \char123a,
f\char125,  \char123x, 0.00001, 1000.\char125,

\ \ \ \ \char123t, -500., 500.\char125,
MaxSteps $\to$ 50000, PrecisionGoal $\to$ 40]

FindRoot[Evaluate[(a[x, t] /.\ x $\to$ 1.134 * 10\^{ }-3) - 4.3103
* 10\^{ }-6 /.\ so1],

\ \ \ \ \char123t, 1\char125]

}\medskip
Here {\tt x} corresponds to $r$, {\tt a} to $\ov\a$, {\tt t} to $\ov C_1$, {\tt f} to
$f$.

\medskip

One gets for $\ov C_1$ ({\tt t} in the programme)
\beq{De429}
\ov C_1=8.8178 \t 10^{-16}.
\end{equation}
Using this value of $\ov C_1$ we get \e s for $\ov\a(r)$ and $f(r)$ solving
them and getting the following results:

On Fig.\ \ref{model1} we give 3D plots of $f(r,\ov C_1)$ and $\ov\a(r,\ov
C_1)$ for some values of $r$ and $\ov C_1$.

\refstepcounter{figure}\label{model1}
\hbox to \textwidth{\ing[width=0.45\textwidth]{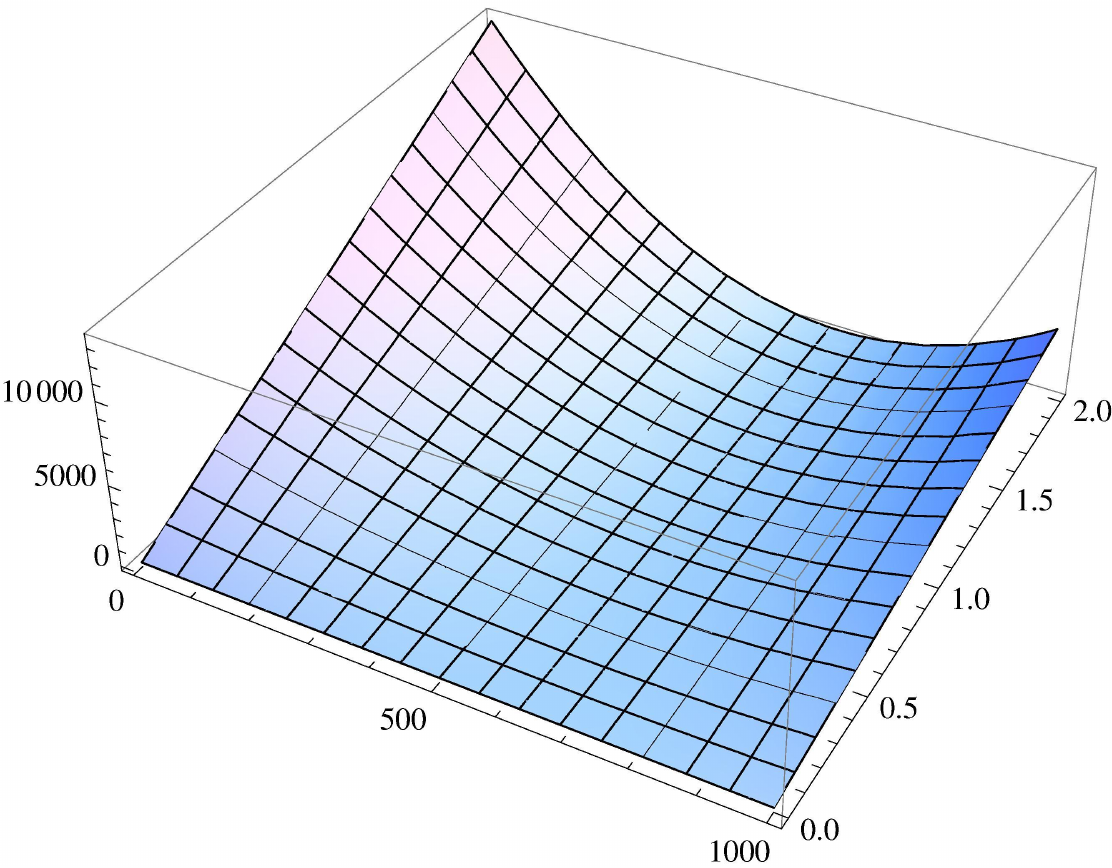}\hfil
\ing[width=0.45\textwidth]{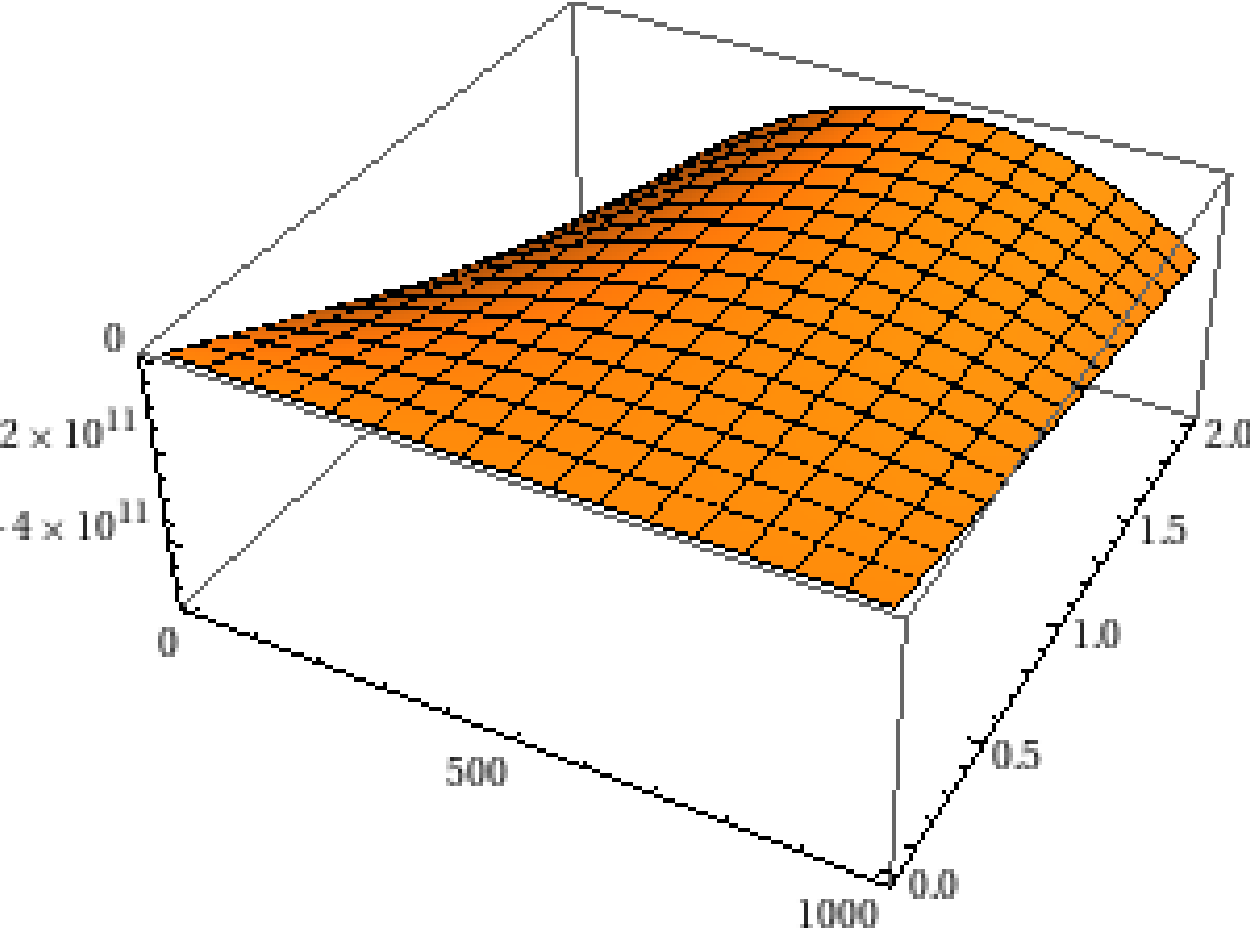}}
\hbox to \textwidth{\hbox to 0.45\textwidth{\hfil(A)\hfil}\hfil
\hbox to 0.45\textwidth{\hfil(B)\hfil}}
\smallskip
\noindent {\small Figure \thefigure:
(A)---a 3D plot of $f(r,\ov C_1)$ for $0.01\le r\le 1000$, $0\le C_1\le 2$,
(B)---a 3D plot of  $\ov\a(r,\ov C_1)$ for $0.01\le r\le 1000$, $0\le C_1\le 2$.}

\bigskip
\refstepcounter{figure}\label{model3}
\hbox to \textwidth{\ing[width=0.45\textwidth]{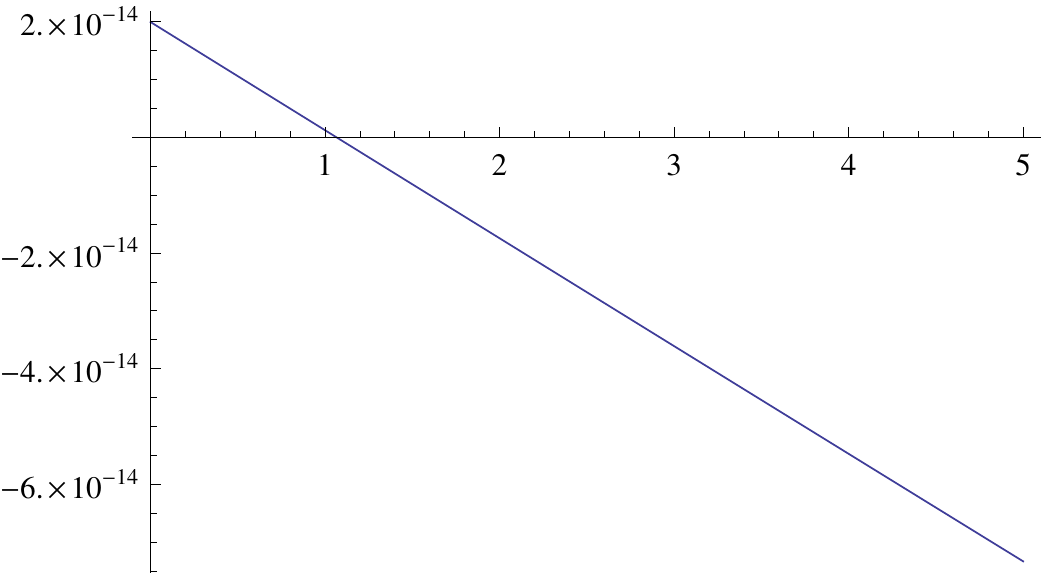}\hfil
\ing[width=0.45\textwidth]{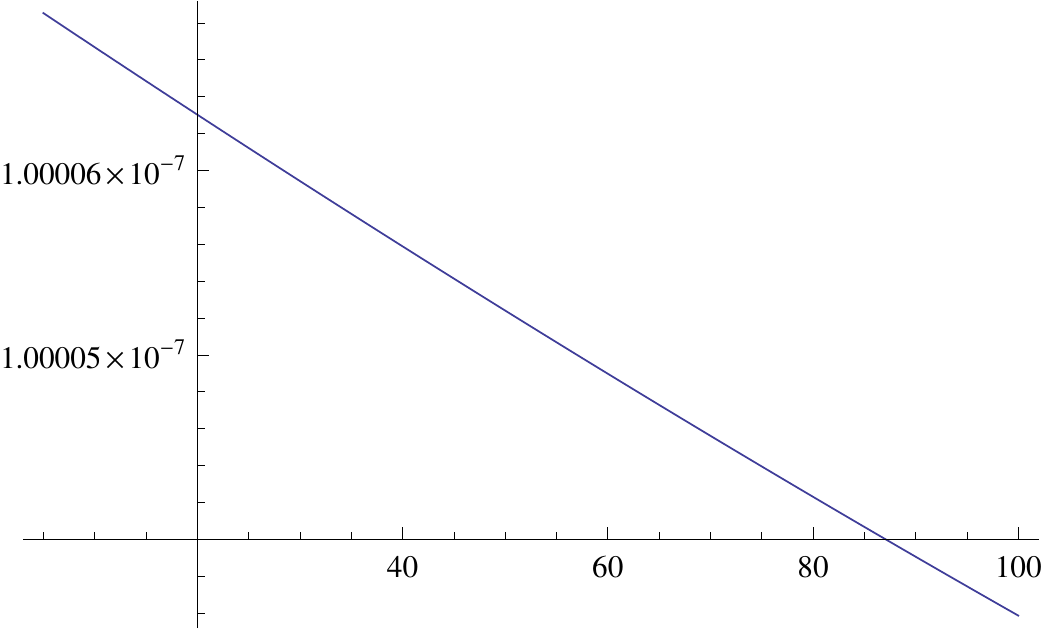}}
\hbox to \textwidth{\hbox to 0.45\textwidth{\hfil(A)\hfil}\hfil
\hbox to 0.45\textwidth{\hfil(B)\hfil}}
\hbox to \textwidth{\ing[width=0.45\textwidth]{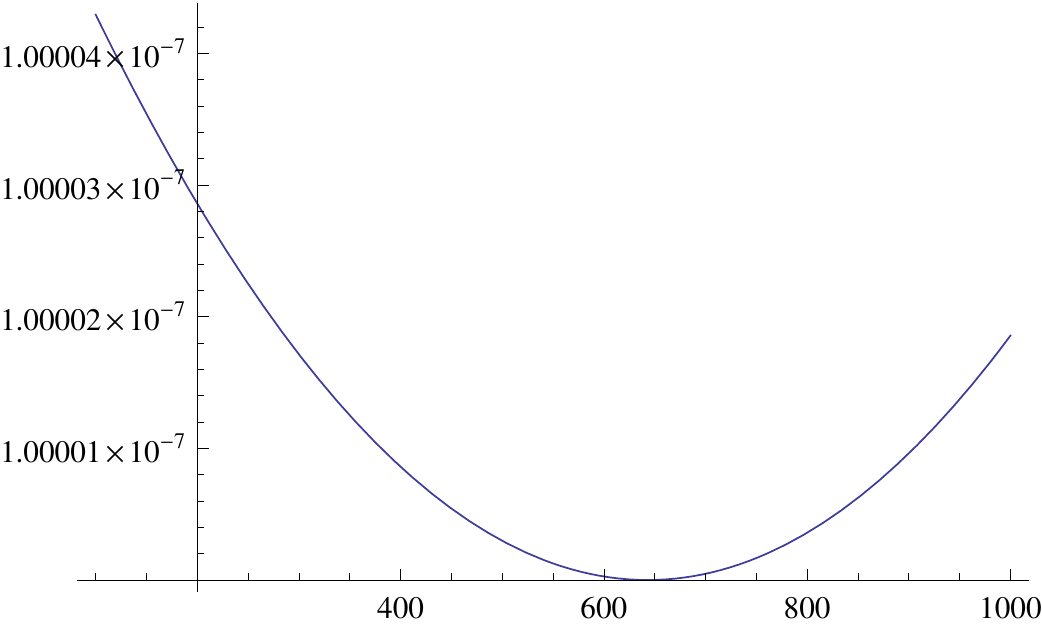}\hfil
\ing[width=0.45\textwidth]{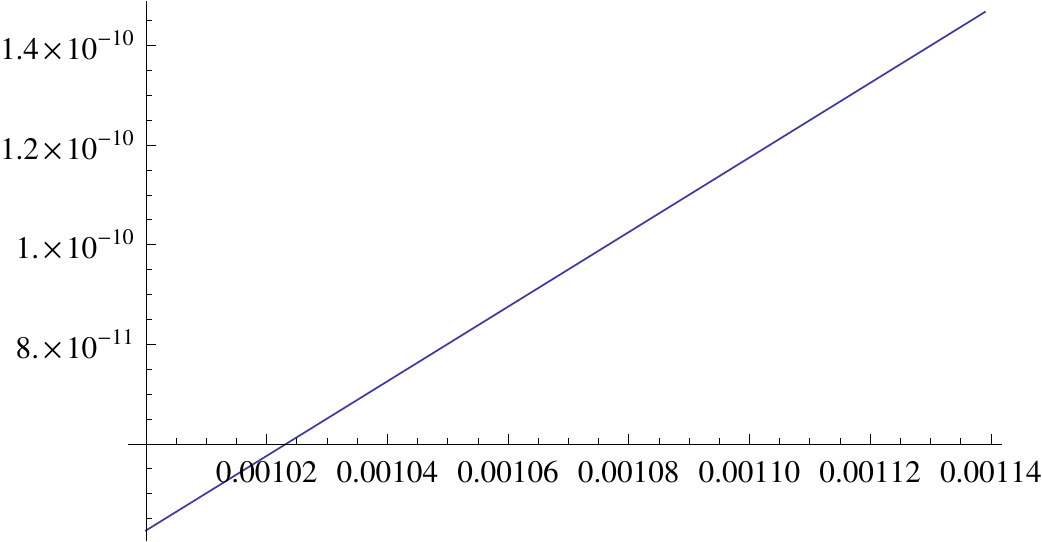}}
\hbox to \textwidth{\hbox to 0.45\textwidth{\hfil(C)\hfil}\hfil
\hbox to 0.45\textwidth{\hfil(D)\hfil}}
\smallskip
\noindent {\small Figure \thefigure:
(A)---a plot of $f(r)-1.00006\t10^{-7}$ for $0.001<r<5$;
(B)---a plot of  $f(r)$ for $5<r<100$;
(C)---a plot of  $f(r)$ for $100<r<1000$;
(D)---a plot of  $\ov\a(r)+0.000240797$ for $0.001<r<0.01139$.}

\bigskip

\hbox to \textwidth{\ing[width=0.45\textwidth]{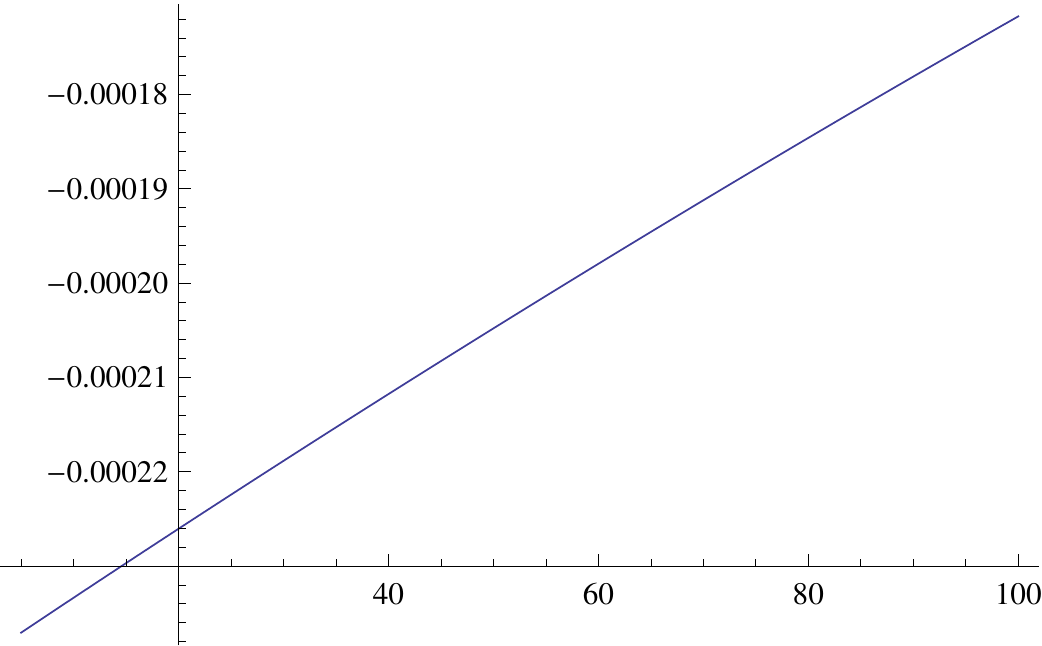}\hfil
\ing[width=0.45\textwidth]{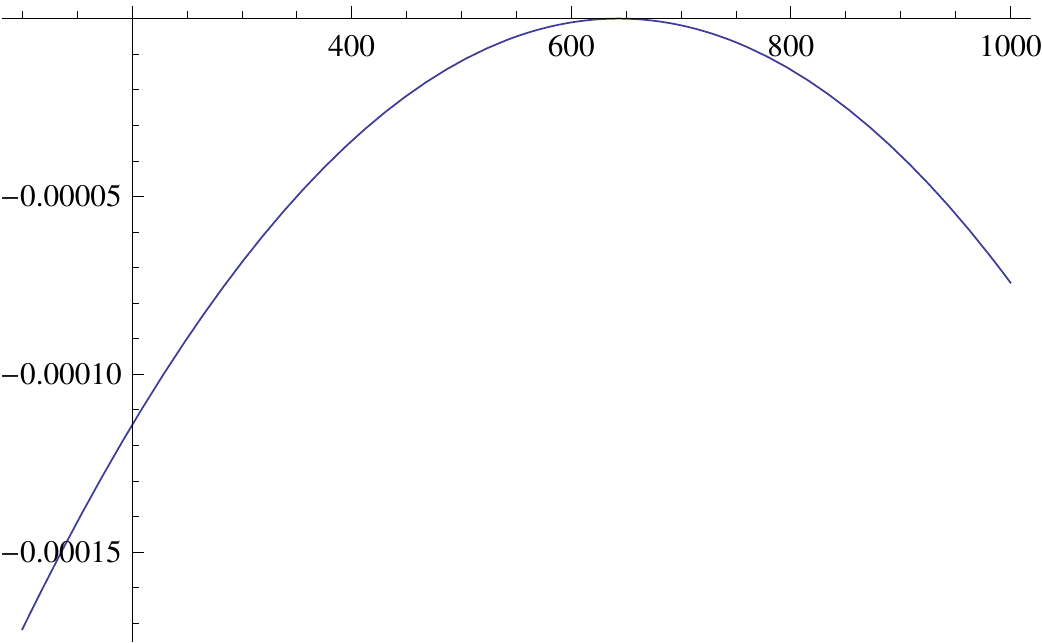}}
\hbox to \textwidth{\hbox to 0.45\textwidth{\hfil(E)\hfil}\hfil
\hbox to 0.45\textwidth{\hfil(F)\hfil}}
\hbox to \textwidth{\ing[width=0.45\textwidth]{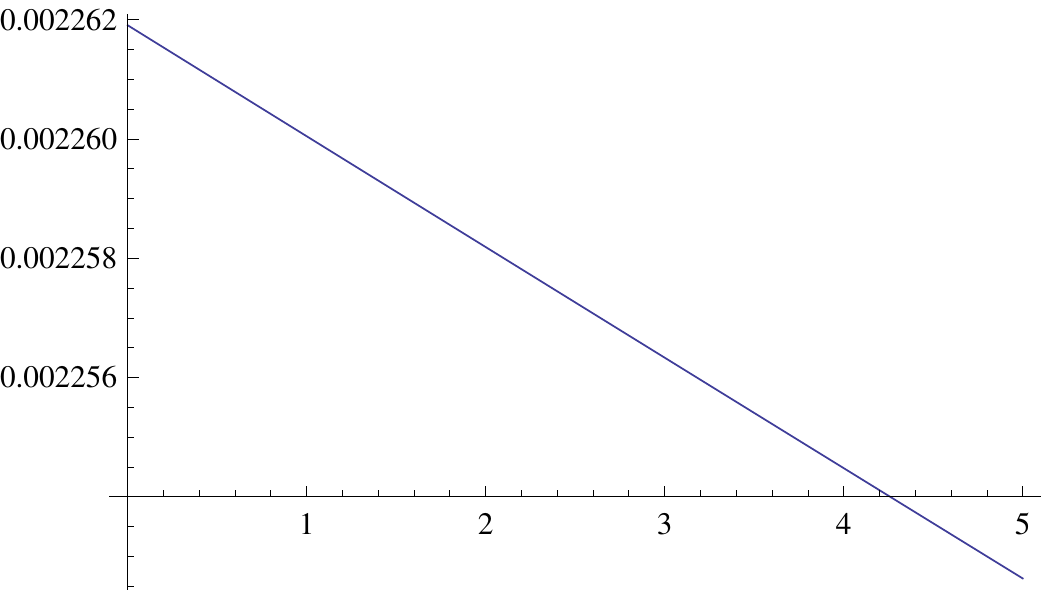}\hfil
\ing[width=0.45\textwidth]{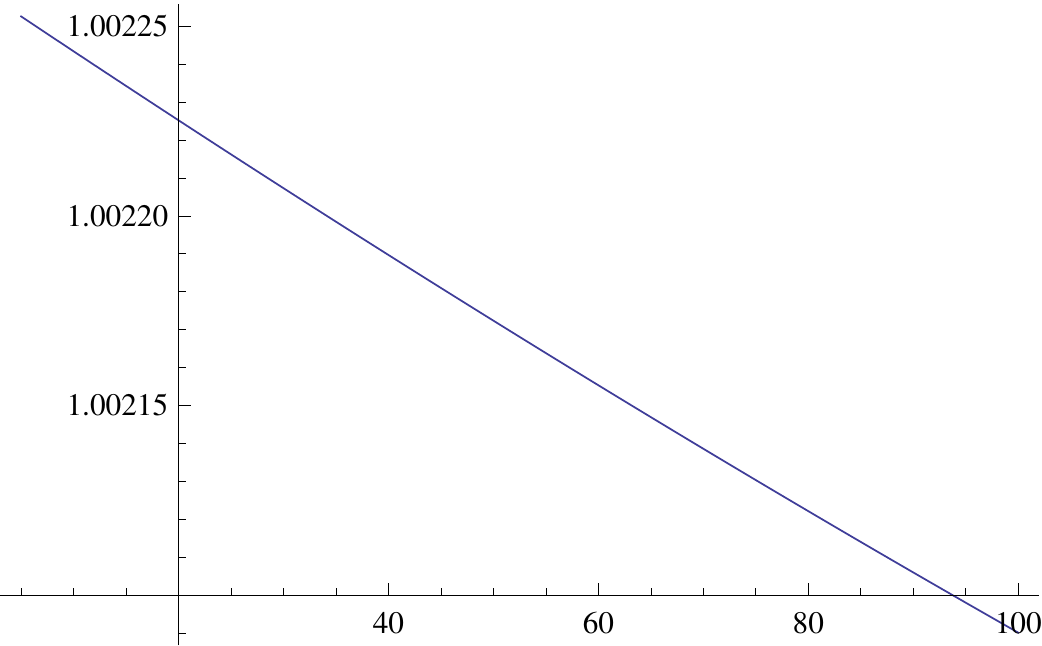}}
\hbox to \textwidth{\hbox to 0.45\textwidth{\hfil(G)\hfil}\hfil
\hbox to 0.45\textwidth{\hfil(H)\hfil}}
\hbox to \textwidth{\hfil \ing[width=0.45\textwidth]{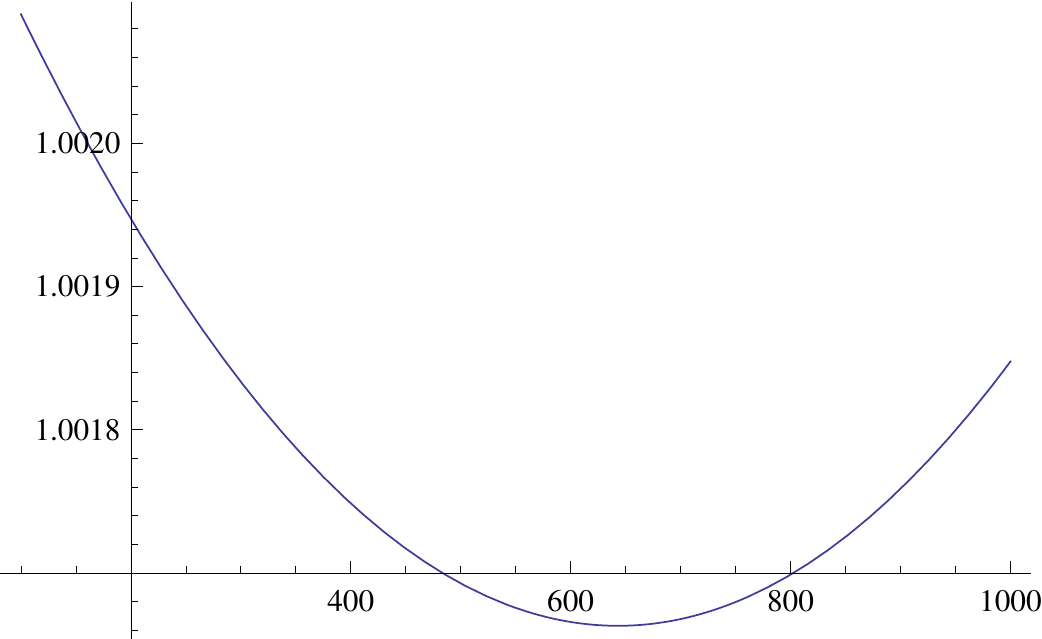}\hfil}
\hbox to \textwidth{\hfil\hbox to 0.45\textwidth{\hfil(I)\hfil}\hfil}
\smallskip
{\small \noindent Figure \thefigure\ (cont.):
(E)---a plot of  $\ov\a(r)$ for $5<r<100$;
(F)---a plot of  $\ov\a(r)$ for $100<r<1000$;
(G)---a plot of  $\g(r)-1$ for $0.001<r<5$;
(H)---a plot of  $\g(r)$ for $5<r<100$;
(I)---a plot of  $\g(r)$ for $100<r<1000$.}

\bigskip

On Fig.\ \ref{model3} we give
plots of $f(r)$, $\ov\a(r)$ and $\g(r)$ for several values of~$r$.

$\g(r)$ is defined according to the formula \er{D.167} ($\frac{r_0}L\simeq
2\t10^{-12}$)
\beq{De430}
\g(r)=\g_0\,\frac{f^6(r)}{\a(r)}\,e^{-2r_0r/L}\simeq\g_0\,\frac{f^6(r)}{\a(r)}
\end{equation}
where
\beq{De431}
\g_0=\frac{e^{2\ov c}}{\ell^4}={\rm const.}
\end{equation}
Using a supplementary condition \er{De428} one can estimate $\g_0$
\beq{De432}
\g_0\simeq 1.0016630608\t 10^{42}.
\end{equation}
Moreover, from these numerical calculations we know only an order of this
\ct. We use this value to plot $\g(r)$ on the Fig.~\ref{model3}.
The \f\ $\o(r)$ (see Eq.\ \er{D.159}) can be rewritten in the following form:
$$
\o(r)=\frac{\ell^2}{k(r)}\,, \qh{where} k(r)=\frac{e^{r_0r/L}}{f^2(r)
\sqrt{\g_0}}\simeq\frac{10^{-21}}{f^2(r)}\,.
$$
Further
calculations demand more precision in numerical calculations and will be done
elsewhere.

The problem of a fine tuning of initial conditions is not the only problem
which we face here. Let us notice that we have two \ct s calculated
numerically: $\ov C_1$ and~$\g_0$. $\ov C_1$~is of order $10^{-16}$ and
$\g_0$ of order $10^{42}$. This means that $\g_0$ should be calculated at
least up to accuracy of $\ov C_1$, i.e.\ $10^{-16}$. From numerical point of
view this is a hard problem. Moreover, $\ov C_1$ should be also calculated
very precisely. It means, up to $10^{-30}$. This results in a higher
precision of the \ct\ $\g_0$, i.e.\ $10^{-30}$. Thus a very large number
($\sim 10^{42}$) should be known with such a precision. From numerical point
of view this is really hard to achieve.

We cannot define an initial condition for $f(r)$ at~$643$, $f(643)=0$,
because we need an infinite accuracy to solve \e s. This is impossible and we
change zero to a sufficiently small number, i.e.\ $10^{-7}$, which is
reasonable from the physical point of view. However, the accuracy to solve \e
s in a sufficiently large region of $r$ demands still high precision of
calculations. A~condition $\pz fr(643)=0$ does not cause such problems. If we
consider more complex \e s \er{Db377}, \er{Dd379} or \er{De379}--\er{De381},
we face the same problems with accuracy and with $f(643)=0$. Moreover,
according to Eqs \er{De379}--\er{De381} we have also additional \ct s (except
$\ov C_1=r_0^2C_1$): $\ell^8 r_0^2$, $\ell^4 r_0^2$, $\ell^4r_0$,
$\ell^8r_0$, $\ell^8$, $\ell^4$. $r_0$ is known up to a factor $\ov \xi$ (of
order~$1$). From practical point of view it is more convenient to define new
set of \ct s: $\ov C_2=r_0^2\ell^4$, $\ov C_3=r_0\ell^4$, $\ell^8r_0^2
=\ell^4\ov C_2$, $\ell^8r_0=\ell^4\ov C_3$ in such a way that
$r_0=\frac{\,\ov C_3\,}{\ell^4}$.

In this way we shift \ct s to $\ov C_1,\ov C_2,\ov C_3$ and $\ell^2$.
$\ell^2$ is an integration \ct, $\ov C_1=r_0\ov C_1$ is also an integration
\ct. Moreover, $r_0^2\ell^8=\ov C_2=\ov C{}_3^2$ and our expansion of the
left hand side of \e s \er{De379}--\er{De381} can be reparametrized to an
expansion \wrt $\ov C_1,\ell^4,\ell^8,\ov C_3,\ov C{}_3^2$. From this point
of view the problem to get $\ov \a(R_\odot)=\frac{r_s}{R_\odot}$ is more
involved. Moreover, we should be very careful with numerical calculations,
taking under consideration that $\ell^2$  is arbitrary (considered as small),
$\ov C_1$~is absolutely arbitrary, $\ov C_3$ is small with an arbitrariness
of order~$1$ ($\ov\xi$~in a definition of $\ov C_3$ and~$r_0$). Thus it is a
problem with two arbitrary \ct s under some boundary condition. We meet also
a problem with $\g_0=\frac{e^{\ov c}}{\ell^4}$, even $\ov c$ is completely
arbitrary.

Let us consider a different parametrization of $\g(r)$ and $\a(r)$, i.e.
\bg{Db380}
\g(r)=\X2(1+\frac{\ell^4}{r^4}\Y2)\X2(1-\frac{r_s}r-2W_1(r)-\frac{\La
r^2}3\Y2) \\
\a(r)=\frac1{1-\frac{r_s}r-2W_2(r)-\frac{\La r^2}3} \label{Db381}
\end{gather}
where
\beq{Db382}
\La=\la_{\rm co}=10^{-52}{\rm\frac1{m^2}}=3.7\t10^{-29}\frac1{r_0^2}
\end{equation}
is a \co ical \ct. In this way
\beq{Db383}
W_1(r)=\frac12\X2(1-\frac{r_s}r-\frac{\g(r)}{1+\frac{\ell^4}{r^4}}-
\frac{\La r^2}3\Y2)
\end{equation}
or
\bg{Db384}
W_1(r)=\frac12\X2(1-\frac{r_s}r-\frac{f^6(r)e^{2(\ov c-r)}}
{\ell^4\a(r)(1+\frac{\ell^4}{r^4})}-\frac{\La r^2}3\Y2)\\
W_2(r)=\frac12\X2(1-\frac1{\a(r)}-\frac{r_s}r-\frac{\La r^2}3\Y2).\label{Db385}
\end{gather}
The term $\frac{\La r^2}3$ is very small and in the \SS\ is negligible. Let us
estimate its influence in comparison to $\frac{r_s}r$ and $\ov V_2(r)
\simeq W_2(r)$. One gets at a distance $r=10^5r_0$
\bea{Db386}
p_1&=&\frac{\,\frac{\La r^2}3}{\frac{r_s}r}\simeq 0.2\t10^{-5}\\
p_2&=&\frac{\,\frac{\La r^2}3}{2\ov V_2(r)}\simeq 0.6\t10^{-4}.\label{Db387}
\end{eqnarray}
Moreover, in Eqs \er{Db377} all the fields, i.e.\ $\a(r)$, $f(r)$, $\vF(r)$ are
coupled in a nontrivial, nonlinear way. Eqs \er{Db377} are not scale
invariant as Eqs \er{D.15}--\er{D.17}. Thus we can expect some nontrivial
effects due to \co ical \ct. We remind to the reader that the field $\vF$
enters \gr al ``\ct'' and \co ical term. In this way $\vF$ is a reason of the
\co ical \ct\ (see Ref.~\cite{10}). We get similarly as before (i.e.\
$\La=0$) for $\ov U_1(r)$ and $\ov U_2(r)$
\bml{Db388}
\X2(\pz u\vf\Y2)^2+u^2=\X2(\frac1{h^2}\X2(\frac{H^2}{c^2}-c^2\Y2)\Y2)
+r_su^3+\frac{c^2r_s}{h^2}\,u- \frac{H^2\ell^4u^4}{h^2c^2(1+u^4\ell^4)}
\\ {}-\frac{H^2}{h^2c^2}\cdot \frac{\D W(\frac1u)}{(1-r_su-W_1(\frac1u)
+\frac1{3u^2})(1+u^4\ell^4)}
+W_2\X2(\frac1u\Y2)\X2(u^2+\frac {c^2}{h^2}\Y2)+\frac{\La c^2}{3h^2u^2}\,.
\end{multline}
Differentiating both sides of Eq.\ \er{Db388} \wrt $\vf$ one gets
\bml{Db389}
\pz{^2u}{\vf^2}+u=\frac{c^2r_s}{h^2}+\frac{3r_su^2}2
- \frac{2H^2\ell^4u^3}{h^2c^2(1+u^4\ell^4)^2}-\frac{\La c^2}{3h^2u}
+uW_2\X2(\frac1u\Y2)\\ {}-\frac12\,\pz{W_2}r\X2(\frac1u\Y2)\X2(1+\frac{c^2}{u^2h^2}
\Y2)- \frac{H^2}{2h^2c^2u^2}\,
\frac{\pz{}r \D W(\frac1u)}{(1-r_su-W_1(\frac1u)-\frac\La{3u^2})
(1+u^4\ell^4)}\\ {} + \frac{H^2}{2h^2c^2}\, \frac{\D W(\frac1u)(r_s+\frac1{u^2}
\,\pz{W_1}r(\frac1u)-\frac{2\La}{3u^2})}
{(1-r_su-W_1(\frac1u)-\frac\La{3u^2})^2(1+u^4\ell^4)}\\
{}-\frac{2H^2\ell^4}{h^2c^2}\,\frac{\D W(\frac1u)u^3}
{(1-r_su-W_1(\frac1u)-\frac\La{3u^2})(1+u^4\ell^4)^2}
\end{multline}
Supposing that $H\simeq c^2$ and using some simplifications one gets
\beq{Db390}
\pz{^2u}{\vf^2}+u = \frac{c^2r_s}{2h^2}+\frac{3r_su}2 - \frac{\La c^2}
{3h^2u^3}-\frac{2c^2\ell^4u^3}{h^2} -\frac{c^2}{2h^2u^2}
\,\wt b_1\X2(\frac1u\Y2)
\end{equation}
where
\bg{Db391}
\D W(r)=W_1(r)-W_2(r)\\
\wt b_1(r)=\pz{W_1(r)}r\,.\label{Db392}
\end{gather}

In the case of a \hy\ orbit one gets
\beq{Db393}
\pz{^2u}{\vf^2}+u = \frac{1}{a_h(e^2-1)}+\frac{3k^2u^2}{c^2}
-\frac{2\ell^2u^3c^2}{k^2a_h(e^2-1)}-\frac{\La c^2}{3h^2u^3}
-\frac{c^2}{2k^2a_h(e^2-1)}\,\wt b_1\X2(\frac1u\Y2).
\end{equation}
The formula above gives us a distortion of a \hy\ orbit.

In the case of an elliptic orbit one gets
\beq{Db394}
u=u_0+\D u_{\rm GR}+\D u_{\rm NGT}+\D u_\La +\D u.
\end{equation}
The formula \er{Db394} gives us an additional \an\ \ph\ movement and
distortion of an elliptic orbit, where $\D u_{\rm GR}$ and $\D u_{\rm NGT}$
are corrections considered above,
\beq{Db395}
\pz{^2\D u_\La}{\vf^2}+\D u_\La = -\frac{\La c^2a_e^2(1-e^2)^2}{2k^2}
\,\frac1{(1+e\cos\vf)^3}\,.
\end{equation}
This formula gives us an \an\ \ph\ movement and distortion of an elliptic
orbit due to a \co ical \ct.
\beq{Db396}
\pz{^2\D u}{\vf^2}+\D u=-\frac{c^2a_e(1-e^2)}{2k^2}
\,\wt b_1\X2(\frac{a_e(1-e^2)}{1+e\cos\vf}\Y2)\frac1{(1+e\cos\vf)^3}\,.
\end{equation}
This formula gives us an \an\ \ph\ movement and distortion of an elliptic
orbit for an \an\ \ac\ $\wt b_1(r)$.

We consider a bending of light getting
\bml{Db397}
\D\vf=\D\vf_{\rm NGT,\La}\\{}+2\X3(\frac b{\,\ov{\ov R}\,}+
\int_{r_1}^{\ov {\ov R}} \frac{W_1(r)+W_2(r)-\frac{2\La}{3r^2}}r
\X2(\frac{r^2}{b^2(1+\frac{\ell^4}{r^4})(1-\frac{r_s}r-\frac{\La r^2}3)}
-1\Y2)^{-1/2}\,dr\Y3)
\end{multline}
where $\D\vf_{\rm NGT,\La}$ is a formula for a bending of light in NGT with a
\co ical \ct\ taken into account,
\bml{Db398}
\D\vf_{\rm NGT,\La}= 2\int_{r_1}^{\ov{\ov R}}
\frac1r\X2(1-\frac{r_s}r-2W_1(r)-\frac{\La r^2}3
\Y2)^{1/2}\\{}\X2(\frac1{b^2}\,
\frac{r^2}{(1+\frac{\ell^4}{r^4})(1-\frac{r_s}r-2W_1(r)-\frac{\La r^2}3)}
-1\Y2)^{-1/2}\,dr
\end{multline}

The relevant formula for Shapiro effect reads
\beq{Db399}
t(r,\ov r_0)=t_{{\rm NGT},\La(r,\ov r_0)}+\wt\d t(r,\ov r_0)
\end{equation}
where
\bea{Db400}
t_{{\rm NGT},\La(r,\ov r_0)}&=&\frac1c\int_{\ov r_0}^r\X2(1+\frac{\ell^4}{r^4}
\Y2)^{1/2}\X2(1-\frac{r_s}r-\frac{\La r^2}3\Y2)\nonumber\\
&&\X2(1-\frac{b^2}{r^2}
\X2(1-\frac{r_s}r-\frac{\La r^2}3\Y2)\X2(1+\frac{\ell^4}{r^4}\Y2)\Y2)^{-1/2}
\,dr\\
\wt\d t(r,\ov r_0)&=&\frac1c\int_{\ov r_0}^r\X2(1+\frac{\ell^4}{r^4}
\Y2)\X2(W_1(r)+W_2(r)+\frac{2\La r^2}3\Y2)\nonumber\\
&&\X2(1-\frac{b^2}{r^2}
\X2(1-\frac{r_s}r-\frac{\La r^2}3\Y2)\X2(1+\frac{\ell^4}{r^4}\Y2)\Y2)^{-1/2}
\,dr.\label{Db401}
\end{eqnarray}

Even an influence of the \co ical \ct\ is negligible inside the \SS, it could
have some important significance on a border of our model, i.e., for
$r>10^9r_0=4.103\t10^9$ \,AU, i.e.\ in a deep space region. In this case
\bea{Db402}
p_1 &\ge& 0.2\t 10^7\\
p_2 &\ge& 0.6\t 10^3. \label{Db403}
\end{eqnarray}
Thus an influence of the \co ical \ct\ could have a significant influence on
a movement of a \sp\ for $r\ge10^9r_0$.

Let us notice the following fact. If we put $\ell=0$ in the formulae
\er{Db388}, \er{Db389}, \er{Db390}, \er{Db393}, \er{Db397},
\er{Db398}, \er{Db399}, \er{Db400}, \er{Db401}, we can consider these \e s
as in Riemannian \s\ case $g_\m=g_{\nu\mu}$ in the following parametrization
\bea{Db404}
e^{2A(r)}&=&1-\frac{r_s}r-2\ov{\ov V}_1(r)-\frac{\La r^2}3 \\
e^{2B(r)}&=&\frac1{1-\frac{r_s}r-2\ov{\ov V}_2(r)-\frac{\La r^2}3} \label{Db405}
\end{eqnarray}
putting $\ov {\ov V}_i(r)$ in the place of $\ov W_i(r)$, $i=1,2$, in such a way that
\bea{Db406}
\ov{\ov V}_1(r)&=&\frac12 \X2(1-\frac{r_s}r-e^{2A(r)}-\frac{\La r^2}3\Y2)\\
\ov{\ov V}_2(r)&=&\frac12 \X2(1-\frac{r_s}r-e^{-2B(r)}-\frac{\La r^2}3\Y2).
\label{Db407}
\end{eqnarray}

\def\oW{\ov{\ov W}}
In order to facilitate a future research in the \nos\ case we should consider
a different parametrization of $\a(r)$ and $\g(r)$:
\bea{Dc407}
\a(r)&=&\X2(1-\frac{r_s}r\,\ef{-\nd}-\frac{\La r^2}3 - 2\oW_2(r)\Y2)^{-1}\\
\g(r)&=&\X2(1+\frac{\ell^4}{r^4}\Y2)\X2(1-\frac{r_s}r\,\ef{-\nd}
-\frac{\La r^2}3 - 2\oW_2(r)\Y2). \label{Dc408}
\end{eqnarray}
It means
\bea{Dc409}
\oW_1(r)&=&\frac12\X2(1-\frac{\g(r)}{1+\frac{\ell^4}{r^4}}
-\frac{r_s}r\,\ef{-\nd}-\frac{\La r^2}3\Y2)\\
\oW_2(r)&=&\frac12\X2(1-\frac1{\a(r)}-\frac{r_s}r\,\ef{-\nd}-\frac{\La r^2}3
\Y2) \label{Dc410}
\end{eqnarray}
or
\beq{Dc411}
\oW_1(r)=\frac12\X3(1-\frac{f^6(r)e^{2(\ov c-r)}}{(1+\frac{\ell^4}{r^4})
\ell^4\a(r)}-\frac{r_s}r\,\ef{-\nd}-\frac{\La r^2}3\Y3).
\end{equation}
In this way we take under consideration an inconstancy of a \gr al ``\ct''
and an influence of a \co ical \ct. In the case of the full \nos\ theory it
is reasonable for the scalar field entering both ``\ct s'' is coupled in a
nontrivial way to $\a,\g$ and~$f$.

We proceed as before considering \e s of motion of massive test particles.
One gets
\bml{Dc412}
\X2(\frac h{r^2}\pz r \vf\Y2)^2 + \frac h{r^2}=
\X2(\frac{H^2}{c^2} - c^2 + \frac{h^2\La}3 \Y2)\\
{}+\frac{H^2}{c^2}\,
\frac{2(\oW_1(r)-\oW_2(r))-\frac{\ell^4}{r^4}(1-\frac{r_s}r\,\ef{-\nd}
-\frac{\La r^2}3 - 2\oW_1(r))}{(1+\frac{\ell^4}{r^4})(1-\frac{r_s}r\,\ef{-\nd}
-\frac{\La r^2}3 - 2\oW_1(r))}\\
{}+\frac{h^2r_s}{r^3}\,\ef{-\nd}+\frac{2\oW_2(r)h^2}{r^2}+\frac{r_sc^2}r
\,\ef{-\nd}+\frac{\La c^2r^2}3 + 2c^2\oW(r).
\end{multline}
Taking as before $u=\frac1r$ and afterwards differentiating both sides of the
\e\ \wrt $\vf$ one gets
\bml{Dc413}
\pz{^2u}{\vf^2}+u = -\frac
{H^2\,\frac1{u^2}\,\pz{\oW}r(\frac1u)}
{c^2h^2(1+\ell^4u^4)\X1(1-r_s u \exp(-\nd \vF(\frac1u))-\frac\La{3u^2}
-2\oW_1(\frac1u)\Y1)}\\
{}-\frac{4H^2\ell^4u^3\oW(\frac1u)}
{c^2h^2(1+\ell^4u^4)^2\X1(1-r_s u \exp(-\nd \vF(\frac1u))-\frac\La{3u^2}
-2\oW_1(\frac1u)\Y1)}\\
{}+\frac{H^2\oW(\frac1u)\X1(r_su\exp({-\nd}\vF(\frac1u))-\frac{r_s}u
\nd \pz\vF r(\frac1u)\exp(-\nd\vF(\frac1u))+\frac{2\La}{3u^3}
-\frac2{u^2}\,\pz{\oW}r(\frac1u)\Y1)}
{c^2h^2(1+\ell^4u^4)\X1(1-r_s u \exp(-\nd \vF(\frac1u))-\frac\La{3u^2}
-2\oW_1(\frac1u)\Y1)^2}\\
{}-\frac{2H^2\ell^4u^3}{h^2c^2(1+\ell^4u^4)}
+\frac{3r_su^2}3\exp\X2(-\nd\vF\X2(\frac1u\Y2)\Y2)
+\frac{r_su}2\,\nd\,\pz\vF r\X2(\frac1u\Y2)\exp\X2(-\nd\vF\X2(\frac1u\Y2)\Y2)\\
{}-\pz{\oW_2}r\X2(\frac1u\Y2)+\frac{r_sc^2}{2h^2}\exp\X2(-\nd\vF\X2(\frac1u
\Y2)\Y2)+\frac{r_sc^2}{2uh^2}\pz\vF r\X2(\frac1u\Y2)\exp\X2(-\nd\vF\X2(\frac1u
\Y2)\Y2)\\
{}-\frac{\La c^2}{3u^3h^2}-\frac{c^2}{u^2h^2}\,\pz{\oW_2}r \X2(\frac1u\Y2)
+2\oW_2\X2(\frac1u\Y2)u
\end{multline}

Formula \er{Dc413} is exact. Let us make some simplifications in the
nonrelativistic case:
\beq{Dc414}
H\simeq c^2, \q \pz\vF r\simeq0
\end{equation}
and let us neglect higher order terms \wrt all corrections to NGT (or to GR).
One gets
\bml{Dc415}
\pz{^2u}{\vf^2}+u=\frac{r_sc^2}{2h^2}+\frac{3r_su^2}2-\frac{\La c^2}
{3u^2h^2}-\frac{2c^2\ell^4u^3}{h^2}\\{}
+\frac12\,r_s\X2(3u^2+\frac{c^2}{h^2}\Y2)
\X2(\exp\X2(-\nd\vF\X2(\frac1u\Y2)-1\Y2)
-\frac{c^2}{h^2u^2}\,\pz{\oW_1}r\X2(\frac1u\Y2)\Y2).
\end{multline}
Thus an \an\ \ac\ term $b_{\rm \an}$ reads
\bml{Dc416}
b_{\rm \an}=c^2\X3(\frac{r_s}{2h^2}\X2(\exp\X2(-\nd\vF\X2(\frac1u\Y2)\Y2)-1\Y2)
-\frac1{h^2c^2}\,\pz{\oW_1}r\X2(\frac1u\Y2)\Y3)\\ {}\simeq
-\frac{c^2}{h^2}\X3(\frac{r_s\nd}2 \,\vF\X2(\frac1u\Y2)+\frac1{u^2}\,
\pz{\oW_2}r\X2(\frac1u\Y2)\Y3)
\end{multline}
(by taking $|\vF(r)|\ll 1$).

In this way an \an\ \ac\ reads
\beq{Dc417}
\ov b_{\rm \an}=-\frac{c^2}2\X3(\frac{r_s\nd}2\,u^2\vF\X2(\frac1u\Y2)
+\pz{\oW_2}r\X2(\frac1u\Y2)\Y3).
\end{equation}
Eq.\ \er{Dc417} generalizes Eq.\ \er{Db374} to the case of nonzero field
$\vF(r)$.

In this parametrization we can also consider bending of light and Shapiro
effect. One gets
\bml{Dc418}
\D\vf=2\int_{\ov r_1}^{\ov{\ov R}}\frac1r
\sqrt{1-\frac{r_s}r\,\ef{-\nd}-\frac{\La r^2}3-2\oW_1(r)}\\
\sqrt{\X3(\frac{r^2}{b^2(1+\frac{\ell^4}{r^4})\X1(
1-\frac{r_s}r\,\ef{-\nd}-\frac{\La r^2}3-2\oW_2(r)\Y1)}-1\Y3)^{-1}}\,dr.
\end{multline}
Taking some simplifications one gets
\bml{Dc419}
\D\vf\simeq \D \vf_{\rm NGT,\La}-2\X3(\frac b{\,\ov{\ov R}\,}+\int_{\ov r_1}
^{\ov{\ov R}} \frac{\oW_1(r)+\oW_2(r)-2\,\frac{r_s}r\,\ef{-\nd}}r\\
\X3(\frac{r^2}{b^2(1+\frac{\ell^4}{r^4})(1-\frac{r_s}r-\frac{\La r^2}3)}-1\Y3)
^{-1/2}\,dr\Y3).
\end{multline}
The consistency of the value of bending of light with $\D\vf_{\rm NGT,\La}$
result can be obtained as before by putting to zero the term in brackets in
Eq.~\er{Dc419} and evaluating a value of~$\ov{\ov R}$.

\def\td{\wt{\wt\d}}
In the case of Shapiro effect we have a relevant formula
\bml{Dc420}
t(r,\ov r_0)=\frac1c \int_{\ov r_0}^r\X3(\X2(1+\frac{\ell^4}{r^4}\Y2)
\X2(1-\frac{r_s}r\,\ef{-\nd}-\frac{\La r^2}3 - 2\oW_2(r)\Y2)\\
\X2(1-\frac{r_s}r\,\ef{-\nd}-\frac{\La r^2}3 - 2\oW_1(r)\Y2)\Y3)^{-1/2}\\
\X3(1-\frac{b^2}{r^2}\X2(1+\frac{\ell^4}{r^4}\Y2)
\X2(1-\frac{r_s}r\,\ef{-\nd}-\frac{\La r^2}3 - 2\oW_1(r)\Y2)\Y3)^{-1/2}dr.
\end{multline}
After some usual approximations one gets
\beq{Dc421}
t(r,\ov r_0)\simeq t_{\rm NGT,\La}(r,\ov r_0)+\td t(r,\ov r_0)
\end{equation}
where $t_{\rm NGT,\La}(r,\ov r_0)$ is a formula for $t(r,\ov r_0)$ in NGT
with an influence of a \co ical \ct\ taken into account, and $\td(r,\ov r_0)$
is a correction
\beq{Dc422}
\td(r,\ov r_0)=\frac1c \int_{\ov r_0}^r
\X2(\oW_1(r)+\oW_2(r)-\frac{2r_s}r\nd \vF(r)\Y2)
\X2(1-\frac{b^2}{r^2}\X2(1-\frac{r_s}r-\frac{\La r^2}3\Y2)\Y2)^{-1/2}dr.
\end{equation}
In the case of a \hy\ orbit we get a similar formula
\bml{Dc423}
\pz{^2u}{\vf^2}+u = \frac1{a_h(e^2-1)}+\frac{3r_su^2}2-
\frac{\La c^2}{3u^2k^2a_h(e^2-1)}\\
{}-\frac{2c^2\ell^4u^3}{k^2a_h(e^2-1)}-\frac{n+2}{a_h(e^2-1)}
\,\vF\X2(\frac1u\Y2)-\frac{c^2}{k^2a_h(e^2-1)u^2}\cdot \pz{\oW_1}r
\X2(\frac1u\Y2).
\end{multline}

If we put $\ell=0$ in the formulae \er{Dc408}--\er{Dc409},
\er{Dc412}--\er{Dc413}, \er{Dc415}--\er{Dc417}, \er{Dc418}--\er{Dc420},
\er{Dc422}--\er{Dc423}, we can consider these \e s as in the Riemannian \s\
case (i.e., $g_\m=g_{\nu\mu}$) in the following parametrization
\bea{Dc424}
e^{2A(r)}&=&1-\frac{r_s}r \,\ef{-\nd}-2\wt V_1(r)-\frac{\La r^2}2\\
e^{2B(r)}&=&\frac1{1-\frac{r_s}r \,\ef{-\nd}-2\wt V_2(r)-\frac{\La r^2}2}
\label{Dc425}
\end{eqnarray}
putting $\wt V_i(r)$ in the place of $\oW_i(r)$, $i=1,2$, in such a way that
\bea{Dc426}
\wt V_1(r)&=&\frac12\X2(1-\frac{r_s}r\,\ef{-\nd}-e^{2A(r)}-\frac{\La r^2}2\Y2)
\\
\wt V_2(r)&=&\frac12\X2(1-\frac{r_s}r\,\ef{-\nd}-e^{-2B(r)}-\frac{\La r^2}2\Y2).
\label{Dc427}
\end{eqnarray}

Let us consider Eq.\ \er{D.30} for $B(r)=B_0=\rm const$. One gets
\beq{Da282}
9-10e^{2B_0}+e^{4B_0}=0.
\end{equation}
Taking
\beq{Da283}
e^{2B_0}=y
\end{equation}
one gets
\beq{Da284}
y^2-10y+9=0
\end{equation}
and
\bea{Da285}
y_1=9 &\qquad& y_2=1\\
B_{0,1}=\log 3&&B_{0,2}=0.\label{Da286}
\end{eqnarray}

Let us consider a small perturbation around \ct\ \so s
\bg{Da287}
e^{2B(r)}=e^{2(B_0+z)}=e^{2B_0}(1+2z)\\
e^{4B(r)}=e^{4(B_0+z)}=e^{4B_0}(1+4z)\label{Da288}
\end{gather}
in such a way that $|z|\ll 1$.

From Eq.\ \er{D.30} one gets
\beq{Da289}
\pz{^2\z(r)}{r^2}=\frac{\z(r)}{r^2}
\end{equation}
where
\bea{Da290}
B_1(r)&=&\log 3 +\frac4{279}\,\z(r)\\
B_2(r)&=&-2\z(r) \label{Da291}\\
B_i&=&B_{0i}+z_i, \q i=1,2,\ |\z|\ll1. \label{Da292}
\end{eqnarray}
Eq.\ \er{Da289} can be easily solved by a substitution $\z(r)=r^\a$.
One gets for $\a$
\beq{Da293}
\a(\a-1)=1
\end{equation}
and we get
\bea{Da294}
\a_1&=&\frac{1+\sqrt5}2\\
\a_2&=&\frac{1-\sqrt5}2\,.\label{Da295}
\end{eqnarray}
In this way
\beq{Da296}
\z(r)=\ov c_1 r^{(1+\sqrt5)/2}+\ov c_2 r^{(1-\sqrt5)/2},
\end{equation}
$\ov c_1$ and $\ov c_2$ are integration \ct s.

Now it is easy to see that if we want to examine a behaviour of the \so\ of
\er{D.30}--\er{D.32} around $r=0$ we should take $\ov c_2=0$ and
\bea{Da297}
B_1(r)&=&\log3 + \frac{4\ov c_1}{279}\,r^{(1+\sqrt5)/2}\\
B_2(r)&=&-2\ov c_1 r^{(1+\sqrt5)/2}\,. \label{Da298}
\end{eqnarray}
In this way $|z|\ll1$ is for $r\ll1$ (around zero). In both cases Eqs
\er{D.30}--\er{D.32} can be solved explicitly. One gets for the first case
($B_1(r)$)
\beq{Da299}
\bal
A_1(r)&=2\log r+\frac{9\ov c_1}{1+\sqrt5}\,r^{(1+\sqrt5)/2}+A_0\\
\wt\vF_1(r)&=18\log r-\frac{153\ov c_1}{2}\,r^{(1+\sqrt5)/2}
-\frac{81\ov c{}_1^2}{5+\sqrt5}\,r^{1+\sqrt5}+\wt\vF_0r+\wt\vF_1.
\eal
\end{equation}
In the second case
\beq{Da300}
\bal
A_2(r)&=\frac{\ov c_1}{1+\sqrt5}\,r^{(1+\sqrt5)/2}+A_0\\
\wt\vF_2(r)&=-\frac{\ov c_1}2\,r^{(1+\sqrt5)/2}-\frac{\ov c{}_1^2}{5+\sqrt5}
\,r^{1+\sqrt5}+\wt\vF_0r +\wt\vF_1.
\eal
\end{equation}
$A_0$, $\wt\vF_0$ and $\wt\vF_1$ are integration \ct s. In this way we can
have a nonsingular behaviour of the \so\ around zero in the second case. In
the second case we have nonsingular behaviour for $B_2(r)$ and $e^{2B_2(r)}$.
However, $A_1(r)$ and $e^{2\bar A_1(r)}$ have a singularity at $r=0$. The
same happens for $\vF_1(r)$ and $G_{\rm eff\,1}(r)$.

One gets
\bea{Da301}
e^{2A_1(r)}&=&e^{2A_0}\exp\X3(\frac{2\ov
c_1}{\sqrt5-1}\,r^{(\sqrt5+1)/2}\Y3) \\
e^{2A_2(r)}&=&\frac{e^{2A_0}}{r^4}\exp\X3(\frac{18\ov c_1}{\sqrt5-1}\,
r^{(\sqrt5+1)/2}\Y3).\label{Da302}
\end{eqnarray}
Moreover
\beq{Da303}
\frac{G_{\rm eff}}{G_N}=\exp\X3(-\frac{(n+2)^2}{\ov M}\,\wt\vF(r)\Y3).
\end{equation}
Thus one gets
\beq{Da304}
\frac{G_{\rm eff\,1}}{G_N}=\exp\X3(\frac{(n+2)^2}{\ov M}\X2(\frac{\,\ov c_1
\,}2\,r^{(\sqrt5+1)/2}+\frac{\ov c_1^2}{5+\sqrt5}\,r^{\sqrt5+1}\Y2)\Y3)
\exp\X1(\wt\vF_0r+\wt\vF_1\Y1)
\end{equation}
and
\beq{Da305}
G_{\rm eff\,1}(0)=G_Ne^{\wt\vF_1},
\end{equation}
\bml{Da305a}
\frac{G_{\rm eff\,2}}{G_N}=r^{-18(n+2)^2/\ov M}\cdot\exp\X3(
\frac{153(n+2)^2\ov c_1}{\ov M}\,r^{(\sqrt5+1)/2}+
\frac{81(n+2)^2\ov c{}_1^2}{(\sqrt5+5)\ov M}\,r^{\sqrt5+1}\Y3)\\
{}\cdot\exp\X1(\wt\vF_0r+\wt\vF_1\Y1).
\end{multline}
Thus if $\ov M>0$
\beq{Da306}
G_{\rm eff\,2}\underset{r\to0}{\longrightarrow}+\iy.
\end{equation}
Using Eq.\ \er{D.5} one can write $\ov \rho(r)$ in both cases (i.e.\ around
$r=0$).
\bml{Da306a}
\ov\rho_1(r)=\frac1{558}\exp\X3((n+2)\X2(\wt\vF_0r+\wt\vF_1
-\frac{9(n+2)}{2\ov Mr^2}\X1(4-17\ov c_1r^{(1+\sqrt5)/2}-18\ov c{}_1^2
r^{1+\sqrt5}\Y1)\Y2)\Y3)\\ {}\cdot
\X3(279\exp\X3(n\X2(\wt\vF_0r+\wt\vF_1-\frac{9(n+2)}{2\ov Mr^2}
\X1(4-17\ov c_1r^{(1+\sqrt5)/2}-18\ov c{}_1^2
r^{1+\sqrt5}\Y1)\Y2)\Y3)\\ {}\cdot
\X3(\frac1n-\frac1{n+2}\exp\X2(2\wt\vF_0r+2\wt\vF_1-\frac{9(n+2)}{\ov Mr^2}
\X1(4-17\ov c_1r^{(1+\sqrt5)/2}-18\ov c{}_1^2
r^{1+\sqrt5}\Y1)\Y2)\Y3)\\
{}+\frac{31\X1(9-\exp\X1(-\frac8{279}\,\ov c_1r^{(1+\sqrt5)/2}\Y1)\Y1)}{r^2}
+\frac49(1+\sqrt5)\ov c_1\exp\X2(-\frac8{279}\,\ov c_1r^{(1+\sqrt5)/2}\Y2)
r^{(\sqrt5-3)/2}\\
{}-31\ov M\exp\X2(-\frac8{279}\,\ov c_1r^{(1+\sqrt5)/2}\Y2)\hskip100pt\\
{}\cdot \X3(2\wt\vF_0-\frac{9(n+2)}{4\ov Mr^3}\X2(-16+17(\sqrt5-3)\ov
c_1r^{(1+\sqrt5)/2} +36(\sqrt5-1)\ov c{}_1^2r^{1+\sqrt5}\Y2)^2\Y3)\Y3)
\end{multline}
and
\bml{Da307}
\ov\rho_2(r)=
\frac12\exp\X3((n+2)\X2(\wt\vF_0r+\wt\vF_1+\frac{n+2}{\ov M}
\X2(-\frac12\,\ov c_1r^{(1+\sqrt5)/2}-(5+\sqrt5)\ov c{}_1^2r^{1+\sqrt5}\Y2)
\Y2)\Y3)\\
{}\cdot\X4(\frac1{n\nd}\exp\X3(n\X2(\wt\vF_0r+\wt\vF_1+\frac{n+2}{\ov M}
\X2(-\frac12\,\ov c_1r^{(1+\sqrt5)/2}-(5+\sqrt5)\ov c{}_1^2r^{1+\sqrt5}\Y2)
\Y2)\Y3)\\
{}\cdot\X3(n+2-n\exp\X3(2\X2(\wt\vF_0r+\wt\vF_1+\frac{n+2}{\ov M}
\X2(-\frac12\ov c_1r^{(1+\sqrt5)/2}-(5+\sqrt5)\ov c{}_1^2r^{1+\sqrt5}\Y2)
\Y2)\Y3)\Y3)\\
{}+\frac{1-\exp\X1(4\ov c_1r^{(1+\sqrt5)/2}\Y1)}{r^2}
-2(1+\sqrt5)\ov c_1\exp\X1(4\ov c_1r^{(1+\sqrt5)/2}\Y1)r^{(\sqrt5-3)/2}\\
-\frac{\exp\X1(4\ov c_1r^{(1+\sqrt5)/2}\Y1)}{16\ov Mr}
\X2(-4\ov M\wt\vF_1\sqrt r +\ov c_1\nd r^{\sqrt5/2}\X1(1+\sqrt5
+8(5+3\sqrt5)\ov c_1r^{(1+\sqrt5)/2}\Y1)\Y2)^2\Y4)
\end{multline}

Let us consider Eq.\ \er{D.30} and let us take the following substitution:
\beq{Da308}
W(r)=\eb2, \q B(r)=\frac12 \log W(r).
\end{equation}
One gets
\beq{Da309}
\pz{^2W}{r^2}=\frac1W\X2(\pz Wr\Y2)^2 + \frac{W^3}{4r^2}-\frac{5W^2}{2r^2}
+\frac{9W}{4r^2}\,.
\end{equation}

Eq.\ \er{Da309} can be extended to a complex plane in such a way that we take
$r\in\C$. From previous investigations we know that around $r=0$
\beq{Da310}
W(r)-1\sim r^{(1+\sqrt5)/2}.
\end{equation}
Thus it is natural to change the independent variable to
\beq{Da311}
\bal
z&=r^{(1+\sqrt5)/2}\\
r&=z^{2/(1+\sqrt5)}.
\eal
\end{equation}
Using \er{Da311} one gets
\beq{Da312}
\pz{^2y}{z^2}=\frac1y\X2(\pz yz\Y2)^2+\frac{\sqrt5-2}z \,\pz
yz+\frac{3-\sqrt5}{8z^2} \,y^3+\frac{5(\sqrt5-3)}{4z^2}\,y^2+
\frac{9(3-\sqrt5)}{8z^2}\,y.
\end{equation}
Using \er{Da311} we are removing a nonalgebraic singularity around $z=0$.

Eqs \er{Da309}, \er{Da312} are \e s of the second order of the following form
\beq{Da313}
\bga
\pz{^2y}{z^2}=F(z,y,p)\\
p=\pz yz\,,
\ega
\end{equation}
where $F$ is an analytic \f\ of $z$, algebraic of $y$ and rational of~$p$.

Such \e s can have movable, essential singularities (they depend on \ct s of
integration). Thus a \so\ of \er{Da313} being a \f\ of~$z$ and \ct s of
integration can have essential singularities which depend on these \ct s.
Moreover, our \e s are more regular than a general case. Our $F(z,y,p)$ is a
rational \f\ \wrt all arguments. The form of $F(z,y,p)$ looks as a standard
form from Painlev\'e analysis, i.e.
\bml{Da314}
\pz{^2y}{z^2}=\frac1y\X2(\pz yz\Y2)^2 + \X2(A(z)y+B(z)+\frac{C(z)}y\Y2)
\pz yz\\ {}+ D(z)y^3+E(z)y^2+F(z)y+G(z)+\frac{H(z)}y\,,
\end{multline}
where
\bg{Da315}
A(z)=C(z)=G(z)=H(z)=0,\\
B(z)=\frac{\sqrt5-2}z\,, \q
D(z)=\frac{3-\sqrt5}{8z^2}\,, \q E(z)=-\frac{9(3-\sqrt5)}{4z^2}\,, \q
F(z)=\frac{9(3-\sqrt5)}{8z^2}\,.\label{Da316}
\end{gather}
Thus this \e\ can belong to the second class of Painlev\'e classification
(see Refs \cite{D83}, \cite{D84} and related Refs \cite{D85}--\cite{D89}) and does
not possess movable essential singularities. In this way it belongs to the type
II--1$^\circ$ ($A=C=0$), i.e.\ to three canonical \e s XI, XII, XIII or
XIII$^1$. For XI, XII a \so\ can be expressed by elliptic \f s. In the case
XIII or XIII$^1$ the \so\ is not integrable in terms of the classical special
\f s. In the case of the last possibility the \so\ can be expresssed by a
transcendent Painlev\'e \f\ which is in this case a meromorphic univalent \f\
(see Ref.~\cite{D84}). Moreover, we should find a transformation of a
dependent variable to get one of these cases. Temporarily we do not know such
a transformation. Let us notice that the \so\ can have movable poles.

Let us consider the transformation
\beq{Dc320}
z \to t=f(z)=\exp\X2(\frac z4(1+\sqrt5)\Y2).
\end{equation}
One gets from \er{Da312}
\beq{Dc321}
\pz{^2y}{t^2}=\frac 1y\X2(\pz yt\Y2)^2+\pz yt+\frac1{16}\,y^3-\frac58\,y^2
+\frac9{16}\,y.
\end{equation}
In this way
\beq{Dc322}
W(r)=y\X2(\exp\X2(\frac{1+\sqrt5}4\,r^{(1+\sqrt5)/2}\Y2)\Y2).
\end{equation}
Moreover, Eq.\ \er{Dc321} does not look as any \e\ from Painlev\'e
classification. It is a little similar to the equation XII (see
Ref.~\cite{D84})
\beq{Dc323}
\pz{^2y}{t^2}=\frac1y\X2(\pz yt\Y2)^2+\a y^3+\b y^2+\g+\frac\d y
\end{equation}
or to XIII (see Ref.~\cite{D84})
\beq{Dc324}
\pz{^2y}{t^2}=\frac1y \X2(\pz yt\Y2)^2 - \frac1y \X2(\pz yt\Y2)
+\frac1t(\a y^2+\b)+\g y^3+\frac\d y.
\end{equation}
However, it is not the same.

Thus we can consider Eq.\ \er{Dc321} as a definition of a new special \f\ and
in terms of this \f\ we describe the \so
\bea{Dc325}
B(r)&=&\frac12 \log\X3(y\X2(\exp\X2(\frac{1+\sqrt5}4\,r^{(1+\sqrt5)/2}\Y2)\Y2)\Y3)\\
A(r)&=&-\frac14\log r+\frac14\int
\frac{y\X1(\exp\X1(\frac{1+\sqrt5}4\,r^{(1+\sqrt5)/2}\Y1)\Y1)}r \,dr
+A_0 \label{Dc326}
\\
\vF(r)&=&\frac{n+2}{4\ov M}\int\X3(\int^t
\frac{(y\X1(\exp\X1(\frac{1+\sqrt5}4\,r^{(1+\sqrt5)/2}\Y1)\Y1)-1)
y\X1(\exp\X1(\frac{1+\sqrt5}4\,r^{(1+\sqrt5)/2}\Y1)\Y1)}
{r^2}\,dr\Y3)\,dt\nonumber\\
&&\qquad  {}+ \vF_0r +\vF_1. \label{Dc327}
\end{eqnarray}
The properties of this interesting \f\ will be examined elsewhere. $A_0$,
$\vF_0$ and $\vF_1$ are integration \ct s.

In the formulae \er{Dc323}--\er{Dc324} $\a,\b,\g,\d$ are \ct s.

According to the suggestions from Refs \cite{D83}, \cite{D84} the
transformation which can help us is a M\"obius transformation
\beq{Dc330}
V(t)=\frac{l(t)y(t)+m(t)}{p(t)y(t)+q(t)}
\end{equation}
where $l,m,p,q$ are analytic \f s of $t\in\C$.

Let us consider Eq.\ \er{Dc321} and let us change a dependent variable into
$u(t)=\frac1{y(t)}$. One gets
\beq{De531}
\pz{^2u}{t^2}=\frac1u\X2(\pz ut\Y2)^2 + \pz ut - \frac9{16}\,u + \frac58
-\frac1{16u}\,.
\end{equation}
$u(t)$, $t\in\C$, is an analytic \f\ in a complex domain. Let us expand
$u(t)$ into a Taylor series around $t=1$. One gets
\beq{De532}
u(t)=\sum_{n=0}^\iy c_n(t-1)^n.
\end{equation}
Using Eq.\ \er{De531} and supposing that \er{De532} absolutely converges we
find recurrence relations
\bea{De533}
c_2&=&\frac1{2c_0}\X2(c_1^2+c_0c_1-\frac9{16}\,c_0^2-\frac58\,c_0+\frac1{16}\Y2)\\
c_{n+2}&=&\frac1{c_0\nd (n+1)}\X3[
-\sum_{i=1}^n c_ic_{n-i+2}(n+2-i)(n+1-i)\nonumber\\
&+&\sum_{i=0}^n c_{i+1}c_{n+1-i}(i+1)(n+1-i)
+\sum_{i=0}^n c_{i}c_{n+1-i}(n+1-i)
-\frac9{16}\sum_{i=0}^n c_ic_{n-i}-\frac58\,c_n\Y3]\hskip30pt\label{De534}
\end{eqnarray}
In this way one gets
\beq{De535}
u(t)=c_0+c_1(t-1)+\sum_{n=2}^\iy c_n(c_0,c_1)(t-1)^n.
\end{equation}
Moreover, there is not any compact formula for $c_n(c_0,c_1)$.

\let\nn\nonumber
We find $c_n(c_0,c_1)$ for $2< n\le 10$ and the result is written down below
\bea{De536}
c_3&=&\frac{-9c_0^3+c_1+16c_1^3-c_0^2(10+11c_1)+c_0(1+4c_1(-5+12c_1))}{96c_0^2}\\
c_4&=&\frac{1}{6144c_0^3}\X1(99c_0^4+c_0^3(290-1184c_1)+6c_0(-5+16c_1)(1+16c_1^2)\nn\\
&+&(1+16c_1^2)^2+4c_0^2(45+8c_1(-40+29c_1))\Y1)\label{De537}\\
c_5&=&\frac1{30720c_0^4}\X2(666c_0^5+c_0^4(1460-2129c_1)+c_1(1+16c_1^2)^2+2c_0(1+16c_1^2)(3+20c_1(-1+4c_1))\nn\\
&-&4c_0^3(-168+5c_1(95+96c_1))+4c_0^2\X1(-35+c_1(149+40c_1(-24+31c_1))\Y1)\Y2)
\label{De538}\\
c_6&=&\frac1{2949120c_0^5}\X2(19161c_0^6+2c_0^5(19195+5576c_1)+c_0^4
(11507+16c_1(4820-14339c_1))\nn\\
&+&10c_0(-5+24c_1)(1+16c_1^2)^2+(1+16c_1^2)^3+4c_0^3\X1(-1875+16c_1(1049
+75c_1(-53+24c_1))\Y1)\nn\\
&+&c_0^2\X1(987+128c_1\X1(-70+c_1(249+10c_1(-104+181c_1))\Y1)\Y1)\Y2)
\label{De539}\\
c_7&=&\frac1{20643840c_0^6}\X2(-6273c_0^7+5c_0^6(-10070+57683c_1)\nn\\
&+&c_0^5(-85267+12c_1(54175-53508c_1))+c_1(1+16c_1^2)^3\nn\\
&+&3c_0(1+16c_1^2)^2(5+4c_1(-5+28c_1))-c_0^4\X1(32100+c_1(-309095+16c_1
(31260+25319c_1))\Y1)\nn\\
&+&c_0^3\X1(9661+8c_1\X1(-7975+16c_1(1991+70c_1(-106+97c_1))\Y1)\Y1)\nn\\
&+&c_0^2\X2(-710+c_1\X1(3209+64c_1\X1(-385+4c_1(310+c_1(-950+2051c_1))\Y1)\Y1)\Y2)\Y2)
\label{De540}
\end{eqnarray}
\bea{nic5}
c_8&=&\frac{1}{2642411520c_0^7}
\X2(-2595735c_0^8+2050c_0^7(-4261+7840c_1)\nn\\
&+&14c_0(-5+32c_1)(1+16c_1^2)^3+(1+16c_1^2)^4-8c_0^6
(1124305+8c_1(-503850+78041c_1))\nn\\
&+&2c_0^5\X1(-933325+48c_1(108766+c_1(226995-635488c_1))\Y1)\nn\\
&+&c_0^4\X1(918214+64c_1\X1(-88240+c_1(442881+8c_1(-139840+49833c_1))\Y1)\Y1)\nn\\
&+&2c_0^3\X2(-47675+32c_1\X1(8991+2c_1\X1(-30665+12c_1(8080+7c_1
(-3825+4768c_1))\Y1)\Y1)\Y2)\nn\\
&+&8c_0^2\X2(475+8c_1\X2(-330+c_1\X1(2601+128c_1\X1(-105+c_1(369+26c_1(-30
+77c_1))\Y1)\Y1)\Y2)\Y2)\Y2)\label{De541}\\
c_9&=&\frac{1}{23781703680c_0^8}
\X3(-9040500c_0^9+c_0^8(-28183600+19095217c_1)+c_1(1+16c_1^2)^4\nn\\
&+&4c_0(1+16c_1^2)^3(7+4c_1(-5+36c_1))+4c_0^7(-6255716+c_1(3493555
+27257952c_1))\nn\\
&-&16c_0^6\X1(83370+c_1(2519699+4c_1(-4356135+3834731c_1))\Y1)\nn\\
&-&8c_0^5\X1(-482315+3c_1\X1(1258535+16c_1(-385707+11c_1(44045+24416c_1))\Y1)\Y1)\nn\\
&+&2c_0^4\X2(-350360+c_1\X1(3041479+96c_1\X1(-165620+c_1(579591+40c_1
(-39940+29491c_1))\Y1)\Y1)\Y2)\nn\\
&+&4c_0^3\X2(12228+c_1\X2(-83665+32c_1\X1(19611+2c_1\X1(-48625+12c_1(11296\nn\\
&&\quad{}+9c_1(-3345+5152c_1))\Y1)\Y1)\Y2)\Y2)
+16c_0^2\X3(-100+c_1\X2(399+4c_1\X2(-1935\nn\\
&&\quad{}+c_1\X1(7059+32c_1\X1(-765
+c_1(2865+8c_1(-595+1767c_1))\Y1)\Y1)\Y2)\Y2)\Y3)\Y3)\label{De542}\\
c_{10}&=&\frac{1}{3805072588800c_0^9}\X3(-171856953c_0^{10}
-2c_0^9(158360075+900848536c_1)\nn\\
&+&90c_0(-1+8c_1)(1+16c_1^2)^4+(1+16c_1^2)^5\nn\\
&+&c_0^8(242189673+16c_1(-435957260+651309781c_1))\nn\\
&+&12c_0^7\X1(57414135-64c_1(10766241+110c_1(-252217+74846c_1))\Y1)\nn\\
&+&c_0^6\X1(206986594-256c_1\X1(8878335+2c_1(-14111572+5c_1(-2436612+8093507c_1
))\Y1)\Y1)\nn\\
&+&16c_0^5\X2(-5501405+2c_1\X1(24973075+24c_1\X1(-4702475+4c_1(4420415+6c_1
(-1453675\nn\\
&&\quad{}+476196c_1))\Y1)\Y1)\Y2)
+6c_0^4\X2(1562017+16c_1\X2(-750850+c_1\X1(5035741+16c_1\X1(-1392280\nn\\
&&\quad{}+c_1(3862581+16c_1(-627390+629657c_1))\Y1)\Y1)\Y2)\Y2)\nn\\
&+&20c_0^3\X3(-19889+32c_1\X2(5706+c_1\X2(-52091+8c_1\X1(19707+c_1\X1(-88859
+96c_1(2693\nn\\
&&\quad{}+c_1(-5985+10888c_1))\Y1)\Y1)\Y2)\Y2)\Y3)
+5c_0^2\X3(1435+128c_1\X3(-209+c_1\X2(1593\nn\\
&&\quad{}+4c_1\X2(-192+c_1\X1(5391+64c_1\X1(
-285+2c_1(562+c_1(-688+2319c_1))\Y1)\Y1)\Y2)\Y2)\Y3)\Y3)\Y3)\label{De543}
\end{eqnarray}

Moreover, for small $(t-1)$ expansion up to the tenth order seems to be
valid. However, we do not know a radius of convergence. In this way if we
want to consider higher expansion and large values of $(t-1)$ we should use a
Borel transformation considering a \f
\beq{De544}
\wt\vf(T(t-1))=\sum_{n=0}^\iy \frac{c_n(t-1)^nT^n}{n!}
\end{equation}
and a transformation
\beq{De545}
\int_0^\iy e^{-T}\wt\vf(T(t-1))\,dT
\end{equation}
which gives us $\wt\vf(t)$ for large $(t-1)$.

\def\nnn{\nonumber\\}
\def\js{(1+16c_1^2)}
In this way
\bml{De546}
B(r)=-\frac12 \log\X3(c_0+c_1\ep\\
{}+\sum_{n=0}^\iy c_n(c_0,c_1)\ep^n\Y3).
\end{multline}
An \ap ion up to the tenth order is given below.
\begin{eqnarray}
\lefteqn{B(r)=\log\X4[c_0+c_1\ep}\nnn
&+&\frac{(\frac1{16}-\frac1{16}c_0(10+9c_0)+c_0c_1+c_1^2)\epm^2}{2c_0}\nnn
&+&\frac{\X1(-9c_0^3+c_1+16c_1^3-c_0^2(10+11c_1)+c_0(1+4c_1(-5+12c_1))\Y1)\epm^3}
{96c_0^2}\nnn
&+&\frac{1}{6144c_0^3}\X1(99c_0^4+c_0^3(290-1184c_1)+6c_0(-5+16c_1)\js\nnn
&+&\js^2+4c_0^2(45+8c_1(-40+29c_1))\Y1)\ep^4\nnn
&+&\frac{1}{30720c_0^4}\X1(666c_0^5+c_0^4(1460-2129c_1)+c_1\js^2
+2c_0\js(3+20c_1(-1+4c_1))\nnn
&-&4c_0^3(-168+5c_1(95+96c_1))+4c_0^2\X1(-35+c_1(149
+40c_1(-24+31c_1))\Y1)\Y1)\nnn
&&\ep^5 + \frac{1}{2949120c_0^5}\X2(19161c_0^6+2c_0^5(19195+5576c_1)\nnn
&+&c_0^4(11507+16c_1(4820-14339c_1))+10c_0(-5+24c_1)(1+16c_1^2)^2\nnn
&+&(1+16c_1^2)^3+4c_0^3\X1(-1875+16c_1(1049+75c_1(-53+24c_1))\Y1)\nn\\
&+&c_0^2\X1(987+128c_1\X1(-70+c_1(249+10c_1(-104+181c_1))\Y1)\Y1)\Y2)
\ep^6\nnn
&+&\frac1{20643840c_0^6}\X2(-6273c_0^7+5c_0^6(-10070+57683c_1)
+c_0^5(-85267+12c_1(54175-53508c_1))\nnn
&+&c_1(1+16c_1^2)^3+3c_0(1+16c_1^2)^2(5+4c_1(-5+28c_1))\nnn
&-&c_0^4\X1(32100+c_1(-309095+16c_1(31260+25319c_1))\Y1)\nn\\
&+&c_0^3\X1(9661+8c_1\X1(-7975+16c_1(1991+70c_1(-106+97c_1))\Y1)\Y1)\nn\\
\noalign{\eject}
&+&c_0^2\X2(-710+c_1\X1(3209+64c_1\X1(-385+4c_1(310+c_1(-950+2051c_1))
\Y1)\Y1)\Y2)\Y2)\nnn
&&\ep^7+\frac{1}{2642411520c_0^7}
\X2(-2595735c_0^8+2050c_0^7(-4261+7840c_1)\nn\\
&+&14c_0(-5+32c_1)(1+16c_1^2)^3+(1+16c_1^2)^4-8c_0^6
(1124305+8c_1(-503850+78041c_1))\nn\\
&+&2c_0^5\X1(-933325+48c_1(108766+c_1(226995-635488c_1))\Y1)\nn\\
&+&c_0^4\X1(918214+64c_1\X1(-88240+c_1(442881+8c_1(-139840+49833c_1))\Y1)\Y1)\nn\\
&+&2c_0^3\X2(-47675+32c_1\X1(8991+2c_1\X1(-30665+12c_1(8080+7c_1
(-3825+4768c_1))\Y1)\Y1)\Y2)\nn\\
&+&8c_0^2\X2(475+8c_1\X2(-330+c_1\X1(2601+128c_1\X1(-105+c_1(369+26c_1(-30
+77c_1))\Y1)\Y1)\Y2)\Y2)\Y2)\nnn
&&\ep^8+\frac{1}{23781703680c_0^8}\X3(-9040500c_0^9\nnn
&+&c_0^8(-28183600+19095217c_1)+c_1(1+16c_1^2)^4\nn\\
&+&4c_0(1+16c_1^2)^3(7+4c_1(-5+36c_1))+4c_0^7(-6255716+c_1(3493555
+27257952c_1))\nn\\
&-&16c_0^6\X1(83370+c_1(2519699+4c_1(-4356135+3834731c_1))\Y1)\nn\\
&-&8c_0^5\X1(-482315+3c_1\X1(1258535+16c_1(-385707+11c_1(44045+24416c_1))\Y1)\Y1)\nn\\
&+&2c_0^4\X2(-350360+c_1\X1(3041479+96c_1\X1(-165620+c_1(579591+40c_1
(-39940+29491c_1))\Y1)\Y1)\Y2)\nn\\
&+&4c_0^3\X2(12228+c_1\X2(-83665+32c_1\X1(19611+2c_1\X1(-48625+12c_1(11296\nn\\
&+&9c_1(-3345+5152c_1))\Y1)\Y1)\Y2)\Y2)
+16c_0^2\X3(-100+c_1\X2(399+4c_1\X2(-1935+c_1\X1(7059\nn\\
&+&32c_1\X1(-765
+c_1(2865+8c_1(-595+1767c_1))\Y1)\Y1)\Y2)\Y2)\Y3)\Y3)\ep^9 \nnn
&+&\frac{1}{3805072588800c_0^9}\X3(-171856953c_0^{10}
-2c_0^9(158360075+900848536c_1)\nn\\
&+&90c_0(-1+8c_1)(1+16c_1^2)^4+(1+16c_1^2)^5\nn\\
&+&c_0^8(242189673+16c_1(-435957260+651309781c_1))\nn\\
&+&12c_0^7\X1(57414135-64c_1(10766241+110c_1(-252217+74846c_1))\Y1)\nn\\
&+&c_0^6\X1(206986594-256c_1\X1(8878335+2c_1(-14111572+5c_1(-2436612+8093507c_1
))\Y1)\Y1)\nn\\
&+&16c_0^5\X2(-5501405+2c_1\X1(24973075+24c_1\X1(-4702475+4c_1(4420415+6c_1
(-1453675\nn\\
&+&476196c_1))\Y1)\Y1)\Y2)
+6c_0^4\X2(1562017+16c_1\X2(-750850+c_1\X1(5035741+16c_1\X1(-1392280\nn\\
&+&c_1(3862581+16c_1(-627390+629657c_1))\Y1)\Y1)\Y2)\Y2)\nn\\
&+&20c_0^3\X3(-19889+32c_1\X2(5706+c_1\X2(-52091+8c_1\X1(19707+c_1\X1(-88859
+96c_1(2693\nn+c_1(-5985\\
&+&10888c_1))\Y1)\Y1)\Y2)\Y2)\Y3)+5c_0^2\X3(1435+128c_1\X3(-209
+c_1\X2(1593+4c_1\X2(-192+c_1\X1(5391+64c_1\X1(-285\nn\\
&+&2c_1(562+c_1(-688+2319c_1))\Y1)\Y1)\Y2)\Y2)\Y3)\Y3)\Y3)\ep^{10}\Y4]\label{De547}
\end{eqnarray}

In the \NK{\JT\ } there are two types of geodesics: extremal and nonextremal.
Up to now we consider only extremal geodesics. Moreover, it is interesting to
consider also nonextremal geodesics as paths of massive particles and photons
(see Eq.\ \er{D.128}). In order to do this we should find a \s\ part of the
\nos\ connection $\ov\G{}^\a_{(\b\g)}$ in the case of our \so. Using results
from Refs \cite4, \cite{D78}, \cite{D79}, \cite{D90}--\cite{D96} we get
\bea{De548}
\ov\G{}^1_{11}&=&\frac12\,\pz\a r\,\frac1{\a(r)} \\
\ov\G{}^1_{(33)}&=&\frac1{\a(r)}\X2(f(r)\ov B-\frac12\,r^2\,\pz{\ov\vf} r\Y2)\sin^2\th
\label{De549}\\
\ov\G{}^1_{22}&=&\frac1{2\a(r)}\X2(f(r)\ov B-\frac12\,r^2\,\pz{\ov\vf} r\Y2) \label{De550}\\
\ov\G{}^1_{44}&=&\frac{\o^2(r)}{\a^2(r)}\,\pz\psi r+\pz\g r\,\frac1{2\a(r)}
\label{De551}\\
\ov\G{}^2_{33}&=&-\frac12\sin2\th, \q \ov\G^3_{(23)}=\cot \th\\
\ov\G{}^2_{(12)}&=&\ov\G{}^3_{(43)}=\frac14\,\pz{\ov\vf} r\label{De552}\\
\ov\G{}^2_{(34)}&=&-\ov\G{}^3_{(24)}\sin^2\th=\frac{\o(r)}{2\a(r)}\,\ov B\sin\th
\label{De552a}\\
\ov\G{}^4_{(14)}&=&\frac{\o^2(r)}{2\a(r)}\,\pz{\ov\vf} r+\pz\g r\,\frac1{2\g(r)}.
\label{De553}
\end{eqnarray}
The remaining connection \cf s are equal to zero and
\beq{De554}
\ov\vf=\log(r^4+f^2(r)), \qquad \psi=\log\X2(1-\frac{\a\g}{\o^2}\Y2),
\qquad \ov B=\frac{2fr-r^2\pz fr}{r^4+f^2(r)}\,.
\end{equation}
Moreover, we have
\beq{De555}
\g(r)=\frac{f^4(r)(f^2(r)+r^4)e^{2(\ov c-r)}}{\ell^4\a(r)}, \qquad \o(r)=f^2(r)e^{\ov c-r}.
\end{equation}
In this way
\beq{De556}
\psi=\log\X2(1-\frac{f^2(r)+r^4}{\ell^4}\Y2).
\end{equation}
Thus we have
\bea{De559}
\ov\G{}^1_{11}&=&\frac1{2\a(r)}\,\pz\a r \\
\ov\G{}^1_{(33)}&=&\frac{2r\sin^2\th}{\a(r)(r^4+f^2(r))}
\X2(r^2(4rf^2(r)-1)-\pz fr\X2(r^4f(r)-\frac1{r^2}\Y2)\Y2) \label{De560}\\
\ov\G{}^1_{22}&=&\frac{2r}{\a(r)(r^4+f^2(r))}
\X2(r^2(4rf^2(r)-1)-\pz fr\X2(r^4f(r)-\frac1{r^2}\Y2)\Y2) \label{De561}\\
\ov\G{}^1_{44}&=&\frac{f^3(r)e^{2(\ov c-r)}}{\a^2(r)}
\X3[\frac{2(f(r)\pz fr+2r^3)}{f^2(r)+r^4-\ell^4}\nn\\
&+&\frac{2\a(r)\bigl(3f^2(r)\pz fr+2r^3f(r)+2r^4\pz fr-f^3(r)
-f(r)r^4\bigr)-\pz\a r\,f(r)(f^2(r)+r^4)}{2\ell^4\a(r)}\Y3]\qquad
\label{De562}\\
\noalign{\eject}
\ov\G{}^2_{(12)}&=&\ov\G{}^3_{(43)}=\frac{2r^3+f(r)\pz fr}{2(r^4+f^2(r))}
\label{De563}\\
\ov\G{}^2_{33}&=&-\frac12\sin2\th, \q \ov\G^3_{(23)}=\cot \th\\
\ov\G{}^2_{(34)}&=&-\ov\G{}^3_{(24)}\sin^2\th=\frac{f^2(r)e^{\ov c-r}
(4f(r)-r\pz fr)r^3}{2\a(r)(r^4+f^2(r))}\cdot \sin\th\label{De564}\\
\ov\G{}^4_{(14)}&=&\frac{2(\pz fr(r)+2r^3)}
{(r^4+f^2(r)-\ell^4)(f^2(r)+r^4)}-\frac1{2\a(r)}\,\pz \a r
+\frac 2{f(r)}\,\pz fr-1+\frac{f\pz fr(r)+2r^3}{f^2(r)+r^4}
\,.\quad\label{De565}
\end{eqnarray}
The remaining connection \cf s $\ov\G{}^\a_{(\b\g)}$ are equal to zero.

Let us consider Eq.\ \er{D.128} using
our connection \cf s $\ov\G{}^\a_{(\b\g)}$\,. In the case of massive point
particles we have
\beq{De566}
c^2\g\X2(\pz t\tau\Y2)^2 - \a\X2(\pz r\tau\Y2)^2 - r^2
\X2(\X2(\pz \th \tau\Y2)^2+\sin^2\th\X2(\pz\vf\tau\Y2)^2\Y2)^2=c^2.
\end{equation}
In the case of a photon
\beq{De567}
c^2\g\X2(\pz t\si\Y2)^2 - \a\X2(\pz r\si\Y2)^2 - r^2
\X2(\X2(\pz \th \si\Y2)^2+\sin^2\th\X2(\pz\vf\si\Y2)^2\Y2)^2=0,
\end{equation}
where $\tau$ means a proper time and $\si$ is an affine parameter along a
photon path.

We remind to the reader that in Einstein Unified Field Theory and in NGT and
in our \NK{\JT\ } there is a first integral of motion for nonextremal
geodesics $g_{\a\b}u^\a u^\b ={\rm const}$ where $u^\a=\pz{x^\a}\tau$.

One gets
\bea{nic6}
\pz{^2r}{\tau^2}&+&\frac1{2\a(r)}\,\pz \a r\X2(\pz r\tau\Y2)^2
+\frac{r^2(4rf^2(r)-1)-\pz fr\X1(r^4f(r)-\frac1{r^2}\Y1)}
{2\a(r)(r^4+f^2(r))}\X3[\X2(\pz \th\tau\Y2)^2+\sin^2\th
\X2(\pz\vf\tau\Y2)^2\Y3]\nnn
&+&\frac{f^3(r)e^{2(\ov c-r)}}{\a^2(r)}
\X3[\frac{2(f(r)\pz fr+2r^3)}{f^2(r)+r^4-\ell^4}
+\frac{2\a(r)\bigl(3f^2(r)\pz fr+2r^3f(r)+2r^4\pz fr-f^3(r)
-f(r)r^4\bigr)}{2\ell^4\a(r)}\nn\\
&-&\frac{\pz\a r\,f(r)(f^2(r)+r^4)}{2\ell^4\a(r)}\Y3]
\X2(\pz t\tau\Y2)^2=0\label{De569}\\
\pz{^2\th}{\tau^2}&+&\frac{2r^3+f(r)\pz fr}{2(r^4+f^2(r))}\,\pz r\tau\,\pz\th\tau\nnn
&+&\frac{r^3f^2(r)e^{\ov c-r}(4f(r)-r\pz fr)\sin \th}
{2\a(r)(r^4+f^2(r))}\,\pz\vf\tau\,\pz t\tau
-\frac12\sin2\th\X2(\pz\vf \tau\Y2)^2=0\label{De570}\\
\pz{^2\vf}{\tau^2}&+&\frac{2r^3+f(r)\pz fr}{2(r^4+f^2(r))}\,\pz t\tau\,\pz\vf\tau\nnn
&-&\frac{r^3f^2(r)e^{\ov c-r}(4f(r)-r\pz fr)\cosec \th}
{2\a(r)(r^4+f^2(r))}\,\pz\th\tau\,\pz t\tau
+\cot\th\X2(\pz \th\tau\Y2)\X2(\pz\vf\tau\Y2)=0\label{De571}\\
\pz{^2t}{\tau^2}&+&\X3[\frac{2\X1(\pz fr\,f(r)+2r^3\Y1)}{(r^4+f^2(r))
(f^2+r^4)}
-\frac1{2\a(r)}\,\pz\a r+\frac2{f(r)}\,\pz fr-1+\frac{f\pz fr+2r^3}
{f^2(r)+r^4}\Y3]\pz r\tau\,\pz t\tau=0\hskip30pt \label{De572}
\end{eqnarray}
for massive point particles. In this case Eq.\ \er{De566} is satisfied. In
the case of a photon we have the same \e s (in place of $\tau$ we have~$\si$)
and Eq.~\er{De567} is satisfied.

Let us consider Eq.\ \er{D.28} or Eq.\ \er{9.22} in the case of
non-Riemannian geometry, i.e.
\beq{De573}
\pz{k^\mu}\si + \ov \G{}^\mu_{(\a\b)}k^\a k^\b=0
\end{equation}
supposing that $k^\mu=(k_r,0,0,\frac {\,\ov\o\,} c)$. One gets
\beq{De576}
\g(r)\ov\o{}^2-\a(r)k_r^2=0
\end{equation}
or
\bea{De577}
k_r&=&\sqrt{\frac{\g(r)}{\a(r)}}\,\ov\o\\
k_r&=&\frac{f^2(r)e^{\ov c-r}}{\ell^2\a(r)}\,\sqrt{f^2(r)+r^4}\,\ov\o \label{De578}\\
\frac1{\,\ov\o{}^2}\,\pz{\ov\o}\si&+&\frac{f^5(r)e^{3(\ov c-r)}}{\ell^2\a^3(r)}
\sqrt{f^2(r)+r^4}\X3[\frac{2(f(r)\pz fr+2r^3)}{f^2(r)+r^4-\ell^4}\nnn
&+&\frac{3f^2(r)\pz fr+2r^3f(r)+2r^4\pz fr-f^3(r)
-f(r)r^4}{\ell^4}-\frac{\pz\a r\,f(r)(f^2(r)+r^4)}{2\ell^4\a(r)}\Y3]
=0 \qquad \label{De574}\\
\pz T\si&=&\frac{f^5(r)e^{3(\ov c-r)}}{\ell^2\a^3(r)}
\sqrt{f^2(r)+r^4}\X3[\frac{2(f(r)\pz fr+2r^3)}{f^2(r)+r^4-\ell^4}\nnn
&+&\frac{3f^2(r)\pz fr+2r^3f(r)+2r^4\pz fr-f^3(r)
-f(r)r^4}{\ell^4}-\frac{\pz\a r\,f(r)(f^2(r)+r^4)}{2\ell^4\a(r)}\Y3]\label{De574a}
\end{eqnarray}
$T=\frac{2\pi}{\ov\o}$ is a period of oscillation of an \elm c wave (a photon).

In this way one gets
\bml{De575}
T(\si)=\frac1{\pi \ell^2}\int_{\si_0}^\si d\si\,
\frac{f^5(r(\si))e^{3(\ov c-r(\si))}}{\a^3(r(\si))}
\sqrt{f^2(r(\si))+r^4(\si)}\X3[\frac{2(f(r(\si))\pz fr(r(\si))+2r^3(\si))}
{f^2(r(\si))+r^4(\si)-\ell^4}\\
{}+\frac{3f^2(r(\si))\pz fr(r(\si))+2r^3(\si)f(r(\si))
+2r^4(\si)\pz fr(r(\si))-f^3(r(\si))-f(r(\si))r^4(\si)}{\ell^4}\\
{}-\frac{\pz\a r(r(\si))f(r(\si))(f^2(r(\si))+r^4(\si))}{2\ell^4\a(r(\si))}\Y3]
\end{multline}
and
\bea{De576a}
\D\ov\o&=&-\frac{\D T(\si)\ov\o{}^2}{2\pi+\ov\o_0\D T(\si)}\\
\ov \o_0&=&\frac{2\pi}{T(\si_0)}\label{De577a}
\end{eqnarray}
where $r(\si)$ is a radial part of a photon path (for a nonextremal
geodesics).

In this way we get a Doppler effect along a photon path.
In the case of nonextremal geodesics an \an\ \ac\ is given by Eqs
\er{De569}--\er{De572}.

Let us consider a top-down approach starting from Eqs \er{D.53}--\er{D.60}
and Eq.~\er{D.66}. Supposing that a skew-\s\ part of $g_{\mu\nu}$,
$g_\[\m]=0$ we get
\beq{Dn244}
\gd\ov\G,\a,\b\g,=\Chr \a{\b\g}
\end{equation}
where $\Chr \a{\b\g}$ is a Christoffel symbol formed from $g_{\a\b}=g_{\b\a}$.
\beq{Dn245}
R_\m=R_{\nu\mu}
\end{equation}
(see Ref.\ \cite6).

$R_\m$ is an ordinary Ricci tensor formed from a Levi-Civit\`a \cn\
$\Chr\a{\b\g}$ for $g_{\a\b}=g_{\b\a}$. In this way we get Eqs
\er{A.116}--\er{A.120} and our theory satisfies a Bohr correspondence
principle. Simultaneously
\beq{Dn246}
\nad{\rm tot}T_{\a\b}=\nad{\rm tot}T_{\b\a}=\ov T_{\a\b}.
\end{equation}

\def\ef#1{e^{#1\vF}}
Now we perform the following simplification. We neglect $\ov\rho$ and $\ov
T_{\a\b}$ in \e s \er{A.117} and \er{A.120}. We get
\beq{Dn247}
R_\m=0
\end{equation}
and
\beq{Dn248}
-2\ov M\,\ov \nabla_\a(g^{\a\b}\pa_\b\vF)+\ef n(\ef2-1)=0.
\end{equation}
Moreover, we still have
\beq{Dn249}
G_{\rm eff}=G_N \ef{-(n+2)}.
\end{equation}

Simultaneously we consider geodetic \e s \er{D.22} in this case for a massive
point body. Doing a Newtonian \ap ion in the case of spherical symmetry and
stationary case we get \e s with an \an\ \ac\ via \er{Dn249}. Eq.\ \er{Dn248}
can be reduced to
\beq{Dn251}
-2\ov M\,\frac1{r^2}\X2(r^2\,\pz\vF r\Y2)+\ef n(\ef2-1)=0,
\end{equation}
i.e.\ to Eq.\ \er{2.26}.

In this way we get our model of an \an\ \ac\ from Section~3. The full
post-Newtonian \ap ion using EIH (Einstein--Infeld--Hoffman) method
(see Ref.~\cite{24}) can be applied if necessary. In this way we get
relativistic corrections from Section~9.

The results from Section~8 can be obtained from Eq.\ \er{Dn188} supposing
$\rho_{\rm tot}=0$ and doing  the same trick as in Section~8.

In this way we get an \an\ \ac\ model considered in Sections 2--10 as a limit
of the \eu\nos\ Jordan--Thiry Theory. The limitation of the model are coming
from the full theory and they are negligible for a \P0/11 case. Let us
notice that our \an\ \ac\ model is coming from the \NK{\JT\ }, in particular
due to an appearence of the scalar field $\vF$ and its self-interaction term.

\def\Pl{Planck}
Let us consider a scale of length from Section~2, Eq.~\er{2.23},
$$
L=\sqrt{\frac{n(n+2)\la_{\rm co}}{2|\ov M|}}
\simeq 10\,{\rm Mpc},
$$
where 
$\la_{\rm co}$ is a \co ical \ct.

$L$ has been calculated for $n=14=\dim G2$ and for $|\ov M|=1$ (see
Ref.~\cite{10}). Moreover, in Ref.~\cite{10} $G2$ has been considered as a
group $H$ in the \NK{\JT\ } (see Refs \cite4, \cite{10}). $G2$ is important
only for Glashow--Weinberg--Salam model of unification of electroweak
interactions. If we want to unify all fundamental interactions we need a
bigger group~$H$, \st $G2\subset H$. We need of course a group $G$ \st
\beq{Dn263}
\SU(2)_L \t \U(1) \t \SU(3)_c \subset G.
\end{equation}
There are a lot of possibilities. One of the most promising is $G=\SO(10)$.
Moreover, we need also a group $G_0$ \st $M=G/G_0$ (see Refs \cite4,
\cite{10} and Section~2).

In our world $G_0=\U(1)_{\rm em} \t \SU(3)_c$. The group $H$ for $G=\SO(10)$
and $G_0=\U(1)_{\rm em}\t \SU(3)_c$ should be \st
\beq{Dn264}
\SO(10)\t \X1(\U(1)_{\rm em}\t \SU(3)_c\Y1) \subset H.
\end{equation}
The simplest choice is $H=\SO(16)$. Why?

First of all $G2\subset \SO(16)$ and $\SO(10)\t\SO(6)\subset\SO(16)$.
Moreover, $\SO(6)\simeq\SU(4)$ and $\U(1)\t \SU(3) \subset\SU(4)$. Thus if we
identify $\U(1)$ with $\U(1)_{\rm em}$ and $\SU(3)$ with $\SU(3)_c$ we get
what we want. In this way
\begin{multline*}
M=\raise6pt\hbox{SO(10)}\bigg/ \lower6pt\hbox{$\U(1)\t\SU(3)$}, \q
S^2\subset M,\\ \q \dim\SO(16)=120, \ \dim\SO(10)=45, \ n_1=\dim M=36.
\end{multline*}
Thus the scale of length should be rescaled and we get for $n=120$
\beq{Dn265}
L=\frac{80.2\,{\rm Mpc}}{\sqrt{|\ov M|}}\,.
\end{equation}
For a scale of time
\beq{Dn266}
T=\frac{258.56}{\sqrt{|\ov M|}}\t10^6\,{\rm yr}.
\end{equation}
Moreover, we still have $L=\xi\cdot 10$\,Mpc, $T=\xi\cdot32\t10^6$\,yr, where
$\xi$ is of order~1.

The uncertainty in $L$ definition (a factor $\xi$) can be shifted to an
uncertainty in $r_0$ definition (a factor $\ov\xi$). This is convenient for
further calculations. Both factors $\xi$ and $\ov\xi$ are of order~$1$.

The very important problem in our theory is a dark matter problem in the \SS.
In our approach it is $\rho(x)$ which is very important in our \so\ in the
case of symmetric metric $g_\m=g_{\nu\mu}$ or in the most general case with 
\nos\ metric $g_\m\ne g_{\nu\mu}$. We can consider extremal and nonextremal
geodetics in this case. It means we consider geodetic with a Levi-Civit\`a 
\cn\ generated
by $g_{(\a\b)}$ (Christoffel symbols) or a general \cn\ from NGT.

In order to get a realistic simulation of \sp\ motion we should satisfy some
constraints coming from known effects in the \SS. In this way we should 
consider a density of dark matter (not only interplanetary matter) and
include it to a total density of a matter considered by us. Thus we should 
satisfy some constraints known in the literature (see Refs \cite{x,y,z,N64a,
v,r}).

\hfuzz0.71pt
\def\theequation{E.\arabic{equation}}
\setcounter{equation}0
\section*{Appendix E}
In this Appendix we give listings of all programmes written in Mathematica~7
important for calculations quoted in Appendix~D. This will facilitate any
future research for any interested reader.

\medskip
{\parindent 0pt
\tt eqn1 = D[D[b[x, b1, t], x], x] == $\dfrac{\tt 9 - 10 * e^{2\,b[x,b1,t]}
+ e^{4\,b[x,b1,t]}}{\tt 8 * x^2}$

\vskip2pt
eqn2 =  D[D[f[x, b1, t], x], x] == $\dfrac{\tt -1 + e^{2\,b[x,b1,t]}}
{\tt 4 * x^2}$

\vskip4pt
eqn3 = D[a[x, b1, t], x] == $\dfrac{\tt -1 + e^{2\,b[x,b1,t]}}{\tt 4 * x}$

\vskip2pt
eqn4 = a[7.36 * 10\^{ }-10, b1, t] == -1.72 * 10\^{ }-9 * 1 / t

eqn5 = b[7.36 * 10\^{ }-10, b1, t] == -1.72 * 10\^{ }-9 * 1 / t

eqn6 = (D[b[x, b1, t], x] /.\ x $\to$ 7.36 * 10\^{ }-10) == b1

ss = NDSolve[\char123 eqn1, eqn3, eqn4, eqn5, eqn6\char125, \char123a, b\char125,

\ \ \ \char123x, 10\^{ }-11, 10\^{ }-3\char125, \char123b1, -100, 100\char125, \char123t, 300, 500\char125, MaxSteps $\to$ 50 000]

eqn8 = D[D[f[x, b1, t], x], x] == $\dfrac{\tt -1 + e^{2\,b[x,b1,t]}}
{\tt 4 * x^2}$ /.\ ss

\vskip2pt
eqn9 = f[7.36 * 10\^{ }-10, b1, t] == 0 /.\ ss

eqn10 = (D[f[x, b1, t], x] /.\ x $\to$ 7.36 * 10\^{ }-10) == 0 /.\ ss

ss1 = NDSolve[\char123eqn8, eqn9, eqn10\char125, f,

\ \ \ \char123x, 10\^{ }-11, 10\^{ }-3\char125,
\char123b1, -100, 100\char125,
\char123t, 300, 500\char125, MaxSteps $\to$ 50 000]

hr1 = Rasterize[Plot3D[Evaluate[f[x, b1, t] /.\ ss1] /.\ b1 $\to$ 14.31,

\ \ \ \char123t, 300, 500\char125,
\char123x, 10\^{ }-11, 10\^{ }-3\char125, ColorFunction $\to$
"RustTones"],

\ \ \ ImageResolution $\to$ 90]

hr2 = Rasterize[Plot3D[Evaluate[D[f[x, 14.31, t], x] /.\ ss1],

\ \ \ \char123t, 300, 500\char125,
\char123x, 10\^{ }-11, 10\^{ }-3\char125], ImageResolution $\to$ 90]

}
\medskip

The above programme is dedicated for a dust filled model.

\medskip

{\parindent 0pt
\tt eqn1 = D[D[b[x, b1, t], x], x] == $\dfrac{\tt 9 - 10 * e^{2\,b[x,b1,t]}
+e^{4\,b[x,b1,t]}}{\tt 8 * x^2}$

\vskip2pt
eqn2 =  D[D[f[x, b1, t], x], x] == $\dfrac{\tt -1 + e^{2\,b[x,b1,t]}}
{\tt 4 * x^2}$

\vskip2pt
eqn3 = D[a[x, b1, t], x] == $\dfrac{\tt -1 + e^{2\,b[x,b1,t]}}{\tt 4 * x}$

\vskip2pt
eqn4 = a[7.36 * 10\^{ }-10, b1, t] == -3.44 * 10\^{ }-12 * Coth[0.003 * t]

eqn5 = b[7.36 * 10\^{ }-10, b1, t] == 3.44 * 10\^{ }-12 * Coth[0.003 * t]

eqn6 = (D[b[x, b1, t], x] /.\ x $\to$ 7.36 * 10\^{ }-10) == b1

ss = NDSolve[\char123eqn1, eqn3, eqn4, eqn5, eqn6\char125, \char123a, b\char125,

\ \ \ \char123x, 10\^{ }-11, 10\^{ }-3\char125, \char123b1, -100, 100\char125, \char123t, 300, 500\char125, MaxSteps $\to$ 50 000]

eqn8 = D[D[f[x, b1, t], x], x] == $\dfrac{\tt -1 + e^{2\,b[x,b1,t]}}
{\tt 4 * x^2}$ /.\ ss

\vskip2pt
eqn9 = f[7.36 * 10\^{ }-10, b1, t] == 0 /.\ ss

eqn10 = (D[f[x, b1, t], x] /.\ x $\to$ 7.36 * 10\^{ }-10) == 0 /.\ ss

ss1 = NDSolve[\char123eqn8, eqn9, eqn10\char125, f,

\ \ \ \char123x, 10\^{ }-11, 10\^{ }-3\char125,
\char123b1, -100, 100\char125,
\char123t, 300, 500\char125, MaxSteps $\to$ 50 000]

hr1 = Rasterize[Plot3D[Evaluate[f[x, 14.31, t] /.\ ss1], \char123t, 300, 500\char125,

\ \ \ \char123x, 10\^{ }-11, 10\^{ }-3\char125, ColorFunction $\to$
"RustTones"], ImageResolution $\to$ 90]

hr2 = Rasterize[Plot3D[Evaluate[D[f[x, 14.31, t], x] /.\ ss1],

\ \ \ \char123t, 300, 500\char125,
\char123x, 10\^{ }-11, 10\^{ }-3\char125], ImageResolution $\to$ 90]

}

\medskip
For a CDM$\La$ model.

\medskip

{\parindent 0pt
\tt eqn1 = D[D[b[x, b1, t], x], x] == $\dfrac{\tt 9 - 10 * e^{2\,b[x,b1,t]}
+ e^{4\,b[x,b1,t]}}{\tt 8 * x^2}$

\vskip2pt
eqn2 =  D[D[f[x, b1, t], x], x] == $\dfrac{\tt -1 + e^{2\,b[x,b1,t]}}
{\tt 4 * x^2}$

\vskip4pt
eqn3 = D[a[x, b1, t], x] == $\dfrac{\tt -1 + e^{2\,b[x,b1,t]}}{\tt 4 * x}$

\vskip2pt
eqn4 = a[7.36 * 10\^{ }-10, b1, t] == -1.72 * 10\^{ }-9 * 1 / t

eqn5 = b[7.36 * 10\^{ }-10, b1, t] == -1.72 * 10\^{ }-9 * 1 / t

eqn6 = (D[b[x, b1, t], x] /.\ x $\to$ 7.36 * 10\^{ }-10) == b1

ss1 = NDSolve[\char123 eqn1, eqn3, eqn4, eqn5, eqn6\char125, \char123a, b\char125,

\ \ \ \char123x, 10\^{ }-11, 10\^{ }-3\char125, \char123b1, -100, 100\char125, \char123t, 300, 500\char125, MaxSteps $\to$ 50 000]

pr1 = Rasterize[Plot3D[Evaluate[Exp[2 * a[x1, 14.31, t]] /.\ ss], \char123t,
300, 500\char125,

\ \ \ \char123x1, 10\^{ }-11, 10\^{ }-3\char125, ColorFunction $\to$
"RustTones"], ImageResolution $\to$ 90]

pr2 = Rasterize[Plot3D[Evaluate[Exp[2 * b[x1, 14.31, t]] /.\ ss], \char123t,
300, 500\char125,

\ \ \ \char123x1, 10\^{ }-11, 10\^{ }-3\char125, ColorFunction $\to$
"RustTones"], ImageResolution $\to$ 90]

aq1 = $\tt 2*e^{b[x[t],b1[t],t]}*(3 + e^{b[x[t],b1[t],t]}x[t])*
(b^{(0,0,1)}[x[t], b1[t], t]
+b1\char19[t] * b^{(0,1,0)}[x[t],b1[t],t])$

\ \ \ +$\tt x\char19[t] * (-7 + e^{2\,b[x[t],b1[t],t]} + 2*e^{b[x[t],b1[t],t]}
* (3 + e^{b[x[t],b1[t],t]}) * b^{(1,0,0)}[x[t],b1[t], t])$ ==

\ \ \ 0 /.\ ss1

aq2 = $\tt 0.000036400000000000004 * x[t] * x\char19[t] - 2*e^{-2\,b[x[t],b1[t],t]}
* (b^{(0,0,1)}[x[t], b1[t], t])$ +

\ \ \ $\tt b1\char19[t] * b^{(0,1,0)}[x[t], b1[t], t] + x\char19[t] * b^{(1,0,0)}
[x[t], b1[t], t])$ == 0 /.\ ss1

aq3 = x[430.]\ == 0.00365368 /.\ ss1

aq4 = b1[430.]\ == 14.3169 /.\ ss1

ss2 = NDSolve[\char123aq1, aq2, aq3, aq4\char125, \char123x, b1\char125,
\char123t, 300, 500\char125, MaxSteps $\to$ 50 000]

hr1 = Plot[Evaluate[x[t] /.\ ss2], \char123t, 300, 500\char125]

hr2 = Plot[Evaluate[b1[t] /.\ ss2], \char123t, 300, 500\char125]

}
\medskip

For a dust filled model.

\medskip
{\parindent 0pt
\tt eqn1 = D[D[b[x, b1, t], x], x] == $\dfrac{\tt 9 - 10 * e^{2\,b[x,b1,t]}
+ e^{4\,b[x,b1,t]}}{\tt 8 * x^2}$

\vskip2pt
eqn2 =  D[D[f[x, b1, t], x], x] == $\dfrac{\tt -1 + e^{2\,b[x,b1,t]}}
{\tt 4 * x^2}$

\vskip4pt
eqn3 = D[a[x, b1, t], x] == $\dfrac{\tt -1 + e^{2\,b[x,b1,t]}}{\tt 4 * x}$

\vskip2pt
eqn4 = a[7.36 * 10\^{ }-10, b1, t] == -3.44 * 10\^{ }-12 * Coth[0.003 * t]

eqn5 = b[7.36 * 10\^{ }-10, b1, t] == 3.44 * 10\^{ }-12 * Coth[0.003 * t]

eqn6 = (D[b[x, b1, t], x] /.\ x $\to$ 7.36 * 10\^{ }-10) == b1

ss = NDSolve[\char123 eqn1, eqn3, eqn4, eqn5, eqn6\char125, \char123a, b\char125,

\ \ \ \char123x, 10\^{ }-11, 10\^{ }-3\char125, \char123b1, -100, 100\char125, \char123t, 300, 500\char125, MaxSteps $\to$ 50 000]

aq1 = $\tt 0.000292 * x[t] * x\char19[t] - 2*e^{-2\,b[x[t],b1[t],t]}
* (b^{(0,0,1)}[x[t], b1[t], t]$ +

\ \ \ $\tt b1\char19[t] * b^{(0,1,0)}[x[t], b1[t], t] + x\char19[t] * b^{(1,0,0)}[x[t],b1[t], t])$ == 0 /.\ ss

aq2 = $\tt \dfrac12 \X3(-0.001 * Coth[0.003 * t] + 3. *\hbox{\^{ }}-6 * t *
 Csch[0.003 * t]^2 +{}$

\ \ \ $\tt \dfrac1{x[t]^2} * e^{-a[x[t],b1[t],t]} * \X2(-x[t]^2 \X1(\X1(-1 +
Coth[b[x[t],b1[t],t]]\Y1) * a^{(0,0,1)}[x[t],b1[t],t] +{}$

\ \ \ \ $\tt (-1 +
Coth[b[x[t],b1[t],t]])\X1(b1\char19[t] * a^{(0,1,0)}[x[t],b1[t],t] + {}$

\ \ \ \ $\tt x\char19[t] * a^{(1,0,0)}[x[t],b1[t],t]\Y1) + Csch[b[x[t],b1[t],t]]^2
(b^{(0,0,1)}[x[t],b1[t],t] + {}$

\ \ \ \ $\tt b1\char19[t] * b^{(0,1,0)}[x[t],b1[t],t] + x\char19[t] * b^{(1,0,0)}
[x[t],b1[t],t]\Y1)\Y1) + {}$

\ \ \ $\tt \X2(4 e^{2\,b[x[t],b1[t],t]}\X1(Sinh[b[x[t],b1[t],t]] * \X1(x[t]
* (a^{(1,0,0)}[x[t],b1[t],t]+{}$

\ \ \ \ $\tt b1\char19[t] * a^{(0,1,0)}[x[t],b1[t],t])
+x\char19[t] * (1 + x[t] * a^{(1,0,0)}[x[t],b1[t],t])\Y1) + {}$

\ \ \ $\tt Cosh[b[x[t],b1[t],t]]x[t] * (b^{(0,0,1)}[x[t],b1[t],t]
+b1\char19[t] * b^{(0,1,0)}[x[t],b1[t],t], t] + {}$

\ \ \ \ $\tt x\char19[t] * b^{(1,0,0)}[x[t],b1[t],t])\Y1)\Y1) \Big/
(-1 + e^{2\,b[x[t],b1[t],t]})^2\Y2)\Y3)$ == 0 /.\ ss

aq3 = x[430.]\ == 0.00365368

aq4 = b1[430.]\ == 14.3169

ss1 = NDSolve[\char123aq1, aq2, aq3, aq4\char125, \char123x, b1\char125,
\char123t, 300, 500\char125, MaxSteps $\to$ 50 000]

Rasterize[Plot3D[Evaluate[Exp[2 * a[x1, b1, t]] /.\ ss] /.\ b1 $\to$ 14.31,

\ \ \char123t, 300, 500\char125, \char123x1, 10\^{ }-11, 10\^{ }-3\char125, ColorFunction $\to$
"RustTones"],

\ \ ImageResolution $\to$ 90]

Rasterize[Plot3D[Evaluate[Exp[2 * b[x1, b1, t]] /.\ ss] /.\ b1 $\to$ 14.31,

\ \ \char123t, 300, 500\char125,
\char123x1, 10\^{ }-11, 10\^{ }-3\char125, ColorFunction $\to$
"RustTones"],

\ \ ImageResolution $\to$ 90]

}
\medskip

For a CDM$\La$ model.

\medskip
{\parindent 0pt
\tt eqn1 = D[D[b[x, b1, t], x], x] == $\dfrac{\tt 9 - 10 * e^{2\,b[x,b1,t]}
+ e^{4\,b[x,b1,t]}}{\tt 8 * x^2}$

\vskip2pt
eqn2 =  D[D[f[x, b1, t], x], x] == $\dfrac{\tt -1 + e^{2\,b[x,b1,t]}}
{\tt 4 * x^2}$

\vskip4pt
eqn3 = D[a[x, b1, t], x] == $\dfrac{\tt -1 + e^{2\,b[x,b1,t]}}{\tt 4 * x}$

\vskip2pt
eqn4 = a[7.36 * 10\^{ }-10, b1, t] == -1.72 * 10\^{ }-9 * 1 / t

eqn5 = b[7.36 * 10\^{ }-10, b1, t] == -1.72 * 10\^{ }-9 * 1 / t

eqn6 = (D[b[x, b1, t], x] /.\ x $\to$ 7.36 * 10\^{ }-10) == b1

ss1 = NDSolve[\char123 eqn1, eqn3, eqn4, eqn5, eqn6\char125, \char123a, b\char125,

\ \ \ \char123x, 10\^{ }-11, 10\^{ }-3\char125, \char123b1, -100, 100\char125, \char123t, 300, 500\char125, MaxSteps $\to$ 50 000]

pr1 = Rasterize[Plot3D[Evaluate[Exp[2 * a[x1, b1, t]] /.\ ss]/.\ b1 $\to$ 14.31,

\ \ \char123t, 300, 500\char125,
\char123x1, 10\^{ }-11, 10\^{ }-3\char125, ColorFunction $\to$
"RustTones"],

\ \ ImageResolution $\to$ 90]

pr2 = Rasterize[Plot3D[Evaluate[Exp[2 * b[x1, b1, t]] /.\ ss]/.\ b1 $\to$ 14.31,

\ \ \char123t, 300, 500\char125, \char123x1, 10\^{ }-11, 10\^{ }-3\char125, ColorFunction $\to$
"RustTones"],

\ \ ImageResolution $\to$ 90]

}

\medskip
For a dust filled model.

\medskip
In all listings {\tt a} means $A$, {\tt b} means $B$, {\tt b1} means $B_1$,
{\tt x} or {\tt x1} mean $r$.

\goodbreak
\medskip
{\parindent 0pt
\tt eqn1 = D[D[b[x], x], x] == $\dfrac{\tt 9 - 10 * e^{2\,b[x]}
+ e^{4\,b[x]}}{\tt 8 * x^2}$

\vskip4pt
eqn3 = D[a[x], x] == $\dfrac{\tt -1 + e^{2\,b[x]}}{\tt 4 * x}$

\vskip2pt
eqn4 = a[7.36 * 10\^{ }-10] == -4 * 10\^{ }-12

eqn5 = b[7.36 * 10\^{ }-10] == 4 * 10\^{ }-12

eqn6 = (D[b[x], x] /.\ x $\to$ 7.36 * 10\^{ }-10) == 14.3117

\goodbreak
ss = NDSolve[\char123 eqn1, eqn3, eqn4, eqn5, eqn6\char125, \char123a,
b\char125, \char123x, 10\^{ }-11, 10\^{ }-2\char125,

\ \ \  \  MaxSteps $\to$ 50 000]

aq1 = $\X3($x\char19[t] == $\tt e^{-b[x[t]]}$ *

\ Sqrt$\X3[\X2(\tt -1 + \dfrac{
\hbox{(4.48 *\^{ }-9\, +\, a1)}^2\ e^{-2\,a[x[t]]}}{a1^2}
-\dfrac{\hbox{2.2089999999999993 *\^{ }-33 a1(-1.\,+\,x1}^2)}{x[t]^2}\Y2)\Y3]$

\  /.\ \char123a1 $\to$ 1.5 * 10\^{ }-6, x1 $\to$ 3\char125 $\Y3)$ /.\ ss

aq2 = $\!\X3($y\char19[f] == $\tt \dfrac{1}{a1^{3/2} \hbox{ Sqrt[(-1.\,+\,x1}^2)]}
\ e^{-(a[y[f]] + b[y[f]])}\ Sqrt[\X1($4.5269352648257145 *\^{ }32

\ \ \ \ (4.4799999610499785 * \^{ }-9 + a1) (4.480000038950022 * \^{ }-9 + a1)
y[f]$\tt ^4$

\ \ \ \ + a1$\tt^2\ e^{2\,a[y[f]]}$ y[f]$\tt^2$
 (-4.5269352648257145 * \^{ }32 * y[f]$\tt^2$ + a1
(-1.\,+\,$\tt x1^2))\Y1)\Y1]$

\ \ \ \ /.\ \char123
a1 $\to$ 1.5 * 10\^{ }-6, x1 $\to$ 3\char125 $\Y2)$ /.\ ss

aq3 = y[0] == 10\^{ }-10

aq4 = y[0] == 10\^{ }-10

ss1 = NDSolve[\char123aq1, aq3\char125, x, \char123t, 0, 10\char125,
MaxSteps $\to$ 50000]

ss2 = NDSolve[\char123aq2, aq4\char125, y, \char123f, 0, 22 * Pi\char125,
MaxSteps $\to$ 50000]

hr1 = PolarPlot[Evaluate[x[t] /.\ ss1], \char123t, 0, 10\char125]

hr2 = PolarPlot[Evaluate[y[f] /.\ ss2], \char123f, 0, 3 * Pi\char125]

aaq1 = (f\char19[t] == x[t]\^{ }-2 * 0.47 * 10\^{ }-16 * Sqrt[(x1\^{ }2
- 1) * a1]

\ \ \  \ /.\ \char123a1 $\to$ 20, x1 $\to$ 7\char125) /.\ ss1

aaq2 = f[0] == 0

ss3 = NDSolve[\char123aaq1, aaq2\char125, f, \char123t, 0, 10\char125,
MaxSteps $\to$ 50000]

hr3 = Plot[Evaluate[f[t] /.\ ss3], \char123t, 0, 0.10\char125]
}
\medskip

The above listing gives a programme in Mathematica 7 to calculate an orbit
for a massive point body. Here {\tt a} means~$A$, {\tt b} means~$B$, {\tt x},
{\tt y} mean~$r$, {\tt a1} means $a_h$, {\tt x1} means~$e$, {\tt f} means~$\vf$.

\medskip

{\parindent 0pt
\tt eqn1 = D[D[b[x], x], x] == $\dfrac{\tt 9 - 10 * e^{2\,b[x]}
+ e^{4\,b[x]}}{\tt 8 * x^2}$

\vskip4pt
eqn3 = D[a[x], x] == $\dfrac{\tt -1 + e^{2\,b[x]}}{\tt 4 * x}$

\vskip2pt
eqn4 = a[7.36 * 10\^{ }-10] == -4 * 10\^{ }-12

eqn5 = b[7.36 * 10\^{ }-10] == 4 * 10\^{ }-12

eqn6 = (D[b[x], x] /.\ x $\to$ 7.36 * 10\^{ }-10) == 14.3117

ss = NDSolve[\char123 eqn1, eqn3, eqn4, eqn5, eqn6\char125, \char123a,
b\char125, \char123x, 10\^{ }-11, 10\^{ }-2\char125,

\ \ \  \  MaxSteps $\to$ 50 000]

aq2 = y\char19[f] == $\tt e^{-b[y[f]]}\ y[f]\ Sqrt\X3[\X3(
-1 + \dfrac{\hbox{\tt 1.4537184246265283 *\^{ }17 * }e^{-2\,a[y[f]]}\
y[f]^2}{a1^3\ (-1. + x1^2)} \Y3)$ /.

\ \ \ \ \char123a1 $\to$ 1.5 * $\tt 10^{-6}$, x1 $\to$ 3\char125$\Y3]$ /.\ ss

aq3 = x[0] == 10\^{ }-10

aq4 = y[0] == 10\^{ }-10

ss2 = NDSolve[\char123aq2, aq4\char125, y, \char123f, 0, 2 * Pi\char125,
MaxSteps $\to$ 50000]

hr2 = PolarPlot[Evaluate[y[f] /.\ ss2], \char123f, 0, 3 * Pi\char125]

}
\medskip

The above listing gives a programme written in Mathematica~7 to calculate an
orbit for a photon. Here {\tt a} means~$A$, {\tt b} means~$B$, {\tt x,
y} mean~$r$, {\tt a1} means $a_h$, {\tt x1} means~$e$, {\tt f} means~$\vf$.

\medskip

{\parindent 0pt
\tt eqn1 = D[D[b[x], x], x] == $\dfrac{\tt 9 - 10 * e^{2\,b[x]}
+ e^{4\,b[x]}}{\tt 8 * x^2}$

\vskip4pt
eqn3 = D[a[x], x] == $\dfrac{\tt -1 + e^{2\,b[x]}}{\tt 4 * x}$

\vskip2pt
eqn4 = a[7.36 * 10\^{ }-10] == -4 * 10\^{ }-12

eqn5 = b[7.36 * 10\^{ }-10] == 4 * 10\^{ }-12

eqn6 = (D[b[x], x] /.\ x $\to$ 7.36 * 10\^{ }-10) == 14.3117

ss = NDSolve[\char123 eqn1, eqn3, eqn4, eqn5, eqn6\char125, \char123a,
b\char125, \char123x, 10\^{ }-11, 10\^{ }-5\char125,

\ \ \  \  MaxSteps $\to$ 50 000]

aaq1 = (D[ty[x], x] == -2 * 0.33 * Pi * D[b[x], x] * Exp[a[x]] *

\ \ \ \ (Sqrt[(0.448 * 10\^{ }-8 * 1 / a1)\^{ }2 * Exp[a[x]] -
(0.47 * 10\^{ }-16)\^{ }2 * a1 *

\ \ \ \ (x1\^{ }2 - 1) * x\^{ }2]\^{ }-1 /.\ ss) /.\ \char123a1 $\to$ 15 * 10\^{ }-5, x1 $\to$ 3\char125

aaq3 = ty[10\^{ }-10] == 0

ss1 = NDSolve[\char123aaq1, aaq3\char125, ty, \char123x, 10\^{ }-11,
10\^{ }-5\char125, MaxSteps $\to$ 50000]

\goodbreak
aq1 = x\char19[t] == $\tt \dfrac{1}{a1}\, e^{-b[x[t]]}$

\ * Sqrt
$\X3[\X3(\tt 2.00704 *\hat{ }-17\ e^{-2\,a[x[t]]} +
\dfrac{2.2089999999999993 * \hat{ }-33\ a1^3 (-1. + x1^2)}{x[t]^2} \Y3)
\Y3]$ /.\

\ \ \ \ \char123a1 $\to$ 15 * 10\^{ }-5, x1 $\to$ 3\char125 /.\ ss

aq3 = x[0] == 10\^{ }=10

ss2 = NDSolve[\char123aq1, aq3\char125, x, \char123t, 0, 10\char125,
MaxSteps $\to$ 50000]

hr1 = Plot[Evaluate[Evaluate[ty[x[t]] /.\ ss1] /.\ ss2], \char123t, 0,
0.0010\char125]

}
\medskip

The above listing gives a programme written in Mathematica~7 to calculate $\D
T(r(\si))$ on a photon orbit. Here {\tt ty[x[t]]} means $\D T(r(\si))$,
{\tt x} means $r$, {\tt x1} means $e$, {\tt a} means $A$, {\tt b} means $B$,
{\tt t} means $\si$.

\medskip

{\parindent 0pt
\tt eq1a0 = a\char19\char19[x] == $\tt\dfrac1{f[x]^2\ (3.1116959999999984\
\hbox{*\^{ }-82 }+\ f[x]^2)^2}$

\ \ \ \ (-0.04371584699453552 \ $\tt e^{122\,q[x]}\ a[x]^2\ f[x]^2$
(3.1116959999999995\ *\^{ }-82 + $\tt f[x]^2$)

\ \  \ \ (4.6675439999999955 *\^{ }-82 + $\tt f[x]^2$) + 0.044444444444444446 $\tt
e^{120\,q[x]}\ a[x]^2$)

\ \  \ \ $\tt f[x]^2$ (3.111695999999998 *\^{ }-82 + $\tt f[x]^2$) (4.667543999999999
*\^{ }-82 + $\tt f[x]^2$) +

\ \ \ \ $\tt f[x]^2$ (3.111695999999998 *\^{ }-82 + $\tt f(x)^2$)
(-2.\ a[x] (1.555847999999999 *\^{ }-82 +

\ \ \ \ $\tt f[x]^2$) - 3.\ (2.074463999999999 *\^{ }-82 +
$\tt f[x]^2$) a\char19[x]) + f[x] (-8.\ a[x]

\ \ \ \ (2.542216107572787 *\^{ }-82 + $\tt f[x]^2$)
(5.2370238924272096 *\^{ }-82 + $\tt f[x]^2$) -

\ \ \ \ 6.\ (3.1116959999999966 *\^{ }-82 + $\tt f[x]^2$)
(5.186159999999999 *\^{ }-82 + $\tt f[x]^2$)

\ \ \ \ a\char19[x]) f\char19[x] + a[x] (-1.7428773593548783 \^{ }-162 -
4.356374399999998 *\^{ }-81

\ \ \ \ $\tt f[x]^2$ - 4.\ $\tt f[x]^4) f\char19[x]^2$ - 4.\ a[x] $\tt
f[x]^2$ (3.1116959682926 *\^{ }-82 + $\tt f[x]^2$)

\ \ \ \ (3.111696031370737 *\^{ }-82 + $\tt f[x]^2$) q\char19[x]$^2$)

\smallskip
eq2a0 = f\char19\char19[x] == $\tt \dfrac 1
{a[x]\ f[x]\ (3.1116959999999984 *\hbox{\^{ }-82 + $\tt f[x]^2$})}$

\ \ \  \ $\X1(\tt e^{120\,q[x]}\ a[x]^2\ f[x]^2$ (-3.457439999999998\^{ }-84
+ 3.4007606557377036 *\^{ }-84 $\tt e^{2\,q[x]}$ +

\ \ \ \ (-0.011111111111111112 + 0.01092896174863388 $\tt e^{2\,q[x]}$)
$\tt f[x]^2$) +

\ \ \ \ f[x] (3.1116959999999984 *\^{ }-82 + $\tt1.\ f[x]^2$) a\char19[x]
f\char19[x] + a[x] f\char19[x] (1.\ f[x]

\ \ \ \ (3.1116959999999984 *\^{ }-82 + $\tt f[x]^2$)
+ (9.335087999999995 *\^{ }-82 +

\ \ \ \ $\tt1.\ f[x]^2$) f\char19[x])$\Y1)$

eq3a0 = q\char19\char19[x] == $\tt \dfrac1
{\hbox{720 a[x] f[x] (3.1116959999999984 *\^{ }-82 + $\tt f[x]^2$)}^2}$

\ \ \ \ $\X1($720.\ f[x] (3.1116959999999984 *\^{ }-82 + $\tt f[x]^2$)$^2$
a\char19[x] q\char19[x] +

\ \ \ \ a[x] (f[x] ($\tt e^{120\,q[x]}$ a[x] $\tt f[x]^2$
(1.879464383999999 *\^{ }-79 - 1.867017599999999

\ \ \ \ *\^{ }-79 $\tt e^{2\,q[x]}$ +
(604.\ - 600.\ $\tt e^{2\,q[x]}$) $\tt f[x]^2$) + 1.3666568831999993 *\^{ }-77
f\char19[x]$^2$) +

\ \ \ \ 720.\ (3.1116959999999984 *\^{ }-82 + $\tt f[x]^2$)
(3.1116959999999984 *\^{ }-82 f[x] +

\ \ \ \ $\tt f[x]^3$ + 6.223391999999997 *\^{ }-82 f\char19[x]) q\char19[x])$\Y1)$

aqq1 = a[0] == 0.1

aqq2 = a\char19[0] == 0.

aqq3 = q[0] == 0.1

aqq4 = q\char19[0] == 0.01

aqq5 = f[0] == 0.1

aqq6 = f\char19[0] == 0

ss = NDSolve[\char123eq1a0, eq2a0, eq3a0, aqq1, aqq2, aqq3, aqq4, aqq5,
aqq6\char125,

\ \ \ \ \char123a, q, f\char125, \char123x, 0, 10\^{ }-2\char125,
MaxSteps $\to$ 5000000]

aqq7 = x\char19[f1] == (x[f1] * Sqrt[x[f1]\^{ }2 *
 0.47 * 10\^{ }-16)\^{ }-2 * (x1\^{ }2 - 1)\^{ }-1 *

\ \ \ \ a1\^{ }-1 * ((1 + 0.448 * 10\^{ }-8 * 1 / a1)\^{ }2 * Exp[2 * x[f1]]
* f[x[f1]]\^{ }-6 *

\ \ \ \ l\^{ }4 - a[x[f1]]\^{ }-1 - a[x[f1]]\^{ }-1 * x[f1]\^{ }2 * (0.47 * 10\^{ }-16)\^{ }-2
*

\ \ \ \ (x1\^{ }2 - 1)\^{ }-1 * a1\^{ }-1] /.\
\char123a1 $\to$ 20, x1 $\to$ 7, l $\to$ 0.42 * 10\^{ }20\char125) /.\ ss

aqq10 = x[0] == 10\^{ }-10

ss1 = NDSolve[\char123aqq7, aqq10\char125, x, \char123f1, 0, 22 * Pi\char125,
MaxSteps $\to$ 5000000]

hr1 = PolarPlot[Evaluate[x[f1] /.\ ss1], \char123f1, 0, 2 * Pi\char125]

}
\medskip
The above programme calculates $r(\vf)$, for an
orbit of a point massive body moving in the background \gr al field from the
\NK{\JT\ }. {\tt x}~means $r$, {\tt f1} means $\vf$,
{\tt x1} means $e$, {\tt a1} means $a_h$, {\tt l} means~$\ell$.

\medskip
{\parindent 0pt
\tt eq1a0 = a\char19\char19[x] == $\tt\dfrac1{f[x]^2\ (3.1116959999999984\
\hbox{*\^{ }-82 }+\ f[x]^2)^2}$

\ \ \ \ (-0.04371584699453552 \ $\tt e^{122\,q[x]}\ a[x]^2\ f[x]^2$
(3.1116959999999995\ *\^{ }-82 + $\tt f[x]^2$)

\ \  \ \ (4.6675439999999955 *\^{ }-82 + $\tt f[x]^2$) + 0.044444444444444446 $\tt
e^{120\,q[x]}\ a[x]^2$

\ \  \ \ $\tt f[x]^2$ (3.111695999999998 *\^{ }-82 + $\tt f[x]^2$) (4.667543999999999
*\^{ }-82 + $\tt f[x]^2$) +

\ \ \ \ $\tt f[x]^2$ (3.111695999999998 *\^{ }-82 + $\tt f(x)^2$)
(-2.\ a[x] (1.555847999999999 *\^{ }-82 +

\ \ \ \ $\tt f[x]^2$) - 3.\ (2.074463999999999 *\^{ }-82 +
$\tt f[x]^2$) a\char19[x]) + f[x] (-8.\ a[x]

\ \ \ \ (2.542216107572787 *\^{ }-82 + $\tt f[x]^2$)
(5.2370238924272096 *\^{ }-82 + $\tt f[x]^2$) -

\ \ \ \ 6.\ (3.1116959999999966 *\^{ }-82 + $\tt f[x]^2$)
(5.186159999999999 *\^{ }-82 + $\tt f[x]^2$)

\ \ \ \ a\char19[x]) f\char19[x] + a[x] (-1.7428773593548783 \^{ }-162 -
4.356374399999998 *\^{ }-81

\ \ \ \ $\tt f[x]^2$ - 4.\ $\tt f[x]^4) f\char19[x]^2$ - 4.\ a[x] $\tt
f[x]^2$ (3.1116959682926 *\^{ }-82 + $\tt f[x]^2$)

\ \ \ \ (3.111696031370737 *\^{ }-82 + $\tt f[x]^2$) q\char19[x]$^2$)

\smallskip
eq2a0 = f\char19\char19[x] == $\tt \dfrac 1
{a[x]\ f[x]\ (3.1116959999999984 *\hbox{\^{ }-82 + $\tt f[x]^2$})}$

\ \ \  \ $\X1(\tt e^{120\,q[x]}\ a[x]^2\ f[x]^2$ (-3.457439999999998\^{ }-84
+ 3.4007606557377036 *\^{ }-84 $\tt e^{2\,q[x]}$ +

\ \ \ \ (-0.011111111111111112 + 0.01092896174863388 $\tt e^{2\,q[x]}$)
$\tt f[x]^2$) +

\ \ \ \ f[x] (3.1116959999999984 *\^{ }-82 + $\tt1.\ f[x]^2$) a\char19[x]
f\char19[x] + a[x] f\char19[x] (1.\ f[x]

\ \ \ \ (3.1116959999999984 *\^{ }-82 + $\tt f[x]^2$)
+ (9.335087999999995 *\^{ }-82 +

\ \ \ \ $\tt1.\ f[x]^2$) f\char19[x])$\Y1)$

\smallskip
eq3a0 = q\char19\char19[x] == $\tt \dfrac1
{\hbox{720 a[x] f[x] (3.1116959999999984 *\^{ }-82 + $\tt f[x]^2$)}^2}$

\ \ \ \ $\X1($720.\ f[x] (3.1116959999999984 *\^{ }-82 + $\tt f[x]^2$)$^2$
a\char19[x] q\char19[x] +

\ \ \ \ a[x] $\X1($f[x] ($\tt e^{120\,q[x]}$ a[x] $\tt f[x]^2$
(1.879464383999999 *\^{ }-79 - 1.867017599999999

\ \ \ \ *\^{ }-79 $\tt e^{2\,q[x]}$ +
(604.\ - 600.\ $\tt e^{2\,q[x]}$) $\tt f[x]^2$) + 1.3666568831999993 *\^{ }-77
f\char19[x]$^2$) +

\ \ \ \ 720.\ (3.1116959999999984 *\^{ }-82 + $\tt f[x]^2$)
(3.1116959999999984 *\^{ }-82 f[x] +

\ \ \ \ $\tt f[x]^3$ + 6.223391999999997 *\^{ }-82 f\char19[x]) q\char19[x])$\Y1)$

aqq1 = a[0] == 0.1

aqq2 = a\char19[0] == 0.

aqq3 = q[0] == 0.1

aqq4 = q\char19[0] == 0.01

aqq5 = f[0] == 0.1

aqq6 = f\char19[0] == 0

ss = NDSolve[\char123eq1a0, eq2a0, eq3a0, aqq1, aqq2, aqq3, aqq4, aqq5,
aqq6\char125,

\ \ \ \ \char123a, q, f\char125, \char123x, 0, 10\^{ }-4\char125,
MaxSteps $\to$ 5000000]

hr1 = Plot[Evaluate[a[x] /.\ ss], \char123x, 10\^{ }-9, 10\^{ }-5\char125]

hr2 = Plot[Evaluate[q[x] /.\ ss], \char123x, 10\^{ }-9, 10\^{ }-5\char125]

hr3 = Plot[Evaluate[f[x] /.\ ss], \char123x, 10\^{ }-9, 10\^{ }-5\char125]

hr4 = Plot[Evaluate[q\char19[x] /.\ ss], \char123x, 10\^{ }-9, 10\^{ }-5\char125]

hr5 = Plot[Evaluate[Exp[-122 * q[x]] /.\ ss], \char123x, 10\^{ }-9, 10\^{ }-5\char125]

hr6 = Plot[Evaluate[f[x]\^{ }2 * Exp[-x] /.\ ss], \char123x, 10\^{ }-9, 10\^{ }-5\char125]

hr7 = Plot[Evaluate[f[x]\^{ }6 * Exp[-2 * x] * a[x]\^{ }-1 *
(0.42 * 10\^{ }-20)\^{ }-4 /.\ ss],

\ \ \ \ \char123x, 10\^{ }-9, 10\^{ }-5\char125]

}
\medskip

The above programme calculates $\a(r)$, $f(r)$, $\vF(r)$, $G_{\rm eff}(r)/G_N$,
$\o(r)$, $\g(r)$, $\pz\vF r(r)$ for initial conditions given by \er{Da245}.
Here {\tt a} means $\a$, {\tt x} means $r$, {\tt q} means $\vF$, $\o(r)$,
$\g(r)$ and $G_{\rm eff}(r)$ are given by the formulae \er{Da254}, \er{Da255},
\er{2.6} for $n=120$, {\tt a1} means $a_h$, {\tt x1}  means $e$.

\medskip
{\parindent 0pt
\tt eqn1 = D[D[b[x], x], x] == $\tt\dfrac{9 - 10 * e^{2\,b[x]}
+ e^{4\,b[x]}}{8 * x^2}$

\vskip4pt
eqn3 = D[a[x], x] == $\dfrac{\tt -1 + e^{2\,b[x]}}{\tt 4 * x}$

\vskip2pt
eqn4 = a[3.7 * 100] == -4 * 10\^{ }-12

eqn5 = b[3.7 * 100] == 4 * 10\^{ }-12

eqn6 = (D[b[x], x] /.\ x $\to$ 3.7 * 100) == 0.

\goodbreak
ss = NDSolve[\char123eqn1, eqn3, eqn4, eqn5, eqn6\char125, \char123a,
b\char125, \char123x, 10\^{ }-3, 10\^{ }5\char125,

\ \ \ \ MaxSteps $\to$ 50000]

hr1 = Plot[Evaluate[a[x] /.\ ss], \char123x, 10\^{ }-3, 10\^{ }5\char125]

hr2 = Plot[Evaluate[b[x] /.\ ss], \char123x, 10\^{ }-3, 10\^{ }5\char125]

hr3 = Plot[Evaluate[Exp[2 * a[x]] /.\ ss], \char123x, 10\^{ }-3, 10\^{ }5\char125]

hr4 = Plot[Evaluate[Exp[2 * b[x]] /.\ ss], \char123x, 10\^{ }-3, 10\^{ }5\char125]

hr5 = Plot[Evaluate[Exp[2 * (a[x] + b[x])] /.\ ss], \char123x, 10\^{ }-3, 10\^{ }5\char125]

hr6 = Plot[Evaluate[D[Exp[2 * a[x]], x] /.\ ss], \char123x, 10\^{ }-3, 10\^{ }5\char125]

hr7 = Plot[Evaluate[D[Exp[2 * b[x]], x] /.\ ss], \char123x, 10\^{ }-3, 10\^{ }5\char125]

hr8 = Plot[Evaluate[D[Exp[2 * (a[x] + b[x])], x] /.\ ss], \char123x, 10\^{ }-3, 10\^{ }5\char125]

hr9 = Plot[Evaluate[Exp[2 * a[x]] /.\ ss], \char123x, 10\^{ }-3, 10\^{ }-2\char125]

hr10 = Plot[Evaluate[Exp[2 * b[x]] /.\ ss], \char123x, 10\^{ }-3, 10\^{ }-2\char125]

hr11 = Plot[Evaluate[D[Exp[2 * (a[x] + b[x])], x] /.\ ss], \char123x, 10\^{ }-3, 10\^{ }-2\char125]

}
\medskip
The above programme written in Mathematica 7 calculates and plots \f s from
Fig. \ref{kkf}, {\tt x} means $r$, {\tt a}---$A$, {\tt b}---$B$.

\medskip
{\parindent 0pt
\tt eqn1 = D[D[b[x], x], x] == $\tt\dfrac{9 - 10 * e^{2\,b[x]}
+ e^{4\,b[x]}}{8 * x^2}$

\vskip4pt
eqn3 = D[a[x], x] == $\dfrac{\tt -1 + e^{2\,b[x]}}{\tt 4 * x}$

\vskip2pt
eqn4 = a[3.7 * 100] == -4 * 10\^{ }-12

eqn5 = b[3.7 * 100] == 4 * 10\^{ }-12

eqn6 = (D[b[x], x] /.\ x $\to$ 3.7 * 100) == 0.

eqn7 = D[D[q[x], x], x] == $-\dfrac{\tt -1 + e^{2\,b[x]}}{\tt 4 * x\hat{\ }2}\,
\tt e^{2\,b[x]}$

eqn8 = q[3.7 * 100] == 0.

eqn9 = (D[q[x], x] /.\ 3.7 * 100) == 0.

\goodbreak
ss = NDSolve[\char123eqn1, eqn3, eqn4, eqn5, eqn6, eqn7, eqn8, eqn9\char125,
\char123a, b, q\char125,

\ \ \ \ \char123x, 10\^{ }-3, 10\^{ }5\char125, MaxSteps $\to$ 50000]

hr1 = Plot[Evaluate[q[x] /.\ ss], \char123x, 10\^{ }-3, 10\^{ }5\char125]

hr2 = Plot[Evaluate[q\char19[x] /.\ ss], \char123x, 10\^{ }-3, 10\^{ }5\char125]

hr3 = Plot[Evaluate[q\char19\char19[x] /.\ ss], \char123x, 10\^{ }-3, 10\^{ }5\char125]

hr4 = Plot[Evaluate[q[x] /.\ ss], \char123x, 10\^{ }-3, 10\^{ }-2\char125]

hr5 = Plot[Evaluate[q\char19[x] /.\ ss], \char123x, 10\^{ }-3, 10\^{ }-2\char125]

hr6 = Plot[Evaluate[q\char19\char19[x], x] /.\ ss], \char123x, 10\^{ }-3, 10\^{ }-2\char125]

hr7 = Plot[Evaluate[Exp[-122\^{ }2 * q[x]] /.\ ss], \char123x, 10\^{ }-3, 10\^{ }-2\char125]

hr8 = Plot[Evaluate[D[Exp[-122\^{ }2 * q[x]], x] /.\ ss], \char123x, 10\^{ }-3, 10\^{ }-2\char125]

hr9 = Plot[Evaluate[Exp[-122\^{ }2 * q[x]] /.\ ss], \char123x, 10\^{ }-3, 10\^{ }5\char125]

hr10 = Plot[Evaluate[D[Exp[-122\^{ }2 * q[x]], x] /.\ ss], \char123x, 10\^{ }-3, 10\^{ }5\char125]

}
\medskip
The above programme written in Mathematica 7 calculates and plots \f s from
Fig. \ref{kkkf}, {\tt x} means $r$, {\tt a}---$A$, {\tt b}---$B$, {\tt
q}---$\wt\vF$.

\medskip
{\parindent 0pt
\tt eqn1 = D[D[b[x], x], x] == $\tt\dfrac{9 - 10 * e^{2\,b[x]}
+ e^{4\,b[x]}}{8 * x^2}$

\vskip4pt
eqn3 = D[a[x], x] == $\dfrac{\tt -1 + e^{2\,b[x]}}{\tt 4 * x}$

\vskip2pt
eqn4 = a[3.7 * 100] == -4 * 10\^{ }-12

eqn5 = b[3.7 * 100] == 4 * 10\^{ }-12

eqn6 = (D[b[x], x] /.\ x $\to$ 3.7 * 100) == 0.

eqn7 = D[D[q[x], x], x] == $-\dfrac{\tt -1 + e^{2\,b[x]}}{\tt 4 * x\hat{\ }2}\,
\tt e^{2\,b[x]}$

eqn8 = q[3.7 * 100] == 0.

eqn9 = (D[q[x], x] /.\ 3.7 * 100) == 0.

\goodbreak
ss = NDSolve[\char123eqn1, eqn3, eqn4, eqn5, eqn6, eqn7, eqn8, eqn9\char125,
\char123a, b, q\char125,

\ \ \ \ \char123x, 10\^{ }-3, 10\^{ }7\char125, MaxSteps $\to$ 50000]

k[x\_] := 3.57 / 2 * 10\^{ }24 * $\X1($-122\^{ }2 * q\char19[x]\^{ }2
* $\tt e^{-2\,b[x]}$ + (x\^{ }-2 * (1 - $\tt e^{-2\,b[x]}$) +

\ \ \ \ 2 / x * $\tt e^{-2\,b[x]}$ * b\char19[x]) - 0.28 * 10\^{ }24 *
(Exp[120 * 122 * q[x]] / 122 -

\ \ \ \ 1 / 120)$\Y1)$ * Exp[122\^{ }2 * q[x]]

hr4 = Plot[Evaluate[k[x] /.\ ss], \char123x, 10\^{ }-3, 10\^{ }-2\char125]

hr5 = Plot[Evaluate[k[x] /.\ ss], \char123x, 10\^{ }-3, 10\^{ }5\char125]

hr6 = Plot[Evaluate[k[x] /.\ ss], \char123x, 10\^{ }-5, 10\^{ }-3\char125]

hr7 = Plot[Evaluate[k[x] /.\ ss], \char123x, 10\^{ }5, 10\^{ }7\char125]

}
\medskip
The above programme written in Mathematica 7 calculates and plots \f s from
Fig. \ref{kkkfx}, {\tt x} means $r$, {\tt a}---$A$, {\tt b}---$B$, {\tt
q}---$\wt\vF$, {\tt k}---$\ov\rho$.

\medskip
{\parindent 0pt
\tt eqn1 = D[D[b[x], x], x] == $\tt\dfrac{9 - 10 * e^{2\,b[x]}
+ e^{4\,b[x]}}{8 * x^2}$

\vskip4pt
eqn3 = D[a[x], x] == $\dfrac{\tt -1 + e^{2\,b[x]}}{\tt 4 * x}$

\vskip2pt
eqn4 = a[3.7 * 100] == -4 * 10\^{ }-12

eqn5 = b[3.7 * 100] == 4 * 10\^{ }-12

eqn6 = (D[b[x], x] /.\ x $\to$ 3.7 * 100) == 0.

\goodbreak
ss = NDSolve[\char123eqn1, eqn3, eqn4, eqn5, eqn6\char125, \char123a,
b\char125, \char123x, 10\^{ }-3, 10\^{ }5\char125,

\ \ \ \ MaxSteps $\to$ 50000]

v1[x\_] := 1 / 2 * (1 - $\tt e^{2\,a[x]}$)

v2[x\_] := 1 / 2 * (1 - $\tt e^{-2\,b[x]}$)

v11[x\_] := 1 / 2 * (1 - $\tt e^{2\,a[x]}$  -4.89078 * 10\^{ }-9 * 1 / x)

v21[x\_] := 1 / 2 * (1 - $\tt e^{-2\,b[x]}$  -4.89078 * 10\^{ }-9 * 1 / x)

bb1[x\_] := D[v11[x], x]

bb2[x\_] := D[v21[x], x]

uu[x\_] := 5.21126 * 10\^{ }-15 * (Exp[-x] + 1 + x\^{ }2 + Sqrt[Pi] *
Erf[x]) * 1 / x

rr1[x\_] := v11[x] / uu[x]

rr2[x\_] := v21[x] / uu[x]

hr1 = Plot[Evaluate[v1[x] /.\ ss], \char123x, 10\^{ }-3, 10\^{ }5\char125]

hr2 = Plot[Evaluate[v2[x] /.\ ss], \char123x, 10\^{ }-3, 10\^{ }5\char125]

hr3 = Plot[Evaluate[v11[x] /.\ ss], \char123x, 10\^{ }-3, 10\^{ }5\char125]

hr4 = Plot[Evaluate[v1[x] /.\ ss], \char123x, 10\^{ }-3, 10\^{ }-2\char125]

hr5 = Plot[Evaluate[v2[x] /.\ ss], \char123x, 10\^{ }-3, 10\^{ }-2\char125]

hr6 = Plot[Evaluate[v11[x] /.\ ss], \char123x, 10\^{ }-3, 10\^{ }-2\char125]

hr7 = Plot[Evaluate[v21[x] /.\ ss], \char123x, 10\^{ }-3, 10\^{ }-2\char125]

hr8 = Plot[Evaluate[bb1[x] /.\ ss], \char123x, 10\^{ }-3, 10\^{ }-2\char125]

hr9 = Plot[Evaluate[v21[x] /.\ ss], \char123x, 10\^{ }-3, 10\^{ }5\char125]

hr10 = Plot[Evaluate[bb1[x] /.\ ss], \char123x, 10\^{ }-3, 10\^{ }5\char125]

hr11 = Plot[Evaluate[bb2[x] /.\ ss], \char123x, 10\^{ }-3, 10\^{ }5\char125]

hr12 = Plot[Evaluate[bb2[x] /.\ ss], \char123x, 10\^{ }-3, 10\^{ }-2\char125]

hr14 = Plot[Evaluate[uu[x] /.\ ss], \char123x, 10\^{ }-3, 10\^{ }-2\char125]

hr15 = Plot[Evaluate[uu[x] /.\ ss], \char123x, 10\^{ }-3, 10\^{ }5\char125]

hr16 = Plot[Evaluate[rr1[x] /.\ ss], \char123x, 10\^{ }-3, 10\^{ }5\char125]

hr17 = Plot[Evaluate[rr2[x] /.\ ss], \char123x, 10\^{ }-3, 10\^{ }5\char125]

hr18 = Plot[Evaluate[rr1[x] /.\ ss], \char123x, 10\^{ }-3, 10\^{ }-2\char125]

hr19 = Plot[Evaluate[rr2[x] /.\ ss], \char123x, 10\^{ }-3, 10\^{ }-2\char125]

}\medskip

The above programme written in Mathematica 7 calculates and plots \f s from
Fig. \ref{kkkb}. {\tt x} means $r$, {\tt a}---$A$, {\tt b}---$B$,
{\tt v1}---$V_1$, {\tt v2}---$V_2$, {\tt v11}---$\ov V_1$, {\tt v21}---$\ov
V_2$, {\tt bb1}---$b_1$, {\tt bb2}---$b_2$, {\tt uu}---$-\ov U$, {\tt
rr1}---$\eta_1$, {\tt rr2}---$\eta_2$.

\medskip

\vskip2pt
{\parindent 0pt
\tt eqn1 = D[D[b[x], x], x] == $\tt\dfrac{9 - 10 * e^{2\,b[x]}
+ e^{4\,b[x]}}{8 * x^2}$

\vskip4pt
eqn3 = D[a[x], x] == $\dfrac{\tt -1 + e^{2\,b[x]}}{\tt 4 * x}$

\vskip4pt
eqn4 = a[7.03] == -6.957012802275961 *\^{ }-10 * (0.5 + 0.00012069886778807667)

eqn5 = b[7.03] == 3.469214116529213 *\^{ }-13

eqn6 = b\char19[7.03] == -2.7206955894132885 *\^{ }-14

\vskip2pt
eqn7 = D[D[q[x], x], x] == - $\dfrac{\tt -1 + e^{2\,b[x]}}{\tt 4 * x\hat{\ }2}
\ \tt e^{2\,b[x]}$

\vskip2pt
eqn8 = q[7.03] == -9.568184581692126 *\^{ }-13

eqn9 = (D[q[x], x] /.\ x $\to$ 7.03) == -5.348841232880765 *\^{ }-15

ss = NDSolve[\char123eqn1, eqn3, eqn4, eqn5, eqn6, eqn7, eqn8, eqn9\char125,

\ \ \ \ \char123a, b, q\char125, \char123x, 10\^{ }-3, 100\char125,
MaxSteps $\to$ 500000]

hr1 = Plot[Evaluate[a[x] /.\ ss], \char123x, 10\^{ }-3, 100\char125]

hr2 = Plot[Evaluate[b[x] /.\ ss], \char123x, 10\^{ }-3, 100\char125]

hr3 = Plot[Evaluate[Exp[2 * a[x]] - 1 /.\ ss], \char123x, 10\^{ }-3, 100\char125]

hr4 = Plot[Evaluate[Exp[2 * b[x]] - 1 /.\ ss], \char123x, 10\^{ }-3, 100\char125]

hr5 = Plot[Evaluate[Exp[2 * (a[x] + b[x])] - 1 /.\ ss], \char123x, 10\^{ }-3, 100\char125]

hr6 = Plot[Evaluate[D[Exp[2 * a[x]], x] /.\ ss], \char123x, 10\^{ }-3, 100\char125]

hr7 = Plot[Evaluate[D[Exp[2 * b[x]], x] /.\ ss], \char123x, 10\^{ }-3, 100\char125]

hr8 = Plot[Evaluate[D[Exp[2 * (a[x] + b[x])], x] /.\ ss], \char123x, 10\^{ }-3, 100\char125]

hr9 = Plot[Evaluate[Exp[2 * a[x]] - 1 + 8.20516 * 10\^{ }-9 /.\ ss], \char123x, 10\^{ }-3,

\ \ \ \ 10\^{ }-2\char125]

hr10 = Plot[Evaluate[Exp[2 * b[x]] - 1 /.\ ss], \char123x, 10\^{ }-3, 10\^{ }-2\char125]

hr11 = Plot[Evaluate[D[Exp[2 * (a[x] + b[x])], x] /.\ ss], \char123x, 10\^{ }-3, 10\^{ }-2\char125]

hr12 = Plot[Evaluate[q[x] /.\ ss], \char123x, 10\^{ }-3, 10\^{ }-2\char125]

hr14 = Plot[Evaluate[q[x] /.\ ss], \char123x, 10\^{ }-3, 100\char125]

hr15 = Plot[Evaluate[q\char19[x] /.\ ss], \char123x, 10\^{ }-3, 10\^{ }-2\char125]

hr16 = Plot[Evaluate[q\char19[x] /.\ ss], \char123x, 10\^{ }-3, 100\char125]

hr17 = Plot[Evaluate[Exp[-122\^{ }2 * q[x]] - 1 /.\ ss], \char123x, 10\^{ }-3, 10\^{ }-2\char125]

hr18 = Plot[Evaluate[Exp[-122\^{ }2 * q[x]] - 1 /.\ ss], \char123x, 10\^{ }-3, 100\char125]

hr19 = Plot[Evaluate[D[Exp[-122\^{ }2 * q[x]], x] /.\ ss], \char123x, 10\^{ }-3, 10\^{ }-2\char125]

hr20 = Plot[Evaluate[D[Exp[-122\^{ }2 * q[x]], x] /.\ ss], \char123x, 10\^{ }-3, 100\char125]

}
\medskip

A programme written in Mathematica 7 to calculate and plot $A(r)$, $B(r)$,
$\ea2$, $\eb2$, $\wt\vF(r)$, $\frac{G\eff}{G_N}(r)$ and their \dv s; {\tt a}
means $A$, {\tt b} means $B$, {\tt q} means $\wt\vF$, {\tt x} means $r$.

\medskip

\vskip2pt
{\parindent 0pt
\tt eqn1 = D[D[b[x], x], x] == $\tt\dfrac{9 - 10 * e^{2\,b[x]}
+ e^{4\,b[x]}}{8 * x^2}$

\vskip4pt
eqn3 = D[a[x], x] == $\dfrac{\tt -1 + e^{2\,b[x]}}{\tt 4 * x}$

\vskip4pt
eqn4 = a[7.03] == -6.957012802275961 *\^{ }-10 * (0.5 + 0.00012069886778807667)

eqn5 = b[7.03] == 3.469214116529213 *\^{ }-13

eqn6 = b\char19[7.03] == -2.7206955894132885 *\^{ }-14

\vskip2pt
eqn7 = D[D[q[x], x], x] == - $\dfrac{\tt -1 + e^{2\,b[x]}}{\tt 4 * x\hat{\ }2}
\ \tt e^{2\,b[x]}$

\vskip2pt
eqn8 = q[7.03] == -9.568184581692126 *\^{ }-13

eqn9 = (D[q[x], x] /.\ x $\to$ 7.03) == -5.348841232880765 *\^{ }-15

ss = NDSolve[\char123eqn1, eqn3, eqn4, eqn5, eqn6, eqn7, eqn8, eqn9\char125,

\ \ \ \ \char123a, b, q\char125, \char123x, 10\^{ }-3, 100\char125,
MaxSteps $\to$ 500000]

k[x\_] := 3.57 / 2 * 10\^{ }-24 * $\X1($-122\^{ }2 * q\char19[x]
\^{ }2 * $\tt e^{-2\,b[x]}\Y1)$ + $\X1($x\^{ }-2 * (1 - $\tt e^{-2\,b[x]}$)

\ \ \ \ + 2 / x * $\tt e^{-2\,b[x]}$ * b\char19[x]$\Y1)$
- 0.28 * 10\^{ }24 * Exp[120 * 122 * q[x]] *

\ \ \ \ (Exp[2 * 122 * q[x]] / 122 - 1 / 120$\Y1)$ * Exp[122\^{ }2 * q[x]]

l[x\_] := k[x] - 0.0000682784

hr4 = Plot[Evaluate[l[x] /.\ ss], \char123x, 10\^{ }-3, 10\^{ }-2\char125]

hr8 = Plot[Evaluate[k\char19[x] /.\ ss], \char123x, 10\^{ }-3, 10\^{ }-2\char125]

hr5 = Plot[Evaluate[l[x] /.\ ss], \char123x, 10\^{ }-3, 100\char125]

hr9 = Plot[Evaluate[k\char19[x] /.\ ss], \char123x, 10\^{ }-3, 100\char125]

hr6 = Plot[Evaluate[l[x] /.\ ss], \char123x, 10\^{ }-5, 2 * 10\^{ }-3\char125]

hr10 = Plot[Evaluate[k\char19[x] /.\ ss], \char123x, 10\^{ }-5, 2 * 10\^{ }-3\char125]

}
\medskip

A programme written in Mathematica 7 to calculate and plot $\wt\rho(r)$ and
its \dv. {\tt k}~means~$\ov\rho$, {\tt l} means $\ov\rho$ minus a \ct,
{\tt x} means $r$, {\tt a} means $A$, {\tt b} means $B$.

\medskip

\vskip2pt
{\parindent 0pt
\tt eqn1 = D[D[b[x], x], x] == $\tt\dfrac{9 - 10 * e^{2\,b[x]}
+ e^{4\,b[x]}}{8 * x^2}$

\vskip4pt
eqn3 = D[a[x], x] == $\dfrac{\tt -1 + e^{2\,b[x]}}{\tt 4 * x}$

\vskip4pt
eqn4 = a[7.03] == -6.957012802275961 *\^{ }-10 * (0.5 + 0.00012069886778807667)

eqn5 = b[7.03] == 3.469214116529213 *\^{ }-13

eqn6 = b\char19[7.03] == -2.7206955894132885 *\^{ }-14

\vskip2pt
eqn7 = D[D[q[x], x], x] == - $\dfrac{\tt -1 + e^{2\,b[x]}}{\tt 4 * x\hat{\ }2}
\ \tt e^{2\,b[x]}$

\vskip2pt
eqn8 = q[7.03] == -9.568184581692126 *\^{ }-13

eqn9 = q\char19[7.03] == -5.348841232880765 *\^{ }-15

ss = NDSolve[\char123eqn1, eqn3, eqn4, eqn5, eqn6, eqn7, eqn8, eqn9\char125,

\ \ \ \ \char123a, b, q\char125, \char123x, 10\^{ }-3, 100\char125,
MaxSteps $\to$ 500000]

v1[x\_] := 1 / 2 * (1 - $\tt e^{2\, a[x]}$)

v2[x\_] := 1 / 2 * (1 - $\tt e^{-2\, b[x]}$)

v3[x\_] := 1 / 2 * (1 - $\tt e^{2\, a[x]}$ - 4.89078 * 10\^{ }-9
* 1 / x * Exp[-122\^{ }2 * q[x]]

\ \ \ \ + 3.7 * 10\^{ }-29 * x\^{ }2 / 3)

v4[x\_] := 1 / 2 * (1 - $\tt e^{-2\, b[x]}$ - 4.89078 * 10\^{ }-9
* 1 / x * Exp[-122\^{ }2 * q[x]]

\ \ \ \ + 3.7 * 10\^{ }-29 * x\^{ }2 / 3)

bb3[x\_] := D[v3[x], x]

bb4[x\_] := D[v4[x], x]

uu[x\_] := 5.21126 * 10\^{ }-15 * (Exp[-x] + 1 + x\^{ }2 + Sqrt[Pi] * Erf[x])
* 1 / x

rr1[x\_] := v3[x] / uu[x]

rr2[x\_] := v4[x] / uu[x]

ba[x\_] := uu\char19[x]

pp1[x] := bb3[x] / ba[x]

pp2[x] := bb4[x] / ba[x]

hr3 = Plot[Evaluate[v3[x] /.\ ss], \char123x, 10\^{ }-3, 100\char125]

hr6 = Plot[Evaluate[v3[x] /.\ ss], \char123x, 10\^{ }-3, 10\^{ }-2\char125]

hr7 = Plot[Evaluate[v4[x] /.\ ss], \char123x, 10\^{ }-3, 10\^{ }-2\char125]

hr8 = Plot[Evaluate[bb3[x] /.\ ss], \char123x, 10\^{ }-3, 10\^{ }-2\char125]

hr9 = Plot[Evaluate[v4[x] /.\ ss], \char123x, 10\^{ }-3, 100\char125]

hr10 = Plot[Evaluate[bb3[x] /.\ ss], \char123x, 10\^{ }-3, 100\char125]

hr11 = Plot[Evaluate[bb4[x] /.\ ss], \char123x, 10\^{ }-3, 100\char125]

hr12 = Plot[Evaluate[bb4[x] /.\ ss], \char123x, 10\^{ }-3, 10\^{ }-2\char125]

hr16 = Plot[Evaluate[rr1[x] /.\ ss], \char123x, 10\^{ }-3, 100\char125]

hr17 = Plot[Evaluate[rr2[x] /.\ ss], \char123x, 10\^{ }-3, 100\char125]

hr18 = Plot[Evaluate[rr1[x] /.\ ss], \char123x, 10\^{ }-3, 10\^{ }-2\char125]

hr19 = Plot[Evaluate[rr2[x] + 233600 /.\ ss], \char123x, 10\^{ }-3, 10\^{ }-2\char125]

hr22 = Plot[Evaluate[pp1[x] /.\ ss], \char123x, 10\^{ }-3, 10\^{ }-2\char125]

hr23 = Plot[Evaluate[pp1[x] /.\ ss], \char123x, 10\^{ }-3, 100\char125]

hr24 = Plot[Evaluate[pp2[x] /.\ ss], \char123x, 10\^{ }-3, 10\^{ }-2\char125]

hr25 = Plot[Evaluate[pp2[x] /.\ ss], \char123x, 10\^{ }-3, 100\char125]

}
\medskip

A programme written in Mathematica 7 to calculate and plot $\wt V_1(r)$,
$\wt V_2(r)$, $\wt b_1(r)$, $\wt b_2(r)$, $\wt \eta_1(r)$, $\wt \eta_2(r)$.
{\tt v3} means $\wt V_1$, {\tt v4} means $\wt V_2$, {\tt bb3} means $\wt b_1$,
{\tt bb4} means $\wt b_2$, {\tt rr1} means $\wt \eta_1$, {\tt rr2}---$\wt
\eta_2$, {\tt x} means $r$, {\tt a}---$A$, and {\tt b}---$B$.

\medskip

\vskip2pt
{\parindent 0pt
\tt eqn1 = D[D[b[x], x], x] == $\tt\dfrac{9 - 10 * e^{2\,b[x]}
+ e^{4\,b[x]}}{8 * x^2}$

\vskip4pt
eqn3 = D[a[x], x] == $\dfrac{\tt -1 + e^{2\,b[x]}}{\tt 4 * x}$

\vskip4pt
eqn4 = a[7.03] == -6.957012802275961 *\^{ }-10 * (0.5 + 0.00012069886778807667)

eqn5 = b[7.03] == 3.469214116529213 *\^{ }-13

eqn6 = b\char19[7.03] == -2.7206955894132885 *\^{ }-14

\vskip2pt
eqn7 = D[D[q[x], x], x] == - $\dfrac{\tt -1 + e^{2\,b[x]}}{\tt 4 * x\hat{\ }2}
\ \tt e^{2\,b[x]}$

\vskip2pt
eqn8 = q[7.03] == -9.568184581692126 *\^{ }-13

eqn9 = (D[q[x], x] /.\ x $\to$ 7.03) == -5.348841232880765 *\^{ }-15

ss = NDSolve[\char123eqn1, eqn3, eqn4, eqn5, eqn6\char125,
\char123a, b, q\char125,

\ \ \ \ \char123x, 10\^{ }-3, 100\char125,  MaxSteps $\to$ 500000]

v1[x\_] := 1 / 2 * (1 - $\tt e^{2\, a[x]}$)

v2[x\_] := 1 / 2 * (1 - $\tt e^{-2\, b[x]}$)

v11[x\_] := 1 / 2 * (1 - $\tt e^{2\, a[x]}$ - 4.89078 * 10\^{ }-9 * 1 / x)

v21[x\_] := 1 / 2 * (1 - $\tt e^{-2\, b[x]}$ - 4.89078 * 10\^{ }-9 * 1 / x)

bb1[x\_] := D[v11[x], x]

bb2[x\_] := D[v21[x], x]

uu[x\_] := 5.21126 * 10\^{ }-15 * (Exp[-x] + 1 + x\^{ }2 + Sqrt[Pi] *
Erf[x]) * 1 / x

rr1[x\_] := v11[x] / uu[x]

rr2[x\_] := v21[x] / uu[x]

ba[x\_] := uu\char19[x]

pp1[x] := bb1[x] / ba[x]

pp2[x] := bb2[x] / ba[x]

hr1 = Plot[Evaluate[v1[x] /.\ ss], \char123x, 10\^{ }-3, 100\char125]

hr2 = Plot[Evaluate[v2[x] /.\ ss], \char123x, 10\^{ }-3, 100\char125]

hr3 = Plot[Evaluate[v11[x] /.\ ss], \char123x, 10\^{ }-3, 100\char125]

hr4 = Plot[Evaluate[v1[x] - 4.10258 * 10\^{ }-9 /.\ ss], \char123x, 10\^{ }-3, 10\^{ }-2\char125]

hr5 = Plot[Evaluate[v2[x] - 1.5 * 10\^{ }-12 /.\ ss], \char123x, 10\^{ }-3, 10\^{ }-2\char125]

hr6 = Plot[Evaluate[v11[x] /.\ ss], \char123x, 10\^{ }-3, 10\^{ }-2\char125]

hr7 = Plot[Evaluate[v21[x] /.\ ss], \char123x, 10\^{ }-3, 10\^{ }-2\char125]

hr8 = Plot[Evaluate[bb1[x] /.\ ss], \char123x, 10\^{ }-3, 10\^{ }-2\char125]

hr9 = Plot[Evaluate[v21[x] /.\ ss], \char123x, 10\^{ }-3, 100\char125]

hr10 = Plot[Evaluate[bb1[x] /.\ ss], \char123x, 10\^{ }-3, 100\char125]

hr11 = Plot[Evaluate[bb2[x] /.\ ss], \char123x, 10\^{ }-3, 100\char125]

hr12 = Plot[Evaluate[bb2[x] /.\ ss], \char123x, 10\^{ }-3, 10\^{ }-2\char125]

hr14 = Plot[Evaluate[uu[x] /.\ ss], \char123x, 10\^{ }-3, 10\^{ }-2\char125]

hr15 = Plot[Evaluate[uu[x] /.\ ss], \char123x, 10\^{ }-3, 100\char125]

hr16 = Plot[Evaluate[rr1[x] /.\ ss], \char123x, 10\^{ }-3, 100\char125]

hr17 = Plot[Evaluate[rr2[x] /.\ ss], \char123x, 10\^{ }-3, 100\char125]

hr18 = Plot[Evaluate[rr1[x] /.\ ss], \char123x, 10\^{ }-3, 10\^{ }-2\char125]

hr19 = Plot[Evaluate[rr2[x] + 233600 /.\ ss], \char123x, 10\^{ }-3, 10\^{ }-2\char125]

hr20 = Plot[Evaluate[ba[x] /.\ ss], \char123x, 10\^{ }-3, 10\^{ }-2\char125]

hr21 = Plot[Evaluate[ba[x] - 5.21136 * 10\^{ }-15 /.\ ss], \char123x, 10\^{ }-3, 100\char125]

hr22 = Plot[Evaluate[pp1[x] /.\ ss], \char123x, 10\^{ }-3, 10\^{ }-2\char125]

hr23 = Plot[Evaluate[pp1[x] /.\ ss], \char123x, 10\^{ }-3, 100\char125]

hr24 = Plot[Evaluate[pp2[x] /.\ ss], \char123x, 10\^{ }-3, 10\^{ }-2\char125]

hr25 = Plot[Evaluate[pp2[x] /.\ ss], \char123x, 10\^{ }-3, 100\char125]

}
\medskip

A programme written in Mathematica 7 to calculate and plot $V_1(r)$,
$V_2(r)$, $\ov V_1(r)$, $\ov V_2(r)$, $b_1(r)$, $b_2(r)$, $\eta_1(r)$,
$\eta_2(r)$, $\ov U(r)$ and their \dv s. {\tt v1} means $V_1$, {\tt v2} means
$V_2$, {\tt v11} means $\ov V_1$, {\tt v21} means $\ov V_2$, {\tt bb1} means
$b_1$, {\tt bb2} means $b_2$, {\tt uu} means $-\ov U$, {\tt x} means $r$,
{\tt rr1}---$\eta_1$, {\tt rr2}---$\eta2$, etc.

\medskip

\vskip2pt
{\parindent 0pt
\tt eqn1 = D[D[b[x], x], x] == $\tt\dfrac{9 - 10 * e^{2\,b[x]}
+ e^{4\,b[x]}}{8 * x^2}$

\vskip4pt
eqn3 = D[a[x], x] == $\dfrac{\tt -1 + e^{2\,b[x]}}{\tt 4 * x}$

\vskip4pt
eqn4 = a[7.03] == -6.957012802275961 *\^{ }-10 * (0.5 + 0.00012069886778807667)

eqn5 = b[7.03] == 3.469214116529213 *\^{ }-13

eqn6 = b\char19[7.03] == -2.7206955894132885 *\^{ }-14

ss = NDSolve[\char123eqn1, eqn3, eqn4, eqn5, eqn6\char125,
\char123a, b\char125,

\ \ \ \ \char123x, 0.001, 1000000\char125,  MaxSteps $\to$ 500000]

v1[x\_] := 1 / 2 * (1 - $\tt e^{2\, a[x]}$)

v2[x\_] := 1 / 2 * (1 - $\tt e^{-2\, b[x]}$)

v11[x\_] := 1 / 2 * (1 - $\tt e^{2\, a[x]}$ - 4.89078 * 10\^{ }-9 * 1 / x)

v21[x\_] := 1 / 2 * (1 - $\tt e^{-2\, b[x]}$ - 4.89078 * 10\^{ }-9 * 1 / x)

bb1[x\_] := D[v11[x], x]

bb2[x\_] := D[v21[x], x]

aq1 = u\char19\char19[f] + u[f] == ((4.104 * a1 \^{ }-1 * (x1 \^{ }2 - 1) \^{ }-1

\ \ \  \ + (2.898 * 10\^{ }-8) * u[f] \^{ }2 - 21.24 * 10\^{ }7 * a1 \^{ }-1

\ \ \ \ * (x1 \^{ }2 - 1) \^{ }-1 * u[f] \^{ }-2 * (bb1[x] /.\ x $\to$ 1 /
u[f])) /.

\ \ \ \ \char123a1 $\to$ 20, x1 $\to$ 5\char125) /.\ ss

aq2 = u[0] == 4.104 * a1 \^{ }-1 * (x1 - 1) \^{ }-1 /.\ \char123a1 $\to$ 20,
x1 $\to$ 5\char125

aq3 = u\char19[0] == 0.

ss1 = NDSolve[\char123aq1, aq2, aq3\char125, u, \char123f, 0, (ArcCos[-1 / x1]
/.\ x1 $\to$ 5)\char125,

\ \ \ \ MaxSteps $\to$ 500000]

hr1 = Plot[Evaluate[u[f] \^{ }-1 /.\ ss1], \char123f, 0, (ArcCos[-1 / x1]
/.\ x1 $\to$ 5)\char125]

hr2 = PolarPlot[Evaluate[u[f] \^{ }-1 /.\ ss1], \char123f, 0, (ArcCos[-1 / x1]
/.\

\ \ \ \ x1 $\to$ 5)\char125]

hr3 = ParametricPlot3D[Evaluate[\char123u[f] \^{ }-1 * Cos[f],

\ \  \ \ u[f] \^{ }-1 * Sin[f] * Cos[3 / 360 * 2 * Pi],

\ \ \ \ u[f] \^{ }-1 * Sin[f] * Sin[3 / 360 * 2 * Pi]\char125 /.\ ss1],

\ \ \ \  \char123f, 0, (ArcCos[-1 / x1] /.\ x1 $\to$ 5)\char125, BoxRatios
$\to$ \char123 1, 1, 2\char125]

}\medskip
A programme written in Mathematica 7 to calculate and plot $r(\vf)$, a polar
plot $r(\vf)$ and a 3D polar plot for $a_h=20$, $e=5$, $I=3^\circ$. {\tt f}
means $\vf$, {\tt u} means $u$, {\tt x} means $r$, {\tt a1} means $a_h$,
{\tt x1} means $e$.

\medskip

All of these programmes use a very powerful instruction of Mathematica~7 {\tt
NDSolve} in several applications.

\medskip

{\parindent 0pt
\tt a\char19\char19[x] = $\tt \dfrac{1}{n\ (2\ +\ n)\ f[x]^6}$
1.3196353633333333 *\^{ }-24 $\X1($-4.523294969090818 *\^{ }12

\ \ \ \  n (2 + n) f[x]$\tt^5$ a\char19[x]
(f[x] + 1.0051766597979596 *\^{ }12 f\char19[x])

\ \ \ \ + a[x] $\X1($8 a[x] f[x]$\tt^5 \ \X1($-3 n (2 + n) s
+ 2 $\tt e^{n\,q[x]}$ (2 + n - $\tt e^{2\,q[x]}$ n) f[x]$\Y1)$

\ \ \ \ + 3 n (2 + n) $\X1($-2 f[x]$\tt^6$ - 4.0207066391918384 *\^{ }12
f[x]$\tt^5$ f\char19[x]

\ \ \ \ - 1.0103801174025828 *\^{ }24 f[x]$\tt^4$ f\char19[x]$\tt^2$

\ \ \ \ - 1.0103801174025828 *\^{ }24 m f[x]$\tt^6$ q\char19[x]$\tt^2\Y1)\Y1)\Y1)$

f\char19\char19[x] = $\tt \dfrac{1}{n\ (2\ +\ n)\ a[x]\ f[x]^3}$
1.3196353633333333 * \^{ }-24 $\X1($2 a[x]$\tt^2$ f[x]$\tt^3$ $\X1($3 n

\ \ \ \ (2 + n) s + 2
$\tt e^{n\,q[x]}$ (-2 + (-1 + $\tt e^{2\,q[x]}$) n) f[x]$\Y1)$
+ 7.577850880519371 *\^{ }23 n

\ \ \ \  (2 + n) f[x]$\tt^3$ a\char19[x]
f\char19[x] + 1.5077649896969395 *\^{ }12

\ \ \ \ n (2 + n) a[x] f\char19[x] $\X1($f[x]$\tt^3$ +
5.025883298989798 *\^{ }11 f[x]$\tt^2$ f\char19[x]$\Y1)\Y1)$

q\char19\char19[x] = $\tt\dfrac{1}{\hbox{m n a[x] f[x]$\tt^5$}}$
6.59817681666666 *\^{ }-25 $\X1($1.5155701761038742 *\^{ }24 m

\ \ \ \ n f[x]$\tt^5$ a\char19[x] q\char 19[x] + a[x]
 $\X1(\tt e^{n\,q[x]}$ (4 - 5 (-1 + $\tt e^{2\,q[x]}$) n)

\ \ \ \ a[x]
f[x]$\tt^5$ + 3.015529979393879 *\^{ }12 m n f[x]$\tt^5$ q\char19[x]$\Y1)\Y1)$

}\medskip
Equations \er{Db377} after substitution $r_0=1.9897\t10^{-12}$,
{\tt a} means $\ov\a$, {\tt f} means usually $f$, {\tt q} means $\vF$,
{\tt l} means $\ell$, {\tt m} means $\ov M$, {\tt s} means $C_1$,
{\tt x} means $r$.

\medskip

{\parindent 0pt
\tt a1\char19\char19[x] = $\tt \dfrac{1}{n\ (2\ +\ n)\ f[x]^6}$
1.3196353633333333 * \^{ }-24 $\X1($-4.523294969090818 *\^{ }12

\ \ \ \ n (2 + n) f[x]$\tt^5$ a1\char19[x]
(f[x] + 1.0051766597979596 *\^{ }12 f\char19[x]) + (1 + a1[x])

\ \ \ \ $\X1($8 (1 + a1[x]) f[x]$\tt^5 \ \X1($-3 n (2 + n) s
+ 2 $\tt e^{n\,q[x]}$ (2 + n - $\tt e^{2\,q[x]}$ n) f[x]$\Y1)$

\ \ \ \ + 3 n (2 + n) $\X1($-2 f[x]$\tt^6$ - 4.0207066391918384 *\^{ }12
f[x]$\tt^5$ f\char19[x]

\ \ \ \ - 1.0103801174025828 *\^{ }24 f[x]$\tt^4$ f\char19[x]$\tt^2$

\ \ \ \ - 1.0103801174025828 *\^{ }24 m f[x]$\tt^6$ q\char19[x]$\tt^2\Y1)\Y1)\Y1)$

f\char19\char19[x] = $\tt \dfrac{1}{n\ (2\ +\ n)\ (1\ +\ a1[x])\ f[x]^3}$
1.3196353633333333 * \^{ }-24

\ \ \ \ $\X1($2 (1 + a1[x])$\tt^2$ f[x]$\tt^3$ $\X1($3 n (2 + n) s + 2
$\tt e^{n\,q[x]}$ (-2 + (-1 + $\tt e^{2\,q[x]}$) n) f[x]$\Y1)$

\ \ \ \ + 7.577850880519371 *\^{ }23 n (2 + n) f[x]$\tt^3$ a1\char19[x]
f\char19[x]

\ \ \ \ + 1.5077649896969395 *\^{ }12 n (2 + n) (1 + a1[x]) f\char19[x]

\ \ \ \ $\X1($f[x]$\tt^3$  + 5.025883298989798 *\^{ }11 f[x]$\tt^2$ f\char19[x]$\Y1)\Y1)$

q\char19\char19[x] = $\tt\dfrac{1}{\hbox{m n (1 + a1[x]) f[x]$\tt^5$}}$
6.59817681666666 *\^{ }-25

\ \ \ \ $\X1($1.5155701761038742 *\^{ }24 m n f[x]$\tt^5$ a1\char19[x]
q\char 19[x]

\ \ \ \ + (1 + a1[x]) $\X1(\tt e^{n\,q[x]}$ (4 - 5 (-1 + $\tt e^{2\,q[x]}$) n) (1 + a1[x])
f[x]$\tt^5$

\ \ \ \ + 3.015529979393879 *\^{ }12 m n f[x]$\tt^5$ q\char19[x]$\Y1)\Y1)$

}\medskip
Equations \er{Dd379} after substitution $r_0=1.9897\t10^{-12}$,
{\tt a1} means $\ov\a$, {\tt f} means usually $f$, {\tt q} means $\vF$,
{\tt l} means $\ell$, {\tt m} means $\ov M$, {\tt s} means $C_1$,
{\tt x} means $r$.

In order to reduce the sizes of our figures in {\tt .eps} we converted
them to PDF files. This can be accomplished by performing the following
at a Linux/Mac command line:
\begin{verbatim}
for i in *eps; do ps2pdf -DEPSCrop $i; done
sed -i 's/\.eps}/}g' *tex
\end{verbatim}
In this way they can be submitted using PDF\LaTeX. Now they are efficient.

\section*{Conclusions, further prospects for research, and remarks}
Our aim is to explain PA-anomaly using \gr al \so\ in alternative theory of
\gr. We want to pass all the \SS\ tests for major bodies in the \SS.

In the paper we consider an \an\ \ac\ of \P0/11 deriving the effect from the
\NK{\JT\ }. We find no possible contradictions with observation data in the
\SS\ caused by the model. We consider a \hy\ orbit of a \sp\ disturbed by this
model giving \e\ for a motion. In Appendix~A we give exact \so\ for a problem
of motion. The \sp\ orbit is bounded. It means it will
never escape from the \SS. It came back after a period estimated about
$10^6$~years. (Moreover, very distant bodies in the \SS\ can influence
attraction from other systems.)
It is easy to give a simple explanation of this, using our model of an
anomalous \ac. In particular, an asymptotic value of a \hy\ velocity of a
\sp\ is $v_P=1.2\t10^4{\rm\frac ms}$. In a deep space region
${}>10^3\,\rm AU$ the only one \ac\ is \an\ one. It is a deceleration. Using
simple formulae for a decelerating movement we get $T\simeq\frac{2v_P}{\ov
b}$ and $L\simeq \frac{v_P^2}{2b}$ (a~range). In both cases we get
$$
T\simeq0.82\t 10^6\,{\rm yr}, \q L\simeq5.5\t 10^5\,\rm AU.
$$
The last value is close to a size of an Oort cloud. See also Appendix~A.

Moreover, a simplified model of motion can work very well up to
a distance of $10^6$\,AU. The interesting point is a connection to
cosmology described in Section~8, where we explained a relation
of~$b$ to a Hubble \ct\ and to a \co ical \ct. In this way $b$ depends on
time via a Hubble parameter. For a scale of a period of a \so\ is of order
$10^6$~years, $b$~does not change significantly during a motion.

Thus the \SS\ is stable. The \an\ \ac\ causes all orbits to be bounded. In
this way the Earth cannot be expelled from the \SS\ far away. There is a body
of work in classical mechanics of point particles interacting via a
newtonian attraction which gives such a possibility. In particular, there is
a \so\ of a 4-body problem \st three point masses are going far away on a cost
of an energy of the remaining one body (see Ref.\ \cite{a63}). One of them is falling on the
second. It is not so surprising if we remind that a newtonian \pt\ is
unbounded from below.
They are approaching infinity at a finite time. One of them $-\iy$
and two of them $+\iy$, \st a distance between the last two is going to zero.
The fourth one bounces back and forth infinitely often between the first
going to $-\iy$ and the closer one from the pair escaping to~$+\iy$. This is
of course a one-dimensional problem, i.e.\ on a straight line.
There is a large body of works on this problem (see Refs \cite{a91},
\cite{a92}), i.e.\ on noncollision singularities.
Moreover, we should remember that Newtonian mechanics is not valid for large
velocities. The velocity cannot be infinite. It means the \so\ should be
considered under Special Relativity Theory. In this case the velocity can be
large but smaller than the velocity of light.
This is a good news. However, there is a bad news. According to a
body of work on $n$-body problem in newtonian mechanics the following theorem
has been proved. Let us suppose we have $n$~point masses interacting
according to a newtonian attraction. The theorem says: a set of all initial
conditions of the problem (it means, a subset of $\R^{6n}$) \st at least two
of those point masses would have the same \cd s in~$\R^3$ (it means, at least
two of them scatter), during an evolution in time, is a set of a zero
Lebesgue measure and of the first Baire class (see Ref.~\cite{a64}, also
Ref.~\cite{a65}).
It means that any scattering for an $n$-body problem in newtonian
mechanics is very rare and really exceptional. Thus any impact of an asteroid
on a planet in the \SS\ is due to finite sizes of these bodies. Moreover, if
we change the newtonian \pt\ into ${}\sim -\frac1{r^2}$ \pt\ the same set
will have a non-zero Lebesgue measure. What does it mean in our case? It
simply means that we can expect also a non-zero Lebesgue measure with an
inclusion of an \an\ \ac. Thus a probability of an impact of a big asteroid
on the Earth can be quite big, not only because of finite sizes of both
bodies.

We work in our paper in the way we call a consecutive approximation method.
We start from simple nonrelativistic model. We find \so\ under some
approximation. Afterwards we find exact \so s with bounded orbits. Next we
extend the model to special Relativity finding also \so s for orbits.
We come next to curved \spt\ (Riemannian) finding \so s for orbits and
trying to pass GR tests in the \SS. Afterwards we consider NGT-like models
(\nos\ metric) in both cases for extremal and nonextremal geodesics (as paths
of massive bodies).

To be honest, it is necessary to mention different approaches to solve an
\an\ \ac\ of \P0/11. First of all it is a Moffat approach using his new
theory of gravity (see Ref.~\cite{52}). This approach has been critized in
Ref.~\cite{N30}. Moreover, it seems that the Moffat theory can resist. The
second approach interesting for us is from Ref.~\cite{54}. There is also an
approach which is far away from our investigations (see Ref.~\cite{N37}).
It is an \elm c-like \so.
Moreover, it is interesting (see Ref.~\cite{Pl}).

The method described by us works in the following way. The first step (first
\ap ion) is a nonrelativistic model. The second step (second \ap ion) is
a special-relativistic model. The third step (third \ap ion) is a Riemannian
model with a dust. The fourth (fourth \ap ion) is a model with non-Riemannian
geometry (\nos\ metric from NGT or Einstein Unified Field Theory) with
extremal geodesics. Both last models are considered with a dust. In the
last step (the last \ap ion) we use nonextremal geodesics (also with a~dust).
Let us notice the following fact. In the first and the second
steps we are using nonrelativistic or relativistic equations of motion for
\sp s supplied with an \an\ \ac. In the third, fourth and last steps we
solve field equations for a \gr al field with scalar and dust sources
(a~scalar field is coming from the \eu\nos\ Jordan--Thiry Theory). The \an\
\ac\ in the first and the second steps is coming also from scalar field in
the \eu\nos\ Jordan--Thiry Theory in order to fit Anderson et al.\ data.
Moreover, further steps use also Anderson et al.\ data in initial conditions
for field \e s. On the background of the \so\ of the field \e s we consider
geodetic \e s which are models of \e s of motion for \sp s and planets.
Solving these \e s under suitable initial conditions we get \e s for orbits.
This can be  achieved by some numerical programs written in Mathematica
and drawn as figures. It seems that our approach with an \ap ion steps method
can give a success in a \so\ of PA by means of pure gravity. Let us
underline the \ap ion steps method is an \ti{\ap ion method on the level of
theories} (of PA anomaly) \ti{not on the level of orbits} (of \sp s or
planets). If someone remembers about it any interpretation of mathematical
formulas in the text can be easily understood. The doubtful conclusion of
being impossible to leave the \SS\ is only on the step of the first \ap ion.
Moreover, the prediction of the first step is \ti{O.K.} up even $10^5$~years.
Even a perturbed hyperbolic orbit is a quite good prediction for a future
position of \P0 and 11. The general \nos\ theory (extremal or nonextremal
geodesics) can predict many additional features but in scales of times larger
than $10^7$~years (this time can be lowered if we construct relativistic
\sp s to go outside of the \SS).

Let us comment the following results. The higher order gravity cannot give a
re\so\ to \P{0/11} anomaly problem (see Ref.~\cite{57}). A~dark matter in the
\SS\ cannot help (see Ref.~\cite{58}) together with models of Kuiper Belt
(see Ref.~\cite{59}). Thus we conclude that our re\so\ together with
mentioned above is the only re\so\ of this problem. There are some prospects
of further research. It is natural to use the model of an \an\ \ac\ described
in the paper to investigate a problem of a flat galaxies rotation curves and
Tully--Fisher law, with some applications to clusters of galaxies and even to
a motion of satellites around galaxies. The globular clusters velocity
dispersions could be also treated in this approach.

The programme of the full formalism of the \eu\nos\ Jordan--Thiry theory for
an \an\ \ac\ problem will be developed in the future (see Appendix~A and~D)
involving numerical \so s of field \e s and an analysis of geodetic \e s.
The nine arbitrary \ct s. i.e.\ six from initial conditions and three
integration \ct s can be very useful in further investigations (see
Appendix~D).
We give in Appendix~D preliminary calculations of orbits for massive test
particles and photons for several values of parameters. We calculate also \gr
al Doppler effect for photons. We give also some examples for orbits in a
full formalism together with a numerical \so\ of field \e s.

The Pioneer anomaly is
still under consideration (see Ref.\ \cite{75} for this effect in Deep Space
Navigation (DSN)).

NASA (National Aeronautics and Space Administration) is planning a dedicated
mission to resolve the Pioneer anomaly. The viable concept uses a \sp\ pair
capable of flying in a flexible formation.  This is a concept of the Deep
Space Gravity Probe. This experiment in a deep space with an accuracy
resolution of $10^{-12}\frac{\rm m}{\rm s^2}$ can resolve an anomaly on an
experimental level.

In this work we consider a bottom-up approach, i.e.\ we are starting from
the nonrelativistic model going to the full formalism of the \NK{\JT\ }. The
top-down approach has been shortly described in the end of Appendix~D.

\eu\rl\ model of an \an\ \ac\ considered in Section~11 does not contradict
our conclusion on a stability of the \SS\ obtained in a simple model from
Section~3. Any stability considerations are along non\rl\ motion of massive
point body, i.e.\ $H\simeq c^2$. In this case all orbits (also \hy-like) are
bounded and the \SS\ is stable. A size of a bounded orbit in \P0/11 case is
about $10^5$\,AU and a time to come back $10^6$\,yr, as in a simple model
from Section~3. The conclusion on a probability of any
scattering of two point massive body seems also to be valid in a \rl\ model
of an \an\ \ac. Moreover, our pessimistic conclusion on an impossibility to
escape from the \SS\ is not valid in the case of \rl\ or almost \rl\ \sp. In
this case a total energy per a unit mass is comparable to its rest energy and
$H=c^2+\frac{k^2}{2a_h}$ cannot be close to $c^2$, even $\frac{k^2}{2a_h}\gg
c^2$. Such a \rl\ \sp\ can escape from the \SS. The orbits in this case can
be unbounded. In the case of light-paths we have also unbounded orbits. The
\rl\ model of an \an\ \ac\ passes three tests of GR (i.e.\ an \an\ \ph\
advance, a bending of light and a Shapiro effect) in the \SS. Moreover, we
need a fine tuning of initial conditions to get a full success. We can also
consider the fourth test, i.e.\ a Doppler effect in a \gr al field. In this
case a prediction of the model does not differ from a GR prediction.

Finally, we give some remarks. There are some misunderstandings connecting
Kaluza--Klein Theory, Einstein's Unified Field Theory, \eu\nos\ \eu\gr\
Theory (NGT), \eu\nos\ Kaluza--Klein Theory (NKKT), \eu\nos\ Jordan--Thiry
Theory (NJTT).

{\bf1.} First of all we comment a \ct\ $\la=\frac{2\sqrt{G_N}}{c^2}$. The \ct\ $\la$
appeared as a free parameter in this theory. Moreover in order to get
Einstein \e s with \elm c sources known from GR it is fixed and it is not free
any more. Why is there not a Planck's length?
I explain it shortly. The Kaluza theory is classical for a paper published by him
is classical as a classical paper in the scientific literature. It is also classical
for this theory is not quantum. For this we cannot get here a Planck's \ct.
This is simply for we need a Planck's \ct\ in order to construct the Planck's
length. \Pl's \ct\ is absent in Kaluza theory for this theory is classical
(non-quantum). The \Pl's length appeared in the further development done by
O.~Klein. O.~Klein considered a Klein--Gordon \e\ in 5-dimensional extension.
The \Pl's \ct\ is present in Klein--Gordon \e.
This \e\ can be considered as an \e\ for a classical scalar field. In
Kaluza--Klein theory \Pl's length appears as a scale of length.

\def\tr{transformation}
{\bf2.}
The classical Kaluza theory has been abandoned by 1950's. Moreover,
due to some mathematical investigations a deep structure has been discovered
behind the theory. Let me describe it shortly. First of all it happens that
behind Maxwell theory of \elm sm there is a principal fibre bundle over a
\spt\ with a structural group $\U(1)$ and a \cn\ defined on this bundle is an
\elm c field. Gauge \tr, four-\pt, the first pair of Maxwell \e\ obtained a
clear geometrical meaning in terms of a fibre bundle approach.

It happens also that a classical Kaluza theory is a theory of metrized (in a
natural way) \elm c fibre bundle (see Ref.~\cite{97}).

This is a true unification of the two fundamental principles of invariance in
physics: a gauge invariance principle and a \cd\  invariance principle, as we
mention in Section~1.

In Section 2 of Ref.~\cite{98} a classical KKT in this setting has been described
(see also the last two lines of page 576 with a fixing of the \ct~$\la$).

Moreover this paper is devoted to the KKT with torsion in such a way that we
put in the place of GR the Einstein--Cartan theory obtaining new features the
so-called ``interference effects'' between gravity and \elm sm going to some
effects which are small, moreover in principle measurable in experiment.

{\bf3.}
Let us consider Einstein Unified Field Theory. A.~Einstein started this
theory in 1920's. In 1950 he came back to this theory describing it in
Appendix~II of the fifth edition of his famous book \ti{The Meaning of
Relativity} (see Ref.~\cite{99}).

It is worth to mention that there are many versions of this theory. The
oldest Einstein--Thomas theory and after that Einstein--Strauss theory,
Einstein--Kaufmann theory. There are also two approaches, weak and strong
field \e s. The Einstein Unified Field Theory can be also considered as a
real theory and Hermitian theory. A slight deviation is the so-called
Bonnor's Unified Field Theory. In all of these approaches there are two
fundamental notions: \nos\ affine \cn\
$\gd\G,\la,\mu\nu,\ne\gd\G,\la,\nu\mu,$ and the \nos\ metric $g_{\mu\nu}
\ne g_{\nu\mu}$. \eu\cn\ and metric can be real or Hermitian. In this theory
there is also a second \cn\ $\gd W,\la,\mu\nu,\ne\gd W,\la,\nu\mu,$. \eu\cn\
$\gd\G,\la,\mu\nu,$ is a so-called constrained \cn, $\gd W,\la,\mu\nu,$ is called
unconstrained. All of these approaches have no free parameters. Some
parameters which appear in \so s of field \e s are \ti{integration \ct s}.

What was an aim to construct such theories? The aim was to find a unified
theory of gravity and \elm sm in such a way that GR and Maxwell theory appear
as some limit of the theory. This approach ended with fiasco. It was
impossible to obtain a Lorentz force.
It was impossible to obtain a Coulomb law too.

One can find all references to all versions of Einstein Unified Field Theory
in Refs \cite{4}, \cite{5}, \cite{7}, \cite{10} and we will not quote them here. Moreover, it
is worth to mention that A.~Einstein considered this theory as a theory of an
extended \gr. Moreover, there is a reference of A.~Einstein's idea to treat
this theory as a theory of an extended gravity only. A.~Einstein published a
paper on it in \ti{Scientific American} (the only one Einstein's paper in
this journal).

Geometrical--mathematical properties of Einstein Unified Field Theory have
been described in a book by Vaclav Hlavat\'y (see Ref.~\cite6).

In those times A.~Einstein started a program of geometrization of physics. Some
notions of this program have been described in Ref.~\cite{100}.

There is also an approach to this theory going in a different direction. It
has been summarized in the book by A.~H.~Klotz (see Ref.~\cite{101}).

{\bf 4.}
Let us comment NGT (\eu\nos\ \eu\gr al Theory) by J.~W.~Moffat. J.~W.~Moffat
reinterpreted Einstein Unified Field Theory as a theory of a pure \gr al
field (see Ref.~\cite9). He introduced material sources to the formalism.
He and his co-workers developed this idea getting many interesting
results which are in principle testable by astronomical observations in the
\SS\ and beyond. He was using both real and Hermitian theory. Simultaneously
he developed a formalism with two \cn s $\gd\G,\la,\mu\nu,$ and $\gd
W,\la,\mu\nu,$. I~refer to some of these papers.

{\bf5.}
Let us comment the \NK{and the \eu\nos\ Jordan--Thiry }. I~posed and
developed these theories using the \nos\ metrization of an \elm c fibre
bundle using differential forms formalism as in my paper (see Ref.~\cite{98}).

Early results concerning the \NK{} have been published (see Refs \cite{102},
\cite{103}, \cite{D79}).

The final result of the theory with some developments has been published in
Ref.~\cite5. The paper contains also an extension to the \eu\nos\ Jordan--Thiry
Theory with a scalar field $\Ps$ (or~$\rho$). In order to get a pure \NK{} it
is enough to put ${\Ps=0}$ (or~${\rho=1}$). All new features as some
``interference effects'' between \elm c fields and \gr\ have been quoted in
Introduction. The theory has no free parameters except integration \ct s in
\so s.

It is possible to get an extension of the theory to the non-Abelian case. In
this case we have one free parameter. Moreover, this parameter can be fixed
by a \co ical \ct. The final version of this theory can be found in
Ref.~\cite7.  In Ref.~\cite4 one can find also an extension to the case with
Higgs' field and spontaneous symmetry breaking. In the last case there are
three free parameters which can be fixed by a \co ical \ct\ and scales of
masses.

I do not refer in my paper to the paper Ref.~\cite{104}, for the authors are using completely different approach (it is
better to say \ti{three approaches}). This approach is far away from
investigations in my work. Moreover, in future both approaches can meet and we
will shake hands. The only one point which is now common is a starting point,
a classical Kaluza Theory. We do not refer to Ref.~\cite{105}.

This paper deals with some problems
in NGT. However, NGT considered by them has only a little touch with NGT
considered here. They introduced a mass for skew-\s\ field
$B_{\mu\nu}=-B_{\nu\mu}$ (in our notation it is $g_\[\mu\nu]$). Moreover, the
$g_\[\mu\nu]$ can obtain a mass in a linear \ap ion of \eu\nos\ Non-Abelian
Kaluza--Klein Theory due to a \co ical \ct\ and it is not necessary to
introduce a mass term. It seems that this is a completely different
approach (see Ref.~\cite{105}). For a cure of NGT by a \co ical \ct\ see also
Ref.~\cite{106}.

Let us notice the following fact. Einstein's Unified Field Theory has been
abandoned for it has been proved using EIH (Einstein--Infeld--Hoffman) method
that there is not a Lorentz force term and Coulomb like law.

These are
disadvantages of Einstein Unified Field Theory but not NGT. This works now
for our advantage, for we do not see any term like Lorentz force and
Coulomb-like law in \gr al physics (I~do not mean a Newton \gr al law which
can be obtained in Einstein Unified Field Theory).
Someone said: ``{\it it is clever to use advantages, moreover, more clever is
to use disadvantages''} and this is a case. Moreover, in the \NK{} we get
Lorentz force term from $(N+4)$-dimensional (5-dimensional in an \elm c case)
geodetic \e s (see Refs~\cite4, \cite5, \cite7, \cite{10}).

All additional notions in the \NK{\JT\ } have been described in Section~2. We
get from $(N+4)$-\di al theory ($N=n+n_1$) four-\di al \e s due to an
invariance of a \nos\ metric and a \cn\ \wrt the right invariance action of
the group (in the \elm c case this is a biinvariance of the group $\U(1)$, see
also Introduction).

Let us notice also the following fact. \eu\e s obtained in the \NK{\JT\ } are
different from these in pure NGT. Due to this we can obtain nonsingular \so s
of field \e s in the \elm c case. These \so s possess a nonsingular metric
$g_\(\a\b)$ and nonsingular electric field. The asymptotic behaviour is as in
the case of Reissner--Nordstrem \so\ (see Refs \cite4, \cite5). This is
impossible to get in pure NGT.

Let us notice the following fact. Our model of an \an\ \ac\ from Section~3 is
non-newtonian and nonrelativistic, for this it has no influence on a photon
path. Moreover, we can apply an \an\ \ac\ for a photon as a particle. We
consider an impact of an \an\ \ac\ on bending of light in nonrelativistic
(newtonian way, Newton \e s of motion). In this way we can expect some
effects which are negligible with comparison to newtonian gravity treatment.
This is the same in the full formalism of \eu\nos\ Jordan--Thiry Theory. The
prediction of the Shapiro effect cannot be sufficiently changed for an
additional \pt\ on the Earth and Venus orbit is close to zero (see Sections 3
and~4). Thus a programme to describe an \an\ \ac\ in the light of the
\eu\nos\ Jordan--Thiry Theory seems to be very promising. Moreover, it causes
serious numerical problems to solve \e s.

Let us give the following remark. In our approach we use for a convenience a
system of units where a time unit $T=\xi\cdot 32\t10^6$\,yr ($\xi$ is a
factor of order~1). We notice that $T$ is close to a value obtained in
Ref.~\cite{11n}. Moreover, there is an evidence in geological sciences for
such a value (see Refs \cite{107}, \cite{108}, \cite{109}). The authors claim that they
found a periodicity of extinctions of species in the geologic past of
$26\t10^6$\,yr. They found also different cycles in fossil diversity of order
$62\t10^6$\,yr. We do not claim that our $T$ has something to do with the
mentioned problems. There is not any physical mechanism to connect our~$T$
with the mentioned ones. This can be absolutely accidental. Moreover, for
someone who believes that ``everything depends on everything'' this
coincidence can be remarkable.

Recently Turyshev et al.\ (see Ref.~\cite{A25}) have analyzed old Doppler data
for \P0 and~11 not included in Ref.~\cite2. They concluded that a Pioneer
anomaly \ac\ depended on time and was temporarily decaying. They constructed
some models of this decaying. In Ref.~\cite{A26} Turyshev et al.\ have found
an explanation of the \an\ \ac\ in terms of ordinary physics, i.e.\ a recoil
force associated with an anisotropic emission of thermal radiation off the
\sp s. J.~D.~Anderson et al. (see Ref.~\cite{A27}) reanalyzed those new data
and found an inverse square component (\wrt the distance from the Sun to \P0
or \P1) of the Pioneer \an\ \ac\ (see Fig.~1 of Ref.~\cite{A27}). They claim
that this component can be explained as an unmodelled reaction force Solar
radiation pressure. According to Section~II of Ref.~\cite{A27} if we subtract
this term the resulting \ac\ is consistent with Refs \cite{1,2,3} which we
use in our model. Simultaneously Anderson et al.\ claim that their results
from Refs \cite{1,2,3} can be successfully applied to the Cassini \sp\ (see
Ref.~\cite{A28}). However the inverse square component mentioned above could
not be explained by a chameleon field effect. Moreover, recently F.~Francisco
et al.\ (see Ref.~\cite{A29}) and B.~Rivers et al.\ (see Ref.~\cite{A30}) claim
to solve the problem of the Pioneer \an\ \ac\ using a thermal radiation
recoil. The idea is very simple. Every \elm c radiation emitted from sources
carries energy and momentum. Thus it is enough to find such sources in the
\sp\ and evaluate the force. Afterwards to compare it with an \an\ \ac.
However the problem is very complex. First of all it is necessary to divide
the \sp\ into some small elements (it means, to use thermal finite element
method). Secondly, to model ways of rays. It means, to use a raytracing
algorithm. This procedure demands a knowledge of some details of an
interaction of a radiation with several surfaces in the \sp. The last point
is very crucial for we need to estimate a momentum transfer. They are using
some phenomenological approaches known in computer graphics which have no
physical justification (see Refs \cite{N20}--\cite{N24}). 

They used results of Anderson et al.\ (see Refs \cite{1,2,3}).
However both claims from Ref.~\cite{A26} and from Refs \cite{A29,A30} cannot be
simultaneously correct for it is impossible to fit different data using the
same physical model. Moreover, they are using in their model a Phong model
(see Ref.~\cite{A31}) in order to include highlights, shading, specular and
diffuse reflections. However the Phong model which is very popular in
computer graphics has not any physical justification. It is physically
inaccurate. Even true photorealism requires more sophisticated models of
surface properties. Thus conclusions based on this model concerning some
mechanical influence of thermal radiation on a surface seem to be premature.
Even more complicated models based on more sound physical foundations
cannot work for our
poor knowledge on properties of Pioneer's external and internal surfaces (see
Ref.~\cite{A32}). How this could be complicated even in everyday applications
is shown in Ref.~\cite{A33}. For this they cannot obtain from their analysis
(which is correct except the last point) reliable results on an \ac\ caused
by thermal effects.

In this way our model presented here of the Pioneer \an\ \ac\ can be still
relevant.

J. D. Anderson and J. R. Morris analyzed massless Brans--Dicke theory in
order to explain \an\ Pioneer \ac\ (see Ref.~\cite{136}). They used exact
spherically symmetric, stationary \so\ of coupled Einstein--massless--scalar
field \e s in four \di s found by B.~C.~Xanthopoulos and T.~Zannias (see
Ref.~\cite{137}). Using this \so\ they analyzed a test mass motion in
Brans--Dicke theory. They found a deviation from Newtonian \ac\ (in non\rl\
limit). Using \SS\ constrain they concluded that an \an\ \ac\ obtained in
this way can explain only a small fraction of an observed \an\ \ac. In this
way our programme of investigations described in this work seems to be very
promising for future. According to our programme it is necessary to use a
full formalism of the \eu\nos\ \JT\ Theory described in Appendix~D of this
work.

The further prospects for research is to use full field \e s derived in the
paper and also \e s of motion for major bodies in the \SS\ to tune initial
conditions (integration \ct s) in such a way that we pass all tests of GR
in the \SS. We
should take under consideration a problem of interplanetary matter together
with dark matter. Such calculations seem to be tedious because we should keep
a high accuracy of calculation on very long distances starting from the
radius of the Mercury orbit or even from the radius of the Sun
up to an Oort cloud distance from the Sun.

The full simulation consists in solving the system of differential \e s
\er{Db377} with initial conditions \er{Db378} or \er{Db379}. Moreover, we can
use equivalently Eqs \er{De379}--\er{De381} and also replacing \er{De379} to
\er{Dd379} and \er{De380}--\er{De381} to \er{Dd379} with initial conditions
\er{Dd380} or \er{Dd381} in such a way that
$$
0 < \widehat{\ov\rho}(r) < \rho_{\rm interplanetary}+\rho_{\rm dark\ matter}
$$
($\wh{\ov\rho}$ from \er{D.160}) and simultaneously all relevant GR tests
in the \SS\ are satisfied. The last means: Shapiro effect, perihelion
movements of Mercury, Icarus, Venus and bending of light. Thus we should
solve geodetic \e s in obtained \gr al background for massive bodies and for
photons (i.e.\ for test particles). This can be achieved by a consecutive
\ap ion methods solving simultaneously two systems of \e s, i.e.\ field \e s
and \e s of motion for test particles. Relevant formulae have been derived
in Appendix~D for any mentioned tests of~GR as \f s of orbits obtained from
geodetic \e s. As geodetic \e s we should take both extremal and nonextremal
to have a comparison. This problem seems to be a very hard task for we
should get \gr al field described in field \e s from a radius of the Sun up
to $10^3$--$10^5$\,AU. This is also very important for an estimation
of~$\wh{\ov\rho}$, $R_\odot \le r\le 10^3-10^5$\,AU. Thus we should start with
some values of integration \ct s in field \e s to simulate required
conditions. Hopefully, after that we can try to find realistic trajectories
for Pioneer 10/11 \sp\ in the \SS\ and beyond.

The second prospect for further research is to apply a first simple model
of an \an\ \ac\ to the systems of satellites of big planets in the \SS\ (we
mention about it in the Conclusions above). The promising project is to apply
it of the satellite system of Uranus (an anomaly reported in Ref.~\cite{B2}).
Moreover, some applications for Jupiter and Saturn systems are also very
interesting. This simple nonrelativistic model can be also applied for
a problem of flat velocity curve of galaxies and can be developed as an
alternative model of dark matter in our Galaxy and other galaxies similar to
MOND approach. We should remember that in our model two \ct s are not
universal physical \ct s but only integration \ct s. Thus the last two
projects (satellite systems and ``dark matter'' in galaxies) are not rigidly
fixed by \ct s as in the MOND approach.

\subsection*{Acknowledgements}
I would like to thank Professor B.~Lesyng for the opportunity to carry out
computations using\break Mathematica\TM~7\footnote{Mathematica\TM\ is the
registered mark of Wolfram Co.} in the Centre of Excellence
BioExploratorium, Faculty of Physics, University of Warsaw. I would like to thank Professor
A.~Krasi\'nski for interesting discussions. I would like to thank Professor
F.~Wilczek for critical remarks. I would like to thank an anonymous referee
for critical remarks to improve my paper.

\newpage

\null\vfill
\centerline{\ing{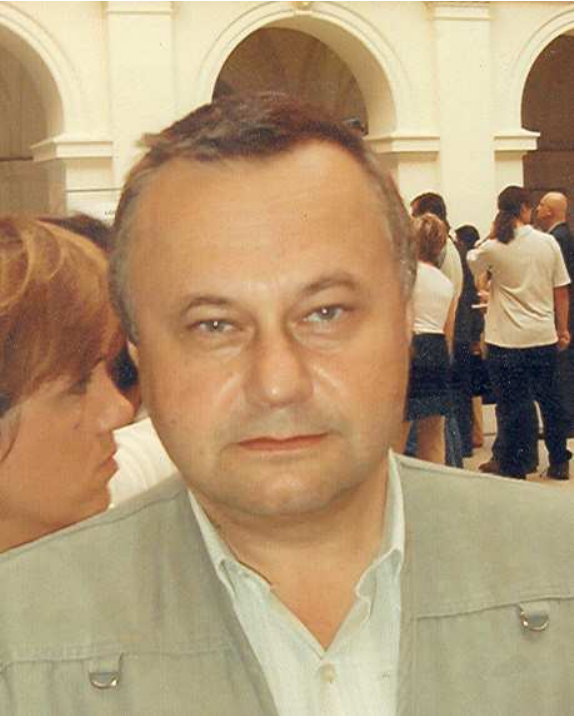}}
\vfill\eject

\end{document}